%% file: Thesis.tex
\documentclass[a4paper, 11pt, oneside]{Thesis}  % Use the "Thesis" style, based on the ECS Thesis style by Steve Gunn
\graphicspath{{Figures/}}  % Location of the graphics files (set up for graphics to be in PDF format)

\hypersetup{linkcolor={blue},citecolor={blue}}



\makeatletter
\newcommand\oneside{\@twosidefalse\@mparswitchfalse}
\newcommand\twoside{\@twosidetrue\@mparswitchfalse}
\makeatother

\renewcommand{\eref}[1]{Eq.~(\ref{#1})}
\newcommand{\erefs}[1]{Eqs.~(\ref{#1})}
\renewcommand{\fref}[1]{Fig.~\ref{#1}}
\newcommand{\pref}[1]{Part~\ref{#1}}
\renewcommand{\cref}[1]{Chapter~\ref{#1}}
\newcommand{\Sref}[1]{Section~\ref{#1}}
\newcommand{\ie}{i.e.}
\newcommand{\qv}[1]{\boldsymbol{#1}}
\renewcommand{\v}[1]{\boldsymbol{#1}}
\renewcommand{\i}{\mathrm{i}}
\newcommand{\m}[1]{\boldsymbol{#1}}
\newcommand{\bfield}{\mathcal{B}}
\newcommand{\efield}{\mathcal{E}}
\newcommand{\bra}[1]{\langle#1\rvert}
\newcommand{\ket}[1]{\lvert#1\rangle}
\newcommand{\expt}[1]{\langle#1\rangle}
\DeclareMathOperator{\Tr}{Tr}
\DeclareMathOperator{\real}{Re}
\DeclareMathOperator{\imag}{Im}
\DeclareMathOperator{\sinc}{sinc}
\DeclareMathOperator{\sgn}{sgn}
\DeclareMathOperator{\grad}{grad}
\DeclareMathOperator{\Lapprox}{\approx}

\newcommand{\rmd}{\mathrm{d}}
\newcommand{\refl}{\mathfrak{r}}
\newcommand{\trans}{\mathfrak{t}}

\newcounter{oldsubsection}
\newcounter{oldsection}
\let\oldsection\thesection
\let\oldsubsection\thesubsection

\newcommand{\subappendicesstart}{%
\setcounter{oldsubsection}{\value{subsection}}%
\setcounter{subsection}{0}%
\renewcommand{\thesubsection}{\thesection.\Alph{subsection}}%
}

\newcommand{\subappendicesend}{%
\setcounter{subsection}{\value{oldsubsection}}%
\renewcommand{\thesubsection}{\oldsubsection}%
}

\newcommand{\appendicesstart}{%
\setcounter{oldsection}{\value{section}}%
\setcounter{section}{0}%
\renewcommand{\thesection}{\thechapter.\Alph{section}}%
}

\newcommand{\appendicesend}{%
\setcounter{section}{\value{oldsection}}%
\renewcommand{\thesection}{\oldsection}%
}

\makeatletter
\newlength \figwidth
\makeatother

\let\oldepigraph\epigraph
\renewcommand{\epigraph}[2]{\setstretch{1.2}\oldepigraph{#1}{#2}\setstretch{1.5}}

\begin{document}

\frontmatter	  % Begin Roman style (i, ii, iii, iv...) page numbering

% Set up the Title Page
\def\fulltitle{Optical Cooling Using the Dipole Force}
\title  \fulltitle
\authors  {\texorpdfstring
            {\href{mailto:andre.xuereb@gmail.com}{Andr\'e Xuereb}}
            {Andr\'e Xuereb}
            }
\addresses  {\groupname\\\deptname\\\univname}  % Do not change this here, instead these must be set in the "Thesis.cls" file, please look through it instead
\date       {\today}
\subject    {}

\maketitle
%% ----------------------------------------------------------------

\setstretch{1.3}  % It is better to have smaller font and larger line spacing than the other way round

% Define the page headers using the FancyHdr package and set up for one-sided printing
\fancyhead{}  % Clears all page headers and footers
\rhead{\thepage}  % Sets the right side header to show the page number
\lhead{}  % Clears the left side page header

\pagestyle{fancy}  % Finally, use the "fancy" page style to implement the FancyHdr headers

%% ----------------------------------------------------------------

% The Abstract Page
\addtotoc{Abstract}  % Add the "Abstract" page entry to the Contents
\abstract{
\addtocontents{toc}{\vspace{1em}}  % Add a gap in the Contents, for aesthetics

The term `laser cooling' is applied to the use of optical means to cool the motional energies of either atoms and molecules, or micromirrors. In the literature, these two strands are kept largely separate; both, however suffer from severe limitations. Laser cooling of atoms and molecules largely relies on the internal level structure of the species being cooled. As a result, only a small number of elements and a tiny number of molecules can be cooled this way. In the case of micromirrors, the problem lies in the engineering of micromirrors that need to satisfy a large number of constraints---these include a high mechanical $Q$-factor, high reflectivity and very good optical quality, weak coupling to the substrate, etc.---in order to enable efficient cooling.\par
\hspace{2em}During the course of this thesis, I will draw these two sides of laser cooling closer together by means of a single, generically applicable scattering theory that can be used to explain the interaction between light and matter at a very general level. I use this `transfer matrix' formalism to explore the use of the retarded dipole--dipole interaction as a means of both enhancing the efficiency of micromirror cooling systems and rendering the laser cooling of atoms and molecules less species selective. In particular, I identify the `external cavity cooling' mechanism, whereby the use of an optical memory in the form of a resonant element (such as a cavity), \emph{outside} which the object to be cooled sits, can potentially lead to the construction of fully integrated optomechanical systems and even two-dimensional arrays of translationally cold atoms, molecules or even micromirrors.
\par
\hspace{2em}The concept of an optical memory is a very general one, and I use it to link together mechanisms that would otherwise appear disparate, including the cavity cooling of atoms and cooling mechanisms based on the non-adiabatic following of atomic populations. A fully vectorial three-dimensional scattering theory including the effects of such a memory is also presented and used to explore several different experimentally-realisable cooling configurations.
}

\clearpage  % Abstract ended, start a new page
%% ----------------------------------------------------------------

\pagestyle{fancy}

%% ----------------------------------------------------------------
\lhead{}  % Set the left side page header to "Contents"
\tableofcontents  % Write out the Table of Contents

%% ----------------------------------------------------------------
\lhead{\emph{List of Figures}}  % Set the left side page header to "List of Figures"
\listoffigures  % Write out the List of Figures
%% ----------------------------------------------------------------

\pagestyle{empty}

%% ----------------------------------------------------------------
% Declaration Page required for the Thesis, your institution may give you a different text to place here
\Declaration{

\addtocontents{toc}{\vspace{1em}}  % Add a gap in the Contents, for aesthetics

I, \textbf{Andr\'e Xuereb}, declare that this thesis titled `\textbf{\fulltitle}' and the work presented in it are my own. I confirm that:

\begin{itemize} 
\item[\tiny{$\blacksquare$}] This work was done wholly or mainly while in candidature for a research degree at this University.
 
\item[\tiny{$\blacksquare$}] Where any part of this thesis has previously been submitted for a degree or any other qualification at this University or any other institution, this has been clearly stated.
 
\item[\tiny{$\blacksquare$}] Where I have consulted the published work of others, this is always clearly attributed.
 
\item[\tiny{$\blacksquare$}] Where I have quoted from the work of others, the source is always given. With the exception of such quotations, this thesis is entirely my own work.
 
\item[\tiny{$\blacksquare$}] I have acknowledged all main sources of help.
 
\item[\tiny{$\blacksquare$}] Where the thesis is based on work done by myself jointly with others, I have made clear exactly what was done by others and what I have contributed myself.
 
\item[\tiny{$\blacksquare$}] Parts of this work have been published---please see list in \aref{ch:Pub}.
\\
\end{itemize}

Signed:\\
\rule[1em]{25em}{0.5pt}  % This prints a line for the signature
 
Date:\\
\rule[1em]{25em}{0.5pt}  % This prints a line to write the date
}
\cleardoublepage  % Declaration ended, now start a new page

%% ----------------------------------------------------------------

\setstretch{1.3}  % Reset the line-spacing to 1.3 for body text (if it has changed)

% The Acknowledgements page, for thanking everyone
\acknowledgements{
\addtocontents{toc}{\vspace{1em}}  % Add a gap in the Contents, for aesthetics
\setcounter{footnote}{1}
\renewcommand{\thefootnote}{\fnsymbol{footnote}}
I started my doctoral research not knowing quite a bit. I will forever be indebted to all those who were there to patiently guide me along the way and, perhaps just as patiently, point me in the right direction in the 1,001 occasions where I interrupted their thoughts to ask some basic question.\\
The Quantum Control group at the University of Southampton has always been a bit of a mystery to me: moments of tremendous insight are generally punctuated by periods where we all seem to be eight years old. I appreciate being thrown into the deep, dark end of experimental atomic physics during the first few weeks of my stay at Southampton when James Bateman and Matthew Himsworth asked me to do some work in their laboratory. My subsequent prediliction for theoretical work should not be interpreted as being due to that experience! James also provided some much needed physical insight at times. Richard Murray, who I am fond of disagreeing with, often provided an extra point of view on whatever it was that we would be discussing at the time. Being an opinionated Mediterranean, I found this rather useful (in retrospect, of course). Hamid Ohadi, the experimentalist \emph{par excellence} and former ion trapper, and his laboratory provided a well-needed distraction whenever I got bored with my sums. I'd also like to thank Nathan Cooper, who is working with Hamid to explore experimentally some of our less crazy ideas and who took over my duty as official tea-maker on Fridays.\\
I must also mention Peter Domokos, whose insight triggered off a sequence of events that led to the results in this thesis that I am perhaps most proud of. Sandro Grech and the various advisers and tutors I had at the University of Malta, especially David Buhagiar and Charles Sammut, always encouraged me to pursue my fascination with mathematics and physics.
\par
Finally, I must thank my family for their support; every one of them\footnote{And there are quite a few: Zak, Rebekah, Luke, Konrad, Charmaine, Maria, and of course my parents.} was an inspiration, and I would have been a very different person were it not for them. Sharon, who always stuck by my side, listening to my grumbling, pushing me along, and generally being there for me, I cannot help but thank from the bottom of my heart.
\par
``Oh, and just one more thing...''---words probably cannot express how exasperated my supervisors, Tim Freegarde and Peter Horak, must have felt at times because of my hard-headedness. Without their help this work would have never gotten \emph{to} the starting blocks and, at the risk of repeating myself, I feel I have to tell them a deeply heartfelt ``thanks!'' Working with both of them proved to be a tremendous learning experience and I have nothing but words of praise for them, both in their role as advisers and in their role as scientists.
\renewcommand{\thefootnote}{\arabic{footnote}}
\setcounter{footnote}{0}
}
\clearpage  % End of the Acknowledgements
%% ----------------------------------------------------------------
% The "Funny Quote Page"
\pagestyle{empty}  % No headers or footers for the following pages

\null\vspace{5em}
% Now comes the "Funny Quote", written in italics
\newcommand{\rawquote}[1]{\raisebox{-0.85em}{\Huge #1}}
\newcommand{\lquote}[1]{\rawquote{#1}}
\newcommand{\rquote}[1]{\rawquote{#1}}
\textit{\llap{\lquote{``}\ }{It} is an important and popular fact that things are not always what they seem. For instance, on the planet Earth, man had always assumed that he was more intelligent than dolphins because he had achieved so much---the wheel, New York, wars and so on---while all the dolphins had ever done was muck about in the water having a good time. But conversely, the dolphins had always believed that they were far more intelligent than man---for precisely the same reasons.\rquote{''}}
\begin{flushright}
Douglas Adams, \emph{The Hitchhiker's Guide to the Galaxy}
\end{flushright}

\vfill
\clearpage  % Funny Quote page ended, start a new page
%% ----------------------------------------------------------------
% End of the pre-able, contents and lists of things
% Begin the Dedication page

\setstretch{2.0}

\pagestyle{empty}  % Page style needs to be empty for this page
\newcommand{\texthbar}{h\hspace{-0.15em}\llap{\raisebox{0.35em}{-}}\hspace{0.15em}}
\dedicatory{For Sharon, for my family, and for a hot cup of tea\\(u lil ommi, li g\texthbar{}allmitni ng\texthbar{}odd)}

\setstretch{1.5}  % Return the line spacing back to 1.5

\addtocontents{toc}{\vspace{2em}}  % Add a gap in the Contents, for aesthetics

%% ----------------------------------------------------------------
\mainmatter	  % Begin normal, numeric (1,2,3...) page numbering
\pagestyle{fancy}  % Return the page headers back to the "fancy" style

% Include the chapters of the thesis, as separate files
% Just uncomment the lines as you write the chapters
\markright{Introduction}
\lhead[\thepage]{\emph{Introduction}}  % Change the left side page header to "Introduction"
\rhead[\emph{Introduction}]{\thepage}  % Change the right side page header to "Introduction"
\input{./Chapters/Introduction} % Introduction
\renewcommand{\chaptermark}[1]{\markright{#1}}
\lhead[\thepage]{\fancyplain{}{\sl{\rightmark}}}
\addtocontents{toc}{\vspace{0.5em}}  % Add a gap in the Contents, for aesthetics
\input{./Chapters/Chapter1} % Atomic physics and cooling methods
\input{./Chapters/Chapter2} % Transfer matrix methods
\input{./Chapters/Chapter3} % Experimental work
\addtocontents{toc}{\vspace{2em}}  % Add a gap in the Contents, for aesthetics
\input{./Chapters/Conclusion} % Conclusions and outlook

%% ----------------------------------------------------------------
% Now begin the Appendices, including them as separate files

\appendix % Cue to tell LaTeX that the following 'chapters' are Appendices
\input{./Chapters/Posters}	% Appendix Title

\addtocontents{toc}{\vspace{2em}}  % Add a gap in the Contents, for aesthetics
\backmatter

%% ----------------------------------------------------------------
\label{Bibliography}
\markright{Bibliography}
\lhead[\thepage]{\emph{Bibliography}}  % Change the left side page header to "Bibliography"
\rhead[\emph{Bibliography}]{\thepage}  % Change the right side page header to "Bibliography"
\bibliographystyle{naturemag}  % Use the "unsrtnat" BibTeX style for formatting the Bibliography

\end{document}

%% file: Chapters/Introduction.tex
\chapter*{Introduction}\label{ch:Introduction}
\addtotoc{Introduction}

\section*{Motivation}
The fundamental quest to know more about the nature of the world around us and the universe of which it is both an incomprehensibly insignificant part, objectively speaking, and a rather important part, subjectively speaking, has driven mankind to the edge of sanity\footnote{I am somewhat fond, however unfairly, of mentioning string theory in this context. See Ref.~\cite{Shapiro2007} for an overview into the birth of string theory that suggests otherwise.} and sometimes beyond. Amidst all of this, cold atoms can be seen as the ideal prototypical system with which to explore nature. In contrast to the sledgehammer approach so typical of high-energy physics experiments, clouds of ultracold matter provide an almost blank canvas that one can use as a microscope to probe the behaviour of matter at the atomic, and even subatomic~\cite{Hudson2002}, scale. Optical cooling is perhaps the only direct way of producing these ultracold clouds.
\par
Therein lies the catch, however. Whereas the most sensitive probes of various physical laws are perhaps the more complicated molecules, most optical cooling methods in use today are applicable to only a small (albeit growing) fraction of the periodic table. Experiments are quickly advancing to the point where a select few ``normal''\footnote{Ultracold di-alkali molecules are reasonably common (see Refs.~\cite{Jones2006} and~\cite{Kohler2006} for reviews), but require ultracold atoms as building blocks.} molecules will soon be optically cooled on demand~\cite{Stuhl2008,Zeppenfeld2009} and the race is now on to find methods of cooling any given molecule to ultracold temperatures.\\
In the related field of optomechanics, and I will in due course examine how the relation with cold atoms runs deeper than the widespread use of the word ``cold'' in the literature, the aim of cooling a macroscopic vibrating object to its ground state has been achieved using cryogenic methods~\cite{OConnell2010}, but several experimental groups~\cite{Groblacher2009a,Schliesser2009,Kippenberg2007} are tirelessly working to achieve the same using optical cooling methods.
\par
New cooling paradigms thus have to be devised, perhaps borrowing a healthy dose of inspiration from those currently in use, to bridge the gaps between the cooling of simple atoms, the cooling of arbitrary molecules, and even the cooling of micro- and mesoscopic structures.

\section*{Brief overview of past work}
Optical cooling of the motion of particles was originally observed in the regime of small dielectric particles, with Ashkin hypothesising~\cite{Ashkin1970} that similar ideas might be applicable to the control of atomic and molecular motion. Wonderful progress was made, both in experiment~\cite{Chu1985,Lett1988} and in theory~\cite{Dalibard1989,Ungar1989} in extending these ideas to three dimensions and in exploiting the internal structure of simple atoms. Cloud temperatures of tens of $\upmu$K have long been considered routine in magneto--optical traps~\cite{Lu1996}. Cooling the motion of dielectric particles proved to be more challenging, especially because the initial explorations~\cite{Ashkin1970} were conducted with the particles in suspension whereas a more desirable configuration~\cite{Chang2009b} would be with the particle suspended in vacuum; cooling of a microscopic particle to millikelvin temperatures was indeed reported very recently~\cite{Li2011}.
\par
Sub-Doppler cooling mechanisms, of the type reported in Ref.~\cite{Dalibard1989}, shift the energy loss process from the decay of the excited state population to the non-adiabatic decay of the population of different hyperfine levels. In the search for more generally-applicable cooling methods, it was realised that one can similarly shift the energy loss process to the decay of a degree of freedom external to the atom. This led to the proposal for the cooling of atomic motion inside a driven cavity~\cite{Horak1997}---demonstrated recently in experiments by the Vuleti\'c~\cite{Leibrandt2009} and Rempe~\cite{Koch2010} groups---by extension of similar mechanisms for cooling atomic motion using the modified vacuum field present in a cavity~\cite{Lewenstein1993}. Following a separate line of research, Braginsky and co-workers~\cite{Braginsky1967} suggested using cavity fields to cool the motion of micro- or mesoscopic mirrors, initially in the context of gravitational wave detectors. Experimentally, this idea lay dormant until technology improved to the point where such effects could be unambiguously demonstrated~\cite{Cohadon1999,Arcizet2006}, for example by cooling the vibrational motion of a micromechanical oscillator from an occupation number of around $53\,000$ quanta ($2.25$\,K) to ca.~$32$ quanta ($1.3$\,mK)~\cite{Groblacher2009b}.
\par
These two streams can be seen as two limiting cases of a general `matter interacting with a cavity field' configuration, with the matter interacting weakly (atom) or strongly (microscopic mirror) with the cavity field. One of my aims in this thesis will be to combine these two fields through the use of a generic matrix-based theoretical approach. The virtues of this approach are that such models are solvable in the fully general case, and that no restriction is placed on the strength of the light--matter interaction.\\
On a more general level, these ideas can be united with Sisyphus--type cooling mechanisms that make use of internal atomic variables following external ones non-adiabatically. This link comes about because a cavity field also introduces a non-adiabatic `delay' element into the dynamics of the situation. This idea of a system with a delay, or memory, as a means to a generic optical cooling mechanism will also be one of the recurring themes throughout this thesis.

\section*{Outline of thesis}
This thesis will begin with a general, if somewhat brief, overview of atomic physics, which will allow me to discuss several of the ``standard'' cooling or trapping methods in current use. In \cref{ch:CoolingMethods:TrapCool} I will look at what can be viewed as a prototypical system for cavity cooling of atoms, which I call `mirror-mediated cooling', and explore it from the semiclassical perspective.\\
The discussion in \pref{part:TMM} will see me exploring two classical theories of light--matter interactions: one based on the transfer matrix method, in \cref{ch:TMM:TMM}, and one based on a delayed fully vectorial three-dimensional scattering theory, in \cref{ch:TMM:Scattering}. The transfer matrix theory is developed in \sref{sec:TMM:Model}, solved in the general case in \sref{sec:TMM:General}, and extended in \sref{sec:TMM:Multilevel} to describe multi-level atoms. It is then applied to the ‘mirror-mediated cooling’ paradigm and, in \cref{ch:TMMApplications}, to another system, `external cavity cooling', which promises to be very important in optomechanical experiments. As a final application of the matrix formalism, I will explore another configuration, in a unidirectional ring cavity with a gain medium, which allows one to dispense with certain (somewhat stringent) requirements in mirror-mediated cooling. The scattering theory developed in \cref{ch:TMM:Scattering} is also initially applied to the one-dimensional mirror-mediated cooling geometry, in order to verify the method, and is then used to describe mirror-mediated cooling in three dimensions and also the ``optical binding'' of a refractive particle to its own reflection.\\
In the final part, I will first discuss the experimental work that is currently underway in the laboratories at the University of Southampton to explore both the consequences of the above, and atom--surface interactions in a more general context. Finally, I will analyse several of the cooling configurations discussed throughout this thesis from an experimental point of view and give an estimate for the performance that can be expected from each configuration under realistic experimental conditions.

\subsection*{Logical connections between chapters}
\begin{figure}[h]
 \centering
    \includegraphics[width=\linewidth]{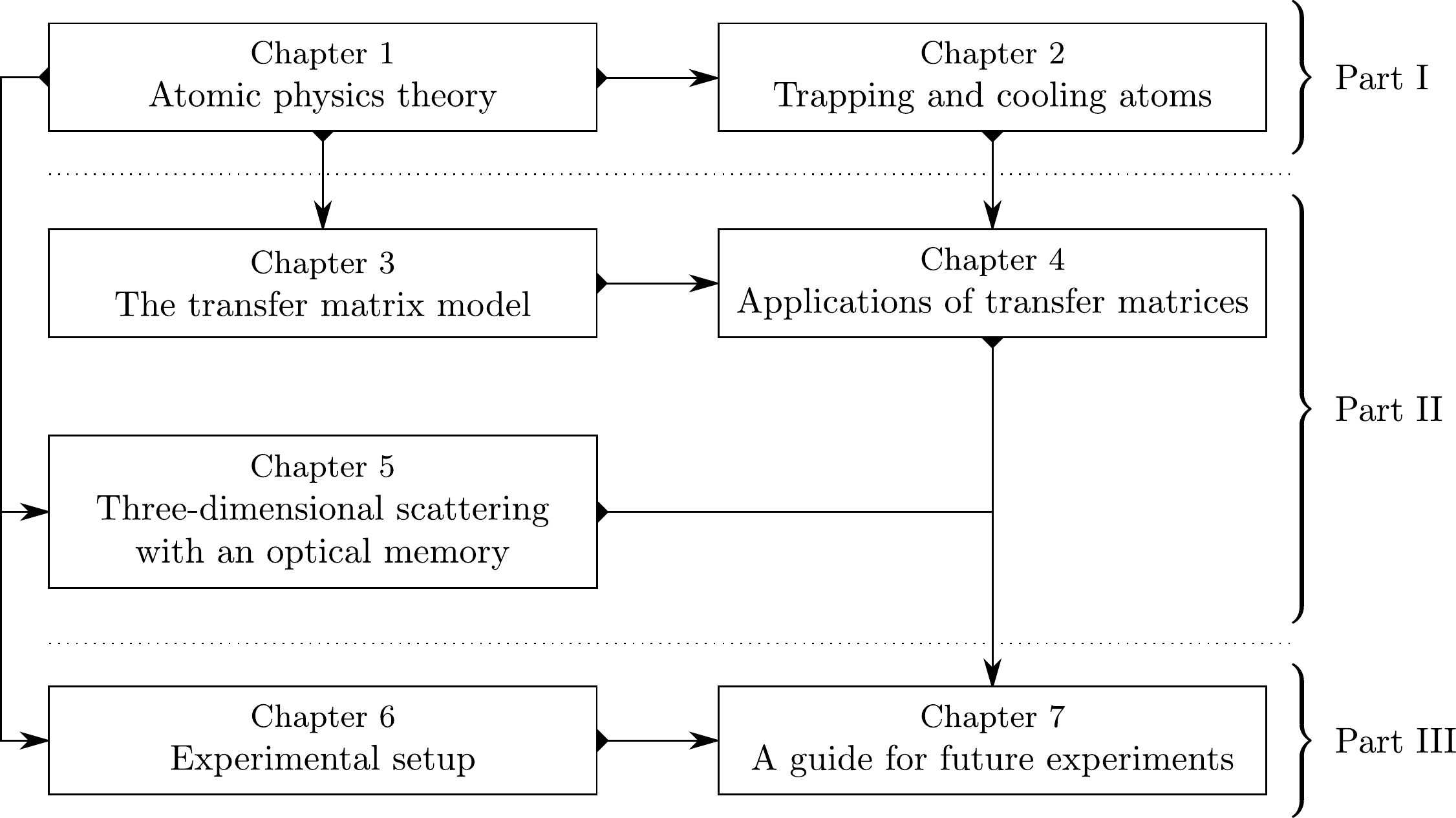}
\end{figure}

%% file: Chapters/Chapter1.tex
\newcommand{\diffn}{\textsl{\textsf{D}}}
\newcommand{\force}{\textsl{\textsf{F}}}
\newcommand{\heatingcoefft}{\varrho}
\newcommand{\xpr}{x^\prime}
\newcommand{\tpr}{t^\prime}
\newcommand{\efpr}{\efield^\prime}
\newcommand{\bfpr}{\bfield^\prime}
\newcommand{\mst}{\boldsymbol{T}}
\newcommand{\im}[1]{\imag\!\left\{#1\right\}}
\newcommand{\re}[1]{\real\!\left\{#1\right\}}
\newcommand{\comm}[2]{\Bigl[#1,#2^\dagger\Bigr]}
\def\totalderivative{\mathrm{D}}

\newpartalt{Atomic Physics Theory \& Cooling Methods}{Atomic Physics Theory\\\&\\Cooling Methods}\label{part:CoolingMethods}
\chapter{Atom--field interactions}\label{ch:CoolingMethods:AFInt}
\epigraph{[...] [T]he semiclassical theory, when extended to take into account both the effect of the field on the molecules and the effect of the molecules on the field, reproduces almost quantitatively the same laws of energy exchange and coherence properties as the quantised field theory, even in the limit of one or a few quanta in the field mode.}{E.\ T.\ Jaynes and F.\ W.\ Cummings, Proceedings of the IEEE \textbf{51}, 89 (1963)}
I begin this chapter with a very brief review of atomic structure, covering the fine and hyperfine structures as well as the magnetic sublevels within the hyperfine manifold; these ideas will be used later to discuss trapping and cooling of atoms. The following sections describe the density matrix approach and build up to a derivation of the Optical Bloch Equations. These equations are the tools necessary to examine the interaction between an atom and the electromagnetic field and, ultimately, to derive an expression for the forces acting on an atom, as parametrised by the polarisability of that atom. The chapter continues with a note on the fluctuation--dissipation theorem and shows how the calculation of the full force acting on the atom allows the prediction of the equilibrium temperature a population of such atoms will tend to, and concludes with a short discussion on multi-level atoms. The reader is referred to Refs.~\cite{CohenTannoudji1978a,CohenTannoudji1978b,Shore1990a,Shore1990b,Woodgate2000,CohenTannoudji2004,Foot2005}, and references therein, for a more in-depth and complete treatment of certain parts of this chapter.
\section{Atomic structure}\label{sec:CoolingMethods:AS}
It has been known for centuries that the spectrum of the light emitted by excited atomic gases is made up of a finite, and usually large, number of distinct spectral lines. These lines are not randomly distributed but can be divided into various closely-spaced groups. Each of these groups can be thought of as being due to the coarse structure of the energy levels of the atom; \ie, each group of lines in the coarse structure originates from a particular transition between two (electron) energy levels. Each energy level has an associated orbital angular momentum, $\qv{L}$.
The total angular momentum $\qv{J}$ of an electron includes a contribution from both $\qv{L}$ and the electron's intrinsic spin angular momentum $\qv{S}$:
\begin{equation}
 \qv{J}=\qv{L}+\qv{S}\,,
\end{equation}
whereby the corresponding quantum number $J$ is an integer or half-integer in the range $\lvert L-S\rvert\leq J\leq L+S$. This is called the $LS$ coupling scheme~\cite{Woodgate2000}. We adopt the convention that a quantised momentum vector $\qv{v}$ has magnitude $\sqrt{v(v+1)}\hbar$. The different possible values of $J$ give rise to the fine structure in the spectrum.
\par
Drilling down further, each line in the fine structure spectrum can be subdivided. This arises from the coupling, or interaction, between the total angular momentum of the electron and the angular momentum $\qv{I}$ of the nucleus. We can then define an atomic angular momentum $\qv{F}$ as~\cite{Woodgate2000}
\begin{equation}
 \qv{F}=\qv{J}+\qv{I}\,,
\end{equation}
where $F$ again takes any integer or half-integer value in the range $\lvert J-I\rvert\leq F\leq J+I$. Finally, we denote the quantisation axis of our system as $z$, whereby the projection of $\qv{F}$ along $z$ is denoted $F_z\equiv m_F\hbar$, with $-F\leq m_F\leq F$, $m_F$ again being an integer or half-integer. In general, $F$ can take several values for each of the fine structure levels; this gives rise to the hyperfine structure in the observed spectrum.\\
We will temporarily constrain ourselves to the transition of perhaps largest experimental interest in cold atoms, the $L=0\rightarrow L=1$ transition in alkali atoms (the so-called D line), where $S=1/2$. First of all, we note that the ground state ($L=0$) has only one fine structure level ($J=1/2$) whereas the excited state ($L=1$) has two: $J=1/2$ and $J=3/2$. The D line is therefore composed of two separate lines: the D$_1$ line ($^2$S$_{1/2}\rightarrow$ $^2$P$_{1/2}$) and the D$_2$ line ($^2$S$_{1/2}\rightarrow$ $^2$P$_{3/2}$). In the preceding sentence we used spectroscopic notation, whereby the levels with $L=0,1,\dots$ are denoted by $\text{S},\text{P},\dots$, and where the superscript is equal to $2S+1$. The Hamiltonian that describes the hyperfine structure for the D line transitions is, to lowest order~\cite{Arimondo1977},
\begin{equation}
 \hat{H}_\text{hfs}=A_\text{hfs}\qv{I}\cdot\qv{J}+B_\text{hfs}\frac{3(\qv{I}\cdot\qv{J})^2+\tfrac{3}{2}(\qv{I}\cdot\qv{J})-I(I+1)J(J+1)}{2I(2I-1)J(2J-1)}\,,
\end{equation}
with $A_\text{hfs}$ being the magnetic dipole constant and $B_\text{hfs}$ the electric quadrupole constant. These are experimentally determined and can be found in standard references, such as Ref.~\cite{Steck2008} for $^{85}$Rb. The two terms in the preceding equation are due to the magnetic dipole and electric quadrupole interaction between $\qv{I}$ and $\qv{J}$, respectively. Higher-order multipole components were ignored in the above expression. The hyperfine levels subsequently incur an energy shift
\begin{equation}
 \Delta E_\text{hfs}=\tfrac{1}{2}A_\text{hfs}K+B_\text{hfs}\frac{\tfrac{3}{2}K(K+1)-2I(I+1)J(J+1)}{4I(2I-1)J(2J-1)}\,,
\end{equation}
where we have defined $K=F(F+1)-I(I+1)-J(J+1)$ for convenience.
\par
In the presence of a static external magnetic field $\v{\bfield}$, it is convenient to define the quantisation axis $z$ as being aligned with the applied field. The different $m_F$ levels (magnetic sublevels) undergo a shift depending on $\bfield=\lvert\v{\bfield}\rvert$. The effect on the atom of its interaction with $\v{\bfield}$ is qualitatively different for various regimes, depending on the size of the energy shifts compared to the splitting between the levels. We will only be concerned with the simplest of these cases, where the energy shift is small compared to the splitting between the hyperfine levels. In such cases, $F$ is a good quantum number (\ie, its value is well-defined as the system evolves in time) and the interaction Hamiltonian can be written
\begin{equation}
\label{eq:MagHamiltonian}
 \hat{H}_\text{B}=\mu_\text{B}g_F\v{F}\cdot\v{\bfield}\,,
\end{equation}
where $\mu_\text{B}$ is the Bohr magneton and $g_F$ the Land\'e $g$-factor given approximately by\footnote{See Ref.~\cite{Woodgate2000,Steck2008}; we approximate $g_S\approx 2$ in their notation.}
\begin{equation}
 g_F\approx\biggl[\frac{3J(J+1)+S(S+1)-L(L+1)}{2J(J+1)}\biggr]\times\biggl[\frac{F(F+1)-I(I+1)+J(J+1)}{2F(F+1)}\biggr]\,.
\end{equation}
The associated energy shift for each magnetic sublevel is then
\begin{equation}
 \Delta E_\text{B}=\mu_\text{B}g_Fm_F\bfield\,,
\end{equation}
which is therefore different for each value of $m_F$, giving rise to the splitting between the magnetic sublevels known as the Zeeman effect.
\par
A similar effect, called the DC Stark effect, occurs in the presence of DC electric fields. The DC Stark effect is generally less pronounced than the Zeeman effect. To lowest order, the energy shift for each level is~\cite{Woodgate2000}
\begin{equation}
 \Delta E_\text{E}\propto\efield_z^2\,,
\end{equation}
where $\efield_z$ is the electric field along the quantisation axis. It is only the higher order terms, omitted in the above expression, that lift the degeneracy for the different levels.

\section{The density matrix}\label{sec:CoolingMethods:DM}
Any physical Hamiltonian that acts on a system is Hermitian; \ie, $\hat{H}^\dagger=\hat{H}$. As such, it is diagonalisable and therefore affords a basis of \emph{eigenstates}. The system acted on by the Hamiltonian is said to be in a pure state if it can be represented as a weighted sum of the eigenstates of the Hamiltonian; in other words, if the Hamiltonian has eigenstates $\ket{\psi_0},\ket{\psi_1},\dots,\ket{\psi_n}$ ($n=0,1,\dots$), a pure state can be described by
\begin{equation}
\label{eq:WavefunctionDecomposition}
 \ket{\Psi}=\sum_i{c_i\ket{\psi_i}}\,,
\end{equation}
with the $c_i$ being complex numbers normalised according to the condition $\sum_i{\lvert c_i\rvert^2}=1$. Often, the above eigenstates are represented as the basis vectors, with
\begin{equation}
 \psi_i=\left(\begin{array}{l}
  \left.\begin{matrix}
    0\\
    \vdots\\
    0\\
  \end{matrix}\right\}i-1\\
  1\\
  \left.\begin{matrix}
    0\\
    \vdots\\
    0\\
  \end{matrix}\right\}n-i\\
              \end{array}
\right)
\end{equation}
being the $i$th basis vector in an $n$-dimensional complex Hilbert space. In general this description does not suffice since the state of the system may be a statistical mixture of the different states $\ket{\Psi_i}$, each of which is itself a weighted sum of the eigenstates of the Hamiltonian, as in \eref{eq:WavefunctionDecomposition}:
\begin{equation}
 \ket{\Psi_i}=\sum_j{c_j^{(i)}\ket{\psi_j}}\,.
\end{equation}
We can conveniently express the state of such a system using the density matrix formalism~\cite{CohenTannoudji1978a}, in which we define the density matrix
\begin{equation}
 \m{\rho}=\sum_i{p_i\ket{\Psi_i}\bra{\Psi_i}}\,,
\end{equation}
where each $p_i$ is interpreted as the probability of the system being in state $\ket{\Psi_i}$; \ie, $\sum_i{p_i}=1$ and $0\leq p_i\leq 1$. We note that the diagonal elements of $\m{\rho}$ are of the form $\m{\rho}_{ii}=\sum_jp_j\big\lvert c_j^{(i)}\big\rvert^2$, where $i$ runs from $1$ to $n$. $\m{\rho}_{ii}$ can be interpreted as the probability of finding the system in the eigenstate $\psi_i$. It is therefore called the population of this state. The off-diagonal elements of this matrix are related to interference effects between pairs of different eigenstates, and are called coherences.\\
The usefulness of the density matrix formalism is readily apparent when calculating expectation values of quantum mechanical operators, for let $\hat{A}$ be such an operator with expectation value $\expt{\hat{A}}$. Then,
\begin{equation}
\label{eq:DensityMatrixTraceRelation}
 \expt{\hat{A}}=\Tr\bigl(\m{\rho}\m{A}\bigr)\,,
\end{equation}
with $\m{A}$ being the matrix representation of $\hat{A}$:
\begin{equation}
 \m{A}\equiv\bigl[\bra{\psi_i}\hat{A}\ket{\psi_j}\bigr]_{i,j}\,.
\end{equation}
We will adopt this notation throughout: an operator $\hat{A}$ and a matrix $\m{A}$ correspond to the same operation in different representations; the $(i,j)$th matrix element of $\m{A}$ will be denoted $\m{A}_{ij}$. Let us see how \eref{eq:DensityMatrixTraceRelation} arises. For a system in pure state $\ket{\Psi}$, the expectation value of $\hat{A}$ is given by $\expt{\hat{A}}=\bra{\Psi}\hat{A}\ket{\Psi}=\sum_{i,j}c_i^\ast c_j\bra{\psi_i}\hat{A}\ket{\psi_j}=\sum_{i,j}\m{\rho}_{ji}\m{A}_{ij}=\Tr\bigl(\m{\rho}\m{A}\bigr)$, where $\m{\rho}=\ket{\Psi}\bra{\Psi}=\sum_{i,j}c_ic_j^\ast\ket{\psi_i}\bra{\psi_j}$. This relation similarly holds in the case of a statistical mixture of states.
\par
The time-dependent Schr\"odinger equation~\cite{Schrodinger1926} for a Hamiltonian $\hat{H}$ acting on a state $\ket{\psi}$, $\hat{H}\ket{\psi}=\i\hbar\tfrac{\partial}{\partial t}\ket{\psi}$, can be used to show that
\begin{equation}
\label{eq:DensityMatrixTimeEvolution}
 \i\hbar\dot{\m{\rho}}=[\m{H},\m{\rho}]\equiv\m{H}\m{\rho}-\m{\rho}\m{H}\,,
\end{equation}
where we use the notation that a dot above a symbol represents its derivative with respect to time:
\begin{align}
\label{eq:QMEPrecursor}
 \dot{\m{\rho}}&=\sum_i{p_i\bigl(\tfrac{\partial}{\partial t}\ket{\Psi_i}\bigr)\bra{\Psi_i}}+\sum_i{p_i\ket{\Psi_i}\bigl(\tfrac{\partial}{\partial t}\bra{\Psi_i}\bigr)}\nonumber\\
&=\sum_i{p_i\bigl(\tfrac{1}{\i\hbar}\hat{H}\ket{\Psi_i}\bigr)\bra{\Psi_i}}+\sum_i{p_i\ket{\Psi_i}\bigl(-\tfrac{1}{\i\hbar}\bra{\Psi_i}\hat{H}^\dagger\bigr)}\nonumber\\
&=\frac{1}{\i\hbar}\Biggl[\hat{H}\Biggl(\sum_i{p_i\ket{\Psi_i}\bra{\Psi_i}}\Biggr)-\Biggl(\sum_i{p_i\ket{\Psi_i}\bra{\Psi_i}}\Biggr)\hat{H}\Biggr]\nonumber\\
&=\tfrac{1}{\i\hbar}[\m{H},\m{\rho}]\,,
\end{align}
since $\hat{H}$ is Hermitian.\footnote{It is perhaps interesting to note that \eref{eq:DensityMatrixTimeEvolution} differs in sign from the Heisenberg equation of motion of an operator~\cite[Complement G$_\text{III}$]{CohenTannoudji1978a}. Mathematically, this is due to the operator ordering, of the form $\hat{U}^\dagger\hat{H}\hat{U}$, used to transform from the Schr\"odinger picture, where the state vector is time-dependent and observables correspond to time-independent operators, to the Heisenberg picture, where only the observables are time-dependent.} \eref{eq:DensityMatrixTimeEvolution} is valid for systems that have no dissipation. Incoherent processes---such as the decay of the electromagnetic field inside a cavity, spontaneous emission, etc.---are a result of the system coupling to an infinity of modes, for example the vacuum electromagnetic field. In most cases~\cite[\textsection 5.1.1]{Gardiner2004}, such processes can be most conveniently described not by directly including them in the Hamiltonian but by adding so-called \emph{Lindblad terms} to the preceding relation~\cite{Lindblad1976}.\footnote{Formally, the Lindblad terms arise upon tracing the master equation \eref{eq:QMEPrecursor} over the bath variables.} These terms are represented by a `superoperator' $\hat{\mathcal{L}}$ acting on $\m{\rho}$, giving the full quantum master equation~\cite[\textsection 5.4.2]{Gardiner2004}:
\begin{equation}
\label{eq:QME}
 \dot{\m{\rho}}=\tfrac{1}{\i\hbar}[\m{H},\m{\rho}]+\hat{\mathcal{L}}\m{\rho}\,.
\end{equation}
If the system described by $\m{\rho}$ is coupled to a zero-temperature environment with a decay constant $\kappa$, then the Lindblad terms take the generic form
\begin{equation}
\label{eq:LindbladDissipativeSystem}
 \hat{\mathcal{L}}\m{\rho} = -\kappa\bigl(\hat{c}^\dagger\hat{c}\m{\rho} + \m{\rho}\hat{c}^\dagger\hat{c} - 2\hat{c}\m{\rho}\hat{c}^\dagger\bigr)\,,
\end{equation}
with $\hat{c}$ being an operator that describes the system coupled to $\m{\rho}$. For example, the decay of a cavity field to which an atom is coupled is described by setting $\kappa$ to be the cavity field decay rate and $\hat{c}$ the annihilation operator, more commonly denoted $\hat{a}$, of the cavity field; $\hat{c}^\dagger$, or $\hat{a}^\dagger$, is then the corresponding creation operator. This description is used in \Sref{sec:CoolingMethods:MMC}, as adapted from the literature (see, in particular, Ref.~\cite{Domokos2001}), in our semiclassical exploration of the mirror-mediated cooling mechanism.\\
The result embodied in \erefs{eq:QME} and~(\ref{eq:LindbladDissipativeSystem}) is surprisingly independent of the properties of the \emph{bath}, or environment, that gives rise to the incoherent processes. The reader is referred to Ref.~\cite{Gardiner2004}, especially Ch.~5, for more discussion about this point. Throughout this thesis, we will assume:~(i)~weak coupling between the system and the bath, (ii)~that the dynamics is Markovian~\cite[Ch.~3]{Gardiner1996}, (iii)~that the rotating wave approximation (see \sref{sec:CoolingMethods:OBE:Inc}) is a valid approximation, and (iv)~that the bath is effectively at zero temperature. These assumptions are crucial in deriving \eref{eq:LindbladDissipativeSystem}.
\par
We now introduce what we shall refer to as the two-level atom (TLA). This is a system that is assumed to have two levels, a ground state $\ket{g}$ labelled `$\text{g}$' and an excited state $\ket{e}$ labelled `$\text{e}$'. In terms of the atomic model we introduced in \sref{sec:CoolingMethods:AS}, a TLA can be emulated by a $J=0\rightarrow J^\prime=1$ transition pumped by purely positive-handed circularly polarised light, such that only the $m_J=0$ and $m_{J^\prime}=1$ sublevels are coupled. In the so-called $D$ transitions of the alkali atoms, atoms can be made to cycle between two hyperfine levels in exactly  the same fashion. Whilst the TLA is an idealisation, therefore, it can be approximated quite well in the laboratory.\\
A justification of \eref{eq:QME} is provided in Ref.~\cite[\textsection 1.5; see Eq.~(1.5.39)]{Gardiner2004} for the case of a TLA coupled to the radiation field. Such a description encompasses stimulated absorption and emission as well as spontaneous emission processes.

\section{The Optical Bloch Equations}\label{sec:CoolingMethods:OBE}
The Optical Bloch Equations (OBE) are a set of equations that explore the behaviour of the elements of $\m{\rho}$ when a TLA interacts with an incident harmonic electric field. To derive the OBE, we start off from the Hamiltonian of a TLA at rest in a radiation field~\cite{CohenTannoudji2004}:
\begin{equation}
 \hat{H}=\hat{H}_\text{A}+\hat{H}_\text{R}+\hat{H}_\text{I}\,.
\end{equation}
The terms on the right-hand side of the above equation are the Hamiltonian of:~the atom, the quantised radiation field, and the interaction between the atom and the field, respectively. Let us now assume that the energy difference between $\ket{g}$ and $\ket{e}$ is equal to $\hbar\omega_0$, including any level shifts due to coupling to the quantised field. We can immediately write down $\hat{H}_\text{A}=\hbar\omega_0\ket{e}\bra{e}$, where we have taken the ground state as defining zero energy. The interaction Hamiltonian has two contributions, $\hat{H}_\text{I,i}$ arising from the incident electric field and $\hat{H}_\text{I,q}$ arising from the quantised field, which we will henceforth assume to be initially in the vacuum state, denoted $\ket{0}$. Let us consider these two contributions separately.

\subsection{Interaction with the quantised field}\label{sec:CoolingMethods:OBE:QEF}
The density matrix of the TLA interacting with the quantised field evolves according to
\begin{equation}
 \dot{\m{\rho}}=\tfrac{1}{\i\hbar}[\m{H}_\text{A}+\m{H}_\text{R}+\m{H}_\text{I,q},\m{\rho}]\,,
\end{equation}
Following the standard procedure outlined in the literature (see, for example, Ref.~\cite[\textsection V.A]{CohenTannoudji2004}), we may derive the following equations for the density matrix elements:
\begin{subequations}
\label{eq:OBEQuantisedField}
\begin{align}
 \dot{\m{\rho}}_\text{gg}&=2\Gamma\m{\rho}_\text{ee}\,,\\
 \dot{\m{\rho}}_\text{ee}&=-2\Gamma\m{\rho}_\text{ee}\,,\\
 \dot{\m{\rho}}_\text{ge}&=\i\omega_0\m{\rho}_\text{ge}-\Gamma\m{\rho}_\text{ge}\,,\text{ and}\\
 \dot{\m{\rho}}_\text{eg}&=-\i\omega_0\m{\rho}_\text{eg}-\Gamma\m{\rho}_\text{eg}\,.
\end{align}
\end{subequations}
The first pair of these equations describes the decay, with a time constant $2\Gamma$,\footnote{This differs from standard notation by a factor of $2$; our $\Gamma$ is frequently denoted $\gamma$ in the literature. We use this notation for simplicity of presentation and consistency with Refs.~\cite{Xuereb2009a} and~\cite{Xuereb2009b}.} of atoms in the excited state to the ground state by means of spontaneous emission. The second pair describes the harmonic evolution, at a frequency $\omega_0$, and decay of the coherences of the atom.\par
Spontaneous emission as an irreversible process is a result of the coupling of the atomic dipole to the infinity of quantised vacuum modes in free space. It can therefore be described as an incoherent process using Lindblad terms. The decay terms in the above equations can be derived by adapting \eref{eq:LindbladDissipativeSystem} for the coupling between the atom and the vacuum field. In this case, $\kappa\rightarrow\Gamma$, and $\hat{c}\rightarrow\ket{g}\bra{e}$ is the lowering operator. \erefs{eq:OBEQuantisedField} then follow if we assume that the TLA, effectively, has infinite mass and suffers no momentum recoil from spontaneous emission.

\subsection{Interaction with the incident field}\label{sec:CoolingMethods:OBE:Inc}
Atoms and molecules, whose sizes are usually in the range of $10^{-10}$--$10^{-9}$\,m, are much smaller than the wavelength of visible electromagnetic radiation (ca.~$400$--$800\times10^{-9}$\,m) and the spatial variation of the electric field over the extent of the electron cloud can therefore be neglected. Several cases violate this criterion, such as the interaction of highly excited Rydberg atoms with visible radiation, but we will only consider cases where we can make this \emph{long-wavelength} assumption. We can thus write the interaction between the atom and the radiation field, by modelling the atom as a point dipole, as $\hat{H}_\text{I,i}=-\m{d}\cdot\v{\efield}$~\cite{CohenTannoudji2004}, where $\m{d}$ is the dipole moment of the atom and $\v{\efield}=\v{\efield}_0\cos(\omega_\text{L}t)$ the incident electric field at the position of the atom. In the basis $\bigl\{\ket{g},\ket{e}\bigr\}$, $\m{d}$ is a Hermitian matrix having only off-diagonal elements.\footnote{The dipole operator has odd parity and its diagonal elements are therefore zero.} We define the Rabi frequency $\Omega$ of the interaction:
\begin{equation}
\label{eq:RabiFrequencyDefinition}
 \Omega=-\tfrac{1}{\hbar}\m{d}_\text{eg}\cdot\v{\efield}_0\,,
\end{equation}
with $\m{d}_\text{eg}=\bra{e}\m{d}\ket{g}$ as usual. By a suitable choice of phase we can ensure that $\Omega$ is real, in which case $\m{d}$ is a symmetric matrix; however, we will keep the rest of this chapter general and do not make this simplification. The evolution of the density matrix under the influence of this interaction is given by
\begin{equation}
 \dot{\m{\rho}}=\tfrac{1}{\i\hbar}[-\m{d}\cdot\v{\efield}_0\cos(\omega_\text{L}t),\m{\rho}]\,,
\end{equation}
which can be expanded to
\begin{subequations}
\label{eq:OBEIncNoRWA}
\begin{align}
 \dot{\m{\rho}}_\text{gg}&=-\i\bigl(\Omega^\ast\m{\rho}_\text{eg}-\Omega\m{\rho}_\text{ge}\bigr)\cos(\omega_\text{L}t)\,,\\
 \dot{\m{\rho}}_\text{ee}&=\i\bigl(\Omega^\ast\m{\rho}_\text{eg}-\Omega\m{\rho}_\text{ge}\bigr)\cos(\omega_\text{L}t)\,,\\
 \dot{\m{\rho}}_\text{ge}&=-\i\Omega^\ast\bigl(\m{\rho}_\text{ee}-\m{\rho}_\text{gg}\bigr)\cos(\omega_\text{L}t)\,,\text{ and}\\
 \dot{\m{\rho}}_\text{eg}&=\i\Omega\bigl(\m{\rho}_\text{ee}-\m{\rho}_\text{gg}\bigr)\cos(\omega_\text{L}t)\,.
\end{align}
\end{subequations}
In the same basis as before, we can write the atomic dipole $\m{d}=\m{d}_\text{eg}\ket{e}\bra{g}+\m{d}_\text{ge}\ket{g}\bra{e}$. We can also rewrite $\cos(\omega_\text{L}t)=\tfrac{1}{2}\bigl[\exp(\i\omega_\text{L}t)+\exp(-\i\omega_\text{L}t)\bigr]$. Then, we have
\begin{multline}
 -\m{d}\cdot\v{\efield}=\tfrac{1}{2}\hbar\bigl[\Omega\ket{e}\bra{g}\exp(-\i\omega_\text{L}t)+\Omega^\ast\ket{g}\bra{e}\exp(\i\omega_\text{L}t)\\+\Omega\ket{e}\bra{g}\exp(-\i\omega_\text{L}t)+\Omega^\ast\ket{g}\bra{e}\exp(\i\omega_\text{L}t)\bigr]\,.
\end{multline}
The first two terms inside the brackets in the above expression are resonant when $\omega_\text{L}$ is close to $\omega_0$. They describe the raising of the atomic energy level from $\ket{g}$ to $\ket{e}$ by absorption of a photon or the lowering of the energy level from $\ket{e}$ to $\ket{g}$ accompanied by the emission of a photon. The other two terms describe nonresonant processes and can be ignored when the incident radiation is sufficiently close to the atomic resonance. This is called the \emph{rotating-wave approximation} (RWA), and allows us to rewrite \erefs{eq:OBEIncNoRWA} as
\begin{subequations}
\label{eq:OBEIncRWA}
\begin{align}
 \dot{\m{\rho}}_\text{gg}&=-\tfrac{1}{2}\i\bigl(\Omega^\ast\m{\rho}_\text{eg}e^{\i\omega_\text{L}t}-\Omega\m{\rho}_\text{ge}e^{-\i\omega_\text{L}t}\bigr)\,,\\
 \dot{\m{\rho}}_\text{ee}&=\tfrac{1}{2}\i\bigl(\Omega^\ast\m{\rho}_\text{eg}e^{\i\omega_\text{L}t}-\Omega\m{\rho}_\text{ge}e^{-\i\omega_\text{L}t}\bigr)\,,\\
 \dot{\m{\rho}}_\text{ge}&=-\tfrac{1}{2}\i\Omega^\ast\bigl(\m{\rho}_\text{ee}-\m{\rho}_\text{gg}\bigr)e^{\i\omega_\text{L}t}\,,\text{ and}\\
 \dot{\m{\rho}}_\text{eg}&=\tfrac{1}{2}\i\Omega\bigl(\m{\rho}_\text{ee}-\m{\rho}_\text{gg}\bigr)e^{-\i\omega_\text{L}t}\,.
\end{align}
\end{subequations}
\ \\
We are now in a position to obtain the full time-independent OBE. The right-hand sides of \erefs{eq:OBEIncRWA} and~(\ref{eq:OBEQuantisedField}) can be added together, according to the approximation of independent rates,\footnote{Since the two systems of equations generally describe time evolution on two vastly different timescales, it is reasonable to assume that the physical processes they describe do not interfere, and that the time derivatives can simply be added.} to give the full time derivative of the elements of $\m{\rho}$. Our final manipulation is to set $\tilde{\m{\rho}}_\text{gg}=\m{\rho}_\text{gg}$, $\tilde{\m{\rho}}_\text{ee}=\m{\rho}_\text{ee}$, $\tilde{\m{\rho}}_\text{ge}=\m{\rho}_\text{ge}\exp(-\i\omega_\text{L}t)$, and $\tilde{\m{\rho}}_\text{eg}=\m{\rho}_\text{eg}\exp(\i\omega_\text{L}t)$, whereby
\begin{subequations}
\label{eq:OBE}
\begin{align}
 \dot{\tilde{\m{\rho}}}_\text{gg}&=-\tfrac{1}{2}\i\bigl(\Omega^\ast\tilde{\m{\rho}}_\text{eg}-\Omega\tilde{\m{\rho}}_\text{ge}\bigr)+2\Gamma\tilde{\m{\rho}}_\text{ee}\,,\\
\label{eq:OBEEE}
 \dot{\tilde{\m{\rho}}}_\text{ee}&=\tfrac{1}{2}\i\bigl(\Omega^\ast\tilde{\m{\rho}}_\text{eg}-\Omega\tilde{\m{\rho}}_\text{ge}\bigr)-2\Gamma\tilde{\m{\rho}}_\text{ee}\,,\\
 \dot{\tilde{\m{\rho}}}_\text{ge}&=-\tfrac{1}{2}\i\Omega^\ast\bigl(\tilde{\m{\rho}}_\text{ee}-\tilde{\m{\rho}}_\text{gg}\bigr)-\bigl(\Gamma+\i\Delta_\text{L}\bigr)\tilde{\m{\rho}}_\text{ge}\,,\text{ and}\\
 \dot{\tilde{\m{\rho}}}_\text{eg}&=\tfrac{1}{2}\i\Omega\bigl(\tilde{\m{\rho}}_\text{ee}-\tilde{\m{\rho}}_\text{gg}\bigr)-\bigl(\Gamma-\i\Delta_\text{L}\bigr)\tilde{\m{\rho}}_\text{eg}\,,
\end{align}
\end{subequations}
with $\Delta_\text{L}=\omega_\text{L}-\omega_0$. These are the Optical Bloch Equations for a TLA at rest~\cite{CohenTannoudji2004}.

\section{Polarisability of a two-level atom}\label{sec:CoolingMethods:PolTLA}
\erefs{eq:OBE} can be expressed in terms of three independent parameters, $u=\tfrac{1}{2}\bigl(\tilde{\m{\rho}}_\text{ge}+\tilde{\m{\rho}}_\text{eg}\bigr)$, $v=\tfrac{1}{2\i}\bigl(\tilde{\m{\rho}}_\text{ge}-\tilde{\m{\rho}}_\text{eg}\bigr)$ and $w=\tfrac{1}{2}\bigl(\tilde{\m{\rho}}_\text{ee}-\tilde{\m{\rho}}_\text{gg}\bigr)$, since the total population of atoms is fixed, \ie, $\tilde{\m{\rho}}_\text{gg}+\tilde{\m{\rho}}_\text{ee}=1$. These three parameters are all real, since $\m{\rho}$ is Hermitian, and can be used to describe the coherent evolution of the state of the atom geometrically, with the `Bloch' vector $(u,v,w)$ describing a path on the so-called Bloch Sphere. We will not be concerned with the Bloch Sphere in this thesis but will use $u$, $v$ and $w$ as a shortcut to deriving some results in this section.\par
If an atom is moving at not too fast a speed, say its velocity $\v{v}$ is such that $\tau_\text{i}\lvert\v{v}\rvert,\tau_\text{L}\lvert\v{v}\rvert\ll\lambda_\text{L}$, with $\tau_\text{L}=2\pi/\omega_\text{L}$ being the duration of an optical period of the incident radiation, $\lambda_\text{L}=2\pi c/\omega_\text{L}$ its wavelength, and $\tau_\text{i}=\min\bigl\{(2\Gamma)^{-1},\Omega^{-1}\bigr\}$ the timescale for the internal evolution of the atom, then we can assume that at each point in time the internal variables of the atom---represented by $\tilde{\m{\rho}}$ or, equivalently, $(u,v,w)$---will reach a state of equilibrium~\cite{CohenTannoudji2004}. In other words, we need only concern ourselves with the steady-state value of $\tilde{\m{\rho}}\equiv\tilde{\m{\rho}}^\text{st}$ (this process is known as \emph{adiabatic elimination}~\cite{Gardiner1984} of the external variables). Let us substitute $u$, $v$ and $w$ for the matrix elements of $\tilde{\m{\rho}}$ in \erefs{eq:OBE}, and let us set all the time derivatives to zero. Then we can write
\begin{subequations}
\begin{align}
 \dot{u}&=0=-2\Gamma u^\text{st}+2\Delta_\text{L}v^\text{st}-\bigl(\Omega^\ast-\Omega\bigr)w^\text{st}\,,\\
 \dot{v}&=0=-2\Delta_\text{L}u^\text{st}-2\Gamma v^\text{st}-\bigl(\Omega^\ast+\Omega\bigr)w^\text{st}\,,\text{ and}\\
 \dot{w}&=0=\tfrac{1}{2}\i\bigl(\Omega^\ast-\Omega\bigr)u^\text{st}+\tfrac{1}{2}\bigl(\Omega^\ast+\Omega\bigr)v^\text{st}-2\Gamma w^\text{st}-\Gamma\,,
\end{align}
\end{subequations}
with the superscript `st' again denoting steady-state values. In particular, then,
\begin{subequations}
\label{eq:OBEFinalST}
\begin{align}
\tilde{\m{\rho}}_\text{gg}^\text{st}&=1-\frac{\lvert\Omega\rvert^2}{4}\frac{1}{\Delta_\text{L}^2+\Gamma^2+\lvert\Omega\rvert^2/2}=\frac{1}{2}\frac{2+s}{1+s}\,,\\
\label{eq:OBEFinalSTEE}
 \tilde{\m{\rho}}_\text{ee}^\text{st}&=\frac{\lvert\Omega\rvert^2}{4}\frac{1}{\Delta_\text{L}^2+\Gamma^2+\lvert\Omega\rvert^2/2}=\frac{1}{2}\frac{s}{1+s}\,,\\
 \tilde{\m{\rho}}_\text{ge}^\text{st}&=\frac{\Omega}{2}\frac{\Delta_\text{L}+\i\Gamma}{\Delta_\text{L}^2+\Gamma^2+\lvert\Omega\rvert^2/2}=\frac{\Omega/2}{\Delta_\text{L}-\i\Gamma}\frac{1}{1+s}\,,\text{ and}\\
 \tilde{\m{\rho}}_\text{eg}^\text{st}&=\frac{\Omega^\ast}{2}\frac{\Delta_\text{L}-\i\Gamma}{\Delta_\text{L}^2+\Gamma^2+\lvert\Omega\rvert^2/2}=\frac{\Omega^\ast/2}{\Delta_\text{L}+\i\Gamma}\frac{1}{1+s}\,,
\end{align}
\end{subequations}
where we have defined the saturation parameter $s=\bigl(\lvert\Omega\rvert^2/2\bigr)\big/\bigl(\Delta_\text{L}^2+\Gamma^2\bigr)$, which is proportional to the intensity of the field interacting with the atom, since $\lvert\Omega\rvert\propto\lvert\v{\efield}_0\rvert$. Much physical insight can be gained by exploring how $\tilde{\m{\rho}}^\text{st}$ behaves when $s$ is varied. For very large $s$, we have $\tilde{\m{\rho}}_\text{gg}^\text{st},\tilde{\m{\rho}}_\text{ee}^\text{st}\approx\tfrac{1}{2}$, in which case the population is evenly distributed between the two states, and $\tilde{\m{\rho}}_\text{ge}^\text{st},\tilde{\m{\rho}}_\text{eg}^\text{st}\approx 0$, so that the coherences between the two states essentially disappear. Conversely, for very small $s$, we have $\tilde{\m{\rho}}_\text{gg}^\text{st}\approx 1$ and $\tilde{\m{\rho}}_\text{ee}^\text{st}\approx 0$; \ie, practically the entire atomic population is in the ground state, and in the limit of small $s$ we have
\begin{align}
 \tilde{\m{\rho}}_\text{ge}^\text{st}=\frac{\Omega/2}{\Delta_\text{L}-\i\Gamma}\,\text{ and }\,\tilde{\m{\rho}}_\text{eg}^\text{st}=\frac{\Omega^\ast/2}{\Delta_\text{L}+\i\Gamma}\,.
\end{align}
The average value of the (induced) atomic dipole moment in the steady state is given by
\begin{align}
\label{eq:AvgDip}
 \expt{\m{d}}&=\Tr\bigl(\m{\rho}^\text{st}\m{d}\bigr)=2\real\bigl\{\m{\rho}^\text{st}_\text{eg}\m{d}_\text{ge}\bigr\}\nonumber\\
&=2\real\biggl\{\frac{\Omega^\ast/2}{\Delta_\text{L}+\i\Gamma}\frac{1}{1+s}\m{d}_\text{ge}e^{-\i\omega_\text{L}t}\biggr\}\,.
\end{align}
We can assume that the induced atomic dipole moment is aligned, in space, with the incident electric field, at which point we can set $\m{d}_\text{ge}=\lvert\m{d}_\text{ge}\rvert e^{i\phi}$, with $\phi$ being some phase, and $\m{d}_\text{ge}\cdot\v{\efield}_0=\lvert\m{d}_\text{ge}\rvert\efield_0 e^{i\phi}$, such that
\begin{equation}
 \expt{\m{d}}=\real\biggl\{-\frac{1}{\hbar}\frac{\lvert\m{d}_\text{ge}\rvert^2e^{2i\phi}}{\Delta_\text{L}+\i\Gamma}\frac{1}{1+s}\efield_0e^{-\i\omega_\text{L}t}\biggr\}\,.
\end{equation}
Let $\chi$ be the (complex) polarisability of the atom, whereby $\expt{\m{d}}=\epsilon_0\real\bigl\{\chi\efield_0e^{-\i\omega_\text{L}t}\bigr\}$~\cite{Jackson1998}, $\epsilon_0$ being the vacuum permittivity. It can be verified by direct calculation that, within the RWA, $\real\{\v{d}\cdot\v{\efield}\}=\real\{\v{d}\}\cdot\real\{\v{\efield}\}$; this justifies the form of the equation in the preceding sentence. Thus,
\begin{equation}
 \chi=-\frac{\lvert\bra{e}\m{d}\ket{g}\rvert^2}{\epsilon_0\hbar}\frac{1}{\Delta_\text{L}+\i\Gamma}\frac{1}{1+s}\,,
\end{equation}
where we have made the simplification $\phi=0$, corresponding to assuming that the matrix elements of $\m{d}$ are real, and have also used the relation $\lvert\m{d}_\text{ge}\rvert=\lvert\m{d}_\text{eg}\rvert=\lvert\bra{e}\m{d}\ket{g}\rvert$. Let us now simplify this expression. First, consider \eref{eq:OBEFinalSTEE}, which describes the population in the excited state, and note that atoms in the excited state decay at a rate $2\Gamma$, as also implied by \eref{eq:OBEEE}. The total scattering rate is therefore given by
\begin{equation}
\label{eq:ScatRate} R=2\Gamma\m{\rho}_\text{ee}^\text{st}=\Gamma\frac{\lvert\Omega\rvert^2/2}{\Delta_\text{L}^2+\Gamma^2+\lvert\Omega\rvert^2/2}=\Gamma\frac{I/I_\text{sat}}{1+\bigl(\Delta_\text{L}/\Gamma\bigr)^2+I/I_\text{sat}}\,,
\end{equation}
which will be justified in \Sref{sec:CoolingMethods:WD}, and where the second equality defines the saturation intensity
\begin{equation}
 I_\text{sat}=\frac{2I\Gamma^2}{\lvert\Omega\rvert^2}\,,
\end{equation}
with $I=\tfrac{1}{2}c\epsilon_0\efield_0^2$~\cite{Hecht2001}. We can use the definitions of the Rabi frequency $\Omega$, \eref{eq:RabiFrequencyDefinition}, and $I_\text{sat}$ to obtain
\begin{equation}
 \lvert\bra{e}\m{d}\ket{g}\rvert^2=\frac{c\epsilon_0\Gamma^2\hbar^2}{I_\text{sat}}\,.
\end{equation}
The scattered power is given by $\hbar\omega_\text{L}R$. Let us now define $\sigma_\text{a}\equiv\hbar\omega_\text{L}R/I=\hbar\omega_\text{L}\Gamma/I_\text{sat}$ on resonance and in the limit of low incident power. Thus, $\sigma_\text{a}$ corresponds to the scattering, or radiative, cross-section (the scattered power divided by the incident energy flux) when $\Delta_\text{L}=0$ and $I\ll I_\text{sat}$. But then,
\begin{equation}
 \frac{\lvert\bra{e}\m{d}\ket{g}\rvert^2}{\epsilon_0\hbar}=\frac{\sigma_\text{a}c}{\omega_\text{L}}\Gamma=\frac{\lambda}{\pi}\frac{\sigma_\text{a}}{2}\Gamma\,,\text{ or}
\end{equation}
\begin{equation}
 \chi=-\frac{\sigma_\text{a}c}{\omega_\text{L}}\Gamma\frac{1}{\Delta_\text{L}+\i\Gamma}\frac{1}{1+s}=-\frac{\lambda}{\pi}\frac{\sigma_\text{a}}{2}\frac{\Gamma}{\Delta_\text{L}+\i\Gamma}\frac{1}{1+s}\,.
\end{equation}
It can also be shown that $\sigma_\text{a}=3\lambda^2/(2\pi)$~\cite[\textsection 68.6]{Drake2005}, which allows us to relate $\chi$ to known quantities. We will use the linear polarisability $\chi$ to define the dimensionless `scattering parameter'\footnote{We will refer to $\zeta$ as `polarisability' throughout---the linear relation between $\chi$ and $\zeta$, as well as the context, allows us to do this without giving rise to any ambiguity. $\chi$ is perhaps more correctly referred to as a `susceptibility'.} $\zeta$. Indeed, let us first look at the 1D wave equation, along the $z$ axis, for a monochromatic plane wave incident normally on a polarisable plane at $z=0$~\cite{Deutsch1995,Jackson1998}:
\begin{equation}
 \Bigl(\tfrac{\partial^2}{\partial z^2}+k^2\Bigr)\v{\efield}=-k^2\eta\chi\delta(z)\v{\efield}\,,
\end{equation}
with $\delta(z)$ being the Dirac $\delta$-function, $k=2\pi/\lambda$ the wavenumber of the wave and $\eta$ the density of the particles making up the plane per unit area. If the wave interacts with one particle, then we can effectively set $\eta=1/\sigma_\text{L}$, $\sigma_\text{L}$ being the mode area of the (possibly focussed) wave. In situations where the atom is at the focus of an extremely tightly-focussed wave, however, this approximation breaks down~\cite{Tey2009} and we effectively have $\eta<1/\sigma_\text{L}$; we will henceforth assume that the atom is not in the centre of such a tightly-focussed beam.\\
By defining $\zeta=\tfrac{1}{2}k\eta\chi$, and making use of the boundary conditions~\cite{Deutsch1995}
\begin{subequations}
\begin{equation}
 \v{\efield}\big|_{z\rightarrow 0^-}=\v{\efield}\big|_{z\rightarrow 0^+}\,, \text{ and}\\
\end{equation}
\begin{equation}
 \bigl(\tfrac{\partial}{\partial z}\v{\efield}\bigr)\big|_{z\rightarrow 0^-}-\bigl(\tfrac{\partial}{\partial z}\v{\efield}\bigr)\big|_{z\rightarrow 0^+}=2k\zeta\Bigl(\v{\efield}\big|_{z=0}\Bigr)\,,
\end{equation}
\end{subequations}
obtained by integrating the wave equation over a small interval centred at $z=0$, we can show that the complex (amplitude) reflection and transmission coefficients are
\begin{equation}
 \refl=\frac{i\zeta}{1-\i\zeta}\,\text{ and }\,\trans=\frac{1}{1-\i\zeta}\,,
\end{equation}
respectively. The polarisability of a TLA can therefore be written~\cite{Xuereb2009b}
\begin{equation}
\label{eq:ZetaDefn}
 \zeta=-\frac{\sigma_\text{a}}{2\sigma_\text{L}}\frac{\Gamma}{\Delta_\text{L}+\i\Gamma}\frac{1}{1+s}\,,\text{ or }\zeta=-\frac{\sigma_\text{a}}{2\sigma_\text{L}}\frac{\Gamma}{\Delta_\text{L}+\i\Gamma}\,,
\end{equation}
where the low-intensity ($s\rightarrow 0$) limit is taken at the end; we will be concerned with this limit in the subsequent work. This form of $\zeta$ is independent of the system of units (SI or CGS) used and expresses the polarisability of the TLA as a dimensionless number. One of the advantages of this formalism is that $\zeta$ can take, at the outset, any complex value. For far-detuned atoms, for example, $\zeta$ is approximately real, while on resonance it is purely imaginary. Going a step further, we can use this parameter as an abstract quantity representing a general 1D `scatterer' with (amplitude) reflectivity $\refl$ and transmissivity $\trans$, noting that for real $\zeta$ the scatterer absorbs none of the incident energy ($\lvert\refl\rvert^2+\lvert\trans\rvert^2=1$). In \pref{part:TMM} we will apply this concept to form a very physical link between the cooling of atoms in radiation fields and optomechanics, which concerns itself with the manipulation of mesoscopic optical elements using light.\par
Let us make some final comments about \eref{eq:ZetaDefn}. First of all, as can easily be verified, this definition for $\zeta$ obeys the Kramers--Kronig relations~\cite{Toll1956}. However, if we rewrite
\begin{equation}
\label{eq:ZetaFreq}
 \zeta(\omega)=-\frac{\sigma_\text{a}}{2\sigma_\text{L}}\frac{\Gamma}{\omega-\omega_0+\i\Gamma}\,,
\end{equation}
whereby $\zeta=\zeta(\omega_\text{L})$ in \eref{eq:ZetaDefn}, we can immediately see that $\zeta(-\omega)\neq\bigl[\zeta(\omega)\bigr]^\ast$. This immediately leads to problems. For let us assume that, in the time domain, $\zeta(t)$ is real, which is equivalent to asserting that $\expt{\v{d}}$ in \eref{eq:AvgDip} is real. Then, using the definition of the Fourier transform,
\begin{equation}
 \zeta(\omega)=\int_{-\infty}^\infty e^{-\i\omega t}\zeta(t)\,\rmd t\,,\text{ so that }\zeta(-\omega)=\int_{-\infty}^\infty e^{\i\omega t}\zeta(t)\,\rmd t=\bigl[\zeta(\omega)\bigr]^\ast\,;
\end{equation}
something is not right in our definition of $\zeta$. Indeed, a more rigorous treatment ignoring the RWA gives~\cite{Wang2009}
\begin{equation}
\label{eq:ZetaFreqFull}
 \zeta(\omega)=-\frac{\sigma_\text{a}}{2\sigma_\text{L}}\biggl(\frac{\Gamma}{\omega-\omega_0+\i\Gamma}-\frac{\Gamma}{\omega+\omega_0+\i\Gamma}\biggr)\,,
\end{equation}
which satisfies the aforementioned equality. In all cases we will be concerned with, however, $\lvert\omega-\omega_0\rvert,\Gamma\lll\omega,\omega_0$---the RWA is valid---in which case the second term in the above expression can be safely neglected and \eref{eq:ZetaFreqFull} reduces to \eref{eq:ZetaFreq}.

\section{Energy balance: Work done on a two-level atom}\label{sec:CoolingMethods:WD}
Consider a small interval between a time $t$ and a time $t+\rmd t$, where $\rmd t$ is infinitesimally small. During this time interval, the electric field incident on the TLA can be taken to be constant, $\v{\efield}_0\cos(\omega_\text{L}t)$, but the dipole moment of the TLA can change, say, due to the more rapid motion of the atomic electrons. The rate of work done, averaged over an optical cycle, by the field on the dipole is then~\cite{Jackson1998}
\begin{align}
\label{eq:AvgPower}
 P&=\overline{\v{\efield}_0\cos(\omega_\text{L}t)\cdot\expt{\dot{\v{d}}}}\nonumber\\
&=2\epsilon_0\omega_\text{L}\lvert\v{\efield}_0\rvert^2\imag\{\chi\}\,.
\end{align}
By making use of the definition of $\chi$, the OBE, and the assumption that $\Omega$ is real, we can show that $P=2\hbar\omega_\text{L}\Gamma\m{\rho}_\text{ee}^\text{st}$. Physically, $P$ is the average power absorbed by the atom. For an atom at rest, this must be equal to the energy lost by the atom per unit time through scattering of photons. Dividing $P$ by the energy per incident photon, $\hbar\omega_\text{L}$, gives the number of photons scattered per unit time by the atom, $R=2\Gamma\m{\rho}_\text{ee}^\text{st}$. This justifies \eref{eq:ScatRate}.\par
Two facts emerge from \eref{eq:AvgPower} that are important for our discussion. First of all, a TLA with real $\chi$ will experience no photon scattering, \ie, scattering is significantly suppressed when $\Delta_\text{L}\gg2\Gamma$. Secondly, and more importantly, suppose that $\v{d}$ is delayed with respect to $\v{\efield}$. Then an extra phase factor, let us say $\varphi$, appears in \eref{eq:AvgPower} such that $P=2\epsilon_0\omega_\text{L}\lvert\v{\efield}_0\rvert^2\imag\bigl\{\chi e^{\i\varphi}\bigr\}$. For $\varphi\neq 0$, the \emph{real} component of $\chi$ can also lead to exchange of energy between the TLA and the electric field. This is exploited in cooling mechanisms that rely on optical pumping in polarisation gradients (see, for example, \Sref{sec:TMM:Multilevel} and Ref.~\cite{Dalibard1989} as well as the concluding remarks in the next section).
\par
A final remark can be made about the work done on a TLA in an optical potential. Indeed, the work done on a TLA undergoing small displacements in the field is not, in general, equal to the change in the potential energy of the atom in the field~\cite{Dalibard1989} because the energy in the electromagnetic field itself changes. The relation ``work done $=$ change of potential energy'' is only valid for closed systems, and because the system we are considering is patently open (the light leaving the system carries away energy) we can not expect these two quantities to be equal. Care must be taken, then, when deriving forces from gradients of optical potentials.

\section{Forces on a two-level atom}\label{sec:CoolingMethods:ForceTLA}
Under most circumstances, excluding of course Bose--Einstein condensation and related phenomena, we can assume that the spread of the wavefunction of a TLA is much smaller than the optical wavelength. The force on a TLA interacting with an incident field is $\v{\force}=\v{\nabla}\bigl(\expt{\v{d}}\cdot\v{\efield}\bigr)$, evaluated at the position of the atom ($\v{r}=\v{0}$); this follows from Ehrenfest's theorem~\cite{CohenTannoudji2004}. Let us set $\v{\efield}=\v{\efield}_0\cos\bigl[\omega_\text{L}t+\phi(\v{r})\bigr]$, with $\phi(\v{0})=0$. Then, we have
\begin{equation}
 \v{\force}=\Biggl[\sum_i\expt{\v{d}_i}\cdot\v{\nabla}\bigl({\v{\efield}_0}\bigr)_i\Biggr]\cos\bigl(\omega_\text{L}t\bigr)-\bigl[\v{\nabla}\phi(\v{r})\bigr]\expt{\v{d}}\cdot\v{\efield}_0\sin\bigl(\omega_\text{L}t\bigr)\,,
\end{equation}
where the subscripted $i$ denotes the spatial dimension and the sum runs over $i=x,y,z$. Moreover, \eref{eq:AvgDip} reduces to
\begin{equation}
 \expt{\v{d}}=2\v{d}_\text{eg}\bigl[u^\text{st}\cos\bigl(\omega_\text{L}t\bigr)-v^\text{st}\sin\bigl(\omega_\text{L}t\bigr)\bigr]\,.
\end{equation}
Putting the last two equations together, we can express the time-averaged force acting on the TLA, assuming real $\Omega$, as
\begin{align}
 \v{\force}&=u^\text{st}\v{\nabla}\bigl(\v{d}_\text{eg}\cdot\v{\efield}_0\bigr)+v^\text{st}\v{d}_\text{eg}\cdot\v{\efield}_0\v{\nabla}\phi(\v{r})\nonumber\\
&=-\hbar u^\text{st}\v{\nabla}\Omega-\hbar\Omega v^\text{st}\v{\nabla}\phi(\v{r})\,,
\end{align}
since $\v{d}_\text{eg}$ has no spatial dependence on this scale. We can give this equation more physical meaning by using \erefs{eq:OBEFinalST} to obtain
\begin{equation}
 u^\text{st}=\frac{\Omega}{2}\frac{\Delta_\text{L}}{\Delta_\text{L}^2+\Gamma^2+\Omega^2/2}\,\text{ and }\,v^\text{st}=\frac{\Omega}{2}\frac{\Gamma}{\Delta_\text{L}^2+\Gamma^2+\Omega^2/2}\,,
\end{equation}
whereby
\begin{subequations}
\begin{align}
\label{eq:FrictionReactDiss}
 \v{\force}&=\underbrace{-\frac{1}{2}\frac{\hbar\Omega\Delta_\text{L}}{\Delta_\text{L}^2+\Gamma^2+\Omega^2/2}\v{\nabla}\Omega}_\text{reactive force}\underbrace{-\frac{1}{2}\frac{\hbar\Omega^2\Gamma}{\Delta_\text{L}^2+\Gamma^2+\Omega^2/2}\v{\nabla}\phi(\v{r})}_\text{dissipative force}\\
\nonumber\\
&=\begin{cases}
  -\dfrac{1}{2}\dfrac{\hbar\Omega\Delta_\text{L}}{\Delta_\text{L}^2+\Gamma^2}\v{\nabla}\Omega-\dfrac{1}{2}\dfrac{\hbar\Omega^2\Gamma}{\Delta_\text{L}^2+\Gamma^2}\v{\nabla}\phi(\v{r})&\text{for }s\rightarrow 0\\
  \\
  -\hbar\Delta_\text{L}\dfrac{\v{\nabla}\Omega}{\Omega}-\hbar\Gamma\v{\nabla}\phi(\v{r})&\text{for }s\rightarrow\infty\,.
 \end{cases}
\end{align}
\end{subequations}
We note that, in the limit of high intensity ($s\rightarrow\infty$), the force acting on the TLA tends towards a limit of a dissipative force whose strength is independent of the intensity. In the following sections we will restrict ourselves to the limit of small $s$. This is useful in considering the interaction of a TLA with a standing wave. For small intensities, the TLA behaves as if it interacts with the two running waves making up the standing wave independently. For very high intensities, the process of absorption of photons from one running wave and stimulated emission into the other running wave can become important and modify the interaction significantly.\\
Another way of looking at this is to recall that the force is the gradient of the product of the induced dipole and the local electric field. The gradient operation is linear, so we need not concern ourselves with its effects. However, suppose that the electric field at the position of the TLA is a sum of two fields, $\v{\efield}_1+\v{\efield}_2$, with the corresponding induced dipole being equal to $\expt{\v{d}_1}+\expt{\v{d}_2}$. Then, the force on the TLA is $\v{\nabla}\bigl[\bigl(\expt{\v{d}_1}+\expt{\v{d}_2}\bigr)\cdot\bigl(\v{\efield}_1+\v{\efield}_2\bigr)\bigr]$, which has two extra terms in addition to the sum of the two independent forces, $\v{\nabla}\bigl(\expt{\v{d}_1}\cdot\v{\efield}_1\bigr)+\v{\nabla}\bigl(\expt{\v{d}_2}\cdot\v{\efield}_2\bigr)$. If the spatial (temporal) average of these extra two terms over a wavelength is zero (this happens, e.g., when there is no coherence between the two electric fields), the two forces can therefore be computed separately and added to give the total spatially (temporally) averaged force.
\par
Let us end this section with a note on the dipole force, which corresponds to the term in the force proportional to $\v{\nabla}\Omega$ (the reactive force). \eref{eq:AvgDip} and \eref{eq:AvgPower} tell us that only $v^\text{st}$ can give rise to a global exchange of energy between the TLA and the light field. The reactive component of $\v{\force}$ above can, in fact, be derived from a potential. In other words, the dipole force is conservative. This is not true in general, however. Employing the same arguments as before, if $\expt{\m{d}}$ is delayed with respect to the field, a component of $v^\text{st}$ enters the dipole force and it is therefore no longer constrained to be strictly conservative.

\section{The Fluctuation--Dissipation Theorem}\label{ch:CoolingMethods:FDT}
In the case of the systems that we will investigate, the fluctuation--dissipation theorem is a relation between the systematic (dissipative) and fluctuating components of a force acting on a particle. We shall make extensive use of this theorem in the form of \eref{eq:FDT} below.\\
Consider a particle interacting with the radiation field, such that the force acting on the particle, as a function of time, is $\force(t)=\force_0(t)+\force_\text{L}(t)$, where the first term describes a continuous force that, for example, acts to dampen the particle's motion and the second term is a Langevin force, which has a zero time average and describes the instantaneous fluctuations of the force about its average. In other words, $\expt{\force(t)}=\expt{\force_0(t)}$.\footnote{A note about notation is due: in this section we use the angle brackets, $\expt{\ \cdot\ }$, to represent the average for a classical force or the expectation value for a quantised force, depending on the nature of the force.} Now, consider
\begin{align}
 \expt{\force(t)\force(t^\prime)}&=\expt{\force_0(t)\force_0(t^\prime)}+\expt{\force_0(t)\force_\text{L}(t^\prime)}+\expt{\force_\text{L}(t)\force_0(t^\prime)}+\expt{\force_\text{L}(t)\force_\text{L}(t^\prime)}\nonumber\\
&=\expt{\force_0(t)}\expt{\force_0(t^\prime)}+\expt{\force_0(t)}\expt{\force_\text{L}(t^\prime)}+\expt{\force_\text{L}(t)}\expt{\force_0(t^\prime)}+\expt{\force_\text{L}(t)\force_\text{L}(t^\prime)}\nonumber\\
&=\expt{\force(t)}\expt{\force(t^\prime)}+\expt{\force_\text{L}(t)\force_\text{L}(t^\prime)}\,,
\end{align}
where the time average is taken on a timescale that is short with respect to the timescale for variations in $\force_0(t)$ but long with respect to that for $\force_\text{L}(t)$, whereby $\expt{\force_0(t)\force_\text{L}(t^\prime)}=\expt{\force_0(t)}\expt{\force_\text{L}(t^\prime)}=0$. The correlation function for $\force_\text{L}(t)$ on this timescale is assumed to have the form
\begin{equation}
 \expt{\force_\text{L}(t)\force_\text{L}(t^\prime)}\approx\diffn\,\delta(t-t^\prime)\,,
\end{equation}
where $\diffn$ is some real number; \ie,
\begin{equation}
\label{eq:FDTDeltaDependence}
 \diffn\,\delta(t-t^\prime)=\expt{\force(t)\force(t^\prime)}-\expt{\force(t)}\expt{\force(t^\prime)}\,,
\end{equation}
cf. Refs.~\cite{CohenTannoudji1992} and~\cite{Gordon1980}, or
\begin{equation}
 \tfrac{1}{2}\diffn=\int_0^\infty\Bigl(\expt{\force(t)\force(t^\prime)}-\expt{\force(t)}\expt{\force(t^\prime)}\Bigr)\rmd t^\prime\,.
\end{equation}
Let us now specify $\force_0=-\heatingcoefft v$, where the \emph{cooling coefficient} $\heatingcoefft$ is some (positive) damping constant and $v$ the velocity of the particle (in 1D). The total force acting on the particle, say of mass $m$, is then
\begin{align}
 \force(t)=\frac{\rmd}{\rmd t}p(t)&=-\heatingcoefft v(t)+\force_\text{L}(t)\nonumber\\
&=m\frac{\rmd}{\rmd t}v(t)\,.
\end{align}
We now multiply this equation by the integrating factor and integrate it from time $t=0$ to some general time $t$. Then:
\begin{align}
 \int_0^t \biggl(\frac{\rmd}{\rmd t}v(t)\biggr)\biggr|_{t=t^\prime}e^{(\heatingcoefft/m)t^\prime}\rmd t^\prime&=-\int_0^t(\heatingcoefft/m)v(t^\prime)e^{(\heatingcoefft/m)t^\prime}\rmd t^\prime+\frac{1}{m}\int_0^t\force_\text{L}(t^\prime)e^{(\heatingcoefft/m)t^\prime}\rmd t^\prime\nonumber\\
&=-\int_0^tv(t^\prime)\biggl(\frac{\rmd}{\rmd t}e^{(\heatingcoefft/m)t}\biggr)\biggr|_{t=t^\prime}\rmd t^\prime+\frac{1}{m}\int_0^t\force_\text{L}(t^\prime)e^{(\heatingcoefft/m)t^\prime}\rmd t^\prime\,,
\end{align}
so that
\begin{align}
v(t)e^{(\heatingcoefft/m)t}&=v(0)+\frac{1}{m}\int_0^t\force_\text{L}(t^\prime)e^{(\heatingcoefft/m)t^\prime}\rmd t^\prime\,,\text{ or}\nonumber\\
v(t)&=v(0)e^{-(\heatingcoefft/m)t}+\frac{1}{m}\int_0^t\force_\text{L}(t^\prime)e^{-(\heatingcoefft/m)(t-t^\prime)}\rmd t^\prime\,.
\end{align}
Let us average this relation over the same timescale as above. Then $\expt{v(t)}=v(0)e^{-(\heatingcoefft/m)t}$, whereby the width of the velocity distribution, defined as
\begin{equation}
 \sigma^2_v(t)=\expt{\bigl[v(t)-\expt{v(t)}\bigr]^2}\,,
\end{equation}
is given by
\begin{align}
 \sigma^2_v(t)&=\frac{1}{m^2}\int_0^t\int_0^t\expt{\force_\text{L}(t^{\prime\prime})\force_\text{L}(t^\prime)}e^{-(\heatingcoefft/m)(t-t^\prime)}e^{-(\heatingcoefft/m)(t-t^{\prime\prime})}\rmd t^\prime\rmd t^{\prime\prime}\nonumber\\
&=\frac{1}{m^2}\int_0^t\int_0^t\diffn\,\delta\bigl(t^\prime-t^{\prime\prime}\bigr)e^{-(\heatingcoefft/m)(t-t^\prime)}e^{-(\heatingcoefft/m)(t-t^{\prime\prime})}\rmd t^\prime\rmd t^{\prime\prime}\nonumber\\
&=\frac{\diffn}{m^2}\int_0^te^{-2(\heatingcoefft/m)(t-t^\prime)}\rmd t^\prime=\frac{\diffn}{2m\heatingcoefft}\Bigl[1-e^{-2(\heatingcoefft/m)t}\Bigr]\,.
\end{align}
For small values of $t$, $\sigma^2_v(t)\approx\diffn t/m^2$. The width of the momentum distribution is similarly given by $\sigma^2_p(t)\approx\diffn t$, whereby $\diffn$ is, by definition, the momentum diffusion coefficient. For very long times we can make two observations. Firstly, $\expt{v(t)}=0$, so that the particle is cooled to zero mean velocity; secondly, $\sigma_v^2(t)=\diffn/(2m\heatingcoefft)$. We will not go into any further depth here, but it can be shown~\cite{Risken1989} that all higher-order correlation functions vanish at long timescales. Moreover, this further implies that the velocity distribution of an ensemble of such particles is Gaussian, of the form $\exp\bigl[-v^2/(\sigma_v/2)^2\bigr]$, with the factor of $\tfrac{1}{2}$ arising due to the symmetry of the distribution. Comparing this with a Maxwell-Boltzmann distribution in thermal equilibrium at temperature $T$, for which $\sigma^2_v=k_\text{B}T/(2m)$ with $k_\text{B}$ being the Boltzmann constant, we can therefore calculate the equilibrium temperature of our particle:
\begin{equation}
\label{eq:FDT}
 k_\text{B}T=\frac{\diffn}{\heatingcoefft}\,.
\end{equation}
This is in fact of the same form as that obtained by Einstein~\cite{Einstein1905} in his discussion of the Brownian motion of particles suspended in a liquid. \eref{eq:FDT} links the fluctuating and dissipating forces experienced by the particle to its equilibrium temperature. Although the justification given here is not completely general, namely in terms of demanding a linear friction force and a constant momentum diffusion constant, \eref{eq:FDT} is a good approximation under most circumstances of interest. See Ref.~\cite{Metcalf1999} for a slightly more general treatment, and Ref.~\cite{Clerk2010} for a treatment based on the theory of quantum noise.\footnote{Considerations related to the general quantum case imply the failure of the Onsager hypothesis, which states that ``the average regression of fluctuations will obey the same laws as the corresponding macroscopic irreversible process''~\cite{Onsager1931}, in cases such as where strong coupling between the system and the noise bath is allowed. Put differently, the quantum regression theorem cannot be called a `theorem' in the mathematical sense. See the discussions in Refs.~\cite{Ford1996} and~\cite{Lax2000} for further details.} The results above can easily be adapted for forces acting in 3D by replacing the forces $\force$, $\force_0$, $\force_\text{L}$ with their vector counterparts $\v{\force}$, $\v{\force}_0$, $\v{\force}_\text{L}$; replacing products of the type $\mathcal{F}(t)\mathcal{F}^\prime(t^\prime)$ with $\v{\mathcal{F}}(t)\cdot\v{\mathcal{F}}^\prime(t^\prime)$ ($\mathcal{F},\mathcal{F}^\prime=\force,\force_0,\force_\text{L}$); and also replacing $v$ with $\v{v}$.

\section{Beyond two-level atoms}\label{sec:CoolingMethods:BeyondTLA}
\begin{figure}[t]
 \centering
    \includegraphics{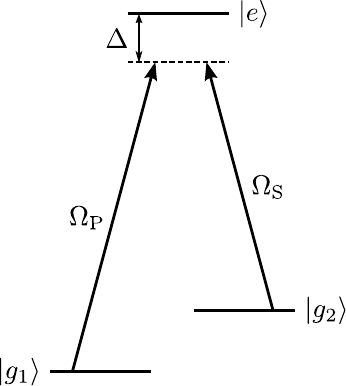}
\caption[Prototypical `$\Lambda$'--type system]{Prototypical `$\Lambda$'--type system used to describe stimulated Raman transitions. The transition between the ground states is taken to be dipole-forbidden.}
 \label{fig:RamanLambdaSystem}
\end{figure}
The two-level atom is a useful, but rather crude, approximation. As noted by Dalibard, Reynaud and Cohen--Tannoudji~\cite{Dalibard1984}, amongst many others, several interesting physical mechanisms can only be explained when one includes the manifold of Zeeman sublevels, \sref{sec:CoolingMethods:AS}, that occur in the energy level structure of real atoms into the model. Indeed, this manifold of sublevels can be included into the definition of the $\zeta$, above, by generalising it into the polarisability tensor, defined as the steady-state expectation value of the polarisability operator $\hat{\boldsymbol{\zeta}}$,
\begin{equation}
\label{eq:ZetaTensorDefn}
 \boldsymbol{\zeta}=\Tr\bigl(\tilde{\m{\rho}}^\text{st}\cdot\hat{\boldsymbol{\zeta}}\bigr)=\sum_{i,j}\langle j\rvert\tilde{\m{\rho}}^\text{st}\lvert i\rangle\langle i\rvert\hat{\boldsymbol{\zeta}}\lvert j\rangle\,,
\end{equation}
where $\tilde{\m{\rho}}^\text{st}$, as defined previously, is the steady-state density matrix describing the system and the summation runs over all the internal sublevels of the atom. The matrix elements of the polarisability operator $\hat{\boldsymbol{\zeta}}$, defined similarly to Eq.~(14.9-24) of Ref.~\cite{Shore1990b}, read in the general $\mu$, $\nu$ basis
\begin{equation}
\label{eq:PolarisabilityOperator}
 \langle i\rvert\hat{\boldsymbol{\zeta}}\lvert j\rangle
                       =\zeta_0\sum_e\begin{bmatrix}
                                    \langle i\rvert\hat{\boldsymbol{d}}_\mu\lvert e\rangle\langle e\rvert\hat{\boldsymbol{d}}_\mu\lvert j\rangle & \langle i\rvert\hat{\boldsymbol{d}}_\mu\lvert e\rangle\langle e\rvert\hat{\boldsymbol{d}}_\nu\lvert j\rangle\\
                                    \langle i\rvert\hat{\boldsymbol{d}}_\nu\lvert e\rangle\langle e\rvert\hat{\boldsymbol{d}}_\mu\lvert j\rangle & \langle i\rvert\hat{\boldsymbol{d}}_\nu\lvert e\rangle\langle e\rvert\hat{\boldsymbol{d}}_\nu\lvert j\rangle
                                    \end{bmatrix}\,,
\end{equation}
where $\zeta_0$ embodies the dimensional constants and the frequency dependence of the polarisability of the atom (see, for example, Ref.~\cite{Wang2009}):
\begin{equation}
\zeta_0=-\frac{k}{2\sigma_\text{L}}\frac{1}{\Delta_\text{L}+i\Gamma}\,,
\end{equation}
In the above equation, the dipole moment operator $\hat{\boldsymbol{d}}_\mu$ ($\hat{\boldsymbol{d}}_\nu$) is related to the $\mu$ ($\nu$) polarised light field and the sum runs over all the internal sublevels, $e$, of the atom. The matrix elements of $\hat{\boldsymbol{d}}_\mu$ ($\hat{\boldsymbol{d}}_\nu$) are given by the appropriate Clebsch--Gordan coefficients. We will explore the implications of this generalised polarisability tensor in \pref{part:TMM}, where it will be used to give an alternative derivation of Sisyphus cooling~\cite{Dalibard1989}.
\par
Stimulated Raman transitions~\cite{Shore1990a} also require multilevel atoms by their very definition. The basic model for an atom undergoing a stimulated Raman transition is a three-level `$\Lambda$'--type level system (\fref{fig:RamanLambdaSystem}): an atom having two `ground' states $\ket{g_1}$ and $\ket{g_2}$, and one excited state $\ket{e}$. The transition between $\ket{g_1}$ and $\ket{g_2}$ is always dipole-forbidden,\footnote{In other words, an atom in one of the $\ket{g_i}$ states will remain in that state, justifying their designation as `ground' states.} due to parity conservation, and the remaining two transitions are coupled by `Pump' ($\ket{g_1}\leftrightarrow\ket{e}$) and `Stokes' ($\ket{g_2}\leftrightarrow\ket{e}$) fields, characterised by strengths $\Omega_\text{P}$ and $\Omega_\text{S}$, respectively. For simplicity, we assume that both coupling fields have a detuning $\Delta$ from resonance. Under these conditions, Rabi oscillations occur between the two ground states at an effective frequency~\cite{Bateman2010b}
\begin{equation}
\frac{\lvert\Omega_\text{P}\Omega_\text{S}^\ast\rvert}{2\Delta}\,.
\end{equation}
It has often been assumed in the literature (see, for example, Ref.~\cite[\S2.1]{kasevich92}), that the hyperfine structure of $\ket{e}$ can easily be taken into account by summing over the appropriate multiple routes, giving an effective Rabi frequency
\begin{equation}
\sum_i\frac{\lvert\Omega_{\text{P};i}\Omega_{\text{S};i}^\ast\rvert}{2\Delta}\,,
\end{equation}
where the label $i$ denotes the coupling to the $i$th sublevel of $\ket{e}$, and $\Delta$ is assumed to be much larger than the energy splittings between the different sublevels. Indeed, it was only in a recent publication by Bateman, Xuereb and Freegarde~\cite{Bateman2010b} that this was proven mathematically to be the case in general, by diagonalising the appropriate multilevel Hamiltonian.\\
For completeness, we will now briefly go through the main arguments in Ref.~\cite{Bateman2010b}. This work was originally published as Bateman, J., Xuereb, A., \& Freegarde, T., Phys.\ Rev.\ A \textbf{81}, 043808 (2010).\footnote{JB set up the problem, solved the three-level system case, provided the interpretation and wrote the paper; AX produced the analytical formulation of the eigensystem of the general $(N+2)$-level Hamiltonian.} The energy structure of the atom we consider is a straightforward extension of that shown in \fref{fig:RamanLambdaSystem}, where we maintain the two ground states $\ket{g_1}$ and $\ket{g_2}$, but where a similar hyperfine structure is taken to exist for the excited state---the state $\ket{e}$ splits into $\ket{e_1},\ket{e_2},\dots,\ket{e_N}$, say. The coupling between the two beams and the different excited sublevels may not be identical, and we therefore set $\Omega_{\text{P};i}$ and $\Omega_{\text{S};i}$ to be, respectively, the coupling strengths for the transitions $\ket{g_1}\leftrightarrow\ket{e_i}$ and $\ket{g_2}\leftrightarrow\ket{e_i}$. The Hamiltonian for this system, after making the RWA and transforming to the interaction picture~\cite{Shore1990a},\footnote{The interaction picture, essentially, removes the fast time evolution from the state vectors and is accomplished by a transformation of the type in \eref{eq:RamanLambdaTransformation} where the transformation matrix is the diagonal matrix of eigenvalues of the time-independent part of the Hamiltonian. It lies in between the Schr\"odinger and the Heisenberg pictures. See Ref.~\cite[Complement G$_\text{III}$]{CohenTannoudji1978a}.} can be written out in state vector notation as
\begin{equation}
\label{eq:RamanLambdaHamiltonian}
\m{H}_\text{A}=\begin{bmatrix}
0 & 0 & \tfrac{1}{2}\Omega_{\text{P};1} & \tfrac{1}{2}\Omega_{\text{P};2} & \dots \\
0 & 0 & \tfrac{1}{2}\Omega_{\text{S};1}e^{\i\delta t} & \tfrac{1}{2}\Omega_{\text{S};3}e^{\i\delta t} & \dots \\
\tfrac{1}{2}\Omega_{\text{P};1}^\ast & \tfrac{1}{2}\Omega_{\text{S};1}^\ast e^{-\i\delta t} & -\Delta & 0 & \dots \\
\tfrac{1}{2}\Omega_{\text{P};2}^\ast & \tfrac{1}{2}\Omega_{\text{S};2}^\ast e^{-\i\delta t} & 0 & -\Delta & \dots \\
\vdots & \vdots & \vdots & \vdots & \ddots
\end{bmatrix}\,,
\end{equation}
which acts on the state vector
\begin{equation}
\begin{pmatrix}
\ket{g_1} \\
\ket{g_2} \\
\ket{e_1} \\
\ket{e_2} \\
\vdots
\end{pmatrix}\,.
\end{equation}
In \eref{eq:RamanLambdaHamiltonian} we made the approximation that the detuning of the Pump beam is $\Delta$ from each of the excited states, and that of the Stokes beam is $\Delta+\delta$. This approximation requires that $\Delta$ be much larger than the splittings between the excited state sublevels. We solve the Schr\"odinger equation with this Hamiltonian by using unitary transformations to find a basis where the time evolution of the states is simple and the transformed Hamiltonian is diagonal. When such a change of basis is done, say by using the transformation matrix $\m{O}_\text{BA}$, then the Hamiltonian in the new basis, $\m{H}_\text{B}$, reads~\cite{Shore1990a}
\begin{equation}
\label{eq:RamanLambdaTransformation}
\m{H}_\text{B}=\m{O}_\text{BA}\biggl(\m{H}_\text{A}\m{O}_\text{BA}^{-1}-\i\frac{\partial}{\partial t}\m{O}_\text{BA}^{-1}\biggr)\,.
\end{equation}
Choosing $\m{O}_\text{BA}$ to be the matrix of eigenvectors of $\m{H}_\text{A}$, the first term in this equation is simply the matrix of its eigenvalues. Upon diagonalising $\m{H}_\text{A}$, it turns out that two of the eigenvectors are superpositions of the two ground states and are decoupled from the rest of the levels in the limit of large detuning. The system can therefore be described as an effective two-level system.
\par
Indeed, we set
\begin{equation}
\lVert\m{H}_\text{B}-\lambda\m{I}\rVert=0\,,
\end{equation}
where $\m{I}$ is the $(N+2)\times(N+2)$ identity matrix and $\lambda$ a parameter. This equation simplifies to the fourth-order polynomial equation
\begin{equation}
\label{eq:RamanLambdaCharacteristicPoly}
16\lambda^2(\Delta-\lambda)^2+4\lambda(\Delta-\lambda)\bigl(\lVert\v{\Omega}_\text{P}\rVert^2+\lVert\v{\Omega}_\text{S}\rVert^2\bigr)+\lVert\v{\Omega}_\text{P}\rVert^2\lVert\v{\Omega}_\text{S}\rVert^2-\lvert\v{\Omega}_\text{P}\cdot\v{\Omega}_\text{S}^\ast\rvert^2=0\,,
\end{equation}
where we have defined the vectors
\begin{equation}
\v{\Omega}_\text{P}=\begin{pmatrix}
\Omega_{\text{P;1}} \\
\Omega_{\text{P;2}} \\
\vdots
\end{pmatrix}\,\text{ and }\,
\v{\Omega}_\text{S}=\begin{pmatrix}
\Omega_{\text{S;1}} \\
\Omega_{\text{S;2}} \\
\vdots
\end{pmatrix}\,.
\end{equation}
The constant term, which is independent of $\lambda$, in \eref{eq:RamanLambdaCharacteristicPoly} is also independent of $\Delta$, so that the product of all the solutions, $\prod_i\lambda_i$, is also independent of $\Delta$. Since not all the $\lambda_i$ are independent of $\Delta$, because the equation is also a second-order polynomial in $\Delta$, then \emph{at least one} eigenvalue must disappear as $\lvert\Delta\rvert\to\infty$. In this limit, we can therefore make the approximation $(\Delta-\lambda)\to\Delta$. The resulting equation has roots
\begin{equation}
\lambda_\pm=\frac{-\bigl(\lVert\v{\Omega}_\text{P}\rVert^2+\lVert\v{\Omega}_\text{S}\rVert^2\bigr)\pm\sqrt{\bigl(\lVert\v{\Omega}_\text{P}\rVert^2-\lVert\v{\Omega}_\text{S}\rVert^2\bigr)^2+4\lvert\v{\Omega}_\text{P}\cdot\v{\Omega}_\text{S}^\ast\rvert^2}}{8\Delta}\,;
\end{equation}
the respective eigenvectors are superpositions of $\ket{g_1}$ and $\ket{g_2}$. We now return to $\m{H}_\text{B}$ and constrain ourselves to the Hilbert space of the two ground states, whereupon we can write
\begin{equation}
\label{eq:RamanLambdaHamiltonian2LevelBare}
\m{H}_\text{B}=\begin{bmatrix}
-\delta\sin^2(\theta) & -\delta e^{\i\delta t}\cos(\theta)\sin(\theta) \\
-\delta e^{-\i\delta t}\cos(\theta)\sin(\theta) & \delta\sin^2(\theta)+\tilde{\Omega}_\text{B}
\end{bmatrix}\,;
\end{equation}
in this restricted Hilbert space, we have
\begin{equation}
\m{O}_\text{BA}=\begin{bmatrix}
\cos(\theta) & e^{\i\delta t}\sin(\theta) \\
-e^{-\i\delta t}\sin(\theta) & \cos(\theta)
\end{bmatrix}\,.
\end{equation}
In the preceding Hamiltonian, we have made use of the definitions, analogous to the two-level Rabi system,
\begin{equation}
\tilde{\Omega}_\text{B}=\sqrt{\Omega_\text{B}^2+\Delta_\text{B}^2}\,,\text{ and }\tan(\theta)=\frac{\Delta_\text{B}-\tilde{\Omega}_\text{B}}{\Omega_\text{B}}\,,
\end{equation}
where
\begin{equation}
\label{eq:RamanLambdaOmegaDelta}
\Omega_\text{B}=\frac{\lvert\v{\Omega}_\text{P}\cdot\v{\Omega}_\text{S}^\ast\rvert}{2\Delta}\,,\text{ and }\Delta_\text{B}=\frac{\lVert\v{\Omega}_\text{P}\rVert^2-\lVert\v{\Omega}_\text{S}\rVert^2}{4\Delta}\,.
\end{equation}
Indeed, we immediately notice that the first relation in \eref{eq:RamanLambdaOmegaDelta} implies that the naive summing over states is formally correct in this limit. In the second relation, $\Delta_\text{B}$ can be identified as the light shift, as justified below.
\par
We further transform this system into a simplified basis by first switching to a time-independent Hamiltonian by means of the matrix
\begin{equation}
\m{O}_\text{CB}=\begin{bmatrix}
1 & 0 \\
0 & e^{\i\delta t}
\end{bmatrix}\,,
\end{equation}
followed by the rotation through an angle $\phi$, through a matrix $\m{O}_\text{DC}$, defined by
\begin{equation}
\tan(2\phi)=\frac{\delta\sin(2\theta)}{\tilde{\Omega}_\text{B}-\delta\cos(2\theta)}\,.
\end{equation}
The difference between the two diagonal elements of the resulting Hamiltonian gives the oscillation frequency for phase evolution of the two dressed states, $\tilde{\Omega}_\text{D}=\sqrt{\Omega_\text{B}^2+\Delta_\text{D}^2}$, where $\Delta_\text{D}=\Delta_\text{B}-\delta$ is the modified detuning, justifying our identification of $\Delta_\text{B}$ with the light shift.
\par
By concatenating the unitary transformations, we can rewrite the bare states in terms of these final dressed states, giving an expression of the form
\begin{equation}
\cos(\theta+\phi)d_1-\sin(\theta+\phi)d_2e^{\i\tilde{\Omega}_\text{D} t}
\end{equation}
for the amplitude of $\ket{g_1}$, where $d_1$ and $d_2$ are the initial amplitudes of the two dressed states. Immediately, we note that the oscillation between $\ket{g_1}$ and $\ket{g_2}$ has a peak-to-peak amplitude that is bounded by
\begin{equation}
\sin[2(\theta+\phi)]=\frac{\Omega_\text{B}}{\sqrt{\Omega_\text{B}^2+\Delta_\text{D}^2}}
\end{equation}
This envelope function describes a power-broadened Lorentzian, centred on the light-shifted frequency difference between the two ground states. It represents the maximum possible population transfer, and any oscillation will be contained within this envelope.

%---

\chapter{Trapping and cooling atoms}\label{ch:CoolingMethods:TrapCool}
\epigraph{There is room for one further general remark. [...] [I]n general one restricts oneself to a discussion of the \emph{energy} exchange, without taking the \emph{momentum} change into account. One feels easily justified in this, because the smallness of the impulses transmitted by the radiation field implies that these can almost always be neglected in practice [...].}{A.\ Einstein, Physikalische Zeitschrift \textbf{18}, 121 (1917)}
The general description given previously of the forces acting on two-level atoms allows the exploration of a number of laser trapping and cooling configurations currently used. In particular, I will look at dipole traps in \Sref{sec:CoolingMethods:DT}, optical molasses in \Sref{sec:CoolingMethods:OM}, and the magneto--optical trap (MOT) in \Sref{sec:CoolingMethods:MOT}. Following these, I will discuss a more recent attempt at a generally applicable laser cooling method, so-called `mirror-mediated cooling', \Sref{sec:CoolingMethods:MMC}, which naturally lends itself to being extended in various ways, as shall be seen in \Sref{sec:CoolingMethods:Other} and \pref{part:TMM}.

\section{Dipole traps}\label{sec:CoolingMethods:DT}
We have already remarked that the `reactive force' in \eref{eq:FrictionReactDiss} is often called the dipole force. The dipole force is proportional to the gradient of the magnitude of the Rabi frequency and is therefore sensitive to spatial nonuniformities in the electric field intensity that the atom is interacting with. In other words, an atom immersed in a tightly focussed light field will experience a force that attracts it to, or repels it from, the focus. Specifically, if $\Delta_\text{L}<0$ the force will point towards increasing light intensity; conversely if $\Delta_\text{L}>0$, the force points away from the focus. In the former case, the atom can be trapped at the focus of the beam, whereas in the latter case configurations can be found having a field minimum at some point in space~\cite{Freegarde2002}, raising the possibility of trapping the atom in such regions.\\
The dipole force is a very general mechanism; it applies not only to TLAs, as explained in the preceding paragraph, but to any object that has a nonzero polarisability. Let us consider an object with an induced dipole moment $\v{d}$ interacting with an electric field $\v{\efield}$. Then, the object will experience a (dipole) potential $U=-\v{d}\cdot\v{\efield}$. This potential then gives rise to the dipole force $\v{\force}=-\v{\nabla}U$ in a manner similar to the force experienced by a TLA. We note that this derivation of $\v{\force}$ implies that the dipole force is conservative; essentially identical arguments to those used when describing the TLA in \Sref{sec:CoolingMethods:ForceTLA} hold in the general case too, whereby a delay between $\v{d}$ and $\v{\efield}$ can give rise to a nonconservative term in the dipole force.\par
This universality inherent in the dipole force makes it a very versatile experimental tool. It has been used to trap atoms in free space~\cite{Dumke2002} and cavities~\cite{Mucke2010}, manipulate microspheres~\cite{Rodrigo2006}, viruses and even living cells~\cite{Ashkin1997}; it is also essential in achieving fully optical Bose--Einstein condensation~\cite{Barrett2001}.

\section{Optical molasses}\label{sec:CoolingMethods:OM}
Optical molasses historically provided the first proof that purely optical mechanisms can be used to slow down the motion of ensembles of atoms~\cite{Chu1985}. Let us see how optical molasses work by looking at the behaviour of a TLA inside a standing wave composed of two weak ($s\ll 1$) counterpropagating travelling waves. \eref{eq:FrictionReactDiss} is again the key to exploring this interaction. For each travelling wave, having a propagation vector $\pm\v{k}$ [\ie, $\phi(\v{r})=\pm\v{k}\cdot\v{r}$], we can easily see that $\v{\nabla}\Omega=0$---we are thus only concerned with the dissipative force---and $\v{\nabla}\phi(\v{r})=\pm\v{k}$. Adding the two forces together thus seems to give a zero net force, but this is only because the Doppler shift has not yet been taken into account. Indeed, the frequency for the $\pm\v{k}$ wave is seen by the atom to be shifted by $\pm\v{k}\cdot\v{v}$, \ie, $\Delta_\text{L}\rightarrow\Delta_\text{L}\pm\v{k}\cdot\v{v}$. In effect, then, the total force seen by the atom is therefore
\begin{align}
\label{eq:OMForce}
\v{\force}_\text{OM}&=-\frac{\hbar\Omega^2\Gamma/2}{\bigl(\Delta_\text{L}+\v{k}\cdot\v{v}\bigr)^2+\Gamma^2}\v{k}+\frac{\hbar\Omega^2\Gamma/2}{\bigl(\Delta_\text{L}-\v{k}\cdot\v{v}\bigr)^2+\Gamma^2}\v{k}\nonumber\\
&\approx\frac{2\hbar\Omega^2\Delta_\text{L}\Gamma}{\bigl(\Delta_\text{L}^2+\Gamma^2\bigr)^2}(\v{k}\cdot\v{v})\v{k}\nonumber\\
&\phantom{\approx}\ =\frac{2\hbar\Omega^2\Delta_\text{L}\Gamma}{\bigl(\Delta_\text{L}^2+\Gamma^2\bigr)^2}(\v{k}\otimes\v{k})\v{v}\,,
\end{align}
with the approximation holding only up to linear order in $\v{k}\cdot\v{v}\Gamma\big/\bigl(\Delta_\text{L}^2+\Gamma^2\bigr)$, and the vector outer product being defined as
\begin{equation}
\v{k}\otimes\v{k}=\begin{pmatrix}
k_1\\
k_2\\
k_3
\end{pmatrix}\otimes\begin{pmatrix}
k_1\\
k_2\\
k_3
\end{pmatrix}=\begin{bmatrix}
k_1^2&k_1k_2&k_1k_3\\
k_1k_2&k_2^2&k_2k_3\\
k_1k_3&k_2k_3&k_3^2
\end{bmatrix}\,.
\end{equation}
Let us now constrain ourselves to one dimension, whereby $\v{\force},\v{v},\v{k}\rightarrow\force,v,k$. For every photon absorption or emission event that the atom undergoes, it experiences a momentum change $\hbar k$; this process occurs at a rate $2R$, with the factor of $2$ arising because of the presence of the two identical beams, and $R$ given by \eref{eq:ScatRate}. Thus, over small times $t$, we have $(\delta p)^2/t=2\hbar^2k^2R$. Identifying $\delta p=\sigma_p(t)$, we therefore have the diffusion and cooling coefficients,
\begin{equation}
\label{eq:OMDiffn}
D=2\hbar^2k^2R=\frac{\hbar^2k^2\Omega^2}{\Delta_\text{L}^2+\Gamma^2}\,,\text{ and }\heatingcoefft=-\frac{2\hbar k^2\Omega^2\Delta_\text{L}\Gamma}{\bigl(\Delta_\text{L}^2+\Gamma^2\bigr)^2}\,,
\end{equation}
respectively (cf. Ref.~\cite{Gordon1980}), whereby the equilibrium temperature of the atom is given by
\begin{equation}
T=\frac{\hbar\Gamma}{2k_\text{B}}\biggl(\frac{\lvert\Delta_\text{L}\rvert}{\Gamma}+\frac{\Gamma}{\lvert\Delta_\text{L}\rvert}\biggr)\,,
\end{equation}
which only makes sense for $\Delta_\text{L}<0$. $T$ attains its lower bound, known as the Doppler temperature $T_\text{D}$, when $\Delta_\text{L}=-\Gamma$, at which point
\begin{equation}
T_\text{D}=\frac{\hbar\Gamma}{k_\text{B}}\,.
\end{equation}
Remarkably, this limiting temperature is independent of the atom's mass $m$ or the intensity of the light beam (recall, however, that we have assumed $s\ll 1$).\par
A more fundamental limit than the Doppler temperature is due to what is called the recoil limit; this gives rise to the recoil temperature $T_\text{R}$. Any time an atom absorbs or emits a photon, it undergoes a momentum change of magnitude $\Delta p=\hbar k$. As such, then, the momentum of the atom can be described as a discrete one-dimensional random walk of step size $\pm\hbar k$. After $n$ steps in such a walk, the momentum $p=S_n\hbar k$ of the atom is described by the quantity $S_n=\sum_{i=1}^n c_i$, where each $c_i=\pm 1$. The statistics of $S_n$ obeys
\begin{equation}
\expt{S_n}=\sum_{i=1}^n \expt{\pm 1}=0\,\text{ and }\,\expt{S_n^2}=\sum_{i=1}^n \expt{(\pm 1)^2}=n\,,
\end{equation}
whereby
\begin{equation}
\sigma_p^2=\bigl(\expt{S_n^2}-\expt{S_n}^2\bigr)(\hbar k)^2=n(\hbar k)^2\,,
\end{equation}
such that the momentum diffusion experienced by the atom is $\diffn=(\hbar k)^2(n/t)$; the quantity $(n/t)$ has to be interpreted as the rate at which photons interact with the atom. The recoil limit applies to any cooling process involving scattering of photons, and therefore at least one scattering event has to occur on average during such a process. The minimum value for $\sigma_p$ is therefore achieved when $n=1$; $\sigma_p=(\hbar k)^2$. Assuming the momentum distribution is Maxwell--Boltzmann,\footnote{This assumption is not very well-founded; ensembles close to, or below, the recoil limit generally obey different statistics (see, for example, Refs.~\cite{Anderson1995} and~\cite{Davis1995}, and Fig.~2 in Ref.~\cite{Aspect1988}). Nonetheless, in the absence of well-definition, assuming Maxwell--Boltzmann statistics allows us to assign a `temperature' to such an ensemble.} this corresponds to a temperature
\begin{equation}
T_\text{R}=\frac{\hbar^2k^2}{2k_\text{B}m}\,.
\end{equation}
For a \textsuperscript{85}Rb atom cycling on the D$_2$ transition, $T_\text{D}=146$\,$\upmu$K and $T_\text{R}=185$\,nK (see Ref.~\cite{Steck2008}, but their definition of the recoil temperature is a factor of $2$ larger than ours; we use a definition consistent with, e.g., Ref.~\cite{Aspect1988}). Not even the recoil limit is a hard limit, though, with several schemes---such as evaporative cooling~\cite{Leanhardt2003}, and even all-optical schemes like velocity-selective coherent population trapping~\cite{Aspect1988}---being devised to overcome it. Such schemes generally work by ensuring that the target state is subject to much less than one scattering event, on the average.

\section{Magneto--optical traps}\label{sec:CoolingMethods:MOT}
To describe MOTs, we have to give up the TLA model we have pursued throughout most of the preceding sections and appeal to the multilevel structure of realistic atoms. In particular, let us consider a $J=1\rightarrow J^\prime=2$ transition.\footnote{We use the standard notation here, with $J$ denoting the ground state and $J^\prime$ the excited state.} Our description of the cooling mechanism that dominates in a MOT will be cursory here---to present a unified approach to laser cooling, we will in \Sref{sec:TMM:Multilevel} treat this type of cooling using an extended form of the transfer matrix method that will be introduced in \pref{part:TMM}.\par
The `sub-Doppler' cooling mechanism in a MOT relies on the phenomenon of optical pumping between the multiple magnetic sublevels in the ground ($J=1$) state. Let us assume that the atom is interacting with two counterpropagating travelling waves of identical frequencies but opposite circular polarisations; this is often referred to as the $\sigma^+$--$\sigma^-$ configuration. The resulting standing wave pattern is linearly polarised everywhere, but this linear polarisation rotates in space on the wavelength scale. Consider now a ground-state static atom in such a field. Optical pumping processes between different magnetic sublevels of the $J=1$ level proceed at rates given by the various Clebsch--Gordan coefficients for the transitions involved. We note that this mechanism relies on pumping between these magnetic sublevels and therefore needs a ground state with $J\geq 1$. The physics behind cooling in this configuration~\cite{Dalibard1989} is rather involved but we can summarise it as follows. In its rest frame, an atom moving with (constant) velocity sees a linear polarisation rotating with a constant angular velocity which, by Larmor's theorem, acts like a fictitious magnetic field parallel to the motion of the atom. This magnetic field acts to effectively couple the magnetic sublevels of the ground state. The motion of the atom also induces an imbalance in the populations of these sublevels, causing it to preferentially scattering photons from one of the two running waves. It can be shown that, under the right conditions, this preferential scattering can lead to a significant friction force acting on the atom, enough to cool a population of such atoms to well below the Doppler temperature~\cite{Lett1988,Dalibard1989,Ungar1989}.\\
We can now describe the theory of operation of MOTs. A MOT is formed at the zero of a quadrupole magnetic field; let us use this magnetic field to define our coordinate system: the $z$-axis points along the axis of symmetry of this magnetic field and the $x$ and $y$ axes define what we will call the `symmetry plane'. The origin of our coordinate system will be taken at the zero of the magnetic field. Traditionally, a MOT requires three pairs of counterpropagating red-detuned circularly-polarised beams, one pair along each of the coordinate axes. The beams in the symmetry plane all have the same helicity, and opposite to that of the beams along the $z$-axis. The polarisation of the beams is as yet undetermined; sub-Doppler cooling occurs for either configuration, depending on the sign of the detuning of the cooling light from the atomic transition. This degree of freedom is therefore fixed by choosing whether to operate above or below atomic resonance.
\par
\begin{figure}
  \centering
  \includegraphics[scale=0.5]{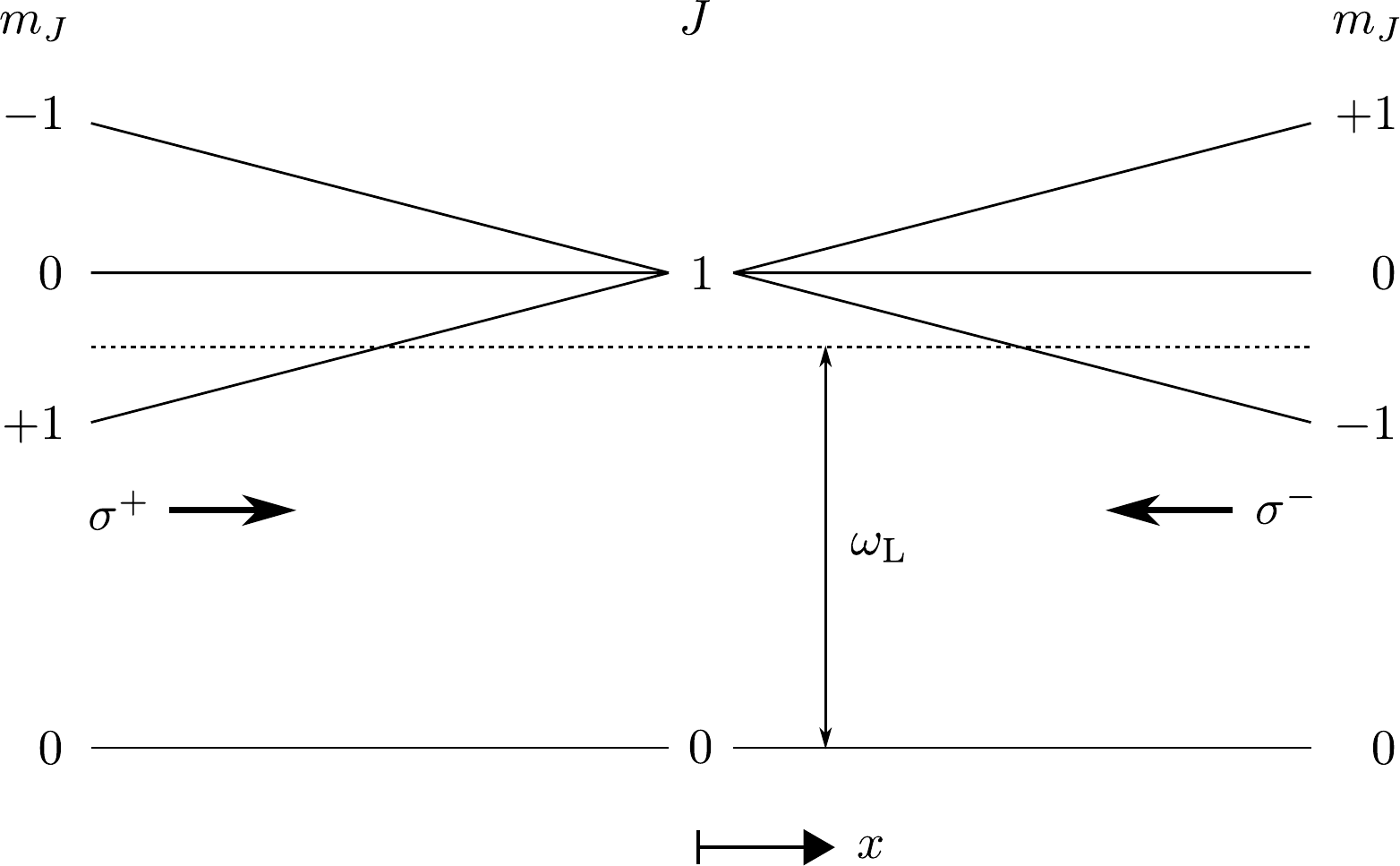}
  \caption[Trapping in a magneto--optical trap]{Close to the centre of a magneto--optical trap, the magnetic field has a constant gradient, taken here to be positive along the positive $x$ direction. An atom to the right of the origin will see the $\sigma^+$ beam as further detuned than the $\sigma^-$ beam, even though both have the same frequency $\omega_\text{L}$, and therefore preferentially scatter from the latter, causing it to experience a restoring force towards $x=0$. A similar principle operates when the atom is to the left of the origin, whereby it will preferentially scatter from the $\sigma^+$ beam. We depict a $J=0\to J^\prime=1$ transition here. The designation $\sigma^\pm$ relates to the effective circular polarisation of the light, and \emph{not} its helicity.}
  \label{fig:MOT_Trapping}
\end{figure}
We have thus far discussed only the cooling mechanism in a MOT. Trapping in a MOT operates as follows. Let us suppose that the magnetic field is such that the gradient is positive in the $x$ direction, and consider a static atom with a positive $x$ coordinate. Due to the nonzero magnetic field, the energy levels of the atom will be split. We will now, for simplicity, consider an atom with a $J=0$ ground state (similar conclusions will hold for $J\neq 0$). The atom is in a region where the magnetic field strength is positive; in other words, the $\Delta m_J=-1$ ($\Delta m_J=+1$) transition is shifted to a lower (higher) energy; this statement follows from \eref{eq:MagHamiltonian}. If the counterpropagating beam (\ie, the beam from the \emph{positive} $x$ direction) is $\sigma^-$ polarised, whereby the opposite beam is $\sigma^+$ polarised, the atom will preferentially scatter photons from the positive $x$ direction and therefore experience a force towards the origin; see \fref{fig:MOT_Trapping}. Conversely, if the magnetic field gradient is negative in the $x$ direction, we require the beam from the positive $x$ direction to be $\sigma^+$ polarised.

\section[Memory-based approach to cooling in laser light]{Memory-based approach to cooling in laser light:~the dipole force delayed\footnote{With sincere apologies to J.\ Dalibard and C.\ Cohen-Tannoudji.}}\label{sec:CoolingMethods:Retarded}
We have stated previously that the dipole force, in its `bare' form, is a conservative force, but if we introduce a delay between the dipole moment of our particle and the field that the particle interacts with, this restriction evaporates and the dipole force gains a non-conservative character~\cite{Braginsky1977}. In the atomic domain, this is generally referred to as a `Sisyphus'--type cooling mechanism, following the nomenclature first applied by Aspect and co-workers~\cite{Aspect1986} with reference to the ancient Greek myth. We will now generalise this idea and see that it leads us to several potentially promising cooling configurations that rely solely upon the dipole force.
\par
In the previous section we considered `standard' Sisyphus cooling. A similar mechanism can be seen to act for atoms inside optical resonators~\cite{Hechenblaikner1998}. Let us consider a simplified model of a resonant optical resonator having almost perfect mirrors. The electric field inside the resonator is, to a very good approximation, a sinusoidal standing wave. If one introduces an atom into the resonator, the effect that the atom has on the resonator will depend greatly on its position in this standing wave. An atom at a node of the field will not affect the field, whereas an atom at an antinode will affect the field rather more strongly.\footnote{A similar point, with implications for photonic crystals, is made about atoms in an optical lattice in Ref.~\cite{Deutsch1995}.} In such a picture, the atom is constantly being pumped into a dressed-state from which it preferentially decays in such a way as to lose kinetic energy. We note that a very similar mechanism is in operation here as in traditional Sisyphus cooling but---crucially---the delay process has been shifted from an internal delay mechanism (optical pumping between the different magnetic sublevels of the atom) to an external one (optical pumping between the atom--cavity dressed states), since the response of the cavity field is necessarily a viscous one.
\par
To illustrate the generality of this idea, let us consider the potential energy $U[x,x_\text{a}(t),v]$ at time $t$ on a test particle at position $x$ due to an atom following a path $x_\text{a}(t)=x_0+vt$. We suppose that the dependence of $U$ on the motion of the atom is mediated entirely through the action of a `memory' in the system, or another time-delayed effect, that can be represented by
\begin{equation}
U[x,x_\text{a}(t),v]=\int_0^\infty U[x,x_\text{a}(t-T),v=0]\,M(T)\,\rmd T\,,
\end{equation}
where $M$ is some function that represents the memory of the system. We can Taylor-expand $U$ in the second variable around $x_\text{a}(t)$ to first order in $v$ and write
\begin{align}
U[x,x_\text{a}(t),v]&=U[x,x_\text{a}(t),v=0]\int_0^\infty M(T)\,\rmd T\nonumber\\
&\phantom{=}\quad-v\biggl[\frac{\partial U(x,y,v=0)}{\partial y}\biggr|_{y=x_0}\biggr]\int_0^\infty T\,M(T)\,\rmd T\,.
\end{align}
Finally, we assume that $M$ is normalised, $\int_0^\infty M(T)\,\rmd T=1$, and set
\begin{equation}
\tau=\int_0^\infty T\,M(T)\,\rmd T\,,
\end{equation}
which has the units of time. Thus, we obtain
\begin{equation}
\label{eq:MemoryFunctionforU}
U[x,x_\text{a}(t),v]=U[x,x_\text{a}(t),v=0]-\tau v\biggl[\frac{\partial U(x,y,v=0)}{\partial y}\biggr|_{y=x_\text{a}(t)}\biggr]\,.
\end{equation}
A similar relation holds for any distributive functional of $U$.\footnote{That is, any $\mathcal{F}$ such that $\mathcal{F}(\alpha U+\beta V)=\alpha\mathcal{F}(U)+\beta\mathcal{F}(V)$ for real $\alpha$ and $\beta$.} In the prototypical Sisyphus mechanism, the system memory lies in the delayed populations. Indeed, if we concentrate on one sublevel having population $\Pi(x,v)\propto -U[x,x_\text{a}(t),v]$, then
\begin{equation}
\Pi[x_\text{a}(t),v]=\Pi^\text{st}[x_\text{a}(t)]-\tau_\text{p}v\biggl[\frac{\partial \Pi^\text{st}(y)}{\partial y}\biggr|_{y=x_\text{a}(t)}\biggr]\,,
\end{equation}
cf. Eq.~(4.22) in Ref.~\cite{Dalibard1989}, with $\tau_\text{p}$ being the \emph{pumping time} associated with our choice of memory function $M(T)=\delta(T-\tau_\text{p})$, or equivalently $M(T)=\exp(-t/\tau_\text{p})/\tau_\text{p}$, and $\Pi^\text{st}(y)\propto -U[x_\text{a}(t),y,v=0]$ the steady-state population when the atom is static.\\
Consider now a situation where the interaction between the particle and the field is mediated through the dipole force, such that the force acting on the particle is
\begin{equation}
\force[x_\text{a}(t),v]=-\frac{\partial U[x,x_\text{a}(t),v]}{\partial x}\biggr|_{x=x_\text{a}(t)}\,.
\end{equation}
By using this relation in \eref{eq:MemoryFunctionforU}, we obtain
\begin{equation}
\label{eq:GeneralTimeDelayedF}
\force[x_\text{a}(t),v]=\force_0[x_\text{a}(t)]+\tau v\biggl[\frac{\partial^2U(x,y,v=0)}{\partial x\,\partial y}\biggr|_{x=x_\text{a}(t),y=x_\text{a}(t)}\biggr]\,,
\end{equation}
with $\force_0$ being the force acting on the particle when it is static. \eref{eq:GeneralTimeDelayedF} may not be trivial to evaluate in practice, and in other circumstances the model itself may not apply; for a given situation $\tau$ may be ill-defined, for example. Nevertheless, it allows us to make a general and powerful prediction that applies to a wide range of systems:~by simply endowing a system with a memory, the dipole force can be used to cool the motion of the particle interacting with the system. It also provides a physical link between models that use an external memory, such as a cavity field, and those that use an internal memory, such as populations in different hyperfine levels.

\section{Cavity fields and atomic motion:~A brief review of current work}
The first few sections of this chapter discussed various ways of capturing atoms or slowing their motion down. In the sections following this, we will explore novel methods that use an optical memory, cf. \sref{sec:CoolingMethods:Retarded}, to achieve cooling. At this stage, therefore, it would be good to briefly summarise the existing work that uses cavities to slow atoms down or otherwise control their motion. Our later work will borrow heavily from the ideas presented here.
\subsection{Cavity-mediated cooling}
The electromagnetic field inside a cavity differs from that in free space in several respects. On the most basic level, the zeroth-order approximation of a (resonant) cavity field is a region in physical space where the electromagnetic field is, for a given driving power, stronger than in other locations. Such a coarse approximation, however, has limited applicability.\footnote{This point is, of course, arguable: the strong coupling of atoms to the electromagnetic field is an interesting field of research with several applications~\cite{Teo2010}.} A first-order approximation to a cavity field would include memory effects which, as we have seen above, imply that the cavity field can be used to control atomic motion. Quantum mechanically, however, the field inside a good cavity is modified in a highly nontrivial way, for even the vacuum fluctuations themselves are modified. This phenomenon is the origin of the Purcell effect~\cite{Purcell1946} and was recognised in the early 1990's as a way of inducing cooling forces on atoms~\cite{Lewenstein1993}. The theory of cavity-mediated cooling of atoms subsequently developed in the direction of driven high-$Q$ cavities~\cite{Cirac1993}. Ref.~\cite{Horak1997} was the first to point out evidence for a novel cooling process taking place inside such cavities that is reminiscent of Sisyphus--type sub-Doppler cooling mechanisms (cf. Ref.~\cite{Dalibard1989}, but see also above):~the basis of the mechanism here being that the photon number in a cavity is, in the strong coupling regime, intimately connected with the position of the atom inside the cavity. By choosing the right system parameters, this connection can be used to slow down the atomic motion. Cavity-mediated cooling of atoms was observed only in recent experiments~\cite{Leibrandt2009,Koch2010}. The potential application of this mechanism to the cooling of the motion of dielectric particles has also been noted~\cite{Barker2010,RomeroIsart2011} and progress is rapidly being made towards the experimental realisation of cavity cooling of microscopic dielectric `particles', both in the form of microspheres~\cite{Barker2010} and in the form of micromirrors~\cite{Kippenberg2008,Aspelmeyer2010}, with profound implications for the study of the foundations of quantum mechanics and even relativity~\cite{Li2011}.\par
Our work in later sections, in particular \sref{sec:TMM:ECCO}, will explore the use of cavities in a different way. By placing a particle outside a cavity, we can still benefit from the presence of the cavity in lengthening the time delay, but without subjecting the particle to the strong field present inside a resonant cavity.
\subsection{Ring cavity cooling}
Investigations of cavity cooling inside Fabry--P\'erot cavities have focussed on the `good-cavity' limit, where the bare cavity has a very high finesse or $Q$-factor. Ring cavities, by their very topology, tend to be larger and with a lower $Q$-factor. This has, perhaps, been one reason why cooling of atoms inside ring cavities has not been explored as thoroughly as cavity-mediated cooling. One distinct advantage of ring cavities is that the system is translationally invariant with respect to the position of the atom; any friction forces developed on the atom are subsequently not dependent on the position of the atom (this will be seen to be a serious limitation of cavity-mediated cooling in \sref{sec:Localisation}). The theory for cooling atoms inside ring cavities~\cite{Gangl2000a,Elsasser2003,Nagy2006,Hemmerling2010,Schulze2010,Niedenzu2010} is similar to that for cavity-mediated cooling. Experimental work on atoms inside ring cavities is rather sparse and tends to focus on recoil-induced effects~\cite{Kruse2003,Slama2007}, which will be discussed next. It has also been proposed (see Refs.~\cite{Vuletic2001a} and~\cite{Salzburger2006}; see also \sref{sec:TMM:AmplifiedOptomechanics}) that the practical limitation of ring cavities as low-$Q$ devices can be lifted by the use of a gain medium inside the ring cavity. This approach holds promise towards improving the performance of such cavity systems to the point that they may be useful in practical realisations of optomechanics experiments.
\subsection{Self-organisation of atoms inside cavities}
The strong feedback between atoms and cavity fields may, under the right conditions, result in self-organisation of an atomic ensemble inside a cavity. We mention here two distinct effects. In Fabry--P\'erot cavities, it was proposed~\cite{Domokos2002b} and subsequently observed~\cite{Baumann2010} that above a threshold pump power, an ensemble of atoms will organise into a checkerboard pattern. Two possible configurations are possible, with the atoms occupying either set of sites in the checkerboard, and transitions occur between the two due to the presence of noise in the system.\par
Collective atomic recoil lasing (CARL)~\cite{Bonifacio1994,Kruse2003,Zimmermann2004,Slama2007} is a related phenomenon that occurs inside ring cavities. CARL occurs when a pump beam inside a ring cavity is reflected off an ensemble of atoms into the counterpropagating mode. The motion of the atoms will Doppler-shift this reflection. An exponential build-up of this latter mode can then occur, accompanied by a spontaneous self-organisation of the ensemble into a grating-like structure. In such a setup, the atomic ensemble acts as a gain medium for the counterpropagating wave, and in its reliance on a gain medium CARL is related to the `amplified optomechanics' idea we discuss in \sref{sec:TMM:AmplifiedOptomechanics}. However, amplified optomechanics uses a gain medium that is spatially separated from the atomic ensemble, in a similar way to the ideas discussed in Ref.~\cite{Vuletic2001a}, whereas in CARL the two are one and the same.

\section{Mirror-mediated cooling}\label{sec:CoolingMethods:MMC}
\begin{figure}
  \centering
  \subfigure[Cavity-Mediated Cooling]{
    \includegraphics[scale=0.5]{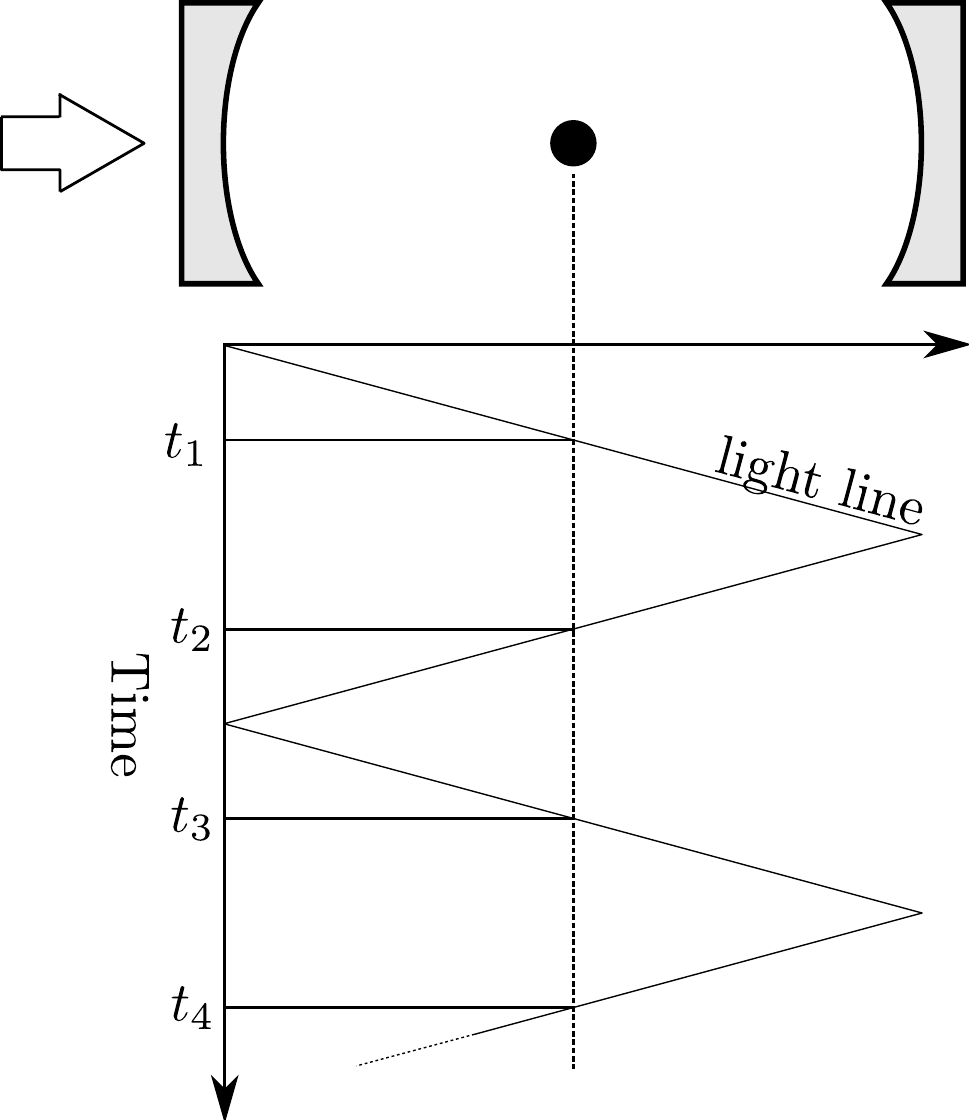}
  }\hspace{1cm}
  \subfigure[Mirror-Mediated Cooling]{
    \includegraphics[scale=0.5]{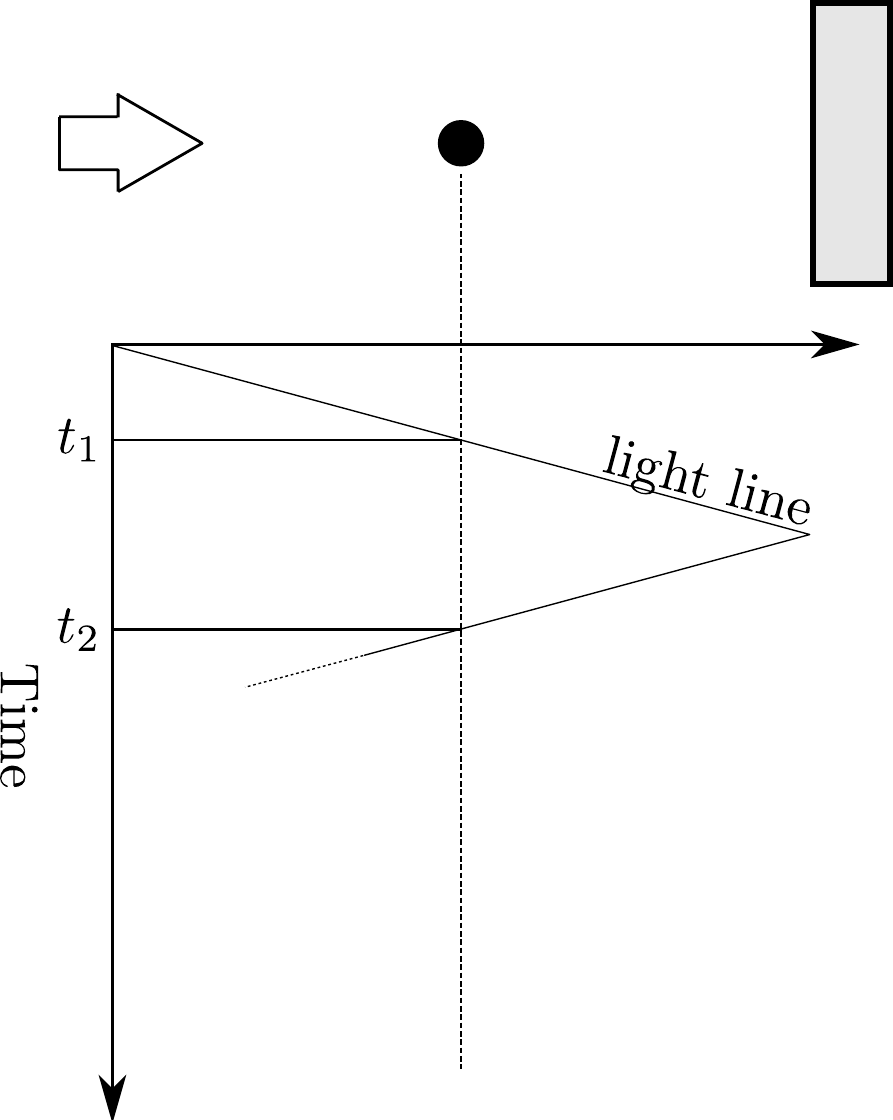}
  }
  \caption[Space--time diagrams for cavity- and mirror-mediated cooling]{Space--time diagrams showing the interaction of an atom with light when the atom is (a)~inside an optical resonator, and (b)~in front of a mirror. It is clear that in case (a) one gets multiple interactions between the two and a `memory time' is not well-defined, whereas in case (b) a definite memory time can be assigned. The blank arrow indicates the pump beam.}
  \label{fig:SpaceTime}
\end{figure}
As explained \sref{sec:CoolingMethods:Retarded}, the dipole force can be invested with a dissipative character by endowing the system as a whole with a memory. Exploring the cooling of atoms inside cavities using this terminology is fine in principle, since a cavity may be viewed as the archetypal optical memory element, but is not easy in practice. Indeed, in such a system, a well-defined ``delay time'' does not exist: a cavity has several memories, each one arising from a separate interaction of \emph{the same light} with the atom. By simplifying the system and removing one of the cavity mirrors, we can once again talk about such a concept as a delay time $\tau$. In this configuration, light interacts with the atom twice, at times $t=0$ and $t=\tau$, say, and no more.\footnote{This is a very `classical' picture, where we describe the light by means of point-like photons. This description is used here to illustrate the physics of the situation. Quantitative calculations can only be made by properly describing the light in terms of the solutions to Maxwell's equations.} These two pictures are illustrated schematically in \fref{fig:SpaceTime}. Whereas we can readily define the system memory time
\begin{equation}
\tau=t_2-t_1
\end{equation}
for \fref{fig:SpaceTime}(b), no such intuitive definition is possible in \fref{fig:SpaceTime}(a) because (i)~if the atom is not at the centre of the cavity, then $t_2-t_1\neq t_3-t_2\neq\dots$, and (ii)~each interaction has a different weighting due to the imperfect reflectivity of the mirrors; it is however possible to assign an effective memory time $\tau=1/(2\kappa)$, where $\kappa$ is the HWHM linewidth of the cavity. Indeed, it will be shown in \aref{sec:TMM:CavityKappaQ} that the finesse of a cavity can be written as
\begin{equation}
\mathcal{F}=\pi N\,,
\end{equation}
where $N$ is the number of round-trips made by the light inside the cavity before the intracavity intensity decays by a factor $1/e^2$. The dipole energy $U$ is proportional to the delayed electric field; so one would expect that the effect of the cavity field on $U$ decays by a factor $1/e$ over $N$ round-trips. It seems reasonable, therefore, to set
\begin{equation}
M(T)=\bigl(1-e^{-1/N}\bigr)\sum_{n\ \text{even}}\delta\bigl(T-nL_1/c)e^{-n/N}+\bigl(1-e^{-1/N}\bigr)\sum_{n\ \text{odd}}\delta\bigl(T-nL_2/c)e^{-n/N}\,,
\end{equation}
with $L_1$ and $L_2$ being the distances from the atom to the two cavity mirrors and $L=L_1+L_2$ the cavity length. This memory function represents an infinite number of discrete interactions, each one being weaker than the previous, of the cavity field with the atom. Thus, in the good-cavity limit of large $N$,
\begin{equation}
\tau=\frac{L_1+L_2}{c}\frac{1}{e^{1/N}-1}\to\frac{LN}{c}=\frac{\mathcal{F}L}{\pi c}\,.
\end{equation}
It will also be shown (see \aref{sec:TMM:CavityKappaQ}) that $\kappa=\pi c/(2\mathcal{F}L)$. It thus follows that $\tau=1/(2\kappa)$ is a good definition for the effective memory time of the system in this model. Throughout the rest of this chapter and in later chapters, we shall see how this memory time $\tau$ governs the interaction of the particle with the field.
\par
The remainder of this section will be devoted to the mathematical description of mirror-mediated cooling and follows closely Ref.~\cite{Xuereb2009a}. It was published as Xuereb, A., Horak, P. \& Freegarde, T. Phys.\ Rev.\ A \textbf{80}, 013836 (2009) and several parts are reproduced \emph{verbatim}. This paper was the first to introduce the `mirror-mediated cooling' mechanism, but was accompanied by another paper~\cite{Horak2010a} that discusses the more computationally-oriented aspects of the Monte--Carlo simulations performed.\\
A semiclassical model of the situation is presented in \Sref{sec:CoolingMethods:MMC:Model}. The model is then explored analytically in a perturbative manner (\Sref{sec:CoolingMethods:MMC:Perturbative}), and numerically using Monte--Carlo simulations (\Sref{sec:CoolingMethods:MMC:Numeric}). Finally, a few brief comments are made about some of the approximations made to describe the system.

\subsection{Mathematical model}\label{sec:CoolingMethods:MMC:Model}
We shall use the TLA model explored in detail in \cref{ch:CoolingMethods:AFInt}. The atom will be assumed to have a transition frequency $\omega_0$ and upper-state decay rate $2\Gamma$, as before, and be coupled to a continuum of quantised electromagnetic modes with frequencies $\omega$ and standing-wave mode functions
\begin{equation}
f(\omega,x)=\sin(\omega x/c)\,,
\end{equation}
with the mirror being at $x=0$. The field modes have annihilation and creation operators $\hat{a}(\omega)$ and $\hat{a}^\dagger(\omega)$, respectively, and the atom--field coupling is described by a single, frequency-independent, coupling coefficient $g$. The mode at frequency $\omega=\omega_\text{L}$ is pumped in a coherent field by a laser, and we assume that the detuning, defined as
\begin{equation}
\Delta_\text{L}=\omega_\text{L}-\omega_0\,,
\end{equation}
obeys the condition $\lvert\Delta_\text{L}\rvert\gg\Gamma$; \ie, we assume far off-resonant operation. The numerical examples we give will be for a \textsuperscript{85}Rb atom cycling in the $5\text{S}_{1/2}\leftrightarrow 5\text{P}_{3/2}$ transition.\\
The starting point for our description is the quantum master equation, \eref{eq:QME}, where the Hamiltonian now reads
\begin{equation}
\label{eq:MMC:Hamiltonian}
\hat{H}=\frac{\hat{p}^2}{2m}-\hbar\Delta\hat{\sigma}^+\hat{\sigma}^-+\int\hbar
(\omega-\omega_\text{L})\hat{a}^{\dagger}(\omega)\hat{a}(\omega)\,\rmd\omega-\i\hbar g\!\!\int[\hat{\sigma}^+\hat{a}(\omega)f(\omega,\hat{x}) - \text{H.c}]\,\rmd\omega\,,
\end{equation}
where $\hat{x}$ and $\hat{p}$ are the position and momentum operators of the atom and `$\text{H.c.}$' denotes the Hermitian conjugate. We also use the Liouvillian terms
\begin{equation}
\label{eq:MMC:Liouvillian}
\mathcal{L}\hat{\rho}=-\Gamma\bigg[\hat{\sigma}^+\hat{\sigma}^-\hat{\rho} +
\hat{\rho}\hat{\sigma}^+\hat{\sigma}^- -2\int_{-1}^{1}N(u)\hat{\sigma}^-e^{-\i u\hat{x}}\hat{\rho}\text{e}^{\i u\hat{
x}}\hat{\sigma}^+\,\rmd u\bigg]\,.
\end{equation}
In this expression, $N(u)$ describes the $1$D projection of the spontaneous emission pattern of the atomic dipole. In the low-saturation regime, we can adiabatically eliminate the internal atomic dynamics and formally express the dipole operator as
\begin{equation}
\hat{\sigma}^-=-\frac{\i\Delta+\Gamma}{\Delta^2+\Gamma^2}\,g\int f(\omega,\hat{x})\hat{a}(\omega)\,\rmd\omega+\hat{\xi}^-\,,
\end{equation}
where $\hat{\xi}^-$ is a noise term~\cite{Gardiner1984}.

\subsection{A perturbative approach to exploring the model}\label{sec:CoolingMethods:MMC:Perturbative}
\subsubsection{Force on a static atom}
The aim of this section is to derive an analytical expression for the friction force acting on the atom when the latter is not moving. To achieve this, we treat the atomic position classically, effecting the replacement $\hat{x}\to x_0$ in \eref{eq:MMC:Hamiltonian} to obtain
\begin{multline}
\hat{H}=\int\hbar(\omega-\omega_\text{L})\hat{a}^{\dagger}(\omega)\hat{a}(\omega)\,
\rmd\omega\\
+\hbar
\frac{g^2\Delta_\text{L}}{\Delta_\text{L}^2+\Gamma^2}\iint\sin(\omega_1x_0/c)\sin(\omega_2x_0/c)\hat{a}^{\dagger}(\omega_1)\hat{a}(\omega_2)\,\rmd\omega_1\,
\rmd\omega_2\,.
\end{multline}
A consequence of assuming $\lvert\Delta_\text{L}\rvert\gg\Gamma$, and operating in the low-saturation regime, is that the population of the excited state is negligible, and therefore last term in \eref{eq:MMC:Liouvillian} does not contribute to the dynamics. With this simplification, the master equation for the annihilation operators reduces to
\begin{equation}
\frac{\rmd}{\rmd t}\hat{a}(\omega,t)=\frac{\i}{\hbar}\bigl[\hat{H},\hat{a}(\omega,t)\bigr]\,.
\end{equation}
By substituting \eref{eq:MMC:Hamiltonian} into this equation, we obtain the integro-differential equation
\begin{equation}
\label{eq:MMC:StartingDE}
\frac{\rmd}{\rmd t}\hat{a}(\omega,t)=-\i(\omega-\omega_\text{L})\hat{a}(\omega,t)-\i\frac{g^2\Delta_\text{L}}{\Delta_\text{L}^2+\Gamma^2}\sin(\omega x_0/c)\int\sin(\omega_1x_0/c)\hat{a}(\omega_1,t)\,\rmd\omega_1\,.
\end{equation}
We now assume coherent states at all times for the fields and replace the operators with their respective expectation values.
Since we are pumping the atom at a single frequency, we take the initial condition $a(\omega,0)=A\,\delta(\omega-\omega_\text{L})$, where $A$ is the amplitude of the pump field, such that $|A|^2$ is the pump power in units of photons per second, and $\delta$ is the Dirac $\delta$-function. We now expand the fields $a(\omega,t)$ in the weak-coupling limit in powers of the coupling constant,
\begin{equation}
a(\omega,t) = \sum_n a_n(\omega,t)\biggl(\frac{g^2\Delta_\text{L}}{\Delta_\text{L}^2+\Gamma^2}\biggr)^n\,,
\end{equation}
with $a_n(\omega,t)$ being the $n$th coefficient of the series expansion. Solving \eref{eq:MMC:StartingDE} to successive orders in $g^2\Delta_\text{L}/\bigl(\Delta_\text{L}^2+\Gamma^2\bigr)$ yields the zeroth order term in this parameter,
\begin{equation}
\label{eq:MMC:a_0}
 a_0(\omega,t)=A\delta(\omega-\omega_\text{L})\,,
\end{equation}
and the first order term
\begin{equation}
 a_1(\omega,t)=A\frac{\exp\left[-\i(\omega-\omega_\text{L})t\right]-1}{\omega-\omega_\text{L}}\sin(\omega x_0/c)\sin(\omega_\text{L}x_0/c)\,.
\end{equation}
To the same level of approximation, we can now find the force acting on the atom to second order in $g^2\Delta_\text{L}/\bigl(\Delta_\text{L}^2+\Gamma^2\bigr)$.\footnote{Mathematically, the fact that the force is at a higher order arises from $\hat{H}$ not having any zeroth order terms [see \eref{eq:MMC:Hamiltonian}]; physically, the force is an interaction of an electric field with an induced dipole moment, and the electric field in this case is itself caused by the induced dipole at an earlier time. Both the induced dipole and this electric field are then, to lowest order, linear in $g^2\Delta_\text{L}/\bigl(\Delta_\text{L}^2+\Gamma^2\bigr)$. This term in the force is therefore quadratic in the same parameter.} The (classical) force acting on the atom is, then,
\begin{equation}
\force(x_0)=-\frac{\partial H}{\partial x}\,,\text{ or}
\end{equation}
\begin{align}
\force(x_0)&=\frac{\hbar}{c}\left|A\right|^2 \frac{g^2\Delta_\text{L}}{\Delta_\text{L}^2+\Gamma^2}\omega_\text{L}\bigg\{\sin(2\omega_\text{L} x_0/c)\nonumber\\
&\ \phantom{=\frac{\hbar}{c}\left|A\right|^2 \frac{g^2\Delta_\text{L}}{\Delta_\text{L}^2+\Gamma^2}\omega_\text{L}\bigg\{}-\frac{\pi}{2}\frac{g^2\Delta_\text{L}}{\Delta_\text{L}^2+\Gamma^2}\sin^2(\omega_\text{L} x_0/c)\big[4\cos^2(\omega_\text{L} x_0/c)-1
\big]\bigg\}\,.
\end{align}
The two terms making up this force have different origins. The first term represents the interaction between the dipole induced in the atom and the (unperturbed) pump field. The second term is the lowest-order correction to the force when the back-action of the atom on the light field is taken into account. The latter term therefore represents the interaction between the dipole induced in the atom by the pump field and the electric field propagated from this same dipole. What we shall see, in the next subsection, is that this propagated electric field can, after being reflected by the mirror, re-interact with the atomic dipole. The motion of the atom between these two interactions gives rise to a velocity-dependent force, \ie, a friction force, on the atom.
\subsubsection{Force on a moving atom: Friction forces}\label{sec:CoolingMethods:MMC:Friction}
Let us now assume that the atom is moving at a constant velocity, $v$. $\hat{x}$ is now a function of time, such that $\hat{x}\to x(t)=x_0+vt$, assuming that by $t=0$ the system has already reached steady-state; in other words, the pump beam has been on for a time longer than $\tau=2x_0/c$, which is the time taken for a disturbance to travel from the position of the atom to the mirror and back again. With this replacement for $\hat{x}$, we proceed to solve \eref{eq:MMC:StartingDE}, keeping terms up to first order in $g^2\Delta_\text{L}/\bigl(\Delta_\text{L}^2+\Gamma^2\bigr)$, as before. The friction force (in the longitudinal direction, \ie, along $x$) can finally be obtained:
\begin{equation}
\label{eq:MMC:AnalyticLongFrictionAll}
 \force_{\parallel}(x_0)=2\pi\hbar k_\text{L}\frac{v}{c}|A|^2\biggl(\frac{g^2\Delta_\text{L}}{\Delta_\text{L}^2+\Gamma^2}\biggr)^2\sin^2(2k_\text{L} x_0)
-\pi\hbar k_\text{L}^2v\tau|A|^2\biggl(\frac{g^2\Delta_\text{L}}{\Delta_\text{L}^2+\Gamma^2}\biggr)^2\sin(4k_\text{L} x_0)\,,
\end{equation}
with $k_\text{L}=\omega_\text{L}/c$ being the wavenumber of the pump field; this force lacks the usual `optical molasses' friction force that is produced by the interaction between the atomic dipole and the Doppler-shifted, unperturbed, pump beam because the atomic excited state has been adiabatically eliminated. The first term in the friction force expression is the velocity-dependent analogue of the second term in $\force(x_0)$ above: it represents the interaction between the induced dipole of the atom and the instantaneous Doppler-shifted field produced by that same dipole. It is the second, `delayed', term, however, that we will be concerned with in the following. Indeed, this term represents the interaction between the atomic dipole and the reflected field it itself produces; the Doppler shift here can effectively be looked at as changing the phase accrued by the reflected field in arriving back at the atom.\\
The delayed friction force is larger than the instantaneous friction force by a factor $k_\text{L}x_0$, and we are therefore fully justified in setting
\begin{equation}
\label{eq:MMC:AnalyticLongFriction}
 \force_{\parallel}(x_0)=-\pi\hbar k_\text{L}^2v\tau|A|^2\biggl(\frac{g^2\Delta_\text{L}}{\Delta_\text{L}^2+\Gamma^2}\biggr)^2\sin(4k_\text{L} x_0)
\end{equation}
if we constrain ourselves to the far-field limit, $k_\text{L}x_0\ggg 1$.
\par
Supposing that the species we are cooling is $^{85}$Rb, and setting $|A|^2=62.5 \Gamma/(2\pi)$, $\Delta_\text{L}=-10\Gamma$,
$\tau=0.5/\Gamma$, and Gaussian beam waist $w=1.4\,\upmu$m, \eref{eq:MMC:AnalyticLongFriction} predicts $1/e$ cooling times of the order of $2$\,ms. The value for $\tau$ that we use implies a separation between the atom and the mirror of the order of several metres. We suggest that this problem can be overcome through the coupling of the light into an optical fibre, thereby avoiding the effects of diffraction. A recent experiment making use of a similar technique is described in Ref.~\cite{Kurtsiefer2009}.
\par
\begin{figure}[t]
 \centering
    \includegraphics[width=1.5\figwidth]{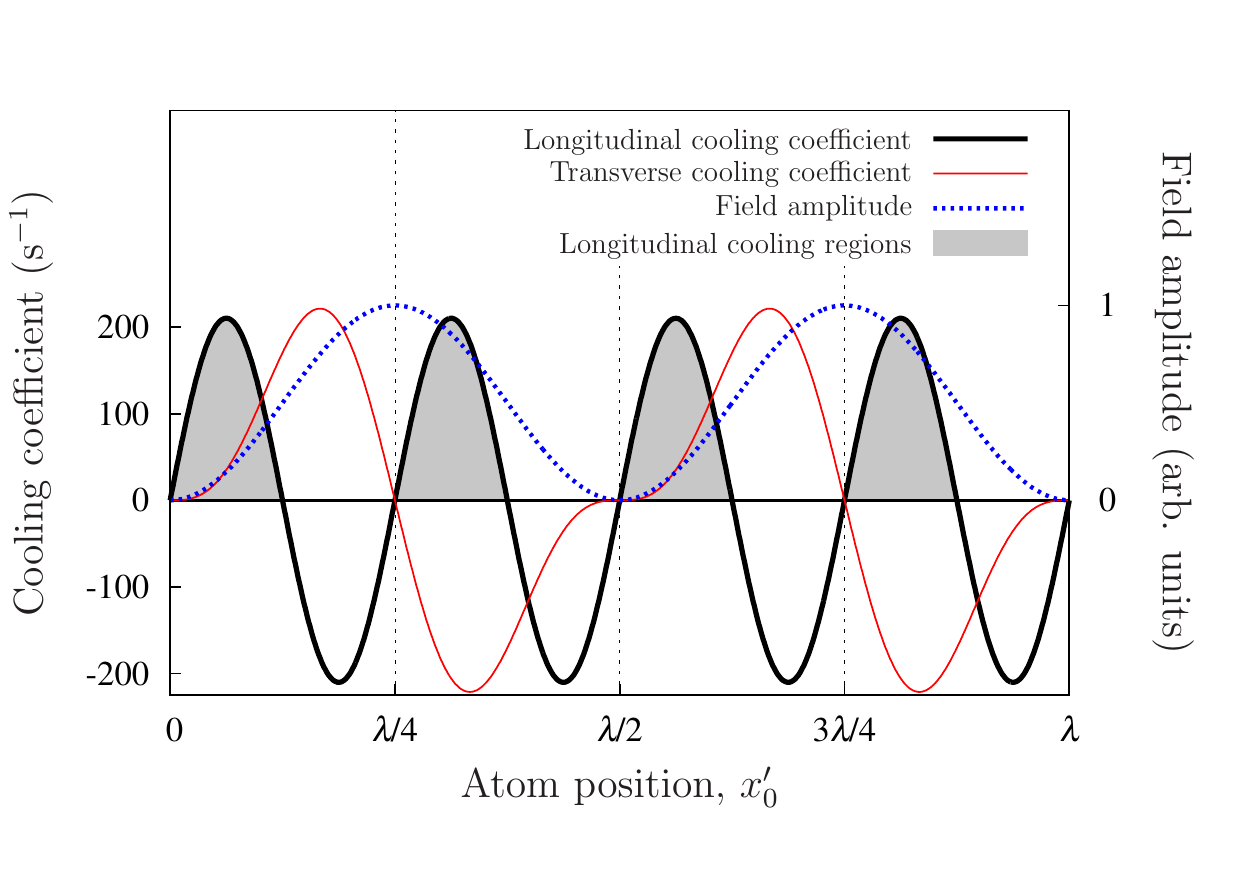}
 \caption[Longitudinal cooling coefficient in mirror-mediated cooling]{Spatial dependence of the longitudinal cooling coefficient $\heatingcoefft$ (thick solid line). The shaded areas promote cooling in the longitudinal direction. Also drawn is the transverse heating coefficient (thin solid line) and the field amplitude (dotted line). Parameters are for $^{85}$Rb atoms and $|A|^2=62.5 \Gamma/(2\pi)$, $\Delta_\text{L}=-10\Gamma$, $\tau=0.5/\Gamma$, $w=1.4~\upmu$m.}
 \label{fig:MMC:Analytic-Spatial}
\end{figure}

\eref{eq:MMC:AnalyticLongFriction} indicates an exponential decay or increase in velocity. We define the \emph{cooling coefficient} as $\heatingcoefft=-\force/v$, whereby we obtain the relation $\rmd p^2/\rmd t=-2\heatingcoefft p^2$, which thus depends on position as $\sin(4k_\text{L}x_0)$. Moreover, since $p^2\propto T$ for a thermal ensemble, we also have $\rmd T/\rmd t=-2\heatingcoefft T$. \fref{fig:MMC:Analytic-Spatial} shows a plot of $\heatingcoefft$ against atomic position, where we introduced the coordinate $x_0^\prime$ relative to the nearest node of the standing wave pump. It is only in certain intervals that we expect the longitudinal force to be a damping force, as indicated in this figure by the shaded regions.
\par
In order to complete the picture, we will now derive the friction force acting in the transverse direction (\ie, in a direction orthogonal to the pump beam). This force arises from the spatial variation of the coupling constant $g$ when the pump field is assumed to be a tightly-focussed Gaussian beam. In this case the coupling constant $g$ becomes a function $g(r)$, where $r$ is the coordinate in the transverse direction. For an atom moving at small constant velocity, we may write $g(r_0+vt)\approx g(r_0)+vtg^{\prime}(r_0)$, where $g^{\prime}(r)=\rmd g/\rmd r$ and time $t=0$, at which $r=r_0$, is defined as before. Substituting this into \eref{eq:MMC:StartingDE} we can derive an expression for the friction force, $\force_{\perp}(x_0)=-\partial H/\partial r$, in the direction of $r$:
\begin{equation}
 \force_{\perp}(x_0)=-2\pi\hbar v\tau|A|^2\left(\frac{2gg^{\prime}\Delta_\text{L}}{\Delta_\text{L}^2+\Gamma^2}\right)^{\!\!2}\sin^3(k_\text{L}x_0)\cos(k_\text{L}x_0)\,,
\end{equation}
with $g$ and $g^\prime$ being evaluated at $r=r_0$. This transverse friction force is also shown in \fref{fig:MMC:Analytic-Spatial}, for comparison with the longitudinal friction force, assuming a Gaussian mode function of waist $w=1.4$\,$\upmu$m. Note that $\force_{\parallel}$ and $\force_{\perp}$ are comparable in magnitude for the parameters chosen here, \ie, where the mode waist is comparable to the optical wavelength. Moreover, we can see that there exist regions where both these forces promote cooling. In the remainder of this section, however, we will concentrate on a one-dimensional treatment of the problem and therefore only consider the longitudinal friction force.
\par
In terms of more familiar parameters, we can rewrite~\eref{eq:MMC:AnalyticLongFriction} in the limit $|\Delta_\text{L}|\gg\Gamma$ as
\begin{equation}
\label{eq:MMC:AnalyticLongFrictionFamiliar}
 \force_{\parallel}(x_0)=-vs\Gamma\frac{\sigma_{\text{a}}}{\sigma_\text{L}}\hbar k_\text{L}^2\tau\sin(4k_\text{L}x_0)\,,
\end{equation}
where $s=g^2|A|^2/(\Delta_\text{L}^2+\Gamma^2)$ is the maximum saturation parameter of the atom in the standing wave, $\sigma_{\text{a}}=3\lambda_\text{L}^2/(2\pi)$ is the atomic radiative cross-section at a wavelength $\lambda_\text{L}=2\pi/k_\text{L}$, as defined in \sref{sec:CoolingMethods:PolTLA}, and where we used the relation $2\pi g^2/\Gamma=4 \sigma_a/(2\sigma_\text{L})$, with $\sigma_\text{L}=\pi w^2/8$ being the mode area of the pump beam of waist $w$, and where the factor of $4$ arises from the standing wave amplitude.
\par
Aside from allowing us to make predictions of cooling times, \erefs{eq:MMC:AnalyticLongFriction}
and~(\ref{eq:MMC:AnalyticLongFrictionFamiliar}) also highlight the dependence of this cooling effect on the variation of important physical parameters. In particular, $\force_{\parallel}$ depends on the square of the detuning, which means that it is possible to obtain cooling with both positive and negative detuning, in stark contrast with the standard Doppler cooling force. The friction force also scales with $w^{-4}$ and $|A|^2$. Hence, for a fixed laser intensity, proportional to $|A|^2/w^2$, \ie, for a fixed atomic saturation, the friction still scales with $w^{-2}$ and thus a tight focus is needed in order to have a sizeable effect. The physical reason for this dependence is that the dipole moment induced in the atom by the pump light is proportional to $|A|^2 g^2\propto|A|^2/w^2$; this polarisation couples to the electric field and, after being reflected by the mirror, polarises the atom again. The size of the force therefore scales as $|A|^2 g^4$, or $|A|^2/w^4$.\\
A very promising feature of these two equations is the linear dependence of the cooling rate on $\tau$: by increasing the distance between the atom and the mirror, we can increase the strength of the friction force acting on the atom. In \Sref{sec:CoolingMethods:MMC:Numeric} we further analyze the dependence of the cooling rate on the various parameters and support the validity of the analytical solution by comparing it with the results of simulations.

\subsubsection{Localising the atom: The effects of adding a harmonic trap}\label{sec:CoolingMethods:MMC:Perturbative:Dipole}
\eref{eq:MMC:AnalyticLongFriction} shows that, in order to observe any cooling effects, we need to localise the particle within around $\lambda_\text{L}/8$. This may be achieved, for example, by an additional far off-resonant and tightly focussed laser beam propagating parallel to the mirror and forming a dipole trap centred at a point $x_0$. In this section, we aim to characterise the effects of this dipole trap, or indeed any harmonic trap, on the atom--field interaction. We characterise this trap by means of its spring constant $k_\text{t}$, such that the trapping force is given by $\force_\text{t}=-k_\text{t}(x-x_0)$, or equivalently by the harmonic oscillator frequency $\omega_\text{t}=\sqrt{k_\text{t}/m}$, where $m$ is the mass of the atom.
\par
With the atom being in a harmonic trap, we can now write down its time-dependent velocity as $v(t)=v_\text{m}\cos(\omega_\text{t}t)$, with a maximum velocity of the atom $v_\text{m}$. Using this expression, it is possible to derive a modified cooling coefficient by performing a perturbative expansion in the dimensionless parameter $v_\text{m}/c$. Proceeding along the lines of the preceding section, we arrive at a modified expression for the friction force,
\begin{equation}
\label{eq:MMC:AnalyticLongFrictionHarm}
 \force_{\parallel}(x_0)=-\pi\hbar
k_\text{L}^2v_\text{m}\tau\sinc(\omega_{\text{t}}\tau)|A|^2\biggl(\frac{g^2\Delta}{\Delta^2+\Gamma^2}\biggl)^2\sin(4k_\text{L}x_0)\,.
\end{equation}
By way of confirmation, it is easy to see that this expression reduces to \eref{eq:MMC:AnalyticLongFriction} in the limit of small $\omega_\text{t}$; \ie, in the free-particle limit for the atomic motion. The sinusoidal dependence on $\omega_\text{t}\tau$ can be justified in an intuitive manner: the effect on the particle is unchanged if the particle undergoes an integer number of oscillations in the round-trip time $\tau$.
\par
While \eref{eq:MMC:AnalyticLongFrictionHarm} was derived for an oscillating atom, it only accounts for the sinusoidal variation of the velocity, and therefore does not include the effect of the finite spatial distribution of the position of the atom. In order to obtain an estimate for the friction force in the presence of this spatial broadening, we calculate the overall energy loss rate experienced by the particle in terms of the time average of \eref{eq:MMC:AnalyticLongFriction}:
\begin{multline}
\label{eq:MMC:AnalyticFrictionTimeAvg}
 \left\langle\frac{\rmd p^2}{\rmd t}\right\rangle=-\frac{\hbar
k_\text{L}^2p_0^2}{m}\tau|A|^2\biggr(\frac{g^2\Delta}{\Delta^2+\Gamma^2}\biggr)^2\int_0^{2\pi}\sin\bigl[4k_\text{L}x_0+4k_\text{L}x_\text{m}\sin(T)\bigr]\cos^2(T)\,\rmd T\,,
\end{multline}
where $p_0=m v_\text{m}$ is the maximum momentum of the particle in the trap, given by $p_0=x_\text{m}\sqrt{m k_\text{t}}$. The value of the integral in \eref{eq:MMC:AnalyticFrictionTimeAvg} can be expressed as
\begin{equation}
\frac{2\pi}{4k_\text{L}x_\text{m}}\bigl[\sin(4k_\text{L}x_0)J_1(4k_\text{L}x_\text{m})+\cos(4k_\text{L}x_0)H_1(4k_\text{L}x_\text{m})\bigr]\,,
\end{equation}
where $J_1$ is the order-$1$ Bessel function of the first kind and $H_1$ is the order-$1$ Struve function~\cite{Gradshteyn1994}. At the point of maximum friction, which occurs at $x_0=-3\lambda_\text{L}/16+n\lambda_\text{L}$ for some integer $n$, the integral in the above equation reduces to $2\pi J_1(4k_\text{L}x_\text{m})/(4k_\text{L}x_\text{m})$; this function can be readily evaluated numerically for a given trap frequency.

For small values of the trap frequency, the effect of the above averaging process is to introduce a factor of $\tfrac{1}{2}$ into \eref{eq:MMC:AnalyticLongFrictionHarm}. The result can be seen as being physically equivalent to the effect of cooling only one degree of freedom when the atom is in a harmonic trap. Finally, we combine the above two ideas to include both the effects of harmonic oscillation and the spatial extent of the atomic motion into  \eref{eq:MMC:AnalyticFrictionTimeAvg}. This can be done, effectively, by replacing $\tau\to\sin(\omega_{\text{t}}\tau)/\omega_{\text{t}}$. The resulting approximate expression for the friction force, taking into account the periodicity in the time delay as well as spatial averaging effects, becomes
\begin{multline}
\label{eq:MMC:AnalyticLongFrictionHarmReduced}
 \left\langle \force_{\parallel}(x_0)\right\rangle=-\tfrac{1}{2}\hbar k_\text{L}^2v_\text{m}\tau|A|^2\biggl(\frac{g^2\Delta}{\Delta^2+\Gamma^2}\biggr)^2\sinc(\omega_\text{t}\tau)\\
\times\int_0^{2\pi}\sin\bigl[4k_\text{L}x_0+4k_\text{L}x_\text{m}\sin(T)\bigr]\cos^2(T)\,\rmd
T\,.
\end{multline}

\subsubsection{Momentum capture range}\label{sec:CoolingMethods:MMC:Perturbative:Cutoff}
\begin{figure}[t]
\centering
   \includegraphics[width=1.5\figwidth]{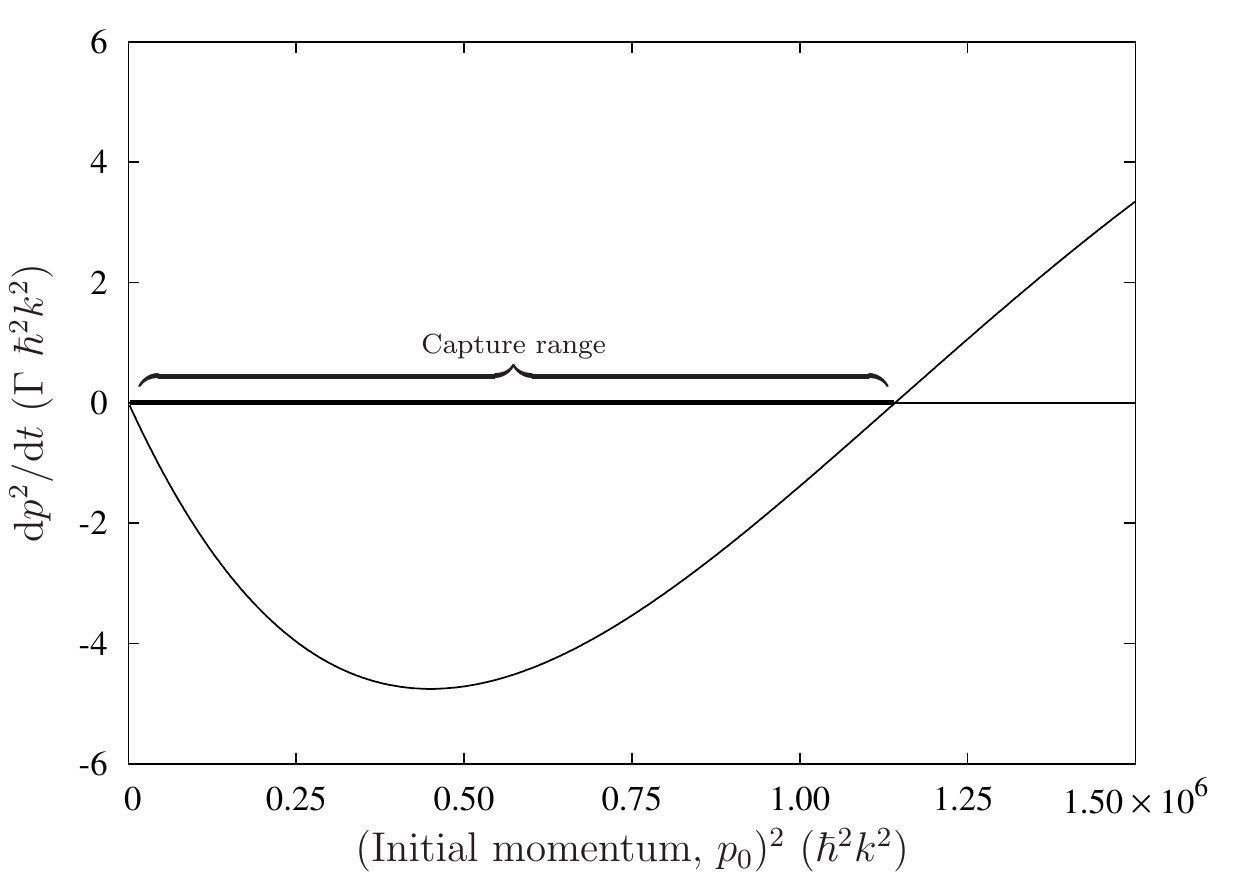}
\caption[Momentum cut-off in mirror-mediated cooling]{Dependence of $\rmd p^2/\rmd t$ on the square of the initial momentum, $p_0^2$, for $\omega_\text{t}=0.45\times 2\pi\Gamma$ and $x_0^\prime=3\lambda_\text{L}/16$. Other parameters are as in \fref{fig:MMC:Analytic-Spatial}. Cooling is achieved only for a finite range of initial momenta.}
\label{fig:MMC:CutoffExpln}
\end{figure}
As discussed above, the addition of the dipole trap introduces several features into the friction force. Plotting the variation of the friction force in \eref{eq:MMC:AnalyticFrictionTimeAvg} with the initial momentum, $p_0$, of the atom, as in \fref{fig:MMC:CutoffExpln}, shows that the amplitude of the force is not monotonic. In fact, it increases from zero, for increasing $p_0$, achieves a maximum, and then decreases again until it reaches zero at some value for $p_0$. Physically, this is due to the broader spatial distribution for faster particles in the harmonic trap. For fast enough velocities, the particle oscillates into the heating regions, as can be deduced from \fref{fig:MMC:Analytic-Spatial}, even if the trap is centred at the position of maximum cooling. This defines a range of initial momenta, starting from zero, within which a particle is cooled by this mechanism; faster particles are heated and ejected from the trap. Note that this result was derived from the friction to lowest order in velocity $v$, and higher-order terms are expected to affect the capture range for high values of $p_0$.
\par
At particular values of $x_0^\prime$, e.g. at $-3\lambda_\text{L}/16$, this capture range can be conveniently estimated by using the location of the first zero of the Bessel function, giving a momentum capture range
\begin{equation}
\label{eq:MMC:CaptureRangeBesselZero}
p_\text{m}\approx 0.958\sqrt{m k_\text{t}}/k_\text{L} = 0.958 m \omega_\text{t}/k_\text{L}\,.
\end{equation}
Thus, $p_\text{m}^2 \propto \omega_\text{t}^2$, for values of $\omega_\text{t}$ that are not too large, and the capture range as defined in \fref{fig:MMC:CutoffExpln} is expected to scale with the square of the trap frequency. We will compare this later, in \Sref{sec:CoolingMethods:MMC:Numeric}, with the results of Monte--Carlo numerical simulations.

\subsubsection{Diffusion and steady-state temperature}\label{sec:CoolingMethods:MMC:SST}
\begin{figure}[t]
\centering
    \includegraphics[width=1.5\figwidth]{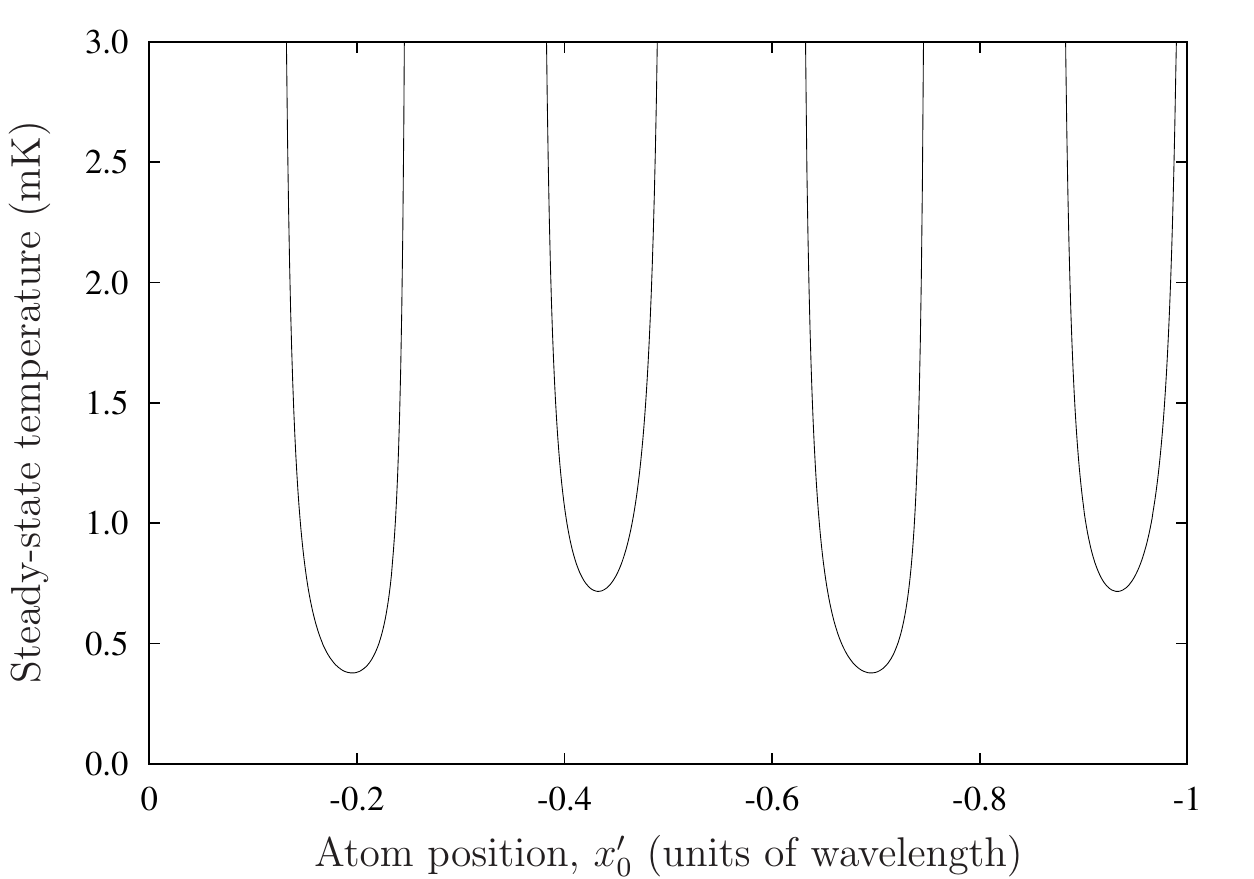}
\caption[Steady-state temperature in mirror-mediated cooling]{Calculated steady-state temperature $T_\text{M}$ for an atom confined in a harmonic trap as a function of position whilst keeping the detuning and pump field constant. $\omega_\text{t}=0.1\times 2\pi\Gamma$; other parameters are as in
\fref{fig:MMC:Analytic-Spatial}.}
\label{fig:MMC:SST_vs_Posn}
\end{figure}
The main result of the preceding discussion was a friction force, which of course cools an atom towards zero momentum in the absence of any other effects. Momentum diffusion due to spontaneous scattering, in a manner similar to that discussed in \sref{ch:CoolingMethods:FDT}, introduces a constant in the equation for $\rmd p^2/\rmd t$ and results in a constant upward shift of the curve in~\fref{fig:MMC:CutoffExpln}. This slightly reduces the capture range for fast particles, but its main effect is to introduce a specific value of the momentum where friction and diffusion exactly compensate each other in the small $p_0$ range. This point corresponds to the steady-state temperature achievable through the cooling mechanism discussed here.
\par
To lowest order in the coupling frequency $g^2$, the momentum diffusion is given by the interaction of the atom with the unperturbed, standing-wave pump field. In this limit, diffusion in our system is therefore identical to that encountered in the usual Doppler cooling mechanism~\cite{Gordon1980,Cook1980,CohenTannoudji1992,BergSorensen1992}, where the diffusion coefficient $\diffn$ is given to lowest order in $s$ by
\begin{equation}
\label{eq:MMC:Diffn}
\diffn=\hbar^2k_\text{L}^2\Gamma s\Big[\cos^2(k_\text{L}x_0)+\tfrac{2}{5}\sin^2(k_\text{L}x_0)\Big]\,.
\end{equation}
The steady-state temperature $T_\text{M}$ of mirror-mediated cooling is then obtained from \eref{eq:FDT} using $\heatingcoefft=-\force_{\parallel}(x_0)/v_\text{m}$, with $\force_{\parallel}(x_0)$ given by \eref{eq:MMC:AnalyticLongFrictionHarm}. For $|\Delta_\text{L}|\gg\Gamma$ we thus find
\begin{equation}
\label{eq:MMC:TempM}
T_\text{M}=\frac{1}{5\pi}\frac{\hbar}{k_B}\frac{\omega_\text{t}\Gamma}{g^2}\frac{2+3\cos^2(k_\text{L}x_0)}{\sin(\omega_\text{t}\tau)\sin(4k_\text{L}x_0)}\,.
\end{equation}
An example of the dependence of $T_\text{M}$ on the trap position is shown in~\fref{fig:MMC:SST_vs_Posn}, predicting a minimum temperature of the order of $400$\,$\upmu$K. Whilst this may seem large in comparison to the Doppler temperature of $141$\,$\upmu$K, one has to keep in mind that $T_\text{M}$, given by \eref{eq:MMC:TempM}, is insensitive to detuning and, for far off-resonant operation of the order of tens of linewidths, it will be the dominant mechanism. This is further discussed in \Sref{sec:CoolingMethods:MMC:Beyond}. We also note that \fref{fig:MMC:SST_vs_Posn} further highlights the importance of the requirement for localising the particle.\\
Using \eref{eq:MMC:AnalyticLongFrictionFamiliar} we can approximate the steady-state temperature at the point of maximum friction by
\begin{equation}
\label{eq:SemiclassicalMMCTemperature}
 T_{\text{M}}\approx\frac{\hbar}{2k_\text{B}\tau}\frac{\sigma_\text{L}}{\sigma_{\text{a}}}\,.
\end{equation}
It is interesting to note that this expression is closely related to the expression for the limiting temperature in Doppler cooling, $k_BT=\hbar\Gamma$, but where the upper state lifetime $1/(2\Gamma)$ is replaced by the atom--mirror delay time, $\tau$, and where a geometrical factor equal to the mode area divided by the atomic radiative cross-section is included.\par
This last point will turn out to be a general trend in the cooling methods we discuss. The steady-state temperature is in general described by a function of the form
\begin{equation}
\label{eq:GeneralTForm}
T=\frac{\hbar}{2k_\text{B}\tilde{\tau}}\phi\,,
\end{equation}
where $\tilde{\tau}$ is the characteristic time---e.g., the memory time---of the system and $\phi$ depends on the geometry of the situation. In Doppler cooling, the characteristic time is the lifetime of the upper state, $\tilde{\tau}=1/(2\Gamma)$, whereas in mirror-mediated cooling it is naturally the delay time $\tau$. In cavity-mediated cooling, where the atom is inside a cavity, the energy loss mechanism is due to the decay of the cavity field, and we thereby have $\tilde{\tau}=1/(2\kappa)$~\cite{Horak1997}, where $\kappa$ is the linewidth of the cavity field.

\subsection{Numerical analysis of mirror-mediated cooling}\label{sec:CoolingMethods:MMC:Numeric}
In this section we now investigate a more accurate numerical model to corroborate the simplified analytical results obtained above.\footnote{The Fokker--Planck equation was obtained, through the use of a suitably extended Wigner transform, by Peter Horak. The basis for the Monte--Carlo code was also written by PH, and then extended by AX. The simulations and data analysis were performed by AX.} In order to render the problem numerically tractable, the continuum of modes is replaced by a discrete set of modes with frequencies $\omega_k$, with $k=1,\dots,N$. The master equation \eref{eq:QME} is then converted by use of the Wigner transform~\cite{Gardiner2004} into a Fokker--Planck equation for the atomic and field variables. Applying a semiclassical approximation and restricting the equation of motion to second-order derivatives, one arrives at an equivalent set of stochastic differential equations for a single atom with momentum $p$ and position $x$ in a discrete multimode field with mode amplitudes $\alpha_k$ \cite{Horak2001},
\begin{subequations}
\label{eq:MMC:SDE}
\begin{align}
 \rmd x=\ &\frac{p}{m}\rmd t\,,\\
 \rmd p=\ &\i\gamma_0\big[\mathcal{E}(x)\tfrac{\rmd}{\rmd x}\mathcal{E}^{\ast}(x)-\mathcal{E}^{\ast}(x)\tfrac{\rmd}{\rmd x}\mathcal{E}(x)\big]\rmd t-U_0\big[\mathcal{E}(x)\tfrac{\rmd}{\rmd x}\mathcal{E}^{\ast}(x)+\mathcal{E}^{\ast}(x)\tfrac{\rmd}{\rmd x}\mathcal{E}(x)\big]\rmd t\nonumber\\
&-k_{\text{t}}(x-x_{\text{t}})\rmd t+\rmd P\,,\text{ and}\\
 \rmd\alpha_k=\ &\i\Delta_k\alpha_k\rmd t-(\i U_0+\gamma_0)\mathcal{E}(x)f_k^{\ast}(x)\rmd t+\rmd A_k\,,
\end{align}
\end{subequations}
where $f_k(x)=\sin(\omega_k x/c)$ are the individual mode functions, $\mathcal{E}(x)=\sum_k \alpha_k f_k(x)$ is the total electric field, $\Delta_k=\omega_k-\omega_\text{L}$ is the detuning of each mode from the pump, $U_0$ is the light shift per photon, and $\gamma_0$ is the photon scattering rate. The terms $\rmd P$ and $\rmd A_k$ are correlated noise terms \cite{Horak2001} responsible for momentum and
field diffusion.
\par
In the following, we set the trap centre to $x_\text{t}=-3\lambda_\text{L}/16$, modulo $\lambda_\text{L}$, which is the point where the analytical solution predicts the maximum of the damping force. We use $N=256$ field modes with a mode spacing of $\Gamma/10$. At the start of every simulation, all field modes are empty with the exception of the pump mode which is initialised at $625$ photons, corresponding to a laser power of around $50$\,pW for our chosen parameters.
\par
The simulations were performed in runs of several thousand trajectories. Each such run was performed at a well-defined initial temperature, with the starting momenta of the particles chosen from a Gaussian distribution, and the starting position being the centre of the trap.

\subsubsection{Friction force and capture range\label{sec:CoolingMethods:MMC:Numeric:Friction}}
\begin{figure}[t]
 \centering
    \subfigure[$\omega_\text{t}=0.3\times 2\pi\Gamma$]{
    \includegraphics[width=1.5\figwidth]{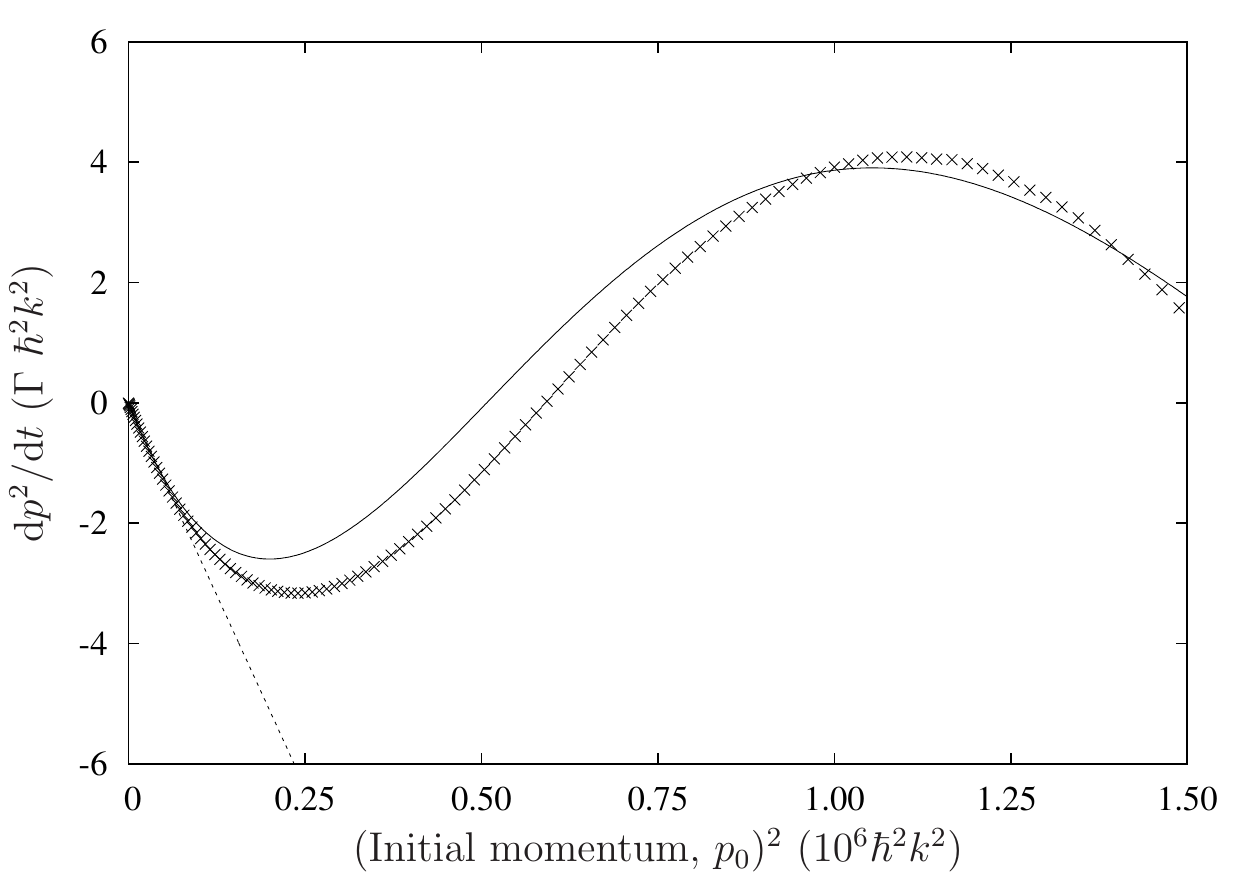}
    }\\
    \subfigure[$\omega_\text{t}=0.5\times 2\pi\Gamma$]{
    \includegraphics[width=1.5\figwidth]{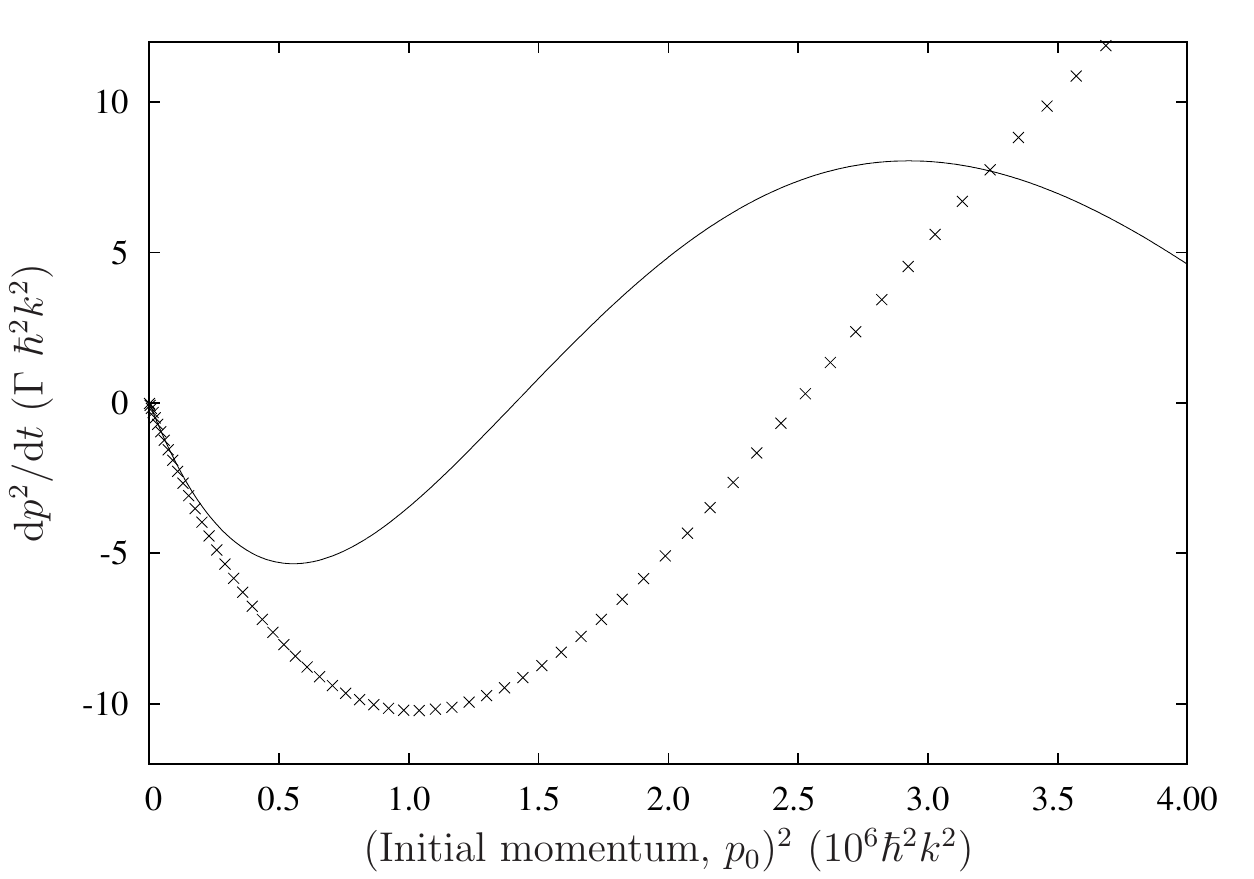}
    }
\caption[Comparison of mirror-mediated cooling simulations with analytical approximation]{Comparison of $\rmd p^2/\rmd t$ for the simulations without noise (data points) with the analytical approximation, \eref{eq:MMC:AnalyticLongFrictionHarmReduced}, including the harmonic trap (solid line). (a)~Weak harmonic
 trap, $\omega_\text{t}=0.3\times 2\pi\Gamma$, showing also the  linear dependence in the limit of small momenta, \eref{eq:MMC:AnalyticLongFrictionHarm} (dotted line). (b)~Stiff trap,  $\omega_\text{t}=0.5\times 2\pi\Gamma$. The trap position $x_0^\prime=-3\lambda/16$ and other parameters are as in \fref{fig:MMC:Analytic-Spatial}.}
 \label{fig:MMC:Comparison}
\end{figure}
\fref{fig:MMC:Comparison} presents the results of a set of simulations performed when setting the noise terms $\rmd P$ and $\rmd A_k$ in equations \eref{eq:MMC:SDE} to zero, \ie, neglecting momentum and photon number diffusion. The simulation data are compared with the result of the perturbative calculations \eref{eq:MMC:AnalyticLongFrictionHarmReduced}. For modest values of $\omega_{\text{t}}$, \fref{fig:MMC:Comparison}(a) justifies the averaging process used to derive \eref{eq:MMC:AnalyticFrictionTimeAvg}, which was based on spatial averaging but neglecting higher order terms in $v$. In contrast, for larger trap frequencies, the numerical simulations diverge significantly from the analytical result, as can be seen in \fref{fig:MMC:Comparison}(b). We expect that the terms in higher powers of the initial speed, which were dropped in the perturbative solution, are responsible for this discrepancy.
\begin{figure}[t]
 \centering
    \includegraphics[width=1.5\figwidth]{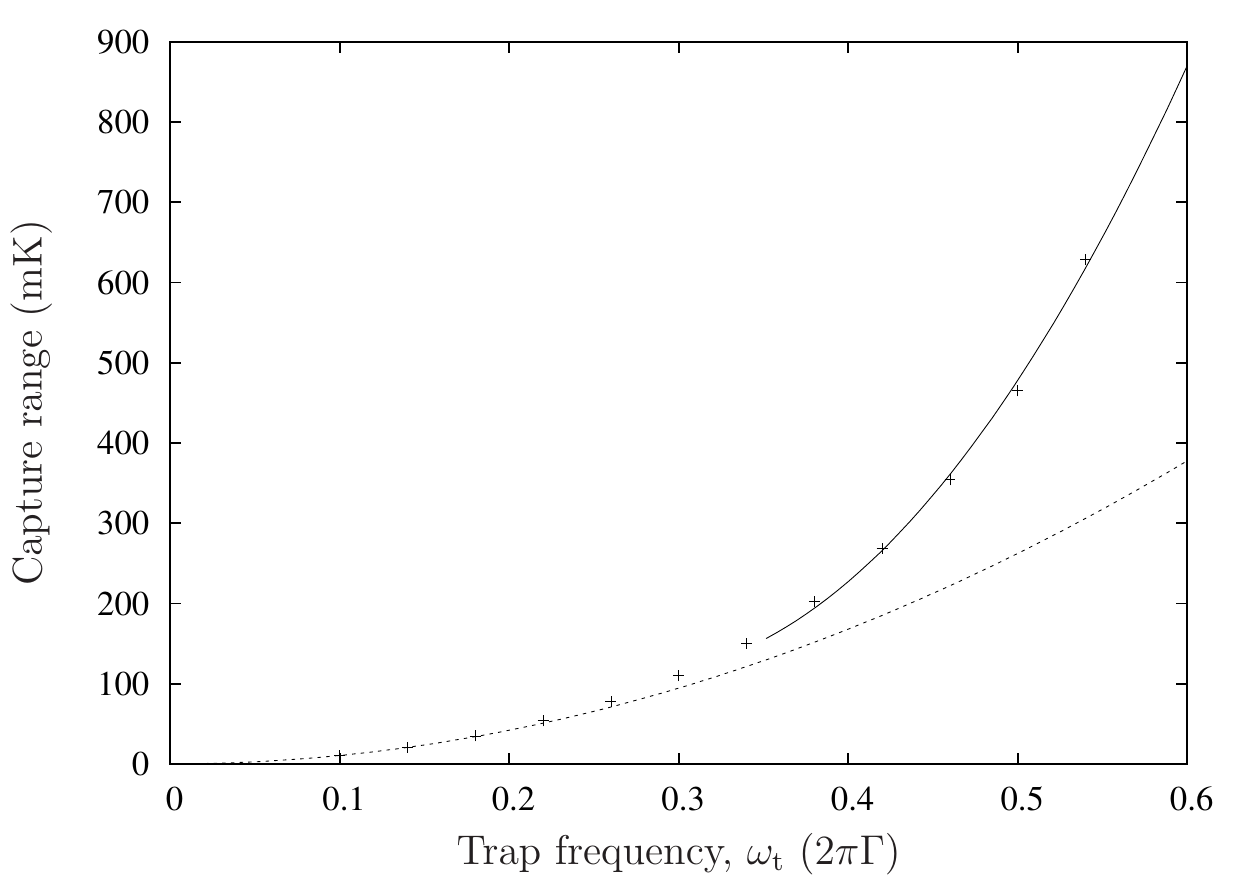}
\caption[Momentum capture range mirror-mediated cooling]{Capture range extracted from the simulations (data points) as compared to the analytical solution (dotted line) for various values of $\omega_\text{t}$. The solid line is a quadratic fit to the data for $\omega_\text{t}\geq 0.3\times 2\pi\Gamma$ and is only intended as a guide to the eye. Other parameters are as in \fref{fig:MMC:Comparison}.}
 \label{fig:MMC:Cutoff}
\end{figure}
\par
We have already seen, in \eref{eq:MMC:CaptureRangeBesselZero}, that the capture range is expected to scale as $\omega_{\text{t}}^2$. For weak traps, as shown in~\fref{fig:MMC:Cutoff}, the numerical simulations agree well with these expectations. For stiffer traps, however, the capture range is consistently larger than that predicted; in fact, the simulations predict a capture range of around $450$\,mK for a trap frequency of $0.5\times 2\pi\Gamma$

\subsubsection{Steady-state temperature\label{sec:CoolingMethods:MMC:Numeric:SST}}
\begin{figure}[t]
 \centering
    \includegraphics[width=1.5\figwidth]{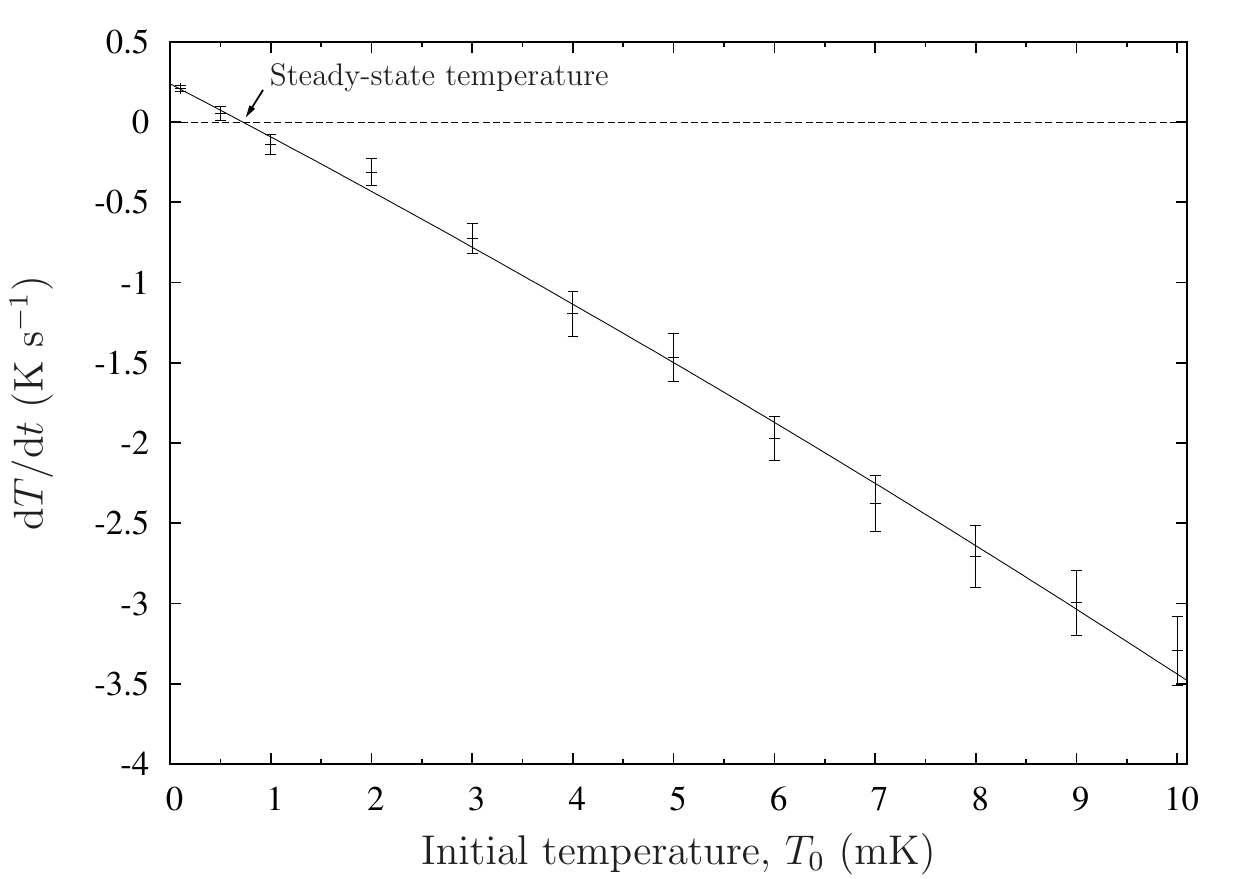}
\caption[Cooling rate for a number of initial temperatures]{Cooling rate ($-\rmd T/\rmd t$) extracted from the simulations starting at a number of initial temperatures. The solid line represents a quadratic fit to the data. $\omega_\text{t}=0.5 \times 2\pi\Gamma$; other parameters are as in \fref{fig:MMC:Comparison}.}
 \label{fig:MMC:Heating-vs-Temp}
\end{figure}
The next step in our investigation was to run simulations involving the full dynamics given by \erefs{eq:MMC:SDE} including the
diffusion terms. Because of the discrete nature of the field modes with uniform frequency spacing used in the simulations, the numerically modelled behaviour is always periodic in time with a periodicity given by the inverse of the frequency spacing. The simulations therefore cannot follow each trajectory to its steady-state unless an unfeasibly large number of modes is used. Instead, simulations were performed in several groups of trajectories, each group forming a thermal ensemble at a well-defined initial temperature. For each such group of trajectories the initial value of $\rmd T/\rmd t$ was calculated. The results for $\omega_\text{t}=0.5\times 2\pi\Gamma$ are shown in \fref{fig:MMC:Heating-vs-Temp}, where the error bars are due to statistical fluctuations for a finite number of stochastic integrations. The steady-state temperature is that temperature at which $\rmd T/\rmd t=0$ as clearly illustrated in this figure. For the chosen parameters, our data suggest a steady-state temperature of $722\pm 54$\,$\upmu$K with a $1/e$ cooling time of around $3.0$\,ms. This compares reasonably well with the steady-state temperature of $597$\,$\upmu$K predicted by \eref{eq:MMC:TempM}.
\begin{figure}[t]
 \centering
    \includegraphics[width=1.5\figwidth]{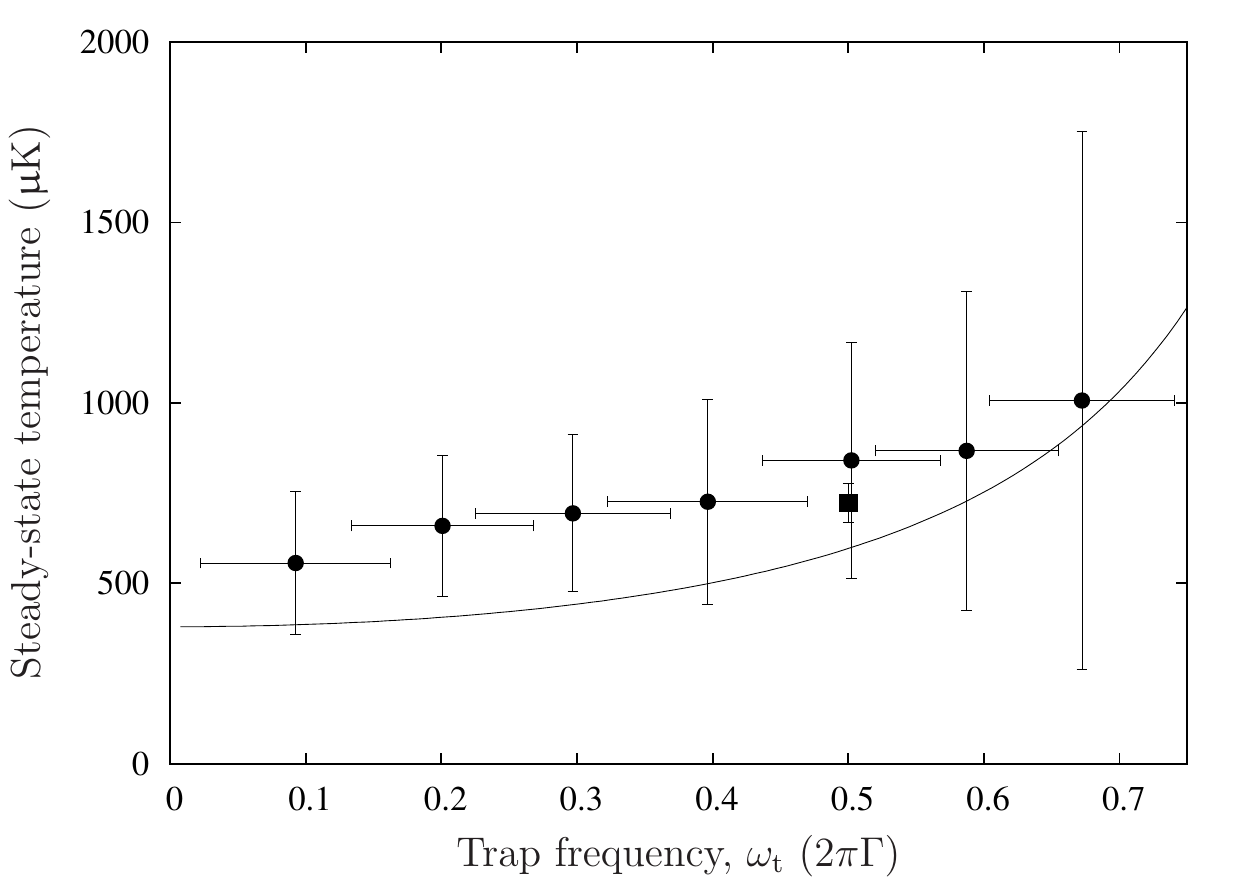}
\caption[Mirror-mediated cooling steady-state temperature: comparison of simulations with analytical approximation]{Steady-state temperature for a number of simulations (circles) compared to the analytical formula \eref{eq:MMC:TempM} (solid line). The solid square represents the equivalent data from~\fref{fig:MMC:Heating-vs-Temp}, resulting from a much larger number of simulations. Parameters are as in \fref{fig:MMC:Comparison}.}
 \label{fig:MMC:SST}
\end{figure}
\par
We finally performed a large number of simulations to investigate the dependence of the steady-state temperature on the trap frequency. \eref{eq:MMC:TempM} indicates that as one decreases $\omega_\text{t}$ the steady state temperature decreases. This is clearly seen in \fref{fig:MMC:SST}, which compares the prediction of \eref{eq:MMC:TempM} with a set of numerical simulations. The trend in the data is reproduced well by the analytical expression. However, the simulated steady-state temperature is consistently a little higher than predicted. We expect that this discrepancy is due mainly to two reasons:~(i)~\eref{eq:MMC:TempM} was derived from the friction \eref{eq:MMC:AnalyticLongFrictionHarm}, \ie, without the spatial averaging of \eref{eq:MMC:AnalyticLongFrictionHarmReduced} which would reduce the friction force; and (ii)~higher order terms in the velocity $v$ are also expected to reduce friction compared to the lowest order analytical result. For both these reasons, therefore, the analytical expression is expected to overestimate the friction force and thus to predict equilibrium temperatures that are too low.

\subsection{Beyond adiabatic theory}\label{sec:CoolingMethods:MMC:Beyond}
\begin{figure}[t]
 \centering
    \includegraphics[width=1.5\figwidth]{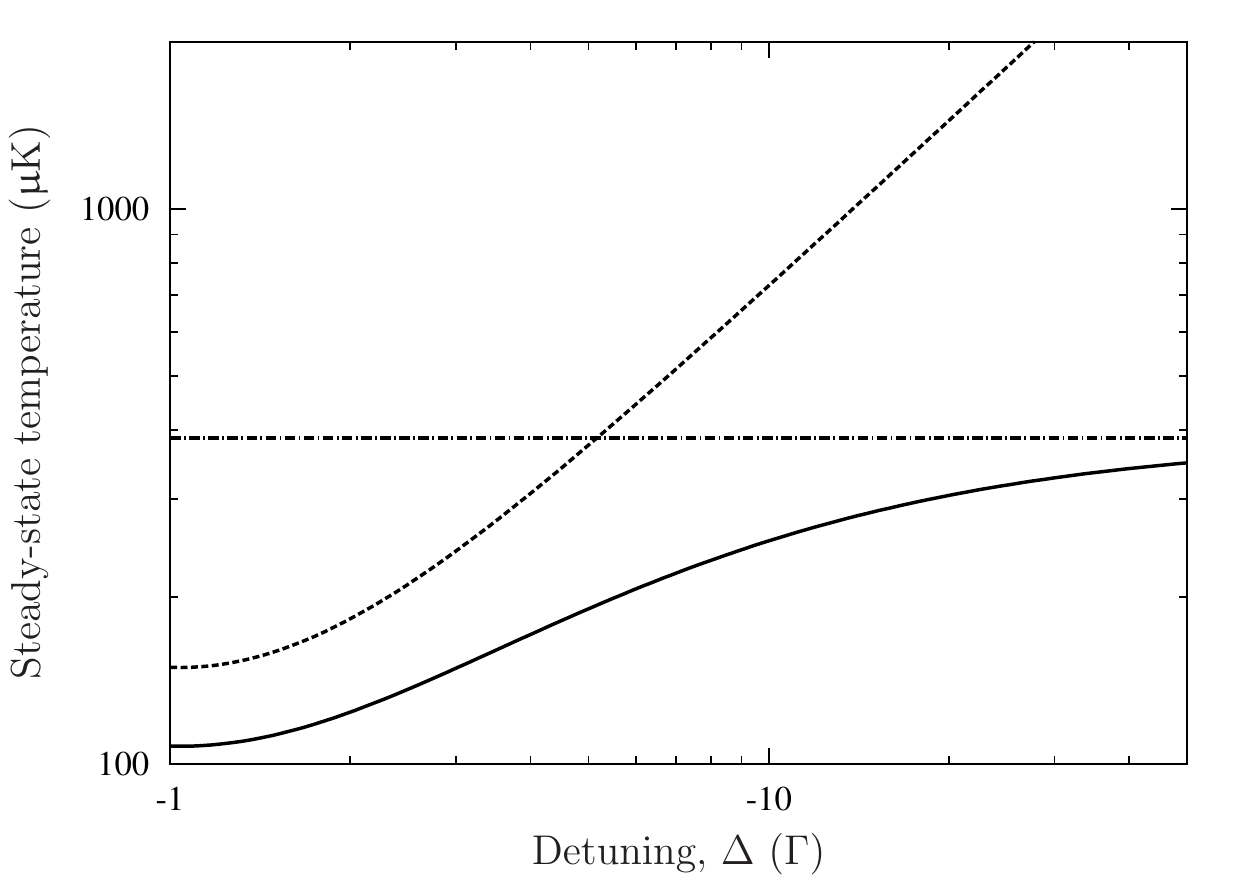}
\caption[Limiting temperatures for Doppler cooling and mirror-mediated cooling]{Comparison between the calculated steady-state temperatures for mirror-mediated cooling $T_\text{M}$ (dash-dotted line), Doppler cooling $T_\text{D}$ (dashed), and in the presence of both effects $T$ (solid), drawn as a function of detuning whilst keeping the saturation parameter constant. $\omega_\text{t}=0.1\times 2\pi\Gamma$; other parameters are as in \fref{fig:MMC:Comparison}.}
 \label{fig:MMC:Doppler_vs_Mirror}
\end{figure}
All the theoretical analysis and simulations discussed so far have been based on adiabatic elimination of the internal atomic degrees of freedom, and therefore neglected Doppler cooling. In~\fref{fig:MMC:Doppler_vs_Mirror}, we explore the variation of
$T_\text{M}$ and the Doppler temperature, $T_\text{D}$, as a function of detuning from resonance when the particle is at the point of greatest friction ($x_0^\prime=-3\lambda/16$), where $T_\text{D}$ is given by
\begin{equation}
T_\text{D} = \hbar\Gamma\,\frac{\Delta_\text{L}^2+\Gamma^2}{2\bigl(-\Delta_\text{L}\bigr)}\,,
\end{equation}
for $\Delta_\text{L}<0$. In the presence of both cooling effects, and assuming that the momentum diffusion terms are identical for both mechanisms, the stationary temperature achieved by the system is given by
\begin{equation}
 T = \biggl(\frac{1}{T_\text{M}}+\frac{1}{T_\text{D}}\biggr)^{-1}\,.
 \label{eq:MMC:fulltemp}
\end{equation}
Thus, for the parameters of \fref{fig:MMC:SST}, the calculated steady-state temperature $T$ reduces to $250$\,$\upmu$K in the limit of vanishing $\omega_\text{t}$.
\par
From \fref{fig:MMC:Doppler_vs_Mirror} one can see that the mirror-mediated force, for our tightly focussed pump, is stronger than the Doppler force for detunings larger than around $10\Gamma$ in magnitude. In practice this has two implications: for large negative detunings, we expect the steady-state temperature of the system to be significantly lower than that predicted by Doppler cooling; whereas for large \emph{positive} detunings, we still predict equilibrium temperatures of the order of mK.
\par
Both our perturbative expressions and our simulations are calculated to lowest orders in the atomic saturation. However, it is well known that in the limit of very large detunings also higher order terms in the saturation parameter $s$ become significant. Using the full expression for the diffusion constant~\cite{Gordon1980}, we can estimate the detuning for which we expect minimum diffusion and temperature. For the value of the saturation parameter $s\lesssim 0.1$ used throughout this section, it can be shown that $T_\text{M}$ attains a minimum at detunings of up to several tens of linewidths. Our chosen parameters are therefore within the range of validity of the model.

\subsection{Concluding remarks}\label{sec:CoolingMethods:MMC:Conclusions}
In conclusion, we have presented a mechanism for cooling particles by optical means which is based fundamentally on the dipole interaction of a particle with a light beam and therefore does not rely on spontaneous emission. The particle is assumed to be trapped and is simultaneously driven by an off-resonant laser beam. After the interaction with the particle the beam is reflected back onto the particle by a distant mirror. The time-delay incurred during the light round-trip to the mirror and back is exploited to create a non-conservative cooling force.
\par
The system was analysed using stochastic simulations of the semiclassical equations of motion representing a single two-level atom coupled to a continuum of electromagnetic modes. The results of these computations were found to agree with the expectations of a perturbative analysis. Our models predict sub-mK steady-state temperatures for \textsuperscript{85}Rb atoms interacting with a tightly focussed laser beam several metres from the mirror, in an arrangement similar to that of Ref.~\cite{Eschner2001}. While most of the theory is presented for a one-dimensional model, results for the friction force in the transverse direction suggest that three-dimensional cooling is possible with this scheme.
\par
The model presented here requires a large separation between the atom and the mirror, of the order of several metres, for an observable cooling effect. This limitation can be overcome in several ways. First, the light could propagate in an optical fibre between the atom and the mirror to avoid the effects of diffraction. Second, the required delayed reflection could be achieved through the use of a cavity instead of a mirror; in contrast to cavity-mediated cooling schemes~\cite{Horak1997,Vuletic2000,Maunz2004,Vilensky2007,Lev2008}, the atom would remain external to the cavity. For a time delay $\tau$ of order $1$\,ns one would require a cavity quality factor $Q=\omega\tau$~\cite{Rempe1992} of the order of $10^6-10^7$, which is achievable with present-day technology~\cite{Mabuchi1994}. This mechanism is explored heuristically in \sref{sec:CoolingMethods:Other:ECCO}, and subsequently investigated in greater depth in \sref{sec:TMM:ECCO}, after we have developed the necessary mathematical tools.

\subappendicesstart

\subsection{Appendix: A note on units}
The units used in the preceding work can perhaps best be called `quantum mechanical'. Here, we provide a number of useful conversions and numerical values, which could be of benefit to readers with a more experimental leaning.
\par
\begin{center}
\begin{tabular}{c | c}
\hline
\hline
Quantity in this work & Experimental value\\
\hline
$A$ & $2\sqrt{2\pi P/\bigl(\hbar k_\text{L}\bigr)}$\\
$g$ & $\sqrt{\Gamma\sigma_\text{a}/\bigl(\pi\sigma_\text{L}\bigr)}$\\
$s$ & $g^2|A|^2/\Delta_\text{L}^2=8\Gamma P\sigma_\text{a}/\bigl(\hbar k_\text{L}\sigma_\text{L}\Delta_\text{L}^2\bigr)$\\
\hline
\end{tabular}
\end{center}
\par
In the above, $P$ is the incident travelling-wave electromagnetic power, related to the incident electric field $\efield$ by
\begin{equation}
P=\tfrac{1}{2}\epsilon_0c\sigma_\text{L}\lvert\efield\rvert^2\,.
\end{equation}
$\Gamma$ is the HWHM linewidth of the upper state. For the D$_2$ line of some common alkali species we have:
\begin{center}
\begin{tabular}{c | c | c | c}
\hline
\hline
Quantity & $^{23}$Na~\cite{Steck2010b} & $^{85}$Rb~\cite{Steck2008}, $^{87}$Rb~\cite{Steck2010a} & $^{133}$Cs~\cite{Steck2010c}\\
\hline
$\Gamma$ & $2\pi\times 4.897$\,MHz & $2\pi\times 3.033$\,MHz & $2\pi\times 2.617$\,MHz\\
$\omega_\text{a}$ & $2\pi\times 508.848$\,THz & $2\pi\times 384.230$\,THz & $2\pi\times 351.726$\,THz\\
$\lambda_\text{a}=2\pi c/\omega_\text{a}$ & $589.158$\,nm & $780.241$\,nm & $852.347$\,nm\\
$k_\text{a}=2\pi/\lambda_\text{a}$ & $10.665\times 10^6$\,m$^{-1}$ & $8.055\times 10^6$\,m$^{-1}$ & $7.372\times 10^6$\,m$^{-1}$\\
$\sigma_\text{a}=3\lambda_\text{a}^2/\bigl(2\pi\bigr)$ & $1.657\times 10^{-13}$\,m$^2$ & $2.905\times 10^{-13}$\,m$^2$ & $3.469\times 10^{-13}$\,m$^2$\\
$T_\text{D}$ & $235.03$\,$\upmu$K & $145.57$\,$\upmu$K & $125.61$\,$\upmu$K\\
\hline
\end{tabular}
\end{center}

\subappendicesend

\section{Exploiting an optical memory in other geometries}\label{sec:CoolingMethods:Other}
\Sref{sec:CoolingMethods:MMC} provided us with a sound theoretical basis, in the case of the simplest possible geometry that permits a memory, for the arguments we put forward in \Sref{sec:CoolingMethods:Retarded}. In the present section, we will explore other geometries with which we can investigate the retarded dipole--dipole interaction. The ideas developed here will be fleshed out in later chapters, after we develop the necessary formalisms, but it is fairly instructive at this early stage to intuitively explore how the different mechanisms arise from very similar physical arguments.
\par
Let us first start by addressing the two major issues with the mirror-mediated cooling mechanism as described in the previous section. To recapitulate, the cooling force in this mechanism arises from the retarded dipole--dipole interaction of a particle with its own reflection in a mirror but (i)~requires a distance of `several metres' between the particle and the mirror for a sizeable effect, and (ii)~the friction force oscillates between cooling and heating on a sub-wavelength scale and has a zero spatial average.\\
Following the discussion of these two points, we will turn our attention to three-dimensional geometries to see how we can exploit the focussing properties of optics to achieve cooling.

\subsection{Lengthening the time delay: External cavity cooling}\label{sec:CoolingMethods:Other:ECCO}
From the general arguments in \Sref{sec:CoolingMethods:Retarded}, as well as the expressions in \Sref{sec:CoolingMethods:MMC}, we can see that the time delay is what governs the overall strength of the friction force in the retarded dipole--dipole interaction. This can be exploited in what we term `external cavity cooling' (\sref{sec:TMM:ECCO}), where the particle to be cooled interacts with a cavity field despite being outside the cavity. The nature of the interaction here is practically identical to mirror-mediated cooling; in particular, it suffers from the same drawback of having the friction force oscillate between a cooling and a heating force several times over the space of a single wavelength. Nevertheless, this is not a problem if what is to be cooled is not the motional energy of an atom but that of a micromirror since~(i)~a micromirror can be positioned, using piezoelectric actuators, with sub-nanometre-scale resolution~\cite{AttocubeSystemsPositioning2010}, and (ii)~the oscillation amplitude for the Brownian motion of a micromirror is also in the picometre to nanometre range. To illustrate the latter point, let us use the effective spring constant of a commercially-available micromirror as calculated in Ref.~\cite{Thompson2008}, $k_\text{M}=28$\,N\,m\textsuperscript{-1}. Then, the RMS displacement for the micromirror at a temperature $T=300$\,K, calculated as
\begin{equation}
\sqrt{\expt{x^2}}=\sqrt{\frac{k_\text{B}T}{k_\text{M}}}\,,
\end{equation}
is of the order of $10^{-11}$\,m.
\begin{figure}[t]
 \centering
    \includegraphics[scale=0.5]{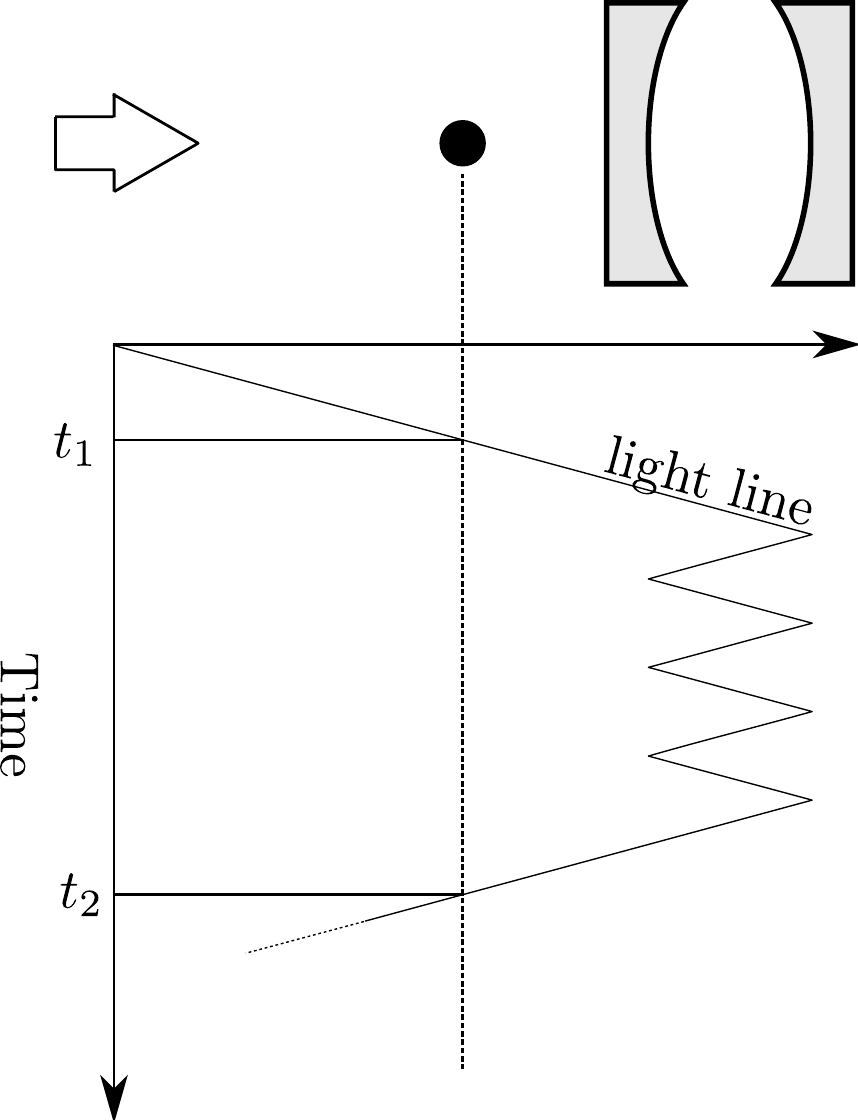}
\caption[Space--time diagram explaining the distance folding mechanism]{Space--time diagram showing the distance folding argument used to explain the action of external cavity cooling. This figure should be compared to \fref{fig:SpaceTime}. The blank arrow indicates the pump beam.}
 \label{fig:SpaceTimeECCO}
\end{figure}
\par
Having established the rationale, then, we can introduce `external cavity cooling' by means of the distance folding argument (see \fref{fig:SpaceTimeECCO}). For suppose we place a particle (atom or micromirror) in front of a cavity. The light that interacts with the particle is then allowed to couple into the cavity. This light then undergoes several round trips inside the cavity, and with each round-trip some light leaks out and re-interacts with the particle. An effective delay time can be defined that is related to the finesse (or, equivalently, the linewidth) of the cavity, and it is this concept that we show in \fref{fig:SpaceTimeECCO}. One notes that this delay time is, for a good cavity, orders of magnitude larger than that due to either mirror separately. Using a good cavity, with a finesse of the order of $10^5$, thereby allows us to squeeze the `several metres' into a sub-millimetre-scale device.\\
This opens the door to several important advances, of which we mention a few here. First of all, one is not constrained to use Fabry--P\'erot-type cavities, whereby the system can be constructed monolithically on a chip-scale device. Secondly, the cavity does not need to have good optical or mechanical access, which allows one to make significantly better and more stable cavities. Finally, having the micromirror or atom outside the cavity means that there is less chance of burning or saturation effects, since the local field surrounding the particle is not amplified by the cavity.

\subsection{Lifting the sub-wavelength dependence: Ring cavity cooling}
\begin{figure}[t]
 \centering
    \subfigure[Mirror-Mediated Cooling]{
    \includegraphics[scale=0.5]{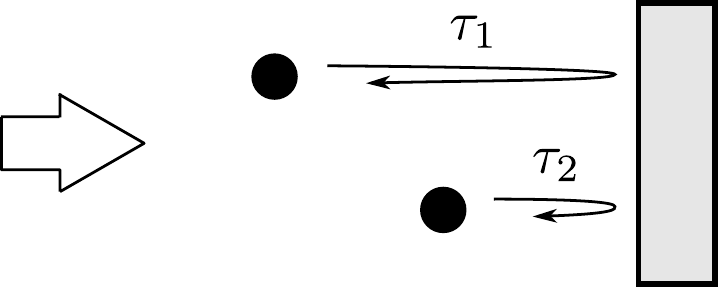}
    }\\[1cm]
    \subfigure[Ring Cavity Cooling, position 1]{
    \includegraphics[scale=0.5]{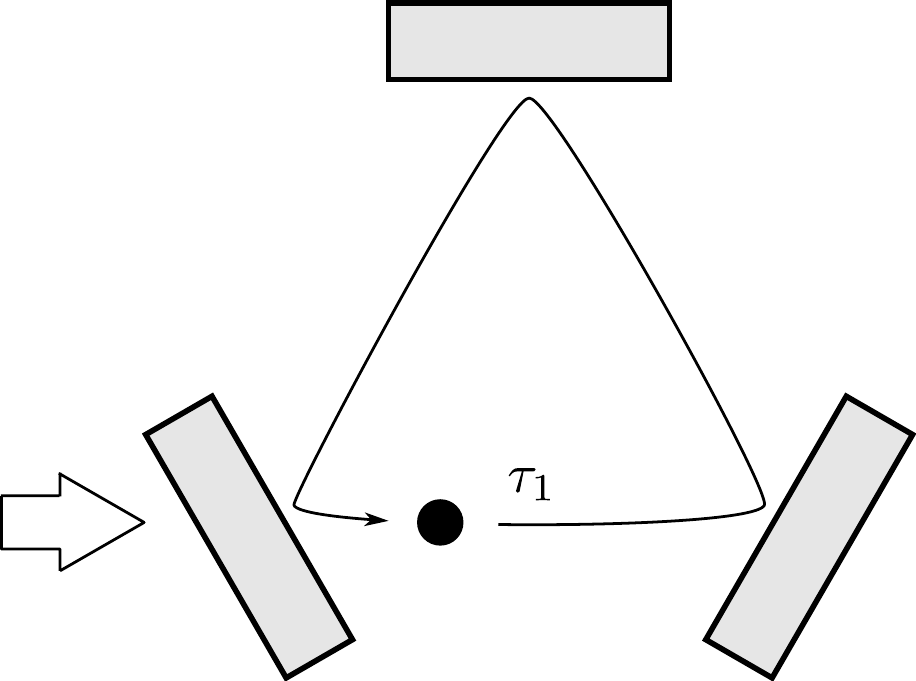}
    }\hspace{1cm}
    \subfigure[Ring Cavity Cooling, position 2]{
    \includegraphics[scale=0.5]{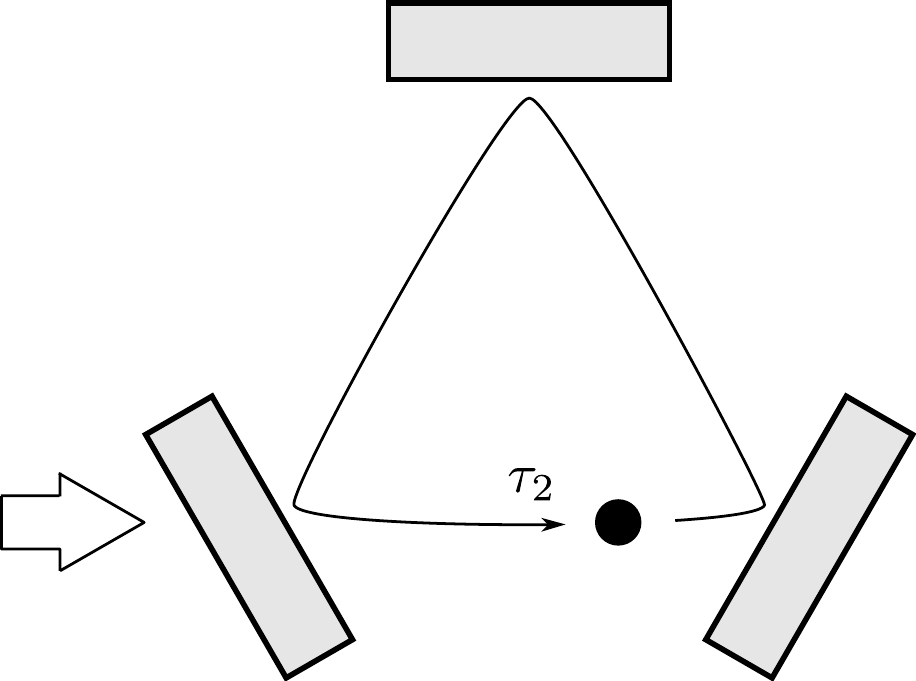}
    }
\caption[Delay times for atom different positions in mirror-mediated cooling and ring cavity cooling]{In (a)~mirror-mediated cooling, the time delay depends on the atomic position ($\tau_1\neq\tau_2$), whereas in (b,~c)~ring cavity cooling, it is independent of the atomic position ($\tau_1=\tau_2$). The blank arrow indicates the pump beam.}
 \label{fig:Invariance}
\end{figure}
A cursory physical analysis of the mirror-mediated cooling mechanism will reveal that the origin of the $\sin(4k_\text{L}x_0)$ dependence lies in the fact that the atom is in a standing wave inside a system that is \emph{not} translationally invariant. To see this latter point, one simply needs to observe that, with the atom at two different places the time delay is of course different. One way of restoring translational invariance into the system is by using a ring cavity, rather than a plane mirror, to introduce a delay; see \fref{fig:Invariance}. The expectation of the physical mechanism being preserved, albeit without the position dependence in the friction force, is borne out when the system is examined in detail, as we shall see in \Sref{sec:TMM:AmplifiedOptomechanics} after we have developed the necessary mathematical model.

\subsection{Exploiting three-dimensional electromagnetism}
\begin{figure}[t]
 \centering
    \subfigure[Optical binding]{
    \includegraphics[scale=0.5]{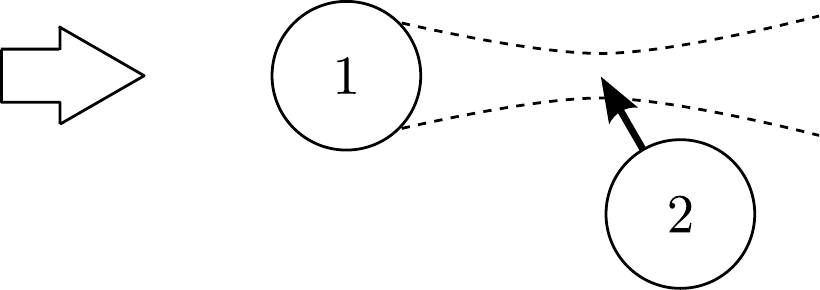}
    }\\[1cm]
    \subfigure[Self--tweezing, static]{
    \includegraphics[scale=0.5]{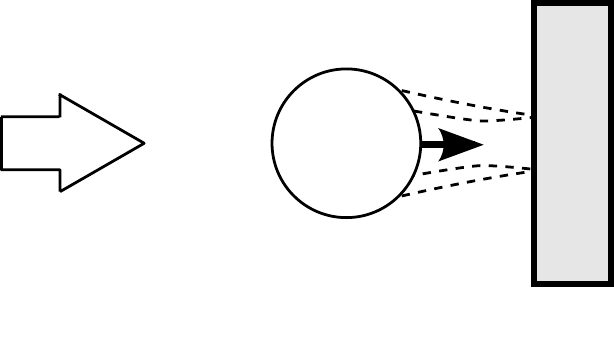}
    }\hspace{1cm}
    \subfigure[Self--tweezing, moving]{
    \includegraphics[scale=0.5]{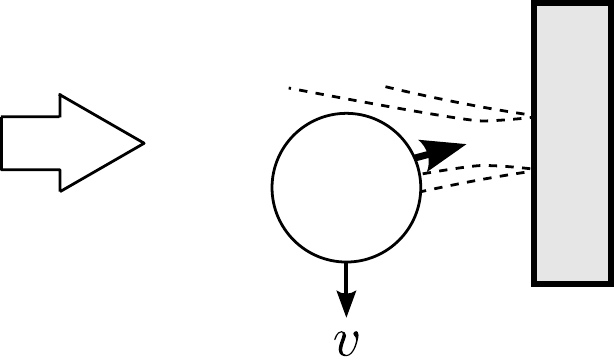}
    }
\caption[Three-dimensional retarded self-binding]{In (a)~optical binding, particle 2 is in a potential well caused by particle 1. Self--tweezing, (b), occurs when a particle is in a potential well caused by itself; if the particle is moving, (c), it will experience a force opposing its motion. The blank arrow indicates the pump beam and the thick black arrows the force.}
 \label{fig:Tweezing}
\end{figure}
Perhaps the two biggest conceptual differences between electromagnetism in one dimension and that in three dimensions are orthogonal polarisations (which can nevertheless be mathematically mimicked in a quasi-one-dimensional geometry) and the spreading out of waves. Indeed, a travelling wave in one dimension can be represented as
\begin{equation}
\cos[\omega(t-x/c)]\,,
\end{equation}
whereby the amplitude of the wave is $1$, for any value of $x$. In three dimensions, however~\cite{Jackson1998} [see, for example, Eq.~(9.19) in this reference], the amplitude of the electric field is dependent on the distance $r$ from the source, and scales as $1/r$ for $r\gg\lambda$. Thus, suppose a particle has a dipole induced by a local electric field $E$. This dipole will subsequently emit a spherical wave, and let us suppose that we reflect this wave back onto the dipole. The potential energy mediated by the interaction between the instantaneous dipole polarisation and the reflected field (which has travelled through a distance $r$) is, up to some phase and constants,
\begin{equation}
U\sim-\frac{E^2}{r}\,.
\end{equation}
We can now apply the result in \eref{eq:GeneralTimeDelayedF}:~the dipole will experience a friction force proportional to $\tau c/r^3=1/r^2$.
\par
The spreading out of waves in three dimensions is but one instance of the more general concept of (de)focussing of an electric field. Let us examine a simple way in which we can exploit this idea, \emph{self-tweezing}. Optical binding~\cite{Karasek2006}, \fref{fig:Tweezing}(a), is an interesting phenomenon whereby light incident on a dielectric sphere is focussed by the sphere itself. This focus, in turn, produces a potential well for a second particle, which is thereby optically bound to the first. Suppose, then that we only have one particle, and that we put a plane mirror `downstream' of that particle, \fref{fig:Tweezing}(b). The mirror will, if the geometry is chosen properly, cause the focus to appear close enough to the particle that it will feel the potential well caused by this electric field. This situation, then is similar to the case with two particles, but the single particle is now optically bound to \emph{itself}. Any motion of the particle around this position will cause its image to lag behind it and a restoring, viscous, force to act on it, \fref{fig:Tweezing}(c).

%% file: Chapters/Chapter2.tex
\newcommand{\mat}[1]{#1}
\def \communit {\frac{\hbar k_0}{2\epsilon_0 \sigma_\text{L}}\delta(t-t^\prime)}
\newcommand{\lre}[1]{\,\real\left\{#1\right.}

\newpartalt{Scattering Models \& Their Applications}{Scattering Models\\\&\\Their Applications}\label{part:TMM}

\chapter{The transfer matrix model}\label{ch:TMM:TMM}
\epigraph{The reader might wonder why it is of interest, physically, to consider $n$-manifolds for which $n$ is larger than $4$, since ordinary spacetime has just four dimensions. In fact many modern theories [...]\ operate within a `spacetime' whose dimension is much larger than $4$.}{R.\ Penrose, \emph{The Road to Reality} (2004)}
\cref{ch:CoolingMethods:AFInt}, and in particular \sref{sec:CoolingMethods:PolTLA}, developed the necessary tools to describe the interaction of an atom with the electromagnetic field, as parametrised by the atom's characteristic polarisability. In one dimension, one can succinctly describe the fields interacting with a linear scatterer through what is called the transfer matrix approach~\cite{Deutsch1995}. Restricting ourselves to one spatial dimension is not an overly restrictive approximation, despite the quote at the beginning of this chapter; the formalism that is discussed in this chapter allows us to describe a wealth of physical situations. The purpose of this chapter is to extend this model significantly, enabling it to account for moving as well as static scatterers; this is done in \sref{sec:TMM:Model} and the model that results is solved generally in \sref{sec:TMM:General}. The extended model discussed here takes into account the first-order Doppler shift but not relativistic effects, and it is therefore correct only up to first order in the velocity of the scatterer.\par
The transfer matrix method is more general than an analysis based on modal decomposition, and is therefore used to describe the optomechanics of scatterers inside cavities in \sref{sec:TMM:Optomechanics}. This method is extended even further in \sref{sec:TMM:Multilevel}, where it is shown that the concept of polarisability can also be applied to atoms having a Zeeman manifold and interacting with circularly polarised light. The results of these sections are confirmed by showing that the standard results for optical molasses, mirror-mediated cooling and cooling of cavity mirrors, as well as sub-Doppler cooling mechanisms~\cite{Dalibard1989}, can be reproduced by simple applications of the transfer matrix method. In \cref{ch:TMMApplications}, the theory developed over the present chapter will then be applied to describe two novel cooling schemes, outside cavities (\sref{sec:TMM:ECCO}) and inside ring cavities (\sref{sec:TMM:AmplifiedOptomechanics}).

\section{An extended scattering theory}\label{sec:TMM:Model}
In this first section we develop and present a scattering theory for optomechanically coupled systems, allowing for the efficient description of the motion of arbitrary combinations of atoms and mirrors interacting through the radiation field. We will restrict the model to one-dimensional motion and small velocities. The main building block is the beamsplitter transfer matrix~\cite{Deutsch1995, Asboth2008}, \ie, the \emph{local relation} between light field amplitudes at the two sides of a scatterer. We will calculate the radiation force acting on a moving scatterer up to linear order in the velocity. The model is completed by including the quantum fluctuations of the radiation force which stem from the quantised nature of the field. We will determine the momentum diffusion coefficient corresponding to the minimum quantum noise level.\\
One system we will consider in some detail is composed of two mirrors; one of them is fixed in space, whilst the other one is mobile. This is the generic scheme for radiation pressure cooling of moving mirrors~\cite{WilsonRae2007, Marquardt2007, Genes2008}. At the same time, in the limit of low reflection the moving mirror can equally well represent the a single atomic dipole interacting with its mirror image in front of a highly reflecting surface (\sref{sec:CoolingMethods:MMC}; see also Refs.~\cite{Eschner2001, Bushev2004, Xuereb2009a}).
\par
The work in this chapter is published as Xuereb, A., Domokos, P., Asb\'oth, J., Horak, P., \& Freegarde, T. Phys.\ Rev.\ A \textbf{79}, 053810 (2009) and Xuereb, A., Freegarde, T., Horak, P., \& Domokos, P., Phys.\ Rev.\ Lett.\ \textbf{105}, 013602 (2010) and is in part reproduced \emph{verbatim}. After our initial exploration of the transfer matrix description of moving scatterers, we subsequently apply this description to several systems, and also extend it in several ways. The resulting model is very general and can be solved to give analytical formulations of the friction forces and momentum diffusion processes acting in a generic optomechanical system.
\subsection{Basic building blocks of the model}\label{sec:TMM:BasicBuildingBlocksModel}
\begin{figure}[t]
 \centering
 \includegraphics[scale=0.75]{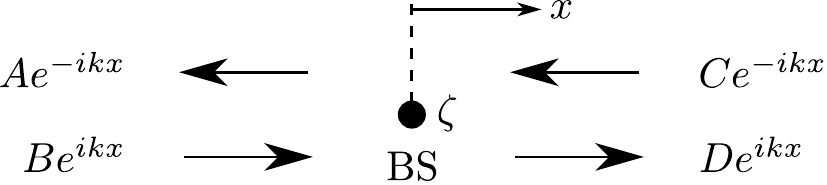}
 \caption[The four different modes that interact through a point-like beamsplitter in one dimension]{The four different modes that interact through a point-like beamsplitter in one dimension.}
 \label{fig:TMM:BS}
\end{figure}

Consider a point-like scatterer (or beamsplitter), $\text{BS}$, moving along the `$x$' axis on the trajectory $x_{\text{BS}}(t)$. Outside the scatterer, the electric field $\efield$ can be expressed in terms of a discrete sum\footnote{This is a simplifying assumption and all our results also hold for a continuum of field modes.} of left- and right-propagating plane wave modes with different wave numbers, $k$, and hence different frequencies, $\omega=kc$:
\begin{equation}
 \label{eq:TMM:Efield}
 \efield=\begin{cases}
    \sum_k\big[A(k)e^{-\i kx-\i\omega t}+B(k)e^{\i kx-\i\omega t}\big]+\rm{c.c.}&x<x_\text{BS}\\
    \sum_k\big[C(k)e^{-\i kx-\i\omega t}+D(k)e^{\i kx-\i\omega t}\big]+\rm{c.c.}&x>x_\text{BS}\,,
   \end{cases}
\end{equation}
where $A(k)$ and $B(k)$ are the mode amplitudes on the left side, $x<x_{\text{BS}}(t)$, while $C(k)$ and $D(k)$ are the amplitudes on the right side, $x>x_{\text{BS}}(t)$, of $\text{BS}$, and where $\text{c.c.}$ denotes the complex conjugate. In accordance, the magnetic field is~\cite{Jackson1998}
\begin{equation}
 \label{eq:TMM:Bfield}
 c \bfield=\begin{cases}
    \sum_k\big[-A(k)e^{-\i kx-\i\omega t}+B(k)e^{\i kx-\i\omega t}\big]+\rm{c.c.}&x<x_\text{BS}\\
    \sum_k\big[-C(k)e^{-\i kx-\i\omega t}+D(k)e^{\i kx-\i\omega t}\big]+\rm{c.c.}&x>x_\text{BS}\,.
   \end{cases}
\end{equation}
As depicted schematically in \fref{fig:TMM:BS}, the scatterer mixes these waves. Our first goal is the derivation of the transfer matrix $M$ connecting the field amplitudes on the right to those on the left side of a beamsplitter moving at a fixed velocity $v$. This relation is well-known~\cite{Deutsch1995} for an immobile scatterer. Therefore, let us first transform the electromagnetic field into a frame moving with the instantaneous velocity $v$ of the $\text{BS}$.

\subsubsection{Transfer matrix for an immobile beamsplitter}\label{sec:TMM:FixedBS}
In the frame co-moving with the $\text{BS}$, the interaction of the field with the scatterer at $\xpr=0$ can be characterised by the single parameter $\zeta$ by means of the one-dimensional wave equation~\cite{Jackson1998, Deutsch1995},
\begin{equation}
\label{eq:TMM:WaveEqn}
\left(\partial_{\xpr}^2-\frac{1}{c^2}\partial_{\tpr}^2\right) \efpr(\xpr,\tpr) = \frac{2}{kc^2}\zeta\,\delta(\xpr)\,\partial^2_{\tpr} \efpr(\xpr,\tpr)\,.
\end{equation}
The electric field can be considered in a modal decomposition similar to~\eref{eq:TMM:Efield}. Since a fixed beamsplitter couples only the plane waves with identical frequency and wave number, the stationary scattering can be fully described within the closed set of modes
\begin{equation}
\efpr(\xpr,\tpr) =\begin{cases} \bigl(A' e^{-\i k\xpr -\i\omega \tpr} + B' e^{\i k\xpr -\i \omega \tpr}\bigr)+\rm{c.c.} &\xpr < 0\\
\bigl(C' e^{-\i k\xpr -\i\omega \tpr} + D' e^{\i k\xpr -\i \omega \tpr}\bigr)+\rm{c.c.} &\xpr > 0\,,
\end{cases}
\end{equation}
where the index $k$ has been dropped. A  linear relation between the field amplitudes on the right of the scatterer and those on the left can be derived from the wave equation~\cite{Deutsch1995}, 
\begin{equation}
\label{eq:TMM:TM_fix}
\begin{pmatrix}
 C^\prime\\
 D^\prime
\end{pmatrix}=M\begin{pmatrix}
 A^\prime\\
 B^\prime
\end{pmatrix}\,\text{, with}
\end{equation}
\begin{equation}
\label{eq:TMM:M0}
M = \begin{bmatrix}
 1-\i\zeta & -\i\zeta\\
 \i\zeta & 1+\i\zeta
\end{bmatrix}
= \frac{1}{\trans}
\begin{bmatrix}
 1 & -\refl   \\
\refl   & \trans^2 - \refl^2
\end{bmatrix}\,.
\end{equation}
In the second form of the transfer matrix $M$, we expressed it in terms of the reflectivity $\refl$ and transmissivity $\trans$ of the beamsplitter. This latter form is more convenient to describe moving mirrors, while for atoms the scattering strength parameter $\zeta$ can be readily expressed in terms of its polarisability $\alpha$ [see \sref{sec:CoolingMethods:PolTLA}, and especially \eref{eq:ZetaDefn}]. In this case the transfer matrix depends on the wave number $k$, which might lead to significant effects, e.g., Doppler cooling, close to resonance with the atom (see \Sref{sec:Force}). 

\subsubsection{Transfer matrix for a moving beamsplitter}\label{sec:TMM:MovBS}
The transformation back into the laboratory-fixed frame involves the change of the coordinates, $\xpr=x-vt$ and $\tpr=t$, and the Lorentz-boost of the electric field up to linear order in $v/c$ \cite[\textsection 11.10]{Jackson1998}:
\begin{equation}
\efield = \efpr + {v} \bfpr\,,
\end{equation}
where we assumed that $\efield$ and $\efpr$ are polarised in the `$y$' direction, $\bfield$ and $\bfpr$ are polarised in the `$z$' direction, and the velocity is along the $x$ axis. The electric field in the laboratory frame becomes
\begin{align}
\efield(x,t) &= \sum_{k^\prime} \Biggl\{A'(k^\prime) e^{-\i k^\prime (x- vt) -\i \omega' t} + B'(k^\prime) e^{\i k^\prime(x-vt) -\i \omega' t}\nonumber\\
&\phantom{=\ \sum_{k^\prime} \Biggl\{}- \frac{v}{c}\Bigl[ A'(k^\prime) e^{-\i k^\prime (x- vt) -\i \omega' t} - B'(k^\prime) e^{\i k^\prime(x-vt) -\i \omega' t}\Bigr]\Biggr\} + \text{c.c.}\nonumber\\
&=\sum_k  \Bigl[\bigl(1-\tfrac{v}{c}\bigr) A'\left(k+kv/c\right) e^{-\i k (1+v/c) x -\i \omega t}\nonumber\\
&\phantom{=\ \sum_k  \bigl(}+ \left(1+\tfrac{v}{c}\right) B'\left(k-kv/c\right) e^{\i k (1-v/c) x -\i \omega t}\Bigr]+\rm{c.c.}\,,
\end{align}
which can be expressed as a linear transformation $\hat L(v)$  of the amplitudes,
\begin{equation}
\begin{pmatrix}
 A(k)\\
 B(k)
\end{pmatrix}=\hat L(-v)\begin{pmatrix}
 A^\prime(k)\\
 B^\prime(k)
\end{pmatrix}\,\text{, with}
\end{equation}
\begin{equation}
\label{eq:TMM:Lmatrix}
\hat L(v) = \begin{bmatrix}
 \left(1+ \frac{v}{c}\right) \hat{P}_{-v} & 0\\
0 & \left(1- \frac{v}{c}\right) \hat{P}_v
\end{bmatrix}\,.
\end{equation}
This construction is explored further in~\aref{sec:POper}. Here we defined the operator $\hat{P}_v:f(k)\mapsto f\left(k+k\tfrac{v}{c}\right)$, which represents the Doppler shift of the plane waves in a moving frame. Obviously, $\hat L^{-1}(v) = \hat L(-v)$ to first order in $v/c$. The total action of the moving $\text{BS}$, 
\begin{equation}
\label{eq:TMM:ABCD}
\begin{pmatrix}
 C(k)\\
 D(k)
\end{pmatrix}= \hat{M}
\begin{pmatrix}
 A(k)\\
 B(k)
\end{pmatrix}\text{,}
\end{equation}
can then be obtained from
\begin{align}
\label{eq:TMM:LML}
 \hat{M} &= \hat L(-v) M \hat L(v)\\
 &=\frac{1}{\trans}\begin{bmatrix}
 1 & -(1-2\tfrac{v}{c})\refl\hat{P}_{2v}\nonumber\\
 (1+2\tfrac{v}{c})\refl\hat{P}_{-2v}& \trans^2-\refl^2
\end{bmatrix}\,,
\end{align}
where we have assumed that $\refl$ and $\trans$ do not depend on the wave number. Compared to $M$ in \eref{eq:TMM:M0}, the difference lies in the off-diagonal terms including the Doppler shift imposed by the reflection on a moving mirror. In other words, the coupled counter-propagating plane wave modes differ in wave number, \ie, $k\left(1+\frac{v}{c}\right)$ right-propagating waves couple to $- k\left(1-\frac{v}{c}\right)$ left-propagating waves. Furthermore, if the polarisability itself depends on the wave number $k$, e.g., as in~\eref{eq:ZetaDefn}, the Doppler shift operator acts also on it. To see this effect explicitly, to linear order in $v/c$, $\hat{M}$ can be written as 
\begin{equation}
\label{eq:TMM:Mzeta}
\begin{bmatrix}
 1-\i\zeta - \i \tfrac{v}{c}{\omega}\tfrac{\partial\zeta}{\partial k} & -\i\zeta\left[1-\tfrac{v}{c}\big(2 - \tfrac{k}{\zeta}\tfrac{\partial\zeta}{\partial k}\big)\right]\hat{P}_{2v}\\
 \i\zeta\left[1+ \tfrac{v}{c}\big(2 - \tfrac{k}{\zeta}\tfrac{\partial\zeta}{\partial k}\big)\right]\hat{P}_{-2v}& 1+\i\zeta -\i\tfrac{v}{c} k\tfrac{\partial\zeta}{\partial k}
\end{bmatrix}\,.
\end{equation}
The transfer matrix in the laboratory frame can thus be conceived as a $2$-by-$2$ supermatrix acting also in the $k$-space. The amplitude $C$ at a given wave number $k$, \ie, $C(k)$, is combined with the amplitudes $A(k)$ and $B\big(k-2k\tfrac{v}{c}\big)$. A similar statement holds for $D(k)$. 

Starting from the knowledge of the incoming field amplitudes, this transfer matrix allows for calculating the total electromagnetic field around a beamsplitter moving with a fixed velocity. In the next step, we derive the force on the moving scatterer through the Maxwell stress tensor.

\subsubsection{Force on a medium in an electromagnetic field}

The Maxwell stress tensor (see~\cite[\textsection 6.7]{Jackson1998}) is defined, for a homogeneous medium in one dimension, $x$, as
\begin{equation}
 \mst_{xx}=-\frac{\epsilon_0}{2}\Big(\big|\efield\big|^2+{c^2} \big|\bfield\big|^2\Big)\,,
\end{equation}
where the electric field $\efield$ and the magnetic field $\bfield$, \eref{eq:TMM:Efield} and \eref{eq:TMM:Bfield}, respectively, have no components along $x$. It is trivial, then, to see that after applying the rotating wave approximation, we obtain
\begin{equation}
\label{eq:TMM:MST}
 \mst_{xx}=-2\epsilon_0\Bigg[\Big|\sum_k A(k)e^{-\i kx-\i\omega t}\Big|^2+\Big|\sum_k B(k)e^{\i kx-\i\omega t}\Big|^2\Bigg]\,,
\end{equation}
since the cross terms in $|\efield|^2$ and $|\bfield|^2$ have opposite signs. Note that $\mst_{xx}$ varies on time scales of the order of the optical period. Let us now introduce a characteristic time, $\tau\gg 2\pi/\omega_0$, over which the variations in $\mst_{xx}$ will be averaged, $\omega_0$ being the central frequency of the pump beam. At $x = 0$, 
\begin{align}
\frac{1}{\tau}\int_0^\tau \Big|\sum_k A(k)e^{-\i\omega t}\Big|^2\rmd t&=\sum_k|A(k)|^2+\sum_{i\neq j}\frac{1}{\tau}\int_0^\tau A(k_i)\big[A(k_j)\big]^\ast e^{-\i(\omega_i-\omega_j)t}\rmd t\nonumber\\
&\approx \sum_k|A(k)|^2+\sum_{i\neq j}A(k_i)\big[A(k_j)\big]^\ast\nonumber\\
&=\Big|\sum_kA(k)\Big|^2\,.
\end{align}
In the approximation we assumed that the frequency bandwidth of the excited modes, $\Delta = \max\left\{ \omega_i-\omega_j\right\}$, around $\omega_0$ is so narrow that $\tau \ll 2 \pi/\Delta$. Since the broadening is due to the Doppler shift,  $\Delta \sim 2\omega_0\tfrac{v}{c}$, where $v$ is the speed of the beamsplitter. For example, taking $v$ to be the typical speed of atoms in a magneto-optical trap, we require $\tau\ll \pi/\big(\omega_0\tfrac{v}{c}\big)\sim 10^{-4}~$s. The time needed to reach the stationary regime of scattering is typically much shorter and thus this condition imposed on the averaging time $\tau$ can be safely fulfilled. 

The force on the medium is given by the surface integral of $\mst_{xx}$ on the surface, $\mathcal{S}$, of a fictitious volume $V=\sigma_\text{L}\,\delta l$ enclosing the medium, where $\sigma_\text{L}$ is the mode area and $\delta l$ the infinitesimal length of the volume along the `$x$' axis. Then, this force is given by
\begin{align}
\label{eq:TMM:Forcedef}
 \force&=\oint_\mathcal{S}\mst_{xx} n_x\rmd\mathcal{S}\nonumber\\
&=\sigma_\text{L}\big[\mst_{xx}({x\rightarrow 0^+})-\mst_{xx}({x\rightarrow 0^-})\big]\,,
\end{align}
where $n_x=\sgn(x)$ is the normal to $\mathcal{S}$. Substituting the relevant expressions for $\mst_{xx}$ into the preceding formula gives
\begin{equation}
\label{eq:TMM:MSTForce}
\force=\frac{\hbar\omega_0}{c}\Big(\big|A\big|^2 +\big|B\big|^2 -\big|C\big|^2-\big|D\big|^2 \Big)\,,
\end{equation}
where $A=[{\hbar\omega_0/(2\sigma_\text{L}\epsilon_0c)}]^{-1/2}\sum_k A(k)$ is the photo-current amplitude, and similarly for $B$, $C$ and $D$, their modulus square giving the number of photons per unit time. Although we considered first the electric field composed of independent modes, in the force expression only the sums of the mode amplitudes occur. An identical result holds when we replace the discrete sum over $k$ by an integral, defining $A=[{\hbar\omega_0/(2\sigma_\text{L}\epsilon_0c)}]^{-1/2}\int\!A(k)\rmd k$, etc.

\subsubsection{Quantum fluctuations of the force}
\label{sec:MSTDiffusion}
In the previous section the force was derived based on the assumption that the field amplitudes are c-numbers. In order to describe the inherent quantum fluctuations of the force, we need to resort to the quantum theory of fields and represent the mode amplitudes by operators: $A(k)\to\hat{A}(k)$. To leading order the fluctuations of the force acting on a beamsplitter amount to a momentum  diffusion process~\cite{Dalibard1989, Castin1990}. The diffusion coefficient will be evaluated in the case of coherent-state fields~\cite{Glauber1963}.
\par
The diffusion coefficient can be deduced from the second-order correlation function of the force operator, \eref{eq:FDTDeltaDependence}. The evaluation of this quantum correlation is system-specific. Quantum correlations, \ie, the operator algebra of the mode amplitudes $\hat A(k)$, $\hat B(k)$, $\hat C(k)$, and $\hat D(k)$, are influenced by multiple scattering and thus depend on the total transfer matrix of the entire system. The simplest case is a single beamsplitter at rest where the ``input'' modes $\hat B(k)$ and $\hat C(k)$ have independent fluctuations. The calculation, delegated to \aref{sec:TMM:AppDiffusion},  includes all the steps needed for the treatment of a general system. The diffusion coefficient for a single beamsplitter is obtained as
\begin{equation}
\label{eq:TMM:DiffCoeff}
 \diffn =  (\hbar k)^2 \Bigl(\big|A\big|^2 +\big|B\big|^2+\big|C\big|^2+\big|D\big|^2 + 2 \re{\refl A^* B -\trans A^* C}+ 2\re{\refl D^*C - \trans D^* B}\Bigr)\,,
\end{equation}
where $A=\expt{\hat{A}}, B=\expt{\hat{B}}, C=\expt{\hat{C}}, D=\expt{\hat{D}}$ are the photo-current amplitudes (their modulus square is of the units of 1/sec), obeying \eref{eq:TMM:ABCD} for $v=0$.
\par
As an example, let us consider the diffusion coefficient for a two-level atom illuminated by counter-propagating monochromatic light waves.  Using the polarisability $\zeta$, the transmission and reflection coefficients can be expressed as $\trans = 1/(1-\i\zeta)$ and $\refl=\i\zeta/(1-\i\zeta)$, respectively [see~\eref{eq:TMM:M0}]. \eref{eq:TMM:DiffCoeff} can then be rewritten in the form
\begin{equation}
\label{eq:TMM:GeneralDiffnBC}
 \diffn =  (\hbar k)^2 \bigg[\frac{2\im{\zeta}}{|1-\i\zeta|^2} \big|B-C\big|^2 + \frac{4|\zeta|^2}{|1-\i\zeta|^2} \Big(\big|B\big|^2 + \big|C\big|^2\Big)\bigg]\,,
\end{equation}
where the first term, apart from the factor $|1-\i\zeta|^2$, corresponds to the result well-known from laser cooling theory. Second- and higher-order terms in $\lvert\zeta\rvert$ correspond to the back-action of the atom on the field and are usually absent in treatments of laser cooling due to the fact that $\lvert\zeta\rvert\ll 1$ is generally implicit in such treatments. Note that the diffusion process due to the recoil accompanying the spontaneous emission of a photon (see Ref.~\cite{Gordon1980}) is missing from this result. The detailed modelling of absorption, \ie, scattering photons into the three-dimensional space, is not included in our approach.

\subsubsection{Example: Force on a moving beamsplitter\label{sec:Force}}
We will now use~\eref{eq:TMM:MSTForce} to derive a general expression for the force on a moving beamsplitter illuminated by two counterpropagating, monochromatic, plane waves with amplitudes $B_0$ and $C_0$. On using~\eref{eq:TMM:ABCD} to express the outgoing field modes in terms of the incoming ones, we note that the outgoing amplitudes comprise two monochromatic terms each:
\begin{equation}
\label{eq:TMM:AD}A= \frac{\i\zeta\left[1-\tfrac{v}{c}\big(2+\tfrac{k}{\zeta}\tfrac{\partial\zeta}{\partial k}\big)\right]}{1-\i\zeta\Big(1-\tfrac{v}{c}\tfrac{k}{\zeta}\tfrac{\partial\zeta}{\partial k}\Big)}B_0 + \frac{1}{1-\i\zeta\Big(1+\tfrac{v}{c}\tfrac{k}{\zeta}\tfrac{\partial\zeta}{\partial k}\Big)}C_0\,.
\end{equation}
and
\begin{equation}
D= \frac{1}{1-\i\zeta\Big(1-\tfrac{v}{c}\tfrac{k}{\zeta}\tfrac{\partial\zeta}{\partial k}\Big)}B_0 + \frac{\i\zeta\left[1+\tfrac{v}{c}\big(2+\tfrac{k}{\zeta}\tfrac{\partial\zeta}{\partial k}\big)\right]}{1-\i\zeta\Big(1+\tfrac{v}{c}\tfrac{k}{\zeta}\tfrac{\partial\zeta}{\partial k}\Big)}C_0\,.
\end{equation}
These relations are substituted into~\eref{eq:TMM:MSTForce}, giving
\begin{multline}
\label{eq:TMM:ForceTLA}
\force=\frac{2\hbar\omega}{\bigl|1-\i\zeta\bigr|^2}\bigg\{\Big(\im{\zeta}+\big|\zeta\big|^2\Big)\Big(\big|B_0\big|^2-\big|C_0\big|^2\Big)-2\re{\zeta}\im{B_0C_0^\ast}\\
-\frac{v}{c}\biggl[\biggl(\omega\im{\frac{1+\i\zeta^\ast}{1-\i\zeta}\frac{\partial\zeta}{\partial\omega}}+2\big|\zeta\big|^2\biggr)\Big(\big|B_0\big|^2+\big|C_0\big|^2\Big)\\+2\im{\omega\frac{1+\i\zeta^\ast}{1-\i\zeta}\frac{\partial\zeta}{\partial\omega}+2\zeta}\re{B_0C_0^\ast}\biggr]\bigg\}\,,
\end{multline}
accurate to first order in $v/c$. For $v=0$ this result reduces to the one in Ref.~\cite{Asboth2008}. Most of the $v$-dependent terms arise from the frequency dependence of the polarisability.  These are the dominant terms in the case of a quasi-resonant excitation of a resonant scatterer, such as a two-level atom, since the prefactor  $\tfrac{k}{\zeta}\tfrac{\partial\zeta}{\partial k} \sim \tfrac{\omega}{\Gamma}$ expresses resonant enhancement. The $v$-dependent terms linear in the polarisability $\zeta$ are in perfect agreement with the friction forces known from standard laser cooling theory, both for propagating and for standing waves. For example, assuming identical laser powers from the two sides, giving a standing wave with wavenumber $k_0$, and averaging spatially gives
\begin{equation}
\label{eq:TMM:MolassesForce}
\force=-4\hbar k_0^2\big|B_0\big|^2\im{\tfrac{\partial\zeta}{\partial\omega}}v\,,
\end{equation}
for small $\lvert\zeta\rvert$ and to first order in $v/c$, which can be immediately recognised as the friction force in ordinary Doppler cooling, \eref{eq:OMForce}, when one uses the definition of $\zeta$ in~\eref{eq:ZetaDefn}. Finally, by making similar substitutions into~\eref{eq:TMM:GeneralDiffnBC}, we obtain
\begin{equation}
\label{eq:TMM:MolassesDiffn}
 \diffn=8(\hbar k_0)^2\im{\zeta}\big|B_0\big|^2\sin^2(k_0 x)\,,
\end{equation}
which, excluding the diffusion effects due to spontaneous emission, matches the standard result in \eref{eq:OMDiffn}. Note, however, that the scattering theory leads to a more general result which is represented by the terms of higher order in $\zeta$. These terms describe the back-action of the scatterer on the field, which we re-iterate is generally neglected in free-space laser cooling theory. The general result in \eref{eq:TMM:ForceTLA} reveals that this velocity-dependent force also acts on a scatterer whose polarisability is independent of the frequency. This is a very general class and we will only focus on such scatterers in the following.

\subsection{General system of a fixed and a mobile scatterer}\label{sec:TMM:GeneralMMC}
\begin{figure}[tb]
 \centering
 \includegraphics[scale=0.75]{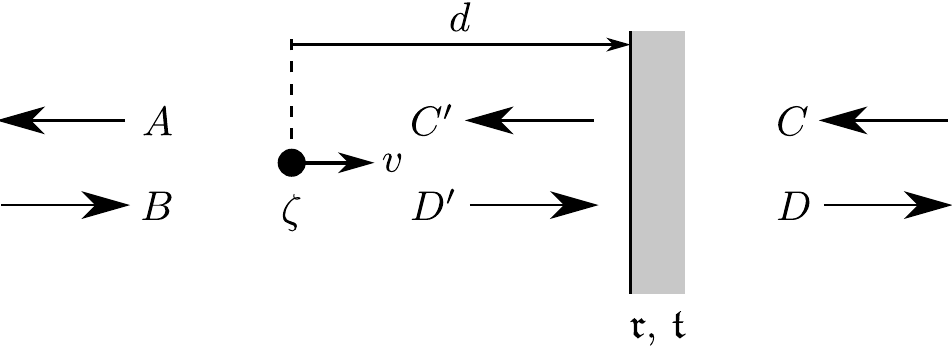}
 \caption[Physical parameters of a model with a fixed and a mobile scatterer]{Physical parameters of our model. $A$, $B$, etc.\ represent the field mode amplitudes.}
 \label{fig:TMM:Model}
\end{figure}
Consider the model in~\fref{fig:TMM:Model} where the scatterer, or `atom', has a polarisability $\zeta$ constant over the frequency range of interest. Letting $M_{\text{a}}$, $M_{\text{p}}$ and $M_{\text{m}}$ be the transfer matrices for the atom, propagation and mirror, respectively, we obtain the relation:
\begin{equation}
\begin{pmatrix}
 A(k)\\
 B(k)
\end{pmatrix} = M_{\text{a}}M_{\text{p}}M_{\text{m}}\begin{pmatrix}
 C(k)\\
 D(k)
\end{pmatrix}\,,
\end{equation}
where
\begin{align}
M_{\text{a}} &= \begin{bmatrix}
 1+\i\zeta & \i\zeta\left(1- 2\tfrac{v}{c}\right)\hat{P}_{2v}\nonumber\\
 -\i\zeta\left(1+2\tfrac{v}{c}\right)\hat{P}_{-2v}& 1-\i\zeta
\end{bmatrix} \\
&= \begin{bmatrix}
 M_{11} & M_{12}\hat{P}_{2v}\\
 M_{21}\hat{P}_{-2v} & M_{22}
\end{bmatrix}\,,
\end{align}
\begin{equation}
M_{\text{p}} = \begin{bmatrix}
 e^{\i kd} & 0\\
 0 & e^{-\i kd}
\end{bmatrix}\,,\text{ and }
M_{\text{m}} = \frac{1}{\trans}\begin{bmatrix}
 \trans^2-\refl^2 & \refl\\
 -\refl & 1
\end{bmatrix}\,.
\end{equation}
The distance between the atom and the mirror is denoted by $d$. Note that the free-propagation transfer matrix $M_\text{p}$ is non-uniform in the $k$-space, and therefore the Doppler shift has an influence on the phase shift accumulated between two scattering events.
\par
The boundary condition is set as follows. We assume that there is no incoming field from the right; therefore $C(k)=0$ for all $k$. The incoming field from the left is assumed to be monochromatic, $B(k)=\mathcal{B}\,\delta(k-k_0)$, with $k_0$ being the pump wavenumber. The resulting field comprises modes with wavenumbers in a narrow region around $k_0$. In the laboratory frame the field mode $A(k)$ interacts with $B(k-2k\tfrac{v}{c})$ and $C^\prime(k)$ through the Doppler shift, and similarly for $D^\prime(k)$. From $C(k)=0$ it directly follows that
\begin{align}
\label{eq:TMM:InputOutput}
A(k)&=\Big(\refl M_{11}e^{\i kd}+M_{12}\hat{P}_{2v}e^{-\i kd}\Big)\Big(\refl M_{21}\hat{P}_{-2v}e^{\i kd}+M_{22}e^{-\i kd}\Big)^{\!-1}B(k)\nonumber\\
&=\frac{1}{M_{22}}\Big(\refl M_{11}e^{\i kd}+M_{12}\hat{P}_{2v}e^{-\i kd}\Big)e^{\i kd}\sum_{n=0}^{\infty}\left(-\refl\frac{M_{21}}{M_{22}}\right)^{\!n} e^{2in kd \left[1-(n+1)\tfrac{v}{c}\right]}B\big(k-2nk\tfrac{v}{c}\big)\,.
\end{align}
We will need the sum of amplitudes, $\mathcal{A}=\int A(k)\rmd k / \mathcal{B}$, defined relative to the incoming amplitude $\mathcal{B}=\int B(k) dk$. Note that $\int \hat{P}_vf(k)\rmd k=\int f(k)\rmd k$. Thus, to first order in $\tfrac{v}{c}$,
\begin{equation}
\label{eq:TMM:IntA}
\mathcal{A}=\frac{M_{12}}{M_{22}}+\left(\frac{M_{12}}{M_{22}}-\frac{M_{11}}{M_{21}}\right)\sum_{n=1}^{\infty}\left(-\refl\frac{M_{21}}{M_{22}}\right)^{\!n} \left[1+2in(n-1)k_0 d\tfrac{v}{c}\right]e^{2in k_0d}\,.
\end{equation}
It is worth introducing the reference point at a distance $L =2 N \pi/k_0$ from the fixed mirror, where the integer $N$ is such that the moving atom's position $x$ is within a wavelength of this reference point. Then the atom--mirror distance can be replaced by $d=L-x$, and $k_0 L$ drops from all the trigonometric functions. The solution, \eref{eq:TMM:IntA}, has a clear physical meaning, in that the reflected field, $\mathcal{A}$, can be decomposed into an interfering sum of fields: the first term is the reflection directly from the atom, whereas the summation is over the electric field undergoing successive atom--mirror round-trips. We can also write the preceding expression in closed form:
\begin{align}
\label{eq:TMM:ReflectedField}
\mathcal{A} =\frac{1}{1-\i\zeta}\Biggl\{&\i\zeta + \refl \frac{e^{-2 \i k_0x}}{1 - \i\zeta -\refl \i\zeta e^{-2 \i k_0x}}\nonumber\\
&-2 \i\frac{v}{c} \zeta \Biggl[1 - \frac{\refl^2 e^{-4 \i k_0x}}{\big(1 - \i\zeta -\refl \i\zeta e^{-2 \i k_0x}\big)^2}-2\i k_0(L-x) \frac{\refl^2 (1-\i\zeta) e^{-4 \i k_0x}}{\big(1 - \i\zeta -\refl \i\zeta e^{-2 \i k_0x}\big)^3}\Biggr] \Biggr\}\,.
\end{align}
This result is valid for arbitrary $\zeta$. The main virtue of our approach is clearly seen, in that we can smoothly move from $\zeta=0$, which indicates the absence of the mobile scatterer, to $|\zeta|\rightarrow\infty$, which corresponds to a perfectly reflecting mirror, \ie, a moving boundary condition for the electromagnetic field.
\par
Let us outline some of the generic features of the above calculation that would be encountered in a general configuration of scatterers. By using the formal Doppler shift operators, we benefit from the transfer matrix method in keeping the description of the system as a whole within $2\times 2$ matrices. The input-output relation for the total system is always obtained in a form similar to that of \eref{eq:TMM:InputOutput}. As long as the Doppler broadening is well below the transient time broadening of the system, the calculation of forces and diffusion requires solely the sum of the mode amplitudes. An important point is that the integrated action of the Doppler shift operator $\hat P_v$ on monochromatic fields is a shift in $k$-space. Therefore, by interchanging the order of terms and putting the $\hat P_v$ terms just to the left of the input field amplitudes, they can be eliminated, such as in \eref{eq:TMM:IntA}. Finally, up to first order in $v/c$, the resulting power series, a trace of multiple reflections, can be evaluated in a closed form, as shown in \eref{eq:TMM:ReflectedField}. In conclusion, the illustrated method lends itself for the description of more complex schemes, for example, the cooling of a moving, partially reflective mirror in a high-finesse Fabry-Perot resonator \cite{Bhattacharya2008}. This scheme, however, rapidly increases in complexity with the summations becoming potentially unmanageable. An alternative is possible, in that the matrix $\hat{M}$ is invertible to first order in $v/c$ so that $\force$ and $\diffn$ can be evaluated in closed form for a mobile scatterer in a generic optical system. We will investigate this solution in \sref{sec:TMM:General} after some illustrative examples.

\subsubsection{Force acting on the mobile scatterer}
To obtain the force on the moving scatterer, we also need to evaluate $C^{\prime}(k)$ and $D^{\prime}(k)$:
\begin{align}
\label{eq:TMM:Cprime}
\begin{pmatrix}
 C^{\prime}(k)\\
 D^{\prime}(k)
\end{pmatrix} &= \begin{bmatrix}
 1-\i\zeta &  -\i\zeta \left(1 - 2 \tfrac{v}{c}\right)\hat{P}_{2v}\\
 \i\zeta \left(1 + 2 \tfrac{v}{c}\right)\hat{P}_{2v}^{-1} & 1+\i\zeta
\end{bmatrix} \begin{pmatrix}
 A(k)\\
 B(k)
\end{pmatrix}\,,
\end{align}
where we applied the inverse of the transfer matrix $M_\text{a}$. Next, we make the following definitions:
\begin{equation}
 \mathbb{A}=|\mathcal{A}|^2\,,\quad \mathbb{B}=|\mathcal{B}|^2\,,
 \mathbb{C}=\frac{1}{\mathbb{B}}\left|\int C^{\prime}(k)\rmd k\right|^2\,,\text{ and }
 \mathbb{D}=\frac{1}{\mathbb{B}}\left|\int D^{\prime}(k)\rmd k\right|^2\,,
\end{equation}
and a simple calculation leads to
\begin{align}
 \mathbb{C}=&\left|1-\i\zeta\right|^2\mathbb{A}+\left|\i\zeta\big(1-2\tfrac{v}{c}\big)\right|^2+2\re{\i\zeta^{\ast}(1-\i\zeta)\big(1-2\tfrac{v}{c}\big)\mathcal{A}}\text{,}\\
 \mathbb{D}=&\left|\i\zeta\big(1+2\tfrac{v}{c}\big)\right|^2\mathbb{A}+\left|1+\i\zeta\right|^2+2\re{\i\zeta(1+\i\zeta^{\ast})\big(1+2\tfrac{v}{c}\big)\mathcal{A}}\,.
\end{align}
Thereby the force acting on the scatterer is obtained as 
\begin{align}
\label{eq:TMM:GeneralForce}
 \force &= (\hbar\omega/c)\mathbb{B}\left(\mathbb{A}+1-\mathbb{C}-\mathbb{D}\right)\nonumber\\
 &= -2\hbar k_0 \mathbb{B}\biggl\{\Bigl[\lvert\zeta\rvert^2\bigl(1+2\tfrac{v}{c}\bigr)+\im{\zeta}\Bigr]\mathbb{A}+\lvert\zeta\rvert^2\bigl(1-2\tfrac{v}{c}\bigr)-\im{\zeta}\nonumber\\
&\phantom{= -2\hbar k_0 \mathbb{B}\biggl\{\ }+2\re{\Bigl(\i\re{\zeta}+\lvert\zeta\rvert^2+2\tfrac{v}{c}\im{\zeta}\Bigr)\mathcal{A}}\biggr\}\,,
\end{align}
where $\mathcal{A}$ has to be substituted from \eref{eq:TMM:ReflectedField}.
\begin{figure}[t]
   \centering
   \includegraphics[width=1.5\figwidth]{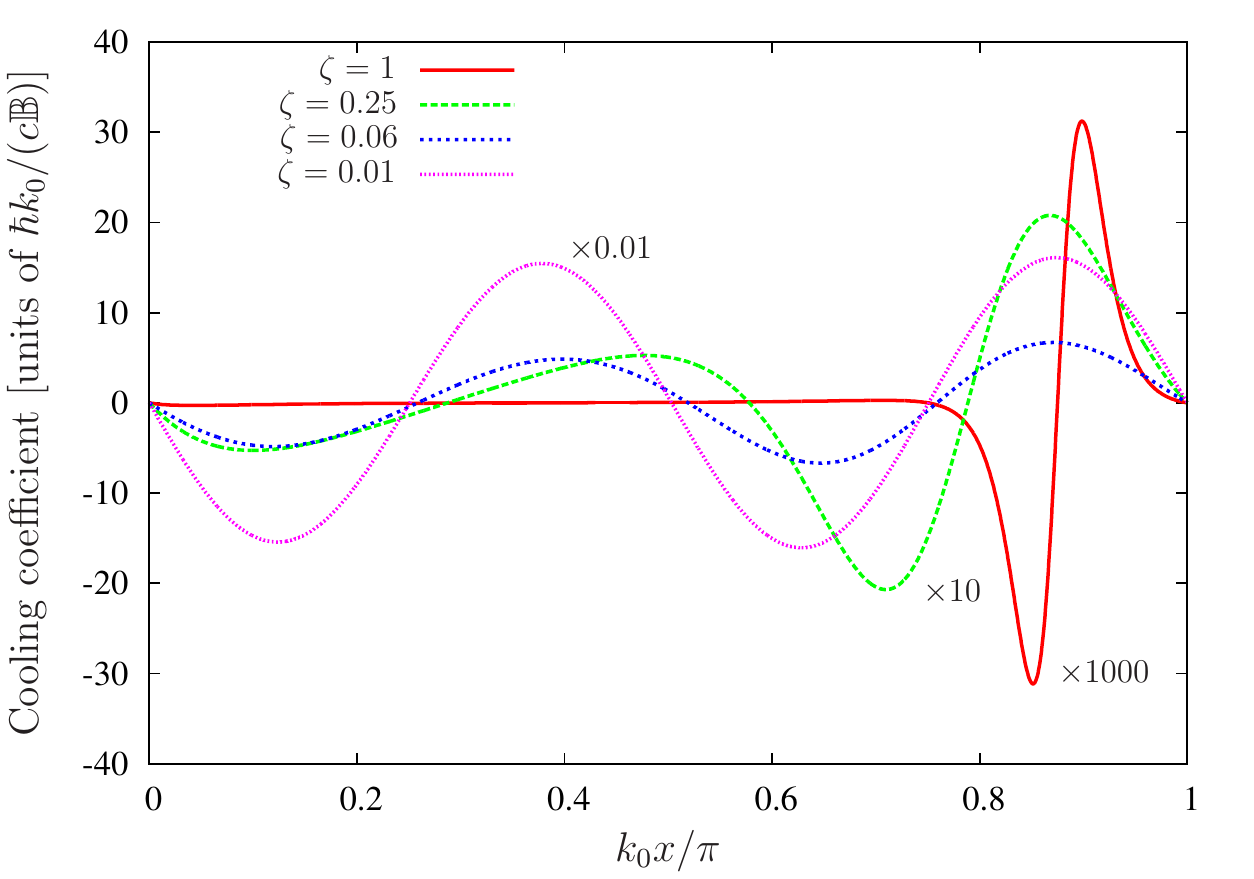}
   \caption[The position dependence of the cooling coefficient]{The position dependence of the cooling coefficient $\heatingcoefft$ in the relation $\force=-\heatingcoefft v$ for the velocity-dependent force acting on the mobile scatterer in \fref{fig:TMM:Model}, for various scattering parameters $\zeta$, evaluated by using \eref{eq:TMM:ReflectedField} and \eref{eq:TMM:GeneralForce} with $k_0 L=2\pi\times 100$. The fixed mirror is assumed to be a perfect mirror. In order to fit all the curves into the same range, they are divided by the factors indicated in the figure. Note that $x$ is defined differently from \fref{fig:MMC:Analytic-Spatial}; the mobile scatterer here is closer to the fixed mirror for \emph{increasing} $x$.}
   \label{fig:TMM:FrictionPosnDependence}
\end{figure}
The cooling coefficient $\heatingcoefft$, defined through the relation $\force=-\heatingcoefft v$, is plotted in \fref{fig:TMM:FrictionPosnDependence} as a function of the position $x$ in a half-wavelength range for various values of $\zeta$. When varying the coupling strength from $\zeta=0.01$ up to $\zeta=1$, the cooling coefficient transforms between two characteristic regimes. For small coupling the linear velocity dependence tends to a simple sinusoidal function while, for large coupling, the friction exhibits a pronounced resonance in a narrow range. This resonance arises from the increased number of reflections between the mobile scatterer and the fixed mirror. It can be observed that the resonance shifts towards $k_0 x =\pi$ on increasing $\zeta$. In the opposite limit of small $\zeta$,  the maximum friction is obtained periodically at $\big(n-\tfrac{1}{4}\big)\pi/2$ according to the sinusoidal function. The position of the maximum friction is plotted in \fref{fig:TMM:Phase}, showing the transition from $7\pi/8$ to $\pi$. The maximum friction force is plotted in \fref{fig:TMM:Maxfric}, showing the two limiting cases of $\zeta^2$ behaviour, in the limit of small $\zeta$, and $\zeta^6$ behaviour, in the limit of large $\zeta$. These two cases are described in \sref{sec:MovingAtom} and \sref{sec:Mirrorcooling}, respectively.
\begin{figure}[t]
   \centering
\subfigure[]{
   \includegraphics[width=1.5\figwidth]{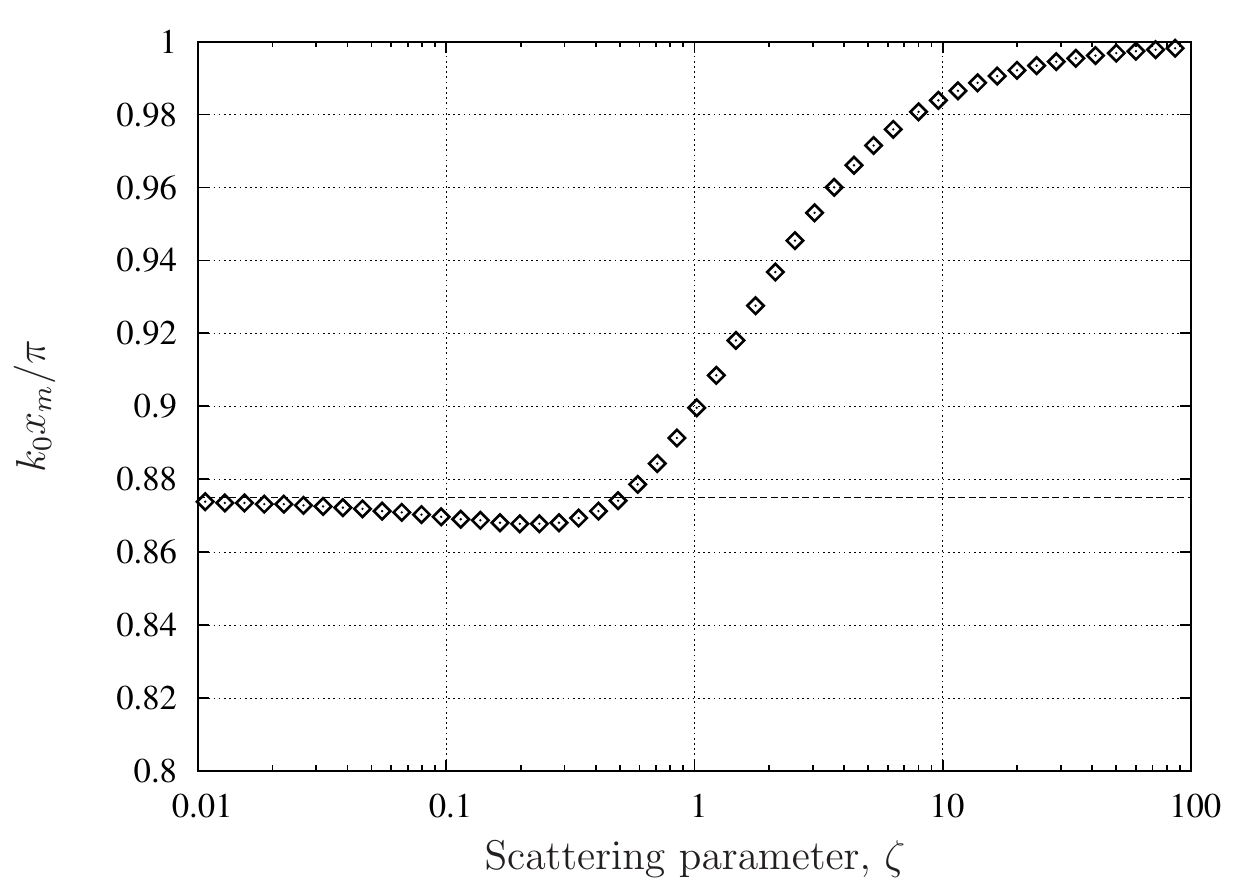}
   \label{fig:TMM:Phase}
}\\
\subfigure[]{
   \includegraphics[width=1.5\figwidth]{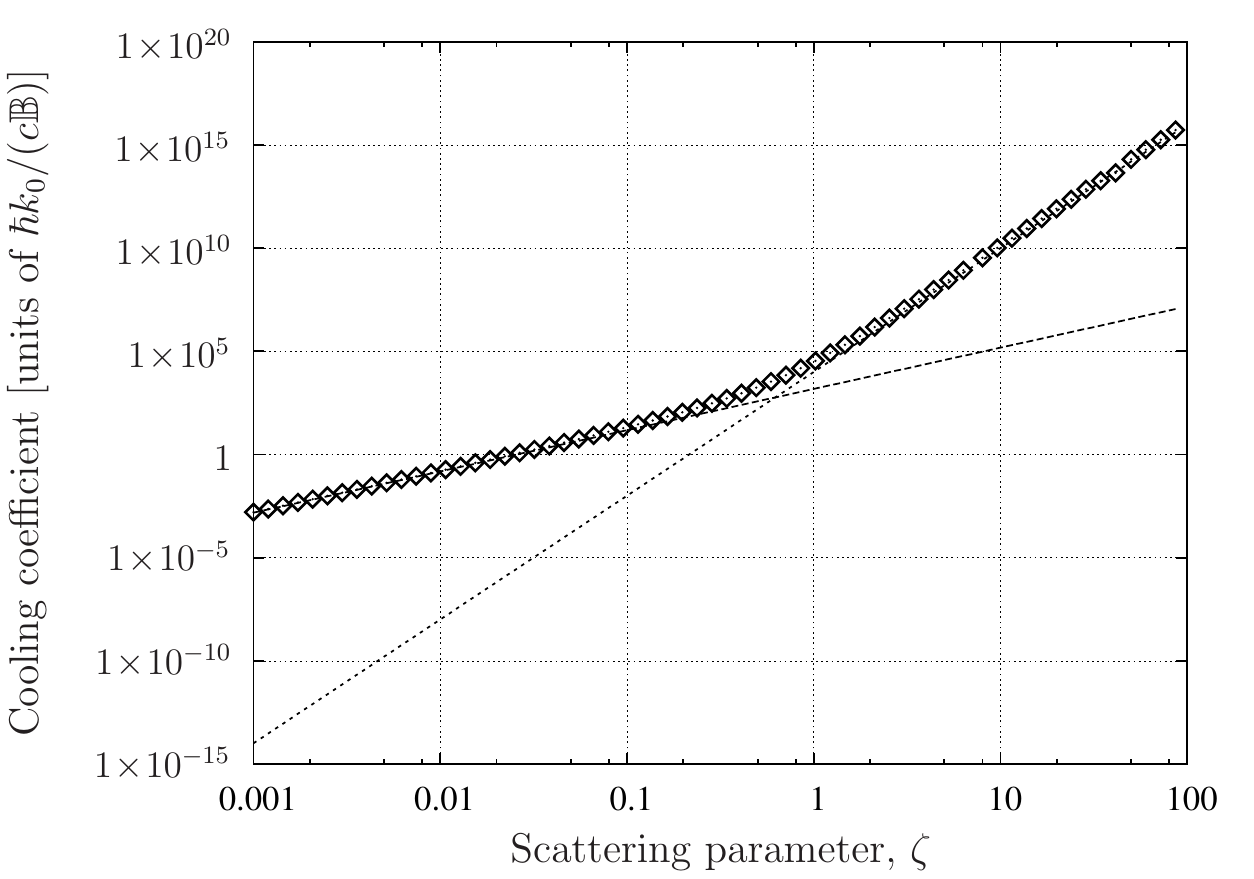}
   \label{fig:TMM:Maxfric}
}
\caption[Position and magnitude of maximum force as a function of $\zeta$]{(a)~The position of the maximum friction force, $k_0x_m$, as a function of the dimensionless scattering parameter $\zeta$ (on a semilog scale) acting on the scatterer in \fref{fig:TMM:Model}, with the fixed mirror being a perfect mirror. This position shifts from $7\pi/8$ to $\pi$ on increasing $\zeta$. (b)~A similar plot, showing the maximum friction force as a function of $\zeta$ (on a log-log scale) with $k_0L=2\pi\times 100$. In the limit of small $\zeta$, the force scales as $\zeta^2$ [cf.~\eref{eq:TMM:MirrorCoolForce}; dashed line] whereas in the limit of large zeta it scales as $\zeta^6$ [cf.~\eref{eq:TMM:FPCoolForce}; dotted line].}
\end{figure}

\subsubsection{Diffusion coefficient}
The calculation of the diffusion coefficient proceeds along the same lines as that corresponding to a single beamsplitter, shown in \aref{sec:TMM:AppDiffusion}. The difference is that the modes $B(k)$ and $C'(k)$ around the mobile scatterer are not independent, for the reflection at the fixed mirror mixes them. Therefore, all the modes $A$, $B$, $C'$, and $D'$ have to be expressed in terms of the leftmost and rightmost incoming modes, $B(k)$ and $C(k)$, respectively. Instead of the derivation of such a general result for the diffusion, here we will restrict ourselves to the special case of $\refl=-1$ ($\Leftrightarrow$ perfect mirror) and real $\zeta$ ($\Leftrightarrow$ no absorption in the moving mirror). In this special case the diffusion calculation simplifies a lot, because (i) the perfect mirror prevents the modes $C$ from penetrating into the interaction region, and (ii) quantum noise accompanying absorption does not intrude in the motion of the scatterer. 
\par
Only the modes $\hat B(k)$ impart independent quantum fluctuations. When all the amplitudes around the scatterer are expressed in terms of $\hat B(k)$, and are inserted into the force correlation function given in \eref{eq:FDTDeltaDependence}, the commutator $[\hat b(t),\hat b^\dagger(t')]$ appears in all the terms (see \aref{sec:TMM:AppDiffusion}). Straightforward algebra leads to
\begin{equation}
\label{eq:TMM:GeneralDiff}
\diffn=\hbar^2 k_0 ^2\mathbb{B}(\mathbb{A} +1-\mathbb{C}-\mathbb{D})^2\,.
\end{equation}
We emphasise that the above result is not general: the diffusion is not necessarily proportional to the square of the force. This simple relation here follows from the assumptions, $\refl=-1$ and $\im{\zeta}=0$, declared above. 
\par
To be consistent with the calculation of the friction force linear in velocity, the diffusion should be evaluated only for $v=0$. From the ratio of these two coefficients,  the steady-state temperature can be deduced. The velocity-independent components of the modes obey the following relations: $\mathbb{A} = 1$ and $\mathbb{C'} = \mathbb{D'}$ (all incoming power is reflected). Therefore the diffusion coefficient further simplifies,
\begin{equation}
\diffn = 4 \hbar^2 k_0 ^2\mathbb{B} \Bigg(1-\frac{1}{\big|1 - \i\zeta + \i\zeta e^{-2 \i k_0x}\big|^2}\Bigg)^2\,.
\end{equation}

\begin{figure}[tbp]
   \centering
   \includegraphics[width=1.5\figwidth]{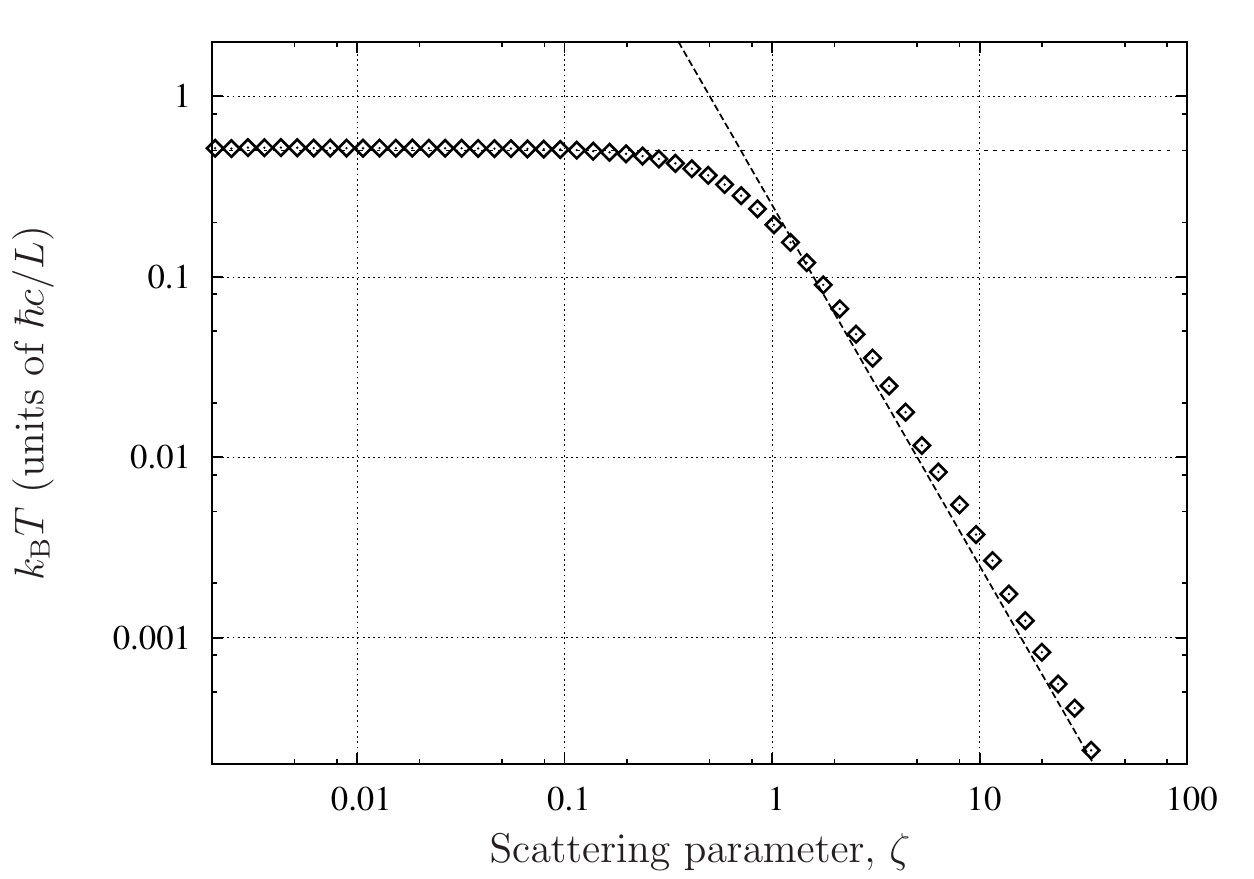}
   \caption[Characteristic temperature for the two-scatterer system]{Characteristic temperature for the two-scatterer system of \fref{fig:TMM:Model}, given by the ratio of the diffusion and cooling coefficients in the points where the friction is maximum, as a function of the dimensionless scattering parameter $\zeta$ on a log-log scale. Constant and $1/\zeta^2$ dependence can be read off in the limits of small and large $\zeta$, respectively. The fixed mirror is a perfect mirror.}
   \label{fig:TMM:Temperature}
\end{figure} 
In \fref{fig:TMM:Temperature}, the temperature $k_\text{B} T= \diffn/\varrho$, is plotted as a function of the scattering parameter $\zeta$. The friction and the diffusion coefficients are taken at the position where the friction is maximum, as shown in \fref{fig:TMM:Phase}.
The two limits of small and large scattering parameter $\zeta$ will be analysed in \sref{sec:MovingAtom} and \sref{sec:Mirrorcooling}, respectively.

\subsection{Atom in front of a perfect mirror}\label{sec:MovingAtom}
An atom pumped with a far off-resonance beam can be modelled as a moving mirror with small and real $\zeta$. In this section we accordingly truncate our expressions to second order in $\zeta$. We also assume that the fixed mirror is perfect; \ie, $\refl=-1$ and $\trans=0$. Thus,
\begin{equation}
\label{eq:TMM:Force}
 \force=2\hbar k_0\mathbb{B}\big[2\zeta\im{\mathcal{A}}-2\zeta^2\re{\mathcal{A}}-\zeta^2\big(1+\tfrac{v}{c}\big)\mathbb{A}-\zeta^2\big(1-\tfrac{v}{c}\big)\big]\,.
\end{equation}
To obtain $\force$ to second order in $\zeta$, we need $\mathcal{A}$ to first order. Using~\eref{eq:TMM:IntA} and~\eref{eq:TMM:Force}, we obtain:
\begin{equation}
\label{eq:TMM:CurlyA}
 \mathcal{A}=-e^{-2\i k_0x}+\zeta(\i-2\i e^{-2\i k_0x}+\i e^{-4\i k_0x})+\zeta\tfrac{v}{c}\big[-2\i+2\i e^{-4\i k_0x}-4 k_0  (L-x) e^{-4\i k_0 x}\big]\,,
\end{equation}
and
\begin{align}
\label{eq:TMM:MirrorCoolForce}
 \force=4\hbar k_0\mathbb{B}\big(&\zeta\sin(2 k_0x)-\zeta^2\big\{2\sin^2( k_0x)\big[4\cos^2( k_0x)-1\big]\big\}\nonumber\\
&-\zeta^2\tfrac{v}{c}\big[4\sin^2(2 k_0x)-4 k_0 (L-x) \sin(4 k_0x)\big]\big)\,,
\end{align}
in agreement with \eref{eq:MMC:AnalyticLongFriction}. In the far field ($x\gg\lambda$), the dominant friction term in the preceding expression is the last term, which renders the $\sin(4 k_0x)$ position dependence shown in \fref{fig:TMM:Friction} for $\zeta=0.01$.
\par
We are now in a position to derive the diffusion coefficient for this system. By substituting~\eref{eq:TMM:CurlyA} into~\eref{eq:TMM:GeneralDiff} and setting $v=0$, we obtain
\begin{equation}
 \diffn=8(\hbar k_0)^2\zeta^2\mathbb{B}\,.
\end{equation}
This allows us to estimate the equilibrium temperature for such a system at a position of maximum friction:
\begin{equation}
\label{eq:TMM:MMCTemp}
 T\approx\frac{\hbar}{k_{\text{B}}\tau}\text{, where }\tau=2(L-x)/c\,,
\end{equation}
which we note is identical in form to the Doppler temperature for a two-level atom undergoing free-space laser cooling~\cite{Metcalf2003}, but where we have replaced the upper state lifetime, $1/\Gamma$, by the round-trip time delay between the atom and the mirror.\footnote{This expression lacks the geometrical factor in \eref{eq:SemiclassicalMMCTemperature}; the reason for this is that the semiclassical treatment of a TLA is, strictly speaking, inconsistent with the assumption $\im{\zeta}=0$; the dominant term of the diffusion is then the first term in \eref{eq:TMM:GeneralDiffnBC}.} Note that this temperature corresponds to the constant value presented in \fref{fig:TMM:Temperature} for $\zeta<0.1$.

\subsection{Optical resonator with mobile mirror}\label{sec:Mirrorcooling}
After the small polarisability case of the previous section, we will now consider the $|\zeta| \rightarrow \infty$ limit. We again assume that the fixed mirror of the resonator is perfect, with $\refl=-1$, and that $C=0$. For simplicity, we assume that the moving mirror has a real polarisability; \ie, it is lossless. We expand the field mode amplitudes as power series in $v/c$, such that $\mathcal{A}=\mathcal{A}_0+\tfrac{v}{c}\mathcal{A}_1+\dots$, and similarly for $\mathcal{C}^{\prime}$.

Let us first calculate the field in the resonator for $v=0$. It follows from \eref{eq:TMM:Cprime} that
\begin{equation}
\mathcal{C}_0^\prime = (1-\i \zeta) \mathcal{A}_0 -\i\zeta = -\frac{e^{-2\i\varphi}}{1-\i\zeta+\i\zeta e^{-2 \i \varphi}}\,,
\end{equation}
with $\varphi = k_0 d$, which has a maximum at $\varphi_0$ obeying
\begin{equation}
\tan(2\varphi_0) = - \frac{1}{\zeta}\,.
\end{equation}
In the limit of $\zeta\rightarrow \infty$, the resonance is Lorentzian:
\begin{equation}
\mathcal{C}_0^\prime = -\frac{e^{-2\i\varphi}}{2 \i (1-\i\zeta) \left[(\varphi-\varphi_0) - \i \tfrac{1}{4 \zeta^2}  \right]}\,,
\end{equation}
with a width of $1/(4\zeta^2)$.

The perfect mirror reflects the total power incoming from the left, $\mathbb{B}$.  Moreover, for real $\zeta$, there is no absorption in the moving mirror, so the outgoing intensity has to be equal to the incoming one: $\mathbb{A} = 1$. This is true if $v=0$; for $v\neq 0$, the field can do work on the mirror. The expansion of the back-reflected intensity  to linear order  in velocity reads $\mathbb{A} = 1 + 2\tfrac{v}{c} \re{\mathcal{A}_0^*\mathcal{A}_1}$.  Extracting the velocity-dependent terms for the general form of the force in \eref{eq:TMM:GeneralForce}, it reduces to 
\begin{equation}
\force_1= \tfrac{v}{c} 4 \hbar k_0 \mathbb{B} \zeta\im{\mathcal{A}_1\big/\left(1 + \i \zeta -\i \zeta e^{2\i\varphi}\right)}\,,
\end{equation}
which, after some algebra, leads to
\begin{equation}
\label{eq:TMM:FPCoolForce}
\force_1 = - \tfrac{1}{2} \tfrac{v}{c} \hbar k_0^2 L \frac{(\varphi-\varphi_0)}{\zeta^4\, \left[\left( \frac{1}{4\zeta^2}\right)^2 + (\varphi-\varphi_0)^2 \right]^3} \mathbb{B}\,.
\end{equation}
On substituting $\kappa = c/\big(4L\zeta^2\big)$, $\Delta_C = - c(\varphi-\varphi_0)/L$, $\eta^2/(2\kappa)=\mathbb{B}$, and $G=c^2k_0^2/L^2$, the friction force renders that derived from the usual radiation pressure Hamiltonian in \aref{sec:RadPressCool}. 

Expressing the field modes interacting with the mobile mirror in terms of the input field mode and performing a calculation similar to that leading to \eref{eq:TMM:DiffCoeff} readily gives
\begin{equation}
\diffn \approx 4(\hbar k_0)^2 |\mathcal{C}^\prime_0|^4 \mathbb{B} \approx \frac{(\hbar k_0)^2\mathbb{B}}{ 4\zeta^4 \left[ \left( \frac{1}{4\zeta^2}\right)^2 + (\varphi-\varphi_0)^2\right]^2}\,.
\end{equation}
The resulting temperature thereby attains a minimum at $4 \zeta^2 (\varphi-\varphi_0)=1$, \ie, $\Delta_C=-\kappa$, in analogy with free-space Doppler cooling, at which point we have
\begin{equation}
\label{eq:TMM:MirrorCoolingTemp}
k_{\text{B}} T \approx  \frac{\hbar c}{4 \zeta^2 L} = \hbar \kappa\,.
\end{equation}
Again, this asymptotic behaviour is reflected in \fref{fig:TMM:Temperature} for large $\zeta$. We note the similarity of the preceding expression with the temperature of an atom cooled in a cavity, in the good-cavity limit~\cite{Horak2001}. We conjecture that this is due to the fact that both systems can be considered to involve the coupling of a laser with a system having a decay rate $\kappa$. This result also holds for the case of an atom undergoing mirror-mediated cooling, as can be seen in \eref{eq:TMM:MMCTemp}.
\par
It is also important to note that the above discussion only treats the effects of the light fields on the scatterer. As such, the temperature limit, \eref{eq:TMM:MirrorCoolingTemp}, is intrinsic to the light forces, and the mechanical damping and heating processes present in a real, macroscopic mirror-cooling setup are not taken into account. In practice, these heating processes may dominate over the heating induced by the quantum noise in the light field \cite{Saulson1990, Cohadon1999}. In such cases, radiation pressure cooling is a possible means to lower the equilibrium temperature owing to the additional, optical, damping process.

\subappendicesstart

\subsection{Appendix: The Doppler shift operator\label{sec:POper}}
\begin{figure}[t]
 \centering
 \includegraphics[scale=0.75]{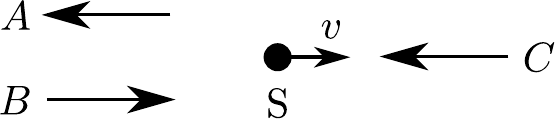}
 \caption[Reflection and transmission of a moving scatterer]{Reflection and transmission of a moving scatterer. $B$ and $C$ are the input field modes, and $A$ is the output field mode. A further output field mode (`$D$') is not drawn because it is not relevant to our discussion here.}
 \label{fig:TMM:POper}
\end{figure}

Consider the situation in~\fref{fig:TMM:POper}, in the laboratory frame, where $\text{S}$ is a scatterer, and suppose that $B$ and $C$ are known. $A(k)$ has contributions arising from both $B\big(k+2k\tfrac{v}{c}\big)$ and $C(k)$, where $k$ is any arbitrary wave number, written separately as:
\begin{align}
 A_B(k)&=a_1 B(k+2k\tfrac{v}{c})\,\text{, and}\\
 A_C(k)&=a_2 C(k)\,.
\end{align}
We can therefore express $A(k)$ as
\begin{equation}
 A(k)=a_1 B(k+2k\tfrac{v}{c})+a_2 C(k)\,.
\end{equation}
Defining $\hat{P}_v$ by $\hat{P}_v:f(k)\mapsto f(k+k\tfrac{v}{c})$, we have
\begin{equation}
 A(k)=\hat{P}_{2v}a_1 B(k)+a_2 C(k)\,.
\end{equation}
A similar expression, involving $\hat{P}_v^{-1}=\hat{P}_{-v}$, holds for $D(k)$. These two operators can then be introduced into~\eref{eq:TMM:TM_fix} as part of the Lorentz transformation, and thus into the transfer matrix for the moving scatterer, giving rise to the form shown in~\eref{eq:TMM:LML}. The resulting transformation, for the transfer matrix $M$, of a scatterer moving with velocity $v$ can be written as:
\begin{equation}
\begin{bmatrix}
(1-\tfrac{v}{c})\hat{P}_v & 0\\
0 & (1+\tfrac{v}{c})\hat{P}_v^{-1}
\end{bmatrix}M\begin{bmatrix}
(1+\tfrac{v}{c})\hat{P}_v^{-1} & 0\\
0 & (1-\tfrac{v}{c})\hat{P}_v
\end{bmatrix}\,,
\end{equation}
to first order in $\tfrac{v}{c}$, where the ordering of the elements of $M$ is as described in the text. Note that this relation is general, in the sense that the elements of $M$ can depend on $k$ (see~\sref{sec:Force}).

For any finite $v$, $\hat{P}_v$ is trivially a bounded operator, having unit norm. This property follows from the important relation $\int \hat{P}^m_v f(k)\rmd k = \int f(k)\rmd k\text{,}$ for any function $f(k)$ and any integer $m$.

This operation can be generalised to $n=2,3$ pairs of modes in $n$ orthogonal dimensions. In two dimensions, this could correspond to a beamsplitter oriented at $45^\circ$ to the $x$-axis; in three dimensions, the beamsplitter would be oriented at $45^\circ$ to all three coordinate axes. We define a new operator by $\hat{S}_i(\v{v}):f(\v{k})\mapsto f(\v{k}+k_i\tfrac{v_i}{c}\mathbf{e}_i)$, where $\mathbf{e}_i$ is the unit vector along the $i$th coordinate axis, $\v{v}$ is the velocity vector of the scatterer, and $\mathbf{x}=(x_1,x_2,\dots)$ for any vector $\mathbf{x}$. In particular, we have $\hat{P}_{v}=\hat{S}_1(v\mathbf{e}_1)$. Now, let $\hat{\m{L}}(\v{v})$ be the $2n\times 2n$ matrix operator:
\begin{equation}
\begin{bmatrix}
(1+\tfrac{v_1}{c})\hat{S}_1^{-1}(\v{v}) & 0 & 0 & \cdots\\
0 & (1-\tfrac{v_1}{c})\hat{S}_1(\v{v}) & 0 & \cdots\\
0 & 0 & (1+\tfrac{v_2}{c})\hat{S}_2^{-1}(\v{v}) & \cdots\\
\vdots & \vdots & \vdots & \ddots
\end{bmatrix}\,.
\end{equation}
Then, the transfer matrix for the scatterer moving with velocity $\v{v}$ is given by
\begin{equation}
\hat{\m{L}}(\mathbf{-v})\,M\,\hat{\m{L}}(\v{v})\,,
\end{equation}
where $M$ is the original transfer matrix for the scatterer, obtained in a manner such as that used to obtain~\eref{eq:TMM:M0}, for example. The ordering of the elements of $M$ is such that it acts on the vector $\big(A_1(\v{k}),B_1(\v{k}),A_2(\v{k}),\dots\big)$:
\begin{equation}
 \begin{pmatrix}
  C_1(\v{k})\\
  D_1(\v{k})\\
  C_2(\v{k})\\
  \vdots
 \end{pmatrix}=\hat{\m{L}}(\mathbf{-v})\,M\,\hat{\m{L}}(\v{v})\begin{pmatrix}
  A_1(\v{k})\\
  B_1(\v{k})\\
  A_2(\v{k})\\
  \vdots
 \end{pmatrix}\,,
\end{equation}
with $A_i(\v{k})$ being the outgoing mode and $B_i(\v{k})$ the incoming mode along the $i$th axis in the negative half-space (assuming that the scatterer is at the origin); and $C_i(\v{k})$ the incoming mode and $D_i(\v{k})$ the outgoing mode in the positive half-space.

\subsection{Appendix: Quantum correlation function of the force operator}\label{sec:TMM:AppDiffusion}
In quantum theory, we need to replace the mode amplitudes $A(k)$ by operators $\hat A(k)$, and similarly for the $B$, $C$, and $D$ modes. 
The cross-correlation of these operators is not trivial because of the boundary condition connecting the mode amplitudes $A(k)$, $B(k)$, $C(k)$ and $D(k)$. The input modes $\hat C(k)$ and $\hat B(k)$ can be considered independent, and the commutator is non-vanishing for the creation and annihilation operators of the same mode, e.g.,
\begin{align}
\Big[ \hat B(k), {\hat B}^\dagger(k^\prime) \Big] & =  \left[ \hat C(k), {\hat C}^\dagger(k^\prime) \right] = \frac{\hbar \omega}{2 \epsilon_0 V} \delta_{k,k^\prime} \text{,}\\
\Big[ \hat B(k), {\hat C}^\dagger(k^\prime) \Big] & =0\,,
\end{align}
assuming a discrete mode index of $k$, and a quantisation volume $V=\sigma_\text{L} l$ with $\sigma_\text{L}$ being the mode area and $l$ a fictitious total length of the space in one dimension. 
\par
In using \eref{eq:FDT}, expressions for $\force$ correct to order $n$ in $v/c$ are compared to $\diffn$ to order $(n-1)$ in this same parameter. Since our force expressions are accurate up to first order in $v/c$, therefore, we need only consider terms that contribute to the diffusion in the $v=0$ case. In the quantum description, the linear relation for the output modes is
 \begin{align}
\label{eq:ADNoise} 
   A(k) = {}& \trans C(k) + \refl B(k) + \sqrt{\epsilon} E \\
   D(k) = {}& \refl C(k) + \trans B(k) + \sqrt{\epsilon} E\,,
\end{align}
where the transmission $\trans=1/M_{22}=1/(1-\i\zeta)$, and reflection $\refl=M_{12}/M_{22} = \i\zeta/(1-\i\zeta)$, as above. The amplitude $E$ represents the quantum noise fed into the system by the absorption. For $\epsilon = 1-\big(|\refl|^2 + |\trans|^2\big)$, this noise ensures that the output modes obey the same commutation relations as the input ones, namely
\begin{align}
\Big[ \hat A(k), {\hat A}^\dagger(k^\prime) \Big] &= \Big[ \hat D(k), {\hat D}^\dagger(k^\prime) \Big] = \frac{\hbar \omega}{2 \epsilon_0 V} \delta_{k,k^\prime}\,,\\
\Big[ \hat A(k), {\hat D}^\dagger(k^\prime) \Big] & =0\,.
\end{align}
However, the linear dependence implies that commutators between input and output mode operators are
\begin{align}
\Big[ \hat A(k), {\hat B}^\dagger(k^\prime) \Big] = \refl \Big[ \hat B(k), {\hat B}^\dagger(k^\prime) \Big] \,,\\
\Big[ \hat A(k), {\hat C}^\dagger(k^\prime) \Big] = \trans \Big[ \hat C(k), {\hat C}^\dagger(k^\prime) \Big] \,,
\end{align}
and similar relations hold for the cross-commutators with $D(k)$.

The proper treatment of quantum fluctuations and the derivation of correlation functions require that the explicit time dependence be considered. Let us introduce the time-varying operators
\begin{equation}
 \hat a(t) = \sum_k \hat A(k) e^{-\i\omega t}\,,
\end{equation}
and similarly for $ \hat b(t)$, $ \hat c(t)$ and $ \hat d(t)$. It follows that
\begin{equation}
\Big[ \hat a(t), {\hat a}^\dagger (t^\prime) \Big] =  \frac{\hbar \omega}{2 \epsilon_0 V} \sum_k  e^{-\i\omega (t-t^\prime)} \approx  \frac{\hbar \omega}{2 c \epsilon_0 \sigma_\text{L}} \delta(t-t^\prime)\,.
\end{equation}
Here we made use of the fact that the non-excited vacuum modes, having a bandwidth that is much broader than that of the pump, also contribute to the summation, allowing us to approximate the summation as an integral over all frequencies. The Fourier-type integral then yields $\delta$-function on the much slower timescale of interest. A similar commutation relation applies to the operators  $ \hat b(t)$, $ \hat c(t)$, and $ \hat d(t)$. The cross-commutators can be derived directly from those concerning the modes, e.g.,
\begin{equation}
\Big[ \hat a(t), {\hat b}^\dagger (t^\prime) \Big] = \refl \frac{\hbar \omega}{2 c \epsilon_0 \sigma_\text{L}}  \delta(t-t^\prime)\,.
\end{equation}
The force operator is
\begin{equation}
\label{eq:TMM:ForceOp}
\hat \force = \sigma_\text{L}\big[\hat \mst_{xx}({x\rightarrow 0^+})-\hat \mst_{xx}({x\rightarrow 0^-})\big]\,,
\end{equation}
as before, where
\begin{align}
\hat \mst_{xx}(x\rightarrow 0) = \begin{cases}
-2\epsilon_0 \Big[\hat a^\dagger(t) \hat a(t) + \hat b^\dagger(t) \hat b(t)\Big]\vspace{0.5em}&\text{for }x\rightarrow 0^-\\
-2\epsilon_0 \Big[\hat c^\dagger(t) \hat c(t) + \hat d^\dagger(t) \hat d(t)\Big]&\text{for }x\rightarrow 0^+
\end{cases}
\end{align}
is the quantised stress tensor. Assuming that the field is in a coherent state, in all normally ordered products, the mode amplitude operators can be replaced by the corresponding coherent state amplitudes, which are c-numbers: e.g.,\ $\hat A(k) \rightarrow A(k)$ and $\hat A^\dagger(k) \rightarrow A^\ast(k)$. The force operator in \eref{eq:TMM:ForceOp} is normally ordered in this way; therefore coherent-state fields render, as a mean value of the quantum expressions, the force \eref{eq:TMM:MSTForce} derived from the classical theory based on the definition \eref{eq:TMM:Forcedef}. Non-trivial quantum effects arise from non-normally ordered products, such as the fourth-order product terms of the second order correlation function of the force \eref{eq:FDTDeltaDependence}.  These terms can be evaluated straightforwardly by invoking the above-derived commutators to rearrange the product into normal order. As an example, consider
\begin{equation}
\big\langle \hat a^\dagger(t) \hat a(t) \hat a^\dagger(t^\prime) \hat a(t^\prime)\big\rangle =\big\langle \hat a^\dagger(t) \hat a^\dagger(t^\prime) \hat a(t)  \hat a(t^\prime)\big\rangle-  \big\langle \hat a^\dagger(t)  \hat a(t^\prime)\big\rangle \frac{\hbar \omega}{2 c \epsilon_0 \sigma_\text{L}} \delta(t-t^\prime)\,.
\end{equation}
For radiation fields in coherent state, the first term is cancelled from the correlation function by the $\langle \hat a^\dagger(t)  \hat a(t^\prime)\rangle^2$ term. The coefficient of $\delta (t-t^\prime)$ in the second term is in normal order and can be replaced by c-numbers and then calculated identically as the force in \sref{sec:Force}, 
\begin{equation}
 \label{eq:TMM:NormalOrderMean}
 \big\langle \hat a^\dagger(t)  \hat a(t)\big\rangle \approx \Big| \sum A (k) \Big|^2 = \frac{\hbar \omega}{2 c \epsilon_0 \sigma_\text{L}} |A|^2\,,
\end{equation}
in terms of the photo-current intensity $|A|^2$.

Assembling all similar contributions, originating from the non-vanishing commutators $[b,b^\dagger]$, $[c,c^\dagger]$, $[d,d^\dagger]$, $[a,b^\dagger]$, etc., one obtains~\eref{eq:TMM:DiffCoeff} presented in \sref{sec:MSTDiffusion}.

\subsection{Appendix: Mirror cooling via the radiation pressure coupling Hamiltonian\label{sec:RadPressCool}}
We describe a generic optomechanical system composed of a single, damped-driven field mode coupled to the motion of a massive particle, whose Hamiltonian is given by~\cite{Courty2001,Vitali2003}
\begin{equation}
\hat{\mathcal{H}} =\hbar \omega_c \hat{a}^\dagger \hat{a}  + \i \hbar \eta (\hat{a}^\dagger e^{-\i \omega t} - \hat{a} e^{\i \omega t}) + \frac{\hat{p}^2}{2m} +V(\hat{x}) +\hbar G \hat{a}^\dagger \hat{a} \hat{x}\,.
\end{equation}
where $\hat{a}$ and $\hat{a}^\dagger$ are the annihilation and creation operators, respectively, of the mode; $\hat{x}$ and $\hat{p}$ are the position and momentum operators associated with the motion; and where we drop the carets to signify expectation values. The mode is driven by a coherent field with an effective amplitude $\eta$ and frequency $\omega$.  This Hamiltonian describes, for example, the radiation pressure coupling of a  moving mirror to the field in a Fabry-Perot resonator. In this case the coupling constant is $G=\omega_c/L$, rendering the cavity mode frequency detuning $\omega_cx/L$ provided the mirror is shifted by an amount $x$. Since the cavity mode is lossy with a photon escape rate of $2 \kappa$, the total system is dissipative. Thereby, with a proper setting of the parameters, in particular the cavity detuning $\Delta_C=\omega-\omega_C$, the mirror motion can be cooled. We will determine the corresponding friction force linear in velocity.

In a frame rotating at frequency $\omega$, the Heisenberg equation of motion for the field mode amplitude reads  
\begin{equation}
\dot{\hat{a}} = \left[ \i (\Delta_C -G \hat{x}) - \kappa\right] \hat{a} + \eta\,.
\end{equation}
where the noise term is omitted. We assume that the mirror moves along the trajectory $x(t) \approx x + v t$ with fixed velocity $v$ during the short time that is needed for the field mode to relax to its steady-state. The variation of $\hat{a}$ arises from the explicit time dependence and from the motion of the mirror.  A steady-state solution is sought in the form of $\hat{a} \approx \hat{a}^{(0)}(x) + v \hat{a}^{(1)}(x)$. On replacing this expansion into the above equation, and using the hydrodynamic derivative $\tfrac{\rmd}{\rmd t} \rightarrow \tfrac{\partial}{\partial t} + v \tfrac{\partial}{\partial x}$, one obtains a hierarchy of equations of different orders of the velocity $v$. To zeroth order the adiabatic field is obtained as
\begin{equation}
a^{(0)} = \frac{\eta}{-\i (\Delta_C-G x) + \kappa}\,.
\end{equation}
The linear response of $a$ to the mirror motion is then
\begin{equation}
a^{(1)} = \frac{1}{\i (\Delta_C-G x) - \kappa} \frac{\partial}{\partial x} a^{(0)} =  \frac{\i \eta G}{\big[-\i ( \Delta_C-G x) + \kappa\big]^3}\,.
\end{equation}

The force acting on the mirror derives from the defining equation $\dot{\hat{p}}= \tfrac{\i}{\hbar}[{\cal \hat{H}}, \hat{p}] =  -\hbar G \hat{a}^\dagger \hat{a}$ and, up to linear order in velocity, is
\begin{equation}
\force_1 =  - 2v \hbar G \re{{a^{(0)\ast}} a^{(1)}} =   4 v \frac{\hbar \eta^2 G^2 \kappa \Delta_C}{ \big( \Delta_C^2 + \kappa^2\big)^3}\,,
\end{equation}
where we used $x=0$ without loss of generality. It can be seen that mirror cooling requires that $\Delta_C<0$, \ie, the cavity resonance frequency is above the pump frequency. In this case, for efficient excitation of the field in the resonator, the frequency of the pump photons is up-shifted at the expense of the mirror's kinetic energy. This cooling force has been derived in \sref{sec:Mirrorcooling}, as a limiting case of the more general scattering theory. To check the perfect agreement between the two results,  the quantity corresponding to $\eta$ can be deduced from the total field energy in the resonator for an immobile mirror, which is $\hbar \omega_C {\hat{a}^{(0)\dagger}}\hat{a}^{(0)}$ here.

\subappendicesend

\section{General solution to the transfer matrix approach}\label{sec:TMM:General}
The approach used to find $\force$ in the previous section illustrates the nature of this quantity, arising from the multiple round-trips of the light between the static and moving mirrors, very well. However, the series summation does not lend itself to easy direct evaluation in a mechanical fashion. By rewriting the Doppler shift operator, however, this limitation can be lifted and $\force$ and $\diffn$ evaluated for a fully general optical system. The basis of this current section was published as Xuereb, A., Freegarde, T., Horak, P., \& Domokos, P. Phys.\ Rev.\ Lett.\ \textbf{105}, 013602 (2010), and will be elaborated on in a manuscript that is currently being prepared. In this section, we will first summarise the general transfer matrix approach and show how solutions in closed form can be found to the friction force, in \Sref{sec:TMM:Force}, and momentum diffusion, in \Sref{sec:Diffusion}, acting on the scatterer. \Sref{sec:TMM:Optomechanics} subsequently explores the linear and nonlinear optomechanical coupling that can be achieved between the field of a cavity and the position of a micromirror placed inside it.\par
Given a static scatterer represented by the transfer matrix $\hat{\mat{M}}$, the effect of the motion of the scatterer on the fields it is interacting with is included by applying the transformation [cf. \eref{eq:TMM:LML}]
\begin{equation}
 \hat{\mat{M}}\rightarrow\hat{\mat{L}}(-v)\hat{\mat{M}}\hat{\mat{L}}(v)\,,
\end{equation}
where the `Doppler shift' operator matrix $\hat{\mat{L}}(v)$ for a scatterer moving with velocity $v$ can be written, from \eref{eq:TMM:Lmatrix}, as
\begin{equation}
\label{eq:TMMFirstordervExplicit}
 \hat{\mat{L}}(v)=\begin{bmatrix}
  1+\tfrac{v}{c}\bigl(1-k_0\partial_k\bigr)&0\\
  0&1-\tfrac{v}{c}\bigl(1-k_0\partial_k\bigr)\\
 \end{bmatrix}\,,
\end{equation}
to linear order in $v/c$ and under the assumption that the pump beam has a very narrow spread of wavenumbers about a central $k_0$. We also use the shorthand notation $\partial_k\equiv\tfrac{\partial}{\partial k}$ throughout and will drop the label $k$ wherever this is not necessary.

\subsection{Force acting on moving scatterer}\label{sec:TMM:Force}
\begin{figure}[tb]
  \centering
  \includegraphics[scale=0.45]{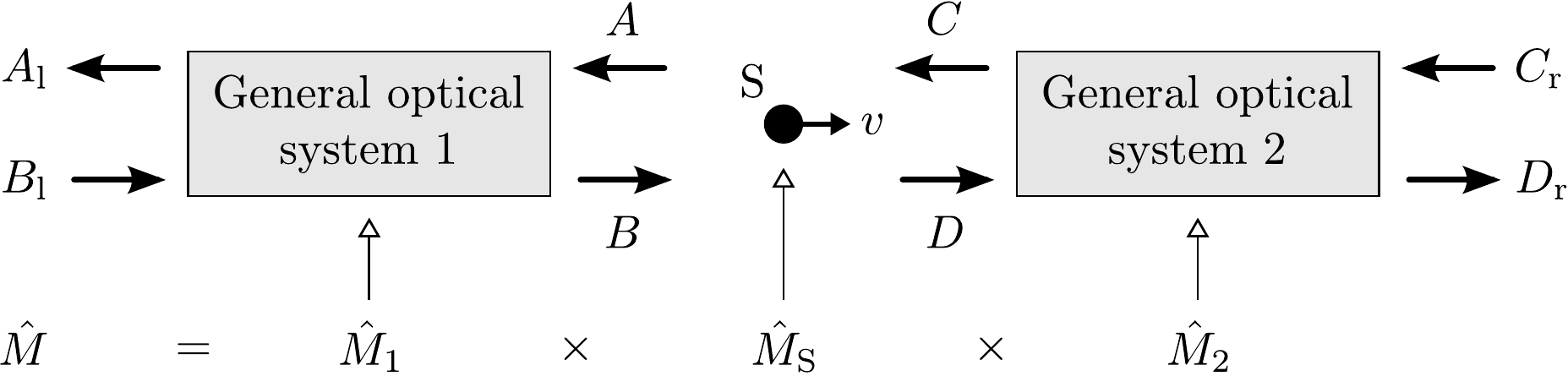}
   \caption[Schematic model of a scatterer interacting with two general optical systems]{The model we consider in this section, drawn schematically. A scatterer $\text{S}$ interacts with two `general optical systems' in one dimension, composed of immobile linear optical elements, one to either side. $\text{S}$ and these two systems are each represented by a $2\times 2$ matrix.}
  \label{fig:Schematic}
\end{figure}
Let us now apply this description to the general model represented by the system in \fref{fig:Schematic}. Our aim is to obtain the force acting on scatterer $\text{S}$; to do this we need to express the fields it interacts with, $A$, $B$, $C$, and $D$, in terms of the input fields $B_\text{l}$ and $C_\text{r}$. As in Ref.~\cite{Xuereb2010b}, we define
\begin{equation}
 \hat{\mat{M}}=\mat{M}_1\times\hat{\mat{M}}_\text{S}\times \mat{M}_2\equiv\begin{bmatrix}
  \hat{\gamma}&\hat{\alpha}\\
  \hat{\delta}&\hat{\beta}
 \end{bmatrix}\,\text{and}\,\bigl(\mat{M}_1\bigr)^{-1}\equiv\bigl[\theta_{ij}\bigr]\,.
\end{equation}
We also define the convenient velocity-independent quantities $\alpha_0$, $\alpha_1^{(0)}$, $\alpha_1^{(1)}$, etc., by
\begin{align}
 \hat{\alpha}&\equiv\alpha_0+\frac{v}{c}\Bigl(\alpha_1^{(0)}+\alpha_1^{(1)}\partial_k\Bigr)\,,\\ 
 \hat{\beta}&\equiv\beta_0+\frac{v}{c}\Bigl(\beta_1^{(0)}+\beta_1^{(1)}\partial_k\Bigr)\,,\\
 \hat{\gamma}&\equiv\gamma_0+\frac{v}{c}\Bigl(\gamma_1^{(0)}+\gamma_1^{(1)}\partial_k\Bigr)\,,\text{ and}\\ 
 \hat{\delta}&\equiv\delta_0+\frac{v}{c}\Bigl(\delta_1^{(0)}+\delta_1^{(1)}\partial_k\Bigr)\,.
\end{align}
A final assumption is needed to be able to express the fields in closed form---we assume that the two input fields are at the same frequency and have only a very small spread in wavenumber:
\begin{equation}
B_\text{l}=B_0\,\delta(k-k_0)\ \text{and}\ C_\text{r}=C_0\,\delta(k-k_0)\,.
\end{equation}
Then, the field amplitudes $\mathcal{A}=\int A(k)\,\rmd k$ and $\mathcal{B}=\int B(k)\,\rmd k$ are given, to first order in $v/c$, by:
\begin{multline}
\label{eq:Afield}
 \mathcal{A}=\biggl(\theta_{11}\frac{\alpha_0}{\beta_0}+\theta_{12}+\frac{v}{c}\biggl\{\frac{\theta_{11}}{\beta_0^2}\Bigl(\alpha_1^{(0)}\beta_0-\alpha_0\beta_1^{(0)}\Bigr)-\frac{1}{\beta_0}\biggl[\partial_k\frac{\theta_{11}}{\beta_0}\Bigl(\alpha_1^{(1)}\beta_0-\alpha_0\beta_1^{(1)}\Bigr)\biggr]\biggr\}\biggr)B_0\\
+\biggl(\theta_{11}\frac{\gamma_0\beta_0-\alpha_0\delta_0}{\beta_0}+\frac{v}{c}\biggl\{\frac{\theta_{11}}{\beta_0^2}\Bigl[\beta_0^2\gamma_1^{(0)}-\alpha_0\beta_0\delta_1^{(0)}-\Bigl(\alpha_1^{(0)}\beta_0-\alpha_0\beta_1^{(0)}\Bigr)\delta_0\Bigr]\\-\biggl[\partial_k\frac{\theta_{11}}{\beta_0}\Bigl(\beta_0\gamma_1^{(1)}-\alpha_0\delta_1^{(1)}\Bigr)\biggr]+\frac{\delta_0}{\beta_0}\biggl[\partial_k\frac{\theta_{11}}{\beta_0}\Bigl(\alpha_1^{(1)}\beta_0-\alpha_0\beta_1^{(1)}\Bigr)\biggr]\biggr\}\biggr)C_0\,,
\end{multline}
and
\begin{multline}
\label{eq:Bfield}
 \mathcal{B}=\biggl(\theta_{21}\frac{\alpha_0}{\beta_0}+\theta_{22}+\frac{v}{c}\biggl\{\frac{\theta_{21}}{\beta_0^2}\Bigl(\alpha_1^{(0)}\beta_0-\alpha_0\beta_1^{(0)}\Bigr)-\frac{1}{\beta_0}\biggl[\partial_k\frac{\theta_{21}}{\beta_0}\Bigl(\alpha_1^{(1)}\beta_0-\alpha_0\beta_1^{(1)}\Bigr)\biggr]\biggr\}\biggr)B_0\\
+\biggl(\theta_{21}\frac{\gamma_0\beta_0-\alpha_0\delta_0}{\beta_0}+\frac{v}{c}\biggl\{\frac{\theta_{21}}{\beta_0^2}\Bigl[\beta_0^2\gamma_1^{(0)}-\alpha_0\beta_0\delta_1^{(0)}-\Bigl(\alpha_1^{(0)}\beta_0-\alpha_0\beta_1^{(0)}\Bigr)\delta_0\Bigr]\\-\biggl[\partial_k\frac{\theta_{21}}{\beta_0}\Bigl(\beta_0\gamma_1^{(1)}-\alpha_0\delta_1^{(1)}\Bigr)\biggr]+\frac{\delta_0}{\beta_0}\biggl[\partial_k\frac{\theta_{21}}{\beta_0}\Bigl(\alpha_1^{(1)}\beta_0-\alpha_0\beta_1^{(1)}\Bigr)\biggr]\biggr\}\biggr)C_0\,,
\end{multline}
where the derivatives are all evaluated at $k=k_0$.
We shall find it useful to express these results in the form $\mathcal{A}=\mathcal{A}_0+\tfrac{v}{c}\mathcal{A}_1$ and $\mathcal{B}=\mathcal{B}_0+\tfrac{v}{c}\mathcal{B}_1$, with $\mathcal{A}_{0,1}$ and $\mathcal{B}_{0,1}$ being independent of $v$. For conciseness, let us now assume that $\zeta$ does not depend on $k$. Then, using the elements of $\hat{\mat{M}}_\text{S}$, we obtain
\begin{align}
 \mathcal{C}&=(1-\i\zeta)\mathcal{A}-\bigl(1-2\tfrac{v}{c}\bigl)\i\zeta\mathcal{B}\nonumber\\&=\bigl[(1-\i\zeta)\mathcal{A}_0-\i\zeta\mathcal{B}_0\bigr]+\tfrac{v}{c}\bigl[(1-\i\zeta)\mathcal{A}_1+2\i\zeta\mathcal{B}_0-\i\zeta\mathcal{B}_1\bigr]\,,
\end{align}
and
\begin{align}
 \mathcal{D}&=\bigl(1+2\tfrac{v}{c}\bigl)\i\zeta\mathcal{A}+(1+\i\zeta)\mathcal{B}\nonumber\\&=\bigl[\i\zeta\mathcal{A}_0+(1+\i\zeta)\mathcal{B}_0\bigr]+\tfrac{v}{c}\bigl[2\i\zeta\mathcal{A}_0-\i\zeta\mathcal{A}_1-(1+\i\zeta)\mathcal{B}_1\bigr]\,.
\end{align}
We denote the velocity-independent parts of $\mathcal{C}$ and $\mathcal{D}$ by $\mathcal{C}_0$ and $\mathcal{D}_0$, respectively. The \emph{friction} force acting on the scatterer can be finally written down as
\begin{align}
\label{eq:FullFriction}
  \force=-4\hbar k_0\frac{v}{c} \Bigl[&\lvert\zeta\rvert^2\bigl(\lvert\mathcal{A}_0\rvert^2-\lvert\mathcal{B}_0\rvert^2\bigr)+\bigl(\lvert\zeta\rvert^2+\im{\zeta}\bigr)\re{\mathcal{A}_0\mathcal{A}_1^\ast}-2\im{\zeta}\re{\mathcal{A}_0\mathcal{B}_0^\ast}\nonumber\\&+\bigl(\lvert\zeta\rvert^2-\im{\zeta}\bigr)\re{\mathcal{B}_0\mathcal{B}_1^\ast}+\im{\zeta}\re{\mathcal{A}_0\mathcal{B}_1^\ast}\nonumber\\&+\re{\Bigl(\lvert\zeta\rvert^2+\i\re{\zeta}\Bigr)\mathcal{A}_1\mathcal{B}_0^\ast}\Bigr]\,.
\end{align}

\subsection{Momentum diffusion experienced by scatterer}\label{sec:Diffusion}
Photon number fluctuations in the input fields $B_\text{l}$ and $C_\text{r}$ give rise to a stochastic force, added to $\force$, that averages out to zero; this stochastic force is, however, responsible for preventing the momentum of the scatter from reaching zero under the action of a friction force~\cite{Metcalf2003}. A complete theory based on the input-output operator formalism can be built along the lines of the previous section. The results we have obtained so far correspond to assuming that the input fields are in a coherent state, see \sref{sec:MSTDiffusion}, whereby the fields $A$, $B$, etc., are the expectation values of analogous bosonic annihilation operators: $\langle\hat{A}\rangle=A$, $\langle\hat{A^\dagger}\rangle=A^\ast$, and so on. The two input modes are the only independent modes in our system, and their corresponding operators obey the usual bosonic commutation relations:
\begin{equation}
 \comm{\hat{B}_\text{l}}{\hat{B}_\text{l}}=\comm{\hat{C}_\text{r}}{\hat{C}_\text{r}}=\communit\,,\text{ and }\comm{\hat{B}_\text{l}}{\hat{C}_\text{r}}=0\,,
\end{equation}
with $\omega=kc$ being the (angular) frequency. Since we are working to first order in $v/c$, we need to evaluate everything in this section to zeroth order in $v/c$.\par
Besides photon number fluctuations, lossy beamsplitters also induce noise into the system; this is equivalent to having a noise mode $\hat{E}$ that modifies the output fields interacting with that component. The noise input in the system by any of its components is independent of the noise input by any other component; in other words, if $\hat{E}_1$ and $\hat{E}_2$ are two noise modes that interact with the system,
\begin{equation}
  \comm{\hat{E}_1}{\hat{E}_1}=\comm{\hat{E}_2}{\hat{E}_2}=\communit\,,\comm{\hat{E}_1}{\hat{E}_2}=0\,,
\end{equation}
and
\begin{equation}
\langle\hat{E}_1\rangle=\langle\hat{E}_2\rangle=0\,,
\end{equation}
with the last pair of equalities being true by construction. We can therefore treat each of the noise modes independently. Moreover, the noise modes are independent of the input modes; for any such noise mode $\hat{E}$:
\begin{equation}
 \comm{\hat{E}}{\hat{B}_\text{l}}=\comm{\hat{E}}{\hat{C}_\text{r}}=0\,.
\end{equation}
In this section, we will generalise the treatment in \sref{sec:TMM:Model} to the situation where the loss-inducing beamsplitter is part of a generic optical system. Let us now assume, for concreteness, that the moving scatterer $\text{S}$ from \Sref{sec:TMM:Force} introduces the only noise mode, $\hat{E}$, in the system. In the case of an isolated scatterer, $\hat{B}$ and $\hat{C}$ represent independent input modes and the above equations guarantee that the two-time self-commutators of $\hat{A}$ and $\hat{D}$ behave as expected; see \aref{sec:TMM:AppDiffusion}. Let us now solve the general problem. One can rewrite \erefs{eq:ADNoise} using a matrix similar to the transfer matrix, obtaining:
\begin{align}
 \begin{pmatrix}
  \hat{A}\\
  \hat{B}
 \end{pmatrix}&=\frac{1}{\trans}\begin{bmatrix}\begin{array}{cc}
 \bigl(\trans^2-\refl^2\bigr)&\refl\\
  -\refl&1
 \end{array}\end{bmatrix}
 \begin{pmatrix}
  \hat{C}\\
  \hat{D}
 \end{pmatrix}+\frac{\sqrt{\epsilon}}{\trans}\begin{pmatrix}
  \hat{E}\\
  \hat{E}
 \end{pmatrix}\nonumber\\
&=\frac{1}{\trans}\begin{bmatrix}\begin{array}{cc|c}
 \bigl(\trans^2-\refl^2\bigr)&\refl&\phantom{+}\sqrt{\epsilon}\\
  -\refl&1&-\sqrt{\epsilon}
 \end{array}\end{bmatrix}
 \begin{pmatrix}
  \hat{C}\\
  \hat{D}\\
  \hat{E}
 \end{pmatrix}\nonumber\\
&=\begin{bmatrix}\begin{array}{c|c}\,\hat{\mat{M}}_\text{S}\,&\begin{matrix}
  \phantom{+}\sqrt{\epsilon}/\trans\\
  -\sqrt{\epsilon}/\trans
 \end{matrix}\end{array}\end{bmatrix}
 \begin{pmatrix}
  \hat{C}\\
  \hat{D}\\
  \hat{E}
 \end{pmatrix}\,,
\end{align}
where $\hat{\mat{M}}_\text{S}$ is included as a $2\times 2$ submatrix in the new $2\times 3$ matrix, which can in turn be embedded in a $3\times 3$ square matrix:\footnote{Note that \eref{eq:ABECDEMatrix} does \emph{not} imply violation of the principle of conservation of energy; indeed, the matrix is not a transfer matrix and the $\hat{E}$ mode is the same mode on either side of the equality.}
\begin{equation}
\label{eq:ABECDEMatrix}
 \begin{pmatrix}
  \hat{A}\\
  \hat{B}\\
  \hat{E}
 \end{pmatrix}=\begin{bmatrix}\begin{array}{c|c}\hat{\mat{M}}_\text{S}&\begin{matrix}
   \phantom{+}\sqrt{\epsilon}/\trans\\
   -\sqrt{\epsilon}/\trans
  \end{matrix}\\\hline
  \begin{matrix}\,0&&0\,\end{matrix}&1
 \end{array}\end{bmatrix}
 \begin{pmatrix}
  \hat{C}\\
  \hat{D}\\
  \hat{E}
 \end{pmatrix}\,.
\end{equation}
This assumes that the noise mode is essentially unaffected by the presence of the scatterer. In summary, then, each transfer matrix in the optical system is replaced by one of two matrices:
\begin{equation}
\hat{\mat{M}}\rightarrow
  \begin{bmatrix}\begin{array}{c|c}\hat{\mat{M}}&\begin{matrix}
   0\\
   0
  \end{matrix}\\\hline
  \begin{matrix}\,0&&0\,\end{matrix}&\,\,1\,\,
 \end{array}\end{bmatrix}\,,
\end{equation}
if the scatterer introduces no noise into the system, or
\begin{equation}
\hat{\mat{M}}\rightarrow
  \begin{bmatrix}\begin{array}{c|c}\hat{\mat{M}}&\begin{matrix}
   \phantom{+}\sqrt{\epsilon}/\trans\\
   -\sqrt{\epsilon}/\trans
  \end{matrix}\\\hline
  \begin{matrix}\,0&&0\,\end{matrix}&1
 \end{array}\end{bmatrix}\,,
\end{equation}
if the scatterer introduces the noise mode; as before, the properties of the scatterer are fully specified by $\refl$ and $\trans$. The third row and column of \emph{every} such matrix refer to the same, singular, noise mode being treated. Applying the above transformation to the matrices representing the generic system in \fref{fig:Schematic}, one obtains the field operators to zeroth order in $v/c$:
\begin{equation}
\label{eq:AfieldNoise}
 \hat{A}=\biggl(\theta_{11}\frac{\alpha_0}{\beta_0}+\theta_{12}\biggr)\hat{B}_\text{l}+\theta_{11}\frac{\gamma_0\beta_0-\alpha_0\delta_0}{\beta_0}\hat{C}_\text{r}+\theta_{11}\frac{\sqrt{\epsilon}}{\trans}\biggl[\bigl(m_{11}-m_{12}\bigr)+\bigl(m_{22}-m_{21}\bigr)\frac{\alpha_0}{\beta_0}\biggr]\hat{E}\,,
\end{equation}
and
\begin{equation}
\label{eq:BfieldNoise}
 \hat{B}=\biggl(\theta_{21}\frac{\alpha_0}{\beta_0}+\theta_{22}\biggr)\hat{B}_\text{l}+\theta_{21}\frac{\gamma_0\beta_0-\alpha_0\delta_0}{\beta_0}\hat{C}_\text{r}+\theta_{21}\frac{\sqrt{\epsilon}}{\trans}\biggl[\bigl(m_{11}-m_{12}\bigr)+\bigl(m_{22}-m_{21}\bigr)\frac{\alpha_0}{\beta_0}\biggr]\hat{E}\,,
\end{equation}
where we have used the shorthand notation $\bigl[m_{ij}\bigr]\equiv \mat{M}_1$. Here, $\hat{E}$ is the noise mode introduced by the scatterer itself. Other noise modes would give rise to contributions that are computed analogously. Now, denote $\hat{A}=b_\text{a}\hat{B}_\text{l}+c_\text{a}\hat{C}_\text{r}+e_\text{a}\hat{E}$ and $\hat{B}=b_\text{b}\hat{B}_\text{l}+c_\text{b}\hat{C}_\text{r}+e_\text{b}\hat{E}$ to this order in $v/c$; $b_\text{a,b}$, $c_\text{a,b}$, and $e_\text{a,b}$ can be read off \erefs{eq:AfieldNoise} and~(\ref{eq:BfieldNoise}). Then
\begin{align}
 \comm{\hat{A}}{\hat{A}}&=\communit\bigl(\lvert b_\text{a}\rvert^2+\lvert c_\text{a}\rvert^2+\lvert e_\text{a}\rvert^2\bigr)\,,\\
 \comm{\hat{B}}{\hat{B}}&=\communit\bigl(\lvert b_\text{b}\rvert^2+\lvert c_\text{b}\rvert^2+\lvert e_\text{b}\rvert^2\bigr)\,,\text{ and}\\
 \comm{\hat{A}}{\hat{B}}&=\communit\bigl(b_\text{a}b^\ast_\text{b}+c_\text{a}c^\ast_\text{b}+e_\text{a}e^\ast_\text{b}\bigr)\,.
\end{align}
We can also express $\hat{C}=b_\text{c}\hat{B}_\text{l}+c_\text{c}\hat{C}_\text{r}+e_\text{c}\hat{E}$ and $\hat{D}=b_\text{d}\hat{B}_\text{l}+c_\text{d}\hat{C}_\text{r}+e_\text{d}\hat{E}$, where
\begin{align}
 &b_\text{c}=b_\text{a}/\trans-b_\text{b}\refl/\trans\,,\ 
 c_\text{c}=c_\text{a}/\trans-c_\text{b}\refl/\trans\,,\text{ and}\ 
 e_\text{c}=e_\text{a}/\trans-e_\text{b}\refl/\trans-\sqrt{\epsilon}/\trans\,;\\
 &b_\text{d}=b_\text{a}\refl/\trans+b_\text{b}\bigl(\trans^2-\refl^2\bigr)/\trans\,,\ 
 c_\text{d}=c_\text{a}\refl/\trans+c_\text{b}\bigl(\trans^2-\refl^2\bigr)/\trans\,,\text{ and}\nonumber\\
 &e_\text{d}=e_\text{a}\refl/\trans+e_\text{b}\bigl(\trans^2-\refl^2\bigr)\refl/\trans+\sqrt{\epsilon}/\trans\,.
\end{align}
Therefore, e.g.,
\begin{align}
 \comm{\hat{C}}{\hat{C}}&=\communit\bigl(\lvert b_\text{c}\rvert^2+\lvert c_\text{c}\rvert^2+\lvert e_\text{c}\rvert^2\bigr)\,,\text{ and }\\
 \comm{\hat{D}}{\hat{D}}&=\communit\bigl(\lvert b_\text{d}\rvert^2+\lvert c_\text{d}\rvert^2+\lvert e_\text{d}\rvert^2\bigr)\,.
\end{align}
Several other nontrivial commutators need to be computed, due to the presence of $\hat{E}$; for example,
\begin{equation}
 \comm{\hat{A}}{\hat{D}}=\communit\bigl(b_\text{a}b^\ast_\text{d}+c_\text{a}c^\ast_\text{d}+e_\text{a}e^\ast_\text{d}\bigr)\,.
\end{equation}
For an isolated scatterer in free space, $b_\text{b}=c_\text{c}=1$, $b_\text{c}=c_\text{b}=0$, $e_\text{b}=e_\text{c}=0$, etc., and one can show that the various commutators behave as expected. Finally, the momentum diffusion constant is given by
\begin{align}
 \diffn\,\delta(t-t^\prime)=2\epsilon_0\sigma_\text{L}\,\hbar k_0\Bigl(&\lvert\mathcal{A}_0\rvert^2\comm{\hat{A}}{\hat{A}}+\lvert\mathcal{B}_0\rvert^2\comm{\hat{B}}{\hat{B}}+\lvert\mathcal{C}_0\rvert^2\comm{\hat{C}}{\hat{C}}+\lvert\mathcal{D}_0\rvert^2\comm{\hat{D}}{\hat{D}}\nonumber\\
&+2\lre{\mathcal{A}_0^\ast\mathcal{B}_0\comm{\hat{A}}{\hat{B}}-\mathcal{A}_0^\ast\mathcal{C}_0\comm{\hat{A}}{\hat{C}}}\nonumber\\
&\phantom{2Re\qquad}\left.-\mathcal{A}_0^\ast\mathcal{D}_0\comm{\hat{A}}{\hat{D}}-\mathcal{B}_0^\ast\mathcal{C}_0\comm{\hat{B}}{\hat{C}}\right.\nonumber\\
&\phantom{2Re\qquad}\left.-\mathcal{B}_0^\ast\mathcal{D}_0\comm{\hat{B}}{\hat{D}}+\mathcal{C}_0^\ast\mathcal{D}_0\comm{\hat{C}}{\hat{D}}\right\}\Bigr)\,,
\end{align}
where, for example, $\mathcal{A}_0^\ast \mathcal{D}_0=\int\langle\hat{A}^\dagger\rangle\rmd k\int\langle\hat{D}\rangle\rmd k=b^\ast_\text{a}b_\text{d}\lvert B_0\rvert^2+c^\ast_\text{a}c_\text{d}\lvert C_0\rvert^2+b^\ast_\text{a}c_\text{d}B^\ast_0C_0+c^\ast_\text{a}b_\text{d}C^\ast_0B_0$, recalling that $\langle\hat{E}\rangle=0$, and analogously for the rest of the terms. This result again reduces to the expected one for an isolated scatterer in free space; it can alternatively be extended to include further noise modes.

\section{Optomechanics of a micromirror inside a cavity}\label{sec:TMM:Optomechanics}
\begin{figure}
  \centering
  \subfigure[\ $\zeta=-0.100$]{
    \includegraphics[width=0.4\textwidth]{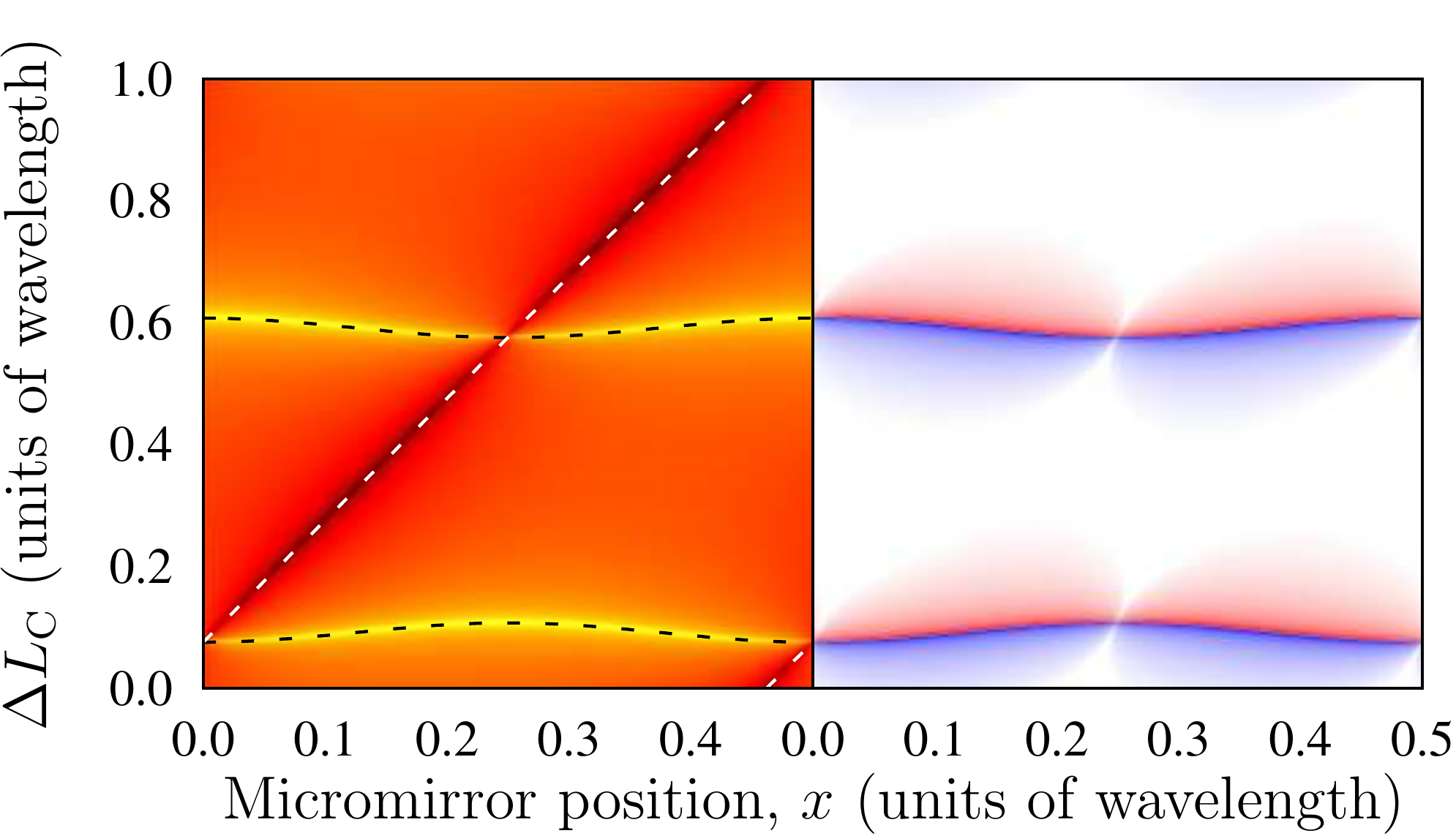}
  }
  \subfigure[\ $\zeta=-0.300$]{
    \includegraphics[width=0.4\textwidth]{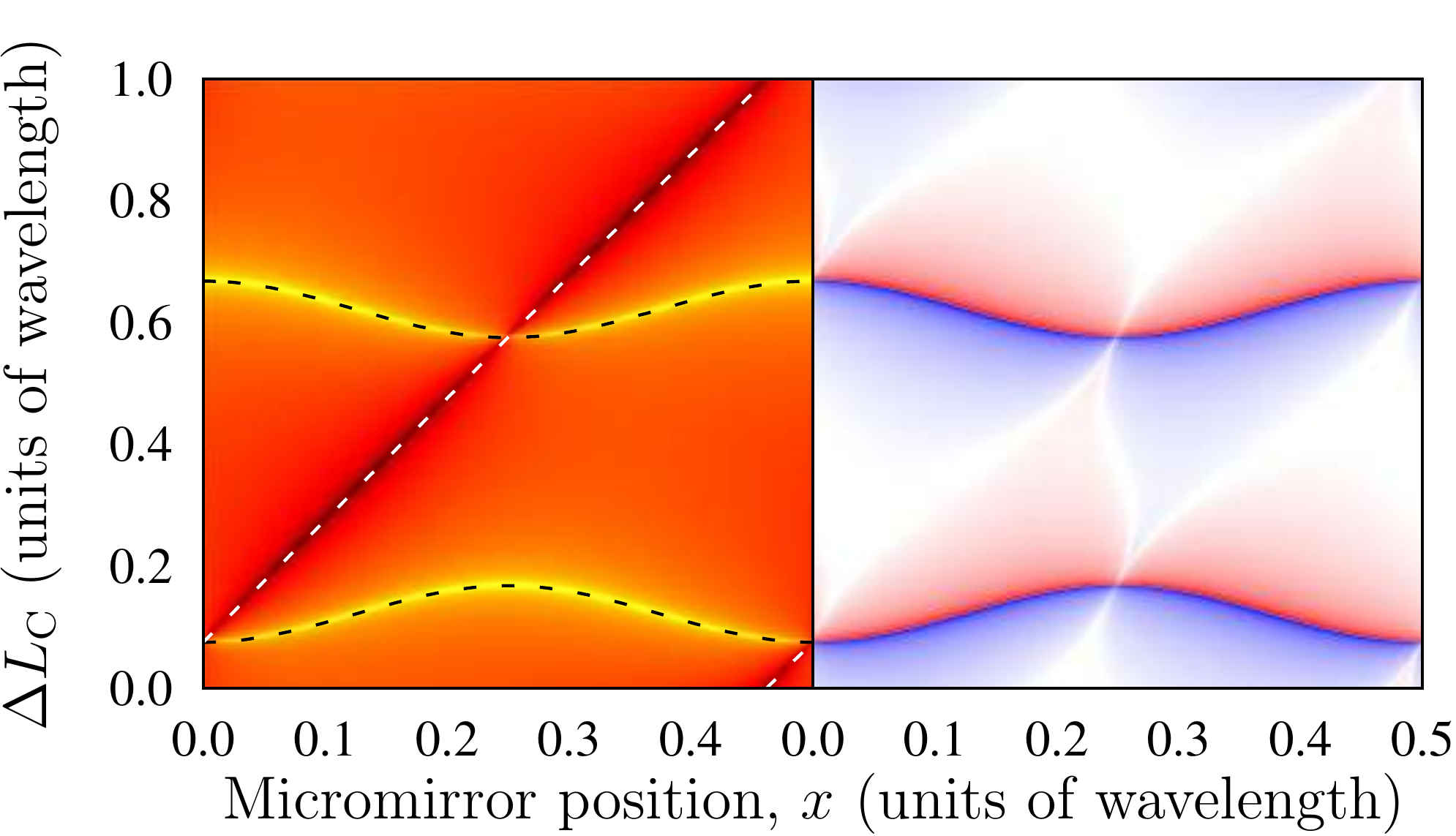}
  }\\
  \subfigure[\ $\zeta=-0.500$]{
    \includegraphics[width=0.4\textwidth]{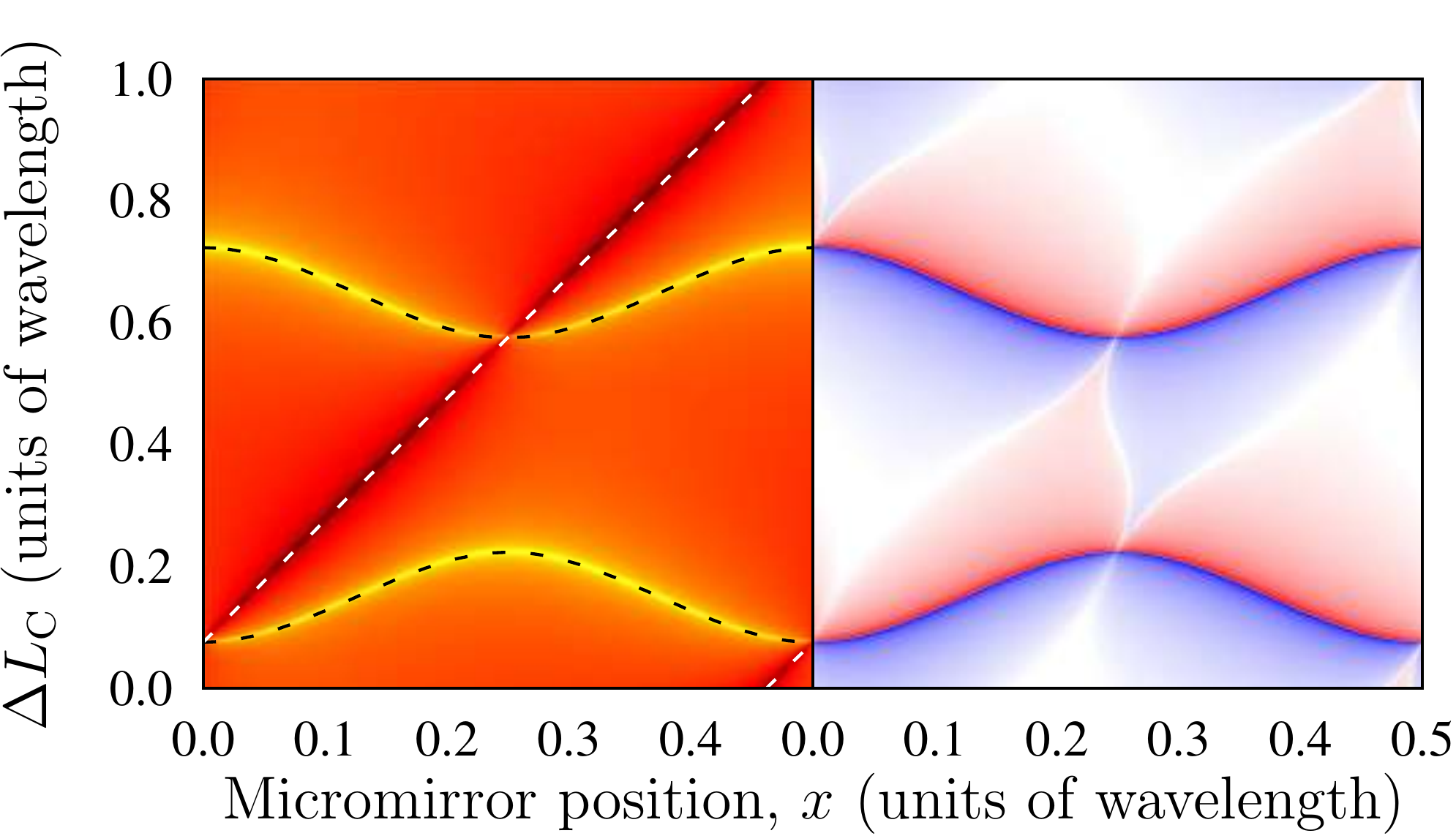}
  }
  \subfigure[\ $\zeta=-0.700$]{
    \includegraphics[width=0.4\textwidth]{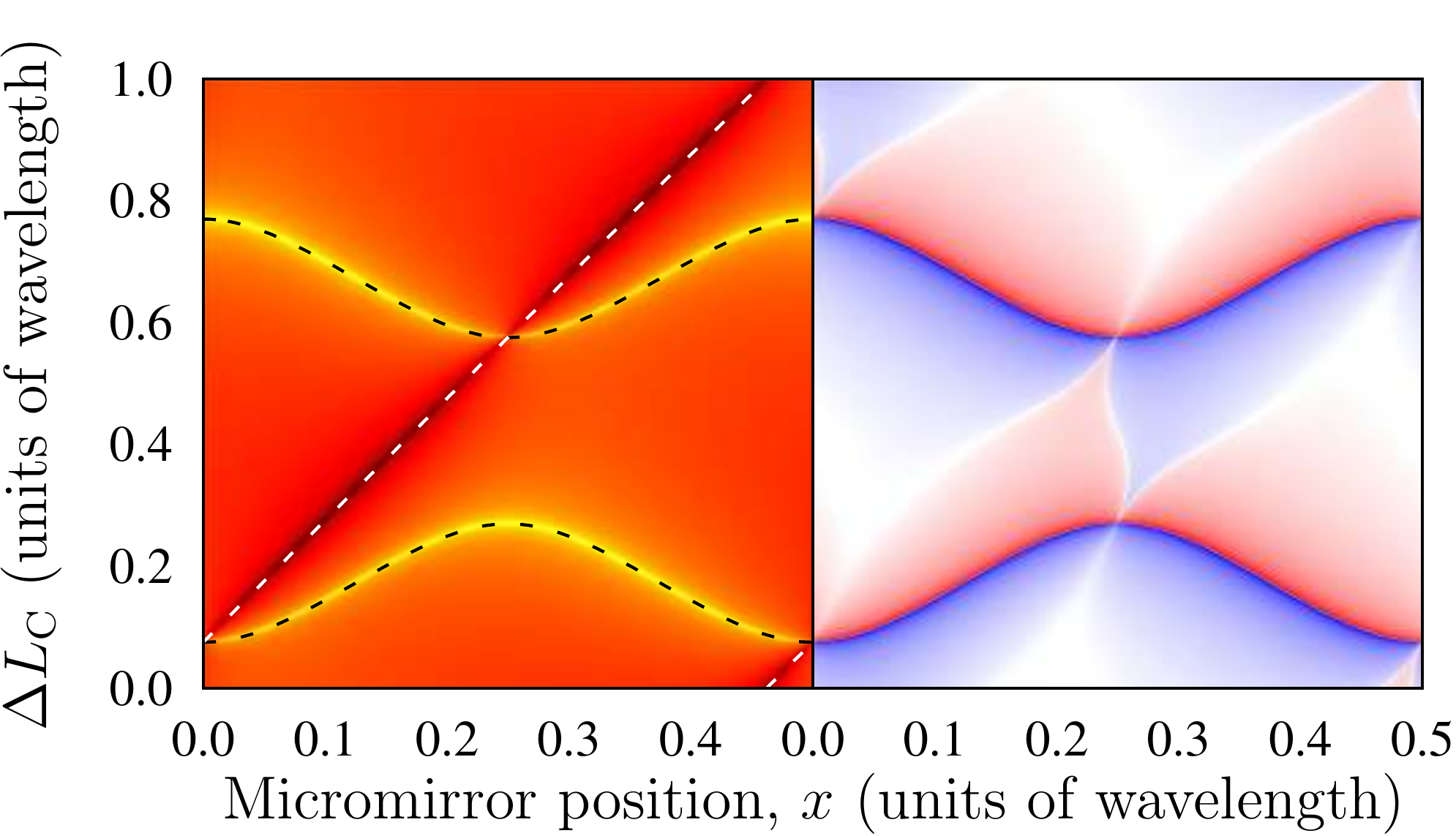}
  }\\
  \subfigure[\ $\zeta=-1.000$]{
    \includegraphics[width=0.4\textwidth]{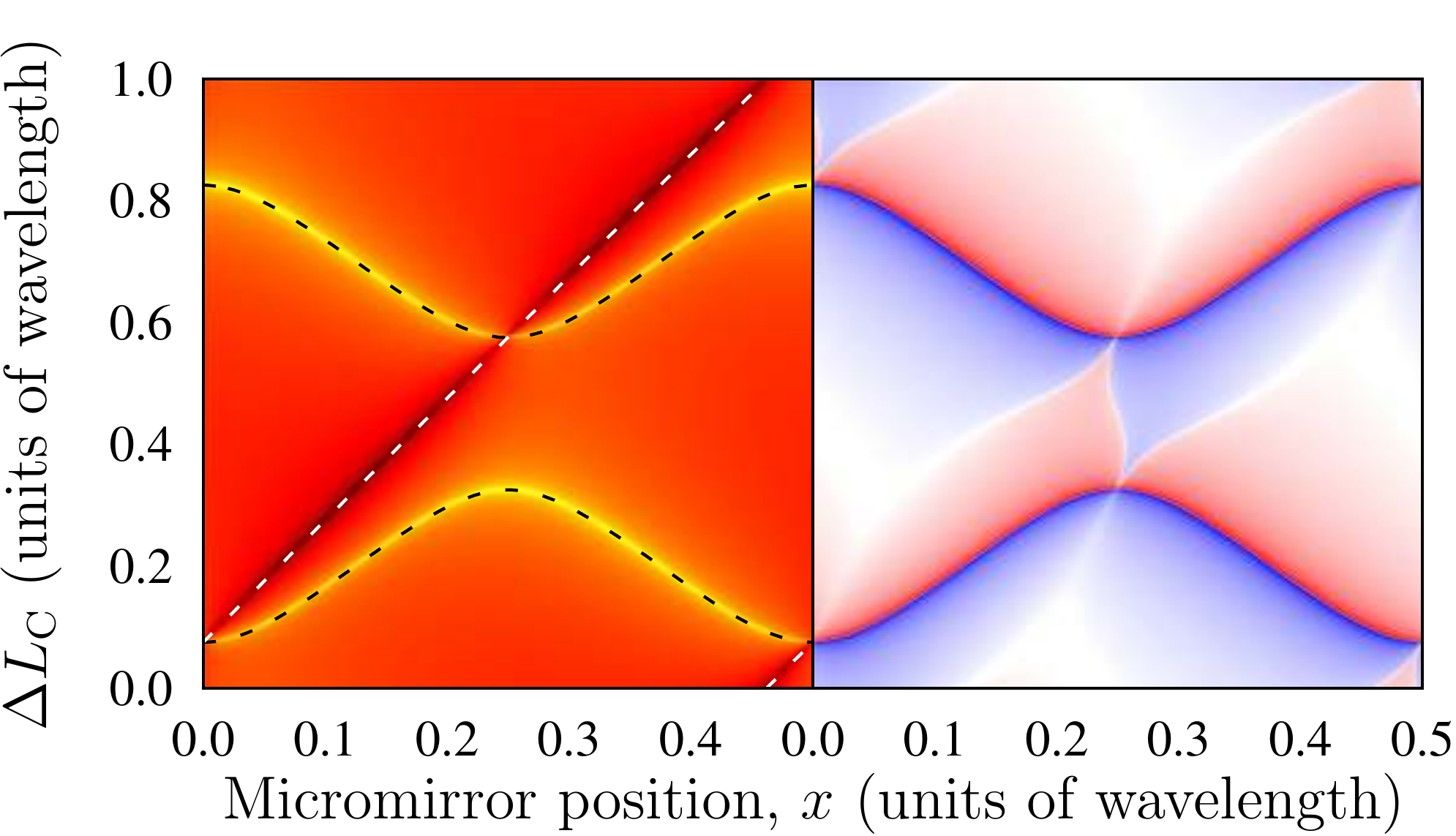}
  }
  \subfigure[\ $\zeta=-2.000$]{
    \includegraphics[width=0.4\textwidth]{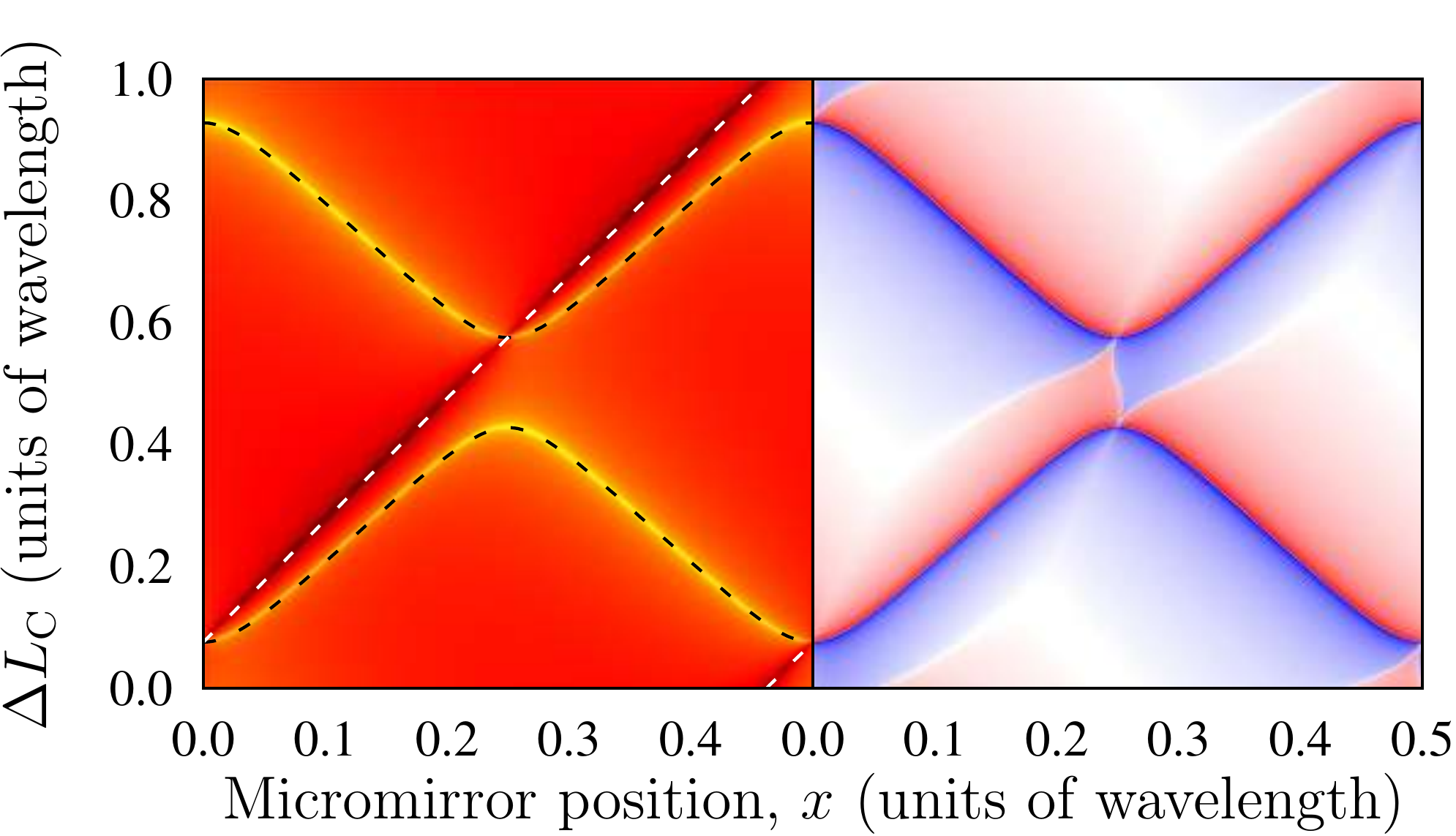}
  }\\
  \subfigure[\ $\zeta=-5.000$]{
    \includegraphics[width=0.4\textwidth]{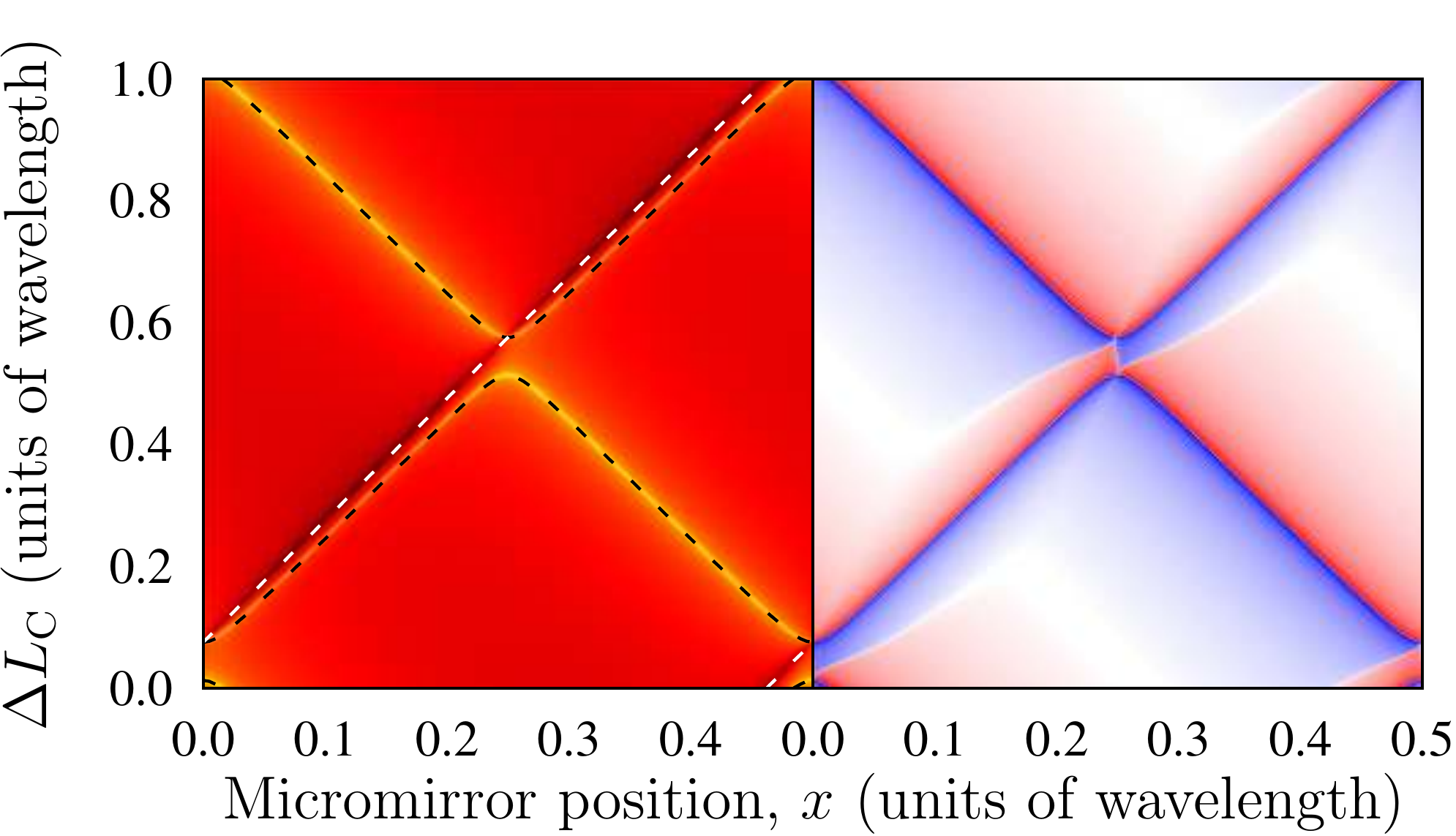}
  }
  \subfigure[\ $\zeta=-10.000$]{
    \includegraphics[width=0.4\textwidth]{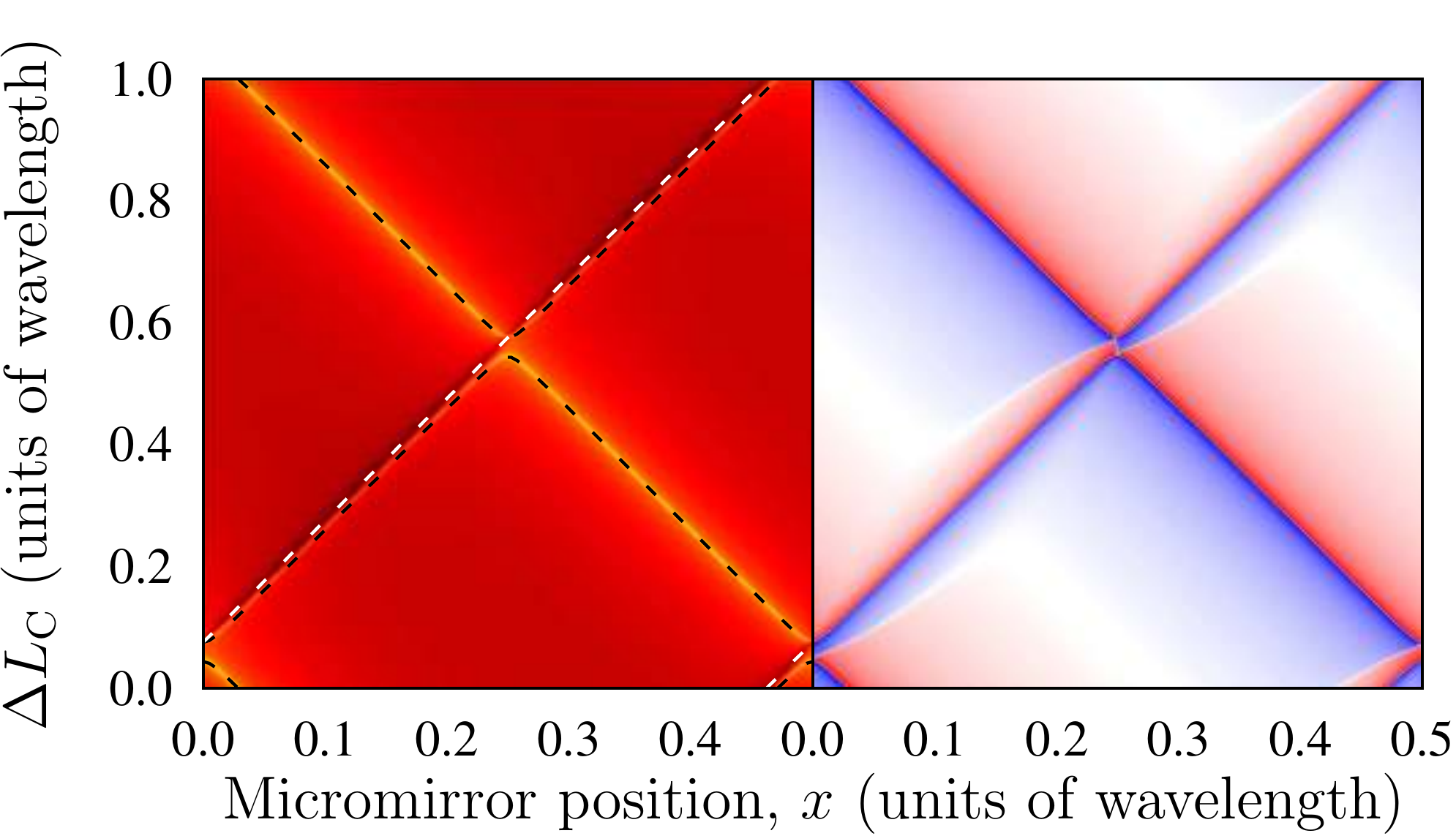}
  }\\
  \subfigure{
    \includegraphics[scale=0.3]{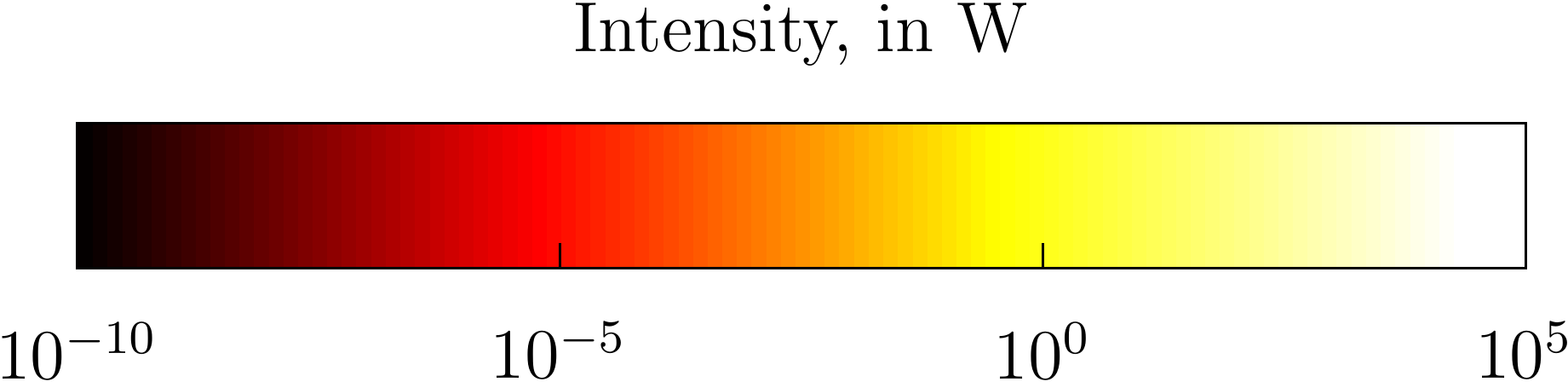}
  }\hspace{1cm}
  \subfigure{
    \includegraphics[scale=0.3]{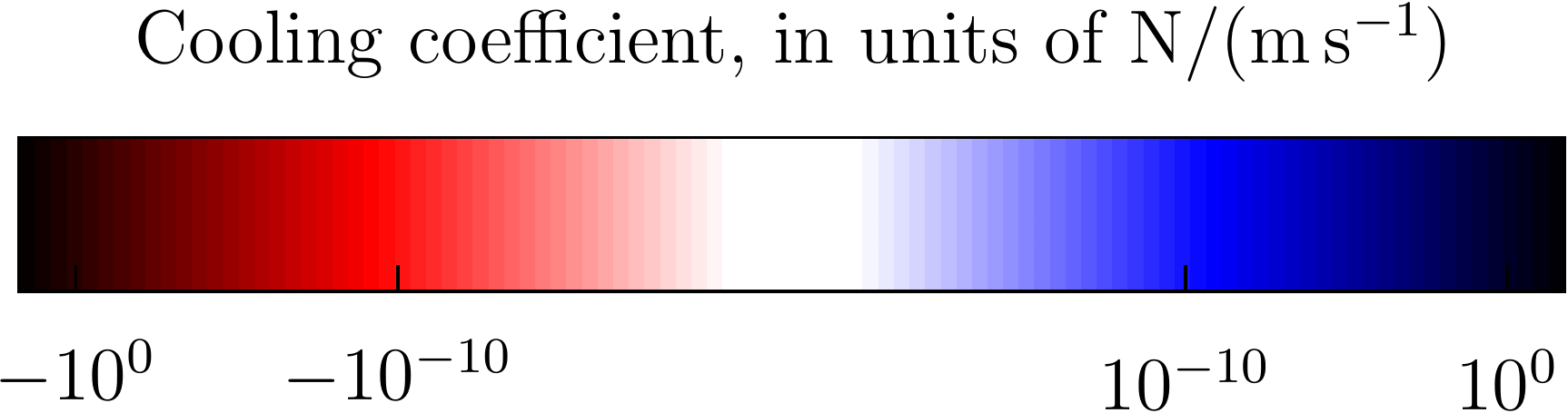}
  }
  \caption[Field intensity and cooling coefficient acting on a micromirror inside a cavity]{Field intensity (left panels) at and cooling coefficient (right panels) acting on the micromirror as the micromirror position ($x$) and cavity length ($L_\text{C}+\Delta L_\text{C}$) are scanned. The subfigures differ only in the polarisability of the mirror, as indicated. The cavity parameters are modelled from Ref.~\cite{Thompson2008}. In the series of left panels, we note the progression from an almost bare cavity situation (a) to a very strong perturbation by the micromirror, leading to avoided crossings (h). The white dashed line traces a cavity node, whereas the black dashed lines [\eref{eq:CavityResonances}] trace the cavity resonances. In the series of right panels, note that the cooling coefficient is---as expected---a cooling force (blue) for red cavity detuning and a heating force (red) for blue detuning. The colourbars are on a logarithmic scale and are for $1$\,W of input power; the large dynamic range is needed to bring out the detail in the panels, due to the very narrow features present.}
  \label{fig:IaF}
\end{figure}
\begin{figure}[t]
  \centering
  \subfigure[]{
  \includegraphics[width=1.5\figwidth]{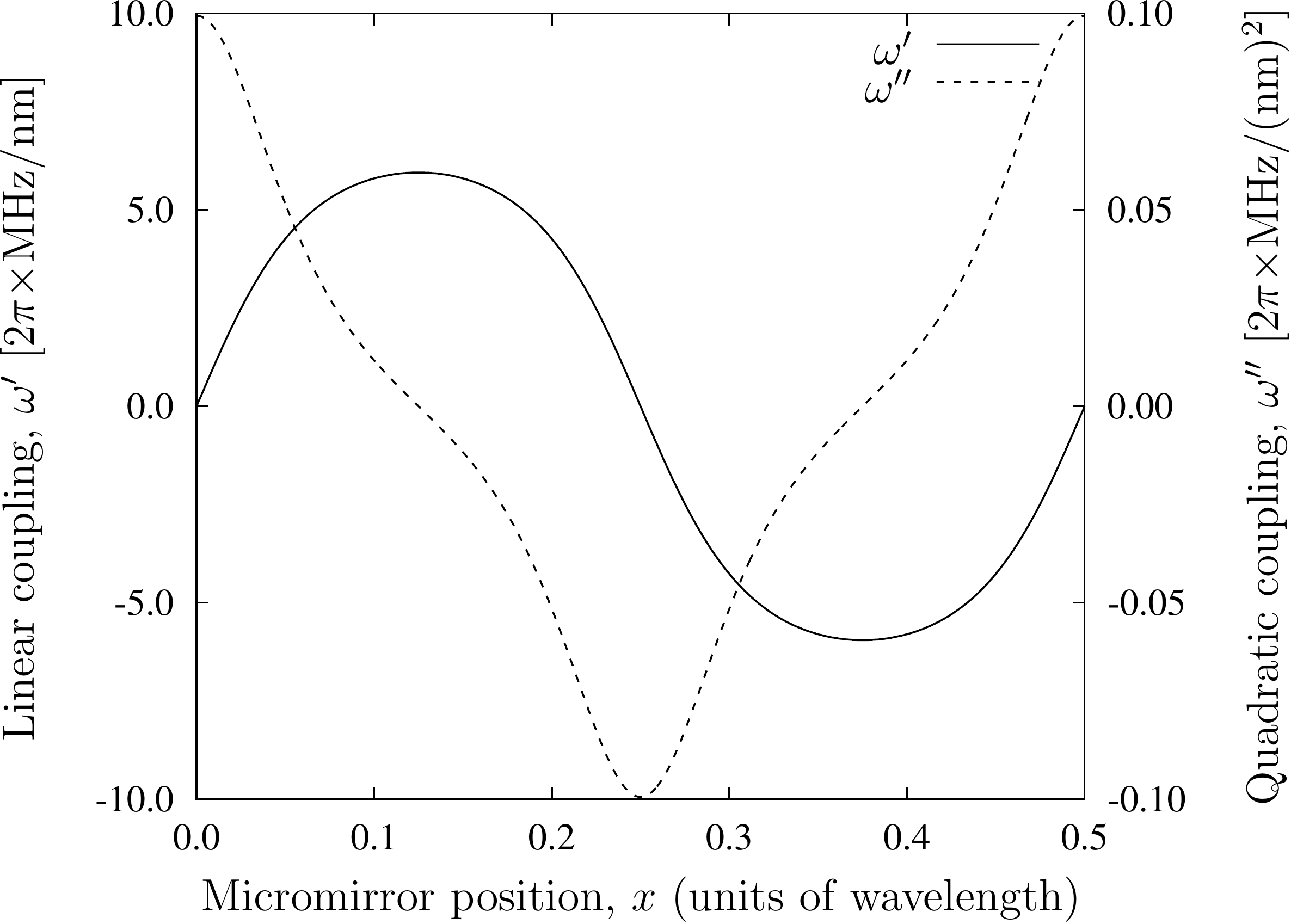}
  }\hspace{1cm}
  \subfigure[]{
  \includegraphics[width=1.5\figwidth]{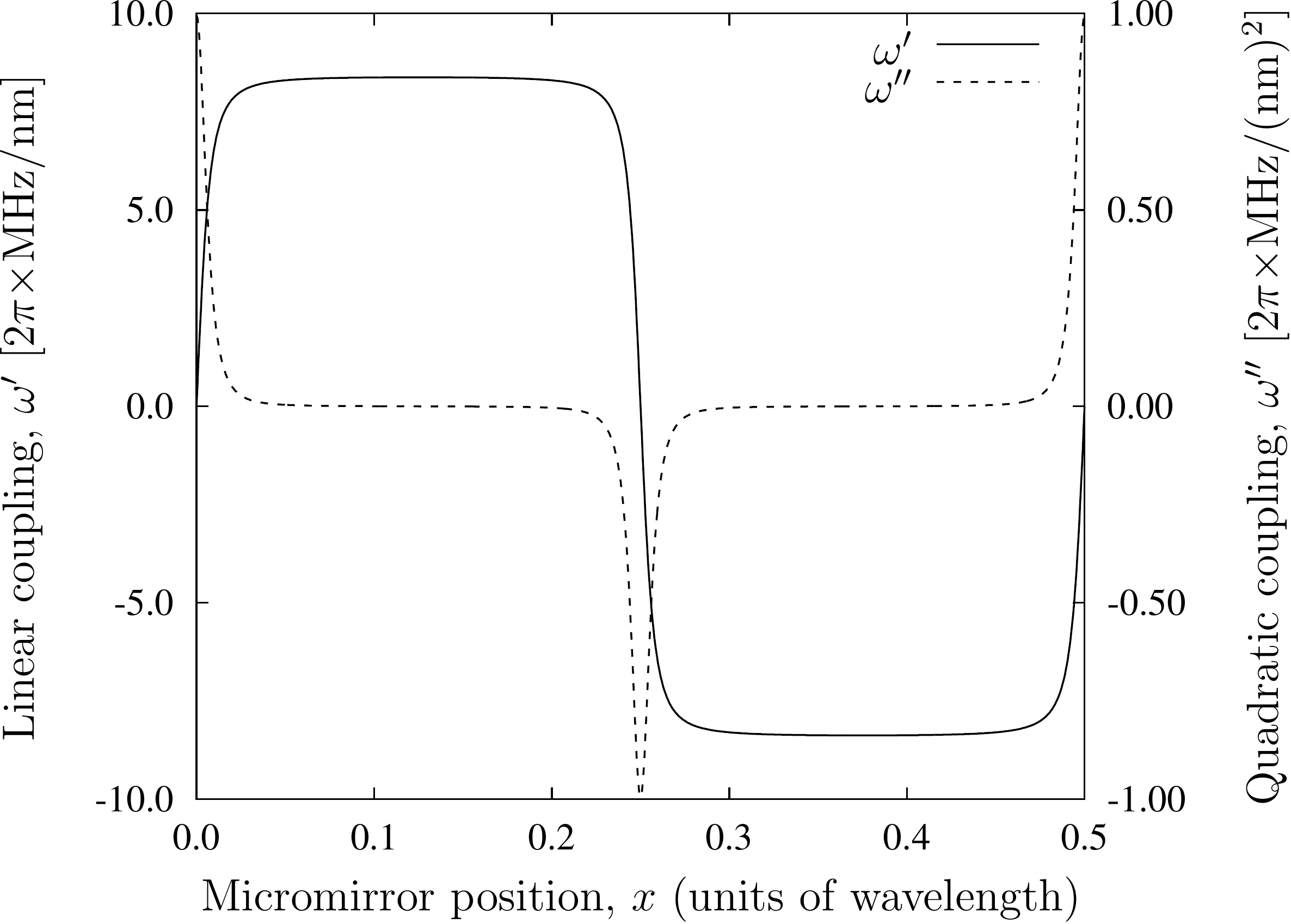}
  }
  \caption[Linear and quadratic optomechanical couplings as a function of mirror position]{Linear and quadratic optomechanical couplings as a function of mirror position for a very good cavity and for (a)~$\zeta=-1$, and (b)~$\zeta=-10$. In each figure we show the linear (solid curve) and quadratic (dashed curve) couplings, from \erefs{eq:LinearCoupling}~and~(\ref{eq:QuadraticCoupling}). Note that the peak value of $\omega^{\prime\prime}$ is roughly proportional to $\zeta$ whereas $\omega^\prime$ is bounded.}
  \label{fig:Couplings}
\end{figure}
\begin{figure}
  \centering
  \subfigure[\ $\zeta=-0.100$]{
    \includegraphics[width=0.4\textwidth]{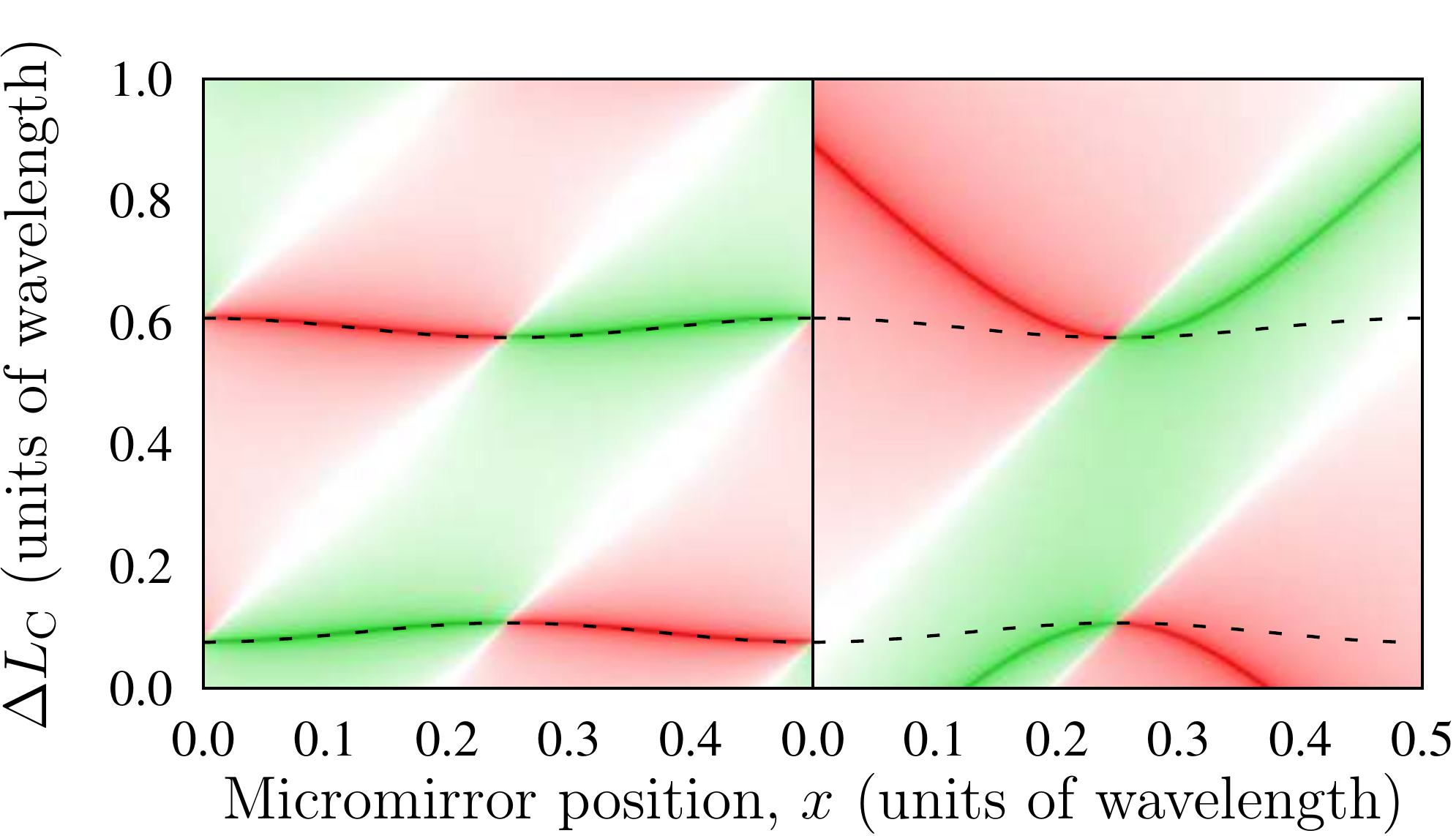}
  }
  \subfigure[\ $\zeta=-0.300$]{
    \includegraphics[width=0.4\textwidth]{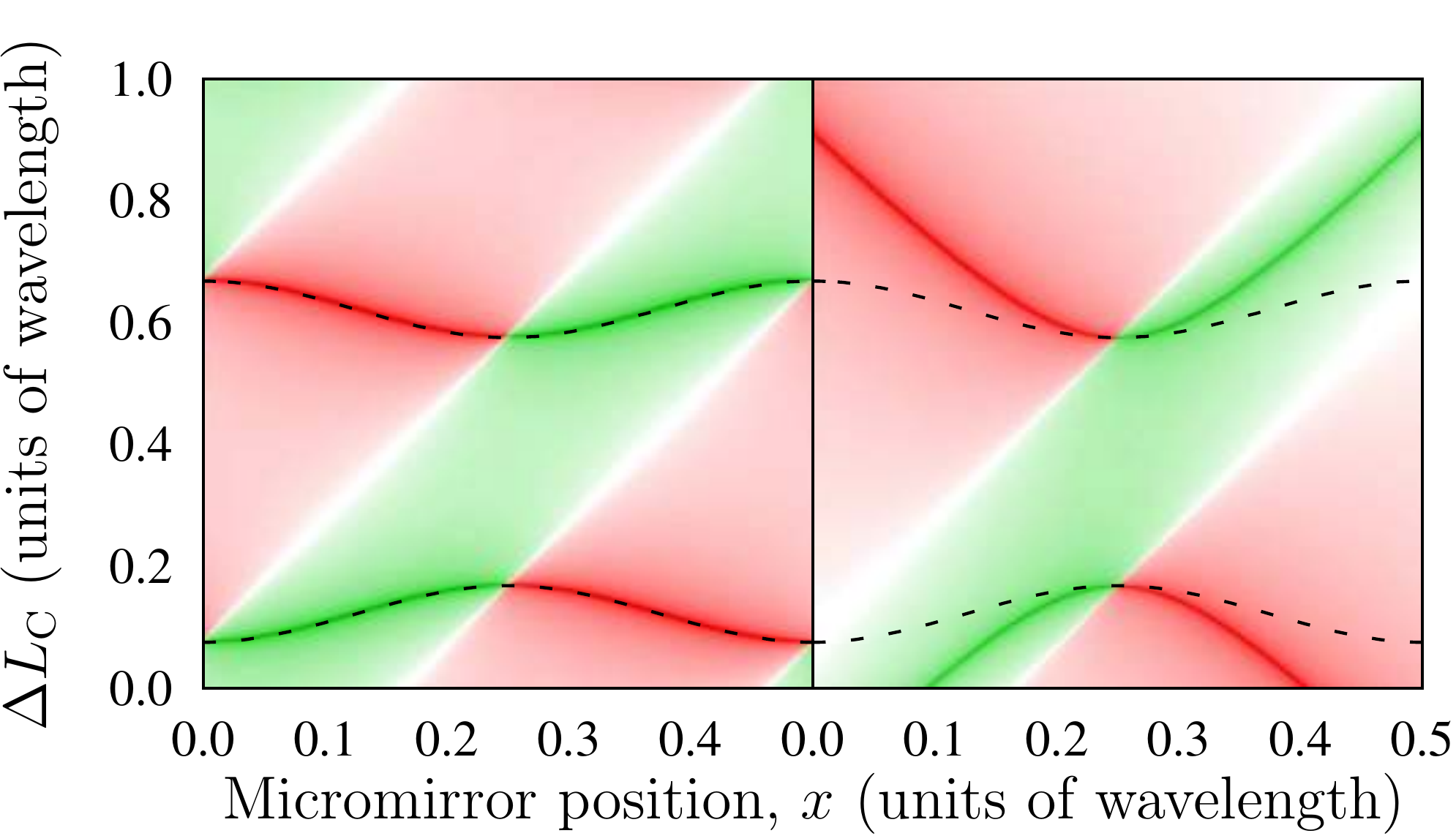}
  }\\
  \subfigure[\ $\zeta=-0.500$]{
    \includegraphics[width=0.4\textwidth]{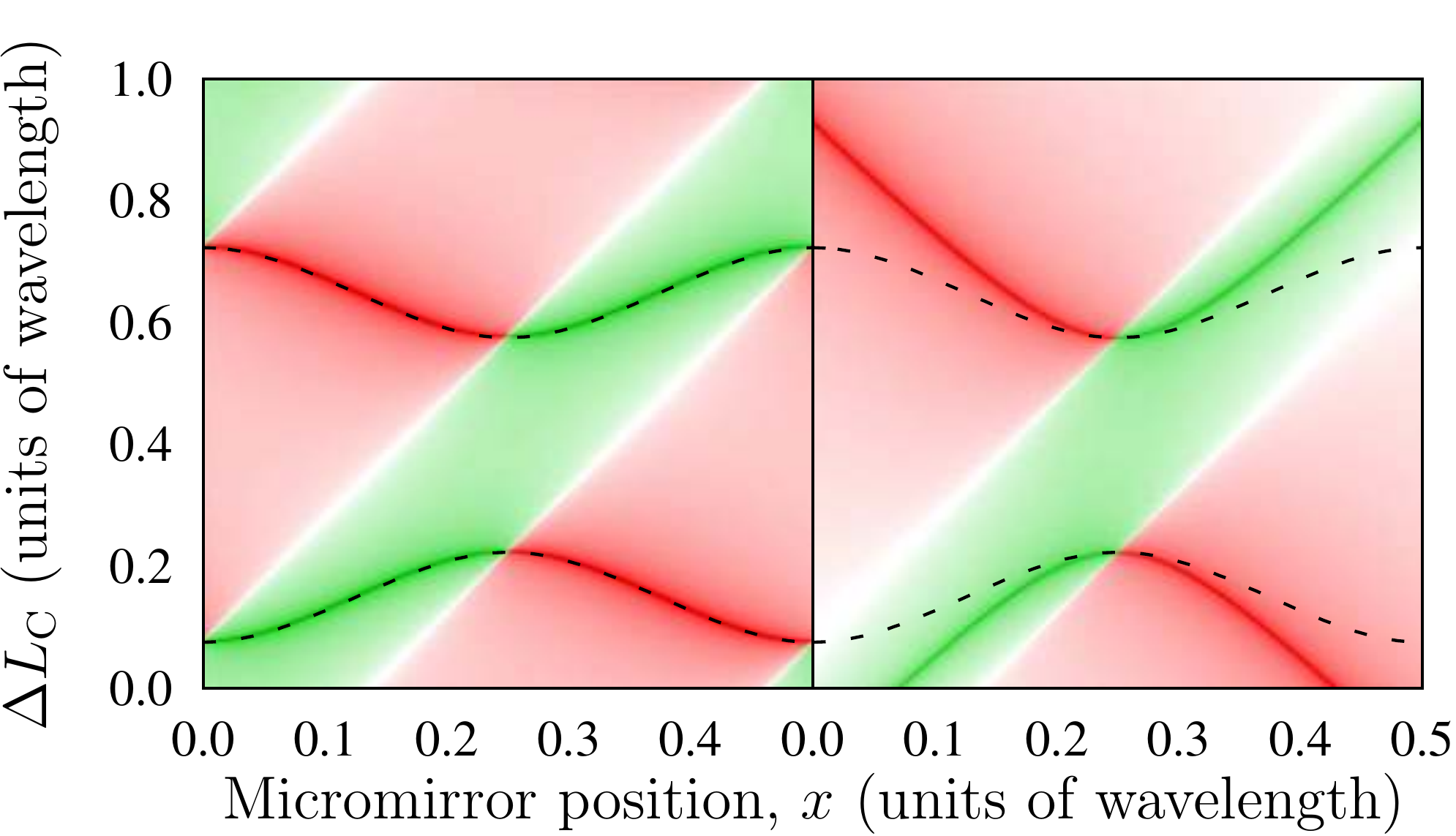}
  }
  \subfigure[\ $\zeta=-0.700$]{
    \includegraphics[width=0.4\textwidth]{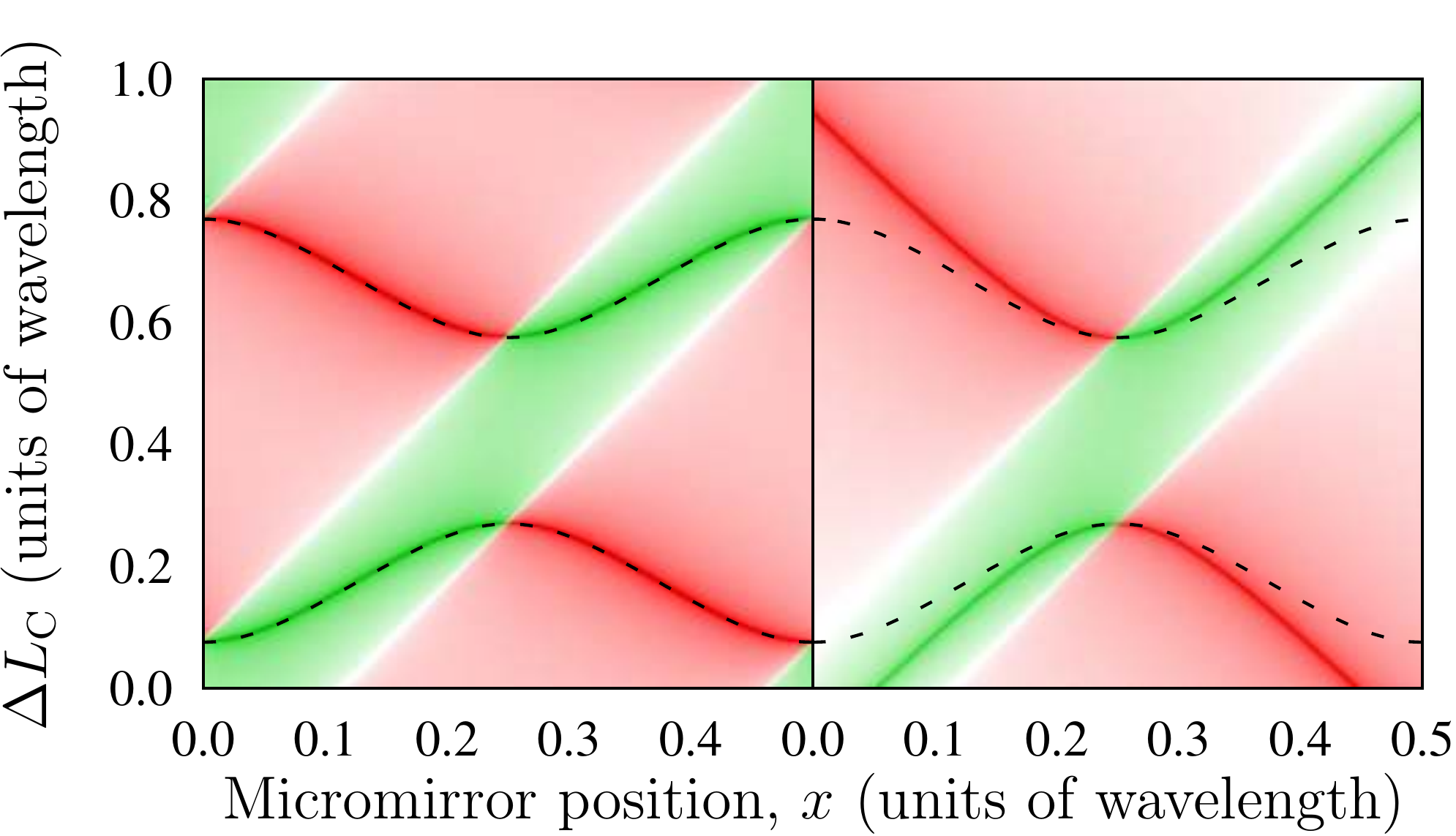}
  }\\
  \subfigure[\ $\zeta=-1.000$]{
    \includegraphics[width=0.4\textwidth]{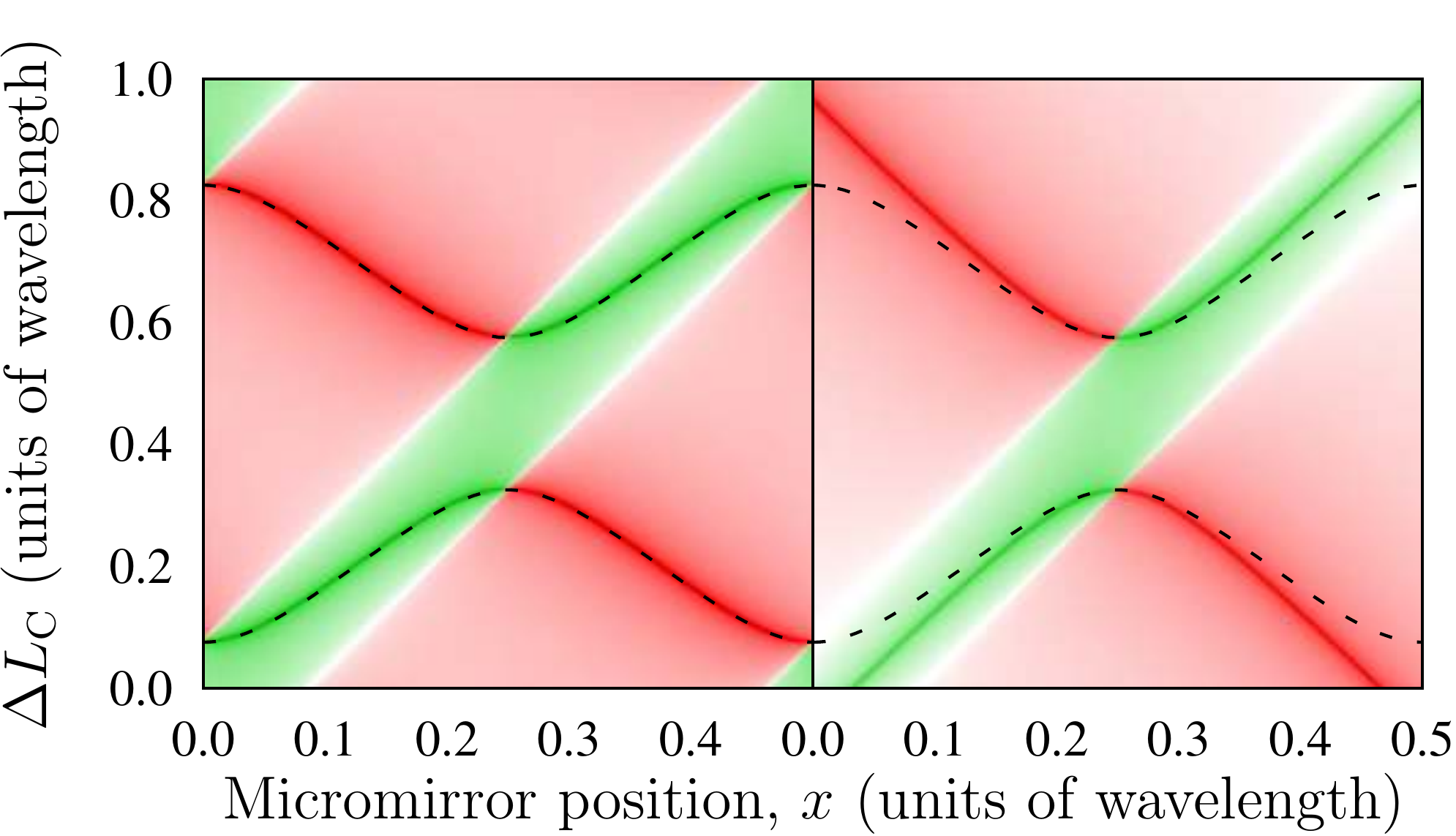}
  }
  \subfigure[\ $\zeta=-2.000$]{
    \includegraphics[width=0.4\textwidth]{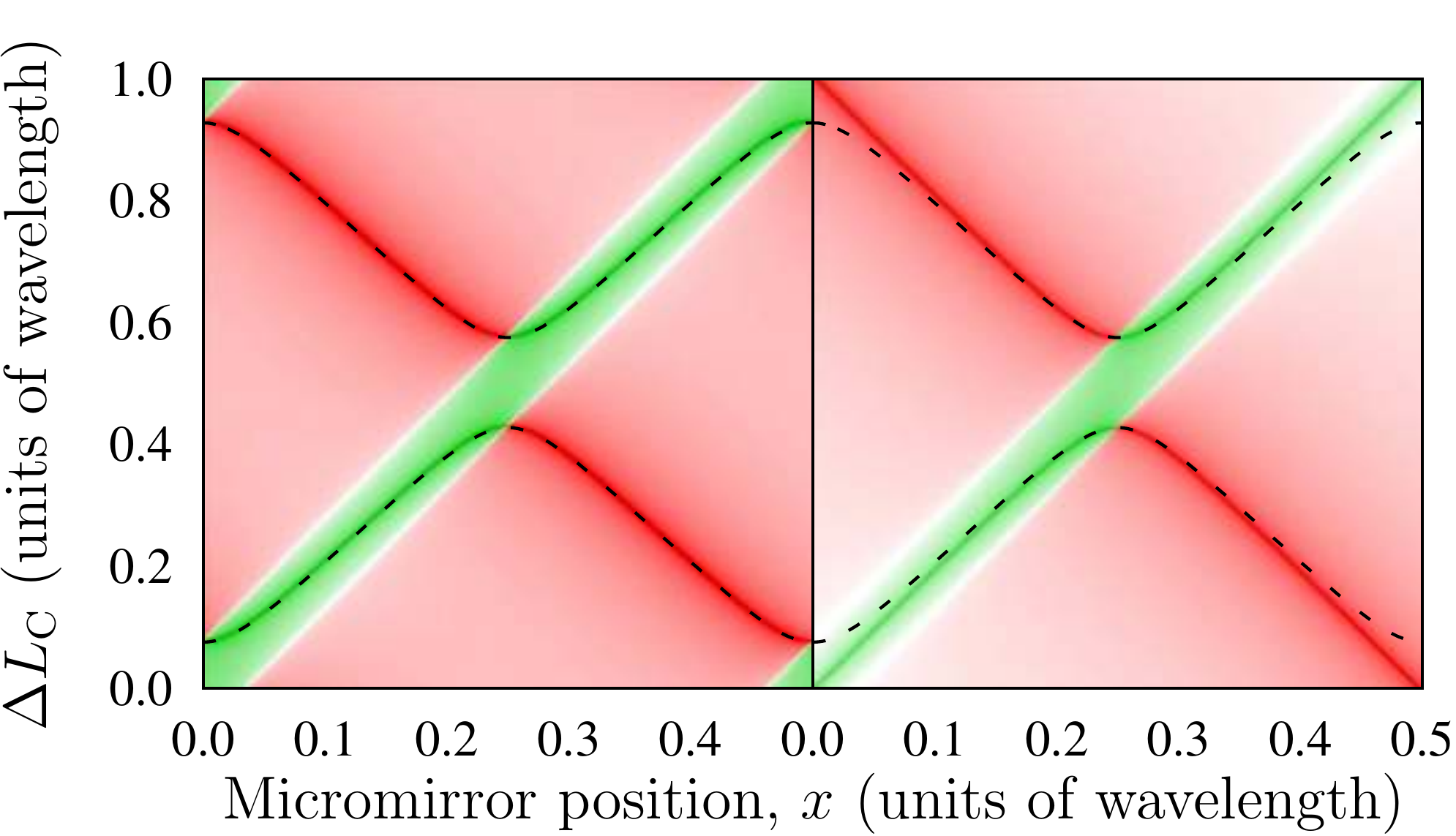}
  }\\
  \subfigure[\ $\zeta=-5.000$]{
    \includegraphics[width=0.4\textwidth]{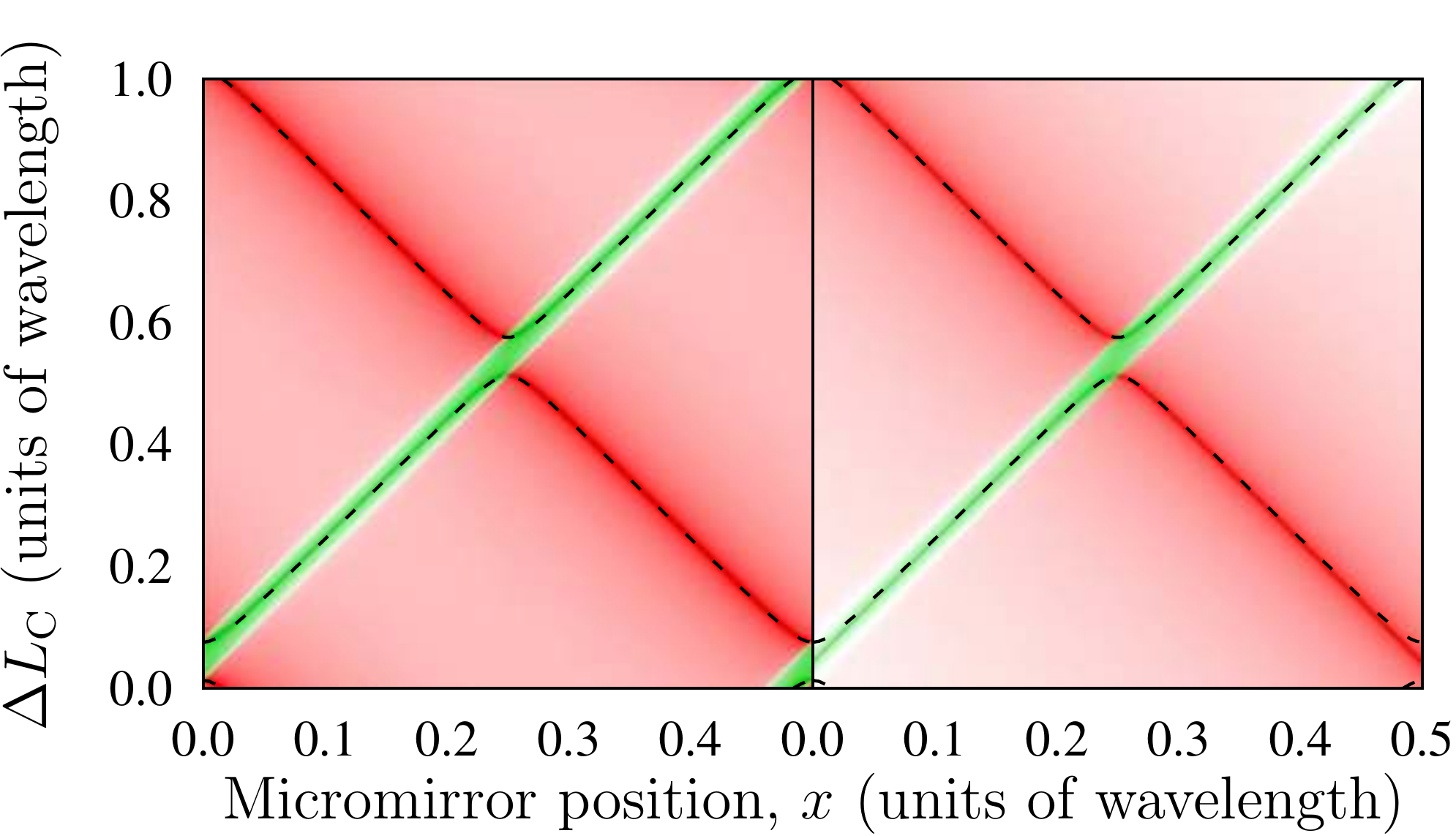}
  }
  \subfigure[\ $\zeta=-10.000$]{
    \includegraphics[width=0.4\textwidth]{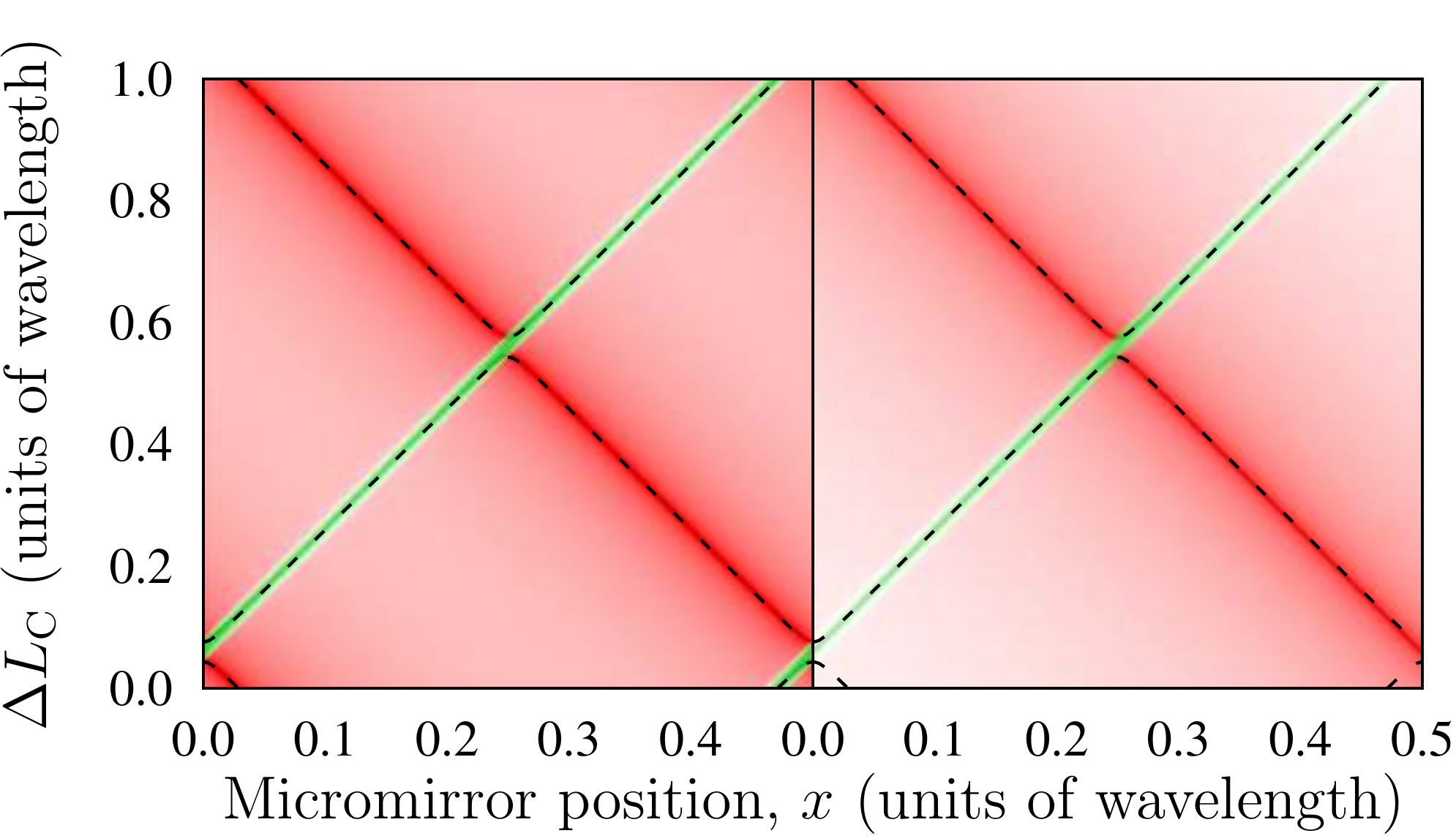}
  }\\
  \subfigure{
    \includegraphics[scale=0.3]{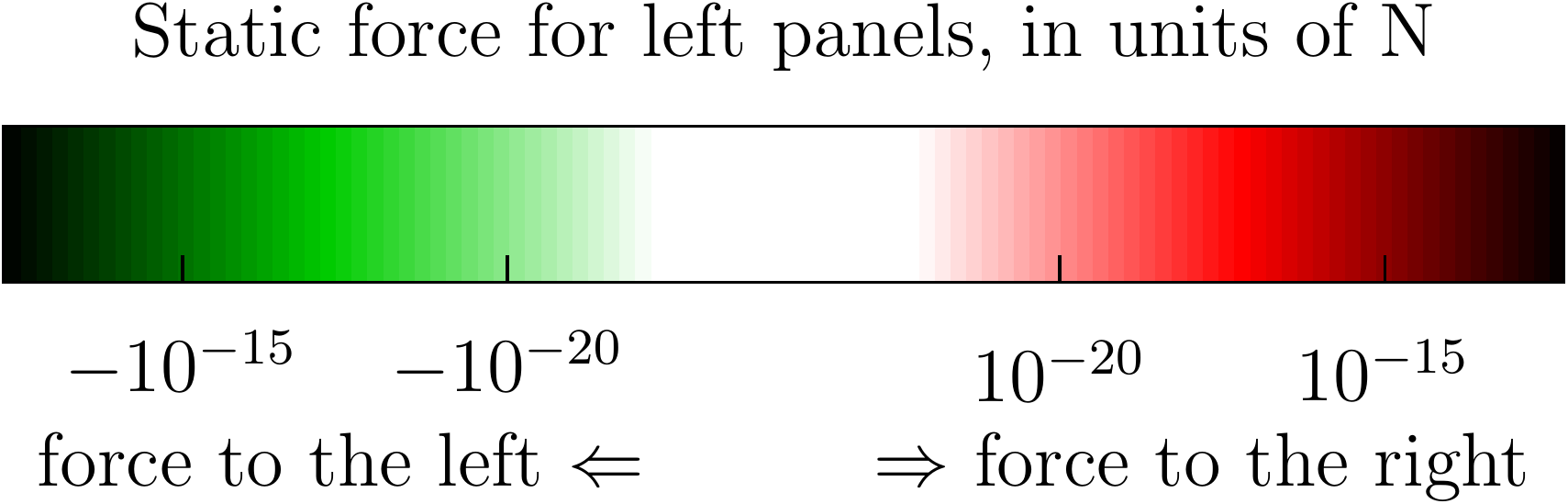}
  }
  \caption[Static friction force acting on a micromirror inside a cavity]{Static friction force (\ie, the force acting on the mirror when $v=0$) computed from the scattering model presented here (left panels) and a model based on a modal decomposition~\cite{Jayich2008} (right panels), showing only one pair of modes. Red and green regions represent forces pointing in opposite directions, as indicated on the colourbar. We note qualitative agreement between the two models for $x\approx 0.25\lambda$ and for $\Delta L_\text{C}$ close to the resonances, especially for large $\lvert\zeta\rvert$. The discrepancies between the two sets of data, that are more pronounced for small polarisability, have significant consequences for any theory based on a coupled-cavity modal decomposition model. The black dashed lines [\eref{eq:CavityResonances}] trace the cavity resonances in the scattering model. The absolute values on the colourbar relate to the left panels.}
  \label{fig:SF}
\end{figure}
Placing a micromirror inside a cavity is actively being explored, both theoretically~\cite{Bhattacharya2007a,Bhattacharya2008} and experimentally~\cite{Thompson2008,Jayich2008,Sankey2010}, as a means of realising strong optomechanical coupling between a movable scatterer and the cavity field. In a recent experiment, Sankey and co-workers~\cite{Sankey2010} looked at both linear and nonlinear coupling between the mechanical degree of freedom of the micromirror and the cavity field. The physical mechanism by which optomechanical coupling, and subsequently cooling, is enhanced by the presence of the cavity can be looked at either using the standard theory of cavity-mediated cooling~\cite{Horak1997,Hechenblaikner1998}, in the case of weakly reflective micromirrors, or by considering the micromirror to form the boundary between two coupled cavities~\cite{Bhattacharya2007a,Bhattacharya2008}, in the case of more strongly reflective micromirrors. The transfer matrix approach detailed in the previous sections provides a natural way to explore this interaction no matter what the optical properties of the micromirror are.
\par
We begin by modelling the system in Ref.~\cite{Thompson2008}: a two--mirror Fabry-P\'erot cavity with a micromirror near its centre, operating at a wavelength $\lambda=1064$\,nm and having a length $L_\text{C}=6.7$\,cm. The micromirror is modelled by its polarisability $\zeta$ which, in light of the small losses observed in practice, is taken to be real and negative. Whereas the real experimental system corresponds to $\zeta\gtrsim-1$, we allow $\zeta$ to vary freely in our model. The two quantities of interest in this section are the intensity of the field close to the micromirror, and the cooling coefficient acting on the micromirror. The former gives us knowledge of the resonant frequencies of the cavity and, therefore, of the optomechanical coupling, to all orders, between the cavity field and the micromirror. The latter is useful in optomechanical cooling experiments; the interest here lies in the fact that cooling the motion of a micromirror is one way towards achieving higher sensitivity in sensing applications, most notably in gravitational-wave detectors~\cite{Braginsky2002}.\\
These quantities are summarised in \fref{fig:IaF}, with the left panels showing the intensity at the mirror and the right panels the cooling coefficient acting on the mirror. Each subfigure (a)--(h) explores a different value for $\zeta$. For $\zeta\approx 0$, the cavity field is similar to the bare-cavity field; in particular, the cavity resonances are only slightly perturbed by the presence and position of the micromirror. The opposite is true of the $\zeta\ll-1$ case, where there is strong coupling between pairs of cavity modes, typified by the avoided crossings in the spectra. The resonance frequencies can be obtained analytically, in the limit of a good bare cavity, as frequency shifts from the bare resonances:
\begin{equation}
\Delta\omega=\frac{c}{L_\text{C}}\tan^{-1}\Biggl\{\frac{\zeta\cos(2k_0x)\mp\zeta\bigl[1+\zeta^2\sin^2(2k_0x)\bigr]}{\zeta^2\cos(2k_0x)\pm\bigl[1+\zeta^2\sin^2(2k_0x)\bigr]}\Biggr\}\,,
\label{eq:CavityResonances}
\end{equation}
with $L_\text{C}$ being the length of the cavity, $x$ the position of the micromirror, and $k_0=2\pi/\lambda$ the wavenumber of the light inside the cavity. The two sets of solutions to \eref{eq:CavityResonances} are, in the $\zeta\rightarrow 0$ limit or at $x=\lambda/8$, separated by a free spectral range. These cavity resonances, plotted as detuned cavity lengths $\Delta L_\text{C}=\bigl(L_\text{C}/\omega\bigr)\Delta\omega$, are traced by means of the dashed black curves in the left panels of \fref{fig:IaF}.\\
In the standard optomechanical coupling Hamiltonian, the mirror--field coupling is represented by a term of the form
\begin{equation}
\hat{H}_\text{OM}^{(1)}\sim\hbar\omega^\prime\hat{x}\hat{a}^\dagger\hat{a}\,,
\end{equation}
where $\hat{x}$ is the position operator of the mirror, and $\omega^\prime\equiv\partial(\Delta\omega)/\partial x$. $\hat{a}$ is the annihilation operator of the field mode that has the dominant interaction with the micromirror; in the $\lvert\zeta\rvert\rightarrow 0$ limit, these field modes are the bare cavity modes of the whole cavity. However, as $\lvert\zeta\rvert$ increases, the micromirror effectively splits the main cavity into two coupled cavities, giving rise to symmetric and antisymmetric modes, seen as the higher (bright) and lower (dark) branches in \fref{fig:IaF}(h) for $0<x<\lambda/4$; in such cases $a$ is the annihilation operator belonging to one of these eigenmodes. We note that similar behaviour was observed in Ref.~\cite{Jayich2008}.\\
Certain effects, such as mechanical squeezing of the mirror position~\cite{Nunnenkamp2010} and quantum non-demolition measurements on the mirror~\cite{Clerk2010b}, require not \emph{linear coupling} to $\hat{x}$ but \emph{quadratic coupling} to $\hat{x}^2$:
\begin{equation}
\hat{H}_\text{OM}^{(2)}\sim\hbar\omega^{\prime\prime}\hat{x}^2\hat{a}^\dagger\hat{a}\,,
\end{equation}
with $\omega^{\prime\prime}\equiv\partial^2(\Delta\omega)/\partial x^2$. In our notation, we have
\begin{equation}
\label{eq:LinearCoupling}
\omega^\prime=\mp\frac{2k_0c}{L_\text{C}}\frac{\zeta\sin(2k_0x)}{\big[1+\zeta^2\sin^2(2k_0x)\bigr]^{1/2}}\,,
\end{equation}
and
\begin{equation}
\label{eq:QuadraticCoupling}
\omega^{\prime\prime}=\mp\frac{4k_0^2c}{L_\text{C}}\frac{\zeta\cos(2k_0x)}{\big[1+\zeta^2\sin^2(2k_0x)\bigr]^{3/2}}\,.
\end{equation}
One thing we note immediately is that there is no value for $x$ such that $\omega^\prime=\omega^{\prime\prime}=0$; in other words, the optomechanical coupling is restricted to be linear or quadratic, to lowest order. Higher-order nonlinearities may be achieved by coupling different transverse modes of the cavity (see, e.g., the experimental results in Ref.~\cite{Sankey2010}) but are overwhelmed by the linear or quadratic couplings in a single-mode cavity. Moreover, the linear coupling $\omega^\prime$ is bounded in the $\zeta\rightarrow\infty$ limit:
\begin{equation}
\lvert\omega^\prime\rvert\leq\frac{2k_0c}{L_\text{C}}\approx 2\pi\times8.42\,\text{MHz/nm}\,,
\end{equation}
with the numeric value corresponding to our parameters. In the same limit, $\omega^{\prime\prime}$ exhibits resonant behaviour (see \fref{fig:Couplings}), indicative of avoided crossings in the spectrum, peaking at a value of:
\begin{equation}
\lvert\omega^{\prime\prime}\rvert\rightarrow\frac{4k_0^2c}{L_\text{C}}\lvert\zeta\rvert\approx 2\pi\times 0.10\,\lvert\zeta\rvert\,\text{MHz/(nm)$^2$}\,.
\end{equation}
We plot the lower ($\mp\rightarrow-$) branches of \erefs{eq:LinearCoupling} and~(\ref{eq:QuadraticCoupling}) in \fref{fig:Couplings} for two values for $\zeta$: $\zeta=-1$, representative of realistic micromirrors, and $\zeta=-10$, representative of highly reflective micromirrors. These correspond to cases (e) and (h) in \fref{fig:IaF}, respectively. Coupling between the pairs of modes is not very strong for the $\zeta=-1$ case; this is manifested by means of the smooth variation with $x$ of $\omega^\prime$ and $\omega^{\prime\prime}$ in \fref{fig:Couplings}(a). The second case shows strong signs of the avoided crossing behaviour seen in \fref{fig:IaF}(h), with $\omega^\prime$ no longer behaving smoothly and $\omega^{\prime\prime}$ acquiring a resonance-like character. Note that, independently of the magnitude of $\zeta$, the strongest quadratic coupling always occurs at the points where $\omega^\prime=0$.\\
The linear frequency shifts, \eref{eq:LinearCoupling} and the solid curves in \fref{fig:Couplings}, describe the same quantity as do the curves in Fig.~5 of Ref.~\cite{Bhattacharya2008}. In other words, our results match those in Ref.~\cite{Bhattacharya2008} in the relevant limit of a high-reflectivity micromirror, but the self-consistent scattering solution presented here is also applicable to the general situation where $\zeta$ can take any complex value. Thus, our `micromirror' could represent an actual micromirror, a poorly reflective membrane, or even an atom, whereas in the latter two cases the decomposition of the system into two coupled cavities is not valid. \fref{fig:SF} presents a comparison of the static force acting on the micromirror, for various values of $\zeta$, as calculated both from the model presented in this work (left panels) and from a modal decomposition model (Ref.~\cite{Jayich2008}, right panels). We note that the two sets of data agree in the large $\lvert\zeta\rvert$ limit, where the `two coupled cavities' model is formally valid, but there are significant qualitative differences when $\lvert\zeta\rvert\ll 1$, where the model used most often is that of a single cavity mode interacting with a scatterer~\cite{Hechenblaikner1998}.
\par
The behaviour of the cooling coefficient acting on the micromirror, as shown on the right panels of \fref{fig:IaF}, also has a number of interesting properties. Its overall trend follows the structure of the field intensity closely, as can be seen from an inspection of this figure. As is expected from earlier investigations~\cite{Hechenblaikner1998,Jayich2008} optomechanical cooling inside cavities proceeds when the light pumped into the cavity is tuned below resonance. This is shown quite clearly in \fref{fig:IaF}, in that the friction force generally acts to cool the micromirror when $\Delta L_\text{C}$ is smaller than the resonant length, and it acts to heat when $\Delta L_\text{C}$ is larger. One notes that for weak mirror--field coupling [\fref{fig:IaF}(a)] the behaviour of the cooling coefficient at a given $\Delta L_\text{C}$ close to resonance is approximately well-defined: the motion of the mirror is either cooled or heated, almost irrespectively of the value of $x$. The situation is qualitatively different for very strong mirror--field coupling [\fref{fig:IaF}(h)], where the cooling coefficient changes sign very rapidly as $x$ is scanned over the resonance at a fixed $\Delta L_\text{C}$. This has profound implications for experimental explorations, since the localisation of the micromirror within the cavity becomes of critical importance to the qualitative behaviour of the system.

\section{Optical pumping and multilevel atoms}\label{sec:TMM:Multilevel}
The work in this section is published as Xuereb, A., Domokos, P., Horak, P., \& Freegarde, T. Phys.\ Scr.\ \textbf{T140}, 014010 (2010) and is reproduced essentially \emph{verbatim}~\cite{Xuereb2010a}. The transfer matrix method described previously is extended here to handle multilevel atoms and arbitrarily polarised incident fields.
\par
The extended model we describe in the present section allows us to treat multi-level atoms as classical scatterers in light fields modified by, in principle, arbitrarily complex optical components such as mirrors, resonators, dispersive or dichroic elements, or filters. After we introduce the general extension in the next section, we verify our formalism for two prototypical sub-Doppler cooling mechanisms---the $J=\tfrac{1}{2}\rightarrow J^\prime=\tfrac{3}{2}$ transition, leading to the Sisyphus cooling mechanism, and the $J=1\rightarrow J^\prime=2$ transition---in \Sref{sec:LinPerpLin} and \Sref{sec:SigmaSigma}, respectively, and show that it agrees with the standard literature.
\subsection{A transfer matrix relating Jones vectors}\label{sec:Extension}
\begin{figure}[t]
 \centering
    \includegraphics[width=0.4\figwidth]{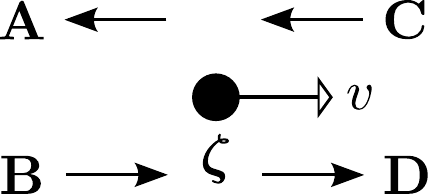}
\caption[Moving scatterer interacting with four field modes]{Moving scatterer interacting with four field modes represented by the Jones vectors $\mathbf{A}$, $\mathbf{B}$, $\mathbf{C}$, and $\mathbf{D}$. The scatterer has velocity $v$ and is described by means of its polarisability tensor $\boldsymbol{\zeta}$. The field mode amplitudes are, in general, functions of the wavenumber $k$.}
 \label{fig:Xuereb2010a:System}
\end{figure}
We investigate the interaction of atoms with light of different polarisations. To this end, we denote the two polarisation basis vectors by $\mu$ and $\nu$, whereby the standard circular polarisation basis is equivalent to setting $\mu=\sigma^+$ and $\nu=\sigma^-$. Starting from the transfer matrix model explored in \sref{sec:TMM:Model} and using the definitions in \fref{fig:Xuereb2010a:System}, we replace each of the field modes by a corresponding Jones vector, similar to the model used in Ref.~\cite{Spreeuw1992}. Thus, for example,
\begin{equation}
 A(k)\rightarrow\mathbf{A}(k)=\begin{pmatrix}
                               A_{\mu}(k)\\
                               A_{\nu}(k)
                              \end{pmatrix}\,,
\end{equation}
and similarly for $B$, $C$ and $D$, which are the mode amplitudes in the positive frequency part of the electric field. The transfer matrix $M$, describing the effect of the scatterer on the four field modes by means of the relation
\begin{equation}
\label{eq:Xuereb2010a:StandardTMM}
 \begin{pmatrix}
    A(k)\\
    B(k)
  \end{pmatrix}=M
 \begin{pmatrix}
    C(k)\\
    D(k)
  \end{pmatrix}\,,
\end{equation}
is now transformed into an order $4$ tensor of the form
\begin{equation}
 \boldsymbol{M}=\begin{bmatrix}
    \boldsymbol{m_{11}} & \boldsymbol{m_{12}}\\
    \boldsymbol{m_{21}} & \boldsymbol{m_{22}}
   \end{bmatrix}\,,
\end{equation}
where each of $\boldsymbol{m_{\alpha\beta}}$ ($\alpha,\beta=1,2$) is a $2\times 2$ matrix relating the respective Jones vector components. A general recipe for transforming the formulae for the field mode amplitudes, as given in \sref{sec:TMM:BasicBuildingBlocksModel}, can be summarised by means of the two replacements
\begin{equation}
 1\rightarrow\mathds{1}=\begin{bmatrix}
                 1 & 0\\
                 0 & 1
                \end{bmatrix}\text{\ and\ }\zeta\rightarrow\boldsymbol{\zeta}\,,
\end{equation}
wherever necessary. In particular, then,
\begin{equation}
\label{eq:Xuereb2010a:NewMatrix}
 M=\begin{bmatrix}
 1-\i\zeta & -\i\zeta\\
 \i\zeta & 1+\i\zeta
\end{bmatrix}\rightarrow\boldsymbol{M}=\begin{bmatrix}
 \mathds{1}-\i\boldsymbol{\zeta} & -\i\boldsymbol{\zeta}\\
 \i\boldsymbol{\zeta} & \mathds{1}+\i\boldsymbol{\zeta}
\end{bmatrix}\,.
\end{equation}
The polarisability tensor $\boldsymbol{\zeta}$ is defined in \eref{eq:ZetaTensorDefn}. We note that this new transfer matrix still allows us to model the interaction of the multilevel atom with an arbitrary system of immobile optical elements such as mirrors, cavities, waveplates, etc. As was done in \sref{sec:TMM:Model}, this interaction is accounted for by the multiplication of the various transfer matrices of the elements making up the system; this model is, in principle, applicable to systems of arbitrary complexity.
\par
Finally, we recall that the diagonal elements, $\langle i\rvert\m{\rho}^\text{st}\lvert i\rangle$, of the steady-state density matrix $\m{\rho}^\text{st}$ are the populations in each of the sublevels, whereas its off-diagonal elements, $\langle i\rvert\m{\rho}^\text{st}\lvert j\rangle$, are the respective coherences. The matrix elements of $\m{\rho}^\text{st}$ are obtained from the appropriate optical Bloch equations (see, for example, the procedure outlined in Ref.~\cite{CohenTannoudji1977b}). We note here that, through its dependence on $\m{\rho}^\text{st}$, $\boldsymbol{M}$ depends on the fields that it helps to determine, and thus \eref{eq:Xuereb2010a:StandardTMM} will in general become a set of nonlinear equations. In cases like the ones considered in the following sections, where only one multilevel atom is interacting with a linear optical system, this problem may be solved using a procedure similar to the one outlined below:~the fields surrounding the atom are obtained from the input fields through linear operations and then used with the optical Bloch equations to obtain the populations and coherences of the atom's various levels. Knowledge of these quantities then determines the fields, and hence the forces acting on the atom, completely.
\par
In the following sections we will restrict our discussion to the case where the input field is not modified by other transfer matrices. We will apply this mechanism to investigate the behaviour of atoms in two cases where the polarisation of the light varies in space on scales of the order of the wavelength to verify the validity of the model given by \eref{eq:Xuereb2010a:NewMatrix}, \eref{eq:ZetaTensorDefn}, and \eref{eq:PolarisabilityOperator}. In the first instance, we illuminate our atom with two counterpropagating linearly polarised beams. We choose the planes of polarisation of the two beams to be orthogonal to each other. The second configuration we will investigate involves illuminating the atom with two circularly polarised beams, choosing opposite handedness for the two beams. These two cases mirror those in Ref.~\cite{Dalibard1989}.

\subsection{Atoms in a gradient of polarisation}\label{sec:LinPerpLin}
\begin{figure}[t]
 \centering
    \includegraphics[scale=0.55]{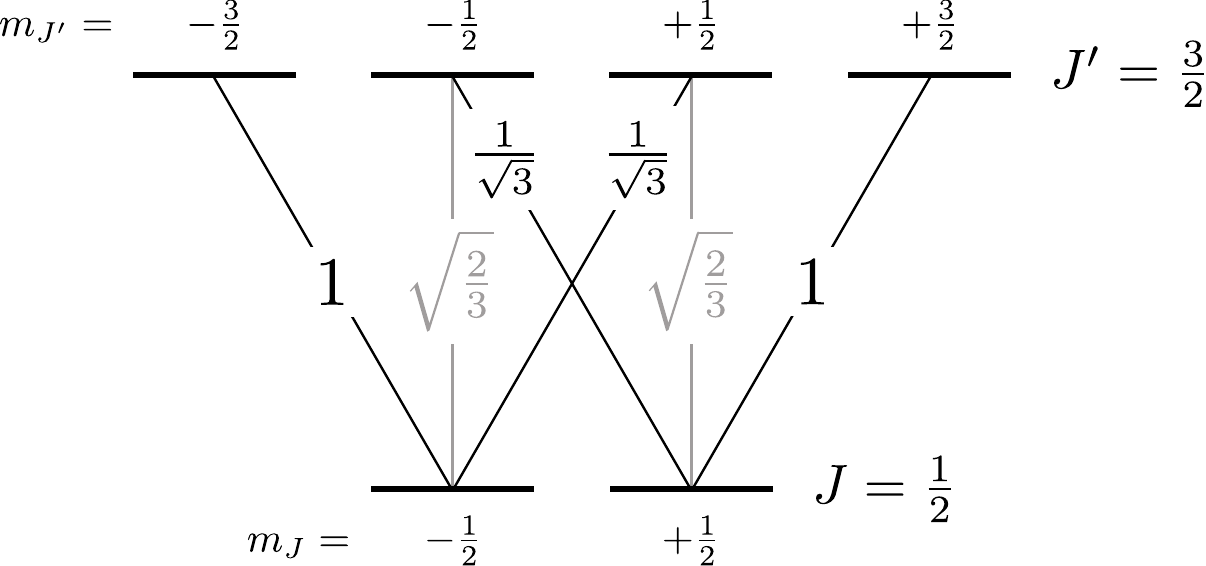}
\caption[Clebsch-Gordan coefficients for a $J=\tfrac{1}{2}\rightarrow J^\prime=\tfrac{3}{2}$ transition]{Clebsch-Gordan coefficients for a $J=\tfrac{1}{2}\rightarrow J^\prime=\tfrac{3}{2}$ transition.}
 \label{fig:Xuereb2010a:1-2_to_3-2}
\end{figure}
In this and the following sections, we will adopt the low-intensity hypothesis. This allows us to simplify the optical Bloch equations and resulting system considerably by neglecting the populations and coherences of the excited state sublevels. We can thus replace $\m{\rho}^\text{st}$ by the ground state steady-state density matrix, $\m{\rho}_\text{g}^\text{st}$. We denote the diagonal element $(i,i)$ of $\m{\rho}_\text{g}^\text{st}$, the population in sublevel $i$, by $\Pi_i$, and the off-diagonal element $(i,j)$, the coherence between sublevels $i$ and $j$, by $C_{i,j}$.
\par
Here we will discuss what is perhaps the simplest transition between two levels with multiple magnetic sublevels: the $J=\tfrac{1}{2}\rightarrow J^\prime=\tfrac{3}{2}$ transition. In this case, we have two ground sublevels so that $\m{\rho}_g^\text{st}$ is a $2\times 2$ matrix. \fref{fig:Xuereb2010a:1-2_to_3-2} tabulates the Clebsch-Gordan coefficients required to evaluate $\boldsymbol{\zeta}$. We thus have:
\begin{equation}
 \m{\rho}_g^\text{st} = \begin{bmatrix}
                     \Pi_{-\tfrac{1}{2}}             & C_{-\tfrac{1}{2},+\tfrac{1}{2}}\\
                     C_{+\tfrac{1}{2},-\tfrac{1}{2}} & \Pi_{+\tfrac{1}{2}}
                    \end{bmatrix}
\end{equation}
and
\begin{equation}
 \hat{\boldsymbol{\zeta}}=\zeta_0\begin{pmatrix}
                 \begin{bmatrix}
                     \tfrac{1}{3} & 0\\
                     0 & 1
                 \end{bmatrix} &
                 \boldsymbol{0}\\
                 \boldsymbol{0}&
                 \begin{bmatrix}
                     1 & 0\\
                     0 & \tfrac{1}{3}
                 \end{bmatrix}
                \end{pmatrix}\,,
\end{equation}
whereby
\begin{equation}
 \boldsymbol{\zeta}=\zeta_0\Biggl(\begin{bmatrix}
                     \tfrac{1}{3} & 0\\
                     0 & 1
                 \end{bmatrix}\Pi_{-\tfrac{1}{2}}+
                 \begin{bmatrix}
                     1 & 0\\
                     0 & \tfrac{1}{3}
                 \end{bmatrix}\Pi_{+\tfrac{1}{2}}\Biggr)\,.
\end{equation}
Suppose, now, that we illuminate the atom with two counterpropagating beams having orthogonal linear polarisation and equal intensity. This can be represented by setting
\begin{equation}
 \mathbf{B}(k)=\tfrac{B}{\sqrt{2}}\begin{pmatrix}
                     1\\
                     1
                   \end{pmatrix}\,\exp(\i kx-\i\pi/4)\end{equation}
and
\begin{equation}
 \mathbf{C}(k)=\tfrac{\i B}{\sqrt{2}}\begin{pmatrix}
                     1\\
                     -1
                   \end{pmatrix}\exp(-\i kx+\i\pi/4)\,,
\end{equation}
where the shift in the $x$ coordinate is introduced to simplify our expressions. Using the optical Bloch equations, we can show that the steady state populations in the ground sublevels at zero atomic velocity are given by
\begin{equation}
\label{eq:Xuereb2010a:LinLinPops}
\Pi_{-\tfrac{1}{2}}=\cos^2(kx)\,\text{ and }\,\Pi_{+\tfrac{1}{2}}=\sin^2(kx)\,,
\end{equation}
noting that the populations do not depend on the field amplitudes in the low intensity regime.
\par
We work to lowest order in $\zeta_0$ and make use of the above relations to find the net force acting on the atom~\cite{Asboth2008}; cf.~\eref{eq:TMM:MSTForce}:
\begin{align}
\label{eq:Xuereb2010a:GeneralForce}
 \force&=\hbar k\Bigl(\lvert\mathbf{A}\rvert^2+\lvert\mathbf{B}\rvert^2-\lvert\mathbf{C}\rvert^2-\lvert\mathbf{D}\rvert^2\Bigr)\nonumber\\
&=2\hbar k\im{\bigl[\boldsymbol{\zeta}\bigl(\mathbf{B}+\mathbf{C}\bigr)\bigr]\cdot\bigl(\mathbf{B}-\mathbf{C}\bigr)^\ast}+4\tfrac{v}{c}\hbar k\im{\bigl(\boldsymbol{\zeta}\mathbf{B}\bigr)\cdot\mathbf{C}^\ast+\bigl(\boldsymbol{\zeta}\mathbf{C}\bigr)\cdot\mathbf{B}^\ast}\nonumber\\
&\phantom{=\ }-2\tfrac{v}{c}\hbar k^2\im{\biggl[\tfrac{\partial\boldsymbol{\zeta}}{\partial k}\bigl(\mathbf{B}+\mathbf{C}\bigr)\biggr]\cdot\bigl(\mathbf{B}+\mathbf{C}\bigr)^\ast}\nonumber\\
&\approx2\hbar k\im{\bigl[\boldsymbol{\zeta}\bigl(\mathbf{B}+\mathbf{C}\bigr)\bigr]\cdot\bigl(\mathbf{B}-\mathbf{C}\bigr)^\ast}+2\tfrac{v}{c}\hbar k^2\im{\biggl[\tfrac{\partial\boldsymbol{\zeta}}{\partial k}\bigl(\mathbf{B}+\mathbf{C}\bigr)\biggr]\cdot\bigl(\mathbf{B}+\mathbf{C}\bigr)^\ast}\,,
\end{align}
where we have assumed that $\lVert k\,\partial\boldsymbol{\zeta}/\partial k\rVert\gg \lVert\boldsymbol{\zeta}\rVert$. The velocity-dependent force terms in the above expression arise through the Doppler shifting of photons both between field modes in the same polarisation and between field modes in different polarisations; these mechanisms are accounted for by the diagonal and off-diagonal terms in $\boldsymbol{\zeta}$,\footnote{We note that, whilst \eref{eq:Xuereb2010a:GeneralForce} is a general expression, the form of $\boldsymbol{\zeta}$ in this section has no nonzero off-diagonal terms, and only the first type of term contributes.} respectively. These terms emerge through the velocity-dependent terms in the generalised transfer matrix.
\\
In the present case, \eref{eq:Xuereb2010a:GeneralForce} simplifies approximately to
\begin{equation}
\label{eq:Xuereb2010a:StaticLinLinForce}
 \force=\tfrac{4}{3}\hbar k\zeta_0\lvert B\rvert^2\sin(2kx)\Bigl(\Pi_{+\tfrac{1}{2}}-\Pi_{-\tfrac{1}{2}}\Bigr)=-\tfrac{2}{3}\hbar k\lvert B\rvert^2\zeta_0\sin(4kx)\,,
\end{equation}
assuming that $\zeta_0$ is real for simplicity.
\\
Apart from the velocity-dependent terms in \eref{eq:Xuereb2010a:GeneralForce}, a second type of friction force emerges from the dynamics of the populations in the ground-state sublevels. Upon solving the optical Bloch equations~\cite{Dalibard1989}, expressions for $\Pi_{\pm\tfrac{1}{2}}$ are obtained that have a velocity-dependent term due to the time $\tau_\text{p}$ it takes for an atom in one sublevel to be pumped to the other. This adds a further velocity-dependent term to~\eref{eq:Xuereb2010a:StaticLinLinForce}, giving an overall force
\begin{equation}
 \force=-\tfrac{2}{3}\hbar k\lvert B\rvert^2\zeta_0\sin(4kx)-\tfrac{8}{3}\hbar k^2\lvert B\rvert^2\zeta_0v\tau_\text{p}\sin^2(2kx),
\end{equation}
which agrees precisely with the standard literature [cf. Eqs.~(4.20) and~(4.23) in Ref.~\cite{Dalibard1989}].

\subsection{Atoms in a gradient of ellipticity}\label{sec:SigmaSigma}
\begin{figure}[t]
 \centering
    \includegraphics[scale=0.55]{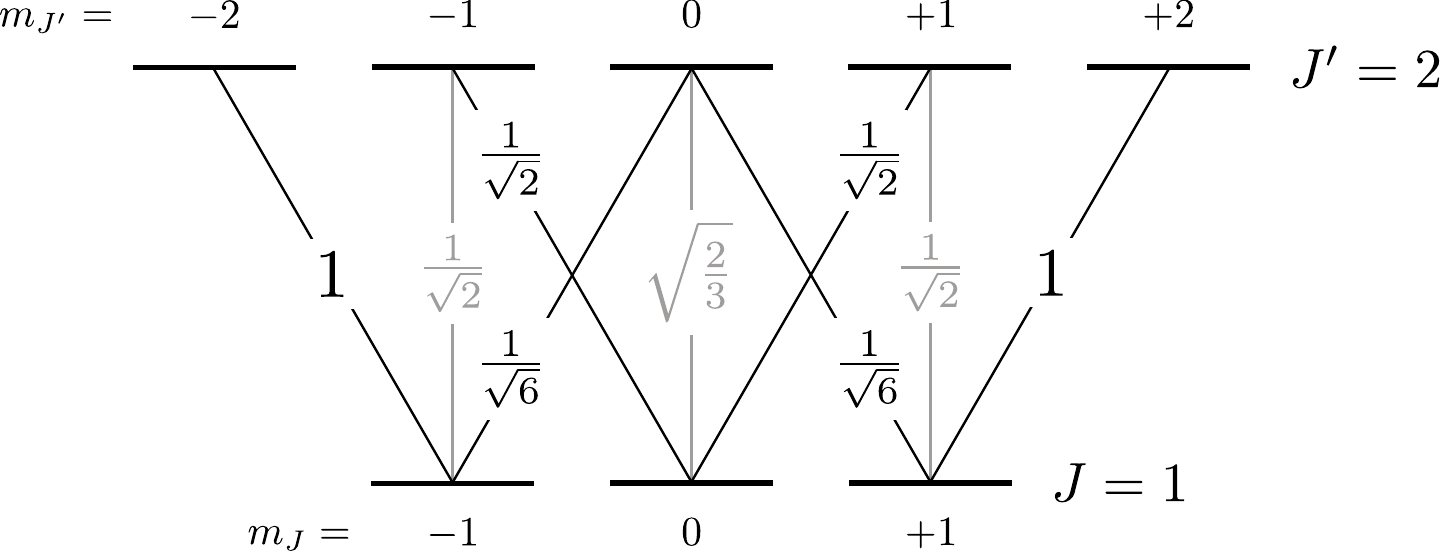}
\caption[Clebsch-Gordan coefficients for a $J=1\rightarrow J^\prime=2$ transition]{Clebsch-Gordan coefficients for a $J=1\rightarrow J^\prime=2$ transition.}
 \label{fig:Xuereb2010a:1_to_2}
\end{figure}
If we illuminate an atom with two counterpropagating beams of light in a $\sigma^+$--$\sigma^-$ configuration, rich dynamics are obtained not in the simplest ($J=\tfrac{1}{2}\rightarrow J^\prime=\tfrac{3}{2}$) case, but in the next simplest, where the ground state has three magnetic sublevels ($J=1$) and the excited state five ($J^\prime=2$). In this case, then, we can express $\m{\rho}_g^\text{st}$ and $\hat{\boldsymbol{\zeta}}$ as
\begin{equation}
 \m{\rho}_g^\text{st} = \begin{bmatrix}
                     \Pi_{-1}  & C_{-1,0} & C_{-1,+1}\\
                     C_{0,-1}  & \Pi_0    & C_{0,+1}\\
                     C_{+1,-1} & C_{+1,0} & \Pi_{+1}
                    \end{bmatrix}
\end{equation}
and
\begin{equation}
 \hat{\boldsymbol{\zeta}}=\zeta_0\begin{pmatrix}
                 \begin{bmatrix}
                     \tfrac{1}{6} & 0\\
                     0            & 1
                 \end{bmatrix} &
                 \boldsymbol{0} &
                 \begin{bmatrix}
                     0 & \tfrac{1}{6}\\
                     0 & 0
                 \end{bmatrix}\\
                 \boldsymbol{0} &
                 \begin{bmatrix}
                     \tfrac{1}{2} & 0\\
                     0            & \tfrac{1}{2}
                 \end{bmatrix} &
                 \boldsymbol{0}\\
                 \begin{bmatrix}
                     0            & 0\\
                     \tfrac{1}{6} & 0
                 \end{bmatrix}&
                 \boldsymbol{0} &
                 \begin{bmatrix}
                     1 & 0\\
                     0 & \tfrac{1}{6}
                 \end{bmatrix}
                \end{pmatrix}\,,
\end{equation}
using the Clebsch-Gordan coefficients in \fref{fig:Xuereb2010a:1_to_2}. Together, these give
\begin{equation}
 \boldsymbol{\zeta}=\zeta_0\Biggl(\begin{bmatrix}
                     \tfrac{1}{6} & 0\\
                     0            & 1
                 \end{bmatrix}\Pi_{-1}+
                 \begin{bmatrix}
                     \tfrac{1}{2} & 0\\
                     0            & \tfrac{1}{2}
                 \end{bmatrix}\Pi_0+
                 \begin{bmatrix}
                     1 & 0\\
                     0 & \tfrac{1}{6}
                 \end{bmatrix}\Pi_{+1}+\begin{bmatrix}
                     0 & \tfrac{1}{6}\\
                     0 & 0
                 \end{bmatrix}C+
                 \begin{bmatrix}
                     0            & 0\\
                     \tfrac{1}{6} & 0
                 \end{bmatrix}C^\ast\Biggr)\,,
\end{equation}
 with $C=C_{+1,-1}=C_{-1,+1}^\ast=\langle+1\vert\m{\rho}_\text{g}^\text{st}\vert\!-\!\!1\rangle$ representing the nonzero coherence between the $m_J=+1$ and the $m_J=-1$ sublevels. Note that we again apply the low intensity hypothesis, thereby replacing $\m{\rho}^\text{st}$ with $\m{\rho}_\text{g}^\text{st}$.
\par
We now illuminate the atom with two counterpropagating beams of equal intensity, $\mathbf{B}$ and $\mathbf{C}$, possessing $\sigma^+$ and $\sigma^-$ polarisation, respectively:
\begin{equation}
 \mathbf{B}(k)=B\begin{pmatrix}
                     1\\
                     0
                   \end{pmatrix}\,\exp(\i kx)\text{\ and\ }
 \mathbf{C}(k)=B\begin{pmatrix}
                     0\\
                     1
                   \end{pmatrix}\exp(-\i kx)\,.
\end{equation}
We again use~\eref{eq:Xuereb2010a:GeneralForce} to derive the force acting on the atom, which is given by
\begin{align}
\label{eq:Xuereb2010a:SigmaSigmaForce}
 \force&=2\hbar k\lvert B\rvert^2\,\imag{\tfrac{5}{6}\zeta_0\bigl(\Pi_{+1}-\Pi_{-1}\bigr)+\tfrac{1}{6}\i\zeta_0\im{C\exp(-2\i kx)}}\nonumber\\
&\phantom{=\ }-2\tfrac{v}{c}\hbar k^2\lvert B\rvert^2\im{\partial\zeta_0/\partial k}\Bigl(\tfrac{7}{6}\bigl(\Pi_{+1}+\Pi_{-1}\bigr)+\Pi_0+\tfrac{1}{3}\re{C\exp(-2\i kx)}\Bigr)\,,
\end{align}
where the populations and coherences are again obtained from the optical Bloch equations, and can be found in Ref.~\cite{Dalibard1989}. By observing the natural correspondence between $\zeta_0$ and $s_\pm$ in this reference, we can see that our expression for the force acting on the atom again agrees with the standard literature to first order in $\tfrac{v}{c}$ [cf. Eq.~(5.9) in Ref.~\cite{Dalibard1989}].
The resulting friction force is thus due to both the Doppler shift, as evident in the terms shown explicitly in~\eref{eq:Xuereb2010a:SigmaSigmaForce}, as well as to the non-adiabatic following of the atomic sublevel populations.

\appendicesstart
\section{Appendix: Cavity properties from the transfer matrix model}
In this section we will derive a consistent set of relations used to describe cavities, based on the transfer matrix formalism. Specifically, we will relate the cavity HWHM linewidth $\kappa$, its finesse $\mathcal{F}$, and its $Q$-factor to one another and to the reflectivity of the cavity mirrors.

\subsection{Cavity finesse}\label{sec:TMM:CavityFinesse}
The finesse of a cavity is defined as
\begin{equation}
\label{eq:FinesseDefn}
\mathcal{F}=\frac{\Delta\lambda}{\delta\lambda}\,,
\end{equation}
where $\delta\lambda$ is the FWHM linewidth of the cavity transmission peak and $\Delta\lambda$ is the free spectral range (FSR) of the cavity. We model the cavity as a Fabry--P\'erot resonator having mirrors of reflectivity $r_1$ and $r_2$ and a length $L$. On resonance, we can find some integer $n$ such that
\begin{equation}
L=\tfrac{1}{2}n\lambda\,.
\end{equation}
The FSR is defined~\cite{Siegman1990} as the wavelength interval such that
\begin{equation}
L=\tfrac{1}{2}(n+1)(\lambda-\Delta\lambda)\,.
\end{equation}
We approximate $\Delta\lambda\ll\lambda$, whereby
\begin{equation}
\label{eq:FSR}
\Delta\lambda=\frac{\lambda^2}{2L}\,.
\end{equation}
The cavity is described by the transfer matrix equation:
\begin{equation}
\begin{pmatrix}
A\\
B
\end{pmatrix}=
\frac{1}{t_1t_2}
\begin{bmatrix}
t_1^2-r_1^2 & r_1\\
-r_1 & 1
\end{bmatrix}
\begin{bmatrix}
e^{-\i kL} & 0\\
0 & e^{\i kL}
\end{bmatrix}
\begin{bmatrix}
t_2^2-r_2^2 & r_2\\
-r_2 & 1
\end{bmatrix}
\begin{pmatrix}
C\\
D
\end{pmatrix}\,.
\end{equation}
We set $C=0$ and write the transmitted field $D$ in terms of the only input field, $B$:
\begin{equation}
\label{eq:TransmissionLorentzian}
\lvert D\rvert^2=\frac{\lvert t_1t_2\rvert^2}{\lvert 1-\lvert r_1r_2\rvert\exp(2\i kL)\rvert^2}\lvert B\rvert^2\,,
\end{equation}
where the phase shifts induced by $r_1$ and $r_2$ have been absorbed in $L$. For a reasonably good cavity ($\lvert t_{1,2}\rvert\ll 1$, $\delta\lambda\ll\lambda$), the transmission is therefore Lorentzian, with a FWHM linewidth
\begin{equation}
\label{eq:FWHMLinewidth}
\delta\lambda=\frac{\lambda}{kL}\frac{1-\lvert r_1r_2\rvert}{\sqrt{\lvert r_1r_2\rvert}}\,.
\end{equation}
Finally, we substitute \eref{eq:FSR} and \eref{eq:FWHMLinewidth} into \eref{eq:FinesseDefn} to obtain
\begin{equation}
\mathcal{F}=\frac{\pi\sqrt{\lvert r_1r_2\rvert}}{1-\lvert r_1r_2\rvert}\,.
\end{equation}
If the approximations $\lvert t_{1,2}\rvert\ll 1$ and $\delta\lambda\ll\lambda$ no longer hold, it can be similarly shown that \eref{eq:TransmissionLorentzian} implies
\begin{equation}
\mathcal{F}=\frac{\pi/2}{\sin^{-1}\biggl(\frac{1-\lvert r_1r_2\rvert}{2\sqrt{\lvert r_1r_2\rvert}}\biggr)}\,.
\end{equation}

\subsection{Physical meaning of the cavity finesse}\label{sec:TMM:CavityFinesseN}
The factor $\rho=\lvert r_1r_2\rvert$ present in the above relations is related to the power lost by the cavity after one round-trip, $1-\rho^2$. Let us set $N=\sqrt{\rho}/(1-\rho)$. The power remaining in the cavity after $N$ round-trips is then
\begin{equation}
\rho^{2N}=\rho^{\frac{2\sqrt{\rho}}{1-\rho}}\to\frac{1}{e^2}\,,
\end{equation}
where we have taken the good-cavity ($\rho\to 1$) limit. In other words, we can write
\begin{equation}
\mathcal{F}=\pi N\,,
\end{equation}
where $N$ is the number of round-trips the light makes inside the cavity before the intensity decays by a factor of $1/e^2$.

\subsection{Cavity linewidth and quality factor}\label{sec:TMM:CavityKappaQ}
The HWHM cavity linewidth in frequency space can be defined in terms of the FWHM linewidth in wavelength space by means of the relation
\begin{equation}
2\kappa=\frac{\omega}{\lambda}\,\delta\lambda\,,
\end{equation}
whereupon
\begin{equation}
\delta\lambda=\frac{\kappa\lambda^2}{\pi c}\,,
\end{equation}
and substituting this expression for the linewidth into \eref{eq:FinesseDefn} gives
\begin{equation}
\mathcal{F}=\frac{\pi c}{2\kappa L}\,,\text{ or }\mathcal{\kappa}=\frac{\pi c}{2\mathcal{F}L}\,.
\end{equation}
The quality factor, or $Q$-factor, is defined as the ratio of the cavity frequency to its FWHM linewidth in frequency space: $Q=\omega/(2\kappa)$, or
\begin{equation}
Q=\frac{2\mathcal{F}L}{\lambda}\,.
\end{equation}
\appendicesend

\chapter{Applications of transfer matrices}\label{ch:TMMApplications}
\epigraph{[...] [T]he sciences do not try to explain, they hardly even try to interpret, they mainly make models. By a model is meant a mathematical construct which, with the addition of certain verbal interpretations, describes observed phenomena. The justification of such a mathematical construct is solely and precisely that it is expected to work [...].}{J.\ von\ Neumann, \emph{Method in the Physical Sciences} (1955)}

In this chapter, I will apply the transfer matrix method developed in \cref{ch:TMM:TMM} to novel cooling geometries outside cavities (\sref{sec:TMM:ECCO} and \sref{sec:TMM:Comparison}), as well as inside active ring cavities (\sref{sec:TMM:AmplifiedOptomechanics}).

\section{External cavity cooling}\label{sec:TMM:ECCO}
The basis for this current section was published as Xuereb, A., Freegarde, T., Horak, P., \& Domokos, P. Phys.\ Rev.\ Lett.\ \textbf{105}, 013602 (2010). We will apply the general solution detailed above to the external cavity cooling configuration, \sref{sec:CoolingMethods:Other:ECCO}, in the case of a micro-mechanical mirror. As a reference system for the analysis of the cooling force in this setup, we also consider the mirror-mediated cooling configuration, which is {the optomechanical} cooling scheme used in many experiments~\cite{Metzger2004,Arcizet2006,Gigan2006,Schliesser2008}. Note that in the external cavity cooling scheme {with a near mirror of complex transmissivity $t$, the limits of small and large $\lvert t\rvert$ render the situation where the cavity is replaced respectively by the near mirror only or the far mirror only.} For intermediate {$t$ compared {with} the transmissivity of the far mirror, $T$}, the moving scatterer interacts with a field reflected back from the {cavity and} is subject to the interference created by the multiple reflections between the two mirrors. Throughout most of this section, although not initially, we {consider in particular} an object having low reflectivity, around $50$\%, which corresponds to a polarisability $\zeta=-1$ and is representative of typical experimental conditions~\cite{Metzger2004}. This ensures that a high-finesse resonator cannot be formed between the object and the near mirror, thereby guaranteeing a parameter range where the cavity formed between the immobile mirrors dominates the interaction. For the sake of simplicity, we restrict ourselves to the special case of scatterers that can be characterised by a real polarisability; this is equivalent to assuming that no absorption takes place in the scatterer. Similar results hold when $\zeta$ is not real.
\par
\begin{figure}
\centering
\includegraphics[width=1.5\figwidth]{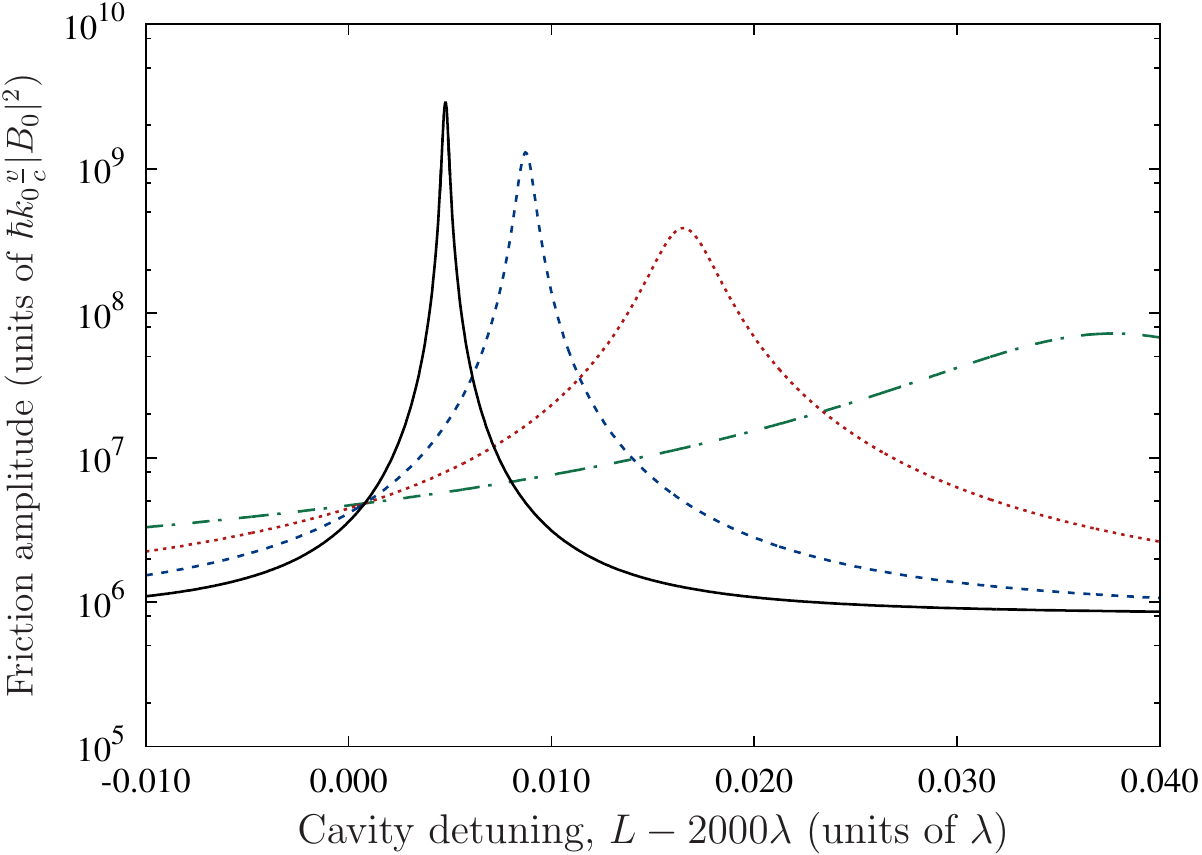}
\caption[The amplitude of the friction force acting on the scatterer, for various near-mirror transmissivities]{The amplitude of the friction force acting on the scatterer, for various near-mirror transmissivities, is shown as a function of the mirror separation in the cavity. The different curves represent different near-mirror transmissivities: $\lvert t\rvert=0.45$ (dashed--dotted curve), $\lvert t\rvert=0.20$ (dotted), $\lvert t\rvert=0.10$ (dashed), $\lvert t\rvert=0.05$ (solid). (Scatterer polarisability $\zeta=-1$, scatterer--cavity separation $x\approx 400\lambda_0$, $\lvert T\rvert=0.01$, $\lambda_0=780$~nm.)}
\label{fig:Detuning}
\end{figure}
A numerical fit to \eref{eq:FullFriction} for $\lvert t\rvert\sim\lvert T\rvert$ and $\lvert\zeta\rvert\ll 1$ renders a friction force of the approximate form
\begin{equation}
\label{eq:ECCOFriction}
\force\approx-8\hbar k_0^2\zeta^2\tfrac{v}{c}(2x+0.17\mathcal{F}L)\sin(4k_0x+\phi)\lvert B_0\rvert^2\,,
\end{equation}
where $\mathcal{F}$ is the cavity finesse, $L$ the cavity length (optimised as discussed below), $x$ the separation between the scatterer and the near mirror, and $\phi$ a phase factor. The gross spatial variation of the friction force is linear in both $L$ and $x$; this is simply because of the linear increase of the retardation time of the reflected field with {the} distance between the scatterer and the mirrors. This dependence is modulated by a wavelength-scale oscillation of the friction force, which thereby follows the same oscillatory dependence as mirror-mediated cooling [cf.~\eref{eq:TMM:MirrorCoolForce}] and constrains cooling to regions of the size of $\lambda_0/8$, where $\lambda_0=2\pi/k_0$. In the case of a micro-mechanical mirror, where the vibrational amplitude is naturally much less than the wavelength, this presents no problem. The form of \eref{eq:ECCOFriction} is dependent on the properties of the scatterer and of the mirrors; for realistic mirrors and $\zeta=-1$, the enhancement factor $0.17\mathcal{F}$ drops to $0.04\mathcal{F}$. With typical experimental parameters this results in an enhancement of $10^3$--$10^4$ over the standard setup.
\par
As shown in \fref{fig:Detuning}, the fine tuning of the cavity {length by varying $L$ on the wavelength scale shows a Lorentzian-like resonant enhancement} of the friction amplitude (\ie, the amplitude of the cooling coefficient), {following that of} the intracavity field intensity. If we denote the complex reflectivities of the near and far mirror by $r$ and $R$, respectively, we can show that the peaks of \fref{fig:Detuning} lie around the cavity resonances, at approximately $L=\tfrac{1}{2}m\lambda_0-\tfrac{1}{2k_0}\arg\big(rR\big)$, with $m$ being an integer, and have approximately the same full-width at half-maximum, $\bigl(1-\lvert r R\rvert\bigr)/\bigl(k_0\sqrt{\lvert r R\rvert}\bigr)=\lambda_0/(2\mathcal{F})$. The enhancement of the friction {force by} the cavity {is due to the multiplication of the retardation time by the number of round trips in the cavity, which thereby acts as a `distance folding' mechanism. For the chosen parameters, the optical path length is effectively $2x+0.04\mathcal{F}L$; \ie, determined predominantly by the cavity length $L$.}
\par
\begin{figure}
\centering
\includegraphics[width=1.5\figwidth]{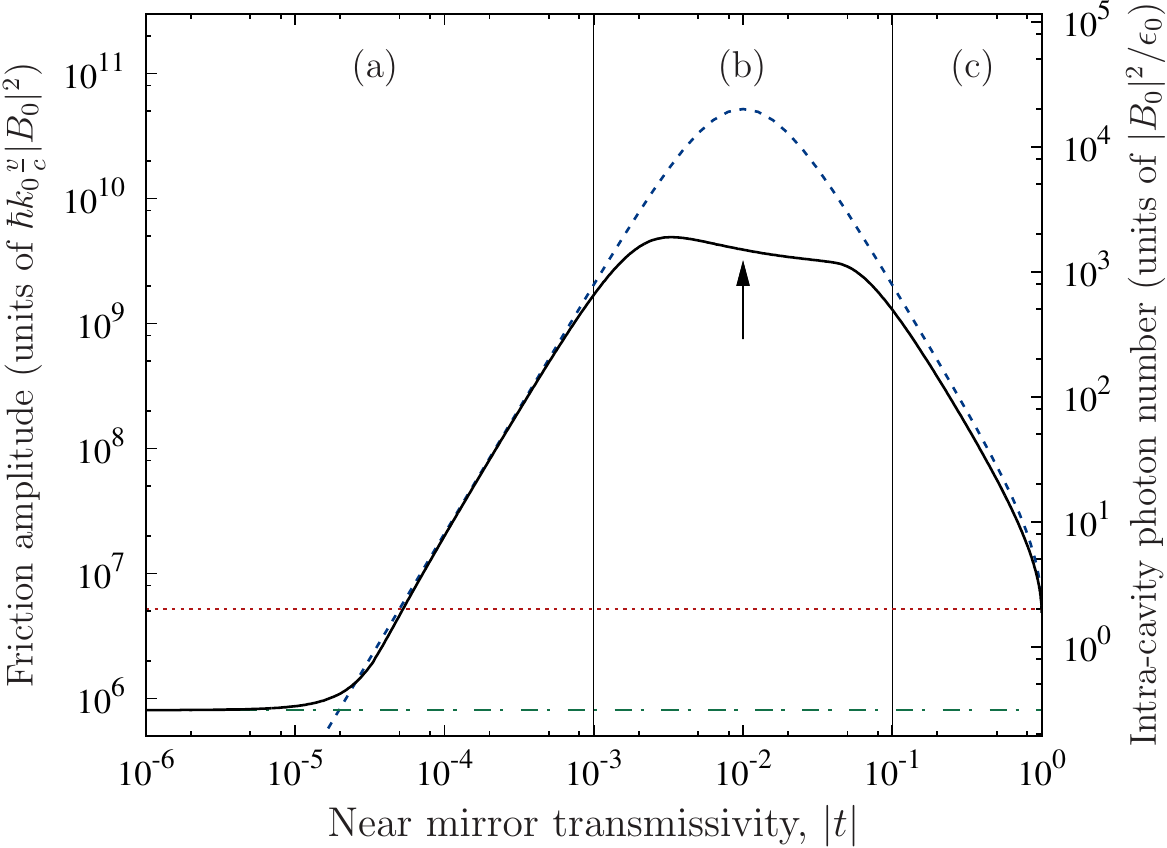}
\caption[Amplitude of the friction acting on a scatterer of polarisability $\zeta=-1$]{Amplitude of the friction acting on a scatterer of polarisability $\zeta=-1$ interacting with a cavity tuned to achieve maximum friction, for varying transmissivity of the near mirror. The friction amplitude (solid curve) approaches that for mirror-mediated cooling using the far (dotted line, $t\rightarrow 1$) or the near (dashed--dotted line, $t\rightarrow 0$) mirror only in the appropriate limits. The arrow indicates the point at which the two cavity mirrors have the same reflectivity. Also shown is the intracavity field (dashed). ($x\approx 400\lambda_0$, $L\approx 2000\lambda_0$, $\lvert T\rvert=0.01$, $\lambda_0=780$~nm, finesse at peak friction $5.0\times 10^4$.)}
\label{fig:TMM:Friction}
\end{figure}
\begin{figure}
\centering
\includegraphics[width=1.5\figwidth]{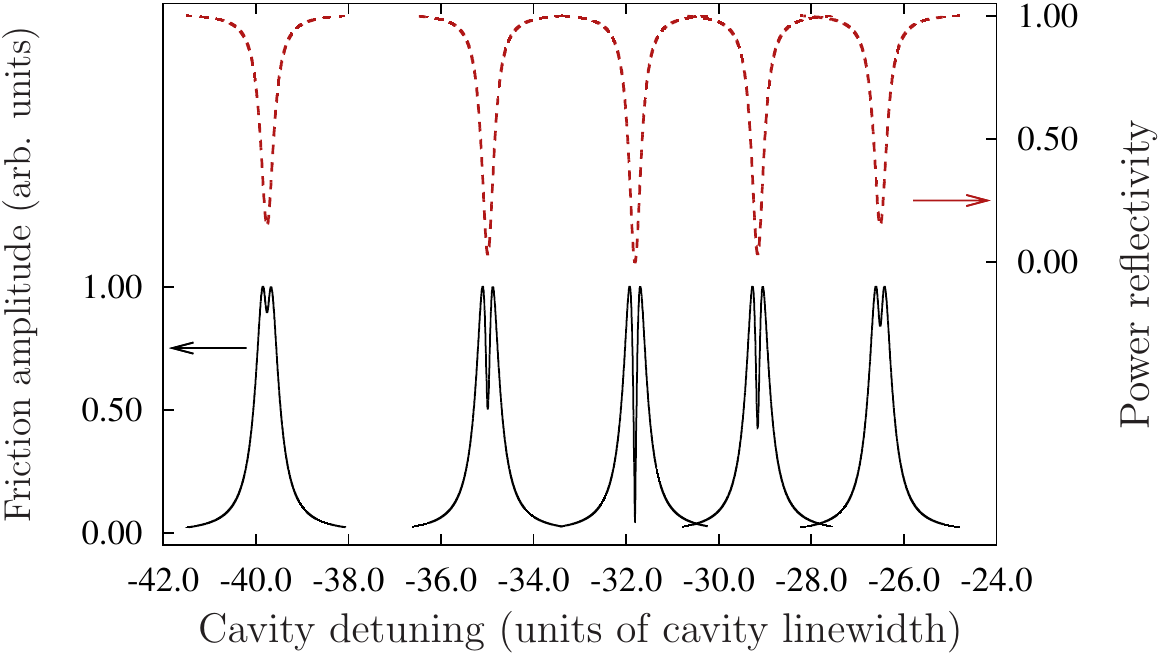}
\caption[Attenuation of the cooling coefficient amplitude for certain parameters]{In region (b) of \fref{fig:TMM:Friction}, the cooling coefficient amplitude (solid curves) is attenuated due to the attenuation in the field reflected from the cavity (dashed). $\lvert T\rvert=0.01$ in every plot; $\lvert t\rvert$ is, from left to right, $6.7\times 10^{-3}$, $8.3\times 10^{-3}$, $1.0\times 10^{-2}$, $1.2\times 10^{-2}$, and $1.5\times 10^{-2}$. (Parameters as in~\fref{fig:TMM:Friction}.)}
\label{fig:CandR}
\end{figure}

{The friction force depends not only upon the retardation but also upon the cavity reflectivity, which drops near resonance in the well-known behaviour of a Fabry--P\'erot resonator. \fref{fig:TMM:Friction} shows the friction amplitude as a function of the near mirror transmissivity $\lvert t \rvert$ for a fixed far mirror transmissivity, $T=1/(1+100\i)$.} {We note that this nonideal reflectivity of the far mirror could equivalently arise from absorption, of ca.\ $0.01$\% with the given parameters, of the incident power by the mirror.}
{For each value of $\lvert t \rvert$, the cavity length $L$ has been adjusted to maximise the friction force, according to curves such as those in \fref{fig:Detuning}. The calculated result follows the intracavity {field (shown {dashed)} except} where the cavity reflectivity drops near resonance [region (b)], and in the extremes of regions (a) and (c), where the geometry is dominated by the near ($\lvert t\rvert\rightarrow 0$) or far ($\lvert t\rvert\rightarrow 1$) mirrors, respectively. \fref{fig:CandR} shows the effect of the drop in reflectivity as the cavity is scanned through resonance for similar mirror reflectivities. When this causes a dip in the friction amplitude peak, the optimum values plotted in \fref{fig:TMM:Friction} occur to either side of the resonance, and the friction force in this region is effectively limited by this interference effect.} {We note that the friction amplitude is not maximised at the point of maximum intracavity field ($t=T$)} because more light is lost through the cavity for larger $\lvert t\rvert$.
\par
{{The external cavity cooling mechanism may prove particularly valuable when the scatterer is a small mirror or other micro-mechanical optical component. In such cases, the advantage gained by using the external cavity over the standard optomechanical cooling scheme depends heavily upon the polarisability or reflectivity of the moving scatterer, which in the above calculations have so far been taken to be modest ($\zeta=-1$; $\lvert r\rvert=0.7$) in comparison with those of the cavity mirrors.} For $\lvert\zeta\rvert\ll 1$, the friction force is enhanced by a factor approximately equal to $\mathcal{F}$} because of the distance folding argument explained above. For larger $\lvert\zeta\rvert$, the system turns into a three--mirror resonator and the advantage of external cavity cooling is not as big, but is still significant. For $\lvert\zeta\rvert\approx 1$ we find enhancement by a factor $0.04\mathcal{F}$, as discussed above. For even larger $\zeta$, when the reflectivity of the moving mirror becomes comparable to that of the fixed mirrors, the scheme behaves similarly to the mirror-mediated cooling {configuration}. The main heating process that counteracts the cooling effect in the case of micromirrors is thermal coupling to the environment, which depends on the geometry. In the case of isolated scatterers that undergo no absorption, the heating is due to quantum fluctuations in the fields (see \aref{sec:TMM:AppDiffusion}); the limit temperature here is $\Lapprox\hbar c/\bigl(0.34k_\text{B}\mathcal{F}L\bigr)=1.87\hbar\kappa/k_\text{B}$ when $\lvert\zeta\rvert\ll 1$, which evaluates to $\Lapprox 0.1$\,mK for the parameters in \fref{fig:TMM:Friction}. This expression for the temperature conforms to the form expected from \eref{eq:GeneralTForm}.
\par
The usual cavity-mediated cooling {mechanism}~\cite{Thompson2008,Favero2008}, where the moving scatterer is inside a two-mirror cavity, can also be described by our general framework in terms of \erefs{eq:Afield} and (\ref{eq:FullFriction}). Compared with this scheme, external cavity cooling has the advantage of always having a sinusoidal spatial dependence; the narrow resonances in the friction force for particles inside a cavity, as seen in the next section, impose more stringent positioning requirements.

\section{Cavity cooling of atoms: within and without a cavity}\label{sec:TMM:Comparison}
In \sref{sec:TMM:ECCO} we explored the cooling of a micromirror outside a high-finesse cavity as a means of enhancing the optomechanical interaction of the mirror with the field. Our formalism is general and, indeed, so are our results: the mechanism works similarly for atoms. Inside a resonator, the enhancement of the interaction is accompanied by an enhancement in the field itself, which precludes using high optical powers in order not to saturate the atom. Outside a cavity, the enhancement in the friction force is lower, but the field itself is not amplified by the presence of a cavity, and therefore one can use much higher powers. The {important question}, then, is whether these two effects compensate for one another in such a way as to render the cooling forces experienced by an atom outside a cavity similar to those it experiences inside.\\
To answer this question we will first {describe} the two models, in \Sref{sec:CMC} and \Sref{sec:ECCO}, respectively, using realistic parameters for state-of-the-art {optical devices}. Taking into account saturation effects, it is seen that the two different models result in similar cooling forces and equilibrium temperatures. An examination of scaling properties of the force acting on the atom {in the two schemes} then follows in \Sref{sec:Scaling}. The work in this section has been accepted for publication in the Eur.\ Phys.\ J.\ D topical issue on \emph{Cold Quantum Matter -- Achievements and Prospects}.

\subsection{Comparison of cavity cooling schemes}\label{sec:ComparisonAtoms}
\subsubsection{Cavity-mediated cooling: Atom inside the cavity}\label{sec:CMC}
\begin{figure}[t]
  \centering
  \subfigure[]{
    \includegraphics[scale=0.5]{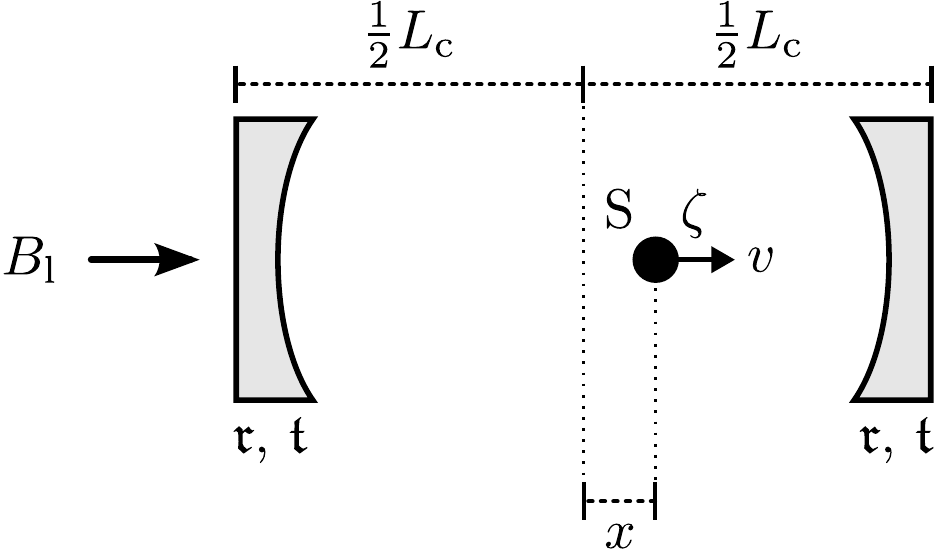}
  }\\
  \subfigure[]{
    \includegraphics[width=1.5\figwidth]{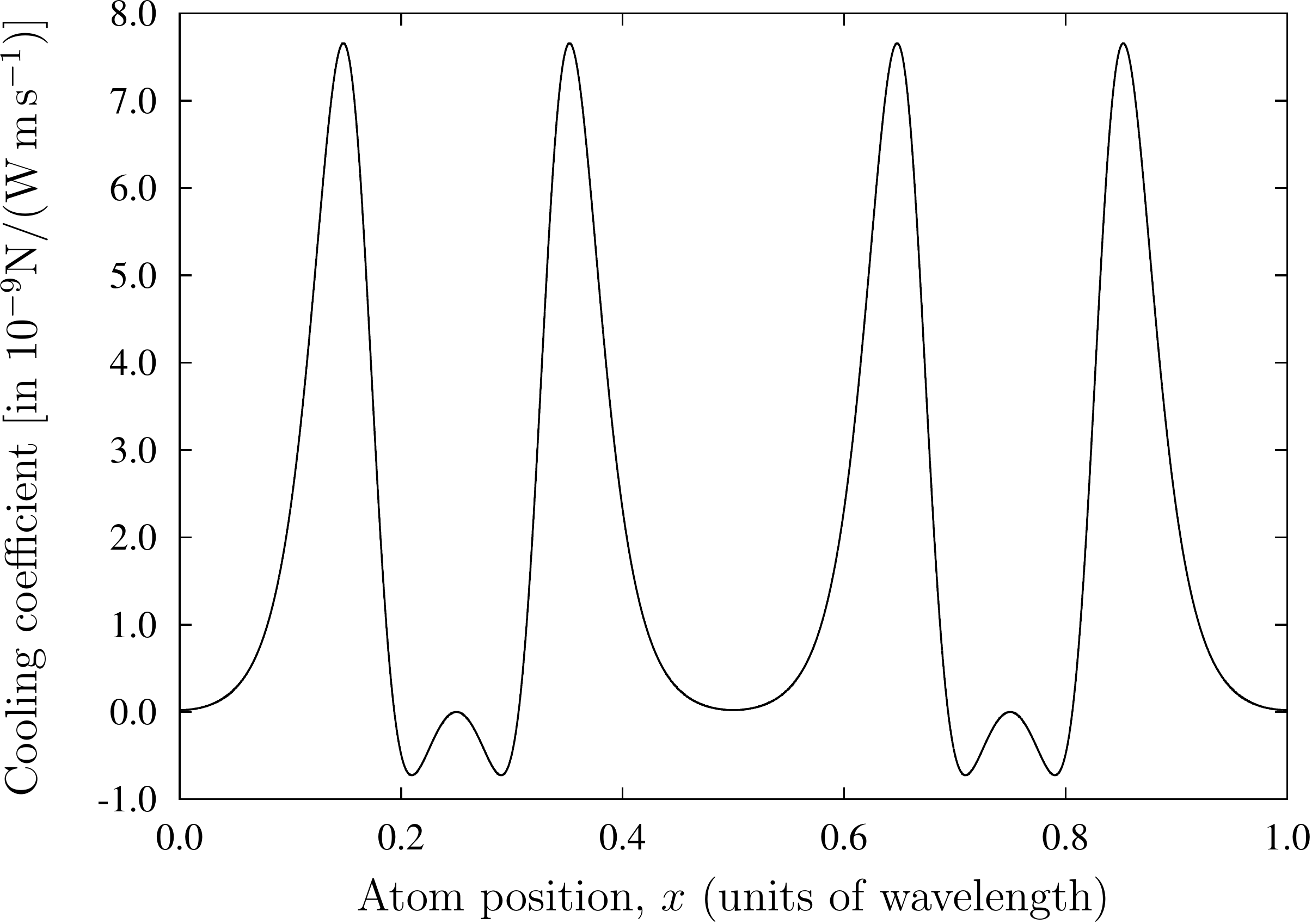}
  }
  \caption[Schematic model for cavity cooling inside cavities]{(a)~{Model} of a {scatterer}, $\text{S}$, inside a symmetric Fabry--P\'erot cavity of length $L_\text{c}$. The {cavity mirrors have} reflection and transmission coefficients, $\refl$ and $\trans$, and $\zeta$ is the polarisability of $\text{S}$. (b)~Cooling coefficient $\heatingcoefft$ per unit input power, experienced by the scatterer at different positions in the cavity for realistic parameters (see text for details).}
  \label{fig:CMC}
\end{figure}
Placing a scatterer---atom~\cite{Horak1997,Leibrandt2009,Koch2010}, micromirror~\cite{Bhattacharya2007a,Thompson2008}, or `point polarisable particle'~\cite{Domokos2003}---inside a cavity has long been pointed out to be a generic means of cooling the translational motion of that scatterer. Cooling of atoms inside resonators has been observed: first~\cite{Leibrandt2009} as a means of counteracting the heating of a trapped ion inside a cavity, leading to a steady-state occupation number for the motion of the atom of around $20$ quanta; and later~\cite{Koch2010} as an increase in the storage time for a neutral atom inside a cavity from $35$\,ms to $1100$\,ms. The generic layout of such an experiment is {shown} in \fref{fig:CMC}(a). For our purposes, we place the scatterer inside a symmetric Fabry--P\'erot cavity of length $L_\text{c}$, which we pump from one side; the dominant field inside the cavity is a standing {wave} field if the reflectivity of the mirrors, $\refl$, is high enough. For a numeric example, we use the same cavity properties as Ref.~\cite{Mucke2010}: {finesse} $\mathcal{F}=56\,000$ {modelled by using mirrors with $\trans=\refl+1=1/(1+133.5\i)$}, {cavity} length $495$\,$\upmu$m, and mode waist $30$\,$\upmu$m{; we use a wavelength $\lambda=780$\,nm. In contrast with Ref.~\cite{Mucke2010}, however}, let us reiterate that our cavity is pumped along its axis.\\
We also take the scatterer to be a {two-level} atom, {with} the cavity {field detuned} $10\Gamma$ to the red of the atom transition frequency. Thus, the polarisability of the atom is $\zeta=4.1\times 10^{-5}+4.1\times 10^{-6}\i$. The maximum cooling coefficient is found at a detuning of $-2.6$\,$\kappa$ from the cavity resonance. As expected~\cite{Domokos2003}, the optimal cooling coefficient occurs for a negative detuning of the pump from the \emph{bare} cavity resonance, but for a positive detuning from the {dressed atom--cavity} resonance.\par
The dependence of the friction force, \eref{eq:FullFriction}, on the position of the {scatterer, scanned over a wavelength, is shown in \fref{fig:CMC}(b)}. {The presence of the cavity manifests itself primarily through a strong enhancement of both the cooling coefficient $-\force/v$ and the intracavity field intensity. The scattering model explored above is only valid in the limit of small saturation. For the $^{85}$Rb D$_2$ transition, assuming that the beam is circularly polarised, the saturation intensity is $1.67$\,mW\,cm$^{-2}$~\cite{Steck2008}. In order to avoid saturation effects, we restrict the power input into the cavity to $2$\,pW; this equates to an intracavity intensity of $23$\,mW\,cm$^{-2}$ and hence a saturation parameter $s=0.14$; this is because $s$ is inversely proportional to the square of the detuning, $-10\Gamma$ in this case, of the pump beam from resonance. In turn, this input power also yields a maximum cooling coefficient of $1.5\times 10^{-20}$\,N/(m\,s$^{-1})$, which corresponds to a $1/e$ velocity cooling time of $9$\,$\upmu$s for the same atom; averaging the friction force over a wavelength gives a cooling time of $37$\,$\upmu$s.}\par
The friction force and the momentum diffusion both scale linearly with the input power, {in the low-saturation regime}. {Therefore, the equilibrium temperature is independent of the pump power in this regime. For the parameters used above}, the equilibrium temperature predicted for a scatterer at the point of maximum friction is $56$\,$\upmu$K; averaging the cooling coefficient, as well as the diffusion coefficient, over a wavelength, gives a higher equilibrium temperature of $220$\,$\upmu$K.

\subsubsection{External cavity cooling: {Atom} outside the cavity}\label{sec:ECCO}
\begin{figure}[t]
  \centering
  \subfigure[]{
    \includegraphics[scale=0.5]{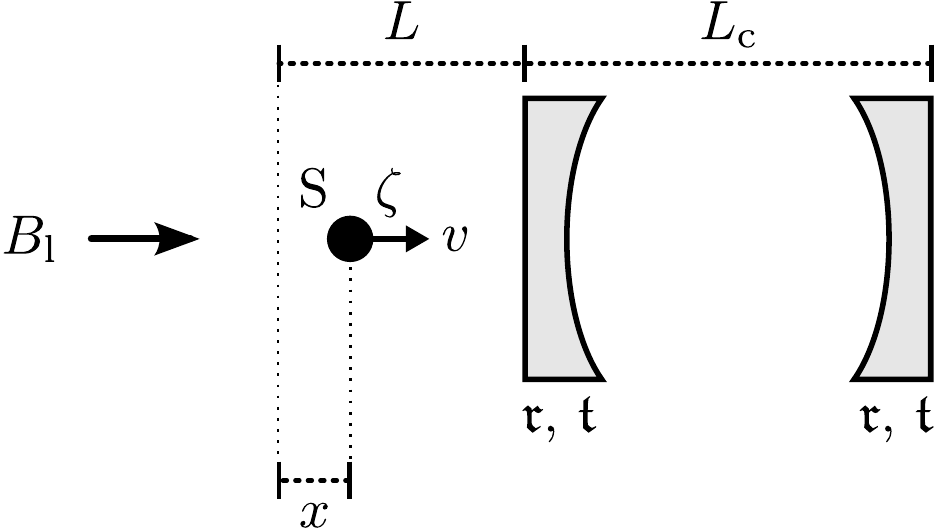}
  }\\
  \subfigure[]{
    \includegraphics[width=1.5\figwidth]{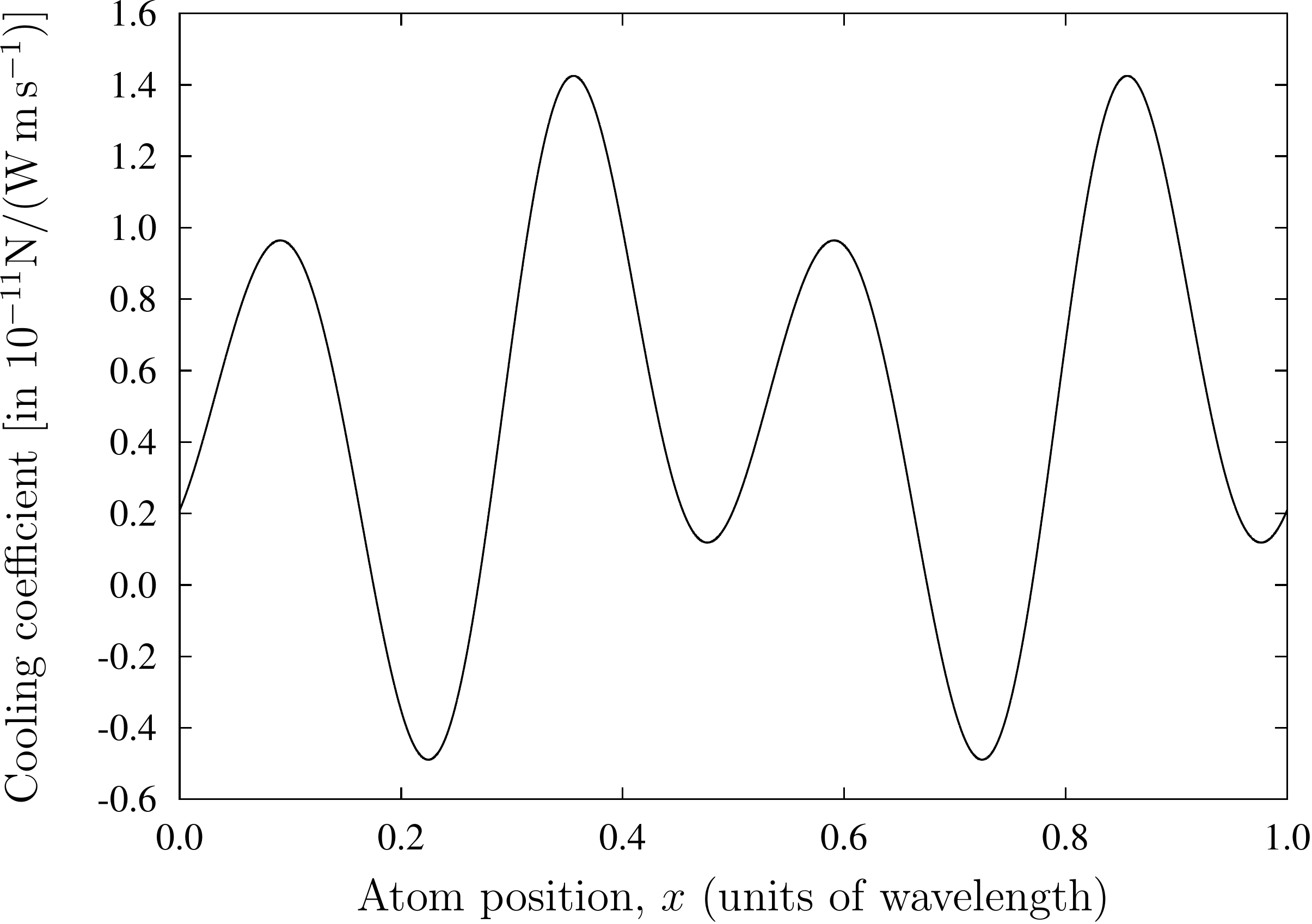}
  }
  \caption[Schematic model for external cavity cooling]{{External cavity cooling. (a)~Model, similar to \fref{fig:CMC}, but with the atom at a distance $L-x$ outside the cavity.} (b)~Cooling coefficient $\heatingcoefft$ per unit input power experienced by the scatterer as $x$ is varied for realistic parameters (see text for details). Note the change of scale, on the vertical axis, from \fref{fig:CMC}(b).}
  \label{fig:ECCO}
\end{figure}
In \sref{sec:TMM:ECCO} above, we also proposed~\cite{Xuereb2010b} that even with the scatterer \emph{outside} the cavity, the cavity's resonance can be exploited to greatly enhance the optomechanical friction experienced by the scatterer \emph{vis-\`a-vis} standard optomechanical cooling setups~\cite{Braginsky1967,Metzger2004,Groblacher2009a}, which place the scatterer in front of a single mirror. It is the aim of this subsection to explore this cooling mechanism, using {experimental parameters similar to those in the previous subsection}, and compare it to the cavity-mediated cooling mechanism discussed {there}.\par
Our mathematical model, \fref{fig:ECCO}, represents the cavity as a standard, symmetric Fabry--P\'erot cavity. However, we emphasise that in principle what is required is simply an optical resonance: the cavity in the model can indeed be replaced by whispering gallery mode resonators~\cite{Schliesser2010} or even solid-state resonators. As a basis for numerical calculations{, and to enable direct comparison}, we model the same resonator as in the previous subsection. It is important to emphasise that the achievable quality factors of the resonators used for external cavity cooling can intrinsically be made larger than the ones in the previous subsection (see, e.g., Ref.~\cite{Rempe1992}) because there does not need to be any form of optical or mechanical access inside the resonator itself.
\par
The pump beam frequency is again taken to be detuned by $10\Gamma$ to the red of the atomic transition. By placing the atom outside the cavity, one is free to use high-numerical-aperture optics to produce a tighter focus than might be possible in a cavity with good optical and mechanical access. Having a tight focus strengthens the atom--field coupling because of the $1/w^2$ dependence of $\zeta$ on the beam waist $w$; whereas the friction force scales linearly with the input power, it also scales as $\zeta^2\sim 1/w^4$ [cf.~\eref{eq:TMM:MirrorCoolForce}]. Focussing the beam therefore increases the atom--field coupling more than the local intensity. Thus, it is now assumed that the beam is focussed down to $1$\,$\upmu$m{, which} gives $\zeta=3.7\times 10^{-2}+3.7\times 10^{-3}\i$.\\
In order to make a fair comparison between the two cases, we choose to set the saturation parameter $s=0.14$, as in the previous subsection. The maximum achievable cooling coefficient is then {$2.9\times 10^{-21}$\,N/(m\,s$^{-1}$) for $200$\,pW} of input power, which is an order of magnitude smaller than the previous result {and leads to a $1/e$ velocity cooling time of $50$\,$\upmu$s and an equilibrium temperature of $280$\,$\upmu$K. The magnitude of the force in this case results from the much smaller pumping beam mode waist and the use of much higher powers, subsequently leading to a} stronger atom--field interaction. {W}ith this beam waist and finesse we would be restricted to input powers several orders of magnitude smaller if the atom were inside such a cavity.
\par
In {summary}, whereas the friction force inside a cavity is much stronger \emph{per unit input power and for the same beam waist}, the restrictions imposed on the magnitude of these quantities {when the atom is inside the cavity} reduce the maximally achievable friction force to a figure comparable to {when it lies} outside the cavity.

\subsection{Scaling properties {of cavity} cooling forces}\label{sec:Scaling}
\subsubsection{Localisation issues}\label{sec:Localisation}
\begin{figure}[t]
  \centering
  \includegraphics[width=1.5\figwidth]{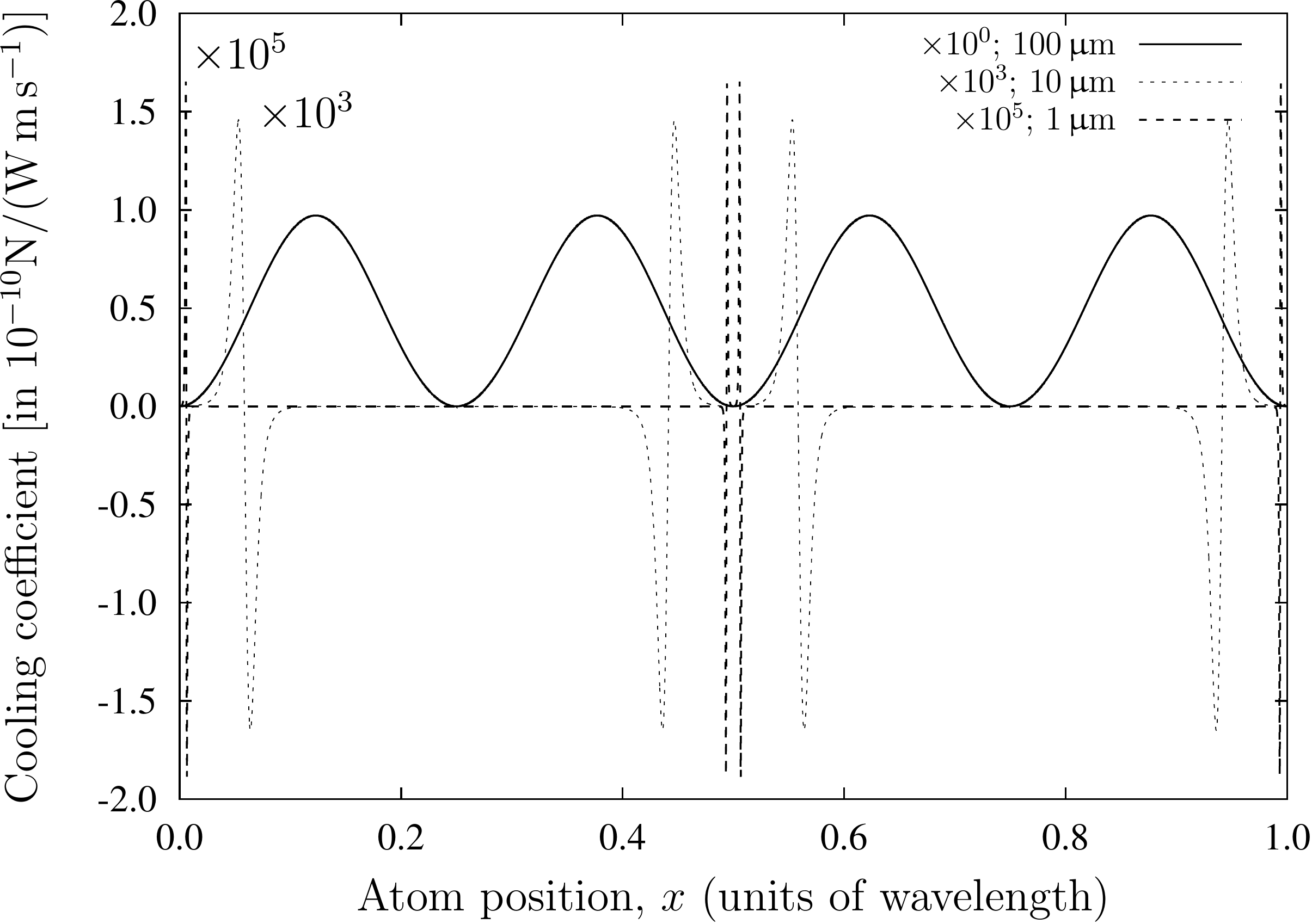}
  \caption[Spatial dependence of the friction force acting on an atom inside a cavity]{Spatial dependence of the friction force acting on an atom \emph{inside} a cavity with different mode waists (\ie, different polarisabilities) but equal detuning from resonance, $10\Gamma$ to the red. The smaller the mode waist the stronger the friction force, by several orders of magnitude, but the more significant localisation issues become. (Parameters as in \Sref{sec:CMC} but with $\partial\zeta/\partial k=0$.)}
  \label{fig:CMC_Waists}
\end{figure}
\begin{figure}[t]
  \centering
  \includegraphics[width=1.5\figwidth]{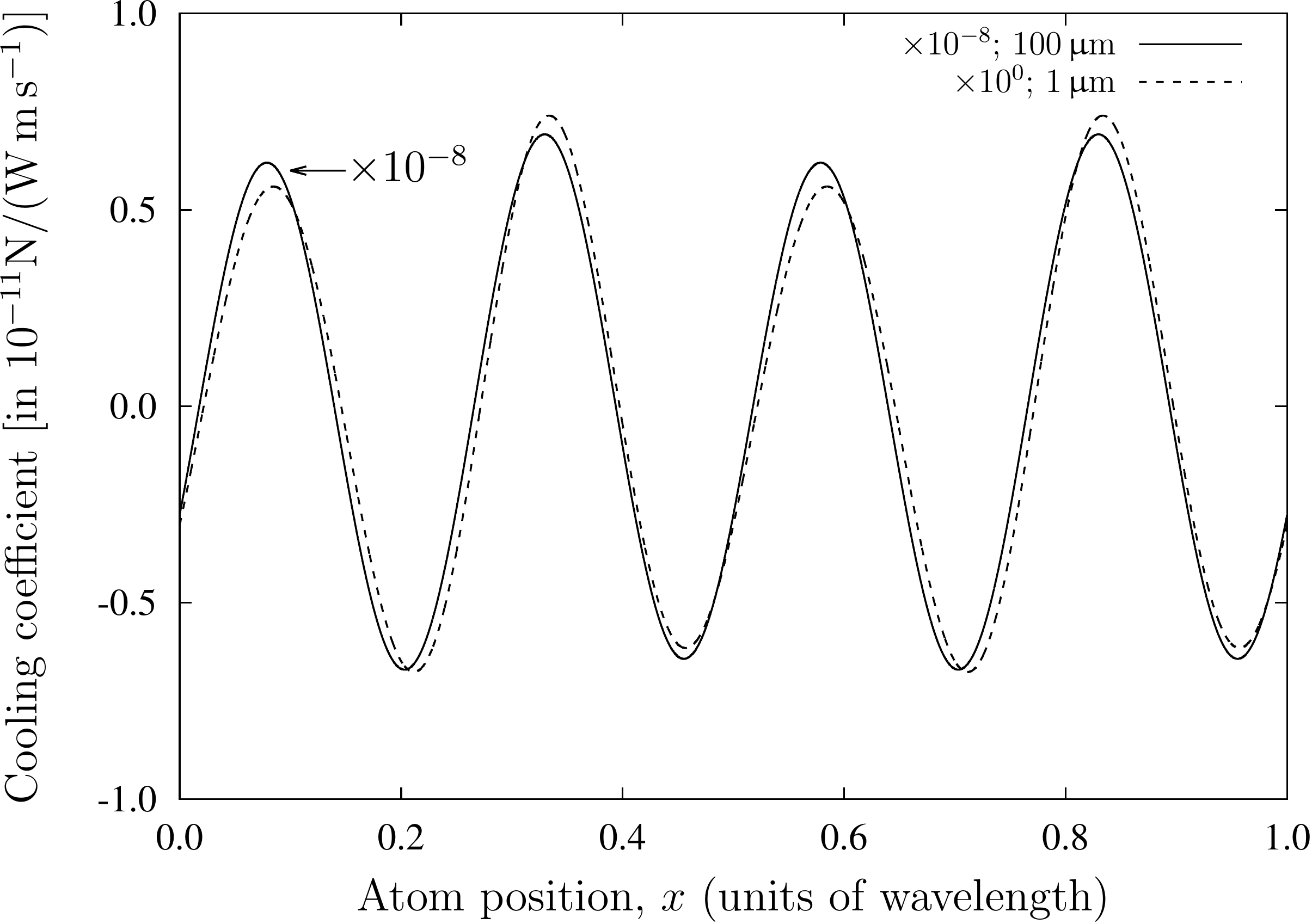}
  \caption[Spatial dependence of the friction force acting on an atom outside a cavity]{Spatial dependence of the friction force acting on an atom \emph{outside} a cavity, with different pumping field waists {but equal} detuning from resonance, $10\Gamma$ to the red. The friction force scales roughly as the inverse fourth power of the waist [cf.~\eref{eq:TMM:MirrorCoolForce}], but the length scale of the cooling and heating regions is unaffected. (Parameters as in \Sref{sec:ECCO} but with $\partial\zeta/\partial k=0$.)}
  \label{fig:ECCO_Waists}
\end{figure}
{The} broad nature of the spatial variations in the force shown in \fref{fig:CMC}(b) is {a consequence} of the small polarisability of the atom in such a cavity. {This is in sharp contrast to the case of large polarisability, achieved by an atom at a tight beam focus, as shown} in \fref{fig:CMC_Waists}, or by a micromirror. A {scatterer of larger polarisability} would experience extremely narrow ($\ll\lambda$) peaks in the friction force inside a cavity but not outside it.\\
{Within the scattering model used in this section, the atom--cavity coupling can be tuned by varying either the beam waist or the laser detuning from atomic resonance. Experimentally, however, atom--cavity coupling is rarely investigated close to resonance, in order to minimise the effects of atomic decorehence through spontaneous emission. In such cases, this coupling can be increased by operating a cavity with a small mode waist; this may in turn be detrimental to the performance of the system due to the strong sub-wavelength nature of the interaction, as explored in \fref{fig:CMC_Waists}.} The net {effect of having a smaller mode waist is} that this not only demands extremely good localisation but also tends to decrease the effective cooling coefficient drastically{---by up to several orders of magnitude---}because of spatial averaging effects.\par
In \Sref{sec:ECCO}, no mention was made of the average friction force acting on the scatterer; indeed this average computes to approximately zero for any case involving far-detuned atoms, or other particles with an approximately constant polarisability, outside cavities. This, then, {also demands strong} localisation of the atom; whilst experimentally challenging this disadvantage is somewhat mitigated by the {easy} mechanical and optical access afforded by external cavity cooling schemes. {In \fref{fig:ECCO_Waists} it is shown that the polarisability of the atom can be varied over a very wide range without affecting the length scale of the cooling and heating regions. The friction force can be seen to vary as $1/w^4$ for a beam waist $w$; in turn, this originates from the $\zeta^2$ scaling of the friction force [cf.~\eref{eq:TMM:MirrorCoolForce}].\\
In contrast with the atomic situation}{, if the scatterer is a micromirror mounted on a cantilever, localisation {does not present a problem}, since such micromirrors naturally undergo small} oscillations {and can be positioned with sub-nm accuracy}.

\subsubsection{Scaling with cavity {finesse and linewidth}}\label{sec:CavityLength}
{Cavity-mediated cooling mechanisms are heavily dependent on the physical properties of the cavity, namely its linewidth $\kappa$ and finesse $\mathcal{F}$. These parameters can be tuned independently by changing the length of the cavity and the reflectivity of its mirrors. This subsection briefly explores how the two mechanisms we are considering scale with $\kappa$ and $\mathcal{F}$.
\par
Expressions for the force acting on an atom inside a good cavity are not simple to write down. Nevertheless, in the good-cavity limit one may obtain an analytical formulation for the limiting temperature~\cite{Horak1997}:
\begin{equation}
T=\frac{\hbar\kappa}{k_\text{B}}\,,
\end{equation}
\ie, making a cavity longer decreases the equilibrium temperature proportionately. This result can be justified by observing that whereas the diffusion constant depends only on the intensity inside the cavity ($\propto\mathcal{F}$), the friction force scales linearly with both the intensity and, if the intensity is kept constant, with the lifetime of the cavity field ($\propto 1/\kappa$). The friction force is therefore proportional to $\mathcal{F}/\kappa$, and the equilibrium temperature proportional to $\kappa$.}
\par
\begin{figure}[t]
  \centering
  \includegraphics[width=1.5\figwidth]{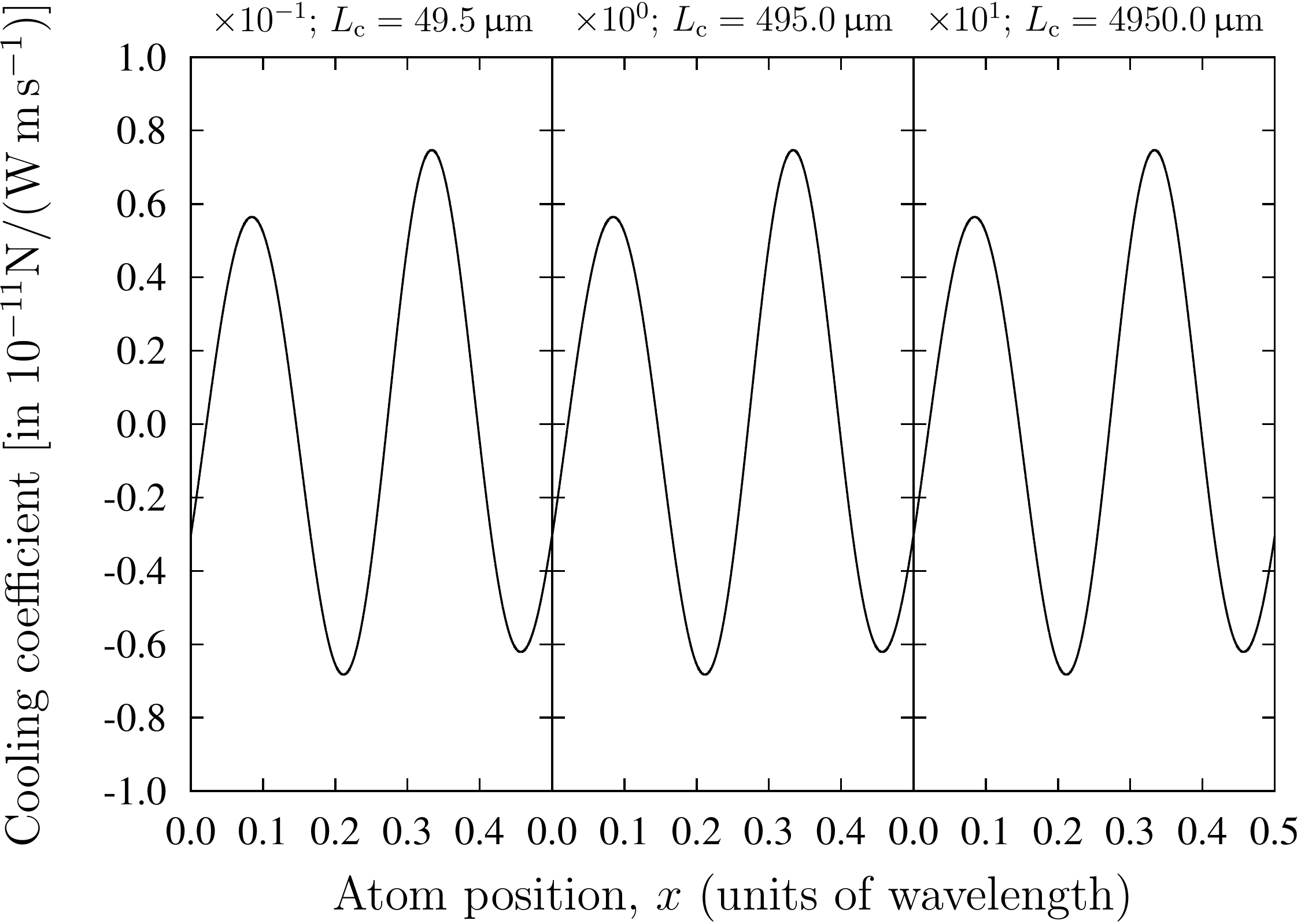}
  \caption[Scaling with cavity length of the friction force acting on an atom outside a cavity]{Spatial dependence of the friction force acting on an atom \emph{outside} a cavity, with detuning $10\Gamma$ to the red of resonance. Three different cavity lengths are shown; the friction force scales almost linearly with the cavity length. Note the different scaling factors and cavity lengths, given above each curve. (Parameters as in \Sref{sec:ECCO} but with $\partial\zeta/\partial k=0$.)}
  \label{fig:ECCO_Lengths}
\end{figure}
As is known from \sref{sec:TMM:ECCO}, the friction force ($\force$) acting on {an atom} outside a cavity scales {approximately} linearly with {both the length and the finesse} of the cavity. This is interpreted in terms of a `distance folding' mechanism: the lifetime of the light inside a cavity scales {inversely} with its {linewidth $\kappa\propto 1/\bigl(\mathcal{F}L_\text{C}\bigr)$} if all other parameters are kept fixed. Within the range of parameters where the light coupled into the cavity dominates the optomechanical effects, this implies that {$\force\propto 1/\kappa$}. One can see this behaviour reproduced in \fref{fig:ECCO_Lengths}, where the cooling coefficient acting on an atom outside each of three cavities having different lengths is shown. This mechanism loses its importance if {the atomic polarisability} is too large, whereby the system behaves more like two coupled cavities, or if the cavity is too long. {The momentum diffusion affecting the atom outside a cavity is essentially independent of the length of the properties of a good cavity, since it depends on the local intensity surrounding the scatterer. Putting these two results together, then, gives (\sref{sec:TMM:ECCO})
\begin{equation}
T\approx 1.9\frac{\hbar\kappa}{k_\text{B}}\,,
\end{equation}
in the limiting case of small polarisability and at the point of maximum friction; \ie, the temperature scales in the same way as for an atom inside the cavity. The numeric factor in the preceding equation depends on $\zeta$ and is larger for $\lvert\zeta\rvert\sim 1$.}

\section{Amplified optomechanics in a ring cavity}\label{sec:TMM:AmplifiedOptomechanics}
In the limit of strong scatterers, friction forces in standing-wave cavities become increasingly position-dependent (cf.~\Sref{sec:TMM:Comparison}), which limits the overall, averaged cooling efficiency. This can be overcome by using ring cavities~\cite{Gangl2000a,Elsasser2003,Kruse2003,Nagy2006,Slama2007,Hemmerling2010,Schulze2010,Niedenzu2010} where the translational symmetry guarantees position-independent forces. On the other hand, ring cavities are usually much larger and of lower $Q$-factor than their standing-wave counterparts. Using a gain medium inside a ring cavity has been proposed~\cite{Vuletic2001a,Salzburger2006} {to offset these losses}, allowing one {to effectively `convert'} a low-$Q$ cavity into a high-$Q$ one, and {thus to} increase the effective optomechanical interaction by orders of magnitude. This {same concept} has also been discussed in theoretical proposals investigating the use of optical parametric amplifiers in standard optomechanical systems~\cite{Huang2009}, or nonlinear media inside cavities~\cite{Kumar2010} as a tool to control the dynamics of a micromechanical oscillator. {A further application of ring cavities is in the investigation of collective atomic recoil lasing (CARL)~\cite{Bonifacio1994}, which exploits the spontaneous self-organisation of an atomic ensemble within a ring cavity, induced by a strong pump beam, to amplify a probe beam through Doppler-shifted reflection of the pump. The gain medium is in this case the atomic ensemble itself.}
\par
{Let us now consider a different system that shares several features with the above mechanisms. In particular, we consider a scatterer inside a ring cavity that includes a gain medium, spatially separated from the scatterer. An isolator is also included in the ring cavity, in such a way as to prevent the pump beam from circulating in the cavity and being amplified; this ensures that the intensity of the field surrounding the scatterer is always low and thereby circumvents any problems caused by atomic saturation or mirror burning. The Doppler-shifted reflection of the pump from the scatterer is, on the other hand, allowed to circulate, and its amplification in turn enhances the velocity-dependent forces acting on the scatterer.}\\
{In such a situation}, one is able to take advantage of properties inherent to the ring cavity system, such as the fact that the forces acting on the particle do not exhibit any sub-wavelength spatial modulation; this is due to the translational symmetry present in the system~\cite{Gangl2000a}. Moreover, modest amplification allows one to use optical fibres to form the ring cavity, opening the door towards increasing the optical length of such cavities. It is necessary to have amplification in such cases to compensate for the losses introduced when coupling to a fibre; such losses would otherwise limit the quality factor of the cavity. The optomechanical force is, {as we will see and} in the parameter domain of interest, linearly dependent on the cavity length; lengthening the cavity thus provides further enhancement of the interaction.
\par
This section is structured as follows. We shall first introduce the physical model, which we proceed to solve using the transfer matrix method to obtain the friction force and momentum diffusion acting on the particle. In the good-cavity limit, \Sref{sec:GoodCavity}, simple expressions for these quantities can be obtained, yielding further insight into the system and allowing us to draw some parallels with traditional cavity cooling. {In this limit, our model becomes equivalent to one based on a standard master equation approach}~\cite{Gangl2000a} {as outlined} in \Sref{sec:Semiclassical}. Realistic numerical values for the various parameters are then used in \Sref{sec:Results} to explore the efficiency and limits of the cooling mechanism. Finally, we will conclude by mentioning some possible extensions to the scheme. The work in this section will be published in the J.\ Mod.\ Opt.\ topical issue on \emph{New cooling mechanisms for atoms and molecules}.

\begin{figure}[t]
  \centering
  \includegraphics[width=\linewidth]{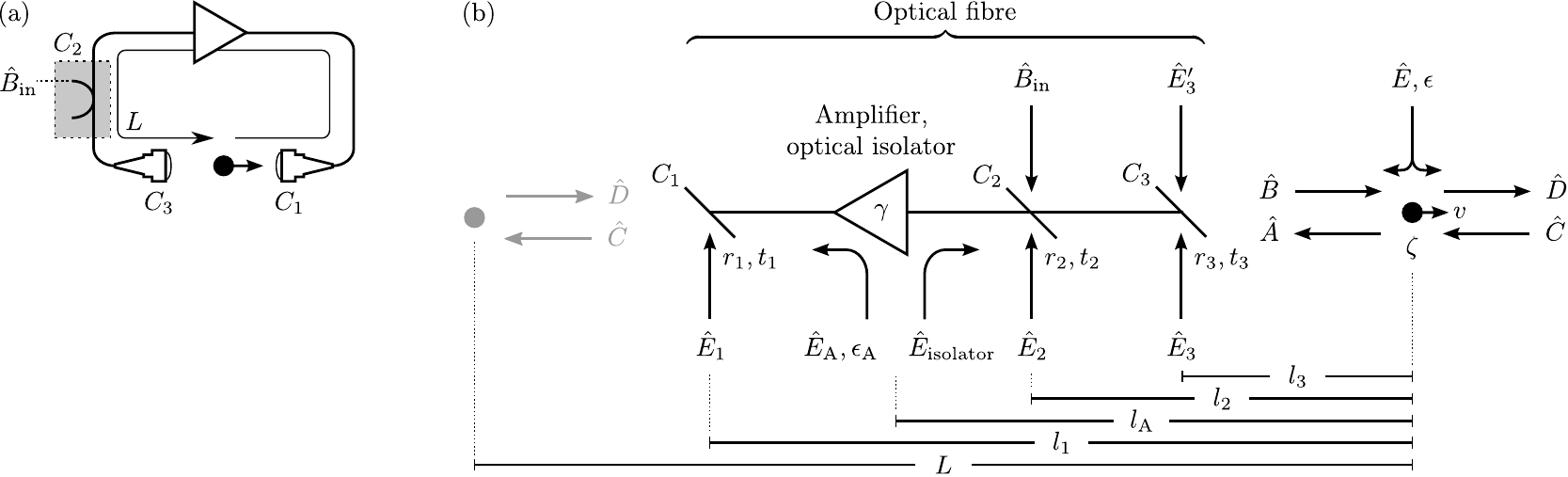}
   \caption[Model for a polarisable particle interacting with the field in a unidirectional ring cavity]{(a)~Physical schematic of a polarisable particle in a unidirectional ring cavity, showing the input field $\hat{B}_\text{in}$. (b)~Equivalent `transfer matrix'--style (unfolded) model; the particle is drawn on both sides of this schematic to illustrate the recursive nature of the cavity. The various components are defined in \Sref{sec:Model}.}
  \label{fig:Model}
\end{figure}
\subsection{General expressions and equilibrium behaviour}\label{sec:Model}
The mathematical model of the ring cavity system, schematically drawn in \fref{fig:Model}(a), is shown in \fref{fig:Model}(b). {A particle, characterised by its polarisability $\zeta$, is in a ring cavity of round-trip length $L$. $C_{1,3}$ are the couplers, between which lies the particle, that terminate the fibre-based cavity, and $C_2$ is the input coupler that injects the pump beam into the cavity. The amplifier, having a gain $\gamma\geq 1$, is assumed to also function as an optical isolator. $r_i$ and $t_i=r_i+1$ are the (amplitude) transmission and reflection coefficients of coupler $C_i$ ($i=1,2,3$). All the ``$\hat{E}$'' modes are noise modes introduced by specific elements. $\epsilon_\text{A}$, the amplitude of $\hat{E}_\text{A}$ introduced into the system, depends on $\gamma$; similarly, $\epsilon$ depends on $\zeta$. The values of $l_1>l_\text{A}>l_2>l_3>0$ are not important.} {One of the travelling wave modes of the cavity is pumped} by light with wavenumber $k_0${, but is prevented from circulating inside the cavity. This avoids resonant enhancement of the pumped mode in the cavity and thus avoids saturating the particle. The backscattered counterpropagating mode, on the other hand, is amplified on every round trip by a factor $\gamma$ by means of the optical amplifier.} The TMM is used to self-consistently solve for the four field amplitudes {at every point in the cavity in the presence of the pump field and the noise modes introduced by the coupler losses and by the amplifier. Note that in the limit where the amplifier is compensating for the ring cavity losses, the amplifier noise is also comparable to the loss-induced noise and must therefore be} taken into account in our model.
\par
Using the notation in \fref{fig:Model}(a) we can relate the expectation values of the amplitudes of the {two input and two scattered} field modes interacting with the particle {in a one-dimensional scheme}, {$A(k)=\langle\hat{A}(k)\rangle$, $B(k)=\langle\hat{B}(k)\rangle$, $C(k)=\langle\hat{C}(k)\rangle$, and $D(k)=\langle\hat{D}(k)\rangle$, to $B_\text{in}(k)=\langle\hat{B}_\text{in}(k)\rangle$} by means of the relations
\begin{equation}
\label{eq:BasicRelations}
B = r_2t_3e^{ik_0l_2}B_\text{in}\,,\ C=\alpha A\,,\ \text{and}\ \begin{pmatrix}A\\B\end{pmatrix}=\hat{M}\begin{pmatrix}C\\D\end{pmatrix}\,,
\end{equation}
where {$\alpha(k)=t_1t_2t_3\gamma\exp(ikL)$} is the factor multiplied to the field amplitude every round trip. {In the preceding equations, as well as in the following, we do not write the index $k$ for simplicity of presentation. The operators $\hat{A}(k)$, etc., denote the annihilation operators of the various field modes. The three equations \erefs{eq:BasicRelations}} have a readily apparent physical significance---respectively, they correspond to: the propagation of $\hat{B}_\text{in}$ to reach the particle; the feeding back of $\hat{A}$ to $\hat{C}$ through the ring cavity; and the usual transfer matrix relation for a particle interacting with the four fields surrounding it. The first two of these relations are substituted into the third, which subsequently simplifies to
\begin{equation}
\label{eq:BaseRelation}
\begin{pmatrix}A\\B_\text{in}\end{pmatrix}=\left(\begin{bmatrix}1&0\\0&r_2t_3e^{\i k_0l_2}\end{bmatrix}-\hat{M}\begin{bmatrix}\alpha&0\\0&0\end{bmatrix}\right)^{-1}\hat{M}\begin{pmatrix}0\\D\end{pmatrix}
\end{equation}
If we assume far off-resonant operation, \ie, $\partial\zeta/\partial k=0$, the velocity-dependent transfer matrix $\hat{M}$ can be written as, cf.~\eref{eq:TMMFirstordervExplicit},
\begin{equation}
\begin{bmatrix}1+\i\zeta&\i\zeta-2\i\zeta\tfrac{v}{c}+2\i k_0\zeta\tfrac{v}{c}\partial_k\\-\i\zeta-2\i\zeta\tfrac{v}{c}+2\i k_0\zeta\tfrac{v}{c}\partial_k&1-\i\zeta\end{bmatrix}\,.
\end{equation}
Note that the partial derivative $\partial_k$ acts not only on $\alpha(k)$ but also on the field mode amplitudes it precedes. \eref{eq:BaseRelation} can be inverted in closed form to first order in $v/c$, similarly to \sref{sec:TMM:General}, and can thus be used to find $\mathcal{A}=\sqrt{2\epsilon_0\sigma_\text{L}/(\hbar k_0)}\int A(k)\rmd k$, $\mathcal{B}=\sqrt{2\epsilon_0\sigma_\text{L}/(\hbar k_0)}\int B(k)\rmd k$, $\mathcal{C}=\sqrt{2\epsilon_0\sigma_\text{L}/(\hbar k_0)}\int C(k)\rmd k$, and $\mathcal{D}=\sqrt{2\epsilon_0\sigma_\text{L}/(\hbar k_0)}\int D(k)\rmd k$, {where the normalisation is with respect to} the pump beam mode area $\sigma_\text{L}$ and {where a monochromatic pump is assumed}: $B_\text{in}(k)=B_0\delta(k-k_0)${. Here, $\lvert\mathcal{A}\rvert^2$, $\lvert\mathcal{B}\rvert^2$, etc., are the photon currents in units of photons per second}. The expectation value of the force acting on the scatterer is then given by \eref{eq:TMM:MSTForce}:
\begin{equation}
\label{eq:RawForce}
\hbar k_0\bigl(\lvert\mathcal{A}\rvert^2+\lvert\mathcal{B}\rvert^2-\lvert\mathcal{C}\rvert^2-\lvert\mathcal{D}\rvert^2\bigr)\,.
\end{equation}
The values of $\mathcal{A}$, $\mathcal{B}$, etc., from the solution of \eref{eq:BaseRelation} are then substituted in \eref{eq:RawForce}, which {we evaluate to} first order in $v/c$, in terms of $B_0$. After some algebra, we obtain the first main result of this section---the friction force acting on the particle:
\begin{equation}
\label{eq:TMMFriction}
\force=-8\hbar k_0^2\tfrac{v}{c}\re{\frac{\bigl(1-\alpha^\ast\bigr)\zeta\re{\zeta}+\i\alpha^\ast\zeta\lvert\zeta\rvert^2}{1-\alpha-\i\zeta}\frac{\partial\alpha}{\partial k}}\frac{\bigl|r_2t_3B_0\bigr|^2}{\bigl|1-\alpha-\i\zeta\bigr|^2}\,.
\end{equation}
By extending the TMM appropriately, one can keep track of the various noise modes interacting with the system. {\erefs{eq:BasicRelations} then become}
\begin{subequations}
\begin{equation}
\hat{A}=\frac{\i\zeta}{1-\i\zeta}\hat{B}+\frac{1}{1-\i\zeta}\hat{C}+\epsilon\hat{E}\,,
\end{equation}
\begin{equation}
\hat{B}=r_2t_3e^{\i k_0l_2}\hat{B}_\text{in}+t_2t_3e^{\i k_0l_\text{A}}\hat{E}_\text{isolator}+r_3e^{\i k_0l_3}\hat{E}_3^\prime\,,\ 
\end{equation}
\begin{equation}
\hat{C}=\alpha\hat{A}+r_1e^{\i k_0(L-l_1)}\hat{E}_1+t_1r_2\gamma e^{\i k_0(L-l_2)}\hat{E}_2+t_1t_2r_3\gamma e^{\i k_0(L-l_3)}\hat{E}_3+t_1\epsilon_\text{A}e^{\i k_0(L-l_\text{A})}\hat{E}_\text{A}\,, \text{ and}
\end{equation}
\begin{equation}
\hat{D}=\frac{1}{1-\i\zeta}\hat{B}+\frac{\i\zeta}{1-\i\zeta}\hat{C}+\epsilon\hat{E}\,,
\end{equation}
\end{subequations}
with $\epsilon=\sqrt{1-\bigl(1+\lvert\zeta\rvert\bigr)/\lvert 1-\i\zeta\rvert^2}$ (see \aref{sec:TMM:AppDiffusion}) and $\epsilon_\text{A}=\sqrt{1-1/\lvert\gamma\rvert^2}$~\cite{Gardiner2004}. These equations can be solved simultaneously for $\hat{A}$, $\hat{B}$, $\hat{C}$, and $\hat{D}$, and the solution used to evaluate the momentum diffusion constant, $\diffn$, {defined as} the two-time autocorrelation function of the force operator (cf. \sref{sec:MSTDiffusion}), keeping in mind that most of the noise modes, as well as $\hat{B}_\text{in}$, obey the commutation relation $\bigl[\hat{E}(t),\hat{E}^\dagger(t^\prime)\bigr]=\hbar k_0/(2\epsilon_0 \sigma_\text{L})\,\delta(t-t^\prime)$. The sole exception is the noise introduced by the amplifier, $\hat{E}_\text{A}$, for which $\bigl[\hat{E}_\text{A}(t),\hat{E}_\text{A}^\dagger(t^\prime)\bigr]=-\hbar k_0/(2\epsilon_0 \sigma_\text{L})\,\delta(t-t^\prime)$; this is due to the model of the amplifier as a negative temperature heat-bath, whereby the creation and annihilation operators effectively switch r\^oles. Further discussion of this model can be found in Ref.~\cite[\textsection 7.2]{Gardiner2004}. All the noise modes are independent from one another and from $\hat{B}_\text{in}$, which simplifies the expressions considerably.

\subsubsection{The good-cavity limit as a simplified case}\label{sec:GoodCavity}
{Before discussing the result of \Sref{sec:Model}, we shall} make several approximations to obtain a transparent set of equations to briefly explore the equilibrium behaviour of the scatterer {and to compare with a standard master equation approach}. In particular, $\zeta$ is assumed to be real, which is tantamount to assuming that the scatterer suffers no optical absorption or, if it is an atom, that it is pumped far off-resonance. Moreover, the cavity is assumed to be very good ($\lvert t_{1,2,3}\rvert\to 1$) {and thus no gain medium is introduced in the cavity ($\gamma=1$)}. With these simplifications, \eref{eq:TMMFriction} reduces to
\begin{align}
\label{eq:TMMFrictionSimple}
\force&\approx -8\hbar k_0^2\tfrac{v}{c}\frac{\zeta^2}{\lvert 1-\alpha\rvert^4}\re{\bigl(1-\alpha^\ast\bigr)^2\frac{\partial\alpha}{\partial k}}\bigl|r_2B_0\bigr|^2\nonumber\\
&\approx 16\hbar k_0^2\zeta^2 v\frac{\kappa\Delta_\text{C}}{\bigl(\Delta_\text{C}^2+\kappa^2\bigr)^2}\frac{1}{\tau}\bigl|r_2B_0\bigr|^2\,.
\end{align}
In the preceding equations, $\Delta_\text{C}$ is the detuning {of the pump} from cavity resonance, $\kappa$ is the HWHM cavity {linewidth,
\begin{equation}
\kappa=\frac{1}{\tau}\frac{1-\lvert t_1t_2t_3\rvert\gamma}{\sqrt{\lvert t_1t_2t_3\rvert\gamma}}\,,
\end{equation}
for $\lvert\zeta\rvert\ll 1$,} {and} $\tau = L/c$ is the round-trip time{.} Using the same approximations {as for \eref{eq:TMMFrictionSimple}}, we also obtain the diffusion constant
\begin{equation}
\label{eq:TMMDiffusionSimple}
\diffn\approx 8\hbar^2 k_0^2\zeta^2\frac{\kappa}{\Delta_\text{C}^2+\kappa^2}\frac{1}{\tau}\bigl|r_2B_0\bigr|^2\,.
\end{equation}
{Note that} $\gamma=1$ {here and therefore} $\hat{E}_\text{A}$ does not contribute to the diffusion constant{.} {\erefs{eq:TMMFrictionSimple} and~(\ref{eq:TMMDiffusionSimple}) hold for the case where $\Delta_\text{C}/\kappa$ is not too large. The cavity can be fully described by means of $\kappa$ and the finesse $\mathcal{F}=\pi c/(2L\kappa)$. Let us now set $\Delta_\text{C}=-\kappa$ in \erefs{eq:TMMFrictionSimple} and~(\ref{eq:TMMDiffusionSimple}), whereby
\begin{equation}
\force=-\tfrac{8}{\pi}\hbar k_0^2\zeta^2 v\frac{\mathcal{F}}{\kappa}\bigl|r_2B_0\bigr|^2\,,\text{ and }
\diffn=\tfrac{8}{\pi}\hbar^2 k_0^2\zeta^2\mathcal{F}\bigl|r_2B_0\bigr|^2\,.
\end{equation}
These two expressions have a readily-apparent physical significance; at a constant finesse, decreasing the cavity linewidth is equivalent to making the cavity longer, whereupon the retardation effects that underlie this cooling mechanism lead to a stronger friction force. At the same time, this has no effect on the intracavity field strength and therefore does not affect the diffusion. On the other hand, improving the cavity finesse by reducing losses at the couplers increases the intracavity intensity, thereby increasing both the friction force and the momentum diffusion.}
\par
Using the above results, we obtain, for $\Delta_\text{C}<0$,
\begin{equation}
\label{eq:TMMTemperatureSimple}
T_\text{A}\approx \frac{\hbar}{k_\text{B}}\biggl(\frac{\lvert\Delta_\text{C}\rvert}{\kappa}+\frac{\kappa}{\lvert\Delta_\text{C}\rvert}\biggr)\frac{\kappa}{2}\geq\frac{\hbar}{k_\text{B}}\kappa\,,
\end{equation}
with the minimum temperature occurring at $\Delta_\text{C}=-\kappa$. One notes that this expression is identical to the corresponding one for standard cavity-mediated cooling~\cite{Horak1997}. Like the corresponding results in \sref{sec:MovingAtom} and \sref{sec:Mirrorcooling}, it can be interpreted in a similar light as the Doppler temperature, albeit with the energy dissipation process shifted from the decay of the atomic excited state to the decay of the cavity field{.}
\par
A particular feature to note in all the preceding expressions is that they are not spatial averages over the position of the particle, but they do not depend on this position either. As a result of this, the force, momentum diffusion and equilibrium temperature do not in any way depend on the position of the particle along the cavity field in a 1D model. The issue of sub-wavelength modulation of the friction force is a major limitation of cooling methods based on intracavity standing fields, in particular mirror-mediated cooling (\sref{sec:TMM:GeneralMMC}) and cavity-mediated cooling (\Sref{sec:TMM:Comparison}).

\subsubsection{Comparison with a semiclassical model}\label{sec:Semiclassical}
In the good-cavity limit {and without gain our TMM model is equivalent to a standard master equation approach} with the Hamiltonian
\begin{align}
\label{eq:Hamiltonian}
\hat{H}=&-\hbar\Delta_\text{a}\hat{\sigma}^+\hat{\sigma}^- -\hbar\Delta_\text{C}\hat{a}_\text{C}^\dagger \hat{a}_\text{C}\nonumber\\
&+\hbar g\bigl(\hat{a}_\text{C}^\dagger\hat{\sigma}^- e^{\i k_0x}+\hat{\sigma}^+\hat{a}_\text{C}e^{-\i k_0x}\bigr)+\hbar g\bigl(a_\text{P}^\ast\hat{\sigma}^- e^{-\i k_0x}+\hat{\sigma}^+a_\text{P}e^{\i k_0x}\bigr)\,,
\end{align}
and the Liouvillian terms
\begin{equation}
\label{eq:Liouvillian}
\mathcal{L}\hat{\rho}=-\Gamma\bigl(\hat{\sigma}^+\hat{\sigma}^-\hat{\rho}-2\hat{\sigma}^-\hat{\rho}\hat{\sigma}^++\hat{\rho}\hat{\sigma}^+\hat{\sigma}^-\bigr)-\kappa\bigl(\hat{a}_\text{C}^\dagger \hat{a}_\text{C}\hat{\rho}-2\hat{a}_\text{C}\hat{\rho} \hat{a}_\text{C}^\dagger+\hat{\rho} \hat{a}_\text{C}^\dagger \hat{a}_\text{C}\bigr)\,,
\end{equation}
as adapted from Ref.~\cite{Gangl2000a} and modified for a unidirectional cavity where only the unpumped mode is allowed to circulate{. Here,} $\hat{\rho}$ is the density matrix of the system, $g$ the atom--field coupling strength, $\hat{a}_\text{C}$ the annihilation operator of the cavity field, $\hat{\sigma}^+$ the atomic dipole raising operator, $\Delta_\text{a}$ the detuning from atomic resonance, $\Gamma$ the atomic upper state HWHM linewidth, and $x$ the coordinate of the atom inside the cavity. The pump field is assumed to be unperturbed by its interaction with the atom, and in the above is replaced by a c-number, $a_\text{P}$. {Calculating the friction force from this model leads again to \eref{eq:TMMFrictionSimple}, thus confirming our TMM results by a more standard technique. The advantage of the TMM approach lies in the simplicity and generality of expressions such as \eref{eq:RawForce}, and the ease with which more optical elements can be introduced into the system. As shown above, the momentum diffusion coefficient is easily calculated from the TMM.}

\begin{figure}[t]
  \centering
  \includegraphics[angle=-90,width=1.5\figwidth]{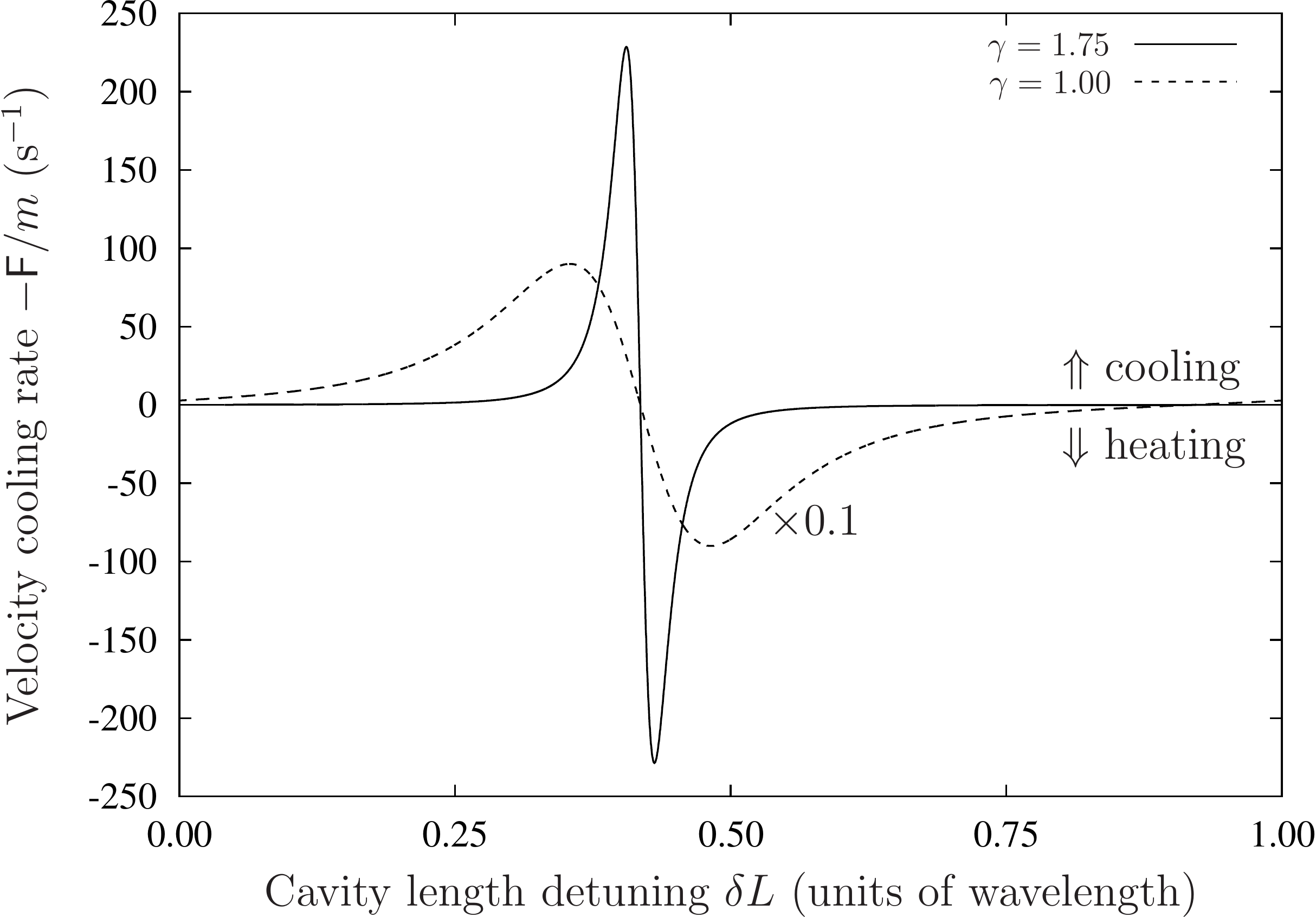}
   \caption[Cooling rate for $^{85}$Rb pumped $-10\Gamma$ from resonance inside a ring cavity]{Cooling rate $-(\text{d}v/\text{d}t)/v$ for $^{85}$Rb pumped $-10\Gamma$ from D$_2$ resonance inside a ring cavity with a round-trip length $L=300$\,m, for two different values of the amplifier gain $\gamma$. Note that the curve for {$\gamma=1$}, as drawn, is scaled \emph{up} by a factor of $10$. The cavity waist is taken to be $10$\,$\upmu$m. ($\lvert t_1\rvert^2=\lvert t_3\rvert^2=0.5$, $\lvert t_2\rvert^2=0.99$, {$B_0$ is chosen such as to give an} atomic saturation $s=0.1$.)}
  \label{fig:Friction}
\end{figure}
\begin{figure}[t]
  \centering
  \includegraphics[angle=-90,width=1.5\figwidth]{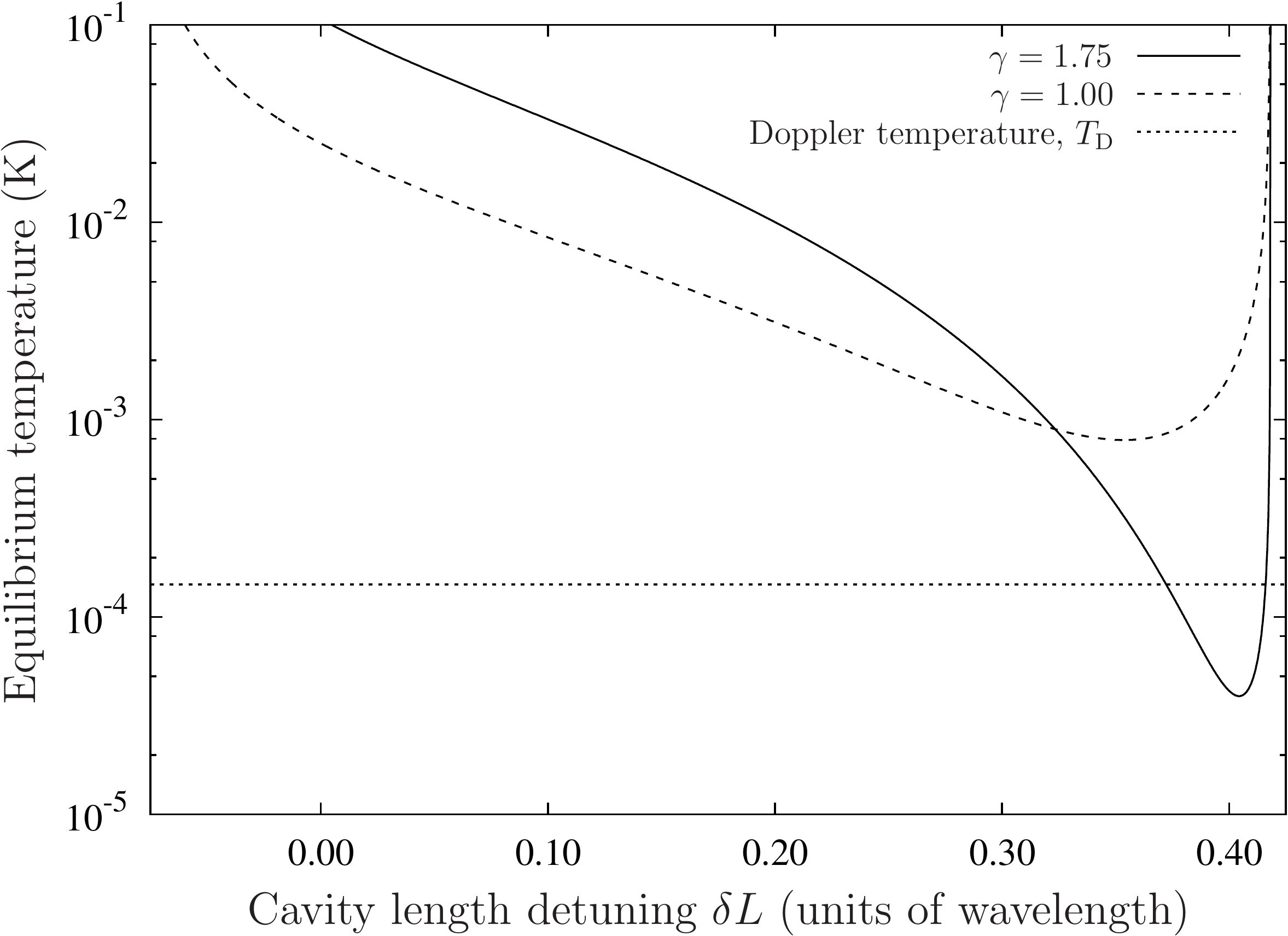}
   \caption[Equilibrium temperature predicted by the transfer matrix model for two values of the amplifier gain $\gamma$]{Equilibrium temperature predicted by the transfer matrix model for two values of the amplifier gain $\gamma$. The Doppler temperature for $^{85}$Rb is also indicated. The horizontal axis differs from that in \fref{fig:Friction} mainly because the temperature is only well-defined for regions where the friction force promotes cooling. (Parameters are as in \fref{fig:Friction}.)}
  \label{fig:Temperature}
\end{figure}
\subsection{Numerical results and discussion}\label{sec:Results}
We can use the conversion factor $\lvert B_0\rvert^2=P/(\hbar k_0c)$, where $P$ is the power of the input beam, to evaluate the above equations numerically in a physically meaningful way. Specifically, the particle is now assumed to be a (two--level) \textsuperscript{85}Rb atom, pumped $-10\Gamma$ from D$_2$ resonance, {where} $\Gamma=2\pi\times 3.03$\,MHz {is} the HWHM linewidth of this same transition at a wavelength of ca.~$780$\,nm; {because} the detuning is much larger than the linewidth, {we simplify the calculations} by setting $\partial\zeta/\partial\omega=0$. The beam waist where the particle interacts with the field is taken to be $10$\,$\upmu$m. With the parameters in \fref{fig:Friction}, the power is reduced by a factor of $1/\lvert t_1t_2t_3\rvert^2=4.04$ with each round-trip, in the presence of no gain in the amplifier. We shall compare this case to the low-gain case; the gain of the amplifier we consider is constrained to be small enough that $\lvert\alpha\rvert^2=\lvert t_1t_2t_3\gamma|^2<1$. Under these conditions, there is no {exponential} build-up of intensity inside the cavity and the system is stable. A cavity with a large enough gain that $\lvert\alpha\rvert^2>1$ would effectively be a laser cavity. Such a system {would have} no stable state in our model, since we assume that the gain medium is not depleted, and will therefore not be considered further in the following.
\par
\fref{fig:Friction} shows the friction force acting on the particle, and \fref{fig:Temperature} the equilibrium temperature, as the length of the cavity is tuned on the scale of one wavelength. In each of these two figures two cases are shown, one representing no gain in the amplifier ({$\gamma=1$}) and one representing a low-gain amplifier ($\gamma=1.75$); note that in both cases the condition $\lvert\alpha\rvert^2<1$ is satisfied.
\par
In order to provide a fair comparison between these two cases, we choose the pump amplitude $B_\text{in}$ such that the saturation of the particle is the same in the two cases. This ensures that any difference in cooling performance is not due to a simple increase in intensity. Since the TMM as presented here is based on a \emph{linear} model of the particle, our results presented above are only valid in the limit of saturation parameter much smaller than $1$. Thus, as a basis for the numerical comparisons between the two different cases, we choose to set the saturation parameter to $0.1$. \fref{fig:Friction} shows that under these conditions the amplified system leads to a significant, approximately $25$-fold, enhancement of the maximum friction force. This can therefore be attributed unambiguously to the effective enhancement of the cavity $Q$-factor by the amplifier.
\par
However, for the parameters considered here, in particular for small particle polarisability $\zeta$ and for $\lvert\alpha\rvert^2<1$, the counterpropagating mode intensity is much smaller than that of the pumped mode, even if the former is amplified. Thus, the intracavity field is always dominated by the pump beam, whereas the friction force is mostly dependent on the Doppler-shifted reflection of the pump from the particle. Specifically, for the parameters used above we find that the total field intensity changes by less than 1\% when the gain is increased from 1 to 1.75. Hence, similar results to those of \fref{fig:Friction} are obtained even \emph{without} pump normalisation.
\par
The steady-state temperature, obtained by the ratio of diffusion and friction, is shown in \fref{fig:Temperature} for the same parameters as above. We observe that the broader resonance in the friction as a function of cavity detuning (i.e., of cavity length), shown in \fref{fig:Friction}, also leads to a wider range of lower temperatures compared to the amplified case. However, as expected, within the narrower resonance of the amplified system where the friction is significantly enhanced, the stationary temperature is also significantly reduced. We see that while the maximum friction force is increased by a factor of $25.4$, the lowest achievable temperature is decreased by a factor of $19.9$ when switching from $\gamma=1$ to $\gamma=1.75$. While the overall cavity intensity is dominated by the pump field, and is therefore hardly affected by the amplifier, the diffusion is actually dominated by the interaction of the weak counterpropagating field with the pump field. This can be seen most clearly by the strong detuning dependence of the analytical expression for $\diffn$ in the good-cavity limit, \eref{eq:TMMDiffusionSimple}. As a consequence, the lowest achievable temperature is improved by a slightly smaller factor than the maximum cooling coefficient. This is consistent with the idea that the amplifier not only increases the cavity lifetime, but also adds a small amount of additional noise into the system. Nevertheless, a strong enhancement of the cooling efficiency is observed in the presence of the amplifier.

\chapter{Three-dimensional scattering with an optical memory}\label{ch:TMM:Scattering}
\epigraph{Homogeneal Rays which flow from several Points of any Object, and fall almost Perpendicularly on any reflecting or refracting Plane or Spherical Surface, shall afterwards diverge from so many other Points, or be Parallel to so many other Lines, or converge to so many other Points, either accurately or without any sensible Error. And the same thing will happen, if the Rays be reflected or refracted successively by two or three or more Plane or spherical Surfaces.\par
[...]\par
Wherever the Rays which come from all the Points of any Object meet again in so many Points after they have been made to converge by Reflexion or Refraction, there they will make a Picture of the Object upon any white Body on which they fall.}{I.\ Newton, \emph{Opticks} (1704)}

The scattering theory presented in the previous chapters can be used to describe a wide variety of one-dimensional, or quasi-one-dimensional, situations involving mobile scatterers and immobile optics. The three-dimensional nature of the electromagnetic field can, however, be exploited to give rise to a different type of retarded dipole-dipole interaction, one mediated not only by the relative phase difference between the successive reflections but also the spreading nature of spherical waves in three dimensions.
\par
I start this chapter with an extension of a self-consistent scattering theory that was used to describe optical binding phenomena~\cite{Depasse1994}. Our treatment essentially identifies the two particles described in Ref.~\cite{Depasse1994} such that the `binding' that takes place really is between a particle and itself, as mediated by a mirror or other delay element. After describing this extension to the theory in \Sref{sec:TMM:Scattering:Theory}, I shall apply it to the mirror-mediated cooling described earlier in this work, both in one dimension (cf.~\sref{sec:TMM:1DMMC}) and in three (\sref{sec:TMM:3DMMC}). I will use these results in \cref{ch:Experimental:Future} to discuss experimentally-accessible configurations for exploring the various mechanisms described in this thesis.

\begin{figure}[t]
  \centering
  \includegraphics{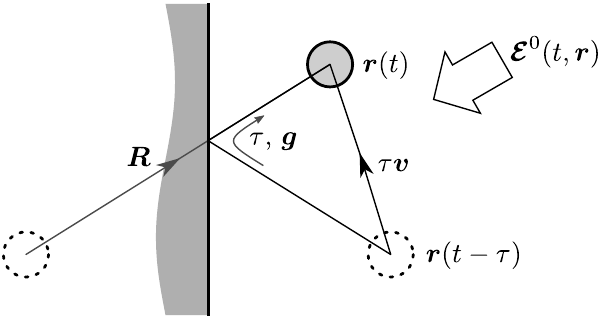}
   \caption[A particle bound to its retarded reflection in a surface]{Polarised by incident radiation $\v{\efield}^0(t,\v{r})$, a particle moving with velocity $\v{v}$ is bound to its retarded reflection in a surface, characterised by round-trip time $\tau$ and field propagator $\m{g}$.}
  \label{fig:TMM:RetardedBinding}
\end{figure}
\section{Optical self-binding of Rayleigh particles}\label{sec:TMM:Scattering:Theory}
The work in this section is being prepared for publication by James Bateman, AX, and Tim Freegarde. Its basis originated from TF and was subsequently elaborated upon by all three authors. The \emph{Mathematica} code used to solve the equations and verify the solutions was written largely by JB.
\par
We first analyse the retarded classical electrostatic interaction between the induced dipoles of a particle, moving along a path $\v{r}(t)$, and its reflection, as shown in \fref{fig:TMM:RetardedBinding}, following the ideas set forth in Ref.~\cite{Depasse1994}. The unperturbed illuminating field is denoted $\v{\efield}^0(t,\v{r})$. A tensor propagator, $\m{g}$, is used to describe the propagation of the scattered field from the particle and back to itself through reflection in the mirror. In one dimension,
\begin{equation}
\label{eq:1DPropagator}
\m{g}(t,\v{r})=-\i e^{\i\omega\tau(t,\v{r})}\mathds{1}\,,
\end{equation}
where the pre-factor $-\i$ accounts for the phase shift upon reflection as well as the Gouy phase shift: in our 1D model we take the particle to be a point-like dipole at the focus of a tightly-focussed beam of mode area $\sigma_\text{L}$. In three dimensions,
\begin{align}
\label{eq:3DPropagator}
\m{g}(t,\mathbf{r})=\frac{\sigma_\text{L}}{R(t)\lambda}e^{\i\omega\tau(t,\v{r})}\Biggl\{&\frac{1-\i kR(t)-\bigl[kR(t)\bigr]^2}{\bigl[kR(t)\bigr]^2}\mathds{1}\nonumber\\
&-\frac{3-3\i kR(t)-\bigl[kR(t)\bigr]^2}{\bigl[kR(t)\bigr]^2}\v{n}(t)\otimes\v{n}(t)\Biggr\}\,.
\end{align}
In the preceding equations, $R(t)=\lVert\v{R}(t)\rVert$, $\v{n}(t)=\v{R}(t)/R(t)$, $\v{R}(t)$ is the position vector of a test particle at time $t$ relative to its image at time $t-\tau[t,\v{r}(t)]$ (see \fref{fig:TMM:RetardedBinding}), and $\otimes$ represents the vector outer product. $\m{g}$ in \eref{eq:3DPropagator} is the tensor Green's function for free-space propagation~\cite{Levine1950,Martin1995} of an electromagnetic wave, normalised to render it dimensionless.
\par
The interaction of the particle with the electric field is assumed to be through the dipole interaction, whereby the particle is described by means of its polarisability $\chi$; this polarisability allows us to define a dimensionless polarisability
\begin{equation}
  \zeta=\chi k/(2\sigma_\text{L})\,,
\end{equation}
as before. The form for $\zeta$ we choose is identical for both one- and three-dimensional cases, despite arising from different considerations; in one dimension $\zeta$ is defined in \sref{sec:CoolingMethods:ForceTLA}, whereas in three dimensions we define $\zeta$ through a particular grouping of constants to yield dimensionless $\zeta$ and $\m{g}$. In this chapter we will, for simplicity, assume a real, scalar value for $\chi$. The resulting equations can be subsequently generalised for complex, vector or tensor polarisabilities. Multiple scattering between the particle and the mirror is taken into account self-consistently by solving for the total electric field $\v{\efield}[t,\v{r}(t)]$, experienced by the particle at position $\v{r}(t)$ and time $t$:
\begin{equation}
 \v{\efield} = \v{\efield}^0+\zeta\m{g}\cdot\v{\efield}\,,
\end{equation}
where all the terms are evaluated at $[t,\v{r}(t)]$. The solution of this equation can be given analytically to lowest order in $\v{v}$ and $\tau$, and for the case when $(\mathds{1}-\zeta\m{g})$ is invertible (\ie, when $\lVert\zeta\m{g}\rVert\ll 1$)~\cite{Depasse1994},
\begin{equation}
\label{eq:ClassicalField}
  \v{\efield}=\left(\begin{array}{c}
  \mathcal{E}_x\\
  \mathcal{E}_y\\
  \mathcal{E}_z
\end{array}\right)=\Bigl[\mathds{1}-\left(\mathds{1}-\zeta\m{g}\right)^{-1}\cdot\tau\zeta\m{g}\cdot\totalderivative_t\Bigr]\bigl(\mathds{1}-\zeta\m{g}\bigr)^{-1}\cdot\v{\efield}^0\,,
\end{equation}
where $\totalderivative_t=\partial_t+\v{v}\cdot\nabla$ is the total time derivative. All the terms in the preceding two equations are evaluated in a frame rotating with the angular frequency $\omega$, in order to remove the fast time variation, at $[t,\v{r}(t)]$. The non-retarded result of Ref.~\cite{Depasse1994}, applied to the particle and its reflection, is thus modified by the appearance of an additional, time- and velocity-dependent, term. The dipole force experienced by the particle may be obtained as in Ref.~\cite{Depasse1994}:
\begin{equation}
  \boldsymbol{\force}=\left(\begin{array}{c}
  \force_x\\
  \force_y\\
  \force_z
\end{array}\right)\,;\ \text{where}\ \force_i=\tfrac{1}{2}\epsilon_0\re{\chi\sum_j\efield_j\partial_i\efield_j^\ast}\,,
\end{equation}
with $i,j$ separately representing the three spatial dimensions $x,y,z$. A series expansion in powers of $\lVert\zeta\mathbf{g}\rVert$ reveals the leading terms for a stationary particle to be the dipole force from the unperturbed field, and then the dipole force upon the polarised particle due to the field propagated from the induced polarisation.

\section{Mirror-mediated cooling in one dimension}\label{sec:TMM:1DMMC}
For a one-dimensional geometry with the particle a distance $x$ from a perfect mirror, the incident illumination combines with its reflection to give an electric field $\v{\efield}^0(x)=\efield_0\hat{\v{y}}\sin(kx)$, and $\tau=2x/c$; $\hat{\v{y}}$ is a unit vector in the $y$ direction. The force in the $x$ direction upon the moving particle is therefore
\begin{align}
\label{eq:Scattering1DMMCForce}
  \force_x=\frac{1}{4}\epsilon_0\chi k\efield_0^2\biggl\{&\sin(2kx)+\frac{\chi k}{\sigma_\text{L}}\Bigl(1-\frac{v}{c}\Bigr)\sin^2(kx)\bigl[4\cos^2(kx)-1\bigr]\nonumber\\
&-\frac{\chi k^2\tau v}{\sigma_\text{L}}\sin(4kx)\biggr\}\,.
\end{align}
The force thus comprises three terms. The first two are the dipole force exerted by the unperturbed field, and a Doppler-shifted optical binding force between the particle and its reflection: the Doppler shift here changes the wavelengths of the Fourier field components and hence the gradient of the field formed by their superposition. The third term, which depends upon the particle velocity, the electric field propagator and the round-trip retardation time, is the velocity-dependent force, and dominates the velocity-dependent part of the second term when the distance from the mirror is many wavelengths. When the sign of this component is such as to oppose the particle velocity, cooling ensues. This third term is qualitatively and quantitatively identical to the equivalent terms derived from a semiclassical approach (\Sref{sec:CoolingMethods:MMC:Friction}) and using a one-dimensional scattering theory (\Sref{sec:MovingAtom}).

\section{Self-binding: mirror-mediated cooling in three dimensions}\label{sec:TMM:3DMMC}
\begin{figure}[t]
  \centering
  \subfigure[Laboratory frame]{
  \includegraphics{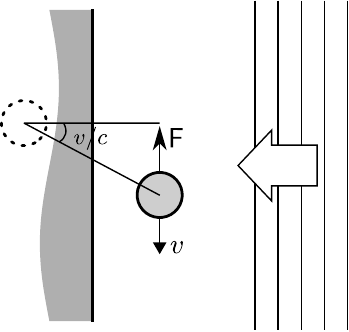}
  }\hspace{1cm}
  \subfigure[Particle frame]{
  \includegraphics{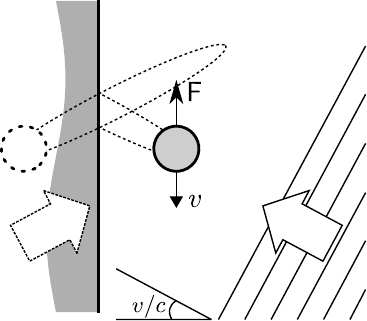}
  }
  \caption[Retarded binding of a normally-illuminated particle to its own reflection]{Retarded binding of a normally-illuminated particle, moving with velocity $v$, to its own reflection, depicted (a)~in the laboratory frame, in which the image lags behind; (b)~in the rest frame of particle, whereby the `wake' trails behind.}
  \label{fig:SelfBinding}
\end{figure}
When a particle or ensemble is strongly coupled to its reflection, we must consider higher-order terms in the expansion of \eref{eq:ClassicalField}, such as the interaction between the propagated particle polarisation and the further polarisation which that induces. This corresponds to the optical binding~\cite{Burns1989,Metzger2006a} of the particle to its reflection, due to the tweezing of the particle by light that it has focussed, as shown in \fref{fig:SelfBinding}(a). The finite time taken for light from the particle to return via the mirror causes the reflected image to trail behind the moving particle, providing a component of the binding force in the direction of the particle velocity and therefore a transverse force even when the geometry shows translational symmetry. \fref{fig:SelfBinding}(b) shows the same geometry in the frame of the particle: it is now the inclination of the transformed incident illumination that causes the focussed `wake' again to lie behind the particle. The sign of the frictional component again alternates with distance from the mirror, but it does so asymmetrically because the apparent field strength described by \eref{eq:ClassicalField} is also modulated. The result is a non-zero force, when averaging over the $x$ coordinate,
\begin{equation}
\left<\left(\begin{array}{c}
  \force_x\\
  \force_y\\
  \force_z
\end{array}\right)\right>=(-v)\frac{\epsilon_0\efield_0^2k^2\chi^2}{128\pi cx^2}\left(\begin{array}{c}
  1\\
  3\\
  3
\end{array}\right)\,,
\end{equation}
under circularly-polarised illumination and taking into account the near-field effects in $\m{g}$. This force becomes comparable with the amplitude of the position-dependent force when $x\lesssim\lambda$, and could therefore be particularly significant for refractive nanoparticles in, for example, colloidal photonic crystals.

%% file: Chapters/Chapter3.tex
\newpartalt{Experimental Work}{Experimental Work}\label{part:Experimental}

\chapter{Experimental setup}\label{ch:Experimental:Past}
\epigraph{Each piece, or part, of the whole nature is always an approximation to the complete truth, or the complete truth so far as we know it. In fact, everything we know is only some kind of approximation, because we know that we do not know all the laws as yet. Therefore, things must be learned only to be unlearned again or, more likely, to be corrected. [...] The test of all knowledge is experiment. Experiment is the sole judge of scientific ``truth''.}{R.\ Feynman, \emph{The Feynman Lectures on Physics} (1964)}
The mechanisms described in the previous chapters, especially mirror-mediated cooling and external cavity cooling, present several exciting avenues not only for theoretical, but also for experimental, research. This chapter presents an overview of the vacuum and laser systems employed by our group in our ongoing investigations into these mechanisms and into MOT miniaturisation and atomic trap arrays. In the first section I describe the physical makeup of the vacuum and laser systems; the second section describes the novel trap geometry and imaging process employed in our system.

\section{Vacuum and laser system}\label{sec:Experimental:Vacuum}
The aim of the current experiment is to investigate atom--surface interactions. A `clean' cloud of ultracold $^{85}$Rb atoms in a magneto-optical trap was chosen as the starting point for these investigations. In designing the experiment, the apparatus had to satisfy a number of criteria, chiefly:
\begin{itemize}
 \item the cold atom cloud needs to be formed close to a surface and the vacuum chamber windows;
 \item the surface, or `sample', may be simply plane or structured on the micro- or nano-scale;
 \item rapid (on the order of one or two weeks) turnaround time for changing the sample; and
 \item very good optical access.
\end{itemize}
Let us look at each of these criteria in turn to explore the design choices they impose on the system.
\subsection{Atom cloud close to surface}
The mirror magneto-optical trap~\cite{Clifford2001} (mirror MOT) was devised as a way to obtain cold atom samples close to a surface. In the present context, however, the standard mirror MOT has a number of disadvantages: the plane of the mirror lies obliquely to the coils that generate the magnetic field necessary to form the MOT, and the beam that forms the MOT illuminates a large cross-section of the mirror directly below the atom cloud---the former will be discussed below in the context of optical access, whereas the latter can be solved by using what we term the `$\Lambda$MOT' beam geometry, as will be discussed in \sref{sec:Experimental:Paper}.

\subsection{Structured surface}\label{sec:Experimental:Vacuum:Structured}
\begin{figure}
\centering
\includegraphics[width=1.5\figwidth]{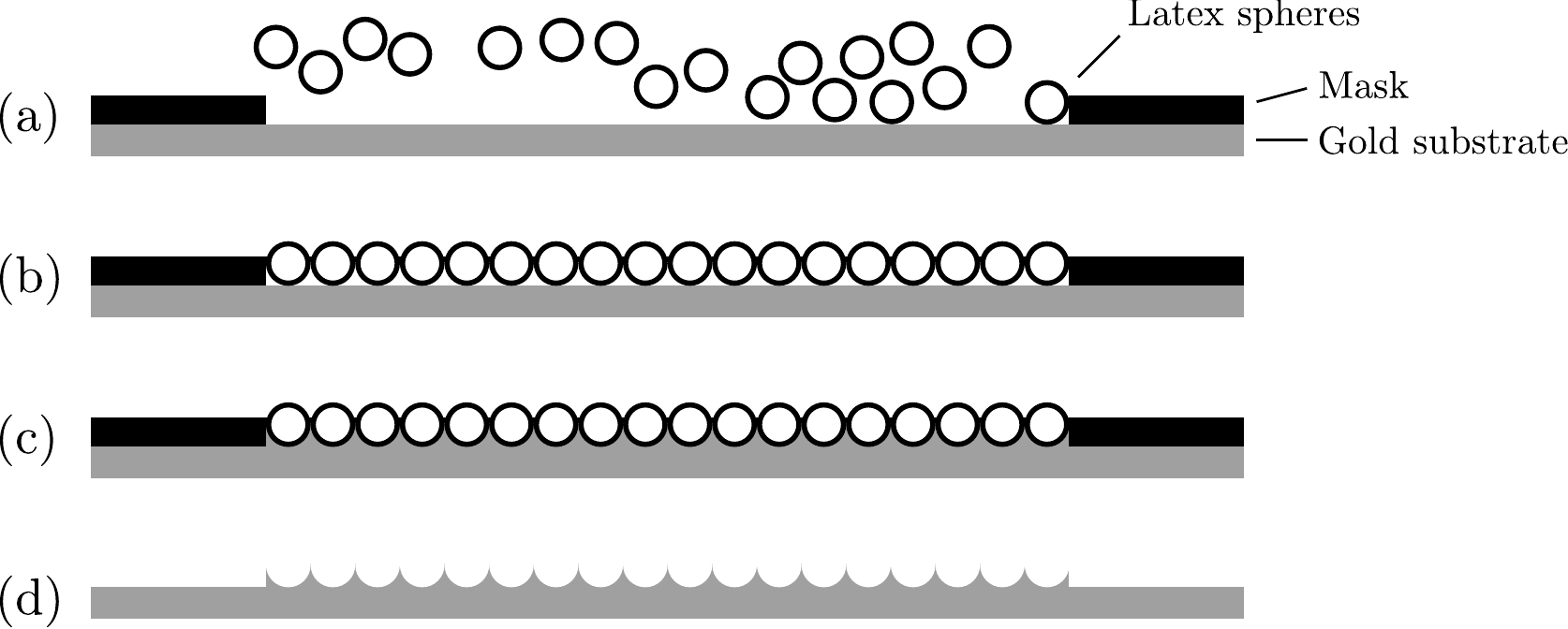}
\caption[Templating process used to make the structured surfaces]{Templating process used to make the structured surfaces. (a) A colloidal suspension of latex spheres in water is allowed to evaporate on a gold substrate within a masked region. (b) An ordered, close-packed monolayer of spheres is formed on the substrate. (c) Gold is electrodeposited and grows from the substrate upwards, around the latex spheres. (d) The spheres and mask are removed by using conventional solvents.}
\label{fig:Templating}
\end{figure}
\begin{figure}[t]
\centering
\fbox{\includegraphics[width=\figwidth]{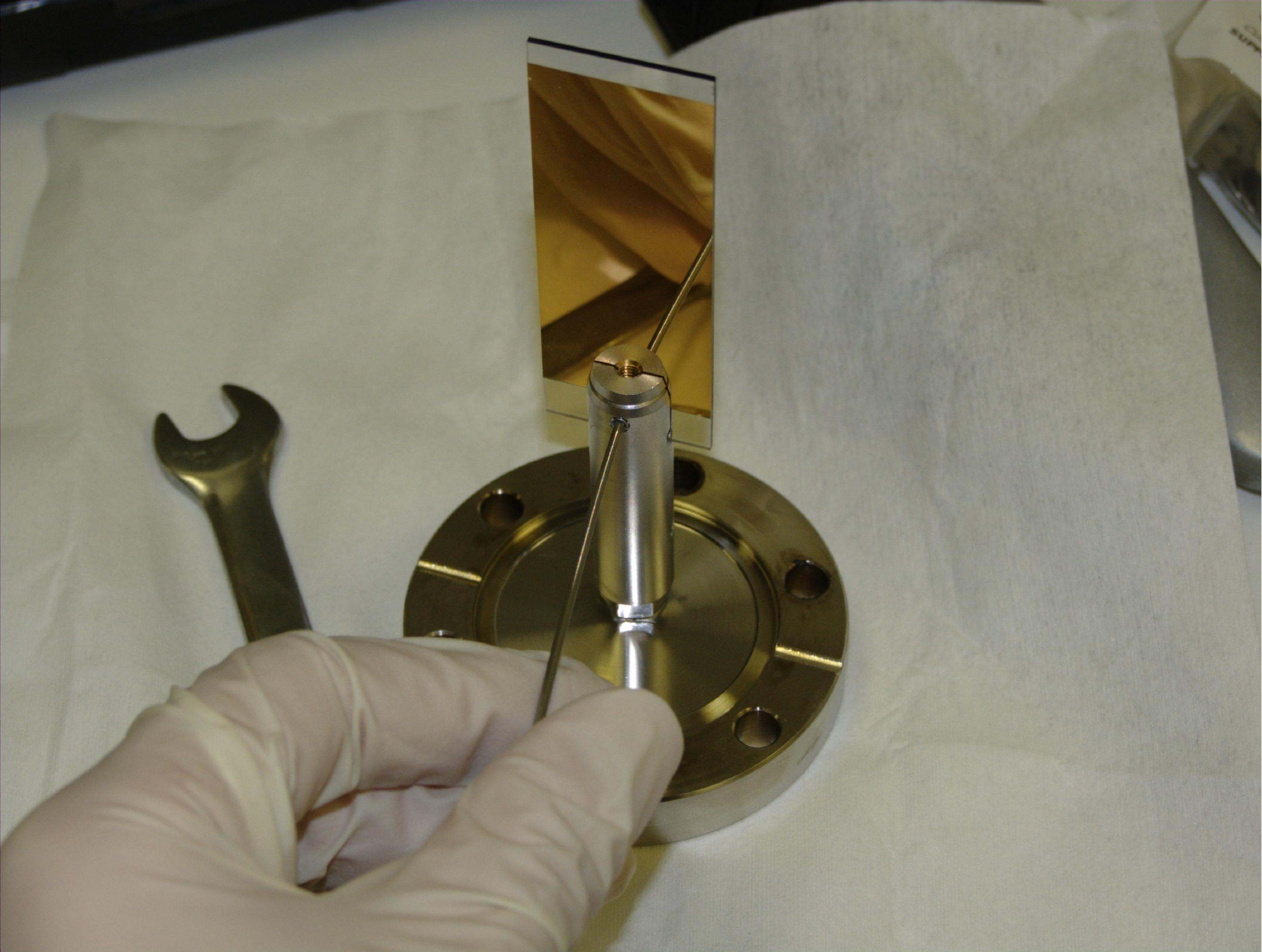}}
\caption[First sample used in the vacuum chamber]{First sample used in the vacuum chamber, pictured being secured into the holder. This sample consisted of a microscope slide onto which a layer of gold was evaporated.}
\label{fig:PlaneSample}
\end{figure}
\begin{figure}[t]
\centering
\fbox{\includegraphics[width=\figwidth]{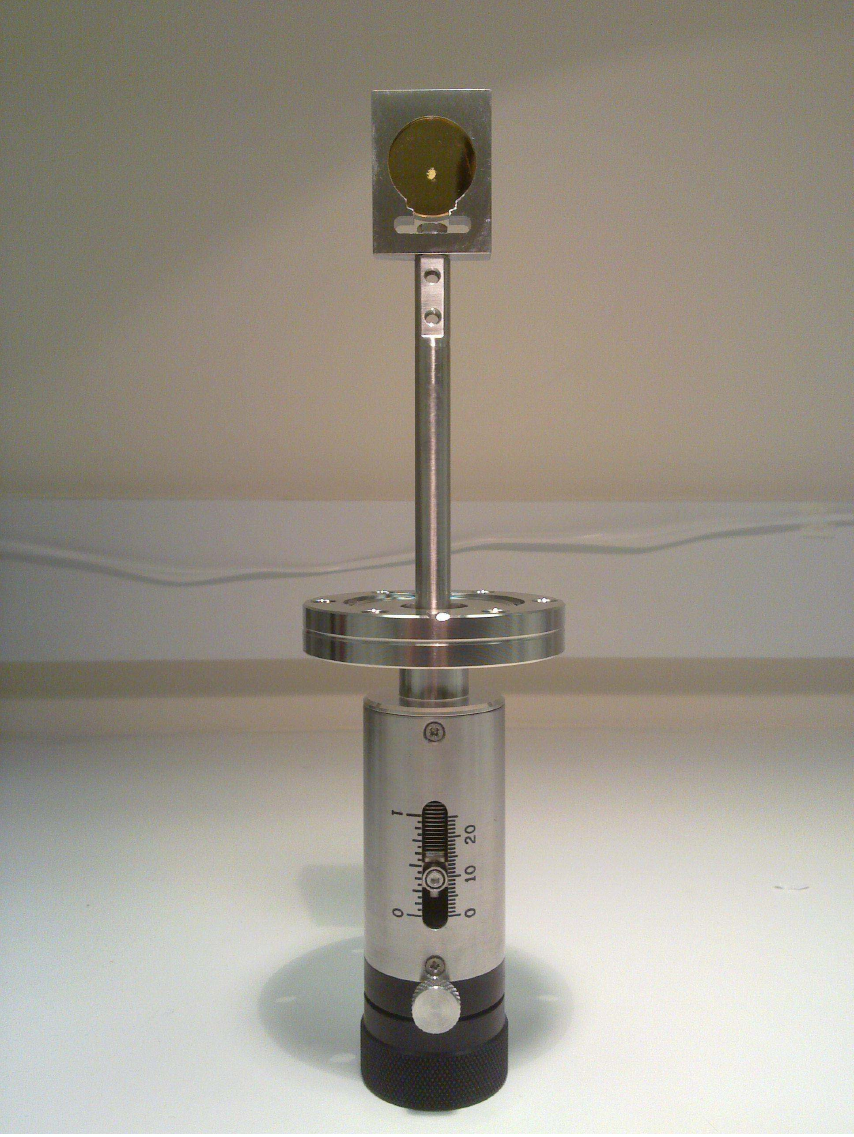}}
\caption[Second sample used, mounted on the vacuum-compatible translation stage]{Second sample used, mounted on the vacuum-compatible translation stage. This sample was templated with hemispherical concave mirrors; the bright dot in the mirror is the reflection of light off the hemispheres.}
\label{fig:TemplatedSample}
\end{figure}
The interaction between atoms and plane surfaces has been investigated both in the context of near field~\cite{Drexhage1968} and far field effects~\cite{Eschner2001}.  The use of structured surfaces allows the coupling of fluorescence from the atom to surface plasmons. A number of advances by Bartlett and co-workers~\cite{Bartlett2000,Bartlett2002} over the past decade have permitted the rapid production of two-dimensional arrays of hemispherical dimples, with radii of $0.1$--$100$\,$\upmu$m, on gold surfaces and with extremely small surface roughness. This process is illustrated schematically in \fref{fig:Templating}; the result is an array of dimples in a small area of an otherwise plane mirror.\\
The initial characterisation of the system in the Southampton Laboratory was conducted using a plane mirror sample, \fref{fig:PlaneSample}. This sample was then replaced by one templated with $100$\,$\upmu$m hemispherical dimples, \fref{fig:TemplatedSample}.

\subsection{Rapid changing of surface}
The physical mechanisms behind the dominant interaction between the atom and the surface depend on the length scale of the structure present on the surface. In the case of a plane sample, the dominant interaction is the Casimir--Polder force in the extreme near field~\cite{Casimir1948,Scheel2009} and the retarded dipole force in the extreme far field (see \sref{sec:CoolingMethods:MMC}; see also Ref.~\cite{Wilson2003}). For a surface with hemispherical dimples having radii of the order of the wavelength, plasmonic effects~\cite{Coyle2001} are expected to dominate. For larger dimples, say those with radii of $10$\,$\upmu$m or greater, the system is approximated better by geometrical optics and one expects the formation of a dipole trap at the focus of each hemisphere. In this regime, one can also envisage depositing magnetically polarised films on the surface that would allow the formation of a microscopic MOT at the focus of each dimple.\\
For these reasons, it is important to be able to switch samples, in order to vary the length scale of the structure, rapidly. This imposes restrictions on the size of the vacuum chamber used: it must be as small as possible without hindering optical access.

\subsection{Good optical access}
\begin{figure}
\centering
\includegraphics[width=\textwidth]{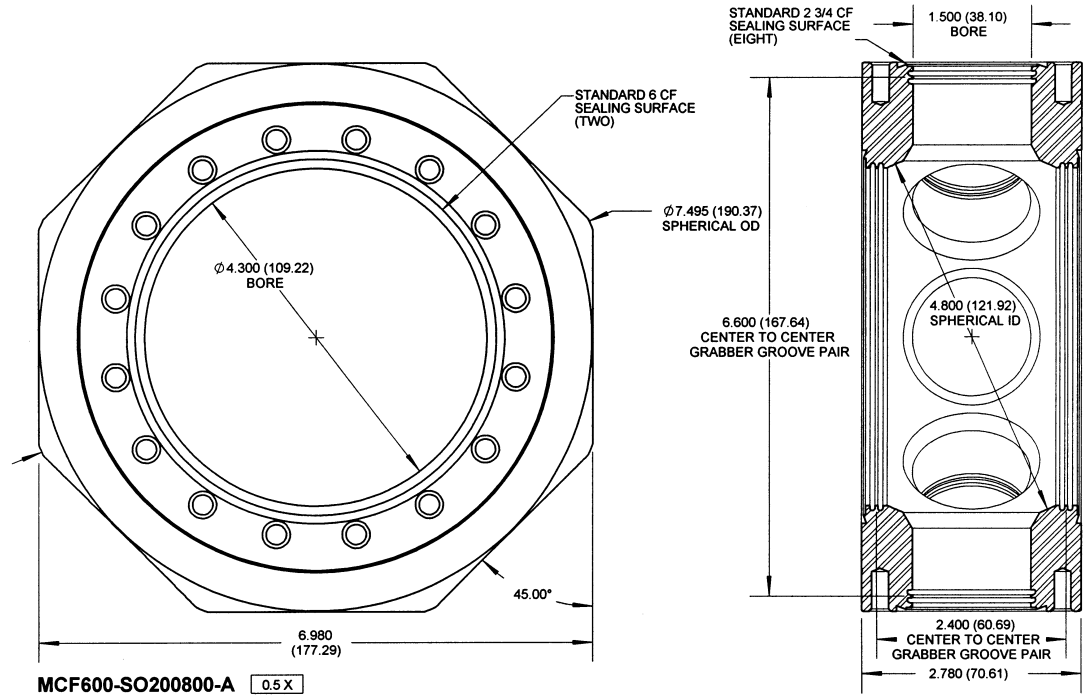}
\caption[Technical drawing for the Kimball Physics spherical octagon]{Technical drawing for the Kimball Physics MCF600-SO200800 spherical octagon. All dimensions are in inches (and millimetres in parentheses). Image reproduced from Ref.~\cite{KimballPhysicsSphOct2010}.}
\label{fig:Chamber}
\end{figure}
A good compromise for a vacuum chamber that satisfies the above restrictions was found in the Kimball Physics Spherical Octagon (part number MCF600-SO200800) chamber, shown in \fref{fig:Chamber}. The two large ports and large aspect ratio enable the use of high numerical aperture fluorescence collection optics. The sample close to which the MOT is formed is placed in the centre of the chamber with its plane parallel to the large ports. This in turn allows the use of a purpose-built high-magnification microscope objective (spatial resolution:~ca.~$2$\,$\upmu$m) to be used with its focal plane coincident with the sample.\\
Two of the smaller ports on the chamber are used for the mount holding the sample and for connecting the chamber to the ion pump and $^{85}$Rb dispensers. The other six are therefore left free, allowing almost unrestricted optical access to the cold atom cloud. The sample, \fref{fig:TemplatedSample}, is mounted on a vacuum-compatible translation stage, allowing the templated surface to be positioned correctly with respect to the MOT beams.

\subsection{Laser system}
The MOT cooling and repump beams are generated by external cavity diode lasers similar to the design in Ref.~\cite{Himsworth2009}. More details of the frequency locking system used are given in the next section.

\section{The \texorpdfstring{$\Lambda$}{`Lambda'-}MOT and multiphoton imaging}\label{sec:Experimental:Paper}
\newcommand{\rb}{$^{85}$Rb}
\newcommand{\name}{\texorpdfstring{$\Lambda$MOT}{`Lambda'-MOT}}

The work in this section is published as Ohadi, H., Himsworth, M., Xuereb, A., \& Freegarde, T. Opt.\ Express \textbf{17}, 23003 (2010) and is reproduced \emph{verbatim}~\cite{Ohadi2009}.\footnote{All authors contributed equally to this paper. Hamid Ohadi and Matthew Himsworth performed the measurements; AX and HO processed the data and wrote the paper. Tim Freegarde supervised the project at all stages.} Here, we describe and characterise the combined magneto--optical trap and imaging system that we developed in our laboratory to be able to explore atom--surface interactions in great detail and with considerable experimental flexibility.
\par
We demonstrate a combined magneto--optical trap and imaging system that is suitable for the investigation of cold atoms near surfaces. In particular, we are able to trap atoms close to optically scattering surfaces and to image them with an excellent signal-to-noise ratio. We also demonstrate a simple magneto--optical atom cloud launching method. We anticipate that this system will be useful for a range of experimental studies of novel atom-surface interactions and atom trap miniaturisation.

\subsection{Introduction and motivation}
Over the past two decades, several configurations for magneto--optical traps have been demonstrated~\cite{Raab1987,Shimizu1991,Emile1992,Lee1996,Reichel1999}. The starting point for most geometries has been the original, `6-beam', configuration~\cite{Raab1987}, where the atom trap is created in the intersection of three counterpropagating laser beams. Despite it having the advantage that the atoms can be trapped far from any surface, thereby reducing spurious scatter in the imaging of such a trap, one cannot easily use this configuration for investigations into atom--surface interactions, for precisely the same reason. Another, more recent, configuration is the so-called `mirror MOT'~\cite{Reichel1999}, where the trap is formed a short distance away from a mirror, which also serves to reduce the number of necessary incident laser beam paths to two. The major drawback of such a configuration is its reduced optical access, due to the oblique angle of the field coils with respect to the mirror. The presence of a reflecting surface close to the trap also presents a problem of an entirely different nature. If the object of one's investigation is to observe the interaction between atoms and surfaces structured at the micrometre scale, for example hemispherical mirrors of the type investigated in~\cite{Coyle2001}, the signal from the atoms will almost certainly be lost due to unwanted scattering of light into the optical system. MOTs on the meso- and microscopic scale, in particular, have received some recent interest~\cite{Folman2000}, but the small atom numbers in such traps have so far hindered their imaging and characterisation~\cite{Pollock2009}. In this section we detail a modified configuration that we call the `\name' and implement an imaging system based on a two-stage excitation process~\cite{Nez1993}, which help us overcome each of these limitations and aid our exploration of different atom--surface interactions.
\par
This section is structured as follows. The next subsection is devoted to the description and characterisation of our trap geometry. We then discuss the mechanism behind our multilevel imaging system and show how it does indeed allow for practically background-free imaging of the atom trap. The subsequent subsection discusses surface loading by magneto-optic launching, which allows us to load atoms onto a surface with a three-dimensional range of motion.

\subsection{The \name}\label{sec:MOT}
\subsubsection{Description}
\begin{figure}[t]
 \centering
    \includegraphics[width=2\figwidth]{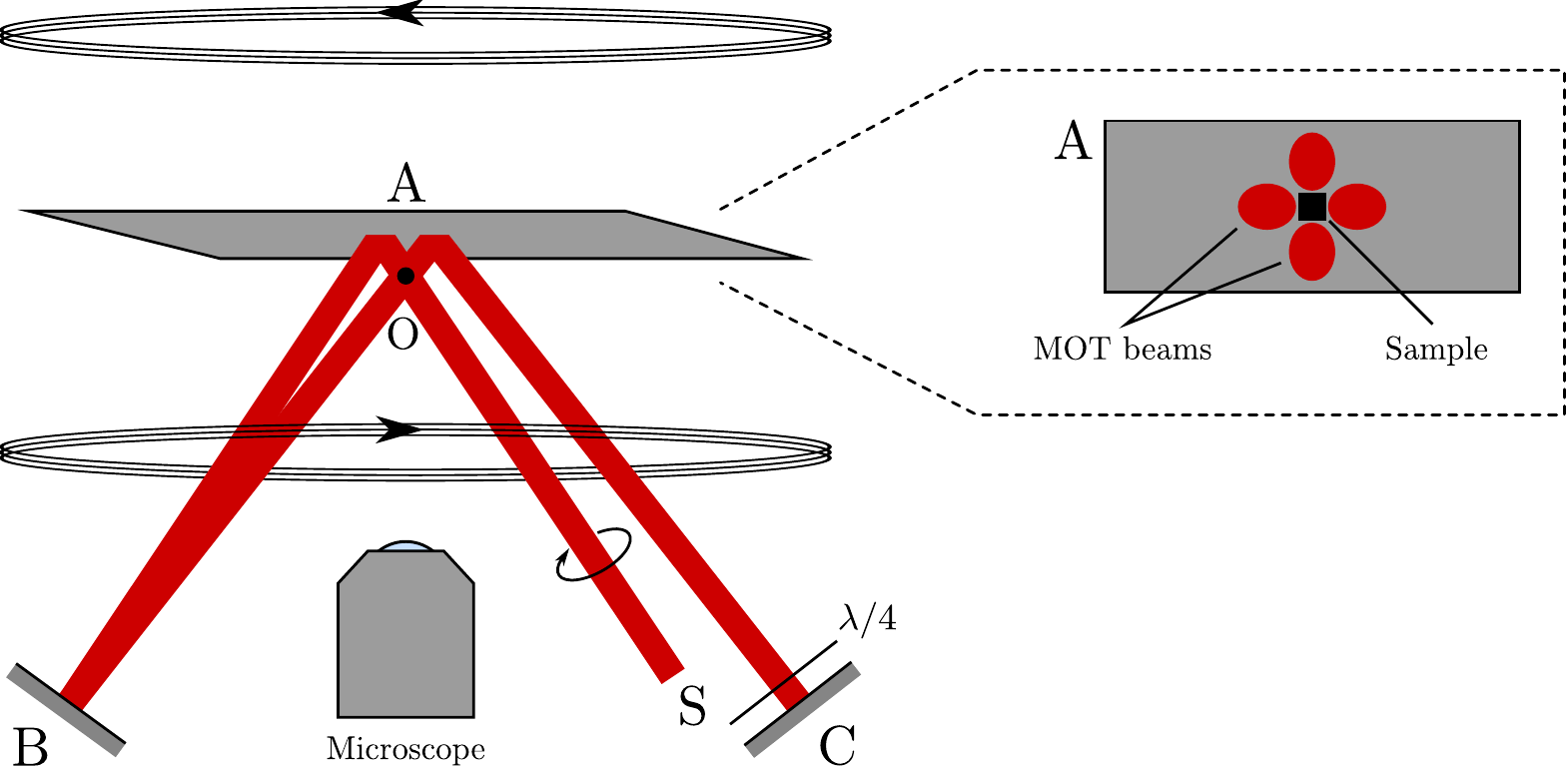}
\caption[Schematic of one of the two beam paths involved in our MOT geometry]{Schematic of one of the two beam paths involved in our MOT geometry. S is the incoming beam; A, B, and C are mirrors. The component marked `$\lambda/4$' is a quarter-wave plate. The cold atom cloud forms in the intersection region, O. In this diagram we do not show a second, identical, beam, which provides trapping and cooling forces in the plane normal to the paper. The area of mirror A immediately adjacent to the trapped atoms is not illuminated, and can therefore be patterned or structured to explore atom--surface interactions. \emph{Inset:} The lower surface of mirror A, showing the MOT beams and the sample area, which is not illuminated by any of the beams.}
 \label{fig:Ohadi2009:VWMOT}
\end{figure}
\begin{figure}[t]
 \centering
    \includegraphics[width=1.5\figwidth]{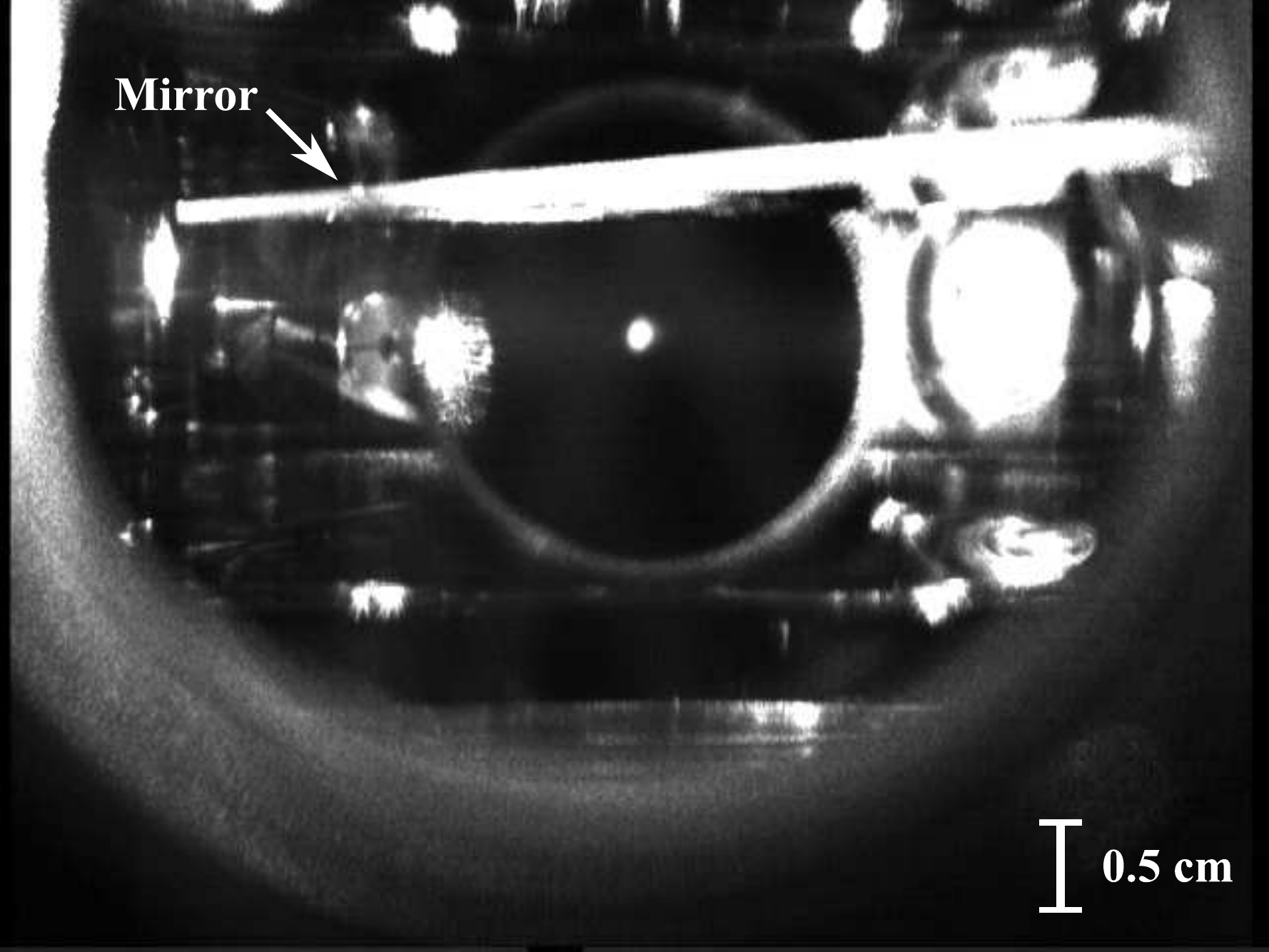}
\caption[Image of our MOT in operation]{Image of our MOT in operation, corresponding to \fref{fig:Ohadi2009:VWMOT}; mirror A is indicated in the picture.}
 \label{fig:Ohadi2009:MOTPic}
\end{figure}
A single beam of circularly polarised light of the right helicity is split using a non-polarising beamsplitter, to generate the two beams that produce the trap, and a half-wave plate is inserted in one of the two resulting beams to achieve the correct polarisations. Each of these beams, denoted S, is then used to construct the geometry shown in \fref{fig:Ohadi2009:VWMOT}. Mirror C is set up so as to retroreflect the beam. Mirrors B and C, together with the quarter-wave plate, allow us to change the polarisation in the retroreflected branch independently of the incoming polarisation. In a normal mirror MOT, the polarisations cannot be modified independently of each other and the quadrupole axis has to be at an oblique angle to the mirror. The four beams travelling towards O thus have the correct polarisations to produce the trapping and cooling forces necessary to form a MOT in this plane. Combined with the second set of beams, this means that the MOT is formed in the intersection region of four pairs of counterpropagating beams. We note that alignment of mirror B such that the beam is retroreflected perfectly will recover the traditional mirror MOT beam geometry, albeit with the incorrect polarisations for a MOT cloud to form.
\par
Several advantages are apparent in the use of this geometry. The trapping volume is the entire overlap of the trapping beams, unlike that in a mirror MOT where half the trapping volume is rendered inaccessible by the presence of the mirror. Optical access is also much improved, both because the coils are oriented in such a way as to be less obstructive, and because we have removed the necessity of having a beam travelling in a plane parallel to mirror A. This allows us to use as much of the $360$\textdegree{} viewing angle in that plane as is necessary for imaging or manipulation beams. If this is not a requirement, a simpler set-up can alternatively be used, where only one set of beams is used in the double-`$\Lambda$' geometry, the trapping and cooling forces in the plane normal to the paper in \fref{fig:Ohadi2009:VWMOT} being produced by means of a separate pair of counterpropagating beams.
\\
An important advantage of this geometry is that the double-`$\Lambda$' shape of the MOT beams affords better imaging of the trap, allowing microscope objectives to be mounted very close to it. With a custom-made objective, we can achieve high-NA imaging ($\mathrm{NA}>0.5$) and a diffraction-limited resolution of $<2$\,$\upmu$m. While a similar degree of optical access may be possible in the traditional 6-beam configuration, we note that this latter configuration is unsuitable for atom--surface interaction studies. In contrast, mirror A in our geometry can be replaced by any other suitable reflecting surface. One candidate for such a reflecting surface would be one of the surfaces of a Dove prism, which could then be used to form a two-dimensional bichromatic evanescent-field trap~\cite{Ovchinnikov1991} close to the mirror surface. This trap would be loaded from the MOT cloud using such techniques as magneto-optic launching, which is explained in \Sref{sec:Launching}.
\par
Aside from this marked increase in optical access, our system is simple to set up and operate. In particular, it requires fewer beam paths than a traditional MOT (two rather than three) and alignment of the beams is also easy: a CCD camera looking up at the mirror can be used to align the beams coarsely; once this is done, optimisation of the cold atom signal provides the fine-tuning of the alignment.

\subsubsection{Characterisation}
A typical trap, shown in \fref{fig:Ohadi2009:MOTPic}, is ellipsoidal in shape with a $1/e$ diameter of the order of $400$\,$\upmu$m along its minor axes and contains around $4\times 10^4$ \rb{} atoms. Combined with a measured trap lifetime $\tau_0\approx 6$\,s, this allows us to infer the trap loading rate, $N_0/\tau_0\approx 6.7\times10^3$\,s$^{-1}$. We measured a cloud temperature of $110\pm 40$\,$\upmu$K, the large uncertainty being due to the imprecision in measuring the cloud size.
\par
Typical parameters for the operation of our trap are: a detuning of $-14.9$\,MHz, or $-5.0$\,$\Gamma$ ($\Gamma\approx 3.0$\,MHz~\cite{Schultz2008}), for the cooling laser and a power of $6$\,mW divided between the two trapping beams (beam diameter: $6$\,mm). The minimum power necessary to produce the MOT was found to be $\approx\!1.3$\,mW in each of the two beams. The trap was loaded from background gas of a natural isotopic mixture of rubidium at a pressure of $10^{-9}$\,mbar. The cooling and repump lasers were locked using the DAVLL technique~\cite{Corwin1998} for long-term stability and flexibility of operation.

\subsection{Multilevel imaging system}\label{sec:Imaging}
\begin{figure}[t]
 \centering
    \includegraphics[width=\figwidth]{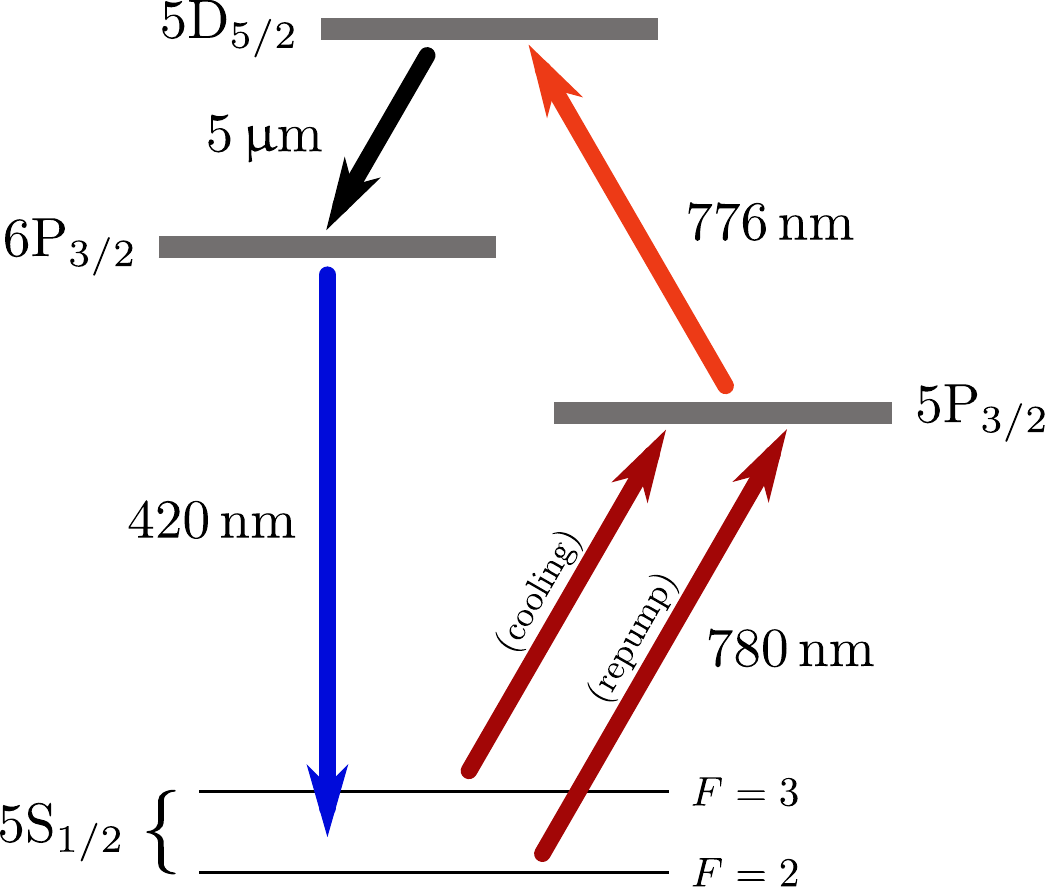}
\caption[The four-level system in \rb{} that we use to image our atoms]{The four-level system in \rb{} that we use to image our atoms. The MOT lasers ($780$\,nm) and a laser at $776$\,nm are used to induce a ladder transition. The population decays back to the ground state, via an intermediate state, and emits a $420$\,nm photon in the process. The hyperfine splitting of the excited states is not drawn for clarity.}
 \label{fig:Ohadi2009:420_Levels}
\end{figure}
The most common method of imaging a cold atom cloud in a MOT is fluorescence imaging. When the cloud is close to a reflecting surface both the cloud and its reflections will be seen by the imaging system (see Ref.~\cite{Clifford2001}, for example). This situation is exacerbated by the presence of surfaces that reflect unwanted light into the imaging optics and thereby decreasing the signal-to-noise ratio of the imaging system. \fref{fig:Ohadi2009:MOTPic}, shows an example of the mirror in our system scattering the MOT beams into the imaging system.
\\
This problem may be overcome using two-stage excitation imaging. We make use of a four-level system in \rb{} (see \fref{fig:Ohadi2009:420_Levels} for details), similarly to Refs.~\cite{Sheludko2008} and~\cite{Vernier2010}; atoms in the $5$S$_{1/2}$ ground state are pumped to the $5$D$_{5/2}$ state via $780$\,nm and $776$\,nm radiation, the former being provided by one of the MOT beams, and then decay back to the ground state via an intermediate $6$P$_{3/2}$ state, emitting $420$\,nm radiation, which we detect. We note that a very similar system was recently used to produce a multiphoton MOT~\cite{SWu2009}. In our system, this process gives a significantly smaller signal than can be obtained through $780$\,nm fluorescence imaging. However, it has the benefit of being entirely background-free: in a well-shielded system, the entire $420$\,nm signal reaching the detector has its origin in the cold atom cloud. Off-the-shelf filters can then be used to remove the $780$\,nm radiation reaching the detector.

\subsubsection*{Generation of the $776$\,nm beam}
\begin{figure}[t]
 \centering
    \includegraphics[width=1.5\figwidth]{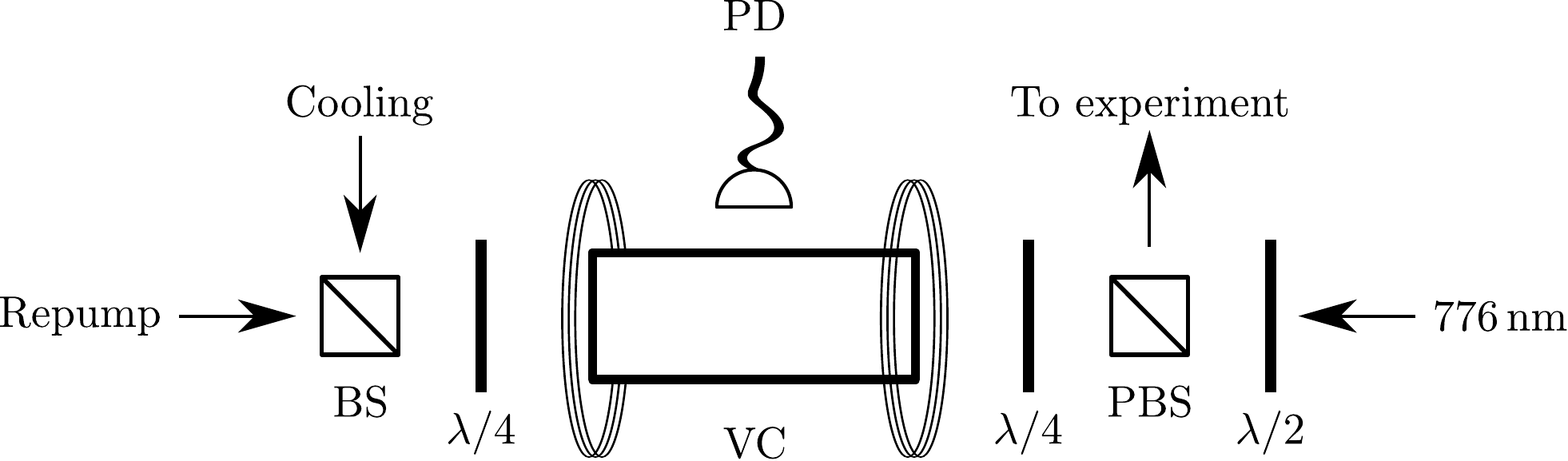}
\caption[$776$\,nm spectroscopy and locking system]{$776$\,nm spectroscopy and locking system. (P)BS: (polarising) beam splitter cube; $\lambda/4$: quarter-wave plate; $\lambda/2$: half-wave plate; VC: heated vapour cell; PD: filtered photodiode.}
 \label{fig:Ohadi2009:776Locking}
\end{figure}
The $776$\,nm beam is produced using a Sanyo DL7140-201S diode and the same external cavity diode laser design used to produce the MOT cooling and trapping beams. Since \rb{} has no spectral features in this wavelength range that are suitable for locking the laser frequency, a multilevel locking system is used (see \fref{fig:Ohadi2009:776Locking}). $5$\,mW from each of the MOT cooling and repump beams ($\approx\!780$\,nm) and $1.5$\,mW from the $776$\,nm beam, all rendered circularly polarised by the quarter-wave plates, enter the heated vapour cell (VC) from opposite ends. A large-area UV-enhanced filtered silicon photodiode (PD), operating in photovoltaic mode, picks up the resulting Doppler-free fluorescence and is amplified by means of a LMP7721 amplifier chip. Magnetic coils surrounding the heated vapour cell control the Zeeman shift of the magnetic sublevels of the atoms inside the cell, shifting this signal, and therefore the lock point, as required. Around $4$\,mW of the $776$\,nm beam is then mixed in with the MOT cooling and repump beams and sent through an optical fibre to the MOT.
\\
\begin{figure}[t]
 \centering
    \includegraphics[width=1.5\figwidth]{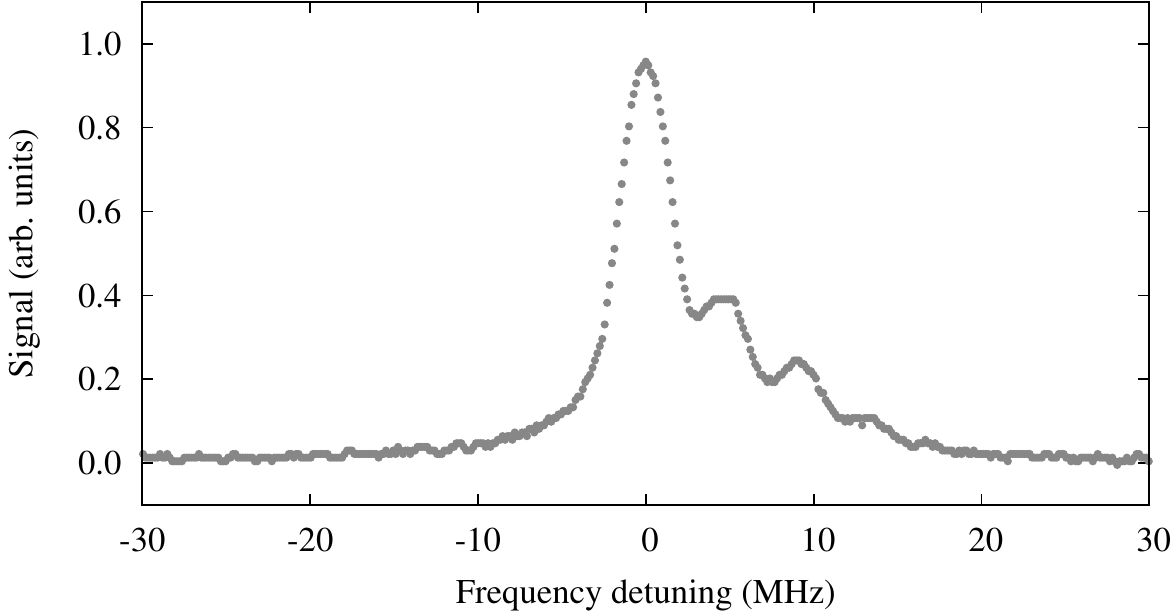}
\caption[$420$\,nm fluorescence from the vapour cell as a function of the detuning of the $776$\,nm beam]{$420$\,nm fluorescence from the vapour cell, observed on PD (see \fref{fig:Ohadi2009:776Locking}) as a function of the detuning of the $776$\,nm beam, with the cooling and repump beams locked and shifted by $80$\,MHz with respect to the frequencies required to make a MOT. The various peaks are due the hyperfine structure in \rb{}. To obtain these data, we removed the quarter-wave plates on either end of the vapour cell, thus having linearly polarised light entering the cell from both ends.}
 \label{fig:Ohadi2009:420Spectrum}
\end{figure}
\begin{figure}[t]
 \centering
    \includegraphics[width=1.5\figwidth]{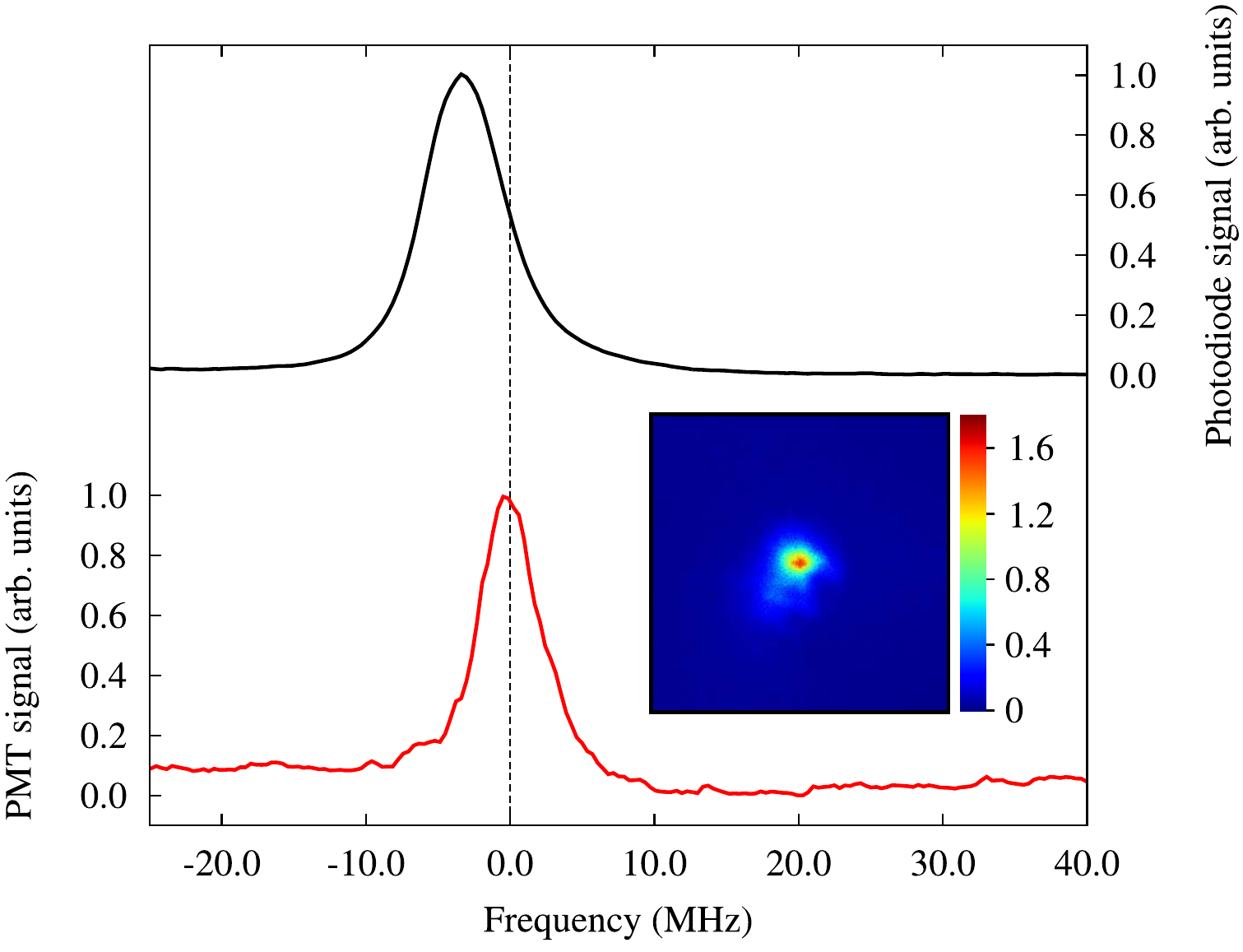}
\caption[$420$\,nm fluorescence observed on PD and on a PMT imaging the MOT cloud as a function of the detuning of the $776$\,nm beam]{$420$\,nm fluorescence observed on PD (solid black line, see \fref{fig:Ohadi2009:776Locking}) and on a PMT imaging the MOT cloud (solid red line) as a function of the detuning of the $776$\,nm beam. The zero on the frequency axis corresponds to the point at which the signal from the MOT cloud is highest; we lock to this point. The magnitude and sign of the shift between the two curves can be set arbitrarily by varying the magnetic field generated by the coils around the vapour cell. \emph{Inset:} MOT cloud imaged at $420$\,nm (scale in $10^3$ counts per second). This image is naturally background-free and has a spatial resolution, limited by the optics used, of ca.~$2$\,$\upmu$m.}
 \label{fig:Ohadi2009:420nmPDPMT}
\end{figure}
\begin{figure}[t]
 \centering
    \includegraphics[width=\linewidth]{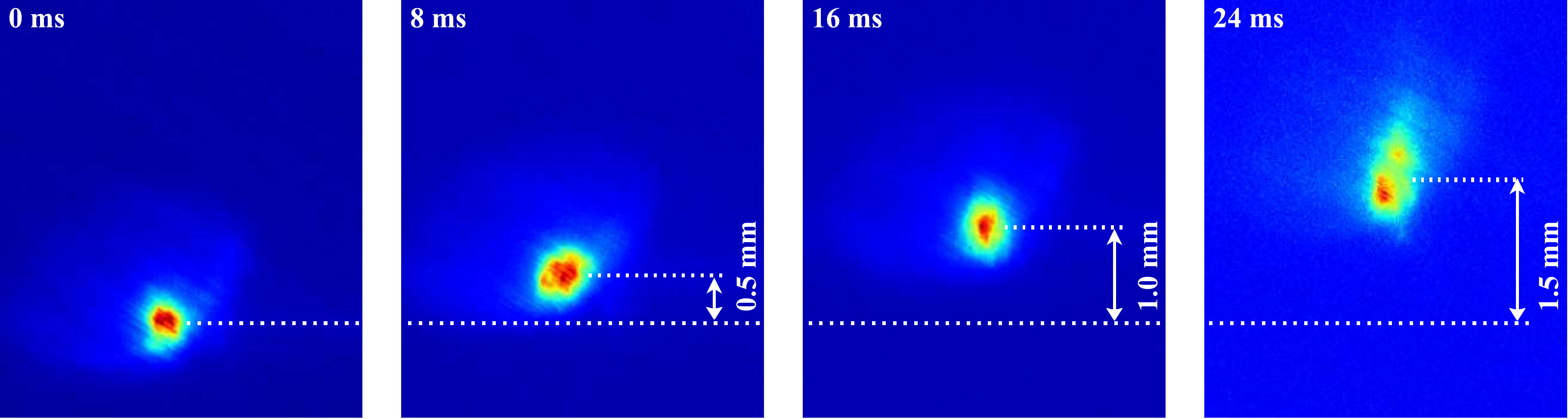}
\caption[A sequence of four false colour fluorescence images of the cloud before and after it has been given a magnetic impulse]{A sequence of four false colour fluorescence images, taken at $8$\,ms intervals, of the cloud before and after it has been given a magnetic impulse. The first shot (leftmost picture) shows the cloud just before the magnetic field is pulsed. The second, and subsequent, shots show the cloud at later times. The transfer efficiency after $24$\,ms is over $40$\%.}
 \label{fig:Ohadi2009:MagLaunch}
\end{figure}
We show a sample $420$\,nm signal, as detected at the photodiode, in \fref{fig:Ohadi2009:420Spectrum}, where the hyperfine splitting of the $5$D$_{5/2}$ level in \rb{} is evident in the shoulders on the right-hand side of the peak in the figure. The $776$\,nm laser diode is locked to the side of one the main peak, at the point indicated by the dashed line in \fref{fig:Ohadi2009:420nmPDPMT}, using a conventional PID circuit. The lock point is found by manually and slowly adjusting the frequency offset of the $776$\,nm laser to maximise the fluorescence from the MOT cloud. Depending on the parameters chosen, the spectra corresponding to the latter figure may exhibit two well-resolved peaks due to the Autler--Townes splitting~\cite{Autler1955}; in the spectrum shown in \fref{fig:Ohadi2009:420nmPDPMT}, however, the presence of the second peak manifests itself as a slight shoulder on the photodiode signal. Locking at a detuning of around $6.5$\,MHz from the peak of absorption in the vapour cell gives the strongest signal in the MOT cloud, as recorded by the photomultiplier tube trace shown in the same figure.

\subsection{Surface loading by magneto-optic launching}\label{sec:Launching}
Transporting cold atoms from the region where the trap naturally forms to the sample is an essential part of many experiments investigating atom--surface effects. Several methods have been devised for moving cold atom clouds, including the use of push beams~\cite{Wohlleben2001} and moving magnetic coils~\cite{Lewandowski2003}. Push beams are easy to set up, requiring either the addition of one extra beam or the switching off of one of the counterpropagating beams, but cannot be used to push atom clouds towards highly reflective surfaces. Using moving magnetic coils requires a rather involved mechanical setup.
\par
We make use of a third method, which we call magneto-optic launching, for transport of the atom cloud by rapidly moving the trap centre and then releasing the cloud, thereby imparting momentum to it. An auxiliary coil is added to the system in~\fref{fig:Ohadi2009:VWMOT}, above the upper MOT coil. After the MOT cloud forms, a long current pulse is applied to this auxiliary coil, which launches the cloud upward with a speed determined by the size and duration of the current pulse, and then the cloud is released from the trap by switching off the MOT beams after $20$\,ms. \fref{fig:Ohadi2009:MagLaunch} shows a series of photographs of the cloud after being launched by a magnetic pulse. It can be seen that the pulse results in an approximately uniform vertical cloud speed of $0.063$\,m\,s$^{-1}$. The physical orientation of our system, with the mirror and sample being \emph{above} the trapping region, allows us to launch the cloud upwards with a much greater degree of control than would be possible if the cloud were merely dropped downwards.
\par
Finally, we note that the equilibrium distance of the MOT cloud from the mirror surface depends on the beam diameter and the size of the `sample area', \ie, the section of the mirror that acts as a sample and is not usable as a plane mirror. With a sample area diameter of $2$\,mm and beam diameter of $4$\,mm, the cloud can be made to form less than $4$\,mm away from the surface, allowing us to use the magneto-optic launching method to move the atoms closer to the surface for interaction studies.

%---

\chapter{A guide for future experiments}\label{ch:Experimental:Future}
\epigraph{[...] [I]t is more important to have beauty in one's equations than to have them fit experiment.}{P.\ A.\ M.\ Dirac, Scientific American \textbf{208}, 5 (1963)}

This chapter aims to provide a guide for experimentalists seeking to observe the effects we predicted in earlier chapters. \sref{sec:Experimental:Future:Overview} presents an overview of the different geometries explored in the previous chapters. For each of these geometries, \sref{sec:Experimental:Future:Forces} calculates and compares the relevant friction forces acting on the particle. \sref{sec:Experimental:Future:Numbers} discusses a number of experimentally-accessible configurations and calculates cooling times and equilibrium temperatures that can be expected in each situation. The first appendix to this chapter is a technical note discussing electric fields inside dielectrics and the origin of the Clausius--Mossotti equation that describes the response of a bulk dielectric to an applied electric field. Finally, two appendices then follow that discuss, respectively, some problems encountered when calculating electric fields inside microscopic hemispherical mirrors, and general expressions for the force acting on an atom inside an arbitrary monochromatic field, ignoring delay effects.

\section{Overview of several different possibilities}\label{sec:Experimental:Future:Overview}
It has been outlined in the preceding chapters that several different geometries exist that allow a memory. Moreover, within each such geometry, one can choose to investigate cooling mechanisms on different classes of particle. The aim of this section is to briefly summarise these different possibilities, noting the advantages and disadvantages of each: \sref{sec:Experimental:Future:Overview:Ions} to \sref{sec:Experimental:Future:Overview:Dielectric} look at species that can be cooled, and \sref{sec:Experimental:Future:Overview:DipoleArrays} to \sref{sec:Experimental:Future:Overview:ConcaveMirror} at the geometries themselves.

\subsection{Trapped ions}\label{sec:Experimental:Future:Overview:Ions}
Ions can be trapped in radio-frequency traps, and laser cooled down to the ground vibrational state of such traps,\footnote{It must be pointed out that such ions would have a translational temperature lower than that which can be achieved through mirror-mediated cooling setups using typical experimental parameters. The aims of such experiments would be (i)~a proof-of-principle demonstration, and (ii)~an exploration of the wavelength-scale variations of the forces.} with remarkable ease. The lifetime of trapped ions can be of the order of hours~\cite{Herskind2008}, which is orders of magnitude longer than the comparable figure for neutral atoms. For these reasons, ions would make ideal test subjects for exploring forces that vary significantly over length scales of the order of a wavelength~\cite{Eschner2001,Hetet2010}. Interest in using trapped ions also arises from their potential applications in quantum information storage and processing~\cite{Cirac1995}.

\subsection{Neutral atoms}\label{sec:Experimental:Future:Overview:Atoms}
The ease of manipulation of trapped ions using electric fields is a double-edged sword, in the sense that this very feature also makes trapped ions highly sensitive to the environment they are immersed in. One can avoid these issues through the use of neutral atoms rather than ions. Neutral atoms, however, cannot easily be confined to sub-wavelength regions without complex experimental systems such as the one used in a recent proof-of-concept experiment presented in Ref.~\cite{Proite2010}.

\subsection{Optomechanics---Cantilevers and micromirrors}\label{sec:Experimental:Future:Overview:Optomechanics}
Recent years have seen a surge in the popularity of optomechanics experiments, mostly with the aim of reaching the ground vibrational state of a vibrating reflective cantilever~\cite{Groblacher2009a}, or of a reflective micro-membrane~\cite{Thompson2008}. The use of such optical elements introduces a number of interesting possibilities:
\begin{itemize}
\item Engineered internal resonances---Ref.~\cite{Karrai2008} looks at using resonances inside a photonic crystal as a means of controlling its motion, in much the same way as one uses atomic resonances in Doppler cooling. In contrast with the case of an atom or ion, however, one can engineer the system to have a wide range of different properties.
\item Strong mirror--field coupling---Single atoms or ions do not have a large polarisability unless the driving field is close to resonance; this is problematic because strong heating effects become important under such conditions. Mirrors, even microscopic ones, consist of vast numbers of atoms, each of which is essentially an individual dipole, and therefore experience correspondingly stronger effects.
\item Positioning---micromirrors enjoy the advantages of both ease of positioning, shared with ions, and the immunity to electrostatic forces, shared with neutral atoms. Moreover, the technology exists to make silicon nitride (SiN) membranes much thinner than an optical wavelength~\cite{Jayich2008}, so such mirrors can indeed be used to explore sub-wavelength structure in the forces.
\end{itemize}

\subsection{Dielectric particles}\label{sec:Experimental:Future:Overview:Dielectric}
Spherical dielectric particles, of sizes on the nanometre~\cite{Chang2009b,Barker2010} or micrometre~\cite{Barker2010b} scales, have been proposed as replacements for individual atoms in cooling experiments. On the small end of the scale, the particles can be suspended using purely optical forces and are therefore not coupled to any physical heat bath. Nevertheless, even such small particles exhibit polarisabilities much larger than that of an individual atom, and therefore correspondingly stronger interactions with the light field. The larger particles would need to be physically supported, perhaps by being mounted on the end of a tapered fibre. This could, in turn, be achieved by ablating the tapered end of the fibre to form a microsphere. The mechanical properties of such a fibre would ensure poor coupling of phonons between the microsphere and the bulk fibre.
\par
The response to the electric field of a dielectric particle on the nanometre scale is related to its complex permeability $\epsilon$ through the Clausius--Mossotti relation:
\begin{equation}
\label{eq:ClausiusMossotti}
\chi=3V\frac{\epsilon-1}{\epsilon+2}\,.
\end{equation}
The derivation of the Clausius--Mossotti equation itself has some interesting subtleties that are discussed in \aref{sec:Experimental:Future:CM}. $\epsilon$ is furthermore related to the real, $\eta$, and imaginary, $\kappa$, parts of the refractive index $n=\sqrt{\epsilon}$ by
\begin{align}
\re{\epsilon}&=\eta^2-\kappa^2\,\text{, and}\\
\im{\epsilon}&=2\eta\kappa\,.
\end{align}
Finally, $\kappa$ is related to the $1/e$ power absorption length of a substance, $1/\alpha$, by $\alpha=2k\kappa$ at wavenumber $k$. For the common dielectric PMMA [poly(methyl methacrylate)] $\eta=1.5$ and, conservatively, $\alpha=50$\,m$^{-1}$ at a wavelength $\lambda=1$\,$\upmu$m in vacuum \cite{Beyer2003}; \ie, $\epsilon=2.2+\bigl(1.2\times 10^{-5}\bigr)\i$. The imaginary part of $\epsilon$ is much smaller than the real part and can generally be neglected when calculating optical forces. It is, however, responsible for absorption of part of the incident light and its effects cannot be neglected when calculating the power absorbed by an illuminated dielectric.

\subsection{Dipole trap arrays}\label{sec:Experimental:Future:Overview:DipoleArrays}
\begin{figure}
\centering
\fbox{\includegraphics[width=\figwidth]{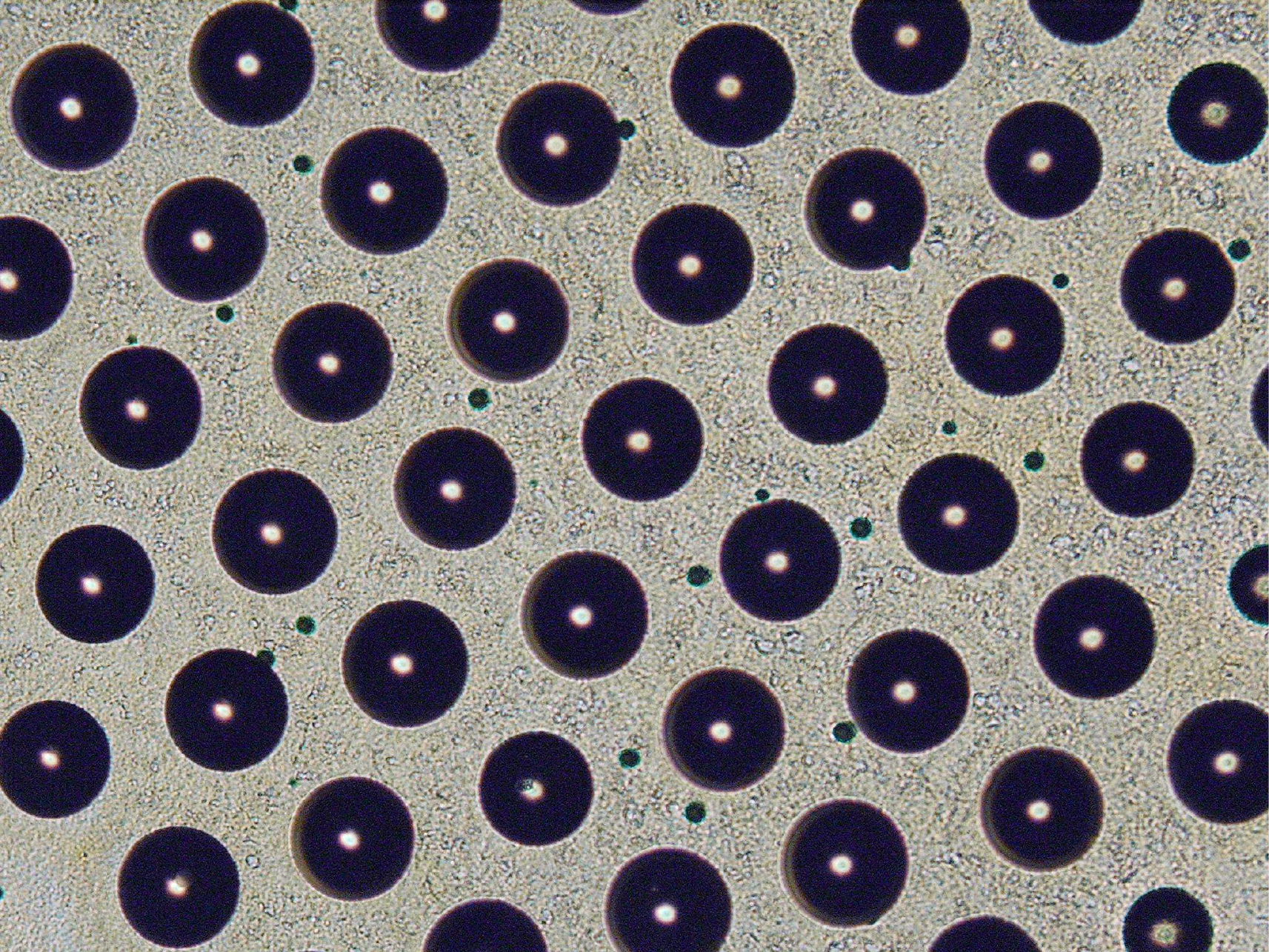}}
\caption[Optical micrograph of a section of a templated gold surface]{Optical micrograph of a section of a templated gold surface. The bright spot in each of the dimples is the focus. The centre--to--centre distance is $100$\,$\upmu$m and the depth of the electrodeposited gold is $15$\,$\upmu$m. (Courtesy Nathan Cooper.)}
\label{fig:Dimples}
\end{figure}
Two-dimensional arrays of dipole traps have been built using microfabricated lens arrays~\cite{Dumke2002} and used to site-selectively address trapped atoms~\cite{Kruse2010}. However, the use of refractive elements such as lens arrays brings with it a number of disadvantages, of which we mention two:
\begin{itemize}
\item Fabrication costs---each array has to be custom-made and it is difficult to mass-produce such optical elements.
\item Integration---lens arrays require optical access from both sides, leading to a larger apparatus and making it difficult to integrate them into so-called atom chips~\cite{Folman2000}.
\end{itemize}
One way to overcome both these issues is to use arrays of reflective concave mirrors, which can be manufactured easily (this was explored in \fref{fig:Templating}). The end product, as shown in \fref{fig:Dimples}, is indeed close to ideal in terms of periodicity and surface quality. These concave mirrors can be used to construct individual dipole traps for either neutral atoms or ions.\par
The interest in using such arrays of mirrors lies not only in their ease of manufacture but also in the surface plasmon resonances that are exhibited by the individual hemispherical mirrors~\cite{Coyle2001}. Such resonances couple to the incident light and give rise the possibility of mechanisms of the ``external cavity cooling'' type, \sref{sec:TMM:ECCO}, using not Fabry--P\'erot cavities but material resonances. The advantages of using such a system are immediately obvious; we mention only that such a setup introduces the possibility of two-dimensional arrays of individual optical resonant elements that require essentially no alignment. Coupled with the fact that external cavity cooling, as with any mechanism based on the dipole force, is not species-selective, this leads to the possibility of producing two-dimensional arrays of cold ions, atoms, or even micromirrors. Such arrays would potentially revolutionise quantum information processing by implementing a scalable two-dimensional register for quantum information.

\subsection{Plane mirror cooling}\label{sec:Experimental:Future:Overview:MMC}
\begin{figure}
\centering
\includegraphics[scale=0.2]{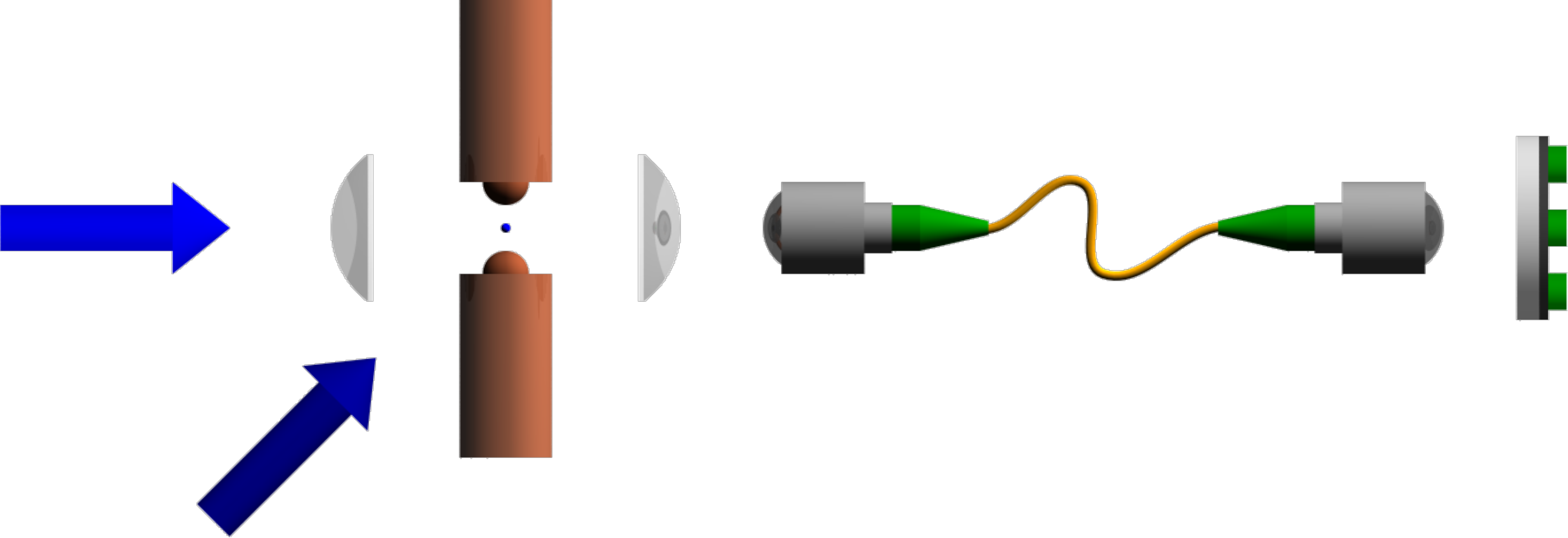}
\caption[Schematic of an experiment to explore mirror-mediated cooling]{Schematic of an experiment to explore the mirror-mediated cooling mechanism. Light is focussed onto an ion trapped in an endcap trap, then coupled into a fibre. This light is retroreflected back into the fibre by means of a mirror whose position can be adjusted through piezoelectric elements.}
\label{fig:Realistic_MMC}
\end{figure}
The mirror-mediated cooling setup would be the most basic proof-of-concept experiment of optical cooling using a memory, seeing as it involves merely one mirror. As has been discussed previously, the friction force in such a setup oscillates on a sub-wavelength scale and, moreover, is only sizeable for atom--mirror distances of the order of metres. A realistic approach to implementing this delay is to couple the light, after interacting with the atom, into a single-mode fibre. The light inside the fibre is then retroreflected and imaged back onto the atom itself. This setup, shown schematically in \fref{fig:Realistic_MMC}, is conceptually similar to the one used in Ref.~\cite{Eschner2001}. In \fref{fig:Realistic_MMC}, the species to be cooled is shown to be a trapped ion rather than a neutral atom. The reason for this is again the small length scale over which the friction force changes from a cooling to a heating force, necessitating very good localisation of the particle to be cooled. One also notes that the delay line length must be stabilised interferometrically.

\subsection{External cavity cooling}\label{sec:Experimental:Future:Overview:ECCO}
The use of an optical resonance to enhance the cooling effect of the retarded dipole--dipole interaction presents a novel way of enhancing the performance of current optomechanical experiments. Indeed, such setups would be less sensitive to misalignment of the mirror to be cooled than traditional optomechanical setups~\cite{Aspelmeyer2010}, which require micromirrors with extremely good optical and mechanical properties. The sub-wavelength modulation of the friction force is an issue with this mechanism too, which necessitates the use of particles---such as thin micromirrors, membranes, or trapped ions---that can be localised to within a small fraction of a wavelength.

\subsection{Ring cavity cooling}\label{sec:Experimental:Future:Overview:RELIC}
\begin{figure}
\centering
\includegraphics[width=\figwidth]{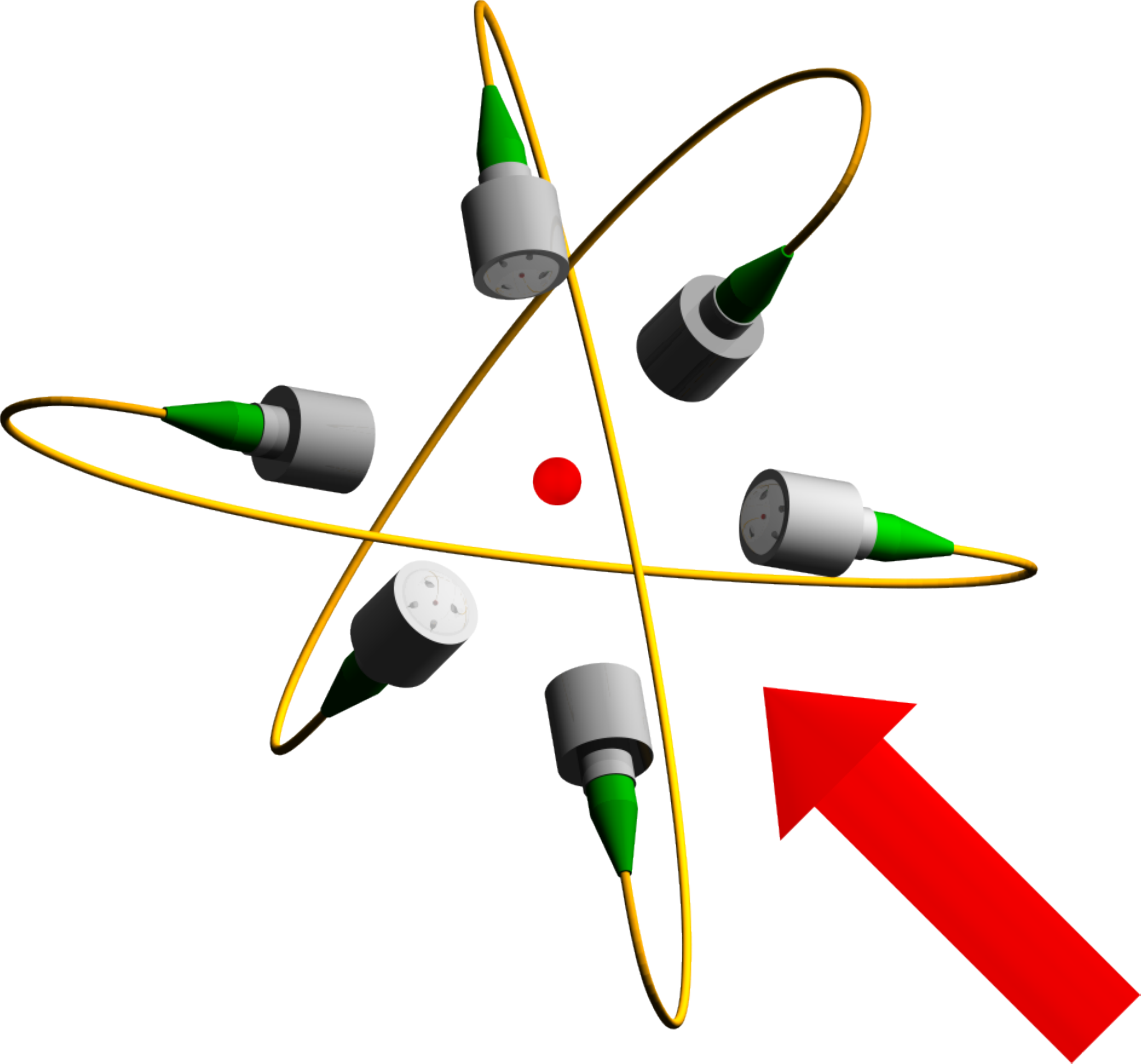}
\caption[Highly schematic illustration of three-dimensional ring cavity cooling]{Highly schematic illustration of a three-dimensional ring cavity cooling setup. The pump light does not couple directly into any cavity, allowing one to use gain media in the fibres similarly to the amplified optomechanics scheme (see \sref{sec:TMM:AmplifiedOptomechanics}).}
\label{fig:3DRELIC}
\end{figure}
One may wish to do away with the sub-wavelength localisation problem inherent in mirror-mediated cooling altogether. Ring cavity cooling, discussed under the guise of ``amplified optomechanics'' in \sref{sec:TMM:AmplifiedOptomechanics}, provides one way of achieving this aim. This mechanism works best with particles that are rather poorly reflective, otherwise the advantages of the amplifier gain are lost, and is therefore more suited towards the cooling of atoms rather than micromirrors. In this instance, both neutral atoms and trapped ions are good candidates; it must also be mentioned that in the case of ions, cooling down to a very low vibrational state is \emph{not} needed in this case. The constraint on the delay line length is not lifted, however:~this must still be stabilised interferometrically.\par
An extension of this scheme can be envisaged where the pump light is not injected into the cavity but is directed off-axis at the particle. This geometry would not require an isolator, since the pump light never enters the cavity, and can easily be extended to three dimensions; see \fref{fig:3DRELIC}.

\subsection{Concave mirror cooling}\label{sec:Experimental:Future:Overview:ConcaveMirror}
\begin{figure}[t]
  \centering
  \subfigure[]{
  \includegraphics{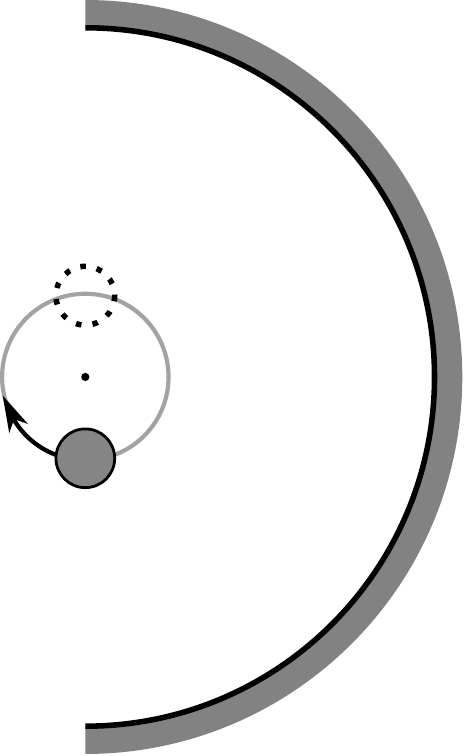}
  }\hspace{1cm}
  \subfigure[]{
  \includegraphics{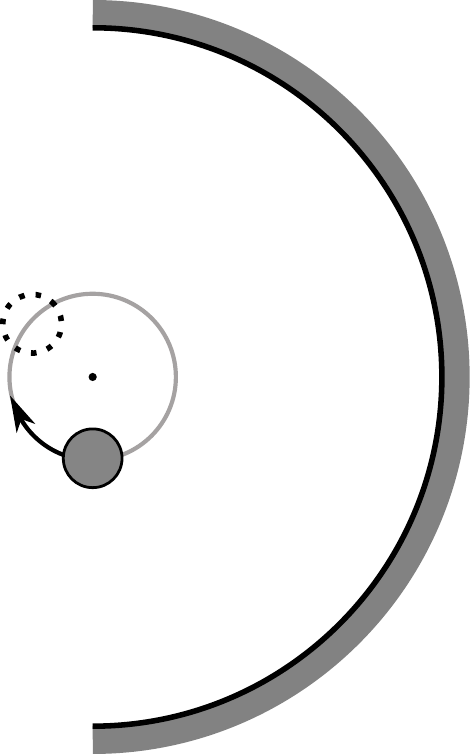}
  }
  \caption[Origin of the friction force for motion of a particle around the centre of a hemispherical mirror]{Origin of the friction force for tangential ``racetrack'' motion of a particle around the centre of curvature of a hemispherical mirror. (a) In the absence of delay, the particle and its image are at diametrically opposite points in the circular locus of the motion of the particle. (b) Due to delay, the image trails behind its equilibrium position slowing the particle down due to the repulsive interaction between the two. Note that the length scales are greatly exaggerated; in reality, the particle, and the path it moves along, would be much smaller than depicted.}
  \label{fig:ConcaveMirrorRacetrack}
\end{figure}
A final possibility is that of using the reflection from a concave hemispherical mirror itself, and not any material resonances supported by the mirror, to cool the motion of a particle at the centre of curvature of the mirror. One would expect that for the effect to be sizeable, the mirror radius would need to be much larger than a wavelength, perhaps of the order of millimetres or centimetres. A brief justification for why a cooling effect is expected to exist in such a geometry is possible using the ideas developed in \sref{sec:CoolingMethods:Retarded} in the case where the particle experiences a repulsive interaction with its image. Such a case could correspond to an atom illuminated by light tuned to the blue of its resonance.\par
When the particle is near the centre of the mirror, a real image is formed that in steady-state is at the same distance from, but on the opposite side of, the centre. Given the symmetry of the situation, it is enough to decompose the motion of the particle into two orthogonal motions, radially and tangentially in a polar coordinate system centred on the centre of curvature of the mirror. In the case of radial motion, we will appeal to \eref{eq:GeneralTimeDelayedF}. The repulsive potential $U$ seen by a motionless particle at a small distance $r$ from the centre can be described by a Gaussian\footnote{To a good approximation, a small particle close to the centre of a spherical mirror produces a spherical Gaussian image upon focussing by the mirror~\cite{Siegman1990}.}, having a peak at $r=0$ and tailing off to zero as $r$ increases. For concreteness, let us take
\begin{equation}
U(\tilde{r},r,v=0)=U_0\exp\bigl[-(r+\tilde{r})^2/w^2\bigr]\,,
\end{equation}
using the notation of \sref{sec:CoolingMethods:Retarded}, where $w$ is the width of the potential, taken to be the same in every direction, and $U_0$ is non-negative. Thus, using \eref{eq:GeneralTimeDelayedF}, we obtain the delayed friction force
\begin{equation}
\force=2U_0\tau v\frac{8r^2-w^2}{w^4}\exp\bigl(-4r^2/w^2\bigr)\,,
\end{equation}
which is a cooling force for small $r$, and where $\tau=2R/c$ is the delay time for a mirror of radius of curvature $R$.\\
The origin of the cooling force for motion in the tangential direction is best explained pictorially. \fref{fig:ConcaveMirrorRacetrack} shows the particle moving in a circle of constant radius around the centre of the mirror. In the absence of delay, \fref{fig:ConcaveMirrorRacetrack}(a), the particle and its image would be in diametrically opposite positions along this circle. Due to the delay, however, the image lags behind its equilibrium position, as shown in \fref{fig:ConcaveMirrorRacetrack}(b). The repulsive interaction between the particle and its image thereby produces a force opposing the motion of the particle. This force can easily be seen to increase with both $v$ and $\tau$, since the distance between the particle and its image decreases with both these quantities, and therefore acts to cool the tangential motion of the particle. Putting these two arguments together implies that a particle undergoing any motion around the centre of curvature of a spherical mirror experiences a friction force mediated by the delayed dipole--dipole interaction.

\section{Cooling forces experienced in different geometries}\label{sec:Experimental:Future:Forces}
In this section we shall use the three-dimensional scattering theory developed in \cref{ch:TMM:Scattering} to obtain the cooling forces experienced by a point-like dipole in a number of different configurations, both (quasi-)one-dimensional and three-dimensional.

\subsection{Longitudinal mirror-mediated cooling}
In one dimension, the mirror-mediated cooling force is given, in agreement with \eref{eq:Scattering1DMMCForce}, as
\begin{equation}
\force=-\frac{1}{2c}\epsilon_0\frac{\efield_0^2}{\sigma_\text{L}}\chi^2k^3x\sin(4kx)v+\mathcal{O}\bigl(\chi^4\bigr)\,,
\end{equation}
where $x$ is the position of the particle with respect to the mirror, and $\chi$ is the polarisability of the particle. Note that this force is oscillatory in the position of the particle and therefore has a zero spatial average on the wavelength scale.\par
A close analogue to the one-dimensional case in three dimensions is that of longitudinal pumping, where the pump field forms a standing wave with the mirror; this is the case considered earlier. When the electric field is assumed to be circularly polarised in the $y$--$z$ plane, there arises a dominant (oscillatory) component of the friction force,
\begin{multline}
\label{eq:Future:MMCLongitudinalNoAverageFriction}
\left(\begin{array}{c}
  \force_x\\
  \force_y\\
  \force_z
\end{array}\right)=\frac{1}{32\pi c}\epsilon_0\efield_0^2\chi^2k^4\frac{1}{kx}\\
\times\left[\begin{array}{ccc}
  4kx\cos(4kx)+\sin(2kx)+3\sin(4kx) & 0 & 0\\
  0 & \cos(kx)\sin^3(kx) & 0\\
  0 & 0 & \cos(kx)\sin^3(kx)
\end{array}\right]\\
\cdot\left(\begin{array}{c}
  v_x\\
  v_y\\
  v_z
\end{array}\right)+\mathcal{O}\bigl[\chi^2/(kx)^2\bigr]+\mathcal{O}\bigl(\chi^3\bigr)\,,
\end{multline}
which averages over a wavelength\footnote{The averaging is done by assuming $kx\gg 1$, which allows us to hold the $1/(kx)^n$ terms constant and average only over the periodic functions.} to give the non-oscillatory terms,
\begin{multline}
\label{eq:Future:MMCLongitudinalAverageFriction}
\left(\begin{array}{c}
  \force_x\\
  \force_y\\
  \force_z
\end{array}\right)=-\frac{1}{128\pi c}\epsilon_0\efield_0^2\chi^2k^4\frac{1}{(kx)^2}\left[\begin{array}{ccc}
  1 & 0 & 0\\
  0 & 3 & 0\\
  0 & 0 & 3
\end{array}\right]\cdot\left(\begin{array}{c}
  v_x\\
  v_y\\
  v_z
\end{array}\right)\\
-\frac{1}{1024\pi c}\epsilon_0\efield_0^2\chi^3k^7\frac{1}{(kx)^3}\left[\begin{array}{ccc}
  4 & 0 & 0\\
  0 & 3 & 0\\
  0 & 0 & 3
\end{array}\right]\cdot\left(\begin{array}{c}
  v_x\\
  v_y\\
  v_z
\end{array}\right)+\mathcal{O}\bigl(\chi^4\bigr)\,.
\end{multline}
It must be noted that these expressions are valid for $kx$ large enough that the point-dipole approximation holds ($kx\gg 1$). In order to explore the physical origin of the above forces, it is helpful to introduce some auxiliary notation. We denote the pump field $\v{\efield}_0$, as before, and the polarisation it induces in the particle $\v{\mathcal{P}}_0$. $\v{\mathcal{P}}_0$ is responsible for producing an electric field $\v{\efield}_1$. Upon reflection by the mirror, $\v{\efield}_1$ is in turn responsible for inducing a polarisation $\v{\mathcal{P}}_1$ in the particle. For each index $i$, $\v{\efield}_i$ is of the order $\chi^i$ and $\v{\mathcal{P}}_i$ of the order $\chi^{i+1}$. The force acting on the particle due to a term of the form $\v{\mathcal{P}}_i^\ast\cdot\v{\efield}_j$ is thereby of order $\chi^{i+j+1}$. Implicit in each of the above friction force expressions is a factor of order $x$ due to the retardation effects. This has to be understood as being physically separate from the factors of $x$ introduced by the spreading of wavefronts in three dimensions; indeed, in one dimension $\force\propto x$ despite the fact that wavefronts do not spread. We will factor this term out in the following.\\
Let us first decompose the friction force in \eref{eq:Future:MMCLongitudinalNoAverageFriction}: there are two sets of terms, of order $\chi^2/(kx)$ and $\chi^2/(kx)^2$, both of which arise from interactions of the form $\v{\mathcal{P}}_0^\ast\cdot\v{\efield}_1$ and $\v{\mathcal{P}}_1^\ast\cdot\v{\efield}_0$. Respectively, these interactions represent the interaction of the polarisation induced by the incident field with the retarded re-radiated field, and that of the polarisation induced by this retarded field with the incident electric field. The first term in the non-zero spatially averaged force, \eref{eq:Future:MMCLongitudinalAverageFriction}, is in this case entirely due to the geometrical spreading out of the wavefronts in the same terms. The second term, which arises from a term that is not written explicitly in \eref{eq:Future:MMCLongitudinalNoAverageFriction}, has a different origin:~it arises from the phase-locked interaction between $\v{\mathcal{P}}_1$ and $\v{\efield}_1$, which has no sub-wavelength spatial dependence, since $\v{\efield}_1$ is a travelling wave and therefore any sub-wavelength dependences are factored out of the product $\v{\mathcal{P}}_1^\ast\cdot\v{\efield}_1$. It is this friction force that therefore arises from a delayed `self-binding' interaction.

\subsection{Transverse mirror-mediated cooling}
In three dimensions one is of course free to choose the direction of illumination. Let us again consider the usual mirror-mediated cooling geometry, but where the pump light is a circularly polarised wave in the $x$--$y$ plane travelling in the $z$-direction, \ie, propagating parallel to the mirror. The resulting friction forces can again be written down as
\begin{multline}
\label{eq:Future:MMCTransverseNoAverageFriction}
\left(\begin{array}{c}
  \force_x\\
  \force_y\\
  \force_z
\end{array}\right)=\frac{1}{32\pi c}\epsilon_0\efield_0^2\chi^2k^4\frac{1}{kx}\\
\times\left[\begin{array}{ccc}
  -2kx\cos(2kx) & 0 & -2kx\cos(2kx)-\sin(2kx)\\
  0 & \sin(2kx) & 0\\
  2kx\cos(2kx) & 0 & 2kx\cos(2kx)+\sin(2kx)
\end{array}\right]\\
\cdot\left(\begin{array}{c}
  v_x\\
  v_y\\
  v_z
\end{array}\right)+\mathcal{O}\bigl[\chi^2/(kx)^2\bigr]+\mathcal{O}\bigl(\chi^3\bigr)\,,
\end{multline}
and averaged over a wavelength to give
\begin{equation}
\label{eq:Future:MMCTransverseAverageFriction}
\left(\begin{array}{c}
  \force_x\\
  \force_y\\
  \force_z
\end{array}\right)=-\frac{1}{512\pi c}\epsilon_0\efield_0^2\chi^3k^7\frac{1}{(kx)^3}\left[\begin{array}{ccc}
  2 & 0 & 0\\
  0 & 1 & 0\\
  0 & 0 & 0
\end{array}\right]\cdot\left(\begin{array}{c}
  v_x\\
  v_y\\
  v_z
\end{array}\right)+\mathcal{O}\bigl(\chi^4\bigr)\,.
\end{equation}
The oscillatory forces have the same origin as those in the longitudinal case. Note, however, that because of the geometry of this situation the only non-oscillatory forces that survive to this order are due to the self-binding terms of the form $\v{\mathcal{P}}_1^\ast\cdot\v{\efield}_1$. In a transverse-pumping geometry, it is this self-binding force that is the dominant effect.

\subsection{Ring cavity cooling}
The mechanism involved in ring cavity cooling is essentially the same as that involved in (one-dimensional) mirror-mediated cooling. Indeed, we can compare \eref{eq:TMMFrictionSimple} to \eref{eq:TMM:MirrorCoolForce} and deduce that the dominant friction force acting on a particle inside a ring cavity will be of the form
\begin{equation}
\force=-\frac{1}{4c}\epsilon_0\frac{\efield_0^2}{\sigma_\text{L}}\chi^2k^3\Lambda Lv\,,
\end{equation}
where $\Lambda$ is a factor due to the properties of the cavity, and $L$ the length of the cavity.

\subsection{Summary: Orders of magnitude}
The table below is a convenient reference for the forms of the dominant friction forces acting on a particle in the geometries discussed in this section. Symbols of the form $\force_{xx}$, for example, denote the force in the $x$ direction proportional to the $x$-component of the velocity.\par
\begin{center}
\begin{tabular}{l l | c c}
\hline
\hline
& & Oscillatory terms & Spatial average\\
\hline
\multicolumn{2}{l|}{Mirror-mediated cooling (MMC; 1D)} & $\chi^2(kx)$ & 0\\
\multirow{2}{*}{MMC (3D)\hspace{1em}$\Biggl\{$} & Longitudinal & $\chi^2/(kx)$ [$\force_{xx}$: $\chi^2$] & $\chi^2/(kx)^2$\\
                          & Transverse   & $\chi^2$ [$\force_{yy}$: $\chi^2/(kx)$] & $\chi^3/(kx)^3$ [$\force_{zz}$: $\chi^2/(kx)^5$]\\
\multicolumn{2}{l|}{Ring cavity cooling (1D)} & --- & $\chi^2(kx)$\\
\hline
\end{tabular}
\end{center}
We reiterate that the $\chi^2$ and $\chi^3$ terms forming the spatial average of the friction forces have different origins, with the former being due to the spreading of wavefronts and the latter the phase-locked interaction between $\v{\mathcal{P}}_1$ and $\v{\efield}_1$.

\section{Cooling times and base temperatures}\label{sec:Experimental:Future:Numbers}
In the previous sections we identified a number of configurations that may be used to show cooling effects mediated by the dipole--dipole force. The aim of this section is to evaluate the cooling forces and base temperatures for a few specific situations.

\subsection{One-dimensional mirror-mediated cooling: Trapped ion}
The `fine structure' in the friction force produced in the mirror-mediated cooling geometry in one dimension is perhaps best explored using trapped ions, as shown schematically in \fref{fig:Realistic_MMC}. We assume that the species being cooled is a Ba$^+$ ion, and that the pump beam is at a wavelength of $\lambda=493$\,nm, detuned by $\Delta=\pm 10\Gamma$ from resonance and focussed down to a $10$\,$\upmu$m spot at the position of the ion. The length of the fibre-based delay line is taken to be $L=2$\,m. In order to ensure operation in the low-saturation regime, we set the pump power to be $P=1$\,nW. The mass of the ion is $m=2.3\times 10^{-25}$\,kg. These numbers result in:
\begin{center}
$1/e$ velocity cooling time: $12$\,ms, and\\
steady-state temperature: $570$\,$\upmu$K.
\end{center}
Both of these numbers decrease linearly with increasing $L$ and include a factor of $\tfrac{1}{2}$ originating from the presence of the harmonic trap confining the motion of the ion, as explained in \sref{sec:CoolingMethods:MMC:Perturbative:Dipole}. Doppler cooling of the ion, after which the position spread of the particle would be smaller than a wavelength (specifically, ca.~$35$\,nm in the setup of Ref.~\cite{Eschner2001}), would be necessary to resolve the sub-wavelength features in the friction force.

\subsection{External cavity cooling: Transmissive membrane}
We have already remarked that external cavity cooling would act to enhance optomechanical cooling mechanisms, such as those used to cool the vibrational motion of reflective mirrors. Let us suppose we use a membrane with power transmissivity of $50$\% and couple the transmitted light into a Fabry-P\'erot cavity, whose mirrors have power transmissivities of $1$\%. A typical commercially-available SiN membrane would have an effective mass $m=5\times 10^{-14}$\,kg~\cite{Thompson2008} for the centre-of-mass mode and negligible absorption. Coupling in $1$\,mW of light at a wavelength $\lambda=780$\,nm thereby gives
\begin{center}
$1/e$ velocity cooling time: $1$\,ms, and\\
steady-state temperature: $35$\,$\upmu$K,
\end{center}
with the steady-state temperature being a lower limit in the absence of any absorbed power or coupling to a substrate.

\subsection{Amplified optomechanics: Neutral atom}
The friction force in amplified optomechanics, or ring cavity cooling in general, is not dependent on the position of the particle to be cooled. To explore this mechanism, we therefore suggest using neutral atoms. Our atom of choice is $^{85}$Rb (mass $m=1.4\times 10^{-25}$\,kg), which can be Doppler-cooled and confined to a small cloud beforehand. We assume that the pump beam, of wavelength $\lambda=780$\,nm and detuning $\Delta=-10\Gamma$, is focussed down to a $10$\,$\upmu$m spot at the position of the atom, and that the power at the input coupler is $13$\,nW to guarantee operation in the low-saturation regime. The fibre-based cavity is taken to be $300$\,m long, the power loss at each of the two couplers terminating the cavity is assumed to be $50$\%, and the input coupler to transmit only $1$\% of the incident power. A gain medium, with gain $1.75$, is assumed to form part of the cavity. We therefore obtain
\begin{center}
$1/e$ velocity cooling time: $4$\,ms, and\\
steady-state temperature: $40$\,$\upmu$K.
\end{center}

\appendicesstart
\section{Appendix: Electric fields inside dielectrics}\label{sec:Experimental:Future:CM}
Within a simple model of the dielectric of volume $V$ as a collection of closely spaced dipoles at random positions, its response to an electric field can be embodied entirely in the relative permittivity $\epsilon$, defined by the relation
\begin{equation}
\v{\mathcal{P}}=V(\epsilon-1)\epsilon_0\v{\efield}\,,
\end{equation}
The Clausius--Mossotti relation, as quoted in \eref{eq:ClausiusMossotti}, connects the susceptibility $\chi$ of the dielectric to the bulk refractive index of the dielectric. Many derivations of this relation (see, for example, Ref.~\cite[\textsection 4.5]{Jackson1998}) divide the dielectric into two regions: a spherical section of the dielectric, large enough to contain several dipoles but small enough that the polarisation is practically constant within; and the rest of the dielectric. One objection to this argument is that it depends critically on the first region chosen as being spherical, an assertion that has no real physical justification and no immediate connection to the main assumption, mentioned below, present in the Clausius--Mossotti relation. Hannay~\cite{Hannay1983} presented an alternative derivation of this same expression that does not make use of this model. Let us briefly recapitulate Hannay's argument.\\
The electric field $\v{\efield}$ produced by an ideal point-like dipole $\v{p}$ is, as a function of the displacement $\v{r}$ from the dipole,
\begin{equation}
\v{\efield}=\frac{1}{4\pi\epsilon_0}\Biggl[\frac{3(\v{p}\cdot\v{r})\v{r}}{\lvert\v{r}\rvert^5}-\frac{\v{p}}{\lvert\v{r}\rvert^3}\Biggr]-\frac{\v{p}}{3\epsilon_0}\delta(\v{r})\,.
\end{equation}
The central assumption that leads to the Clausius--Mossotti equation is that a ``test'' dipole inserted at a random position in the dielectric experiences an electric field that is free from the influence of the $\delta$-spikes that occur at each of the dipoles making up the dielectric. These $\delta$-spikes, of which there are $N$, contribute a spatially-averaged field
\begin{equation}
-\frac{N\v{p}}{3V\epsilon_0}=-\frac{\v{\mathcal{P}}}{3V\epsilon_0}\,,
\end{equation}
defining the macroscopic polarisation of the medium by $\v{\mathcal{P}}\equiv N\v{p}$. Thus, the field experienced by the test dipole---and, therefore, the field inside the medium---is equal to the ``normalised'' field
\begin{equation}
\v{\efield}_\text{norm}=\v{\efield}+\frac{\v{\mathcal{P}}}{3V\epsilon_0}\,,
\end{equation}
where $\v{\efield}$ is the local microscopic spatial average of the electric field that the dielectric is immersed in. Finally, $\v{\efield}_\text{norm}$ is related to $\v{\mathcal{P}}$ through the definition of $\chi$~\cite[\textsection 4.5]{Jackson1998},
\begin{equation}
\v{\mathcal{P}}=\epsilon_0\chi\v{\efield}_\text{norm}=\epsilon_0\chi\Biggl(\v{\efield}+\frac{\v{\mathcal{P}}}{3V\epsilon_0}\Biggr)\,.
\end{equation}
Thus,
\begin{equation}
\v{\mathcal{P}}=V(\epsilon-1)\epsilon_0\v{\efield}=\frac{\epsilon_0\chi\v{\efield}}{1-\chi/\bigl(3V\bigr)}\,,
\end{equation}
which can be rearranged to give the Clausius--Mossotti equation:
\begin{equation}
\chi=3V\frac{\epsilon-1}{\epsilon+2}\,.
\end{equation}
For a particle of volume $V$ made from a typical dielectric with a refractive index $n=1.5$, $\chi=0.9V\approx V$. An intuitive understanding of $\chi$ is therefore possible as the volume of dielectric that is polarised by an incoming field. It is perhaps interesting to note that this means that each single molecular dipole in the dielectric has an effective susceptibility
\begin{equation}
\chi_\text{eff}\approx\frac{4}{3}\pi a^3\,,
\end{equation}
where $2a$ is the mean distance between the molecules making up the dielectric; $a\sim 10^{-10}$\,m is typically several orders of magnitude larger than the molecular radius itself. However, $\chi_\text{eff}$ is of the same order as the susceptibility of a free molecule~\cite{Miller1990}. In other words, the polarisation of an individual dipole due to an off-resonant electric field is of about the same order whether that dipole is isolated or in a bulk solid.
\par
This $\chi$ can be used to calculate the power dissipated by a small dielectric sphere, modelled as a single point dipole, due to blackbody radiation. At a temperature $T$ there are $n_k=1/\bigl\{\exp\bigr[\hbar ck/\bigl(k_\text{B}T\bigr)\bigr]-1\bigr\}$ photons with a wavevector $\v{k}$ and wavenumber $k$. These photons produce an equivalent electric field of optical power $P_k=\hbar kc^2n_k\sigma_\text{L}/V_\text{q}$, $V_\text{q}$ being the quantisation volume, and therefore lead to an absorbed power due to that mode
\begin{equation}
P_{\text{abs},k}\approx\frac{\hbar k^2c^2n_k}{V_\text{q}}\im{\chi}\,,
\end{equation}
for small $\lvert\chi\rvert$. We must now sum over every mode to obtain the total absorbed power $P_\text{abs}=\sum_{\v{k}}P_{\text{abs},k}$, which quickly becomes cumbersome since the number of modes becomes infinite as the quantisation volume grows indefinitely. By assuming that the modes are evenly distributed in $\v{k}$-space, with a density $(2\pi)^3/V_\text{q}$, we can transform this sum into a three-dimensional integral
\begin{align}
P_\text{abs}&=\frac{2V_\text{q}}{(2\pi)^3}\iiint\frac{\hbar k^2c^2n_k}{V_\text{q}}\im{\chi}\,\rmd^3\v{k}\nonumber\\
&=\frac{\hbar c^2}{4\pi^3}\im{\chi}\int_0^{2\pi}\int_0^\pi\int_0^\infty\frac{k^4}{\exp\bigr[\hbar ck/\bigl(k_\text{B}T\bigr)\bigr]-1}\sin(\theta)\,\rmd k\,\rmd\theta\,\rmd\phi\nonumber\\
&=\frac{24\zeta(5)}{\pi^2 c^3\hbar^4}\im{\chi}\bigl(k_\text{B}T\bigr)^5\,,
\end{align}
where the extra factor of $2$ accounts for the two polarisations, where $\chi$ was assumed to be independent of $k$, and where $\zeta(5)$ is the Riemann zeta function. The $k$-integral is performed by appealing to the definition of the $\zeta(z)\equiv\zeta(z,1)$~\cite[\textsection 9.51]{Gradshteyn1994}. Our final step is to note that in thermal equilibrium, $P_\text{abs}$ is equal to the power dissipated by the sphere, $P_\text{diss}$.
%\par
%Another subtlety related to the effect of electric fields within dielectrics has given rise to the so-called ``Abraham--Minkowski controversy''. Essentially, the momentum of a photon propagating in a dielectric of (real) refractive index $n$ is either $n$ (Minkowski), or $1/n$ (Abraham), times its momentum in free space. Theoretical arguments and experimental evidence for both these alternatives have been put forward (see Ref.~\cite{Hinds2009} and references therein). Indeed, the authors of Ref.~\cite{Hinds2009} forward a surprisingly elegant argument that lays to rest, in some ways, this century-old controversy. The crux of their argument is that the difference between the two forms for the photon momentum, for light interacting with a dipole moment $\v{d}$, arises from the $\v{d}\times\v{\bfield}$ term in the momentum transferred by the Lorentz force. This term is generally ignored since the magnetic field $\v{\bfield}$ is often regarded as being small, but may in fact have significant effects for fields that vary rapidly on the timescale of the motion of an atom. Consequently, it can be shown that the measured (kinetic) momentum of a photon in a dielectric has the Abraham form. A surprising conclusion of this argument is that the dispersive light force in a red-detuned pulse \emph{repels}, rather than attracts, a two-level atom.

\begin{figure}
\centering
\includegraphics[width=\figwidth]{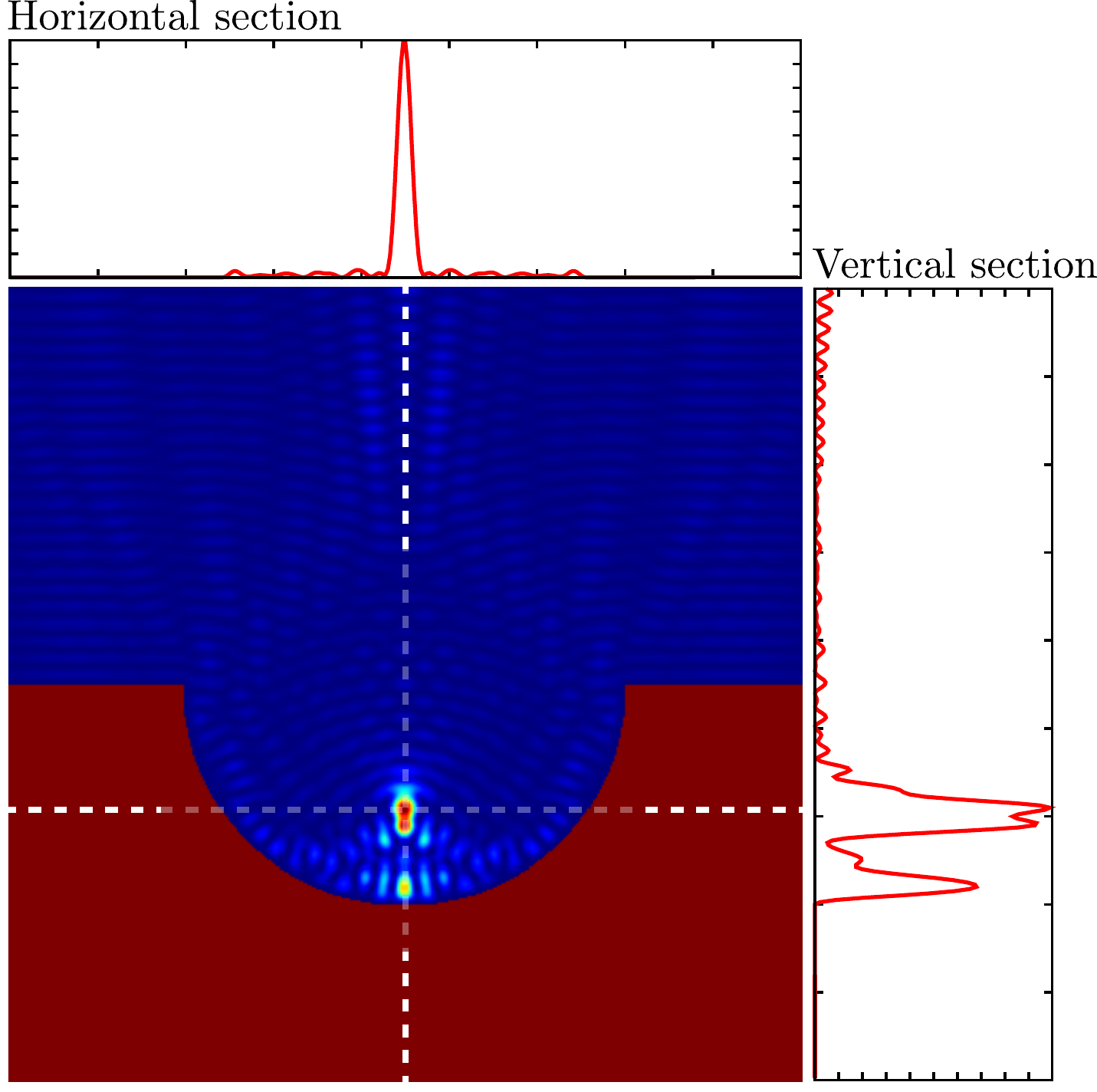}
\caption[Electric field intensity inside a $10$\,$\upmu$m diameter hemispherical void in an ideal metal substrate]{Finite-difference time-domain analysis of the electric field intensity on a 2D slice of a $10$\,$\upmu$m diameter hemispherical void in an ideal metal substrate. The incident field is a linearly polarised plane wave, with a wavelength of $780$\,nm, propagating downwards. Two sections, both intersecting the focus, are also shown: the vertical (horizontal) section runs along the vertical (horizontal) dashed line.}
\label{fig:10um}
\end{figure}
\section{Appendix: Calculating the electric field inside hemispherical mirrors}\label{sec:Future:Dipole}
Hemispherical voids templated on gold surfaces are a good system to work with experimentally: once made, they require no further alignment; and regular, close-packed arrays can be made with several tens (for large diameters) up to several hundreds (for diameters of the order of $1$\,$\upmu$m~\cite{Bartlett2004}) of dimples. However, the analysis of the electromagnetic fields inside the dimples presents a challenge. The most direct route to exploring these fields is through the numerical solution of Maxwell's equations. A large number of software packages are available, with several operating either on finite element method (FEM) or finite-difference time-domain (FDTD) principles or employing Mie theory (see Ref.~\cite{Parsons2010} for a recent review of such techniques). Analyses of scattering of electromagnetic radiation off spherical particles are usually performed using Mie theory~\cite{Jackson1998}, which exploits the fact that the vector spherical harmonics form a complete orthogonal set of modes on the sphere. In the case of a hemisphere, however, no such set of modes exists---and the situation is even worse for truncated hemispheres---and FEM or FDTD techniques are more desirable. It is interesting to note that this problem can be formally circumvented in certain situations. For example, the authors of Ref.~\cite{Hetet2010} implicitly assume knowledge of the field outside the spherical region defined by a perfect truncated hemispherical mirror to explore the behaviour of the vacuum field inside this same spherical region. In the absence of a compact analytical solution, we use an open-source software package called \textsc{MEEP}~\cite{Oskooi2010} for our analysis.\\
\begin{figure}
\centering
\includegraphics[width=\figwidth]{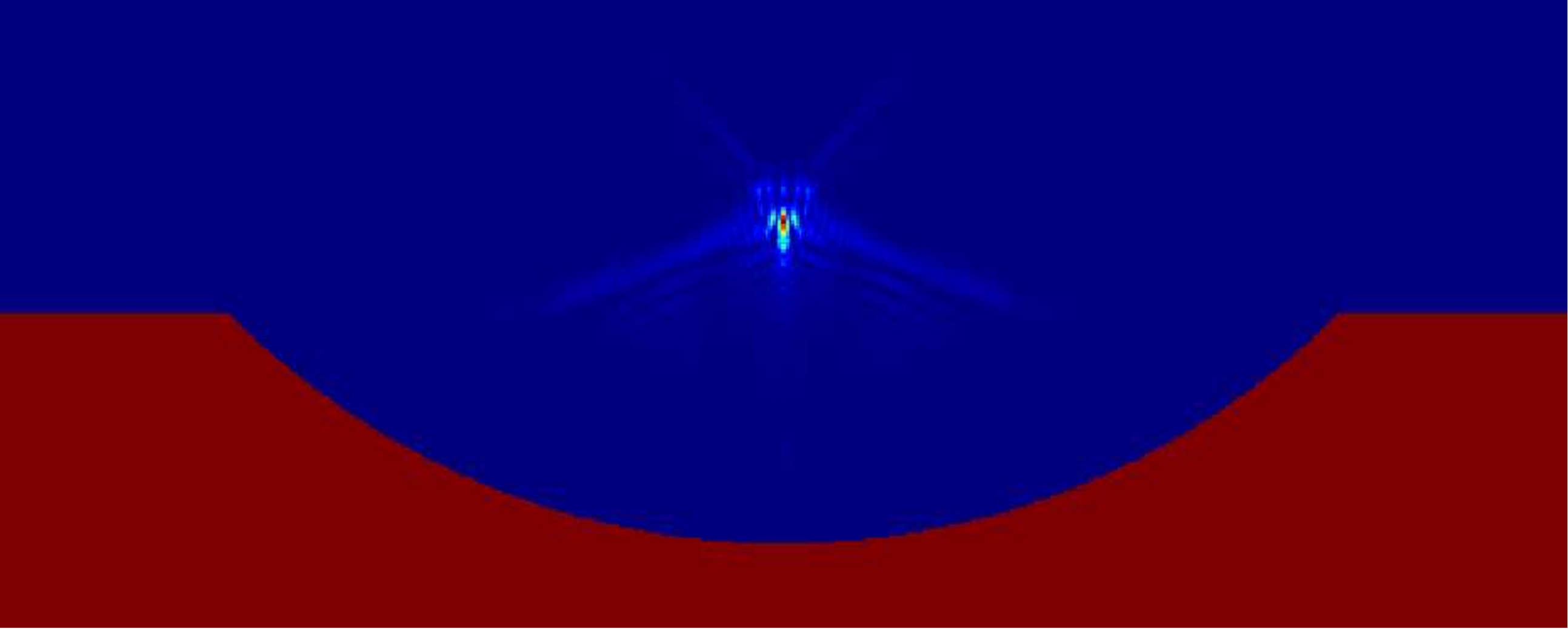}
\caption[Electric field intensity inside a $100$\,$\upmu$m diameter hemispherical void in an ideal metal substrate]{Electric field intensity inside a $100$\,$\upmu$m diameter hemispherical void in an ideal metal substrate. The size of the hemispherical template and the depth of the void ($15$\,$\upmu$m vertically from the bottom to the lip) match the surface in \fref{fig:Dimples}.}
\label{fig:100um}
\end{figure}
We show an example of such a simulated field in \fref{fig:10um}. Close to the geometrical focus of the dimple, a strong focus is found; the size and peak intensity of this focus can be used to explore the possibility of  producing single- or few-atom dipole traps inside such cavities. 2D arrays of dipole traps were demonstrated~\cite{Dumke2002} a number of years ago, and recently used to perform site-selective manipulation of atoms~\cite{Kruse2010}---both these experiments relied on a purpose-built refractive micro-lens array. Using templated surfaces offers a number of advantages over the micro-lens array, not least ease of manufacture and the possibility of integration into so-called atom chips~\cite{Folman2000}; in this regard, the use of reflective rather than refractive optical elements is of paramount importance.\\
The dimples in the first surface produced for the experimental investigations in Southampton were not grown to full hemispheres. Rather, latex spheres with a diameter of $100$\,$\upmu$m were used and gold was only templated up to a depth of $15$\,$\upmu$m. Part of the resulting surface is shown in \fref{fig:Dimples}, with the field inside one such dimple simulated in \fref{fig:100um}. Note that aliasing artifacts are more apparent in this figure than in \fref{fig:10um}, the reason being that the $0.05$\,$\upmu$m resolution possible in the case of the latter was not possible in simulating the larger sample, due to computer memory constraints. In the case of \fref{fig:100um}, a resolution of $0.17$\,$\upmu$m was used. In both cases, the incoming field was a plane wave with a wavelength of $\lambda=780$\,nm.

\section{Appendix: Force acting on an atom inside an arbitrary monochromatic field}
In a remarkable piece of work dating to 1980, Gordon and Ashkin~\cite{Gordon1980} give several useful expressions for the force and diffusion experienced by an atom inside what they called a `radiation trap'---essentially an arbitrary (monochromatic) electric field. Implicit in their work is the assumption that the system has no `memory'; we cannot directly apply their equations to mirror-mediated or external cavity cooling systems, for example. Nonetheless, such a model is perhaps the easiest way of exploring the behaviour of atoms inside fields as complex as those in hemispherical voids on a metal surface. Unfortunately, the authors of Ref.~\cite{Gordon1980} do not give explicit general formulae for the velocity-dependent force acting on an atom; we will now generalise their expressions [specifically, Eqs.~(14) and~(15)] to the case when the atom is not motionless. Let us first briefly introduce the notation we will use in this section:\footnote{Our notation will be identical to Ref.~\cite{Gordon1980} wherever possible.} $\v{\force}$ is the force acting on the atom; $D$ is the difference in the populations of the upper and lower states; $\sigma$ is the atomic lowering operator; $\Delta=\omega-\omega_\text{a}$, with $\omega$ the frequency of the driving field and $\omega_\text{a}$ the atomic resonance frequency; $\gamma=\Gamma-\i\Delta$; $\v{\mu}_{21}$ the atomic dipole operator; $\v{v}$ the atomic velocity; and $\efield e^{-\i\omega t}$ the classical incident electric field. We also define
\begin{equation}
 g=\tfrac{\i}{\hbar}\v{\mu}_{21}\cdot\efield\,,
\end{equation}
and the saturation parameter $s=2\lvert g\rvert^2/\lvert\gamma\rvert^2$. It is also convenient to define the vectors $\v{\alpha}$ and $\v{\beta}$ by $\grad g=(\v{\alpha}+\i\v{\beta})g$. In practice, numerical simulation gives us knowledge of $\efield$, assuming that the perturbation of the atom on the field is quite small. By specifying the magnetic field, we can determine $\omega_\text{a}$, and therefore $\Delta$. Given the atomic species, we also know $\Gamma$ and $\v{\mu}_{21}$; knowledge of $\gamma$, $g$, $s$, $\v{\alpha}$, $\v{\beta}$ and $\expt\sigma$ then follows, as we will see. Finally, this determines $\expt{\v{\force}}$, the classical force acting on the atom.\\
To first order in $\v{v}$, we have
\begin{align}
\label{eq:ForceGordon1980}
 \expt{\v{\force}}&=-\i\hbar\bigl[\expt{\sigma}^\ast\grad g-\expt{\sigma}\grad g^\ast\bigr]\,,\\
 \expt{\dot{\sigma}}+\gamma\expt{\sigma}&=\expt{D}g\,,\\
 \expt{\dot{D}}+2\Gamma\expt{D}&=2\Gamma-2(g^\ast\expt{\sigma}+g\expt{\sigma}^\ast)\,,\\
 \expt{\dot{D}}&=-\tfrac{2s}{1+s}(\v{v}\cdot\v{\alpha})\expt{D}\,,\\
 \expt{\dot{\sigma}}&=\bigl[(\v{v}\cdot\v{\alpha})\tfrac{1-s}{1+s}+\i(\v{v}\cdot\v{\beta})\bigr]\expt{\sigma}\,\text{, and}\\
 \dot{g}&=\v{v}\cdot(\v{\alpha}+\i\v{\beta})g\,,
\end{align}
directly from Ref.~\cite{Gordon1980}. Thus, to the same order,
\begin{equation}
 \expt{D}=\bigl[1-\tfrac{1}{\Gamma}(g^\ast\expt{\sigma}+g\expt{\sigma}^\ast)\bigr]+\tfrac{1}{\Gamma}\tfrac{s}{1+s}(\v{v}\cdot\v{\alpha})\bigl[1-\tfrac{1}{\Gamma}(g^\ast\expt{\sigma}+g\expt{\sigma}^\ast)\bigr]\,,
\end{equation}
whereby
\begin{multline}
 \expt{\sigma}\bigl[\gamma+(\v{v}\cdot\v{\alpha})\tfrac{1-s}{1+s}+\i(\v{v}\cdot\v{\beta})\bigr]=\bigl[1-\tfrac{1}{\Gamma}(g^\ast\expt{\sigma}+g\expt{\sigma}^\ast)\bigr]g\\+\tfrac{1}{\Gamma}\tfrac{s}{1+s}(\v{v}\cdot\v{\alpha})\bigl[1-\tfrac{1}{\Gamma}(g^\ast\expt{\sigma}+g\expt{\sigma}^\ast)\bigr]\,.
\end{multline}
We can solve this for $\expt{\sigma}$ to obtain
\begin{multline}
 \expt{\sigma}=\frac{g}{\gamma(1+s)}+\frac{g}{\lvert\gamma\rvert^2(1+s)}\bigl(\tfrac{1}{2\Gamma}\gamma^\ast+\tfrac{1-s}{1+s}\bigr)(\v{v}\cdot\v{\alpha})\\
-2\bigl[\Gamma(1-s)+\tfrac{1}{\Gamma}\lvert g\rvert^2\bigr]\frac{g}{\gamma\lvert\gamma\rvert^2(1+s)^3}(\v{v}\cdot\v{\alpha})\\
+\frac{\bigl[2\Delta-\i\gamma(1+s)\bigr]g}{\gamma\lvert\gamma\rvert^2(1+s)^2}(\v{v}\cdot\v{\beta})\,,
\end{multline}
which we can plug into \eref{eq:ForceGordon1980}, together with $g$, to obtain the (velocity-dependent) force acting on the atom.

\appendicesend

%% file: Chapters/Conclusion.tex
\chapter*{Conclusions and outlook}\label{ch:Conclusion}
\addtotoc{Conclusions and outlook}

Cold atom experiments have progressed significantly over the $25$ years since the first cold atoms were observed in optical molasses. Increasingly, research is focussing more and more towards the use of cold atoms in applications such as sensing, metrology and information processing. This trend has made it ever more important to find means of cooling more general species of atoms and molecules; current methods are simply too species-selective. Throughout the course of this thesis I have shown how the retarded dipole--dipole interaction can be used to achieve this aim; by investing the dipole force with a non-conservative nature, one can transfer energy from a moving polarisable particle to or from the light field---the nature of the particle itself is largely irrelevant. Several key theoretical results were reported in this thesis; these are summarised below:
\begin{itemize}
\item By introducing a time delay, in a very general sense, into an optical system, one can endow the dipole interaction with a non-conservative nature.
\item This delayed dipole--dipole interaction is a very general mechanism and applies to anything that is acted upon by the dipole force.
\item In order to explore these interactions, we devised an extended formalism based on the transfer matrix method and applied it successfully to various cooling schemes.
\item The prototypical mirror-mediated cooling mechanism was explored in great detail and its shortcomings addressed through external cavity cooling and amplified optomechanics.
\item Apart from rendering the exploration of the above cooling mechanisms possible, the formalisms we extended also naturally lead to the unification of cooling mechanisms for atomic motion, and optomechanical cooling mechanisms for micromirror vibrations.
\end{itemize}
The basis for an experimental investigation of these cooling mechanisms was also laid in the latter parts of this thesis, the main results of which are
\begin{itemize}
\item an experimental apparatus that simplifies the investigation of the interactions between atoms and surfaces structured on the nano- or micro-scale;
\item a multi-level imaging system for exploring cold atom clouds near highly reflective and highly scattering surfaces; and
\item the exploration of a number of physical configurations that promise experimentally-observable effects under realisable conditions.
\end{itemize}
\par
Work is currently underway to better understand the nature of the retarded dipole--dipole interaction in three dimensions and in a more general geometric setting; it is hoped that this understanding will greatly facilitate the production of systems, perhaps of an integrated nature, that can be tailored to generate cold samples of any arbitrary species. Such systems would revolutionise sensing and metrology applications and also provide an unprecedented means of interfacing with the quantum nature of matter at the micro- and mesoscopic scales.\\
Several questions are raised by the work in this thesis that are as yet unanswered, but which can form the basis of theoretical work in the future, are:
\begin{itemize}
\item Can the vectorial electromagnetic field inside a possibly truncated, micro- or mesoscopic, hemispherical mirror be expressed succinctly? Small hemispherical mirrors violate the approximations usually assumed in optics, namely the paraxial and ray approximations. The resulting behaviour of the electric field inside such deeply curved surfaces is very intricate. Some work has been done using scattering theories (see, for example, Refs.~\cite{Balian1970} and~\cite{Balian1971}), but these theories lead to numerically-intensive computations and do not give the insight required to explore the system from the point of view of the retarded dipole--dipole interaction.
\item A natural generalisation of the transfer matrix method for static scatterers in three dimensions is the scattering matrix theory; can this latter theoretical framework be extended to account for the motion of scatterers? The transfer matrix method explored in this thesis provides a very general formalism for exploring light--matter interactions. It is, however, limited to systems that are inherently in one dimension, systems that can be reduced to one dimension (such as ring cavities), or systems interacting with an electric field that can be described in terms of plane waves along orthogonal dimensions. Systems such as colloidal crystals in two or three dimensions cannot be explored using the matrix theory described in this thesis, and yet potentially present highly interesting dynamics mediated by light.
\item The transfer matrix approach presented is based on the assumption of a quasi-static system. Within such a framework, the motion of any scatterer must be slow enough for the system to reach optical steady-state at every point. Thus, any motion must be slow on the timescale defined by the decay lifetime of any cavity interacting with the scatterer. This condition specifically rules out operation in the so-called `resolved sideband' regime which, in the case of both micromirrors~\cite{Aspelmeyer2010} and ions~\cite{Ohadi2008}, has been shown to lead to the most efficient cooling processes. In particular, it is not yet known whether the external cavity cooling mechanism would be as effective in this regime.
\end{itemize}

%% file: Chapters/Posters.tex
\newpartalt{Appendices: Publications \& Talks}{\emph{Appendices:}\\Publications\\\&\\Talks}

\newcommand{\years}[1]{{\tiny\sc{#1}\ }}
\newcommand{\markattached}{\rlap{\hspace{-0.7em}$\star$}}
\chapter{Publications}\label{ch:Pub}
Wherever published work was used in this thesis, this was clearly indicated. The following is a complete list of my work published in peer--reviewed journals or on preprint archives, in chronological order. Preprints of the peer-reviewed publications, marked with a `$\star$' in the following list, that are most relevant to this thesis are then reproduced in the same order over the following several pages.
\section{Peer-reviewed journal articles}
\noindent
\markattached\years{2009}\textbf{Andr\'e Xuereb}, Peter Domokos, Janos Asb\'oth, Peter Horak, and Tim Freegarde; \emph{Scattering theory of heating and cooling in optomechanical systems}; Phys.\ Rev.\ A \textbf{79}, 053810 (2009); \href{http://www.arxiv.org/abs/0903.3132}{\texttt{arXiv:0903.3132}}. Given a synopsis in the APS journal \emph{Physics} and mentioned in the Research Highlights section of Nature Photonics \textbf{3}, 7 (2009)\\
\markattached\years{2009}\textbf{Andr\'e Xuereb}, Peter Horak, and Tim Freegarde; \emph{Atom cooling using the dipole force of a single retroreflected laser beam}; Phys.\ Rev.\ A \textbf{80}, 013836 (2009); \href{http://www.arxiv.org/abs/0903.2945}{\texttt{arXiv:0903.2945}}\\
\markattached\years{2009}Hamid Ohadi, Matthew Himsworth, \textbf{Andr\'e Xuereb}, and Tim Freegarde; \emph{Magneto-optical trapping and background-free imaging for atoms near nanostructured surfaces}; Opt.\ Express \textbf{17}, 25, 23003 (2009); \href{http://www.arxiv.org/abs/0910.5003}{\texttt{arXiv:0910.5003}}\\
\years{2010}\textbf{Andr\'e Xuereb}, Mathias Groth, Karl Krieger, Otto Asunta, Taina Kurki-Suonio, Jari Likonen, David P Coster, ASDEX Upgrade Team; \emph{DIVIMP-B2-EIRENE modelling of $^{13}$C migration and deposition in ASDEX Upgrade L-mode plasmas}; J.\ Nucl.\ Mater.\ \textbf{396}, 2--3, 228 (2010). Based on work done whilst on an IAESTE traineeship at the Helsinki University of Technology (TKK), July--August 2006\\
\markattached\years{2010}James Bateman, \textbf{Andr\'e Xuereb}, and Tim Freegarde; \emph{Stimulated Raman transitions via multiple atomic levels}; Phys.\ Rev.\ A \textbf{81}, 043808 (2010); \href{http://www.arxiv.org/abs/0908.2389}{\texttt{arXiv:0908.2389}}\\
\markattached\years{2010}\textbf{Andr\'e Xuereb}, Tim Freegarde, Peter Horak and Peter Domokos; \emph{Optomechanical cooling with generalized interferometers}; Phys.\ Rev.\ Lett.\ \textbf{105}, 013602 (2010); \href{http://www.arxiv.org/abs/1002.0463}{\texttt{arXiv:1002.0463}}\\
\years{2010}James Bateman, Richard Murray, Matthew Himsworth, Hamid Ohadi, \textbf{Andr\'e\linebreak Xuereb}, and Tim Freegarde; \emph{H\"ansch--Couillaud locking of Mach--Zehnder interferometer for carrier removal from a phase-modulated optical spectrum}; J.\ Opt.\ Soc.\ Am.\ B \textbf{27}, 1530 (2010); \href{http://www.arxiv.org/abs/0911.1695}{\texttt{arXiv:0911.1695}}\\
\markattached\years{2010}Peter Horak, \textbf{Andr\'e Xuereb}, and Tim Freegarde; \emph{Optical cooling of atoms in microtraps by time--delayed reflection}; J.\ Comput.\ Theor.\ Nanosci.\ \textbf{7}, 1747 (2010); \href{http://www.arxiv.org/abs/0911.4805}{\texttt{arXiv:0911.4805}}\\
\markattached\years{2010}\textbf{Andr\'e Xuereb}, Peter Domokos, Peter Horak, and Tim Freegarde; \emph{Scattering theory of multilevel atoms interacting with arbitrary radiation fields}; Phys.\ Scr.\ \textbf{T140}, 014010 (2010); \href{http://www.arxiv.org/abs/0910.0802}{\texttt{arXiv:0910.0802}}\\
\markattached\years{In press}\textbf{Andr\'e Xuereb}, Peter Horak, and Tim Freegarde; \emph{Amplified optomechanics in a unidirectional ring cavity}; to appear in J.\ Mod.\ Opt.; \href{http://www.arxiv.org/abs/1101.0130}{\texttt{arXiv:1101.0130}}\\
\markattached\years{In press}\textbf{Andr\'e Xuereb}, Peter Domokos, Peter Horak, and Tim Freegarde; \emph{Cavity cooling of atoms: Within and without a cavity}; to appear in Eur.\ Phys.\ J.\ D; \href{http://www.arxiv.org/abs/1101.2739}{\texttt{arXiv:1101.2739}}

\section{Conference proceedings}
\years{2011}Peter Domokos, \textbf{Andr\'e Xuereb}, Peter Horak, and Tim Freegarde; \emph{Efficient optomechanical cooling in one-dimensional
interferometers}; presented at SPIE Photonics West 2011 (invited contribution)

\oneside
\includepdf[addtotoc={1,section,2,Scattering theory of cooling and heating in opto-mechanical systems,preprint:TMM},pages=-,offset=1in -1in,scale=0.9]{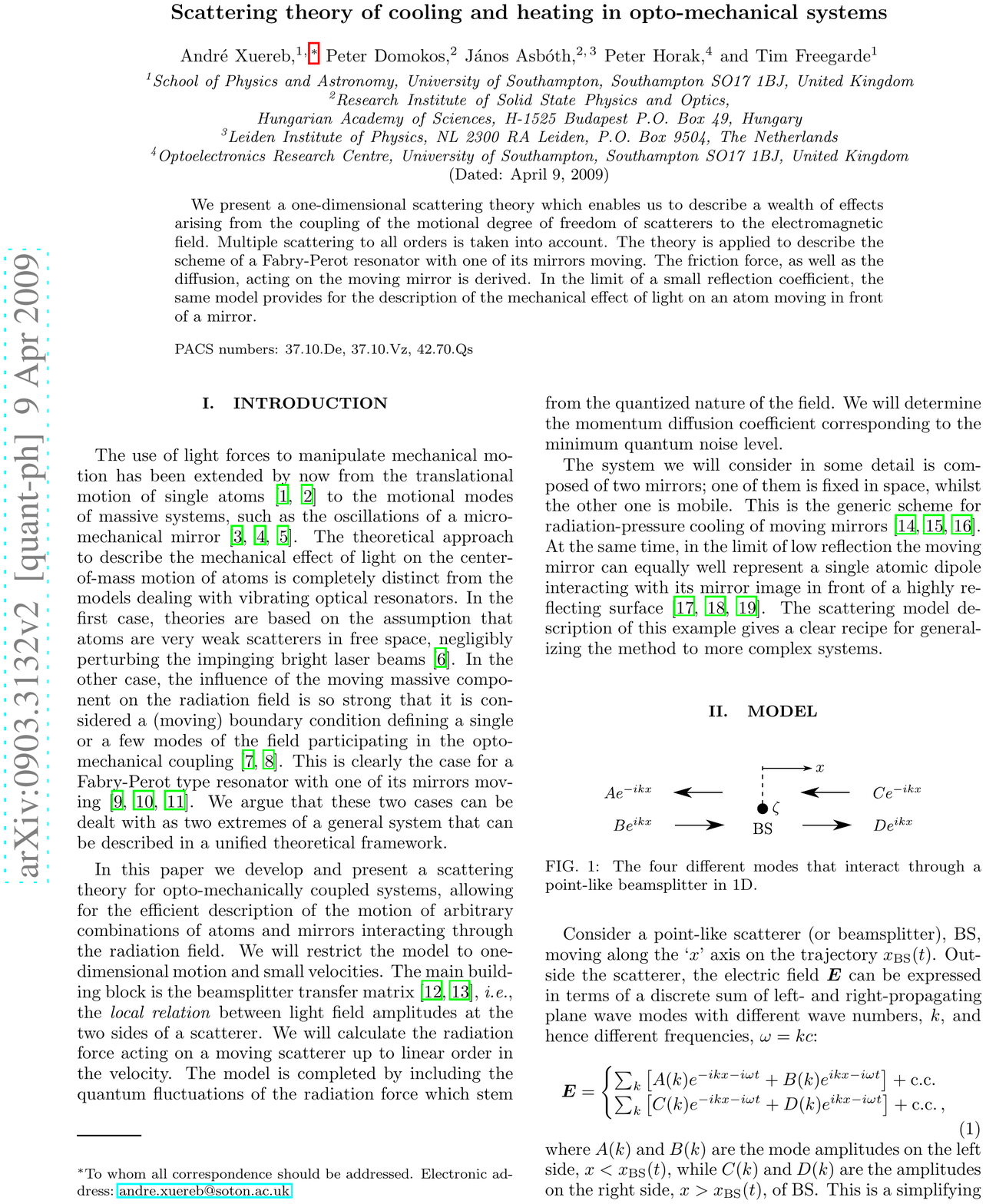}
\includepdf[addtotoc={1,section,2,Atom cooling using the dipole force of a single retroflected laser beam,preprint:MMC},pages=-,offset=1in -1in,scale=0.9]{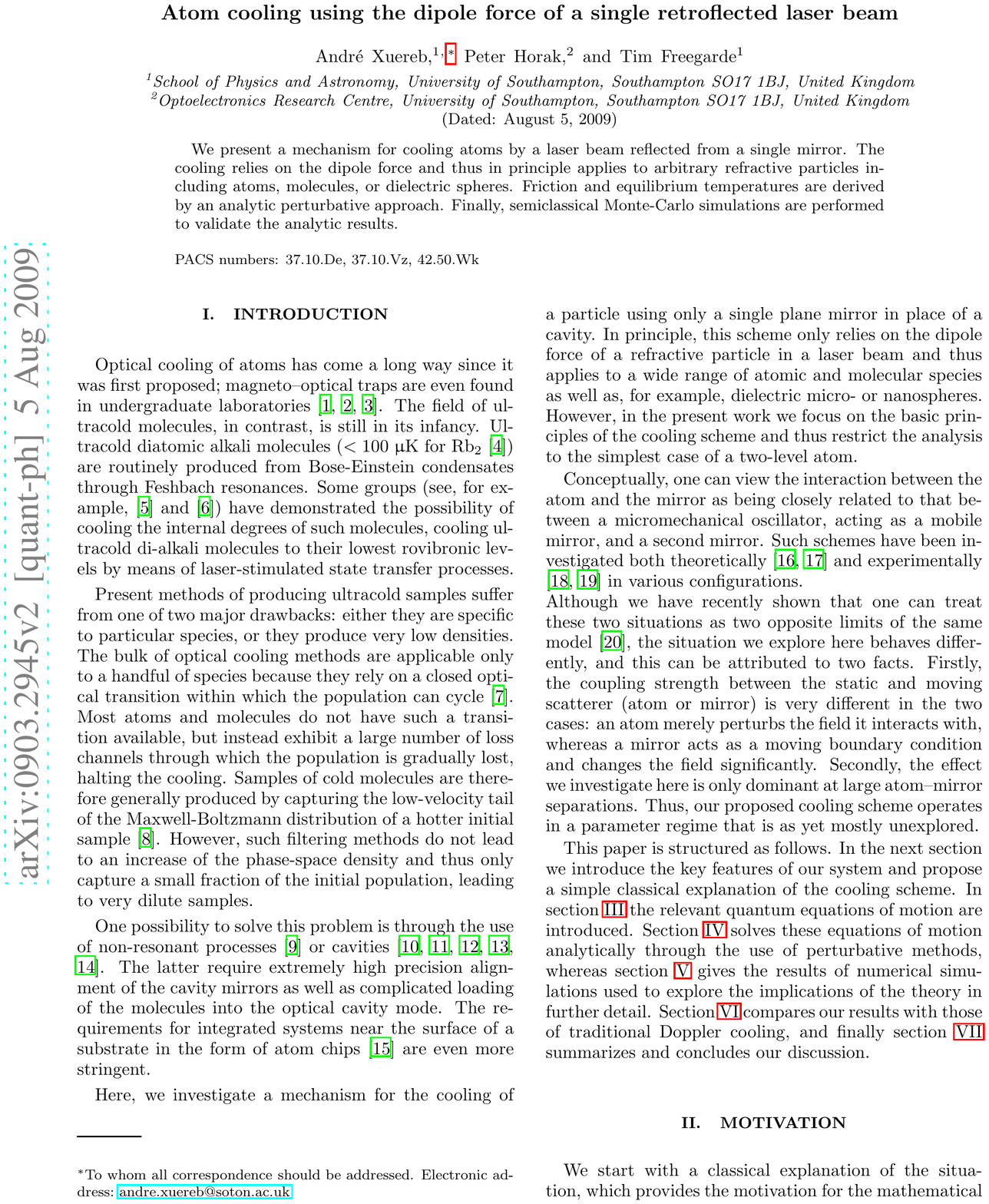}
\includepdf[addtotoc={1,section,2,Magneto-optical trapping and background-free imaging for atoms near nanostructured surfaces,preprint:LambdaMOT},pages=-,offset=1in -1in,scale=0.9]{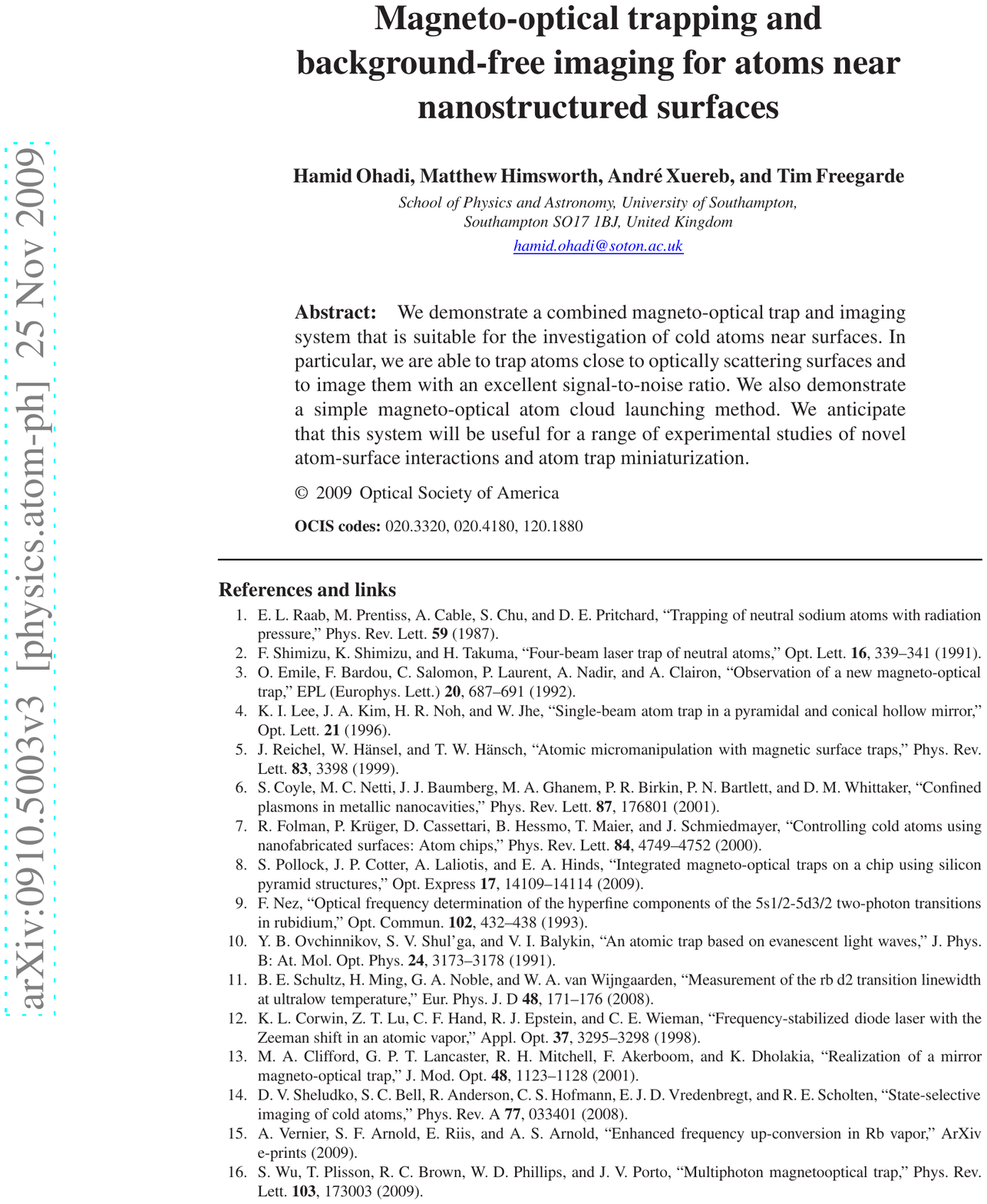}
\includepdf[addtotoc={1,section,2,Stimulated Raman transitions via multiple atomic levels,preprint:RamanLambda},pages=-,offset=1in -1in,scale=0.9]{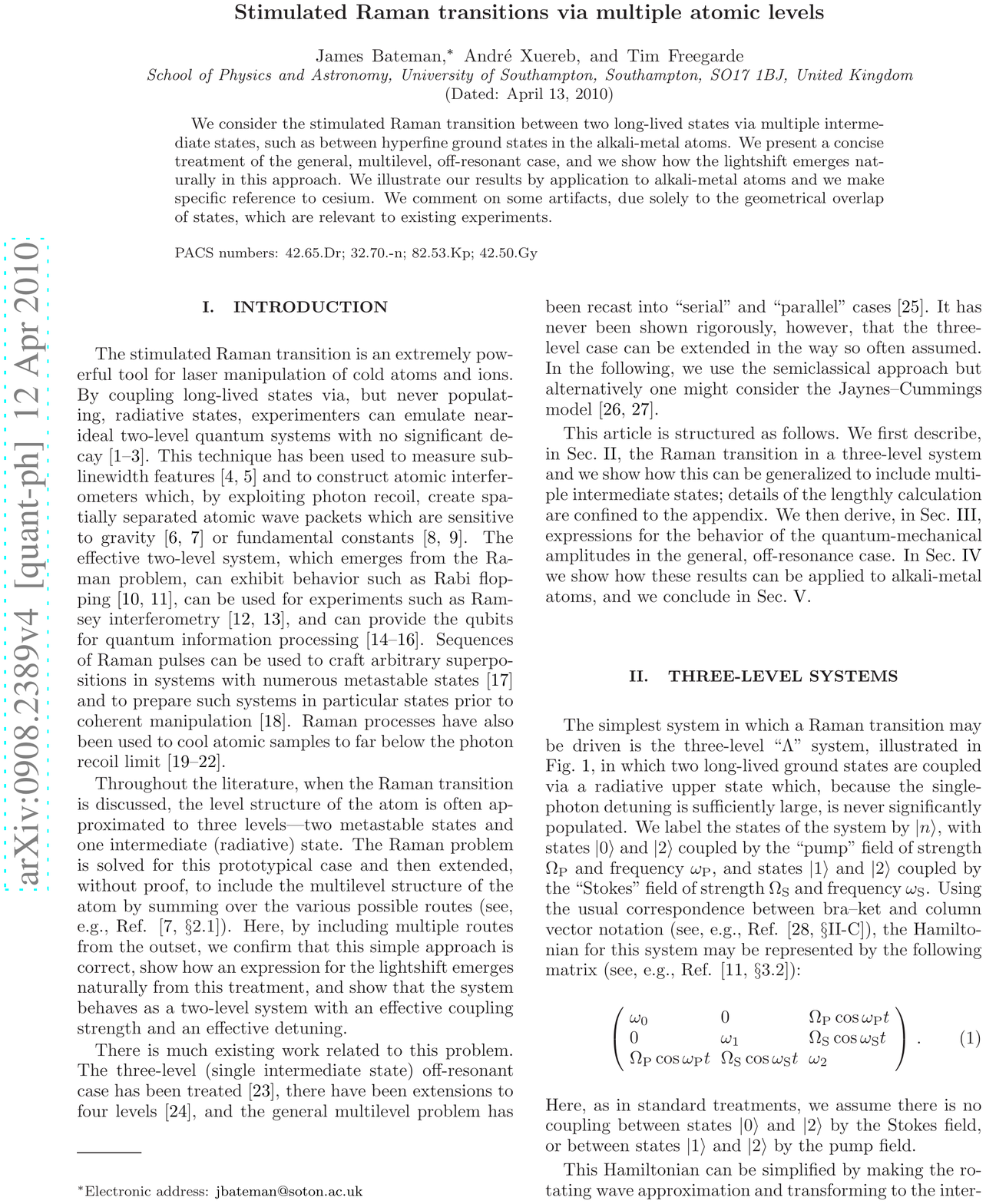}
\includepdf[addtotoc={1,section,2,Optomechanical cooling with generalized interferometers,preprint:PRL},pages=-,offset=1in -1in,scale=0.9]{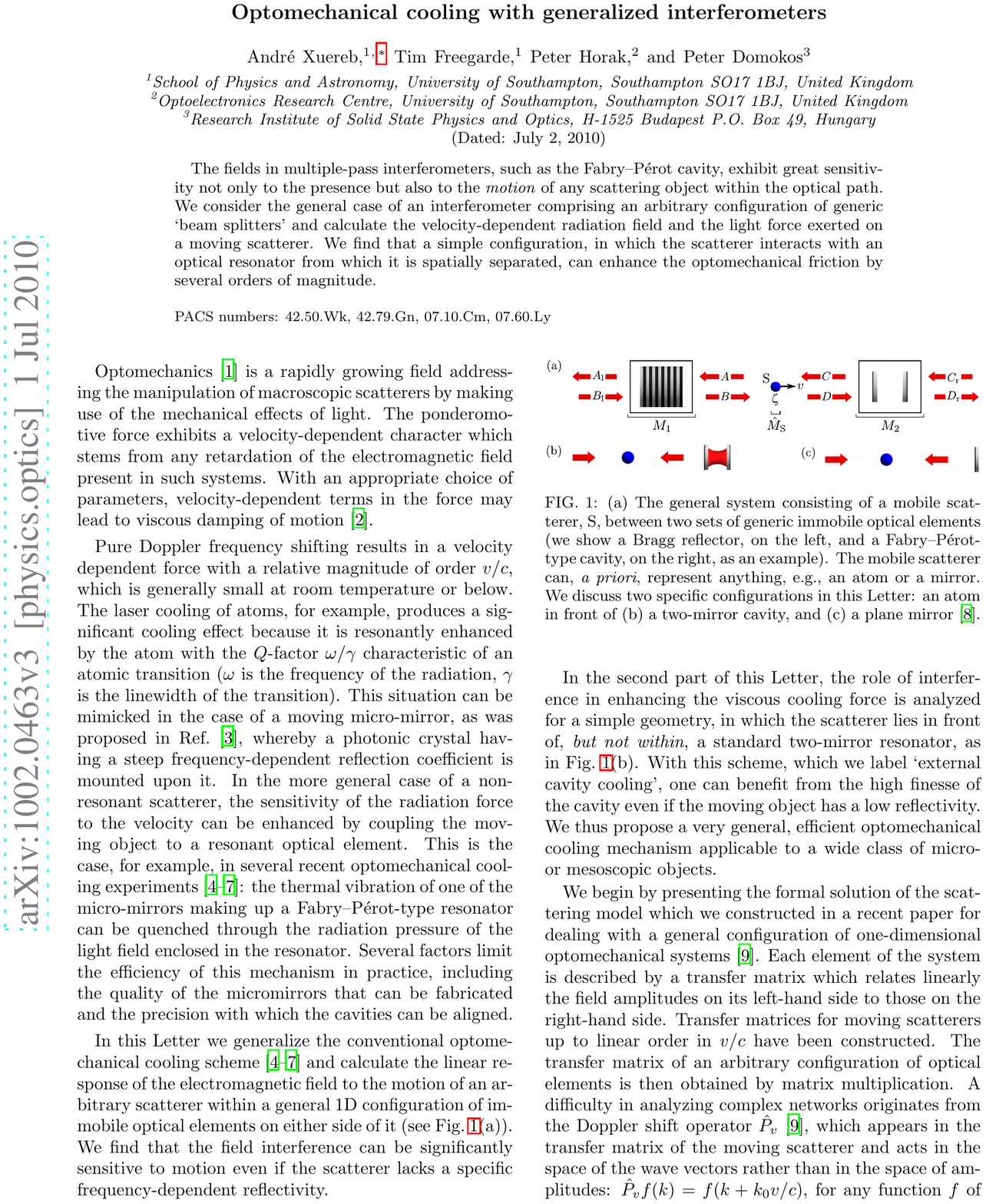}
\includepdf[addtotoc={1,section,2,Optical cooling of atoms in microtraps by time-delayed reflection,preprint:MonteCarlo},pages=-,offset=1in -1in,scale=0.9]{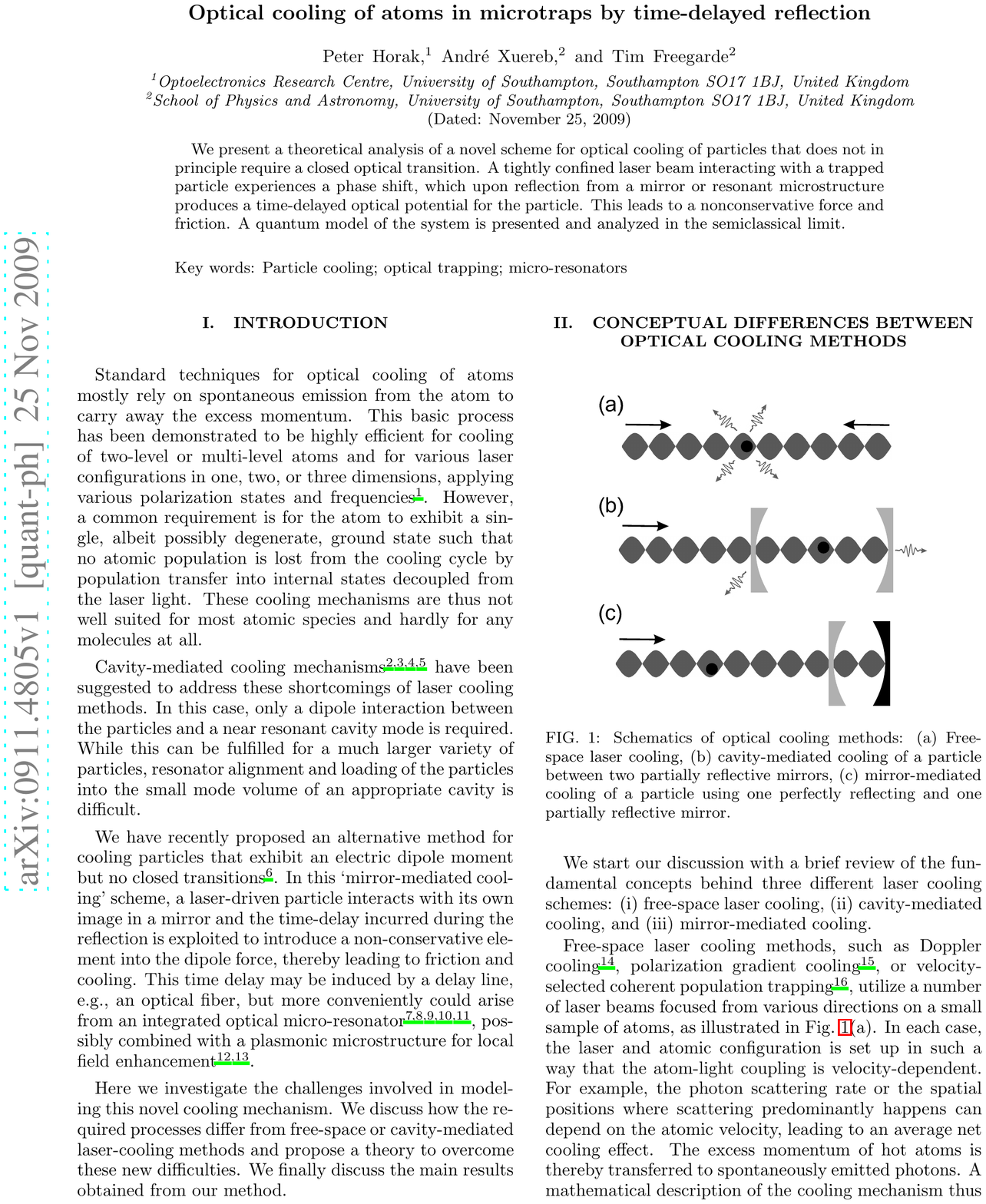}
\includepdf[addtotoc={1,section,2,Scattering theory of multilevel atoms interacting with arbitrary radiation fields,preprint:MultiLevel},pages=-,offset=1in -1in,scale=0.9]{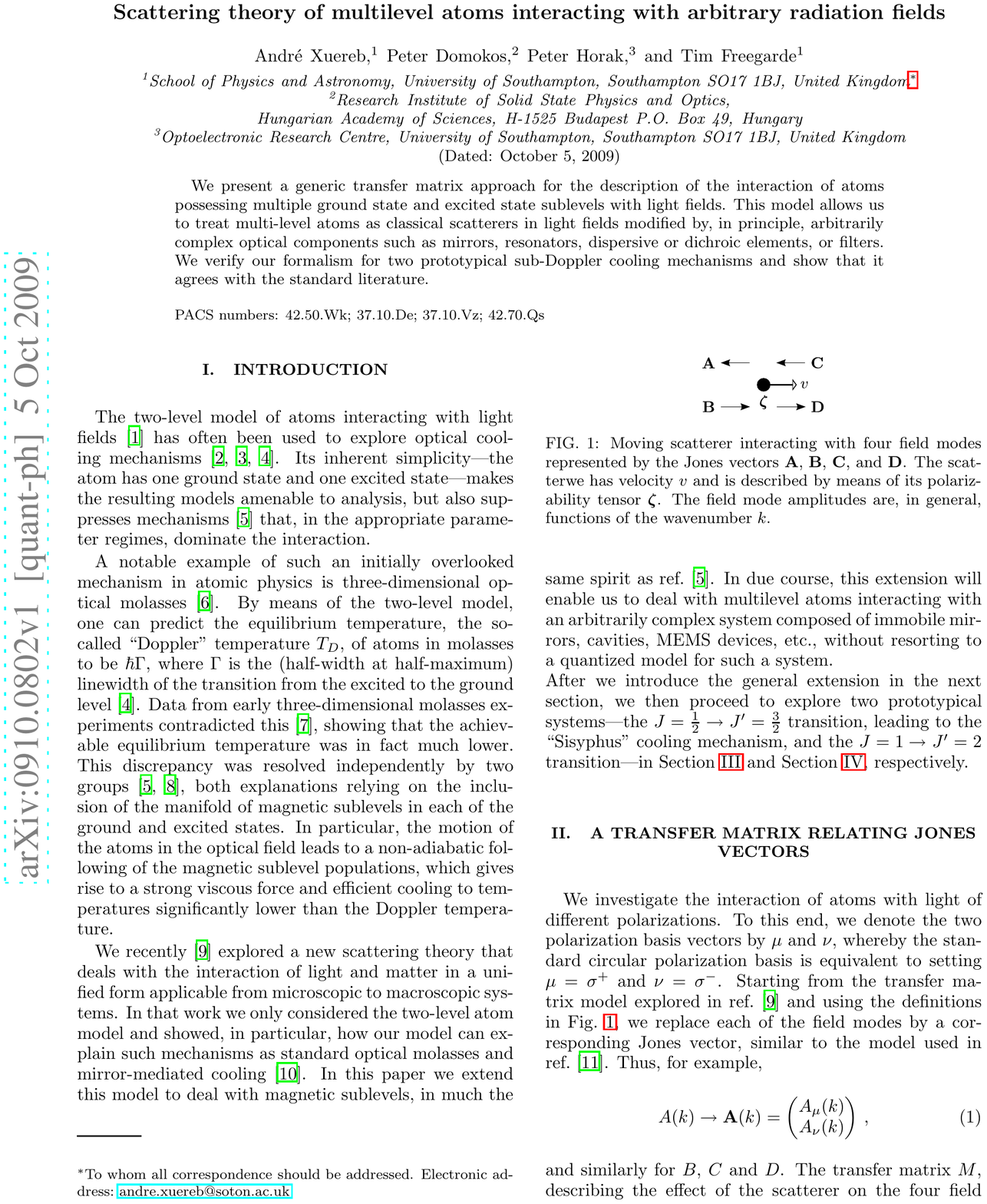}
\includepdf[addtotoc={1,section,2,Amplified optomechanics in a unidirectional ring cavity,preprint:AOM},pages=-,offset=1in -1in,scale=0.9]{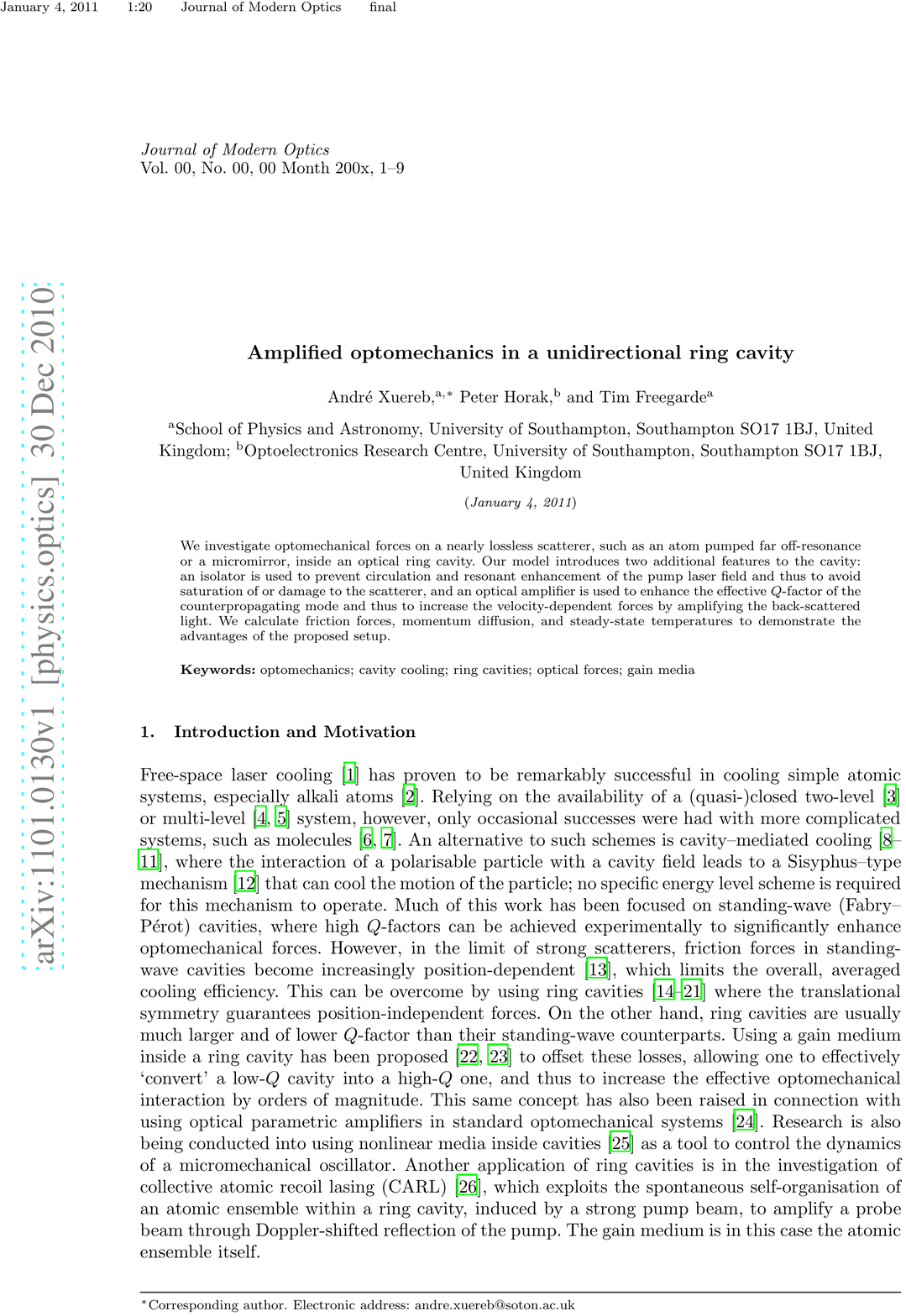}
\includepdf[addtotoc={1,section,2,Cavity cooling of atoms: Within and without a cavity,preprint:CavityCooling},pages=-,offset=1in -1in,scale=0.9]{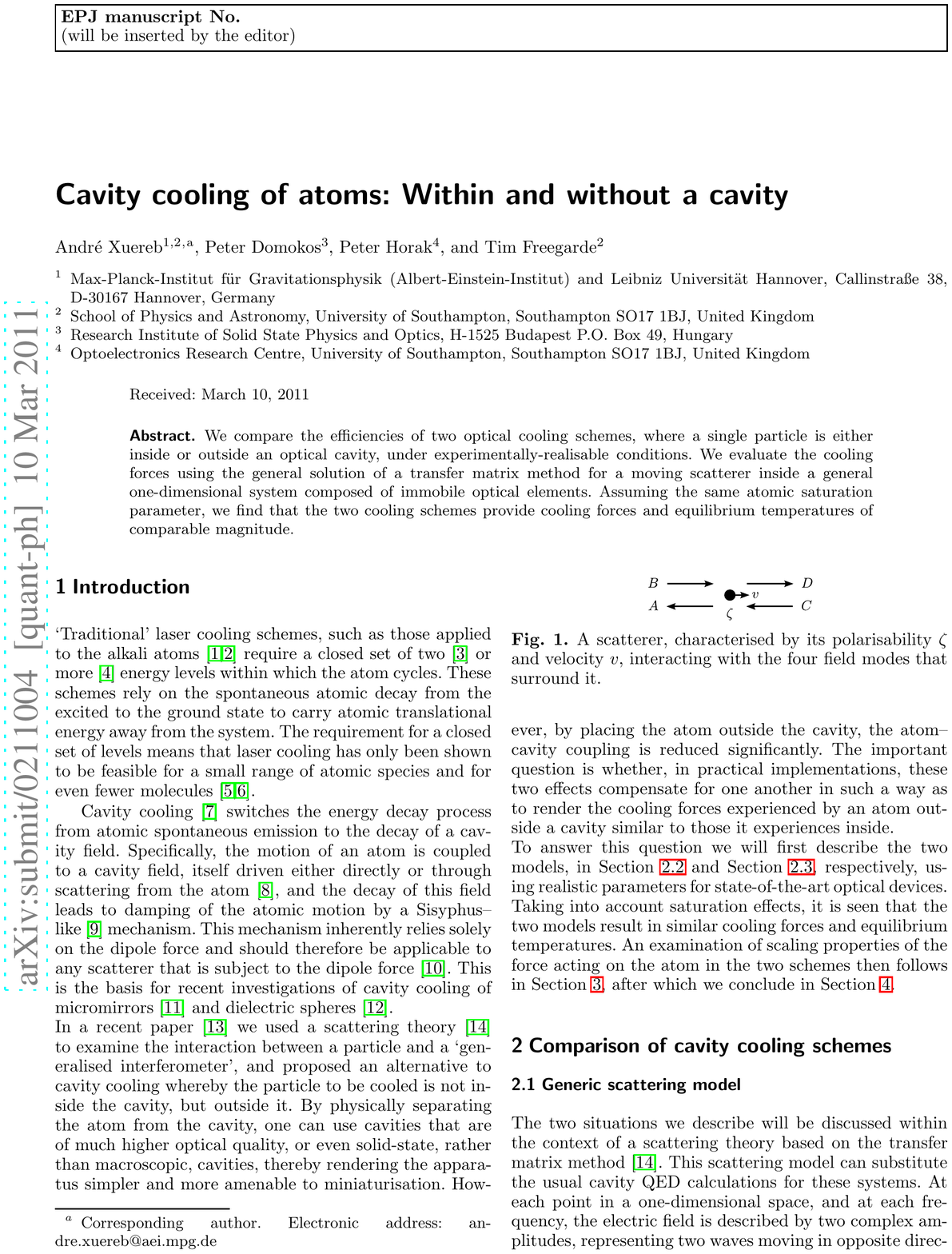}
\twoside

\chapter{Posters and Presentations}\label{ch:Posters}
\section{Talks}
Parts of this thesis were presented orally in the conferences and seminars listed below.
\par
\years{2010, Invited}\emph{Cooling polarisable particles with an optical memory}; University of Malta (Malta, November)\\
\years{2010, Invited}---; University of Hannover (Hannover, November)\\
\years{2010, Invited}---; Institute for Quantum Optics and Quantum Information (IQOQI; Innsbruck, July)\\
\years{2010, Invited}---; University of Vienna (Vienna, June)\\
\years{2010}\emph{Laser cooling using the dissipative dipole force}; University of Southampton School of Physics and Astronomy weekly QLM Seminar (Southampton, February)\\
\years{2009}\emph{Scattering theory of light--matter interactions}; 2nd UK Atom--Cavity Network Meeting (Leeds, December)\\
\years{2009}\emph{Scattering theory of cooling in optomechanical systems}; CMMC09 Workshop (Obergurgl, February)\\
\years{2008}\emph{Cooling of atoms using nanostructured surfaces}; University of Southampton School of Phy\-sics and Astronomy weekly QLM Seminar (Southampton, December)

\section{Posters}
A number of posters were also presented, both at specialist conferences and at research showcases aimed at a lay audience. These posters are listed below and reprinted over the next several pages.
\par
\years{2010}\emph{Cooling atoms, particles and polarisable objects using the dissipative dipole force}; International Conference on Quantum Optics (Obergurgl, 2010), Final CMMC Meeting (Herrsching, 2010) and ICAP 2010 (Cairns, 2010)\\
\years{2010}\emph{Mirror-mediated cooling of a particle by coupling to its own reflection}; Final CMMC Meeting (Herrsching, 2010) and ICAP 2010 (Cairns, 2010)\\
\years{2009}\emph{Scattering theory of cooling in optomechanical systems}; ICOLS09 (Hokkaido, 2009)\\
\years{2009}\emph{Novel Optical Cooling Methods for Atoms and Molecules}; University of Southampton FESM Research Showcases 2009 and 2010 (Southampton, 2009 and 2010) and explains author's research to non-specialists\\
\years{2008}\emph{Novel Optical Cooling Methods for Atoms and Molecules}; Photon08 (Edinburgh, 2008) and Les Houches School (Les Houches, 2008)\\
\years{2008}\emph{Semiclassical Theory of Coherent Atom Cooling with a Single Mirror}; EuroQUAM Inaugural Conference (Barcelona, 2008)
\newpage
\begin{figure*}
\centering
\fbox{\includegraphics[width=\textwidth]{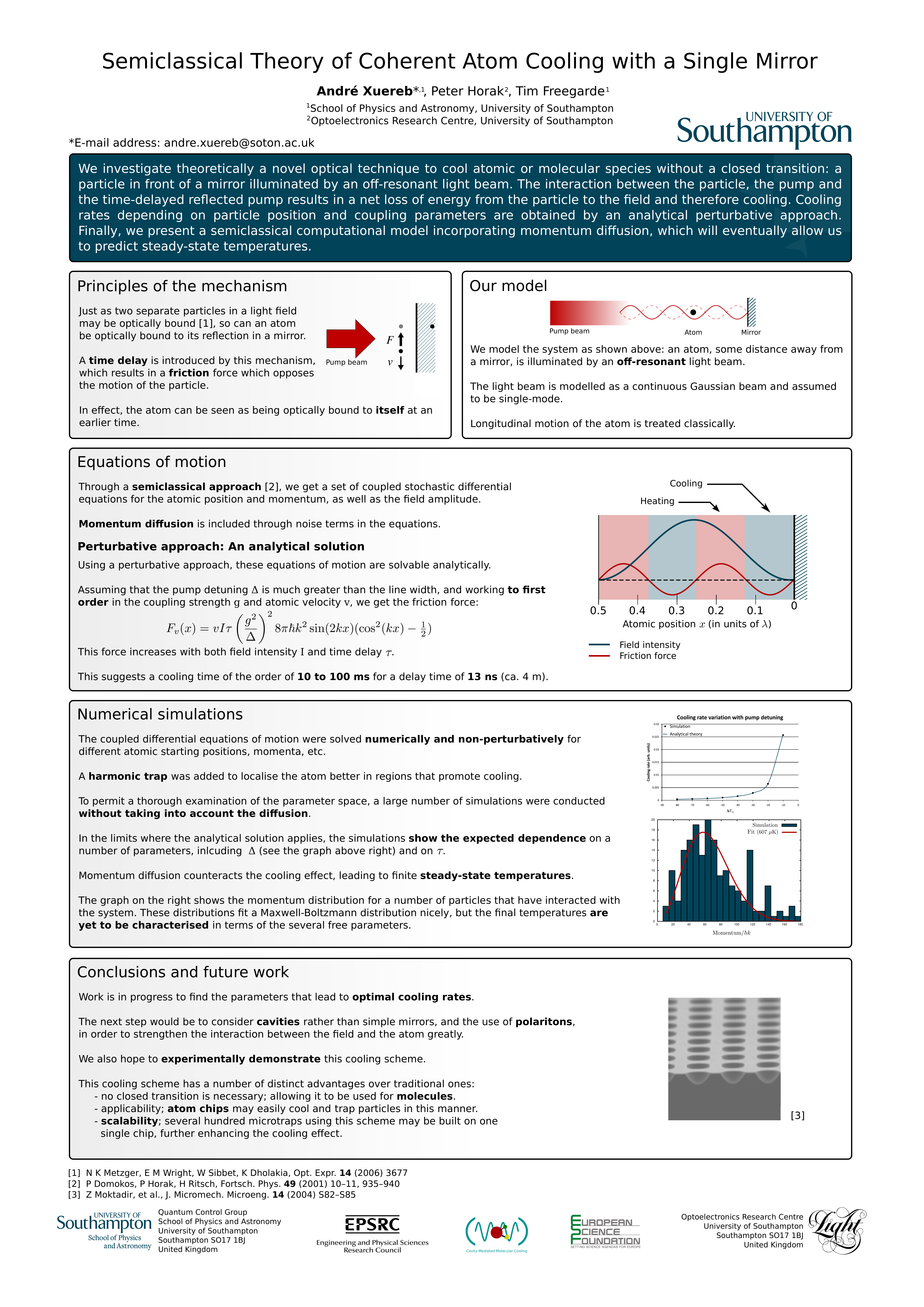}}
\caption[Poster 1]{Explanation of the mirror-mediated cooling mechanism and some initial investigations into the model.}
\end{figure*}
\newpage
\begin{figure*}
\centering
\fbox{\includegraphics[width=\textwidth]{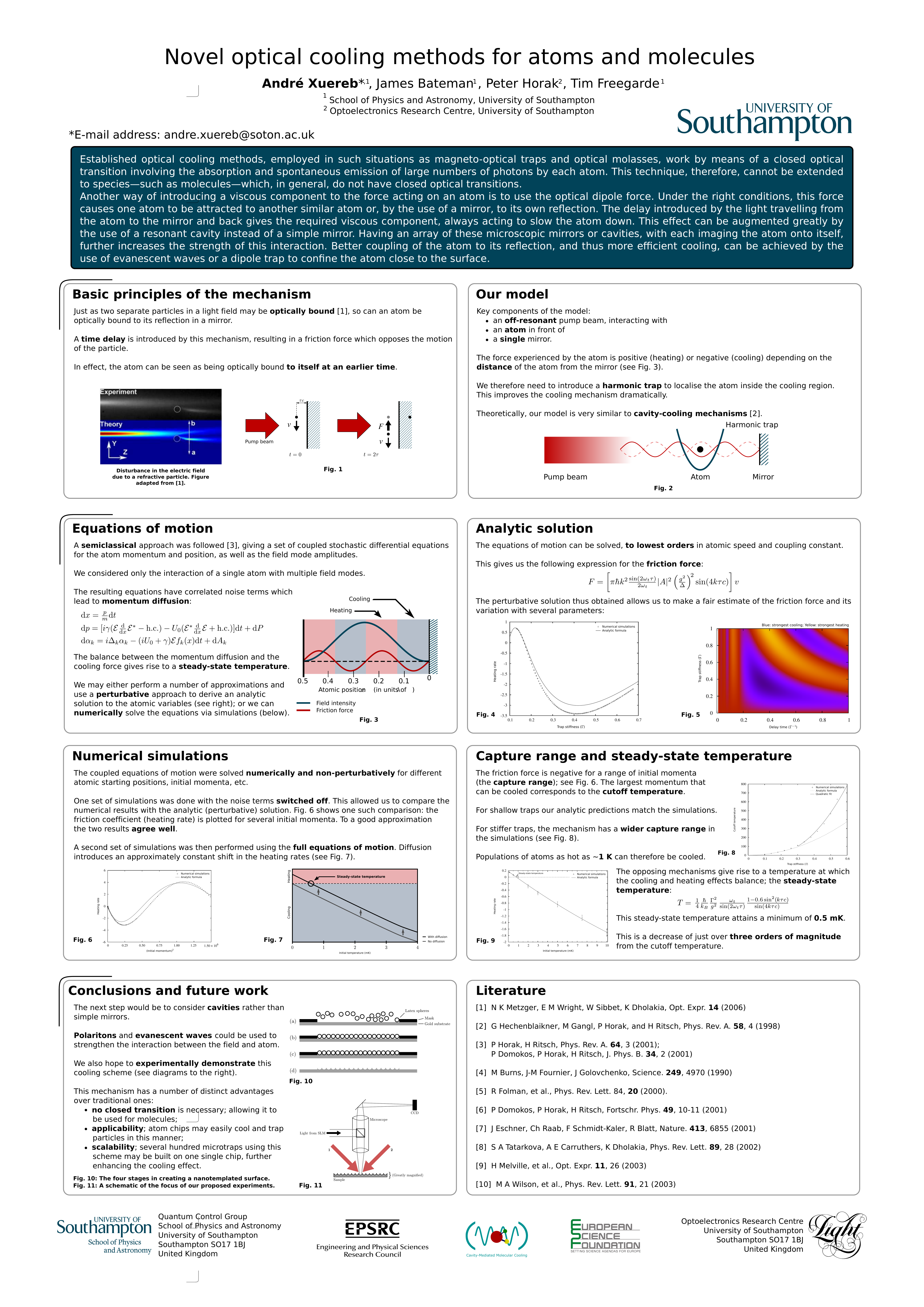}}
\caption[Poster 2]{A more in-depth treatment of mirror-mediated cooling and some speculation about possible experimental realisations.}
\end{figure*}
\newpage
\begin{figure*}
\centering
\fbox{\includegraphics[width=\textwidth]{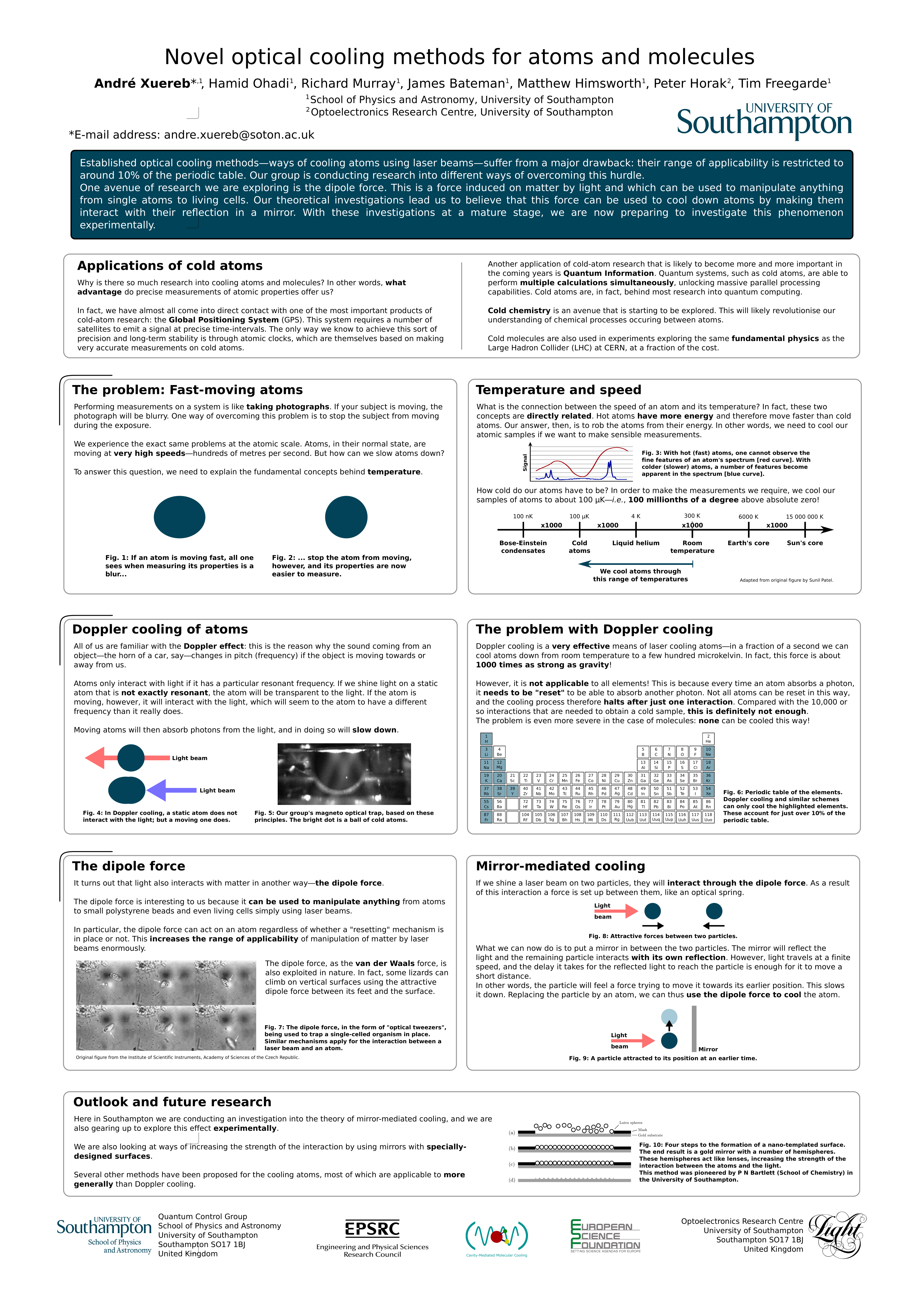}}
\caption[Poster 3]{An explanation of laser cooling intended for a lay audience.}
\end{figure*}
\newpage
\begin{figure*}
\centering
\fbox{\includegraphics[width=\textwidth]{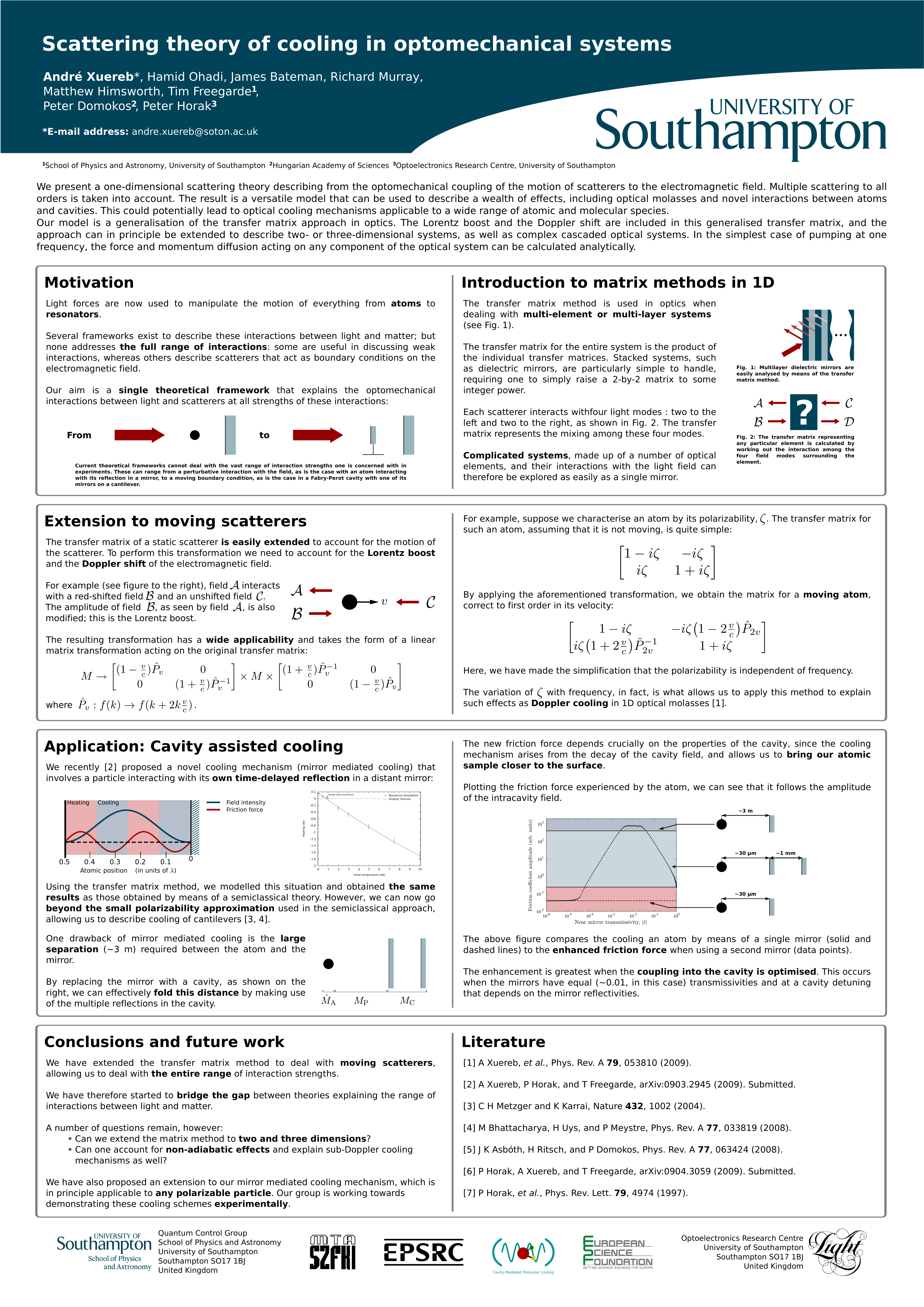}}
\caption[Poster 4]{Introduction to our transfer matrix theory and its use in explaining both mirror-mediated cooling and external cavity cooling.}
\end{figure*}
\newpage
\begin{figure*}
\centering
\fbox{\includegraphics[width=\textwidth]{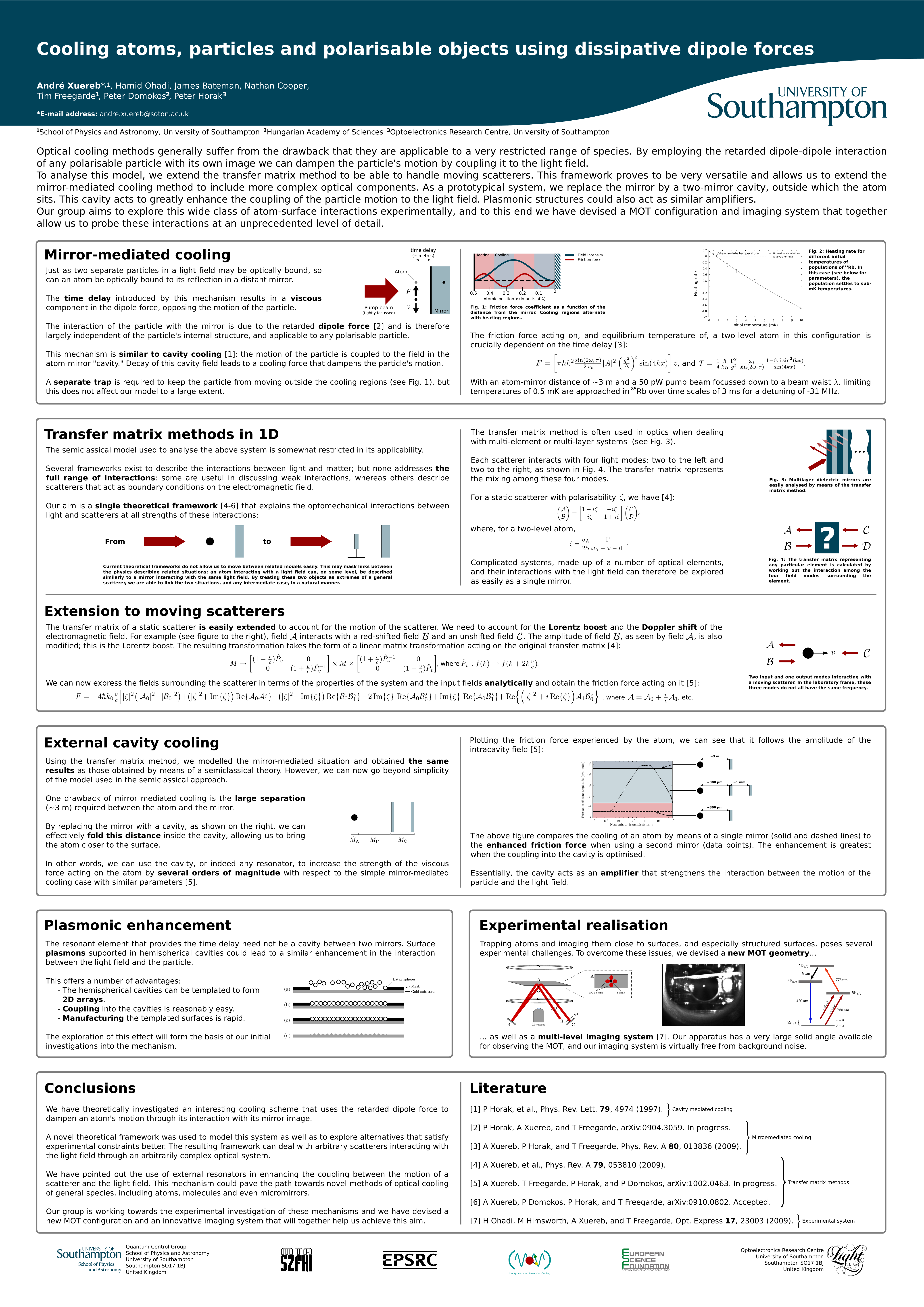}}
\caption[Poster 5]{An exploration of mirror-mediated cooling from the semi-classical and transfer matrix approaches, as well as external cavity cooling. Also some comments on the experimental apparatus used to conduct these investigations.}
\end{figure*}
\newpage
\begin{figure*}
\centering
\fbox{\includegraphics[width=\textwidth]{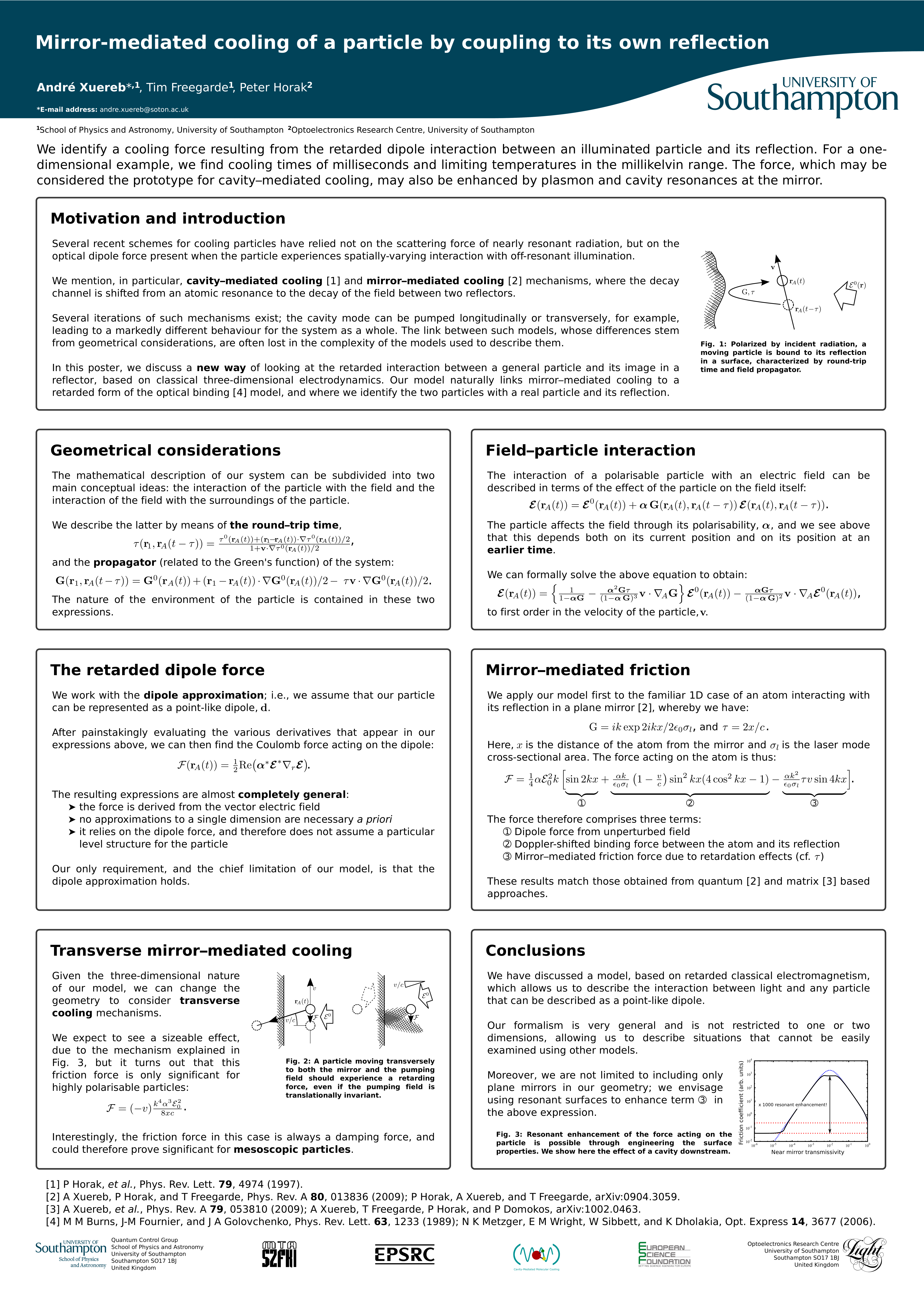}}
\caption[Poster 6]{Introduction of a classical, fully 3D and vectorial scattering theory to explain the interaction of a polarisable particle with its own reflection.}
\end{figure*}

%% file: Thesis.bbl
\begin{thebibliography}{100}
\expandafter\ifx\csname url\endcsname\relax
  \def\url#1{\texttt{#1}}\fi
\expandafter\ifx\csname urlprefix\endcsname\relax\def\urlprefix{URL }\fi
\providecommand{\bibinfo}[2]{#2}
\providecommand{\eprint}[2][]{\url{#2}}

\bibitem{Shapiro2007}
\bibinfo{author}{Shapiro, J.~A.}
\newblock \bibinfo{title}{Reminiscence on the birth of string theory}.
\newblock \emph{\bibinfo{journal}{arXiv e-prints}}  (\bibinfo{year}{2007}).
\newblock \eprint{arXiv:0711.3448}.

\bibitem{Hudson2002}
\bibinfo{author}{Hudson, J.~J.}, \bibinfo{author}{Sauer, B.~E.},
  \bibinfo{author}{Tarbutt, M.~R.} \& \bibinfo{author}{Hinds, E.~A.}
\newblock \bibinfo{title}{Measurement of the electron electric dipole moment
  using {YbF} molecules}.
\newblock \emph{\bibinfo{journal}{Phys. Rev. Lett.}}
  \textbf{\bibinfo{volume}{89}}, \bibinfo{pages}{023003}
  (\bibinfo{year}{2002}).

\bibitem{Jones2006}
\bibinfo{author}{Jones, K.~M.}, \bibinfo{author}{Tiesinga, E.},
  \bibinfo{author}{Lett, P.~D.} \& \bibinfo{author}{Julienne, P.~S.}
\newblock \bibinfo{title}{Ultracold photoassociation spectroscopy: Long-range
  molecules and atomic scattering}.
\newblock \emph{\bibinfo{journal}{Rev. Mod. Phys.}}
  \bibinfo{pages}{\textbf{\bibinfo{volume}{78}}, 483} (\bibinfo{year}{2006}).

\bibitem{Kohler2006}
\bibinfo{author}{K{\"{o}}hler, T.}, \bibinfo{author}{G{\'{o}}ral, K.} \&
  \bibinfo{author}{Julienne, P.~S.}
\newblock \bibinfo{title}{Production of cold molecules via magnetically tunable
  {F}eshbach resonances}.
\newblock \emph{\bibinfo{journal}{Rev. Mod. Phys.}}
  \bibinfo{pages}{\textbf{\bibinfo{volume}{78}}, 1311} (\bibinfo{year}{2006}).

\bibitem{Stuhl2008}
\bibinfo{author}{Stuhl, B.~K.}, \bibinfo{author}{Sawyer, B.~C.},
  \bibinfo{author}{Wang, D.} \& \bibinfo{author}{Ye, J.}
\newblock \bibinfo{title}{Magneto-optical trap for polar molecules}.
\newblock \emph{\bibinfo{journal}{Phys. Rev. Lett.}}
  \textbf{\bibinfo{volume}{101}}, \bibinfo{pages}{243002}
  (\bibinfo{year}{2008}).

\bibitem{Zeppenfeld2009}
\bibinfo{author}{Zeppenfeld, M.}, \bibinfo{author}{Motsch, M.},
  \bibinfo{author}{Pinkse, P. W.~H.} \& \bibinfo{author}{Rempe, G.}
\newblock \bibinfo{title}{Optoelectrical cooling of polar molecules}.
\newblock \emph{\bibinfo{journal}{Phys. Rev. A}} \textbf{\bibinfo{volume}{80}},
  \bibinfo{pages}{041401} (\bibinfo{year}{2009}).

\bibitem{OConnell2010}
\bibinfo{author}{O'Connell, A.~D.}, \bibinfo{author}{Hofheinz, M.},
  \bibinfo{author}{Ansmann, M.}, \bibinfo{author}{Bialczak, R.~C.},
  \bibinfo{author}{Lenander, M.}, \bibinfo{author}{Lucero, E.},
  \bibinfo{author}{Neeley, M.}, \bibinfo{author}{Sank, D.},
  \bibinfo{author}{Wang, H.}, \bibinfo{author}{Weides, M.},
  \bibinfo{author}{Wenner, J.}, \bibinfo{author}{Martinis, J.~M.} \&
  \bibinfo{author}{Cleland, A.~N.}
\newblock \bibinfo{title}{Quantum ground state and single-phonon control of a
  mechanical resonator}.
\newblock \emph{\bibinfo{journal}{Nature}}
  \bibinfo{pages}{\textbf{\bibinfo{volume}{464}}, 697} (\bibinfo{year}{2010}).

\bibitem{Groblacher2009a}
\bibinfo{author}{Gr{\"{o}}blacher, S.}, \bibinfo{author}{Hammerer, K.},
  \bibinfo{author}{Vanner, M.~R.} \& \bibinfo{author}{Aspelmeyer, M.}
\newblock \bibinfo{title}{Observation of strong coupling between a
  micromechanical resonator and an optical cavity field}.
\newblock \emph{\bibinfo{journal}{Nature}}
  \bibinfo{pages}{\textbf{\bibinfo{volume}{460}}, 724} (\bibinfo{year}{2009}).

\bibitem{Schliesser2009}
\bibinfo{author}{Schliesser, A.}, \bibinfo{author}{Arcizet, O.},
  \bibinfo{author}{Rivi{\`{e}}re, R.}, \bibinfo{author}{Anetsberger, G.} \&
  \bibinfo{author}{Kippenberg, T.~J.}
\newblock \bibinfo{title}{Resolved-sideband cooling and position measurement of
  a micromechanical oscillator close to the {H}eisenberg uncertainty limit}.
\newblock \emph{\bibinfo{journal}{Nat. Phys.}}
  \bibinfo{pages}{\textbf{\bibinfo{volume}{5}}, 509} (\bibinfo{year}{2009}).

\bibitem{Kippenberg2007}
\bibinfo{author}{Kippenberg, T.~J.} \& \bibinfo{author}{Vahala, K.~J.}
\newblock \bibinfo{title}{Cavity opto-mechanics}.
\newblock \emph{\bibinfo{journal}{Opt. Express}}
  \bibinfo{pages}{\textbf{\bibinfo{volume}{15}}, 17172} (\bibinfo{year}{2007}).

\bibitem{Ashkin1970}
\bibinfo{author}{Ashkin, A.}
\newblock \bibinfo{title}{Acceleration and trapping of particles by radiation
  pressure}.
\newblock \emph{\bibinfo{journal}{Phys. Rev. Lett.}}
  \bibinfo{pages}{\textbf{\bibinfo{volume}{24}}, 156} (\bibinfo{year}{1970}).

\bibitem{Chu1985}
\bibinfo{author}{Chu, S.}, \bibinfo{author}{Hollberg, L.},
  \bibinfo{author}{Bjorkholm, J.~E.}, \bibinfo{author}{Cable, A.} \&
  \bibinfo{author}{Ashkin, A.}
\newblock \bibinfo{title}{Three-dimensional viscous confinement and cooling of
  atoms by resonance radiation pressure}.
\newblock \emph{\bibinfo{journal}{Phys. Rev. Lett.}}
  \bibinfo{pages}{\textbf{\bibinfo{volume}{55}}, 48} (\bibinfo{year}{1985}).

\bibitem{Lett1988}
\bibinfo{author}{Lett, P.~D.}, \bibinfo{author}{Watts, R.~N.},
  \bibinfo{author}{Westbrook, C.~I.}, \bibinfo{author}{Phillips, W.~D.},
  \bibinfo{author}{Gould, P.~L.} \& \bibinfo{author}{Metcalf, H.~J.}
\newblock \bibinfo{title}{{Observation of Atoms Laser Cooled below the Doppler
  Limit}}.
\newblock \emph{\bibinfo{journal}{Phys. Rev. Lett.}}
  \bibinfo{pages}{\textbf{\bibinfo{volume}{61}}, 169} (\bibinfo{year}{1988}).

\bibitem{Dalibard1989}
\bibinfo{author}{Dalibard, J.} \& \bibinfo{author}{Cohen-Tannoudji, C.}
\newblock \bibinfo{title}{Laser cooling below the {D}oppler limit by
  polarization gradients: simple theoretical models}.
\newblock \emph{\bibinfo{journal}{J. Opt. Soc. Am. B}}
  \bibinfo{pages}{\textbf{\bibinfo{volume}{6}}, 2023} (\bibinfo{year}{1989}).

\bibitem{Ungar1989}
\bibinfo{author}{Ungar, P.~J.}, \bibinfo{author}{Weiss, D.~S.},
  \bibinfo{author}{Riis, E.} \& \bibinfo{author}{Chu, S.}
\newblock \bibinfo{title}{Optical molasses and multilevel atoms: theory}.
\newblock \emph{\bibinfo{journal}{J. Opt. Soc. Am. B}}
  \bibinfo{pages}{\textbf{\bibinfo{volume}{6}}, 2058} (\bibinfo{year}{1989}).

\bibitem{Lu1996}
\bibinfo{author}{Lu, Z.~T.}, \bibinfo{author}{Corwin, K.~L.},
  \bibinfo{author}{Renn, M.~J.}, \bibinfo{author}{Anderson, M.~H.},
  \bibinfo{author}{Cornell, E.~A.} \& \bibinfo{author}{Wieman, C.~E.}
\newblock \bibinfo{title}{Low-velocity intense source of atoms from a
  magneto-optical trap}.
\newblock \emph{\bibinfo{journal}{Phys. Rev. Lett.}}
  \textbf{\bibinfo{volume}{77}} (\bibinfo{year}{1996}).

\bibitem{Chang2009b}
\bibinfo{author}{Chang, D.~E.}, \bibinfo{author}{Regal, C.~A.},
  \bibinfo{author}{Papp, S.~B.}, \bibinfo{author}{Wilson, D.~J.},
  \bibinfo{author}{Ye, J.}, \bibinfo{author}{Painter, O.},
  \bibinfo{author}{Kimble, H.~J.} \& \bibinfo{author}{Zoller, P.}
\newblock \bibinfo{title}{{Cavity opto-mechanics using an optically levitated
  nanosphere}}.
\newblock \emph{\bibinfo{journal}{Proc. Natl. Acad. Sci.}}
  \bibinfo{pages}{\textbf{\bibinfo{volume}{107}}, 1005} (\bibinfo{year}{2010}).

\bibitem{Li2011}
\bibinfo{author}{Li, T.}, \bibinfo{author}{Kheifets, S.} \&
  \bibinfo{author}{Raizen, M.~G.}
\newblock \bibinfo{title}{Millikelvin cooling of an optically trapped
  microsphere in vacuum}.
\newblock \emph{\bibinfo{journal}{arXiv e-prints}}  (\bibinfo{year}{2011}).
\newblock \eprint{arXiv:1101.1283}.

\bibitem{Horak1997}
\bibinfo{author}{Horak, P.}, \bibinfo{author}{Hechenblaikner, G.},
  \bibinfo{author}{Gheri, K.~M.}, \bibinfo{author}{Stecher, H.} \&
  \bibinfo{author}{Ritsch, H.}
\newblock \bibinfo{title}{Cavity-induced atom cooling in the strong coupling
  regime}.
\newblock \emph{\bibinfo{journal}{Phys. Rev. Lett.}}
  \bibinfo{pages}{\textbf{\bibinfo{volume}{79}}, 4974} (\bibinfo{year}{1997}).

\bibitem{Leibrandt2009}
\bibinfo{author}{Leibrandt, D.~R.}, \bibinfo{author}{Labaziewicz, J.},
  \bibinfo{author}{Vuleti{\'{c}}, V.} \& \bibinfo{author}{Chuang, I.~L.}
\newblock \bibinfo{title}{Cavity sideband cooling of a single trapped ion}.
\newblock \emph{\bibinfo{journal}{Phys. Rev. Lett.}}
  \textbf{\bibinfo{volume}{103}}, \bibinfo{pages}{103001}
  (\bibinfo{year}{2009}).

\bibitem{Koch2010}
\bibinfo{author}{Koch, M.}, \bibinfo{author}{Sames, C.},
  \bibinfo{author}{Kubanek, A.}, \bibinfo{author}{Apel, M.},
  \bibinfo{author}{Balbach, M.}, \bibinfo{author}{Ourjoumtsev, A.},
  \bibinfo{author}{Pinkse, P. W.~H.} \& \bibinfo{author}{Rempe, G.}
\newblock \bibinfo{title}{Feedback cooling of a single neutral atom}.
\newblock \emph{\bibinfo{journal}{Phys. Rev. Lett.}}
  \textbf{\bibinfo{volume}{105}}, \bibinfo{pages}{173003}
  (\bibinfo{year}{2010}).

\bibitem{Lewenstein1993}
\bibinfo{author}{Lewenstein, M.} \& \bibinfo{author}{Roso, L.}
\newblock \bibinfo{title}{Cooling of atoms in colored vacua}.
\newblock \emph{\bibinfo{journal}{Phys. Rev. A}}
  \bibinfo{pages}{\textbf{\bibinfo{volume}{47}}, 3385} (\bibinfo{year}{1993}).

\bibitem{Braginsky1967}
\bibinfo{author}{Braginsky, V.~B.} \& \bibinfo{author}{Manukin, A.~B.}
\newblock \bibinfo{title}{Ponderomotive effects of electromagnetic radiation}.
\newblock \emph{\bibinfo{journal}{Sov. Phys. JETP}}
  \bibinfo{pages}{\textbf{\bibinfo{volume}{25}}, 653} (\bibinfo{year}{1967}).

\bibitem{Cohadon1999}
\bibinfo{author}{Cohadon, P.~F.}, \bibinfo{author}{Heidmann, A.} \&
  \bibinfo{author}{Pinard, M.}
\newblock \bibinfo{title}{Cooling of a mirror by radiation pressure}.
\newblock \emph{\bibinfo{journal}{Phys. Rev. Lett.}}
  \bibinfo{pages}{\textbf{\bibinfo{volume}{83}}, 3174} (\bibinfo{year}{1999}).

\bibitem{Arcizet2006}
\bibinfo{author}{Arcizet, O.}, \bibinfo{author}{Cohadon, P.~F.},
  \bibinfo{author}{Briant, T.}, \bibinfo{author}{Pinard, M.} \&
  \bibinfo{author}{Heidmann, A.}
\newblock \bibinfo{title}{Radiation-pressure cooling and optomechanical
  instability of a micromirror}.
\newblock \emph{\bibinfo{journal}{Nature}}
  \bibinfo{pages}{\textbf{\bibinfo{volume}{444}}, 71} (\bibinfo{year}{2006}).

\bibitem{Groblacher2009b}
\bibinfo{author}{Gr{\"{o}}blacher, S.}, \bibinfo{author}{Hertzberg, J.~B.},
  \bibinfo{author}{Vanner, M.~R.}, \bibinfo{author}{Cole, G.~D.},
  \bibinfo{author}{Gigan, S.}, \bibinfo{author}{Schwab, K.~C.} \&
  \bibinfo{author}{Aspelmeyer, M.}
\newblock \bibinfo{title}{Demonstration of an ultracold micro-optomechanical
  oscillator in a cryogenic cavity}.
\newblock \emph{\bibinfo{journal}{Nat Phys}}
  \bibinfo{pages}{\textbf{\bibinfo{volume}{5}}, 485} (\bibinfo{year}{2009}).

\bibitem{CohenTannoudji1978a}
\bibinfo{author}{Cohen-Tannoudji, C.}, \bibinfo{author}{Diu, B.} \&
  \bibinfo{author}{Laloe, F.}
\newblock \emph{\bibinfo{title}{Quantum Mechanics, Volume 1}}
  (\bibinfo{publisher}{Wiley-Interscience}, \bibinfo{year}{1978}).

\bibitem{CohenTannoudji1978b}
\bibinfo{author}{Cohen-Tannoudji, C.}
\newblock \emph{\bibinfo{title}{Quantum Mechanics, Volume 2}}
  (\bibinfo{publisher}{Wiley-Interscience}, \bibinfo{year}{1978}).

\bibitem{Shore1990a}
\bibinfo{author}{Shore, B.~W.}
\newblock \emph{\bibinfo{title}{The Theory of Coherent Atomic Excitation:
  Simple Atoms and Fields, Volume 1}} (\bibinfo{publisher}{Wiley VCH},
  \bibinfo{year}{1990}).

\bibitem{Shore1990b}
\bibinfo{author}{Shore, B.~W.}
\newblock \emph{\bibinfo{title}{The Theory of Coherent Atomic Excitation:
  Simple Atoms and Fields, Volume 2}} (\bibinfo{publisher}{Wiley VCH},
  \bibinfo{year}{1990}).

\bibitem{Woodgate2000}
\bibinfo{author}{Woodgate, G.~K.}
\newblock \emph{\bibinfo{title}{Elemetary Atomic Structure}}
  (\bibinfo{publisher}{Oxford University Press}, \bibinfo{year}{2000}),
  \bibinfo{edition}{seventh} edn.

\bibitem{CohenTannoudji2004}
\bibinfo{author}{Cohen-Tannoudji, C.}, \bibinfo{author}{Dupont-Roc, J.} \&
  \bibinfo{author}{Grynberg, G.}
\newblock \emph{\bibinfo{title}{Atom--Photon Interactions: Basic Processes and
  Applications}} (\bibinfo{publisher}{Wiley-Interscience},
  \bibinfo{year}{1992}).

\bibitem{Foot2005}
\bibinfo{author}{Foot, C.~J.}
\newblock \emph{\bibinfo{title}{Atomic Physics (Oxford Master Series in Atomic,
  Optical and Laser Physics)}} (\bibinfo{publisher}{{Oxford University Press,
  USA}}, \bibinfo{year}{2005}).

\bibitem{Arimondo1977}
\bibinfo{author}{Arimondo, E.}, \bibinfo{author}{Inguscio, M.} \&
  \bibinfo{author}{Violino, P.}
\newblock \bibinfo{title}{Experimental determinations of the hyperfine
  structure in the alkali atoms}.
\newblock \emph{\bibinfo{journal}{Rev. Mod. Phys.}}
  \bibinfo{pages}{\textbf{\bibinfo{volume}{49}}, 31} (\bibinfo{year}{1977}).

\bibitem{Steck2008}
\bibinfo{author}{Steck, D.~A.}
\newblock \bibinfo{title}{{Rubidium 85 D Line Data}} (\bibinfo{year}{2008}).
\newblock \urlprefix\url{http://steck.us/alkalidata/rubidium85numbers.pdf}.

\bibitem{Schrodinger1926}
\bibinfo{author}{Schr{\"{o}}dinger, E.}
\newblock \bibinfo{title}{An undulatory theory of the mechanics of atoms and
  molecules}.
\newblock \emph{\bibinfo{journal}{Phys. Rev.}}
  \bibinfo{pages}{\textbf{\bibinfo{volume}{28}}, 1049} (\bibinfo{year}{1926}).

\bibitem{Gardiner2004}
\bibinfo{author}{Gardiner, C.~W.} \& \bibinfo{author}{Zoller, P.}
\newblock \emph{\bibinfo{title}{Quantum Noise}} (\bibinfo{publisher}{Springer},
  \bibinfo{year}{2004}), \bibinfo{edition}{third} edn.

\bibitem{Lindblad1976}
\bibinfo{author}{Lindblad, G.}
\newblock \bibinfo{title}{On the generators of quantum dynamical semigroups}.
\newblock \emph{\bibinfo{journal}{Commun. Math. Phys.}}
  \bibinfo{pages}{\textbf{\bibinfo{volume}{48}}, 119} (\bibinfo{year}{1976}).

\bibitem{Domokos2001}
\bibinfo{author}{Domokos, P.}, \bibinfo{author}{Horak, P.} \&
  \bibinfo{author}{Ritsch, H.}
\newblock \bibinfo{title}{Semiclassical theory of cavity-assisted atom
  cooling}.
\newblock \emph{\bibinfo{journal}{J. Phys. B}}
  \bibinfo{pages}{\textbf{\bibinfo{volume}{34}}, 187} (\bibinfo{year}{2001}).

\bibitem{Gardiner1996}
\bibinfo{author}{Gardiner, C.~W.}
\newblock \emph{\bibinfo{title}{{Handbook of stochastic methods: for physics,
  chemistry and the natural sciences}}} (\bibinfo{publisher}{Springer},
  \bibinfo{year}{1996}), \bibinfo{edition}{2} edn.

\bibitem{Xuereb2009a}
\bibinfo{author}{Xuereb, A.}, \bibinfo{author}{Horak, P.} \&
  \bibinfo{author}{Freegarde, T.}
\newblock \bibinfo{title}{Atom cooling using the dipole force of a single
  retroflected laser beam}.
\newblock \emph{\bibinfo{journal}{Phys. Rev. A}} \textbf{\bibinfo{volume}{80}},
  \bibinfo{pages}{013836} (\bibinfo{year}{2009}).

\bibitem{Xuereb2009b}
\bibinfo{author}{Xuereb, A.}, \bibinfo{author}{Domokos, P.},
  \bibinfo{author}{Asb{\'{o}}th, J.}, \bibinfo{author}{Horak, P.} \&
  \bibinfo{author}{Freegarde, T.}
\newblock \bibinfo{title}{Scattering theory of cooling and heating in
  optomechanical systems}.
\newblock \emph{\bibinfo{journal}{Phys. Rev. A}} \textbf{\bibinfo{volume}{79}},
  \bibinfo{pages}{053810} (\bibinfo{year}{2009}).

\bibitem{Gardiner1984}
\bibinfo{author}{Gardiner, C.~W.}
\newblock \bibinfo{title}{Adiabatic elimination in stochastic systems. {I}.
  {F}ormulation of methods and application to few-variable systems}.
\newblock \emph{\bibinfo{journal}{Phys. Rev. A}} \textbf{\bibinfo{volume}{29}},
  \bibinfo{pages}{2814} (\bibinfo{year}{1984}).

\bibitem{Jackson1998}
\bibinfo{author}{Jackson, J.~D.}
\newblock \emph{\bibinfo{title}{Classical Electrodynamics}}
  (\bibinfo{publisher}{Wiley}, \bibinfo{year}{1998}), \bibinfo{edition}{third}
  edn.

\bibitem{Hecht2001}
\bibinfo{author}{Hecht, E.}
\newblock \emph{\bibinfo{title}{Optics}} (\bibinfo{publisher}{Addison Wesley},
  \bibinfo{year}{2001}), \bibinfo{edition}{4th} edn.

\bibitem{Drake2005}
 \emph{\bibinfo{title}{{Springer Handbook of Atomic, Molecular, and Optical
  Physics}}} (\bibinfo{publisher}{Springer}, \bibinfo{year}{2005}),
  \bibinfo{edition}{2nd} edn.

\bibitem{Deutsch1995}
\bibinfo{author}{Deutsch, I.~H.}, \bibinfo{author}{Spreeuw, R. J.~C.},
  \bibinfo{author}{Rolston, S.~L.} \& \bibinfo{author}{Phillips, W.~D.}
\newblock \bibinfo{title}{Photonic band gaps in optical lattices}.
\newblock \emph{\bibinfo{journal}{Phys. Rev. A}}
  \bibinfo{pages}{\textbf{\bibinfo{volume}{52}}, 1394} (\bibinfo{year}{1995}).

\bibitem{Tey2009}
\bibinfo{author}{Tey, M.~K.}, \bibinfo{author}{Maslennikov, G.},
  \bibinfo{author}{Liew, T. C.~H.}, \bibinfo{author}{Aljunid, S.~A.},
  \bibinfo{author}{Huber, F.}, \bibinfo{author}{Chng, B.},
  \bibinfo{author}{Chen, Z.}, \bibinfo{author}{Scarani, V.} \&
  \bibinfo{author}{Kurtsiefer, C.}
\newblock \bibinfo{title}{Interfacing light and single atoms with a lens}.
\newblock \emph{\bibinfo{journal}{New J. Phys.}} \textbf{\bibinfo{volume}{11}},
  \bibinfo{pages}{043011} (\bibinfo{year}{2009}).

\bibitem{Toll1956}
\bibinfo{author}{Toll, J.~S.}
\newblock \bibinfo{title}{Causality and the dispersion relation: Logical
  foundations}.
\newblock \emph{\bibinfo{journal}{Phys. Rev.}}
  \bibinfo{pages}{\textbf{\bibinfo{volume}{104}}, 1760} (\bibinfo{year}{1956}).

\bibitem{Wang2009}
\bibinfo{author}{Wang, D.-w.}, \bibinfo{author}{Li, A.-j.},
  \bibinfo{author}{Wang, L.-g.}, \bibinfo{author}{Zhu, S.-y.} \&
  \bibinfo{author}{Zubairy, M.~S.}
\newblock \bibinfo{title}{Effect of the counterrotating terms on polarizability
  in atom-field interactions}.
\newblock \emph{\bibinfo{journal}{Phys. Rev. A}} \textbf{\bibinfo{volume}{80}},
  \bibinfo{pages}{063826} (\bibinfo{year}{2009}).

\bibitem{CohenTannoudji1992}
\bibinfo{author}{Cohen-Tannoudji, C.}
\newblock \bibinfo{title}{Atomic motion in laser light}.
\newblock \bibinfo{pages}{In \bibinfo{editor}{Dalibard, J.},
  \bibinfo{editor}{Zinn-Justin, J.} \& \bibinfo{editor}{Raimond, J.~M.} (eds.)
  \emph{\bibinfo{booktitle}{Fundamental Systems in Quantum Optics, Proceedings
  of the Les Houches Summer School, Session LIII}}1} (\bibinfo{publisher}{North
  Holland}, \bibinfo{year}{1992}).

\bibitem{Gordon1980}
\bibinfo{author}{Gordon, J.~P.} \& \bibinfo{author}{Ashkin, A.}
\newblock \bibinfo{title}{Motion of atoms in a radiation trap}.
\newblock \emph{\bibinfo{journal}{Phys. Rev. A}}
  \bibinfo{pages}{\textbf{\bibinfo{volume}{21}}, 1606} (\bibinfo{year}{1980}).

\bibitem{Risken1989}
\bibinfo{author}{Risken, H.}
\newblock \emph{\bibinfo{title}{The Fokker-Planck Equation: Methods of
  Solutions and Applications}}.
\newblock Springer Series in Synergetics (\bibinfo{publisher}{Springer},
  \bibinfo{year}{1996}), \bibinfo{edition}{third} edn.

\bibitem{Einstein1905}
\bibinfo{author}{Einstein, A.}
\newblock \bibinfo{title}{{\"{U}}ber die von der molekularkinetischen {T}heorie
  der {W}{\"{a}}rme geforderte {B}ewegung von in ruhenden {F}l{\"{u}}ssigkeiten
  suspendierten {T}eilchen}.
\newblock \emph{\bibinfo{journal}{Ann. Phys.}}
  \bibinfo{pages}{\textbf{\bibinfo{volume}{322}}, 549} (\bibinfo{year}{1905}).

\bibitem{Metcalf1999}
\bibinfo{author}{Metcalf, H.~J.} \& \bibinfo{author}{van~der Straten, P.}
\newblock \emph{\bibinfo{title}{Laser Cooling and Trapping}}
  (\bibinfo{publisher}{Springer}, \bibinfo{year}{1999}),
  \bibinfo{edition}{first} edn.

\bibitem{Clerk2010}
\bibinfo{author}{Clerk, A.~A.}, \bibinfo{author}{Devoret, M.~H.},
  \bibinfo{author}{Girvin, S.~M.}, \bibinfo{author}{Marquardt, F.} \&
  \bibinfo{author}{Schoelkopf, R.~J.}
\newblock \bibinfo{title}{Introduction to quantum noise, measurement, and
  amplification}.
\newblock \emph{\bibinfo{journal}{Rev. Mod. Phys.}}
  \bibinfo{pages}{\textbf{\bibinfo{volume}{82}}, 1155} (\bibinfo{year}{2010}).

\bibitem{Onsager1931}
\bibinfo{author}{Onsager, L.}
\newblock \bibinfo{title}{Reciprocal relations in irreversible processes. ii.}
\newblock \emph{\bibinfo{journal}{Phys. Rev.}}
  \bibinfo{pages}{\textbf{\bibinfo{volume}{38}}, 2265} (\bibinfo{year}{1931}).

\bibitem{Ford1996}
\bibinfo{author}{Ford, G.~W.} \& \bibinfo{author}{O'Connell, R.~F.}
\newblock \bibinfo{title}{There is no quantum regression theorem}.
\newblock \emph{\bibinfo{journal}{Phys. Rev. Lett.}}
  \bibinfo{pages}{\textbf{\bibinfo{volume}{77}}, 798} (\bibinfo{year}{1996}).

\bibitem{Lax2000}
\bibinfo{author}{Lax, M.}
\newblock \bibinfo{title}{The {L}ax--{O}nsager regression `theorem' revisited}.
\newblock \emph{\bibinfo{journal}{Opt. Commun.}}
  \bibinfo{pages}{\textbf{\bibinfo{volume}{179}}, 463} (\bibinfo{year}{2000}).

\bibitem{Dalibard1984}
\bibinfo{author}{Dalibard, J.}, \bibinfo{author}{Reynaud, S.} \&
  \bibinfo{author}{Cohen-Tannoudji, C.}
\newblock \bibinfo{title}{Potentialities of a new $\sigma_+$-$\sigma_-$ laser
  configuration for radiative cooling and trapping}.
\newblock \emph{\bibinfo{journal}{J. Phys. B}}
  \bibinfo{pages}{\textbf{\bibinfo{volume}{17}}, 4577} (\bibinfo{year}{1984}).

\bibitem{Bateman2010b}
\bibinfo{author}{Bateman, J.}, \bibinfo{author}{Xuereb, A.} \&
  \bibinfo{author}{Freegarde, T.}
\newblock \bibinfo{title}{Stimulated raman transitions via multiple atomic
  levels}.
\newblock \emph{\bibinfo{journal}{Phys. Rev. A}} \textbf{\bibinfo{volume}{81}},
  \bibinfo{pages}{043808} (\bibinfo{year}{2010}).

\bibitem{kasevich92}
\bibinfo{author}{Kasevich, M.} \& \bibinfo{author}{Chu, S.}
\newblock \bibinfo{title}{{Measurement of the gravitational acceleration of an
  atom with a light-pulse atom interferometer}}.
\newblock \emph{\bibinfo{journal}{Applied Physics B: Lasers and Optics}}
  \bibinfo{pages}{\textbf{\bibinfo{volume}{54}}, 321} (\bibinfo{year}{1992}).

\bibitem{Freegarde2002}
\bibinfo{author}{Freegarde, T.} \& \bibinfo{author}{Dholakia, K.}
\newblock \bibinfo{title}{Cavity-enhanced optical bottle beam as a mechanical
  amplifier}.
\newblock \emph{\bibinfo{journal}{Phys. Rev. A}} \textbf{\bibinfo{volume}{66}},
  \bibinfo{pages}{013413} (\bibinfo{year}{2002}).

\bibitem{Dumke2002}
\bibinfo{author}{Dumke, R.}, \bibinfo{author}{Volk, M.},
  \bibinfo{author}{M{\"{u}}ther, T.}, \bibinfo{author}{Buchkremer, F. B.~J.},
  \bibinfo{author}{Birkl, G.} \& \bibinfo{author}{Ertmer, W.}
\newblock \bibinfo{title}{Micro-optical realization of arrays of selectively
  addressable dipole traps: A scalable configuration for quantum computation
  with atomic qubits}.
\newblock \emph{\bibinfo{journal}{Phys. Rev. Lett.}}
  \textbf{\bibinfo{volume}{89}}, \bibinfo{pages}{097903}
  (\bibinfo{year}{2002}).

\bibitem{Mucke2010}
\bibinfo{author}{M{\"{u}}cke, M.}, \bibinfo{author}{Figueroa, E.},
  \bibinfo{author}{Bochmann, J.}, \bibinfo{author}{Hahn, C.},
  \bibinfo{author}{Murr, K.}, \bibinfo{author}{Ritter, S.},
  \bibinfo{author}{Villas-Boas, C.~J.} \& \bibinfo{author}{Rempe, G.}
\newblock \bibinfo{title}{Electromagnetically induced transparency with single
  atoms in a cavity}.
\newblock \emph{\bibinfo{journal}{Nature}}
  \bibinfo{pages}{\textbf{\bibinfo{volume}{465}}, 755} (\bibinfo{year}{2010}).

\bibitem{Rodrigo2006}
\bibinfo{author}{Rodrigo, P.~J.}, \bibinfo{author}{Perch-Nielsen, I.~R.},
  \bibinfo{author}{Alonzo, C.~A.} \& \bibinfo{author}{Gl{\"{u}}ckstad, J.}
\newblock \bibinfo{title}{{GPC}-based optical micromanipulation in {3D}
  real-time using a single spatial light modulator}.
\newblock \emph{\bibinfo{journal}{Opt. Express}}
  \bibinfo{pages}{\textbf{\bibinfo{volume}{14}}, 13107} (\bibinfo{year}{2006}).

\bibitem{Ashkin1997}
\bibinfo{author}{Ashkin, A.}
\newblock \bibinfo{title}{Optical trapping and manipulation of neutral
  particles using lasers}.
\newblock \emph{\bibinfo{journal}{Proc. Natl. Acad. Sci.}}
  \bibinfo{pages}{\textbf{\bibinfo{volume}{94}}, 4853} (\bibinfo{year}{1997}).

\bibitem{Barrett2001}
\bibinfo{author}{Barrett, M.~D.}, \bibinfo{author}{Sauer, J.~A.} \&
  \bibinfo{author}{Chapman, M.~S.}
\newblock \bibinfo{title}{All-optical formation of an atomic {B}ose-{E}instein
  condensate}.
\newblock \emph{\bibinfo{journal}{Phys. Rev. Lett.}}
  \textbf{\bibinfo{volume}{87}}, \bibinfo{pages}{010404}
  (\bibinfo{year}{2001}).

\bibitem{Anderson1995}
\bibinfo{author}{Anderson, M.~H.}, \bibinfo{author}{Ensher, J.~R.},
  \bibinfo{author}{Matthews, M.~R.}, \bibinfo{author}{Wieman, C.~E.} \&
  \bibinfo{author}{Cornell, E.~A.}
\newblock \bibinfo{title}{Observation of {B}ose-{E}instein condensation in a
  dilute atomic vapor}.
\newblock \emph{\bibinfo{journal}{Science}}
  \bibinfo{pages}{\textbf{\bibinfo{volume}{269}}, 198} (\bibinfo{year}{1995}).

\bibitem{Davis1995}
\bibinfo{author}{Davis, K.~B.}, \bibinfo{author}{Mewes, M.~O.},
  \bibinfo{author}{Andrews, M.~R.}, \bibinfo{author}{van Druten, N.~J.},
  \bibinfo{author}{Durfee, D.~S.}, \bibinfo{author}{Kurn, D.~M.} \&
  \bibinfo{author}{Ketterle, W.}
\newblock \bibinfo{title}{{B}ose-{E}instein condensation in a gas of sodium
  atoms}.
\newblock \emph{\bibinfo{journal}{Phys. Rev. Lett.}}
  \bibinfo{pages}{\textbf{\bibinfo{volume}{75}}, 3969} (\bibinfo{year}{1995}).

\bibitem{Aspect1988}
\bibinfo{author}{Aspect, A.}, \bibinfo{author}{Arimondo, E.},
  \bibinfo{author}{Kaiser, R.}, \bibinfo{author}{Vansteenkiste, N.} \&
  \bibinfo{author}{Cohen-Tannoudji, C.}
\newblock \bibinfo{title}{Laser cooling below the one-photon recoil energy by
  velocity-selective coherent population trapping}.
\newblock \emph{\bibinfo{journal}{Phys. Rev. Lett.}}
  \bibinfo{pages}{\textbf{\bibinfo{volume}{61}}, 826} (\bibinfo{year}{1988}).

\bibitem{Leanhardt2003}
\bibinfo{author}{Leanhardt, A.~E.}, \bibinfo{author}{Pasquini, T.~A.},
  \bibinfo{author}{Saba, M.}, \bibinfo{author}{Schirotzek, A.},
  \bibinfo{author}{Shin, Y.}, \bibinfo{author}{Kielpinski, D.},
  \bibinfo{author}{Pritchard, D.~E.} \& \bibinfo{author}{Ketterle, W.}
\newblock \bibinfo{title}{Cooling {B}ose-{E}instein condensates below 500
  picokelvin}.
\newblock \emph{\bibinfo{journal}{Science}}
  \bibinfo{pages}{\textbf{\bibinfo{volume}{301}}, 1513} (\bibinfo{year}{2003}).

\bibitem{Braginsky1977}
\bibinfo{author}{Braginsky, V.~B.} \& \bibinfo{author}{Manukin, A.~B.}
\newblock \emph{\bibinfo{title}{Measurement of Weak Forces in Physics
  Experiments}} (\bibinfo{publisher}{University of Chicago},
  \bibinfo{year}{1977}), \bibinfo{edition}{1st} edn.

\bibitem{Aspect1986}
\bibinfo{author}{Aspect, A.}, \bibinfo{author}{Dalibard, J.},
  \bibinfo{author}{Heidmann, A.}, \bibinfo{author}{Salomon, C.} \&
  \bibinfo{author}{Cohen-Tannoudji, C.}
\newblock \bibinfo{title}{Cooling atoms with stimulated emission}.
\newblock \emph{\bibinfo{journal}{Phys. Rev. Lett.}}
  \bibinfo{pages}{\textbf{\bibinfo{volume}{57}}, 1688} (\bibinfo{year}{1986}).

\bibitem{Hechenblaikner1998}
\bibinfo{author}{Hechenblaikner, G.}, \bibinfo{author}{Gangl, M.},
  \bibinfo{author}{Horak, P.} \& \bibinfo{author}{Ritsch, H.}
\newblock \bibinfo{title}{Cooling an atom in a weakly driven high-{Q} cavity}.
\newblock \emph{\bibinfo{journal}{Phys. Rev. A}}
  \bibinfo{pages}{\textbf{\bibinfo{volume}{58}}, 3030} (\bibinfo{year}{1998}).

\bibitem{Teo2010}
\bibinfo{author}{Teo, C.} \& \bibinfo{author}{Scarani, V.}
\newblock \bibinfo{title}{Lenses as an atom-photon interface: A semiclassical
  model}.
\newblock \emph{\bibinfo{journal}{arXiv e-prints}}  (\bibinfo{year}{2010}).
\newblock \eprint{arXiv:1012.0630}.

\bibitem{Purcell1946}
\bibinfo{author}{Purcell, E.~M.}
\newblock \bibinfo{title}{Spontaneous emission probabilities at radio
  frequencies}.
\newblock \emph{\bibinfo{journal}{Phys. Rev.}} \textbf{\bibinfo{volume}{69}},
  \bibinfo{pages}{681} (\bibinfo{year}{1946}).

\bibitem{Cirac1993}
\bibinfo{author}{Cirac, J.~I.}, \bibinfo{author}{Parkins, A.~S.},
  \bibinfo{author}{Blatt, R.} \& \bibinfo{author}{Zoller, P.}
\newblock \bibinfo{title}{Cooling of a trapped ion coupled strongly to a
  quantized cavity mode}.
\newblock \emph{\bibinfo{journal}{Opt. Commun.}}
  \bibinfo{pages}{\textbf{\bibinfo{volume}{97}}, 353} (\bibinfo{year}{1993}).

\bibitem{Barker2010}
\bibinfo{author}{Barker, P.~F.} \& \bibinfo{author}{Shneider, M.~N.}
\newblock \bibinfo{title}{Cavity cooling of an optically trapped nanoparticle}.
\newblock \emph{\bibinfo{journal}{Phys. Rev. A}} \textbf{\bibinfo{volume}{81}},
  \bibinfo{pages}{023826} (\bibinfo{year}{2010}).

\bibitem{RomeroIsart2011}
\bibinfo{author}{Romero-Isart, O.}, \bibinfo{author}{Pflanzer, A.~C.},
  \bibinfo{author}{Juan, M.~L.}, \bibinfo{author}{Quidant, R.},
  \bibinfo{author}{Kiesel, N.}, \bibinfo{author}{Aspelmeyer, M.} \&
  \bibinfo{author}{Cirac, J.~I.}
\newblock \bibinfo{title}{Optically levitating dielectrics in the quantum
  regime: Theory and protocols}.
\newblock \emph{\bibinfo{journal}{Phys. Rev. A}} \textbf{\bibinfo{volume}{83}},
  \bibinfo{pages}{013803} (\bibinfo{year}{2011}).

\bibitem{Kippenberg2008}
\bibinfo{author}{Kippenberg, T.~J.} \& \bibinfo{author}{Vahala, K.~J.}
\newblock \bibinfo{title}{{Cavity Optomechanics: Back-Action at the
  Mesoscale}}.
\newblock \emph{\bibinfo{journal}{Science}}
  \bibinfo{pages}{\textbf{\bibinfo{volume}{321}}, 1172} (\bibinfo{year}{2008}).

\bibitem{Aspelmeyer2010}
\bibinfo{author}{Aspelmeyer, M.}, \bibinfo{author}{Gr{\"{o}}blacher, S.},
  \bibinfo{author}{Hammerer, K.} \& \bibinfo{author}{Kiesel, N.}
\newblock \bibinfo{title}{Quantum optomechanics---throwing a glance}.
\newblock \emph{\bibinfo{journal}{J. Opt. Soc. Am. B}}
  \bibinfo{pages}{\textbf{\bibinfo{volume}{27}}, A189} (\bibinfo{year}{2010}).

\bibitem{Gangl2000a}
\bibinfo{author}{Gangl, M.} \& \bibinfo{author}{Ritsch, H.}
\newblock \bibinfo{title}{Cold atoms in a high-${Q}$ ring cavity}.
\newblock \emph{\bibinfo{journal}{Phys. Rev. A}} \textbf{\bibinfo{volume}{61}},
  \bibinfo{pages}{043405} (\bibinfo{year}{2000}).

\bibitem{Elsasser2003}
\bibinfo{author}{Els{\"{a}}sser, T.}, \bibinfo{author}{Nagorny, B.} \&
  \bibinfo{author}{Hemmerich, A.}
\newblock \bibinfo{title}{Collective sideband cooling in an optical ring
  cavity}.
\newblock \emph{\bibinfo{journal}{Phys. Rev. A}} \textbf{\bibinfo{volume}{67}},
  \bibinfo{pages}{051401} (\bibinfo{year}{2003}).

\bibitem{Nagy2006}
\bibinfo{author}{Nagy, D.}, \bibinfo{author}{Asb{\'{o}}th, J.~K.} \&
  \bibinfo{author}{Domokos, P.}
\newblock \bibinfo{title}{Collective cooling of atoms in a ring cavity}.
\newblock \emph{\bibinfo{journal}{Acta Physica Hungarica B}}
  \bibinfo{pages}{\textbf{\bibinfo{volume}{26}}, 141} (\bibinfo{year}{2006}).

\bibitem{Hemmerling2010}
\bibinfo{author}{Hemmerling, M.} \& \bibinfo{author}{Robb, G. R.~M.}
\newblock \bibinfo{title}{Slowing atoms using optical cavities pumped by
  phase-modulated light}.
\newblock \emph{\bibinfo{journal}{Phys. Rev. A}} \textbf{\bibinfo{volume}{82}},
  \bibinfo{pages}{053420} (\bibinfo{year}{2010}).

\bibitem{Schulze2010}
\bibinfo{author}{Schulze, R.~J.}, \bibinfo{author}{Genes, C.} \&
  \bibinfo{author}{Ritsch, H.}
\newblock \bibinfo{title}{Optomechanical approach to cooling of small
  polarizable particles in a strongly pumped ring cavity}.
\newblock \emph{\bibinfo{journal}{Phys. Rev. A}} \textbf{\bibinfo{volume}{81}},
  \bibinfo{pages}{063820} (\bibinfo{year}{2010}).

\bibitem{Niedenzu2010}
\bibinfo{author}{Niedenzu, W.}, \bibinfo{author}{Schulze, R.},
  \bibinfo{author}{Vukics, A.} \& \bibinfo{author}{Ritsch, H.}
\newblock \bibinfo{title}{Microscopic dynamics of ultracold particles in a
  ring-cavity optical lattice}.
\newblock \emph{\bibinfo{journal}{Phys. Rev. A}} \textbf{\bibinfo{volume}{82}},
  \bibinfo{pages}{043605} (\bibinfo{year}{2010}).

\bibitem{Kruse2003}
\bibinfo{author}{Kruse, D.}, \bibinfo{author}{Ruder, M.},
  \bibinfo{author}{Benhelm, J.}, \bibinfo{author}{von Cube, C.},
  \bibinfo{author}{Zimmermann, C.}, \bibinfo{author}{Courteille, P.~W.},
  \bibinfo{author}{Els{\"{a}}sser, T.}, \bibinfo{author}{Nagorny, B.} \&
  \bibinfo{author}{Hemmerich, A.}
\newblock \bibinfo{title}{Cold atoms in a high-${Q}$ ring cavity}.
\newblock \emph{\bibinfo{journal}{Phys. Rev. A}} \textbf{\bibinfo{volume}{67}},
  \bibinfo{pages}{051802} (\bibinfo{year}{2003}).

\bibitem{Slama2007}
\bibinfo{author}{Slama, S.}, \bibinfo{author}{Bux, S.}, \bibinfo{author}{Krenz,
  G.}, \bibinfo{author}{Zimmermann, C.} \& \bibinfo{author}{Courteille, P.~W.}
\newblock \bibinfo{title}{Superradiant rayleigh scattering and collective
  atomic recoil lasing in a ring cavity}.
\newblock \emph{\bibinfo{journal}{Phys. Rev. Lett.}}
  \textbf{\bibinfo{volume}{98}}, \bibinfo{pages}{053603}
  (\bibinfo{year}{2007}).

\bibitem{Vuletic2001a}
\bibinfo{author}{Vuleti{\'{c}}, V.}
\newblock \emph{\bibinfo{title}{{L}aser {P}hysics at the
  {L}imits}}\bibinfo{pages}{, chap. \bibinfo{chapter}{Cavity Cooling with a Hot
  Cavity}, 305} (\bibinfo{publisher}{Springer}, \bibinfo{year}{2001}).

\bibitem{Salzburger2006}
\bibinfo{author}{Salzburger, T.} \& \bibinfo{author}{Ritsch, H.}
\newblock \bibinfo{title}{Lasing and cooling in a finite-temperature cavity}.
\newblock \emph{\bibinfo{journal}{Phys. Rev. A}} \textbf{\bibinfo{volume}{74}},
  \bibinfo{pages}{033806} (\bibinfo{year}{2006}).

\bibitem{Domokos2002b}
\bibinfo{author}{Domokos, P.} \& \bibinfo{author}{Ritsch, H.}
\newblock \bibinfo{title}{Collective cooling and self-organization of atoms in
  a cavity}.
\newblock \emph{\bibinfo{journal}{Phys. Rev. Lett.}}
  \textbf{\bibinfo{volume}{89}}, \bibinfo{pages}{253003}
  (\bibinfo{year}{2002}).

\bibitem{Baumann2010}
\bibinfo{author}{Baumann, K.}, \bibinfo{author}{Guerlin, C.},
  \bibinfo{author}{Brennecke, F.} \& \bibinfo{author}{Esslinger, T.}
\newblock \bibinfo{title}{Dicke quantum phase transition with a superfluid gas
  in an optical cavity}.
\newblock \emph{\bibinfo{journal}{Nature}}
  \bibinfo{pages}{\textbf{\bibinfo{volume}{464}}, 1301} (\bibinfo{year}{2010}).

\bibitem{Bonifacio1994}
\bibinfo{author}{Bonifacio, R.}, \bibinfo{author}{De~Salvo, L.},
  \bibinfo{author}{Narducci, L.~M.} \& \bibinfo{author}{D'Angelo, E.~J.}
\newblock \bibinfo{title}{Exponential gain and self-bunching in a collective
  atomic recoil laser}.
\newblock \emph{\bibinfo{journal}{Phys. Rev. A}}
  \bibinfo{pages}{\textbf{\bibinfo{volume}{50}}, 1716} (\bibinfo{year}{1994}).

\bibitem{Zimmermann2004}
\bibinfo{author}{Zimmermann, C.}, \bibinfo{author}{Kruse, D.},
  \bibinfo{author}{Cube, C.~V.}, \bibinfo{author}{Slama, S.},
  \bibinfo{author}{Deh, B.} \& \bibinfo{author}{Courteille, P.}
\newblock \bibinfo{title}{Collective atomic recoil lasing}.
\newblock \emph{\bibinfo{journal}{J. Mod. Opt.}}
  \bibinfo{pages}{\textbf{\bibinfo{volume}{51}}, 957} (\bibinfo{year}{2004}).

\bibitem{Horak2010a}
\bibinfo{author}{Horak, P.}, \bibinfo{author}{Xuereb, A.} \&
  \bibinfo{author}{Freegarde, T.}
\newblock \bibinfo{title}{Optical cooling of atoms in microtraps by
  time-delayed reflection}.
\newblock \emph{\bibinfo{journal}{J. Comput. Theor. Nanosci.}}
  \bibinfo{pages}{\textbf{\bibinfo{volume}{7}}, 1747} (\bibinfo{year}{September
  2010}).

\bibitem{Kurtsiefer2009}
\bibinfo{author}{Aljunid, S.~A.}, \bibinfo{author}{Tey, M.~K.},
  \bibinfo{author}{Chng, B.}, \bibinfo{author}{Chen, Z.}, \bibinfo{author}{Lee,
  J.}, \bibinfo{author}{Liew, T.}, \bibinfo{author}{Maslennikov, G.},
  \bibinfo{author}{Scarani, V.} \& \bibinfo{author}{Kurtsiefer, C.}
\newblock \bibinfo{title}{Substantial scattering of a weak coherent beam by a
  single atom}.
\newblock In \emph{\bibinfo{booktitle}{2009 Conference on Lasers and
  Electro-Optics and the XIth European Quantum Electronics Conference
  (CLEO\textregistered/Europe-EQEC 2009), Munich, Germany}},
  \bibinfo{pages}{89} (\bibinfo{publisher}{IEEE}, \bibinfo{year}{2009}).

\bibitem{Gradshteyn1994}
\bibinfo{author}{Gradshteyn, I.~S.} \& \bibinfo{author}{Ryzhik, I.~M.}
\newblock \emph{\bibinfo{title}{Table of integrals, series and products}}
  (\bibinfo{publisher}{Academic Press}, \bibinfo{year}{1994}),
  \bibinfo{edition}{fifth} edn.

\bibitem{Cook1980}
\bibinfo{author}{Cook, R.~J.}
\newblock \bibinfo{title}{Theory of resonance-radiation pressure}.
\newblock \emph{\bibinfo{journal}{Phys. Rev. A}}
  \bibinfo{pages}{\textbf{\bibinfo{volume}{22}}, 1078} (\bibinfo{year}{1980}).

\bibitem{BergSorensen1992}
\bibinfo{author}{Berg-S{\o}rensen, K.}, \bibinfo{author}{Castin, Y.},
  \bibinfo{author}{Bonderup, E.} \& \bibinfo{author}{M{\o}lmer, K.}
\newblock \bibinfo{title}{Momentum diffusion of atoms moving in laser fields}.
\newblock \emph{\bibinfo{journal}{J. Phys. B}}
  \bibinfo{pages}{\textbf{\bibinfo{volume}{25}}, 4195} (\bibinfo{year}{1992}).

\bibitem{Horak2001}
\bibinfo{author}{Horak, P.} \& \bibinfo{author}{Ritsch, H.}
\newblock \bibinfo{title}{Scaling properties of cavity-enhanced atom cooling}.
\newblock \emph{\bibinfo{journal}{Phys. Rev. A}} \textbf{\bibinfo{volume}{64}},
  \bibinfo{pages}{033422} (\bibinfo{year}{2001}).

\bibitem{Eschner2001}
\bibinfo{author}{Eschner, J.}, \bibinfo{author}{Raab, C.},
  \bibinfo{author}{Schmidt-Kaler, F.} \& \bibinfo{author}{Blatt, R.}
\newblock \bibinfo{title}{Light interference from single atoms and their mirror
  images}.
\newblock \emph{\bibinfo{journal}{Nature}}
  \bibinfo{pages}{\textbf{\bibinfo{volume}{413}}, 495} (\bibinfo{year}{2001}).

\bibitem{Vuletic2000}
\bibinfo{author}{Vuleti{\'{c}}, V.} \& \bibinfo{author}{Chu, S.}
\newblock \bibinfo{title}{Laser cooling of atoms, ions, or molecules by
  coherent scattering}.
\newblock \emph{\bibinfo{journal}{Phys. Rev. Lett.}}
  \bibinfo{pages}{\textbf{\bibinfo{volume}{84}}, 3787} (\bibinfo{year}{2000}).

\bibitem{Maunz2004}
\bibinfo{author}{Maunz, P.}, \bibinfo{author}{Puppe, T.},
  \bibinfo{author}{Schuster, I.}, \bibinfo{author}{Syassen, N.},
  \bibinfo{author}{Pinkse, P. W.~H.} \& \bibinfo{author}{Rempe, G.}
\newblock \bibinfo{title}{Cavity cooling of a single atom}.
\newblock \emph{\bibinfo{journal}{Nature}}
  \bibinfo{pages}{\textbf{\bibinfo{volume}{428}}, 50} (\bibinfo{year}{2004}).

\bibitem{Vilensky2007}
\bibinfo{author}{Vilensky, M.~Y.}, \bibinfo{author}{Prior, Y.} \&
  \bibinfo{author}{Averbukh, I.~S.}
\newblock \bibinfo{title}{Cooling in a bistable optical cavity}.
\newblock \emph{\bibinfo{journal}{Phys. Rev. Lett.}}
  \textbf{\bibinfo{volume}{99}}, \bibinfo{pages}{103002}
  (\bibinfo{year}{2007}).

\bibitem{Lev2008}
\bibinfo{author}{Lev, B.~L.}, \bibinfo{author}{Vukics, A.},
  \bibinfo{author}{Hudson, E.~R.}, \bibinfo{author}{Sawyer, B.~C.},
  \bibinfo{author}{Domokos, P.}, \bibinfo{author}{Ritsch, H.} \&
  \bibinfo{author}{Ye, J.}
\newblock \bibinfo{title}{Prospects for the cavity-assisted laser cooling of
  molecules}.
\newblock \emph{\bibinfo{journal}{Phys. Rev. A}} \textbf{\bibinfo{volume}{77}},
  \bibinfo{pages}{023402} (\bibinfo{year}{2008}).

\bibitem{Rempe1992}
\bibinfo{author}{Rempe, G.}, \bibinfo{author}{Thompson, R.~J.},
  \bibinfo{author}{Kimble, H.~J.} \& \bibinfo{author}{Lalezari, R.}
\newblock \bibinfo{title}{Measurement of ultralow losses in an optical
  interferometer}.
\newblock \emph{\bibinfo{journal}{Opt. Lett.}}
  \bibinfo{pages}{\textbf{\bibinfo{volume}{17}}, 363} (\bibinfo{year}{1992}).

\bibitem{Mabuchi1994}
\bibinfo{author}{Mabuchi, H.} \& \bibinfo{author}{Kimble, H.~J.}
\newblock \bibinfo{title}{Atom galleries for whispering atoms: binding atoms in
  stable orbits around an optical resonator}.
\newblock \emph{\bibinfo{journal}{Opt. Lett.}}
  \bibinfo{pages}{\textbf{\bibinfo{volume}{19}}, 749} (\bibinfo{year}{1994}).

\bibitem{Steck2010b}
\bibinfo{author}{Steck, D.~A.}
\newblock \bibinfo{title}{{Sodium D Line Data}} (\bibinfo{year}{2010}).
\newblock \urlprefix\url{http://steck.us/alkalidata/sodiumnumbers.pdf}.

\bibitem{Steck2010a}
\bibinfo{author}{Steck, D.~A.}
\newblock \bibinfo{title}{{Rubidium 87 D Line Data}} (\bibinfo{year}{2010}).
\newblock \urlprefix\url{http://steck.us/alkalidata/rubidium87numbers.pdf}.

\bibitem{Steck2010c}
\bibinfo{author}{Steck, D.~A.}
\newblock \bibinfo{title}{{Cesium D Line Data}} (\bibinfo{year}{2010}).
\newblock \urlprefix\url{http://steck.us/alkalidata/cesiumnumbers.pdf}.

\bibitem{AttocubeSystemsPositioning2010}
\bibinfo{organization}{attocube systems AG}.
\newblock \emph{\bibinfo{title}{ANSxy50 Data Sheet}} (\bibinfo{year}{2010}).

\bibitem{Thompson2008}
\bibinfo{author}{Thompson, J.~D.}, \bibinfo{author}{Zwickl, B.~M.},
  \bibinfo{author}{Jayich, A.~M.}, \bibinfo{author}{Marquardt, F.},
  \bibinfo{author}{Girvin, S.~M.} \& \bibinfo{author}{Harris, J. G.~E.}
\newblock \bibinfo{title}{Strong dispersive coupling of a high-finesse cavity
  to a micromechanical membrane}.
\newblock \emph{\bibinfo{journal}{Nature}}
  \bibinfo{pages}{\textbf{\bibinfo{volume}{452}}, 72} (\bibinfo{year}{2008}).

\bibitem{Karasek2006}
\bibinfo{author}{Kar{\'{a}}sek, V.}, \bibinfo{author}{Dholakia, K.} \&
  \bibinfo{author}{Zem{\'{a}}nek, P.}
\newblock \bibinfo{title}{Analysis of optical binding in one dimension}.
\newblock \emph{\bibinfo{journal}{Appl. Phys. B}}
  \bibinfo{pages}{\textbf{\bibinfo{volume}{84}}, 149} (\bibinfo{year}{2006}).

\bibitem{Asboth2008}
\bibinfo{author}{Asb{\'{o}}th, J.~K.}, \bibinfo{author}{Ritsch, H.} \&
  \bibinfo{author}{Domokos, P.}
\newblock \bibinfo{title}{Optomechanical coupling in a one-dimensional optical
  lattice}.
\newblock \emph{\bibinfo{journal}{Phys. Rev. A}} \textbf{\bibinfo{volume}{77}},
  \bibinfo{pages}{063424} (\bibinfo{year}{2008}).

\bibitem{WilsonRae2007}
\bibinfo{author}{Wilson-Rae, I.}, \bibinfo{author}{Nooshi, N.},
  \bibinfo{author}{Zwerger, W.} \& \bibinfo{author}{Kippenberg, T.~J.}
\newblock \bibinfo{title}{Theory of ground state cooling of a mechanical
  oscillator using dynamical backaction}.
\newblock \emph{\bibinfo{journal}{Phys. Rev. Lett.}}
  \textbf{\bibinfo{volume}{99}}, \bibinfo{pages}{093901}
  (\bibinfo{year}{2007}).

\bibitem{Marquardt2007}
\bibinfo{author}{Marquardt, F.}, \bibinfo{author}{Chen, J.~P.},
  \bibinfo{author}{Clerk, A.~A.} \& \bibinfo{author}{Girvin, S.~M.}
\newblock \bibinfo{title}{Quantum theory of cavity-assisted sideband cooling of
  mechanical motion}.
\newblock \emph{\bibinfo{journal}{Phys. Rev. Lett.}}
  \textbf{\bibinfo{volume}{99}} (\bibinfo{year}{2007}).

\bibitem{Genes2008}
\bibinfo{author}{Genes, C.}, \bibinfo{author}{Vitali, D.},
  \bibinfo{author}{Tombesi, P.}, \bibinfo{author}{Gigan, S.} \&
  \bibinfo{author}{Aspelmeyer, M.}
\newblock \bibinfo{title}{Ground-state cooling of a micromechanical oscillator:
  Comparing cold damping and cavity-assisted cooling schemes}.
\newblock \emph{\bibinfo{journal}{Phys. Rev. A}} \textbf{\bibinfo{volume}{77}},
  \bibinfo{pages}{033804} (\bibinfo{year}{2008}).

\bibitem{Bushev2004}
\bibinfo{author}{Bushev, P.}, \bibinfo{author}{Wilson, A.},
  \bibinfo{author}{Eschner, J.}, \bibinfo{author}{Raab, C.},
  \bibinfo{author}{Schmidt-Kaler, F.}, \bibinfo{author}{Becher, C.} \&
  \bibinfo{author}{Blatt, R.}
\newblock \bibinfo{title}{Forces between a single atom and its distant mirror
  image}.
\newblock \emph{\bibinfo{journal}{Phys. Rev. Lett.}}
  \textbf{\bibinfo{volume}{92}}, \bibinfo{pages}{223602}
  (\bibinfo{year}{2004}).

\bibitem{Castin1990}
\bibinfo{author}{Castin, Y.} \& \bibinfo{author}{M{\o}lmer, K.}
\newblock \bibinfo{title}{Atomic momentum diffusion in a $\sigma_+$--$\sigma_-$
  laser configuration: influence of an internal sublevel structure}.
\newblock \emph{\bibinfo{journal}{J. Phys. B}}
  \bibinfo{pages}{\textbf{\bibinfo{volume}{23}}, 4101} (\bibinfo{year}{1990}).

\bibitem{Glauber1963}
\bibinfo{author}{Glauber, R.~J.}
\newblock \bibinfo{title}{Coherent and incoherent states of the radiation
  field}.
\newblock \emph{\bibinfo{journal}{Phys. Rev.}}
  \bibinfo{pages}{\textbf{\bibinfo{volume}{131}}, 2766} (\bibinfo{year}{1963}).

\bibitem{Bhattacharya2008}
\bibinfo{author}{Bhattacharya, M.}, \bibinfo{author}{Uys, H.} \&
  \bibinfo{author}{Meystre, P.}
\newblock \bibinfo{title}{Optomechanical trapping and cooling of partially
  reflective mirrors}.
\newblock \emph{\bibinfo{journal}{Phys. Rev. A}} \textbf{\bibinfo{volume}{77}},
  \bibinfo{pages}{033819} (\bibinfo{year}{2008}).

\bibitem{Metcalf2003}
\bibinfo{author}{Metcalf, H.~J.} \& \bibinfo{author}{van~der Straten, P.}
\newblock \bibinfo{title}{Laser cooling and trapping of atoms}.
\newblock \emph{\bibinfo{journal}{J. Opt. Soc. Am. B}}
  \bibinfo{pages}{\textbf{\bibinfo{volume}{20}}, 887} (\bibinfo{year}{2003}).

\bibitem{Saulson1990}
\bibinfo{author}{Saulson, P.~R.}
\newblock \bibinfo{title}{Thermal noise in mechanical experiments}.
\newblock \emph{\bibinfo{journal}{Phys. Rev. D}}
  \bibinfo{pages}{\textbf{\bibinfo{volume}{42}}, 2437} (\bibinfo{year}{1990}).

\bibitem{Courty2001}
\bibinfo{author}{Courty, J.~M.}, \bibinfo{author}{Heidmann, A.} \&
  \bibinfo{author}{Pinard, M.}
\newblock \bibinfo{title}{Quantum limits of cold damping with optomechanical
  coupling}.
\newblock \emph{\bibinfo{journal}{Eur. Phys. J. D}}
  \bibinfo{pages}{\textbf{\bibinfo{volume}{17}}, 399} (\bibinfo{year}{2001}).

\bibitem{Vitali2003}
\bibinfo{author}{Vitali, D.}, \bibinfo{author}{Mancini, S.},
  \bibinfo{author}{Ribichini, L.} \& \bibinfo{author}{Tombesi, P.}
\newblock \bibinfo{title}{Macroscopic mechanical oscillators at the quantum
  limit through optomechanical coupling}.
\newblock \emph{\bibinfo{journal}{J. Opt. Soc. Am. B}}
  \bibinfo{pages}{\textbf{\bibinfo{volume}{20}}, 1054} (\bibinfo{year}{2003}).

\bibitem{Xuereb2010b}
\bibinfo{author}{Xuereb, A.}, \bibinfo{author}{Freegarde, T.},
  \bibinfo{author}{Horak, P.} \& \bibinfo{author}{Domokos, P.}
\newblock \bibinfo{title}{Optomechanical cooling with generalized
  interferometers}.
\newblock \emph{\bibinfo{journal}{Phys. Rev. Lett.}}
  \textbf{\bibinfo{volume}{105}}, \bibinfo{pages}{013602}
  (\bibinfo{year}{2010}).

\bibitem{Jayich2008}
\bibinfo{author}{Jayich, A.~M.}, \bibinfo{author}{Sankey, J.~C.},
  \bibinfo{author}{Zwickl, B.~M.}, \bibinfo{author}{Yang, C.},
  \bibinfo{author}{Thompson, J.~D.}, \bibinfo{author}{Girvin, S.~M.},
  \bibinfo{author}{Clerk, A.~A.}, \bibinfo{author}{Marquardt, F.} \&
  \bibinfo{author}{Harris, J. G.~E.}
\newblock \bibinfo{title}{Dispersive optomechanics: a membrane inside a
  cavity}.
\newblock \emph{\bibinfo{journal}{New J. Phys.}} \textbf{\bibinfo{volume}{10}},
  \bibinfo{pages}{095008} (\bibinfo{year}{2008}).

\bibitem{Bhattacharya2007a}
\bibinfo{author}{Bhattacharya, M.} \& \bibinfo{author}{Meystre, P.}
\newblock \bibinfo{title}{Trapping and cooling a mirror to its quantum
  mechanical ground state}.
\newblock \emph{\bibinfo{journal}{Phys. Rev. Lett.}}
  \textbf{\bibinfo{volume}{99}}, \bibinfo{pages}{073601}
  (\bibinfo{year}{2007}).

\bibitem{Sankey2010}
\bibinfo{author}{Sankey, J.~C.}, \bibinfo{author}{Yang, C.},
  \bibinfo{author}{Zwickl, B.~M.}, \bibinfo{author}{Jayich, A.~M.} \&
  \bibinfo{author}{Harris, J. G.~E.}
\newblock \bibinfo{title}{Strong and tunable nonlinear optomechanical coupling
  in a low-loss system}.
\newblock \emph{\bibinfo{journal}{Nat. Phys.}}
  \bibinfo{pages}{\textbf{\bibinfo{volume}{6}}, 707} (\bibinfo{year}{2010}).

\bibitem{Braginsky2002}
\bibinfo{author}{Braginsky, V.}
\newblock \bibinfo{title}{Low quantum noise tranquilizer for {F}abry--{P}erot
  interferometer}.
\newblock \emph{\bibinfo{journal}{Phys. Lett. A}}
  \bibinfo{pages}{\textbf{\bibinfo{volume}{293}}, 228} (\bibinfo{year}{2002}).

\bibitem{Nunnenkamp2010}
\bibinfo{author}{Nunnenkamp, A.}, \bibinfo{author}{B{\o}rkje, K.},
  \bibinfo{author}{Harris, J. G.~E.} \& \bibinfo{author}{Girvin, S.~M.}
\newblock \bibinfo{title}{Cooling and squeezing via quadratic optomechanical
  coupling}.
\newblock \emph{\bibinfo{journal}{Phys. Rev. A}} \textbf{\bibinfo{volume}{82}},
  \bibinfo{pages}{021806} (\bibinfo{year}{2010}).

\bibitem{Clerk2010b}
\bibinfo{author}{Clerk, A.~A.}, \bibinfo{author}{Marquardt, F.} \&
  \bibinfo{author}{Harris, J. G.~E.}
\newblock \bibinfo{title}{Quantum measurement of phonon shot noise}.
\newblock \emph{\bibinfo{journal}{Phys. Rev. Lett.}}
  \textbf{\bibinfo{volume}{104}}, \bibinfo{pages}{213603}
  (\bibinfo{year}{2010}).

\bibitem{Xuereb2010a}
\bibinfo{author}{Xuereb, A.}, \bibinfo{author}{Domokos, P.},
  \bibinfo{author}{Horak, P.} \& \bibinfo{author}{Freegarde, T.}
\newblock \bibinfo{title}{Scattering theory of multilevel atoms interacting
  with arbitrary radiation fields}.
\newblock \emph{\bibinfo{journal}{Phys. Scr.}} \textbf{\bibinfo{volume}{2010}},
  \bibinfo{pages}{014010} (\bibinfo{year}{2010}).

\bibitem{Spreeuw1992}
\bibinfo{author}{Spreeuw, R. J.~C.}, \bibinfo{author}{Beijersbergen, M.~W.} \&
  \bibinfo{author}{Woerdman, J.~P.}
\newblock \bibinfo{title}{Optical ring cavities as tailored four-level systems:
  An application of the group {U}(2,2)}.
\newblock \emph{\bibinfo{journal}{Phys. Rev. A}}
  \bibinfo{pages}{\textbf{\bibinfo{volume}{45}}, 1213} (\bibinfo{year}{1992}).

\bibitem{CohenTannoudji1977b}
\bibinfo{author}{Cohen-Tannoudji, C.}
\newblock \bibinfo{title}{Atoms in strong resonant fields}.
\newblock \bibinfo{pages}{In \bibinfo{editor}{Balian, R.},
  \bibinfo{editor}{Haroche, S.} \& \bibinfo{editor}{Liberman, S.} (eds.)
  \emph{\bibinfo{booktitle}{Frontiers in laser spectroscopy, Proceedings of the
  Les Houches Summer School, Session XXVII}}1} (\bibinfo{publisher}{North
  Holland}, \bibinfo{year}{1977}).

\bibitem{Siegman1990}
\bibinfo{author}{Siegman, A.~E.}
\newblock \emph{\bibinfo{title}{Lasers}} (\bibinfo{publisher}{University
  Science Books}, \bibinfo{address}{Sausalito, CA}, \bibinfo{year}{1990}).

\bibitem{Metzger2004}
\bibinfo{author}{Metzger, C.~H.} \& \bibinfo{author}{Karrai, K.}
\newblock \bibinfo{title}{Cavity cooling of a microlever}.
\newblock \emph{\bibinfo{journal}{Nature}}
  \bibinfo{pages}{\textbf{\bibinfo{volume}{432}}, 1002} (\bibinfo{year}{2004}).

\bibitem{Gigan2006}
\bibinfo{author}{Gigan, S.}, \bibinfo{author}{Bohm, H.~R.},
  \bibinfo{author}{Paternostro, M.}, \bibinfo{author}{Blaser, F.},
  \bibinfo{author}{Langer, G.}, \bibinfo{author}{Hertzberg, J.~B.},
  \bibinfo{author}{Schwab, K.~C.}, \bibinfo{author}{Bauerle, D.},
  \bibinfo{author}{Aspelmeyer, M.} \& \bibinfo{author}{Zeilinger, A.}
\newblock \bibinfo{title}{Self-cooling of a micromirror by radiation pressure}.
\newblock \emph{\bibinfo{journal}{Nature}}
  \bibinfo{pages}{\textbf{\bibinfo{volume}{444}}, 67} (\bibinfo{year}{2006}).

\bibitem{Schliesser2008}
\bibinfo{author}{Schliesser, A.}, \bibinfo{author}{Rivi{\`{e}}re, R.},
  \bibinfo{author}{Anetsberger, G.}, \bibinfo{author}{Arcizet, O.} \&
  \bibinfo{author}{Kippenberg, T.~J.}
\newblock \bibinfo{title}{Resolved-sideband cooling of a micromechanical
  oscillator}.
\newblock \emph{\bibinfo{journal}{Nat. Phys.}}
  \bibinfo{pages}{\textbf{\bibinfo{volume}{4}}, 415} (\bibinfo{year}{2008}).

\bibitem{Favero2008}
\bibinfo{author}{Favero, I.} \& \bibinfo{author}{Karrai, K.}
\newblock \bibinfo{title}{Cavity cooling of a nanomechanical resonator by light
  scattering}.
\newblock \emph{\bibinfo{journal}{New J. Phys.}} \textbf{\bibinfo{volume}{10}},
  \bibinfo{pages}{095006} (\bibinfo{year}{2008}).

\bibitem{Domokos2003}
\bibinfo{author}{Domokos, P.} \& \bibinfo{author}{Ritsch, H.}
\newblock \bibinfo{title}{Mechanical effects of light in optical resonators}.
\newblock \emph{\bibinfo{journal}{J. Opt. Soc. Am. B}}
  \bibinfo{pages}{\textbf{\bibinfo{volume}{20}}, 1098} (\bibinfo{year}{2003}).

\bibitem{Schliesser2010}
\bibinfo{author}{Schliesser, A.} \& \bibinfo{author}{Kippenberg, T.~J.}
\newblock \bibinfo{title}{Cavity optomechanics with whispering-gallery-mode
  optical micro-resonators}.
\newblock \emph{\bibinfo{journal}{arXiv e-prints}}  (\bibinfo{year}{2010}).
\newblock \eprint{arXiv:1003.5922}.

\bibitem{Huang2009}
\bibinfo{author}{Huang, S.} \& \bibinfo{author}{Agarwal, G.~S.}
\newblock \bibinfo{title}{Enhancement of cavity cooling of a micromechanical
  mirror using parametric interactions}.
\newblock \emph{\bibinfo{journal}{Phys. Rev. A}} \textbf{\bibinfo{volume}{79}},
  \bibinfo{pages}{013821} (\bibinfo{year}{2009}).

\bibitem{Kumar2010}
\bibinfo{author}{Kumar, T.}, \bibinfo{author}{Bhattacherjee, A.~B.} \&
  \bibinfo{author}{Man{M}ohan}.
\newblock \bibinfo{title}{Dynamics of a movable micromirror in a nonlinear
  optical cavity}.
\newblock \emph{\bibinfo{journal}{Phys. Rev. A}} \textbf{\bibinfo{volume}{81}},
  \bibinfo{pages}{013835} (\bibinfo{year}{2010}).

\bibitem{Depasse1994}
\bibinfo{author}{Dapasse, F.} \& \bibinfo{author}{Vigoureux, J.-M.}
\newblock \bibinfo{title}{{Optical binding force between two Rayleigh
  particles}}.
\newblock \emph{\bibinfo{journal}{J. Phys. D}}
  \bibinfo{pages}{\textbf{\bibinfo{volume}{27}}, 914} (\bibinfo{year}{1994}).

\bibitem{Levine1950}
\bibinfo{author}{Levine, H.} \& \bibinfo{author}{Schwinger, J.}
\newblock \bibinfo{title}{On the theory of electromagnetic wave diffraction by
  an aperture in an infinite plane conducting screen}.
\newblock \emph{\bibinfo{journal}{Comm. Pure Appl. Math.}}
  \bibinfo{pages}{\textbf{\bibinfo{volume}{3}}, 355} (\bibinfo{year}{1950}).

\bibitem{Martin1995}
\bibinfo{author}{Martin, O. J.~F.}, \bibinfo{author}{Girard, C.} \&
  \bibinfo{author}{Dereux, A.}
\newblock \bibinfo{title}{Generalized field propagator for electromagnetic
  scattering and light confinement}.
\newblock \emph{\bibinfo{journal}{Phys. Rev. Lett.}}
  \bibinfo{pages}{\textbf{\bibinfo{volume}{74}}, 526} (\bibinfo{year}{1995}).

\bibitem{Burns1989}
\bibinfo{author}{Burns, M.~M.}, \bibinfo{author}{Fournier, J.~M.} \&
  \bibinfo{author}{Golovchenko, J.~A.}
\newblock \bibinfo{title}{Optical binding}.
\newblock \emph{\bibinfo{journal}{Phys. Rev. Lett.}}
  \bibinfo{pages}{\textbf{\bibinfo{volume}{63}}, 1233} (\bibinfo{year}{1989}).

\bibitem{Metzger2006a}
\bibinfo{author}{Metzger, N.~K.}, \bibinfo{author}{Wright, E.~M.},
  \bibinfo{author}{Sibbett, W.} \& \bibinfo{author}{Dholakia, K.}
\newblock \bibinfo{title}{Visualization of optical binding of microparticles
  using a femtosecond fiber optical trap}.
\newblock \emph{\bibinfo{journal}{Opt. Express}}
  \bibinfo{pages}{\textbf{\bibinfo{volume}{14}}, 3677} (\bibinfo{year}{2006}).

\bibitem{Clifford2001}
\bibinfo{author}{Clifford, M.~A.}, \bibinfo{author}{Lancaster, G. P.~T.},
  \bibinfo{author}{Mitchell, R.~H.}, \bibinfo{author}{Akerboom, F.} \&
  \bibinfo{author}{Dholakia, K.}
\newblock \bibinfo{title}{Realization of a mirror magneto-optical trap}.
\newblock \emph{\bibinfo{journal}{J. Mod. Opt.}}
  \bibinfo{pages}{\textbf{\bibinfo{volume}{48}}, 1123} (\bibinfo{year}{2001}).

\bibitem{Drexhage1968}
\bibinfo{author}{Drexhage, K.~H.}, \bibinfo{author}{Kuhn, H.} \&
  \bibinfo{author}{Sch{\"{a}}fer, F.~P.}
\newblock \bibinfo{title}{Variation of the fluorescence decay time of a
  molecule in front of a mirror}.
\newblock \emph{\bibinfo{journal}{Ber. Bunsen Phys. Chem.}}
  \textbf{\bibinfo{volume}{72}}, \bibinfo{pages}{329} (\bibinfo{year}{1968}).

\bibitem{Bartlett2000}
\bibinfo{author}{Bartlett, P.~N.}, \bibinfo{author}{Birkin, P.~R.} \&
  \bibinfo{author}{Ghanem, M.~A.}
\newblock \bibinfo{title}{Electrochemical deposition of macroporous platinum,
  palladium and cobalt films using polystyrene latex sphere templates}.
\newblock \bibinfo{pages}{\emph{\bibinfo{journal}{Chem. Commun.}} 1671}
  (\bibinfo{year}{2000}).

\bibitem{Bartlett2002}
\bibinfo{author}{Bartlett, P.~N.}, \bibinfo{author}{Baumberg, J.~J.},
  \bibinfo{author}{Birkin, P.~R.}, \bibinfo{author}{Ghanem, M.~A.} \&
  \bibinfo{author}{Netti, M.~C.}
\newblock \bibinfo{title}{Highly ordered macroporous gold and platinum films
  formed by electrochemical deposition through templates assembled from
  submicron diameter monodisperse polystyrene spheres}.
\newblock \emph{\bibinfo{journal}{Chem. Mater.}}
  \bibinfo{pages}{\textbf{\bibinfo{volume}{14}}, 2199} (\bibinfo{year}{2002}).

\bibitem{Casimir1948}
\bibinfo{author}{Casimir, H. B.~G.} \& \bibinfo{author}{Polder, D.}
\newblock \bibinfo{title}{{The Influence of Retardation on the London-van der
  Waals Forces}}.
\newblock \emph{\bibinfo{journal}{Phys. Rev.}}
  \bibinfo{pages}{\textbf{\bibinfo{volume}{73}}, 360} (\bibinfo{year}{1948}).

\bibitem{Scheel2009}
\bibinfo{author}{Scheel, S.} \& \bibinfo{author}{Buhmann, S.~Y.}
\newblock \bibinfo{title}{Casimir-polder forces on moving atoms}.
\newblock \emph{\bibinfo{journal}{Phys. Rev. A}} \textbf{\bibinfo{volume}{80}},
  \bibinfo{pages}{042902} (\bibinfo{year}{2009}).

\bibitem{Wilson2003}
\bibinfo{author}{Wilson, M.~A.}, \bibinfo{author}{Bushev, P.},
  \bibinfo{author}{Eschner, J.}, \bibinfo{author}{Kaler, S.~F.},
  \bibinfo{author}{Becher, C.}, \bibinfo{author}{Blatt, R.} \&
  \bibinfo{author}{Dorner, U.}
\newblock \bibinfo{title}{Vacuum-field level shifts in a single trapped ion
  mediated by a single distant mirror}.
\newblock \emph{\bibinfo{journal}{Phys. Rev. Lett.}}
  \textbf{\bibinfo{volume}{91}}, \bibinfo{pages}{213602}
  (\bibinfo{year}{2003}).

\bibitem{Coyle2001}
\bibinfo{author}{Coyle, S.}, \bibinfo{author}{Netti, M.~C.},
  \bibinfo{author}{Baumberg, J.~J.}, \bibinfo{author}{Ghanem, M.~A.},
  \bibinfo{author}{Birkin, P.~R.}, \bibinfo{author}{Bartlett, P.~N.} \&
  \bibinfo{author}{Whittaker, D.~M.}
\newblock \bibinfo{title}{Confined plasmons in metallic nanocavities}.
\newblock \emph{\bibinfo{journal}{Phys. Rev. Lett.}}
  \textbf{\bibinfo{volume}{87}}, \bibinfo{pages}{176801}
  (\bibinfo{year}{2001}).

\bibitem{KimballPhysicsSphOct2010}
\bibinfo{organization}{Kimball Physics, inc.}
\newblock \emph{\bibinfo{title}{Spherical Octagon Catalogue}}
  (\bibinfo{year}{2010}).

\bibitem{Himsworth2009}
\bibinfo{author}{Himsworth, M.}
\newblock \emph{\bibinfo{title}{Coherent Manipulation of Ultracold Rubidium}}.
\newblock Ph.D. thesis, \bibinfo{school}{University of Southampton}
  (\bibinfo{year}{2009}).

\bibitem{Ohadi2009}
\bibinfo{author}{Ohadi, H.}, \bibinfo{author}{Himsworth, M.},
  \bibinfo{author}{Xuereb, A.} \& \bibinfo{author}{Freegarde, T.}
\newblock \bibinfo{title}{Magneto-optical trapping and background-free imaging
  for atoms near nanostructured surfaces}.
\newblock \emph{\bibinfo{journal}{Opt. Express}}
  \bibinfo{pages}{\textbf{\bibinfo{volume}{17}}, 23003} (\bibinfo{year}{2009}).

\bibitem{Raab1987}
\bibinfo{author}{Raab, E.~L.}, \bibinfo{author}{Prentiss, M.},
  \bibinfo{author}{Cable, A.}, \bibinfo{author}{Chu, S.} \&
  \bibinfo{author}{Pritchard, D.~E.}
\newblock \bibinfo{title}{Trapping of neutral sodium atoms with radiation
  pressure}.
\newblock \emph{\bibinfo{journal}{Phys. Rev. Lett.}}
  \textbf{\bibinfo{volume}{59}}, \bibinfo{pages}{2631} (\bibinfo{year}{1987}).

\bibitem{Shimizu1991}
\bibinfo{author}{Shimizu, F.}, \bibinfo{author}{Shimizu, K.} \&
  \bibinfo{author}{Takuma, H.}
\newblock \bibinfo{title}{Four-beam laser trap of neutral atoms}.
\newblock \emph{\bibinfo{journal}{Opt. Lett.}}
  \bibinfo{pages}{\textbf{\bibinfo{volume}{16}}, 339} (\bibinfo{year}{1991}).

\bibitem{Emile1992}
\bibinfo{author}{Emile, O.}, \bibinfo{author}{Bardou, F.},
  \bibinfo{author}{Salomon, C.}, \bibinfo{author}{Laurent, P.},
  \bibinfo{author}{Nadir, A.} \& \bibinfo{author}{Clairon, A.}
\newblock \bibinfo{title}{Observation of a new magneto-optical trap}.
\newblock \emph{\bibinfo{journal}{Europhys. Lett.}}
  \bibinfo{pages}{\textbf{\bibinfo{volume}{20}}, 687} (\bibinfo{year}{1992}).

\bibitem{Lee1996}
\bibinfo{author}{Lee, K.~I.}, \bibinfo{author}{Kim, J.~A.},
  \bibinfo{author}{Noh, H.~R.} \& \bibinfo{author}{Jhe, W.}
\newblock \bibinfo{title}{Single-beam atom trap in a pyramidal and conical
  hollow mirror}.
\newblock \emph{\bibinfo{journal}{Opt. Lett.}} \textbf{\bibinfo{volume}{21}},
  \bibinfo{pages}{1177} (\bibinfo{year}{1996}).

\bibitem{Reichel1999}
\bibinfo{author}{Reichel, J.}, \bibinfo{author}{H{\"{a}}nsel, W.} \&
  \bibinfo{author}{H{\"{a}}nsch, T.~W.}
\newblock \bibinfo{title}{Atomic micromanipulation with magnetic surface
  traps}.
\newblock \emph{\bibinfo{journal}{Phys. Rev. Lett.}}
  \textbf{\bibinfo{volume}{83}}, \bibinfo{pages}{3398} (\bibinfo{year}{1999}).

\bibitem{Folman2000}
\bibinfo{author}{Folman, R.}, \bibinfo{author}{Kr{\"{u}}ger, P.},
  \bibinfo{author}{Cassettari, D.}, \bibinfo{author}{Hessmo, B.},
  \bibinfo{author}{Maier, T.} \& \bibinfo{author}{Schmiedmayer, J.}
\newblock \bibinfo{title}{Controlling cold atoms using nanofabricated surfaces:
  Atom chips}.
\newblock \emph{\bibinfo{journal}{Phys. Rev. Lett.}}
  \bibinfo{pages}{\textbf{\bibinfo{volume}{84}}, 4749} (\bibinfo{year}{2000}).

\bibitem{Pollock2009}
\bibinfo{author}{Pollock, S.}, \bibinfo{author}{Cotter, J.~P.},
  \bibinfo{author}{Laliotis, A.} \& \bibinfo{author}{Hinds, E.~A.}
\newblock \bibinfo{title}{Integrated magneto-optical traps on a chip using
  silicon pyramid structures}.
\newblock \emph{\bibinfo{journal}{Opt. Express}}
  \bibinfo{pages}{\textbf{\bibinfo{volume}{17}}, 14109} (\bibinfo{year}{2009}).

\bibitem{Nez1993}
\bibinfo{author}{Nez, F.}
\newblock \bibinfo{title}{Optical frequency determination of the hyperfine
  components of the 5{S}$_{1/2}$--5{D}$_{3/2}$ two-photon transitions in
  rubidium}.
\newblock \emph{\bibinfo{journal}{Opt. Commun.}}
  \bibinfo{pages}{\textbf{\bibinfo{volume}{102}}, 432} (\bibinfo{year}{1993}).

\bibitem{Ovchinnikov1991}
\bibinfo{author}{Ovchinnikov, Y.~B.}, \bibinfo{author}{Shul'ga, S.~V.} \&
  \bibinfo{author}{Balykin, V.~I.}
\newblock \bibinfo{title}{An atomic trap based on evanescent light waves}.
\newblock \emph{\bibinfo{journal}{J. Phys. B}}
  \bibinfo{pages}{\textbf{\bibinfo{volume}{24}}, 3173} (\bibinfo{year}{1991}).

\bibitem{Schultz2008}
\bibinfo{author}{Schultz, B.~E.}, \bibinfo{author}{Ming, H.},
  \bibinfo{author}{Noble, G.~A.} \& \bibinfo{author}{van Wijngaarden, W.~A.}
\newblock \bibinfo{title}{Measurement of the {Rb} {D}2 transition linewidth at
  ultralow temperature}.
\newblock \emph{\bibinfo{journal}{Eur. Phys. J. D}}
  \bibinfo{pages}{\textbf{\bibinfo{volume}{48}}, 171} (\bibinfo{year}{2008}).

\bibitem{Corwin1998}
\bibinfo{author}{Corwin, K.~L.}, \bibinfo{author}{Lu, Z.~T.},
  \bibinfo{author}{Hand, C.~F.}, \bibinfo{author}{Epstein, R.~J.} \&
  \bibinfo{author}{Wieman, C.~E.}
\newblock \bibinfo{title}{Frequency-stabilized diode laser with the {Z}eeman
  shift in an atomic vapor}.
\newblock \emph{\bibinfo{journal}{Appl. Opt.}}
  \bibinfo{pages}{\textbf{\bibinfo{volume}{37}}, 3295} (\bibinfo{year}{1998}).

\bibitem{Sheludko2008}
\bibinfo{author}{Sheludko, D.~V.}, \bibinfo{author}{Bell, S.~C.},
  \bibinfo{author}{Anderson, R.}, \bibinfo{author}{Hofmann, C.~S.},
  \bibinfo{author}{Vredenbregt, E. J.~D.} \& \bibinfo{author}{Scholten, R.~E.}
\newblock \bibinfo{title}{State-selective imaging of cold atoms}.
\newblock \emph{\bibinfo{journal}{Phys. Rev. A}} \textbf{\bibinfo{volume}{77}},
  \bibinfo{pages}{033401} (\bibinfo{year}{2008}).

\bibitem{Vernier2010}
\bibinfo{author}{Vernier, A.}, \bibinfo{author}{Franke-Arnold, S.},
  \bibinfo{author}{Riis, E.} \& \bibinfo{author}{Arnold, A.~S.}
\newblock \bibinfo{title}{Enhanced frequency up-conversion inrb vapor}.
\newblock \emph{\bibinfo{journal}{Opt. Express}}
  \bibinfo{pages}{\textbf{\bibinfo{volume}{18}}, 17020} (\bibinfo{year}{2010}).

\bibitem{SWu2009}
\bibinfo{author}{Wu, S.}, \bibinfo{author}{Plisson, T.},
  \bibinfo{author}{Brown, R.~C.}, \bibinfo{author}{Phillips, W.~D.} \&
  \bibinfo{author}{Porto, J.~V.}
\newblock \bibinfo{title}{Multiphoton magnetooptical trap}.
\newblock \emph{\bibinfo{journal}{Phys. Rev. Lett.}}
  \textbf{\bibinfo{volume}{103}}, \bibinfo{pages}{173003}
  (\bibinfo{year}{2009}).

\bibitem{Autler1955}
\bibinfo{author}{Autler, S.~H.} \& \bibinfo{author}{Townes, C.~H.}
\newblock \bibinfo{title}{Stark effect in rapidly varying fields}.
\newblock \emph{\bibinfo{journal}{Phys. Rev.}}
  \bibinfo{pages}{\textbf{\bibinfo{volume}{100}}, 703} (\bibinfo{year}{1955}).

\bibitem{Wohlleben2001}
\bibinfo{author}{Wohlleben, W.}, \bibinfo{author}{Chevy, F.},
  \bibinfo{author}{Madison, K.} \& \bibinfo{author}{Dalibard, J.}
\newblock \bibinfo{title}{An atom faucet}.
\newblock \emph{\bibinfo{journal}{Eur. Phys. J. D}}
  \bibinfo{pages}{\textbf{\bibinfo{volume}{15}}, 237} (\bibinfo{year}{2001}).

\bibitem{Lewandowski2003}
\bibinfo{author}{Lewandowski, H.~J.}, \bibinfo{author}{Harber, D.~M.},
  \bibinfo{author}{Whitaker, D.~L.} \& \bibinfo{author}{Cornell, E.~A.}
\newblock \bibinfo{title}{Simplified system for creating a {B}ose-{E}instein
  condensate}.
\newblock \emph{\bibinfo{journal}{J. Low Temp. Phys.}}
  \bibinfo{pages}{\textbf{\bibinfo{volume}{132}}, 309} (\bibinfo{year}{2003}).

\bibitem{Herskind2008}
\bibinfo{author}{Herskind, P.}, \bibinfo{author}{Dantan, A.},
  \bibinfo{author}{Langkilde-Lauesen, M.}, \bibinfo{author}{Mortensen, A.},
  \bibinfo{author}{S{\o}rensen, J.} \& \bibinfo{author}{Drewsen, M.}
\newblock \bibinfo{title}{Loading of large ion coulomb crystals into a linear
  paul trap incorporating an optical cavity}.
\newblock \emph{\bibinfo{journal}{Appl. Phys. B}}
  \bibinfo{pages}{\textbf{\bibinfo{volume}{93}}, 373} (\bibinfo{year}{2008}).

\bibitem{Hetet2010}
\bibinfo{author}{H{\'{e}}tet, G.}, \bibinfo{author}{Slodicka, L.},
  \bibinfo{author}{Gl{\"{a}}tzle, A.}, \bibinfo{author}{Hennrich, M.} \&
  \bibinfo{author}{Blatt, R.}
\newblock \bibinfo{title}{{QED} with a spherical mirror}.
\newblock \emph{\bibinfo{journal}{Phys. Rev. A}} \textbf{\bibinfo{volume}{82}},
  \bibinfo{pages}{063812} (\bibinfo{year}{2010}).

\bibitem{Cirac1995}
\bibinfo{author}{Cirac, J.~I.} \& \bibinfo{author}{Zoller, P.}
\newblock \bibinfo{title}{Quantum computations with cold trapped ions}.
\newblock \emph{\bibinfo{journal}{Phys. Rev. Lett.}}
  \bibinfo{pages}{\textbf{\bibinfo{volume}{74}}, 4091} (\bibinfo{year}{1995}).

\bibitem{Proite2010}
\bibinfo{author}{Proite, N.~A.}, \bibinfo{author}{Simmons, Z.~J.} \&
  \bibinfo{author}{Yavuz, D.~D.}
\newblock \bibinfo{title}{Observation of atomic localization using
  electromagnetically induced transparency}.
\newblock \emph{\bibinfo{journal}{arXiv e-prints}}  (\bibinfo{year}{2010}).
\newblock \eprint{arXiv:1011.2754}.

\bibitem{Karrai2008}
\bibinfo{author}{Karrai, K.}, \bibinfo{author}{Favero, I.} \&
  \bibinfo{author}{Metzger, C.}
\newblock \bibinfo{title}{Doppler optomechanics of a photonic crystal}.
\newblock \emph{\bibinfo{journal}{Phys. Rev. Lett.}}
  \textbf{\bibinfo{volume}{100}}, \bibinfo{pages}{240801}
  (\bibinfo{year}{2008}).

\bibitem{Barker2010b}
\bibinfo{author}{Barker, P.~F.}
\newblock \bibinfo{title}{Doppler cooling a microsphere}.
\newblock \emph{\bibinfo{journal}{Phys. Rev. Lett.}}
  \textbf{\bibinfo{volume}{105}}, \bibinfo{pages}{073002}
  (\bibinfo{year}{2010}).

\bibitem{Beyer2003}
\bibinfo{author}{Beyer, O.}, \bibinfo{author}{Nee, I.},
  \bibinfo{author}{Havermeyer, F.} \& \bibinfo{author}{Buse, K.}
\newblock \bibinfo{title}{Wavelength division multiplexing with {B}ragg
  gratings in poly (methyl methacrylate) ({PMMA})}.
\newblock In \emph{\bibinfo{booktitle}{Photorefractive Effects, Materials, and
  Devices}}, \bibinfo{pages}{577} (\bibinfo{publisher}{Optical Society of
  America}, \bibinfo{year}{2003}).

\bibitem{Kruse2010}
\bibinfo{author}{Kruse, J.}, \bibinfo{author}{Gierl, C.},
  \bibinfo{author}{Schlosser, M.} \& \bibinfo{author}{Birkl, G.}
\newblock \bibinfo{title}{Reconfigurable site-selective manipulation of atomic
  quantum systems in two-dimensional arrays of dipole traps}.
\newblock \emph{\bibinfo{journal}{Phys. Rev. A}} \textbf{\bibinfo{volume}{81}},
  \bibinfo{pages}{060308} (\bibinfo{year}{2010}).

\bibitem{Hannay1983}
\bibinfo{author}{Hannay, J.~H.}
\newblock \bibinfo{title}{The {C}lausius-{M}ossotti equation: an alternative
  derivation}.
\newblock \emph{\bibinfo{journal}{Eur. J. Phys.}} \textbf{\bibinfo{volume}{4}},
  \bibinfo{pages}{141} (\bibinfo{year}{1983}).

\bibitem{Miller1990}
\bibinfo{author}{Miller, K.~J.}
\newblock \bibinfo{title}{Calculation of the molecular polarizability tensor}.
\newblock \emph{\bibinfo{journal}{J. Am. Chem. Soc.}}
  \bibinfo{pages}{\textbf{\bibinfo{volume}{112}}, 8543} (\bibinfo{year}{1990}).

\bibitem{Bartlett2004}
\bibinfo{author}{Bartlett, P.~N.}, \bibinfo{author}{Baumberg, J.~J.},
  \bibinfo{author}{Coyle, S.} \& \bibinfo{author}{Abdelsalam, M.~E.}
\newblock \bibinfo{title}{Optical properties of nanostructured metal films}.
\newblock \emph{\bibinfo{journal}{Faraday Disc.}}
  \bibinfo{pages}{\textbf{\bibinfo{volume}{125}}, 117} (\bibinfo{year}{2004}).

\bibitem{Parsons2010}
\bibinfo{author}{Parsons, J.}, \bibinfo{author}{Burrows, C.~P.},
  \bibinfo{author}{Sambles, J.~R.} \& \bibinfo{author}{Barnes, W.~L.}
\newblock \bibinfo{title}{A comparison of techniques used to simulate the
  scattering of electromagnetic radiation by metallic nanostructures}.
\newblock \emph{\bibinfo{journal}{J. Mod. Opt.}}
  \bibinfo{pages}{\textbf{\bibinfo{volume}{57}}, 356} (\bibinfo{year}{2010}).

\bibitem{Oskooi2010}
\bibinfo{author}{Oskooi, A.~F.}, \bibinfo{author}{Roundy, D.},
  \bibinfo{author}{Ibanescu, M.}, \bibinfo{author}{Bermel, P.},
  \bibinfo{author}{Joannopoulos, J.~D.} \& \bibinfo{author}{Johnson, S.~G.}
\newblock \bibinfo{title}{{MEEP}: {A} flexible free-software package for
  electromagnetic simulations by the {FDTD} method}.
\newblock \emph{\bibinfo{journal}{Comput. Phys. Commun.}}
  \bibinfo{pages}{\textbf{\bibinfo{volume}{181}}, 687} (\bibinfo{year}{2010}).

\bibitem{Balian1970}
\bibinfo{author}{Balian, R.} \& \bibinfo{author}{Bloch, C.}
\newblock \bibinfo{title}{Distribution of eigenfrequencies for the wave
  equation in a finite domain {I}. {T}hree-dimensional problem with smooth
  boundary surface}.
\newblock \emph{\bibinfo{journal}{Ann. Phys.}}
  \bibinfo{pages}{\textbf{\bibinfo{volume}{60}}, 401} (\bibinfo{year}{1970}).

\bibitem{Balian1971}
\bibinfo{author}{Balian, R.} \& \bibinfo{author}{Bloch, C.}
\newblock \bibinfo{title}{Distribution of eigenfrequencies for the wave
  equation in a finite domain. {II}. {E}lectromagnetic field. {R}iemannian
  spaces}.
\newblock \emph{\bibinfo{journal}{Ann. Phys.}}
  \bibinfo{pages}{\textbf{\bibinfo{volume}{64}}, 271} (\bibinfo{year}{1971}).

\bibitem{Ohadi2008}
\bibinfo{author}{Ohadi, H.}
\newblock \emph{\bibinfo{title}{Single {Ca}$^+$ Ions in a {P}enning Trap for
  Applications in Quantum Information Processing}}.
\newblock Ph.D. thesis, \bibinfo{school}{Imperial College}
  (\bibinfo{year}{2008}).

\end{thebibliography}
